# Calibration of the IceCube Neutrino Observatory

Von der Fakultät für Mathematik, Informatik und Naturwissenschaften der RWTH Aachen University zur Erlangung des akademischen Grades eines Doktors der Naturwissenschaften genehmigte Dissertation

vorgelegt von

Master of Science Martin Rongen

aus Düren.

Berichter: Universitätsprofessor Dr. Christopher Wiebusch
Privatdozent Dr. Oliver Pooth

# ABSTRACT

The IceCube Neutrino Observatory instruments roughly one cubic kilometer of deep, glacial ice below the geographic South Pole with 5160 optical sensors to register the Cherenkov light of passing relativistic, charged particles. Since its construction was completed in 2010, a wide range of analyses has been performed. Those include, among others, the discovery of a high energetic astrophysical neutrino flux[1] [2], competitive measurements of neutrino oscillation parameters[3] and world-leading limits on dark matter detection[4]. With ever-increasing statistics the influence of insufficiently known aspects of the detector performance start to limit the potential gain of future analyses. This thesis presents calibration studies on both the hardware characteristics as well as the optical properties of the instrumented ice. Improving the knowledge of the detector systematics and the methods to study them does not only aid IceCube but also inform the design of potential future IceCube extensions.

[1] *Aartsen et al., "Observation of High-Energy Astrophysical Neutrinos in Three Years of IceCube Data"*

[2] *Aartsen et al., "Observation and Characterization of a Cosmic Muon Neutrino Flux from the Northern Hemisphere using six years of IceCube data"*

[3] *Aartsen et al., "Measurement of Atmospheric Neutrino Oscillations at 6-56 GeV with IceCube DeepCore"*

[4] *Aartsen et al., "Improved limits on dark matter annihilation in the Sun with the 79-string IceCube detector and implications for supersymmetry"*

# ZUSAMMENFASSUNG

Das IceCube Neutrino Observatory instrumentiert etwa einen Kubikkilometer tiefen Gletschereises unterhalb des geographischen Südpols mit 5160 optischen Sensoren, um das Cherenkov-Licht relativistischer, geladener Teilchen zu registrieren. Seit der Fertigstellung des Detektors im Jahr 2010, wurde eine breite Palette von Analysen durchgeführt. Zu den Ergebnissen gehören unter anderem die Entdeckung eines hochenergetischen astrophysikalischen Neutrinoflusses[1] [2], kompetitive Messungen von Neutrinooszillationsparametern[3] und weltweit führende Grenzen bei der Detektion dunkler Materie[4]. Mit immer größer werdenden Statistiken beginnt der Einfluss von nicht ausreichend bekannten Aspekten der Detektoreigenschaften den potentiellen Gewinn zukünftiger Analysen zu begrenzen. Diese Arbeit präsentiert Kalibrationsstudien sowohl zu den Hardwareeigenschaften als auch den optischen Eigenschaften des instrumentierten Eises. Die verbesserte Kenntnis der Detektorsystematiken und der Methoden zu deren Untersuchung unterstützt nicht nur IceCube, sondern ist auch entscheidend für das Design möglicher zukünftiger IceCube-Erweiterungen.

*The first principle is that you must not fool yourself and you are the easiest person to fool.*

*— Richard Feynman*

# ACKNOWLEDGMENTS

The work presented in this thesis would not have been possible without the support of numerous people. Some of whom I wish to specifically thank at this occasion:

My adviser Christopher Wiebusch for granting me the support and freedom to pursue independent scientific goals.

The members of IceCube calibration working group, in particular Dawn Williams, Christopher Wendt and Dmitry Chirkin, for their input and patience throughout countless calls, meetings and mail discussions.

The Aachen IceCube group, in particular the numerous office colleague throughout the years, for providing a very open-minded and comfortable work environment.

Gwen de Wasseige and Devyn Rysewyk for being exceptional travel companions on two very memorable deployments to Antarctica.

My wife Sarah and my parents for their enduring support.

# CONTENTS

# LIST OF FIGURES

# ACRONYMS

**ADC** . . . . . . Analog-to-digital converter

**FADC** . . . . Fast ADC

**AMANDA** . . . Antarctic Muon And Neutrino Detector Array

**Antares** . . . . . Astronomy with a Neutrino Telescope and Abyss environmental Research

**APD** . . . . . . . Avalanche photodiode

**ATWD** . . . . . Analog transient waveform digitizer

**c-axis** . . . . . . crystal axis

**CDF** . . . . . . . Constant fraction discriminator

**DAC** . . . . . . Digital-to-analog converter

**DARD** . . . . . Data Acquisition foR a flashing Dom

**DAQ** . . . . . . data acquisition

**D-Egg** . . . . . . Dual optical sensor in an Ellipsoid Glass for Gen2

**DOM** . . . . . . Digital Optical Module

**DUMAND** . . . Deep Underwater Muon And Neutrino Detector Project

**EDML** . . . . . European Project for Ice Coring in Antarctica (EPICA) at Dronning Maud Land (DML)

**EHWD** . . . . . Enhanced Hot Water Drill

**FOM** . . . . . . Fiber optical module

**FPGA** . . . . . . Field-programmable gate array

**FWHM** . . . . . Full width half maximum

**GVD** . . . . . . Gigaton Volume Detector in Lake Baikal

**ICL** . . . . . . . IceCube laboratory

**KM3Net** . . . . Cubic Kilometre Neutrino Network

**LC** . . . . . . . . Local coincidence

**HLC** . . . . Hard local coincidence

**SLC** . . . . . Soft local coincidence

**LED** . . . . . . . Light emitting diode

**LLH** . . . . . . . (negative) log-likelihood, sometimes also denoted GOF

**LPO** . . . . . . . lattice-preferred orientation

**MC** . . . . . . . Monte Carlo (simulation)

**mDOM** . . . . . multi-PMT digital optical module

**PE** . . . . . . . . Photoelectron

**SPE** . . . . . Single photoelectron
**PINGU** . . . . . Precision IceCube Next Generation Upgrade
**PMT** . . . . . . Photomultiplier tube
**POCAM** . . . . Precision Optical Calibration Module
**QE** . . . . . . . . quantum efficiency
**SiPM** . . . . . . Silicon photomultiplier
**Spice** . . . . . . South Pole ice
**SPTR** . . . . . . Single photon timing resolution
**TCD** . . . . . . . Time-to-digital converter
**VCSEL** . . . . . Vertical cavity surface emitting laser
**WOM** . . . . . . Wavelength shifting optical module

# 1

# SETTING THE STAGE

**The following chapter is an adaptation of the introduction to my Masters Thesis[5]. It was chosen here as a very condensed overview.**

[5] Rongen, *"Measurement of laser-induced thermoacoustic signals in ice for the detection of high-energy neutrinos"*

Mankind has always been looking to the stars trying to understand its place in nature, and soon the annual transit of constellations could be used to accurately keep track of time.

The invention of the telescope in the early 17$^{\text{th}}$ century[6] enabled the systematic observation of individual objects. Since then the sky has been mapped on the entire electromagnetic spectrum ranging from radio-waves to highly energetic gamma-rays. As a result, our view of the universe has slowly evolved from the rejection of the geocentric model to the Standard Model of Cosmology, which today gives the best description of the universe at large. On the way numerous fundamental laws of physics from Newtons Universal Law of Gravity to General Relativity and quantum mechanics had to be developed in order to explain astronomical and lab based observations alike.

[6] *King,* The History of the Telescope

In 1912, Victor Hess conducted a series of balloon based experiments to measure the decay of Earth's natural ionizing radiation with respect to the height over ground[7]. In contrast to his expectation he observed an increase in radiation with height. This discovery was interpreted and later confirmed to be a result of highly energetic, charged particles hitting the atmosphere. Until today the energy spectrum, spanning 14 orders of magnitude, and composition have been studied in great detail[8]. Nevertheless the acceleration mechanisms or distinct sources of high energy cosmic rays could not be experimentally clearly identified yet.

[7] *Hess, "Über Beobachtungen der durchdringenden Strahlung bei sieben Freiballonfahrten"*

[8] *Tanabashi, Hagiwara, et al., "Review of Particle Physics"*

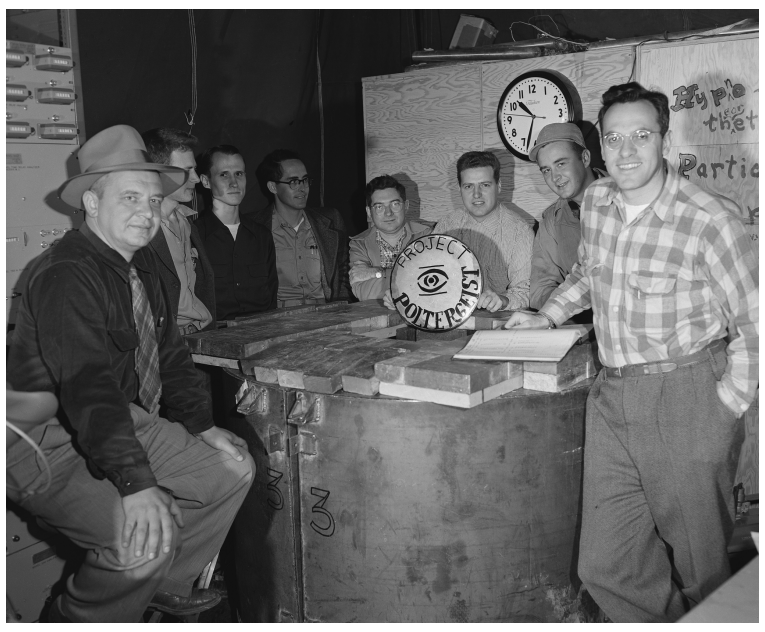

Figure 1.1: Clyde Cowan and Frederick Reines with their team of Project Poltergeist.
[Los Alamos National Laboratory, *Poltergeist experiment*]

In 1930, Wolfgang Pauli proposed the existence of a new family of light and neutral particles, later called neutrinos, in order to resolve the unexplained, continuous energy spectrum of the beta-decay[9]. By their very nature these particles can not be detected directly, as they only undergo weak interactions with a very low cross-section. In 1956, Project Poltergeist was able to identify neutrinos from a nearby nuclear reactor using the distinct timing signature of the inverse beta decay[10].

[9] *Pauli,* Open Letter

[10] *Cowan, Reines, et al., "Detection of the Free Neutrino: A Confirmation"*

Neutrinos are a by-product of hadronic interactions mediated by the weak force and thus should also be an abundant constituent of the cosmic particle flux. Charged cosmic ray particles are deflected by the interstellar and intergalactic magnetic fields. High-energy photons are quickly absorbed in interstellar clouds or interact with residual photon fields. In contrast, neutrinos preserve the original direction and intensity of their source.

Coincident with a close supernova in 1987, 25 neutrinos[11] above background level were detected in various experiments[12]. Since then a number of collaborations have tried to build dedicated experiments for the detection of cosmic neutrinos. At GeV energies and above many experiments rely on the measurement of Cherenkov radiation emitted by secondary particles traveling through a clear medium (water or ice) at more than the speed of light in the medium. In 2013, IceCube has been the first experiment to present a statistically significant flux of neutrinos which originates from outside our atmosphere[13].

[11] *between Kamiokande-II, IMB and Baksan*

[12] *K. Hirata, T. Kajita, et al. (1987). "Observation of a neutrino burst from the supernova SN1987A"; R. M. Bionta et al. (Apr. 1987). "Observation of a neutrino burst in coincidence with supernova 1987A in the Large Magellanic Cloud"*

[13] *M. G. Aartsen et al. (2013b). "Evidence for High-Energy Extraterrestrial Neutrinos at the IceCube Detector"*

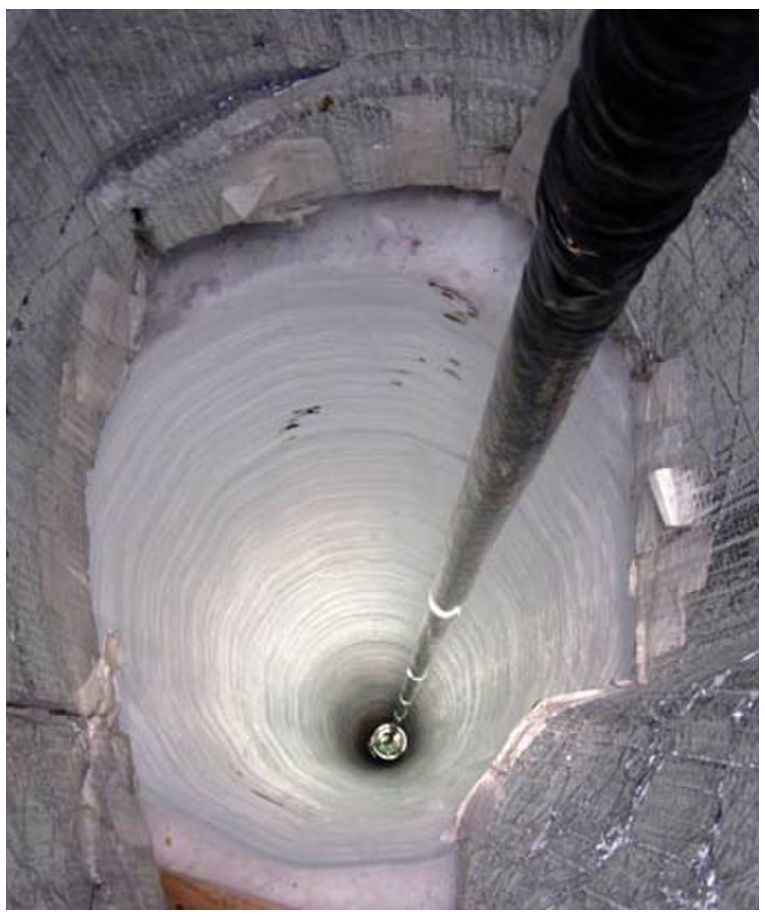

Figure 1.2: An IceCube sensor descending into the ice. A complex and challenging medium to characterize.
[IceCube collaboration, "Internal graphics resource"]

As experimental live time increases, many analyses performed within IceCube are now transitioning from being limited by the available statistics to being limited by systematic uncertainties related to the detector response.

Usually in a particle physics experiments great care is taken to design and characterize the detection material. Deploying into a naturally occurring medium, the Antarctic glacier, IceCube has no control of the material properties and can only calibrate the material properties using in-situ data from the experiment itself. This makes the calibration particularly challenging.

This thesis explores a number of systematics related both to the detector hardware as well as to the instrumented ice. In the process the understanding of the IceCube detector response is improved and new insights and appreciation for the detector medium is gained.

# 2

# NEUTRINO ASTRONOMY

**This chapter introduces the concept of neutrino astronomy and the experimental techniques involved. After a short review of neutrinos as part of the Standard Model of Particle Physics in the first section, their connection to cosmic rays and other astronomical messengers will be discussed in the second section. The final section presents water Cherenkov detectors as the primary experimental tool.**

## 2.1 NEUTRINOS IN THE STANDARD MODEL OF PARTICLE PHYSICS

The Standard Model of Particle Physics aims to describe matter through its non-divisible, fundamental constituents, also called elementary particles, and their interactions.

To-date, 17 such elementary particles, as shown in Figure 2.1, and their anti-particles have been identified. They can be categorized into three generations of elementary fermions, four gauge bosons required to describe the electromagnetic, weak and strong force[14] and the mass inducing, scalar Higgs boson[15].

[14] *the fourth known fundamental force, gravity, is currently not being described by the Standard Model*

[15] *Tanabashi, Hagiwara, et al., "Review of Particle Physics"*

Fermions are further sub-divided into strongly-interacting, charged quarks, which are found as constituents of hadrons, and leptons. For each charged lepton, such as the electron, an associated, neutral neutrino can be identified. While the charged leptons also participate in the electromagnetic force, neutrinos can only interact weakly.

The three generations of quarks and charged leptons differ in their mass, with a member of the later generation[16] being heavier but otherwise generally sharing the same properties as its lighter counterpart.

[16] *such as the muon with respect to the electron*

As neutrinos interact only via the weak force, their total interaction probability with matter is generally small. Above 10 GeV, as relevant for neutrino telescopes, deep inelastic nucleon scattering[17] through the exchange of a charged current W-boson (CC) or a neutral current Z boson (NC) as seen in Figures 2.2 and 2.3 dominate the cross section. In the case of a NC reaction the resulting hadronic cascade is the only observable, while in the case of a CC reaction the neutrino gets converted into a charged lepton of the same generation, which can also be detected.

[17] *this means the target nucleon is destroyed in the process*

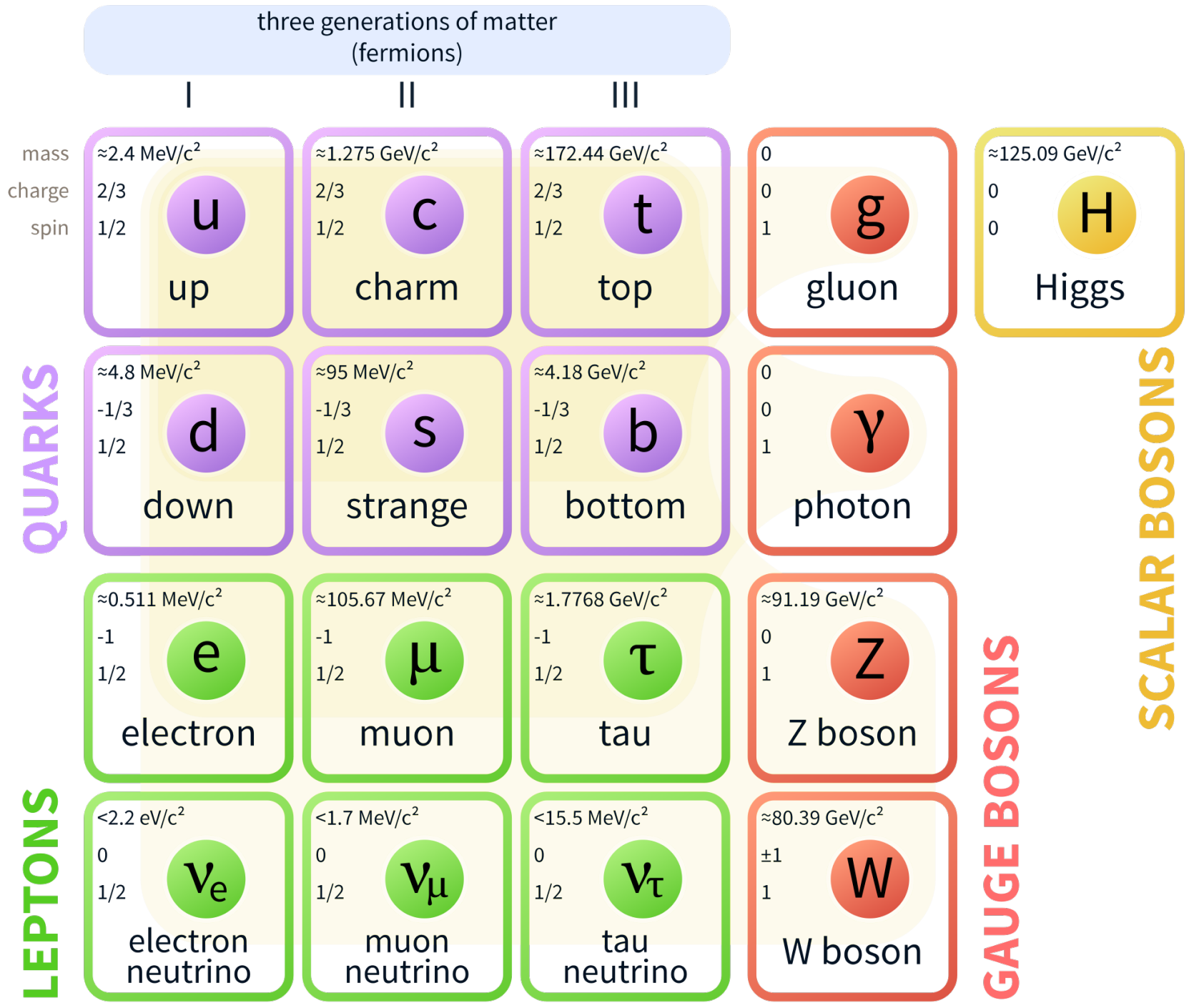

Figure 2.1: Schematic arrangement of the elementary particles comprising the Standard Model. [Wikimedia Commons (2018d). *Standard Model of Elementary Particles*]

The total cross section of neutrinos and anti-neutrinos with energies above 100 GeV is shown in Figure 2.4. It rises at all energies, with a reduced slope above several TeVs where the momentum transfer becomes larger than the mass of the exchanged boson. One can get a more natural understanding of the cross section by estimating the neutrino energy where the interaction length becomes equivalent to the density weighted Earth diameter. This is the case for energies above $\sim$100 TeV.

At 6.325 PeV, the anti-electron neutrino can resonantly produce a $W^-$ boson in the s-channel through the interaction with an electron at rest. This so called Glashow Resonance[18] locally exceeds the inelastic scattering cross section by more than an order of magnitude and offers an interesting target for PeV neutrino searches.

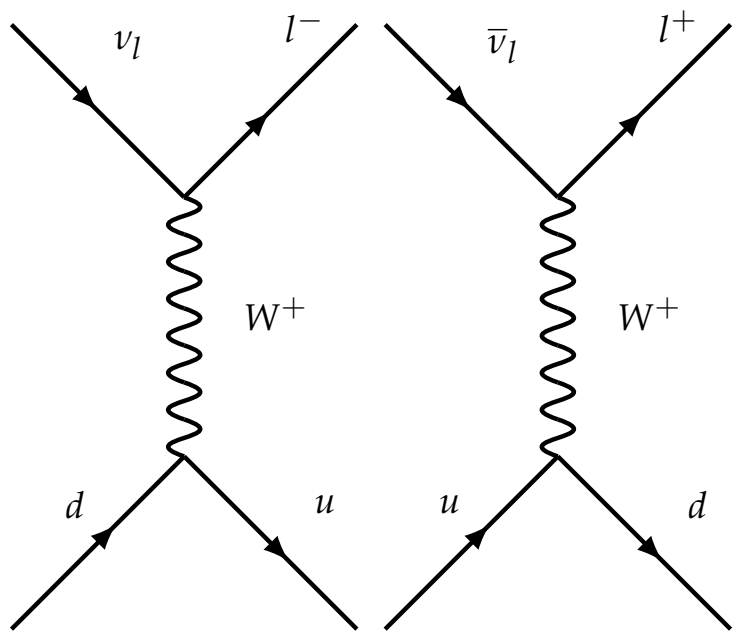

Figure 2.2: Feynman diagrams for neutrino charged current interactions.

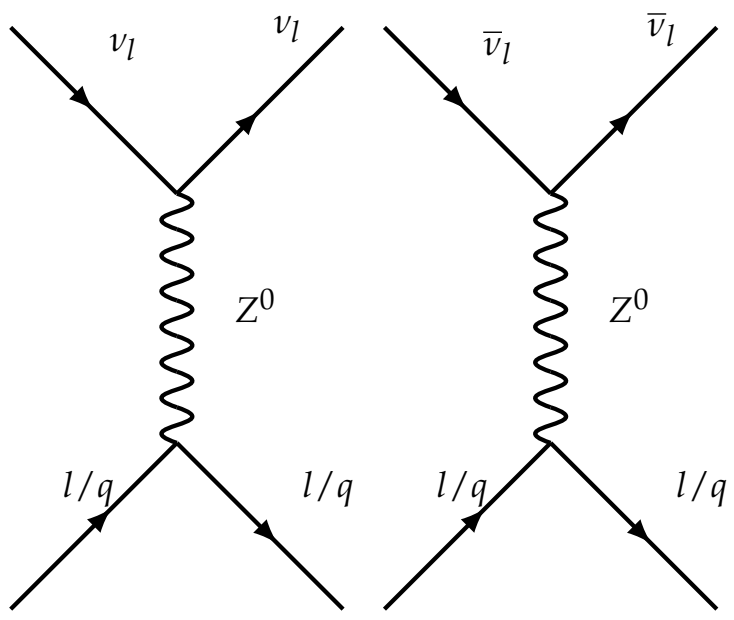

Figure 2.3: Feynman diagrams for neutrino neutral current interactions.

### 2.1.1 *Neutrino oscillations*

Neutrinos exhibit an effect called oscillation, where the flavor composition of a neutrino beam changes during propagation. While not strictly part of the Standard Model, the effect is theoretically well described and has in recent years been measured in great detail.

18 *Sheldon L. Glashow (Apr. 1960). "Resonant Scattering of Antineutrinos"*

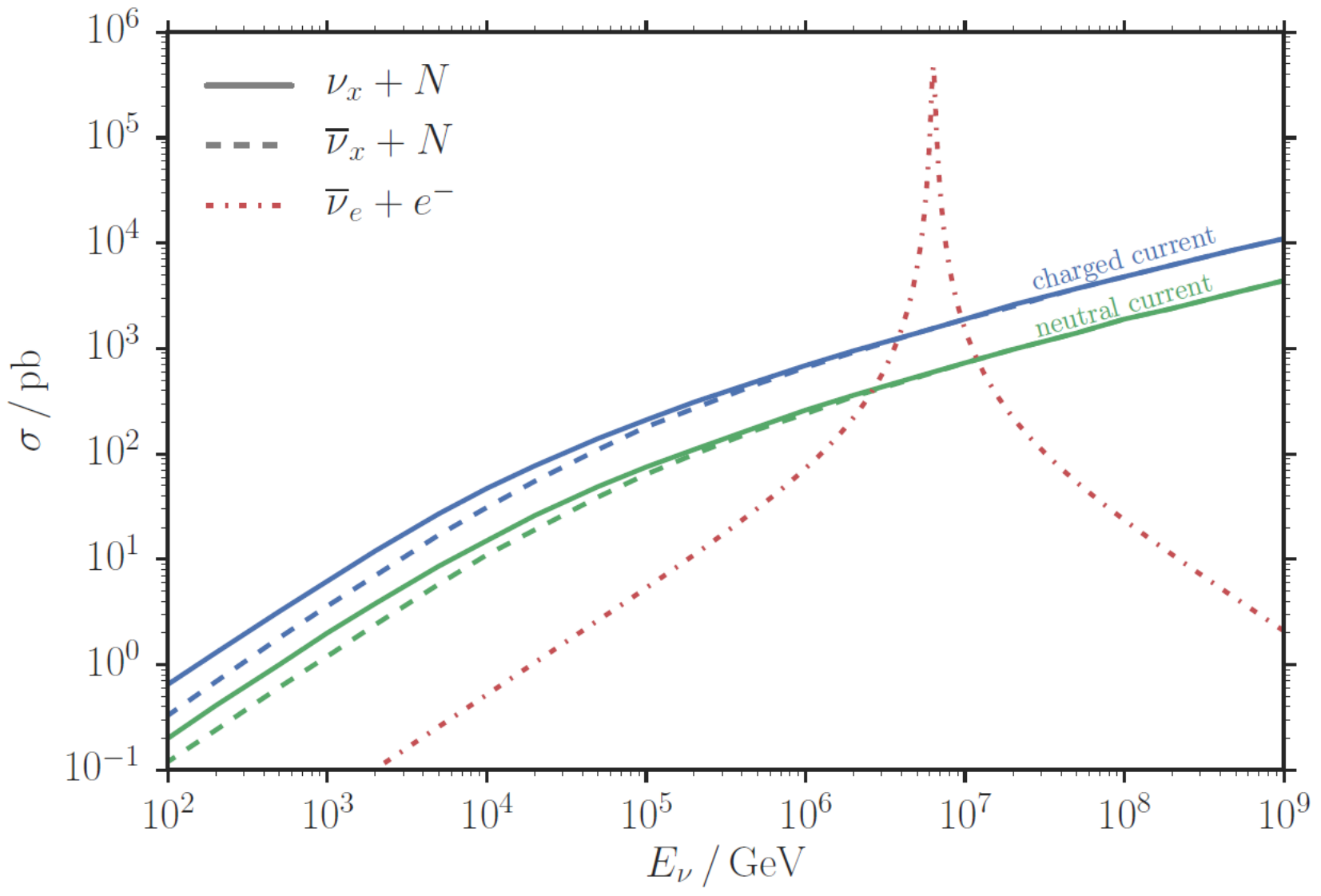

Figure 2.4: Cross sections for charged and neutral current neutrino interactions as well as the Glashow resonance.
[Leif Rädel (2017). "Measurement of High-Energy Muon Neutrinos with the IceCube Neutrino Observatory." RWTH Aachen University] based on:
[J. A. Formaggio and G. P. Zeller (May 31, 2013). "From eV to EeV: Neutrino Cross Sections Across Energy Scales"] and
[Amanda Cooper-Sarkar et al. (June 19, 2011). "The high energy neutrino cross-section in the Standard Model and its uncertainty"]

Usually elementary particles are defined as the eigenstates of a Hamiltonian. A minimal set of orthogonal eigenstates forms a basis. For quarks and charged leptons the flavors are defined by ordering the mass eigenstates[19] of particles.

[19] *as for example identified via a cross section resonance*

In the case of neutrinos, the flavor is defined through the observation of an associated charged lepton in a weak interaction. It is not a-priory given that the neutrino interaction / flavor basis is identical to the vacuum propagation / mass basis. In general the transformation between two sets of bases can be expressed through a unitary rotation[20] whose associated matrix is called *Pontecorvo-Maki-Nakagawa-Sakat (PMNS) matrix*[21] and is given in equation 2.1:

[20] $|\nu_\alpha\rangle = \sum_i U^*_{\alpha_i} |\nu_i\rangle$, *where* $|\nu_\alpha\rangle$ *is a flavor and* $|\nu_i\rangle$ *a mass eigenstate*

[21] *Maki et al., "Remarks on the Unified Model of Elementary Particles"*

$$U = \begin{pmatrix} c_{12}c_{13} & s_{12}c_{13} & s_{13}e^{-i\delta} \\ -s_{12}c_{23} - c_{12}s_{23}s_{13}e^{i\delta} & c_{12}c_{23} - s_{12}s_{23}s_{13}e^{i\delta} & s_{23}c_{13} \\ s_{12}s_{23} - c_{12}s_{23}s_{13}e^{i\delta} & -c_{12}c_{23} - s_{12}c_{23}s_{13}e^{i\delta} & c_{23}c_{13} \end{pmatrix} \tag{2.1}$$

With $c_{ij}$ and $s_{ij}$ denoting the cosine and sine of the three mixing angles $\theta_{12} \approx 34°$, $\theta_{13} \approx 34°$ and $\theta_{23} \approx 9°$[22]. An additional complex phase $\delta$ is introduced to facilitate potential charge conjugation parity symmetry violation[23].

[22] *Tanabashi, Hagiwara, et al., "Review of Particle Physics"*

[23] *If neutrinos are not Dirac particles but Majorana particles two additional phases need to be introduced. [Tanabashi, Hagiwara, et al., "Review of Particle Physics"]*

Through a weak interaction, a neutrino flavor eigenstate, for example a muon neutrino as identified through the associated production of a muon, is created. This flavor eigenstate can also be expressed as a superposition of mass eigenstates. Mass eigenstates propagate according to the Schrödinger equation[24]:

[24] *Schrödinger, "An Undulatory Theory of the Mechanics of Atoms and Molecules"*

$$i\frac{\delta}{\delta t}|\nu_i(t)\rangle = H|\nu_i(t)\rangle \tag{2.2}$$

This, for energies significantly larger than the rest mass which is usually given for neutrinos, results in a free-space propagation of:

$$|\nu_i(t)\rangle = e^{-i(E\cdot t - p\cdot x)}|\nu_i(0)\rangle \approx e^{-i(m_i^2\frac{L}{2E})}|\nu_i(0)\rangle, \tag{2.3}$$

where L is the propagation distance given by $L = t \cdot \beta c$. Therefore, the individual mass eigenstates propagate at distinct characteristic frequencies $\frac{m_i^2}{2E}$. The superposition beats[25], with the flavor composition continuously changing over the propagation distance.

[25] *as in German "Schwebung"*

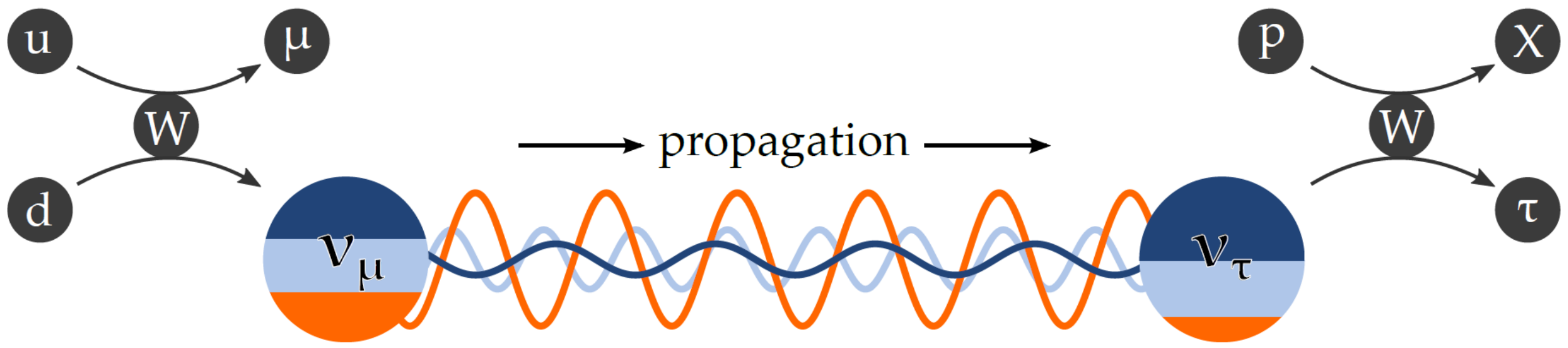

Figure 2.5: Schematic depiction of the neutrino oscillation mechanism. Neutrino interactions are governed by the flavor state, while the mass eigenstates propagate.
[Sebastian Euler (2014). "Observation of oscillations of atmospheric neutrinos with the IceCube Neutrino Observatory." Aachen: RWTH Aachen University]

A measurement projects the local flavor mixing through a weak interaction. In a simplified two flavor approximation the probability to observe an initial flavor state $\nu_i$ of energy $E$ as a flavor $\nu_j$ at a distance L is given by:

$$P(\nu_i \rightarrow \nu_j) = \sin^2(2\theta ij) \cdot \sin(1.267 \cdot \Delta m_{ij}^2/eV^2 \cdot \frac{L/km}{E/GeV}) \tag{2.4}$$

Obviously the oscillation effect relies on non-vanishing mass differences. The observation of oscillation effects proves that neutrinos have a mass, although it can currently not be measured directly.

The current best knowledge on the neutrino mass differences is[26]:

[26] *Tanabashi, Hagiwara, et al., "Review of Particle Physics"*

$$|\Delta m_{21}^2| \approx 7.4 \cdot 10^{-5} eV^2 \qquad |\Delta m_{23}^2| \approx 2.5 \cdot 10^{-3} eV^2 \tag{2.5}$$

This leaves the neutrino mass ordering, that is whether $m_3$ is heavier or lighter than $m_1$ and $m_2$, as well as the absolute neutrino mass scale to be discovered.

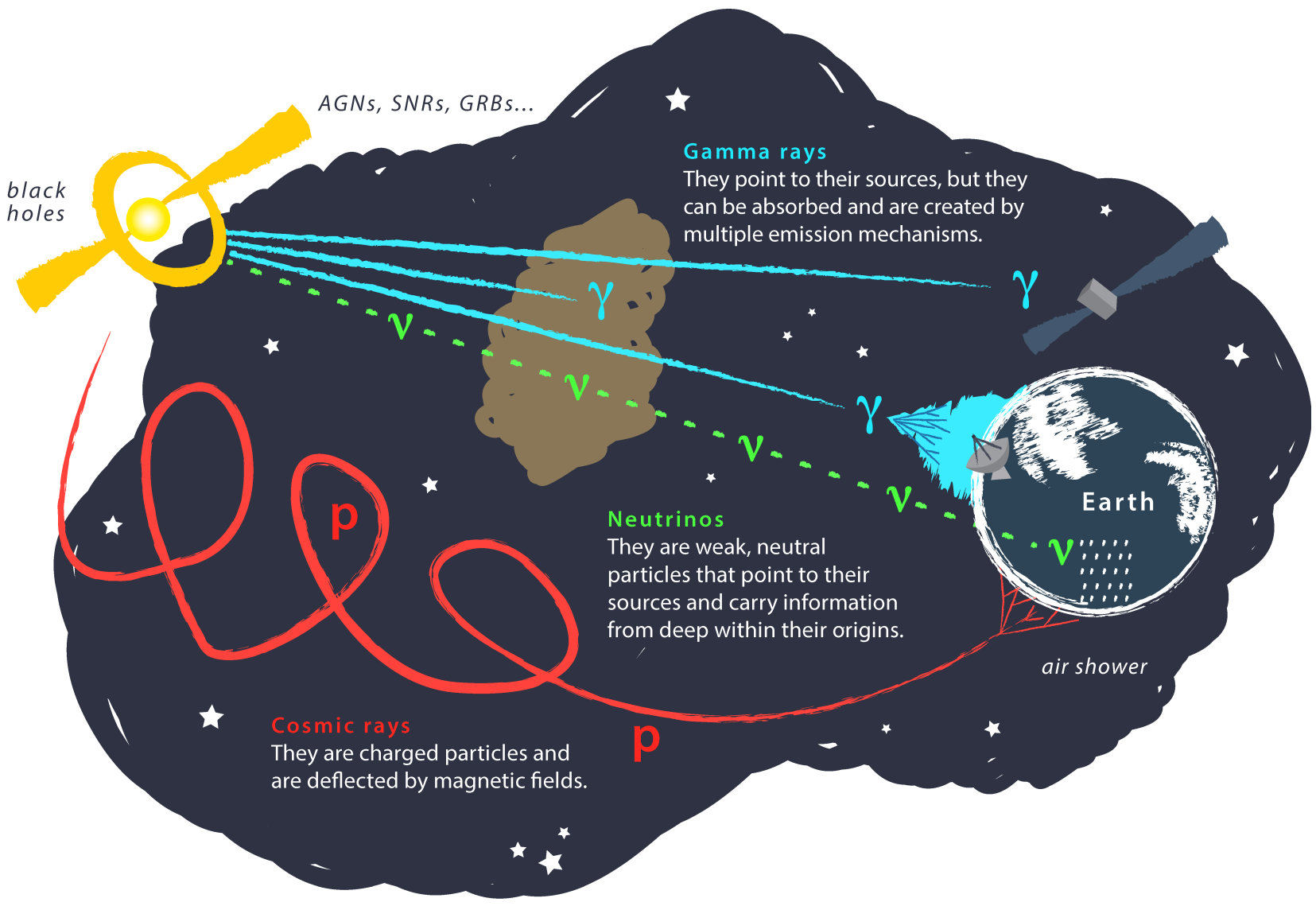

Figure 2.6: Schematic depiction of the different astronomical messengers and their propagation properties. Neutrinos are the only messenger to arrive at Earth unattenuated and undeflected.
[IceCube collaboration (2019). "Internal graphics resource"]

## 2.2 MULTI-MESSENGER ASTRONOMY

Astronomy aims to catalog and understand the celestial objects populating the sky. As a modern extension, cosmology aims to understand the structure and evolution of the universe at large. Given the vast extend of space, the only practical mean to observe the universe is the study of radiation reaching an earth-bound observer.

Any particle stable at cosmological distances can reach Earth from its source and serve as a messenger. Given our understanding of particle physics, photons, cosmic rays and neutrinos satisfy this requirement and will be discussed in more detail in the following.

Though not a messenger particle in the traditional sense, the onset of gravitational wave detections[27] and the successful correlation to gamma-rays in the recent case of a neutron star merger[28], has in a dramatic way highlighted the potential of the field.

[27] *Abbasi et al., "Observation of Gravitational Waves from a Binary Black Hole Merger"*

[28] *Abbott, Abbott, et al., "Multi-messenger Observations of a Binary Neutron Star Merger"*

### 2.2.1 *Photon Astronomy*

Until the mid-twentieth century the sky could only be observed through the visible portion of the electromagnetic spectrum using classical telescopes. With advances in sensor technology and the availability of space flight, the electromagnetic spectrum is now accessible from radio waves to high energy x-rays and gamma rays. As these different wavelengths originate in different production mechanisms, each observation offers a unique insight.

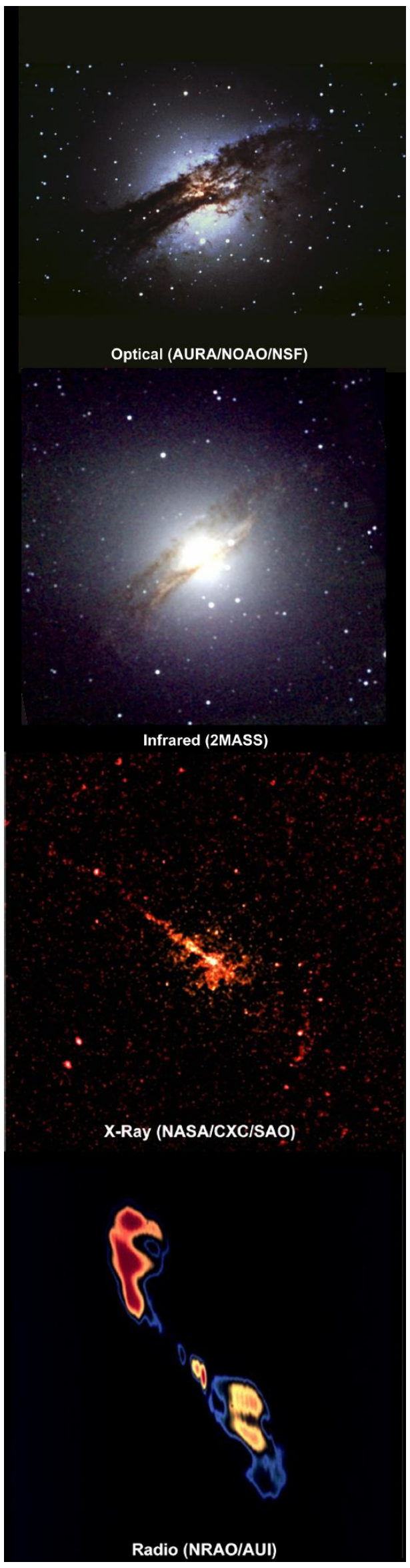

Figure 2.7: Centaurus A as seen by X-ray, radio, infrared and optical telescopes.
[Harvard-Smithsonian Center for Astrophysics, *Centaurus A Multi-wavelength images*]

This is probably most famously illustrated through the composite observation of Centaurus A, as seen in Figure 2.7. Centaurus A is a close by radio galaxy with an active galactic nucleus (AGN). A postulated, recent merger with a smaller spiral galaxy[29] triggered both an elevated star formation and feeds matter to the central super-massive black hole. The black hole in turn ejects relativistic jets perpendicular to the galactic plane as seen in radio and x-ray observations. The core of the galaxy is in the visible obscured by an opaque dust band.

### 2.2.2 *Cosmic rays*

While photons are abundant and comparatively easy to detect, low-energy photons primarily probe thermal sources. At energies above a few-dozen GeVs photons get absorbed quickly[30]. Therefore, other particles, namely cosmic rays and the associated neutrinos are required to probe the universe at high energies.

Cosmic rays are charged atomic nuclei, that are at low energies dominated by protons. The measured energy spectrum of charged cosmic radiation is shown in Figure 2.8. From 10 GeV to around 10 EeV the spectrum follows a broken power law, with a possible cut-off above[31]. Up to the so called knee at $2.7 \cdot 10^6$ GeV the spectral index is 2.7. Above, the index softens to around 3. At $4 \cdot 10^8$ GeV (second knee) the spectrum softens again to 3.1, before returning to the initial 2.7 above $4 \cdot 10^9$ GeV (the ankle).

Changes in the spectral index are attributed to a changing composition or source region, with the ankle denoting the energy above which the galactic magnetic field can no longer trap particles to be accelerated further[32].

Above $\sim 3 \cdot 10^{10}$ GeV the measured flux drops off rapidly, although the interpretation is still limited by the statistical uncertainty of the experiments[33]. A rapid drop off can potentially be explained by a maximum energy of the cosmic ray accelerators. Alternatively, above $\sim 5 \cdot 10^{10}$ GeV cosmic ray protons and photons from the cosmic-microwave background[34] reach a center-of-mass energy sufficient to resonantly produce a $\Delta^+$ resonance. This so called GZK[35] effect[36] severely limits the propagation range and gives an upper bound to the energy spectrum.

#### 2.2.2.1 *Cosmic ray interactions in the atmosphere*

Direct detection of cosmic rays is only possible in space. Due to the steeply falling flux and the size and weight limitations of satellites, the maximum feasible energy is about 100 TeV[37].

[29] *F. Aharonian et al. (Mar. 2009). "Discovery of very high energy gamma ray emission from Centaurus A with H.E.S.S."*

[30] *Imre Bartos and Marek Kowalski (2017). "Multimessenger Astronomy." IOP Publishing*

[31] *Thomas K. Gaisser et al. (2016).* Cosmic Rays and Particle Physics. *Cambridge University Press*

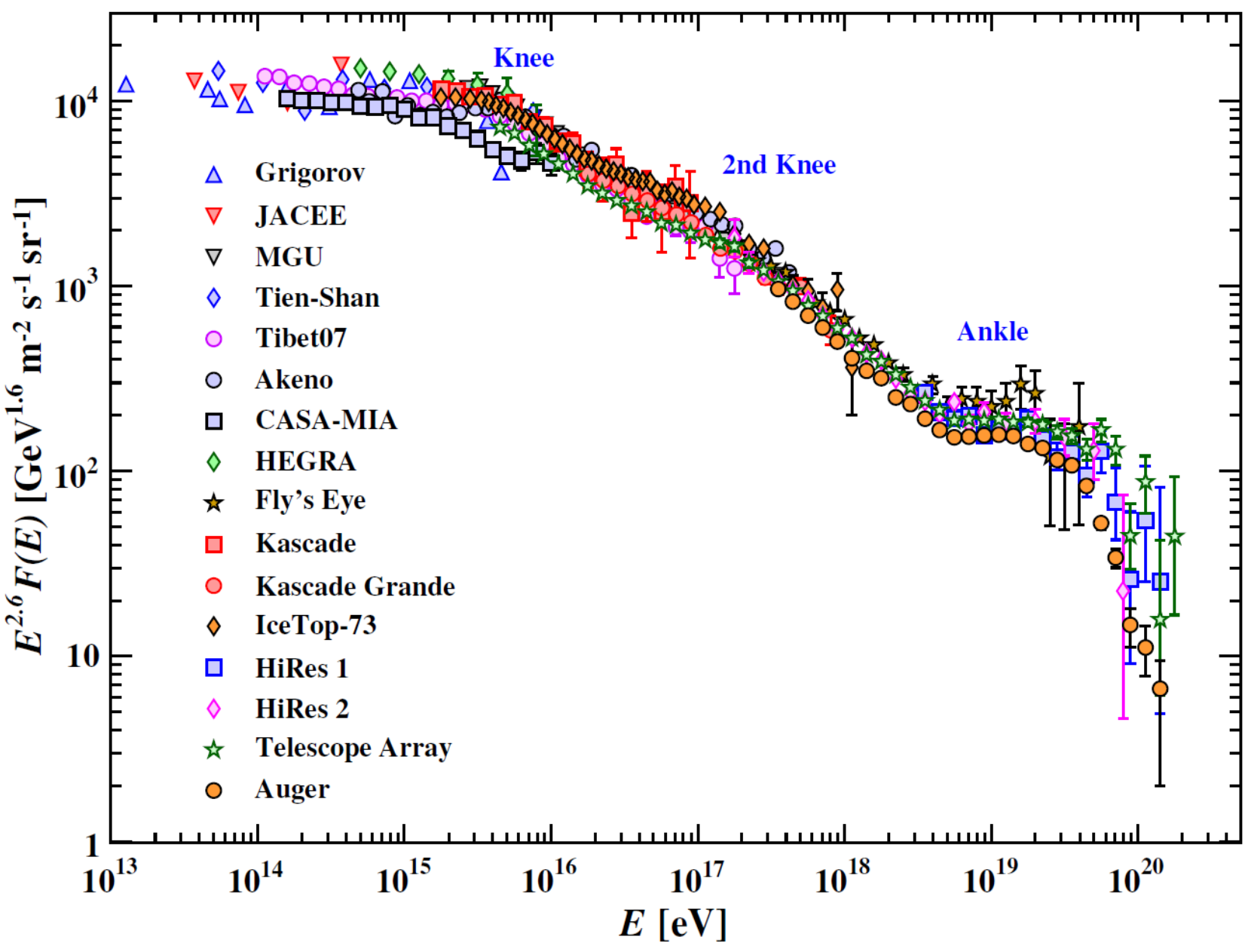

Figure 2.8: All-particle cosmic ray energy spectrum as measured by various air shower experiments. Transitions in the spectral index indicate a changing origin or composition.
[C. Patrignani et al. (2016). "Review of Particle Physics"]

[32] *Gaisser, "The Cosmic-ray Spectrum: from the knee to the ankle"*

[33] *Fenu et al., "The cosmic ray energy spectrum measured using the Pierre Auger Observatory"*

[34] *following a Maxwell-Boltzmann distribution*

[35] *acronym for: Greisen-Zatsepin-Kuzmin*

[36] *Greisen, "End to the Cosmic-Ray Spectrum?"*

[37] *Gaisser et al.,* Cosmic Rays and Particle Physics

For higher energies the properties of the primary cosmic rays are inferred from observations of the air showers they produce when hitting the atmosphere. The first interaction with an air molecule initiates a hadronic cascade. The decay of neutral pions in this cascade into high energy photons then initiates a simultaneously developing electromagnetic cascade, driven by Bremsstrahlung and pair production. In addition, charged mesons decay into long ranged muons. These three principal components of an air shower are also shown in Figure 2.9.

As the air shower develops, the energy of the primary particle is divided among a growing number of secondary particles. The secondary particles in the electromagnetic cascade at some point reach a critical energy and the multiplication processes stop. At this point, the so called shower maximum is reached and the electromagnetic secondary particles just keep propagating to ground, unless they decay or are absorbed.

An observer may either detect the air shower "footprint" of particles reaching ground level or monitor the electromagnetic radiation, usually radio emission, Cherenkov radiation or fluorescence light, produced during the shower development. The first interaction height as well as the position of the shower maximum depends on the mass of the cosmic ray primary, therefore the observation of the shower development as well as the ratio of the electromagnetic to the muonic component at ground can be used to infer the primary mass[38].

[38] *Gaisser et al.,* Cosmic Rays and Particle Physics

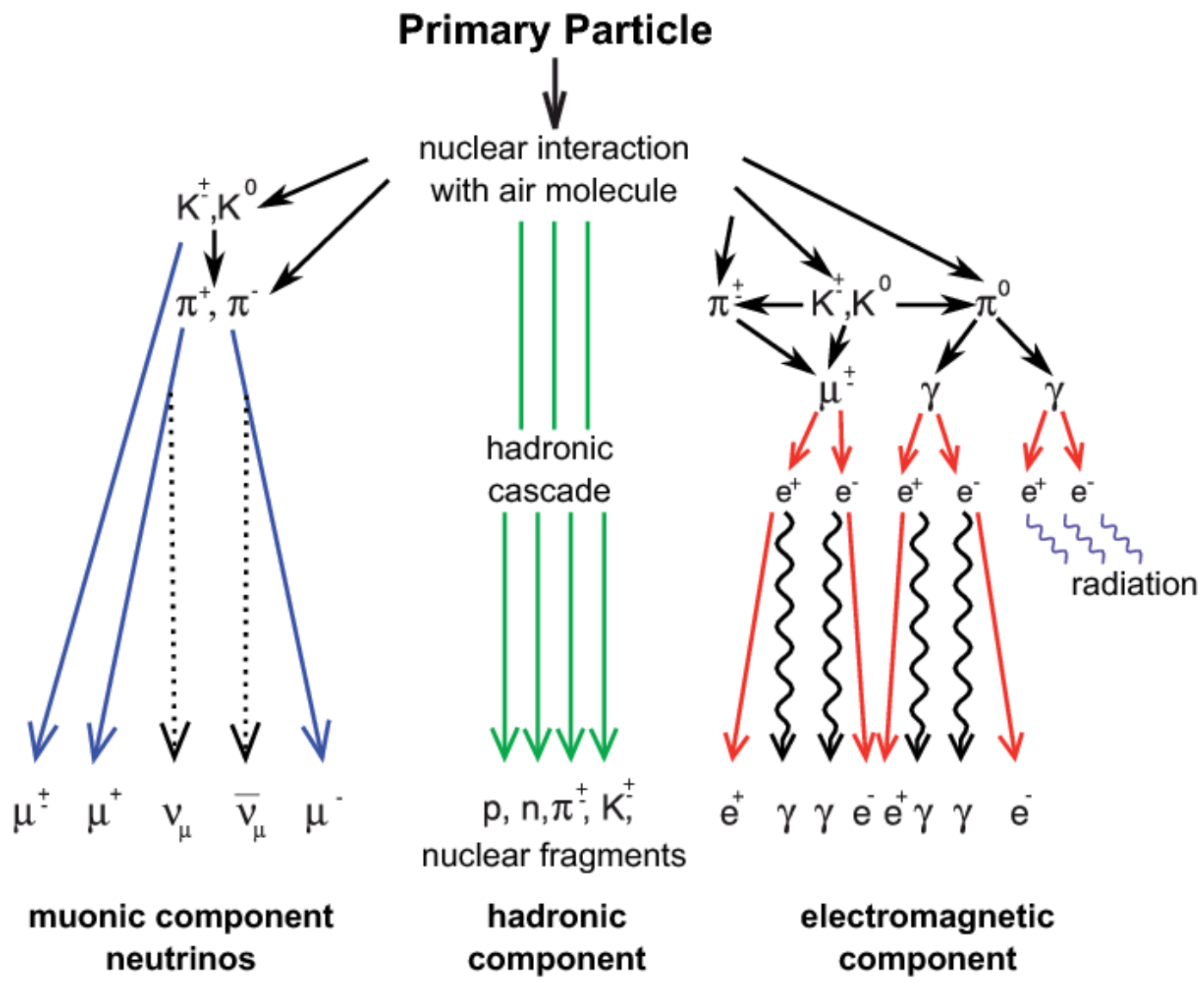

Figure 2.9: Schematic view of the different components of an extensive air shower developing in the atmosphere.
[A. Haungs et al. (2015). "KCDC - The KASCADE Cosmic-ray Data Centre"]

### 2.2.2.2 *Atmospheric neutrinos*

During the development of an air shower, neutrinos are primarily produced through the decay of pions and muons. As the decay probability competes against the re-interaction probability of these particles with air molecules, the neutrino yield depends on the energy and the local air density.

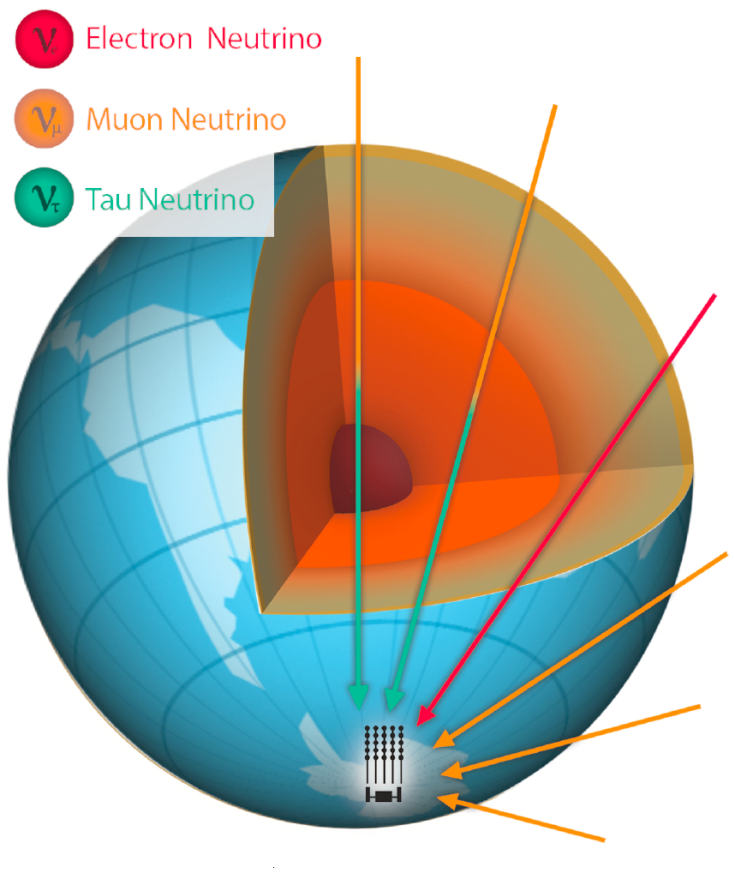

Figure 2.10: Different propagation baselines of atmospheric neutrinos as detected by IceCube enable the measurement of atmospheric neutrino oscillations
[IceCube collaboration, "Internal graphics resource"]

Unlike other air shower components, neutrinos, once produced, do not re-interact or decay. It follows that the Earth atmosphere at large can be thought of as a reasonably homogeneous neutrino source. For a stationary observer the measured flavor composition in any direction will not only depend on the energy but also on the propagation distance and traversed Earth density profile. This concept is also illustrated in Figure 2.10 and is used for all oscillation analyses with IceCube.

### 2.2.2.3 *Acceleration mechanisms*

While the sources of CRs and their acceleration mechanisms remain unknown, the Fermi mechanism[39] combined with shock acceleration[40] offers a generic description which results in a power-law energy spectrum with the correct spectral index and is applicable to a large variety of potential source classes.

[39] *Enrico Fermi (Apr. 1949). "On the Origin of the Cosmic Radiation"*

[40] *Roger Blandford and David Eichler (Oct. 1987). "Particle acceleration at astrophysical shocks: A theory of cosmic ray origin"*

Consider particles repeatedly undergoing an interaction, each time gaining a constant fraction $\xi$ of their energy. In addition, each interaction shall introduce a probability $P_{esc}$ of a particle to be lost from the acceleration region.

After $n$ interactions each particle has an energy $E_n = E_0 \cdot (1+\xi)^n$, but the number of particles reaching that energy decreases as

$$N(>E_n) \propto (1-P_{esc})^n = (1-P_{esc})^{\frac{\ln(E_n/E_0)}{\ln(1+\xi)}} = \left(\frac{E_n}{E_0}\right)^{\frac{\ln(1-P_{esc})}{\ln(1+\xi)}} . \quad (2.6)$$

If $P_{esc}$ and $\xi$ are assumed to be energy independent this results in a power-law spectrum

$$N(>E_n) \propto \left(\frac{E_n}{E_0}\right)^{-\alpha} . \quad (2.7)$$

Given $P_{esc} \ll 1$ and $\xi \ll 1$ yields $\alpha \approx \frac{P_{esc}}{\xi}$. Correspondingly, the differential spectral index $\gamma$ for the energy spectrum $d\Phi/dE$ is:

$$\gamma = \alpha + 1 \approx 1 + \frac{P_{esc}}{\xi}. \quad (2.8)$$

Such a scenario can for astronomical objects be realized in so called shock fronts. These are the magnetic boundary surfaces between colliding plasma fields, as are for example found in supernova blast waves or near active galactic nuclei[41].

For an ideal shock[42], the relative energy gain as well as the escape probability are found to be the shock velocity $\xi = P_{esc} = \beta_S$. As such, shock acceleration predicts a CR spectral index of $\gamma \approx 2$ at the source. The difference in spectral index compared to the observed value of $2.3 - 2.7$ at Earth is accounted for through the energy dependent diffusion of cosmic rays in our galaxy[43].

[41] *Biermann and Strittmatter, "Synchrotron emission from shock waves in active galactic nuclei"*

[42] *large Mach number*

[43] *Stanev,* High Energy Cosmic Rays

As the acceleration is a gradual process, a limited size or lifetime of the accelerator will introduce an energy cutoff. Independent of the Fermi mechanism, Hillas approximated the maximum energy simply based on magnetic confinement.

The size of the acceleration region has to be at least two times the particle Larmor radius[44], yielding a maximum energy of

$$E_{max} \leq 10^{19} eV \cdot Z \cdot \beta_S \cdot \frac{R}{kpc} \cdot \frac{B}{\mu G}. \quad (2.9)$$

[44] *Hillas, "The Origin of Ultra-High-Energy Cosmic Rays"*

Figure 2.11 shows that dependency compared to the usual size and magnetic field strength of many potential source classes.

Alternative scenarios for the generation of high energy cosmic rays include beyond-the-standard model theories, as for example top-down scenarios, where heavy dark matter particles decay or annihilate[45]. This allows for close by sources, which are not excluded by the GZK effect.

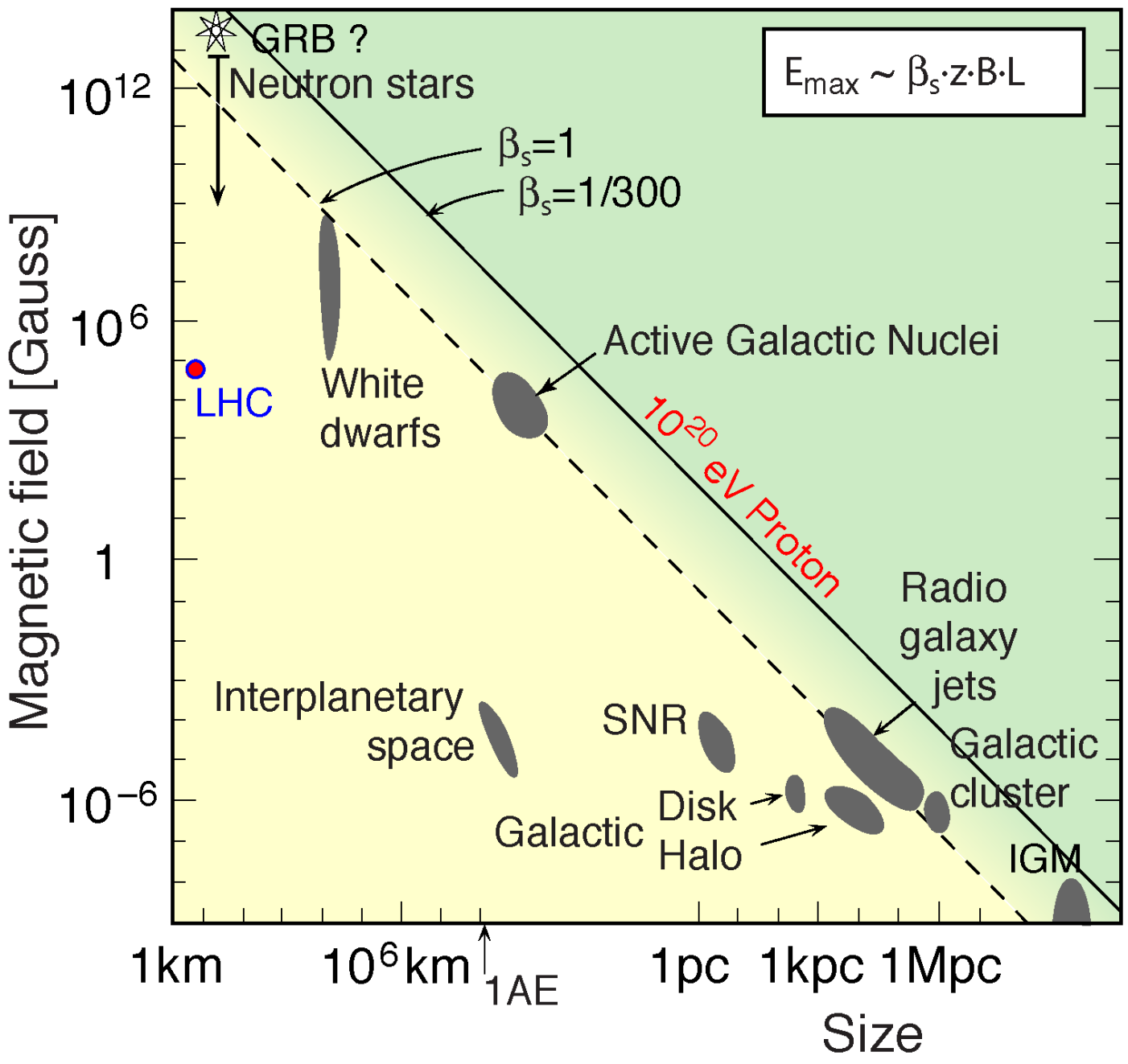

Figure 2.11: Hillas plot, depicting the typical size and magnetic field of potential source classes of ultra high energy cosmic rays and their resulting maximum energy as expected from the gyroradius. [Ralph Engel et al. (Oct. 2009). "Cosmic rays from the knee to the highest energies"]

#### 2.2.2.4 *Propagation and source searches*

While the energy spectrum of cosmic rays has been measured in great detail and the arrival directions can be measured to sub-degree accuracy[46], the sources of high energy cosmic rays remain essentially unknown.

Being charged particles, cosmic rays get deflected in magnetic fields. While intra-galactic field strengths are not well understood, they are believed to be smaller than $10^{-9}$ G, leading to a maximum deflection of $\sim 2°$ for particles up to $10^{20}$ eV and originating from at most 50 Mpc away[47].

Within our Milky Way, the field orientations and strengths are strongly position dependent, but generally on the order of micro-Gauss. A common approximation to estimate the mean deflection is [48]:

$$\sigma = 3 - 6° \cdot \frac{100\,\text{EeV}}{E} \cdot Z \qquad (2.10)$$

It follows that the measured arrival directions do not point directly back to the source position even for the highest energy cosmic rays and arrival directions are largely randomized below several EeV.

46 *Auger Collaboration and M. Ave (Sept. 13, 2007). "Reconstruction accuracy of the surface detector array of the Pierre Auger Observatory"*

47 *M. G. Aartsen et al. (Nov. 30, 2015b). "Search for correlations between the arrival directions of IceCube neutrino events and ultrahigh-energy cosmic rays detected by the Pierre Auger Observatory and the Telescope Array"*

48 *Aartsen et al., "Search for correlations between the arrival directions of IceCube neutrino events and ultrahigh-energy cosmic rays detected by the Pierre Auger Observatory and the Telescope Array"*

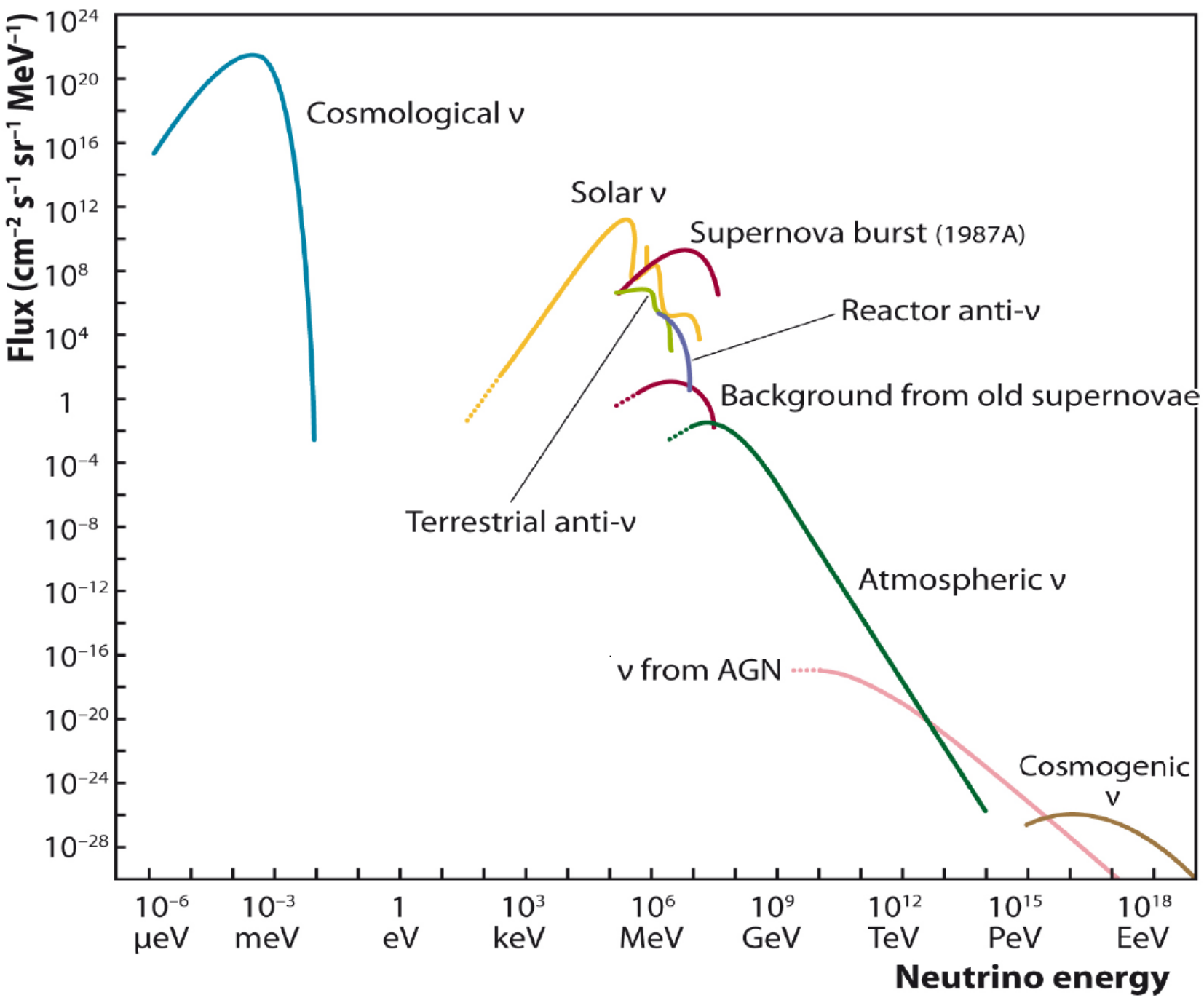

Figure 2.12: Measured and expected fluxes of natural and reactor neutrinos. Astrophysical neutrinos (denoted here as $\nu$ from AGN) can only be distinguished from atmospheric neutrinos at very high energies.
[Christian Spiering (July 2012). "Towards high-energy neutrino astronomy"]

### 2.2.3 *Neutrinos*

Astrophysical neutrinos are produced as secondary products of cosmic rays interacting with radiation and matter fields around the source. As in an air shower[49], this produces mainly pions and kaons decaying into neutrinos as well as high-energy photons. Therefore, high-energy photons are believed to be strongly correlated to neutrinos and cosmic rays, unless these photons originate from leptonic sources for example through synchrotron radiation.

49 *see section 2.2.2.2*

Neutrinos are not deflected in magnetic fields and do generally not interact as they leave even optically thick sources. Given these properties, they retain their original direction and energy distributions, making them the ideal messenger particles to identify the sources of cosmic rays.

Recent highlights with regards to the measurement of an astrophysical neutrino flux can be found in section 6.2.1.1.

## 2.3 CHERENKOV DETECTORS AS NEUTRINO TELESCOPES

Neutrino telescopes, such as IceCube, are large scale particle detectors designed to observe the natural flux of astrophysical neutrinos. As atmospheric neutrinos and underground muons result in nearly identical detector signatures, they are also being recorded by neutrino telescopes. During analysis, they are treated as background or used to expand the physics scope.

As the flux of astrophysical neutrinos at Earth is small, volumes of the order of $km^3$ need to be instrumented in order to observe a few astrophysical events per year. It is not feasible to provide artificial detection media such as scintillators at this scale and instead naturally occurring materials are instrumented.

### 2.3.1 *Detection principle*

Natural materials limit the applicable detection techniques. Transparent di-electric media allow for cubic-kilometer sized Cherenkov detectors, while thermo-acoustic and radio signals are being investigated for higher energies.

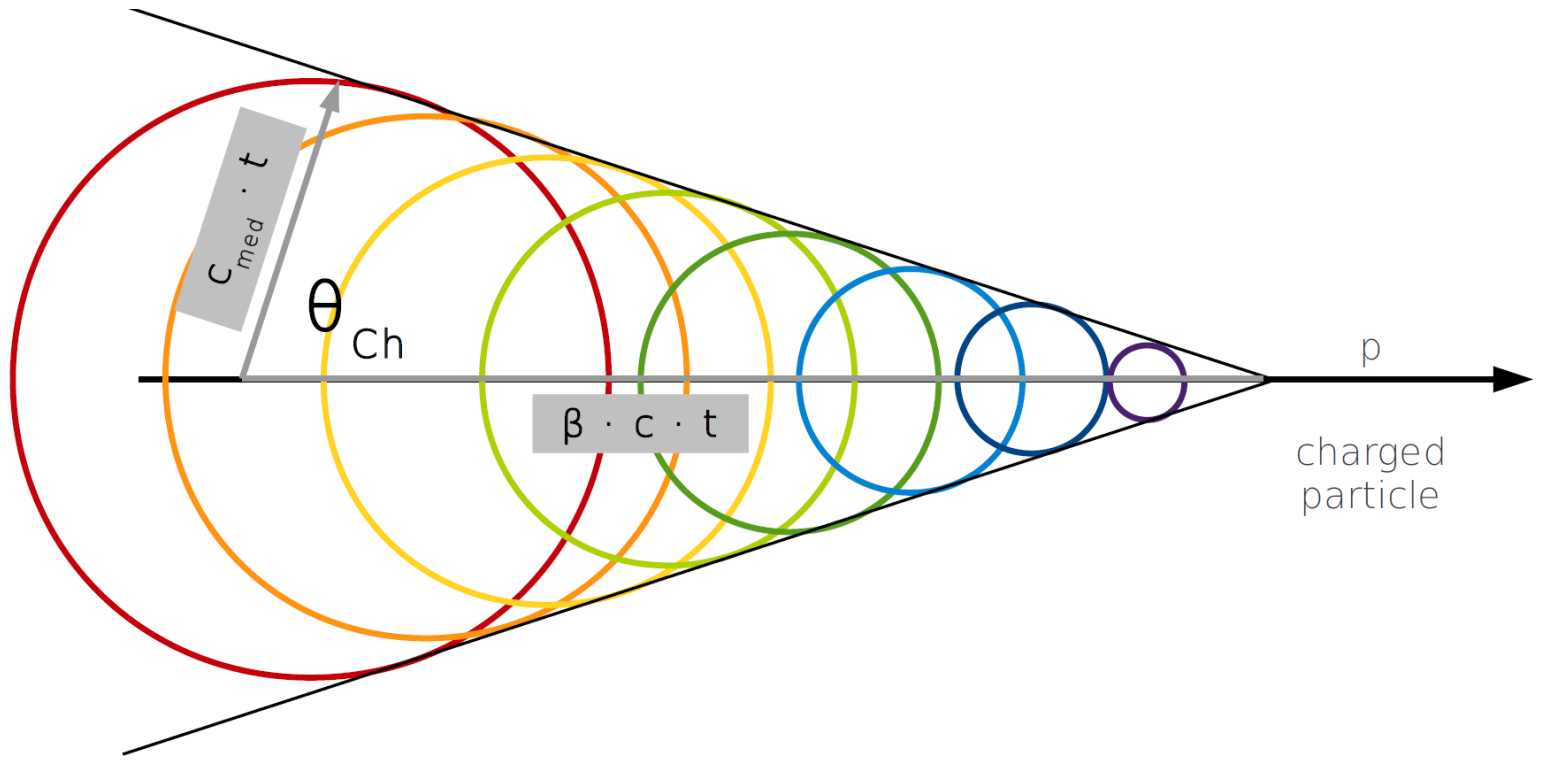

Figure 2.13: The principle of Cherenkov light emission. Wavelets get emitted at each position along the particle trajectory. For particle velocities above the speed of light in the medium, they interfere constructively at a cone described by the characteristic Cherenkov angle.
[Anne Schukraft (2013). "Search for a diffuse flux of extragalactic neutrinos with the IceCube neutrino observatory." Aachen: RWTH Aachen University]

Charged particles traversing a di-electric medium polarize the surrounding atoms, which in turn relax emitting electromagnetic dipole radiation[50]. While the particle is slower than the speed of light in the medium ($\beta < 1/n$) the polarization cloud can keep up with the particle and the resulting radiation interferes destructively, leading to no observable light.

[50] *P. A. Čerenkov (Aug. 1937). "Visible Radiation Produced by Electrons Moving in a Medium with Velocities Exceeding that of Light"*

Once the particle is faster than the speed of light in the medium, the wavefronts interfere constructively. Figure 2.13 shows the Huygens-Fresnel wavelet construction around the particle trajectory. The resulting light is called Cherenkov radiation[51] and

[51] *Čerenkov, "Visible Radiation Produced by Electrons Moving in a Medium with Velocities Exceeding that of Light"*

is emitted in a characteristic cone with an opening angle of

$$\cos(\theta) = \frac{1}{\beta \cdot n} \overset{\beta=1;\, n_{ice}=1.31}{=} 41^\circ. \tag{2.11}$$

The wavelength spectrum is described by the Frank-Tamm formula[52,53]

$$\frac{\mathrm{d}N}{\mathrm{d}\lambda \mathrm{d}x} = \frac{2\pi\alpha Z^2}{\lambda^2}\left(1 - \frac{1}{n^2\beta^2}\right) \tag{2.12}$$

and increases towards smaller wavelengths, giving a blue appearance to the human eye.

[52] *Frank and Tamm, "Coherent Visible Radiation of Fast Electrons Passing Through Matter"*

[53] *$\alpha$ being the fine-structure constant and Z being the particle's charge*

In order to be able to detect the emitted light, it is advantageous for the medium to be transparent for the emitted radiation. Abundant natural media which fulfill this requirement are water and ice.

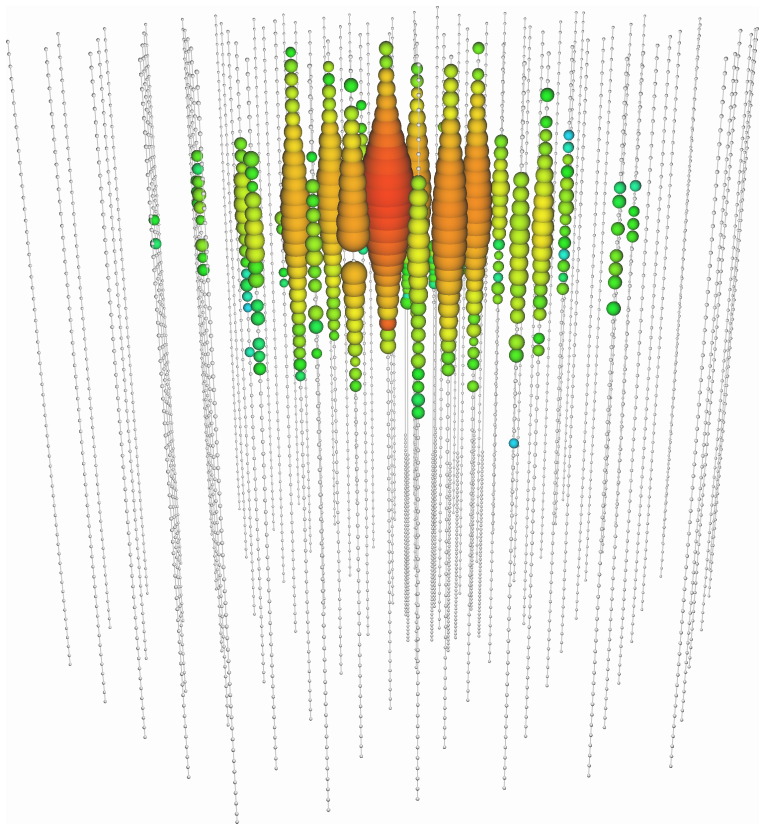

Figure 2.14: Signature of a high energy cascade event in IceCube. Grey dots denote optical sensors. Hit sensors are indicated by colored spheres. Size indicates the total received charge, while the color denotes time from red being early to blue being late.
[Rädel, "Measurement of High-Energy Muon Neutrinos with the IceCube Neutrino Observatory"]

Cherenkov neutrino telescopes instrument these media with arrays of photo-sensors, usually photomultiplier tubes, to register the Cherenkov radiation of secondary particles produced in neutrino interactions.

## 2.3.2 *Event signatures*

As described in section 2.1, a neutrino undergoing deep-inelastic scattering creates a hadronic cascade and in the case of a CC interaction an additional lepton according to the neutrino flavor. In a CC interaction, the neutrino energy is converted to visible secondaries, while in a NC reaction, where the neutrino is not fully converted and carries away a fraction of the total energy, on average only ~30% of the neutrino energy can be observed[54].

[54] *Aartsen et al., "Energy Reconstruction Methods in the IceCube Neutrino Telescope"*

In the following, the resulting generic detector signatures are discussed and their impact on the reconstruction performance is presented using the IceCube detector as a representative example.

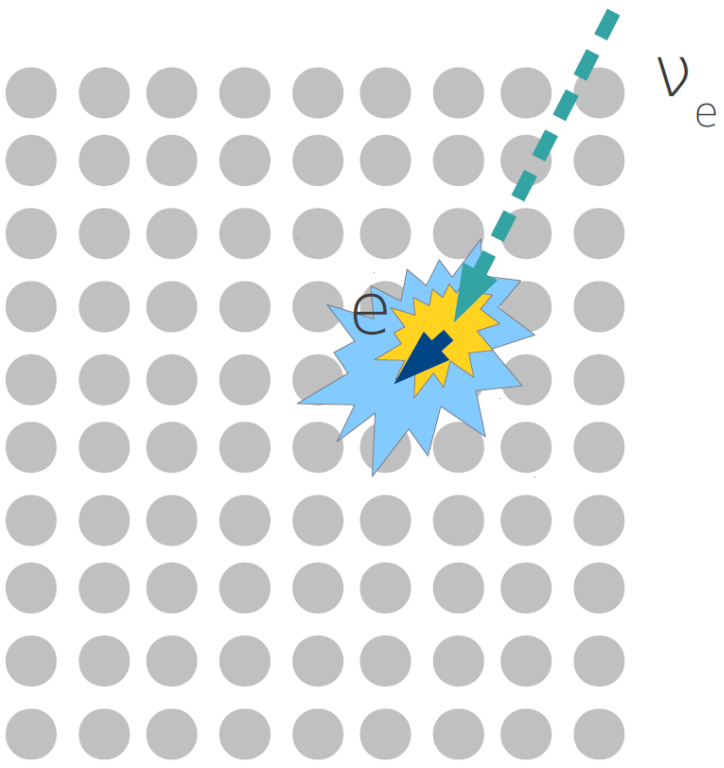

Figure 2.15: Observable superposition of a hadronic (yellow) and an electromagnetic (blue) cascade resulting from an electron neutrino charged current interaction inside the detection volume.
[Schukraft, "Search for a diffuse flux of extragalactic neutrinos with the IceCube neutrino observatory"]

### 2.3.2.1 *NC and electron neutrino CC interactions*

Both hadronic and electromagnetic cascades, resulting from electrons created in a $\nu_e$ CC interaction, are short[55] compared to the usual sensor spacing. Through light diffusion in the detector medium, they appear as almost point-like sources of light, creating a nearly spherical signatures in the detector.

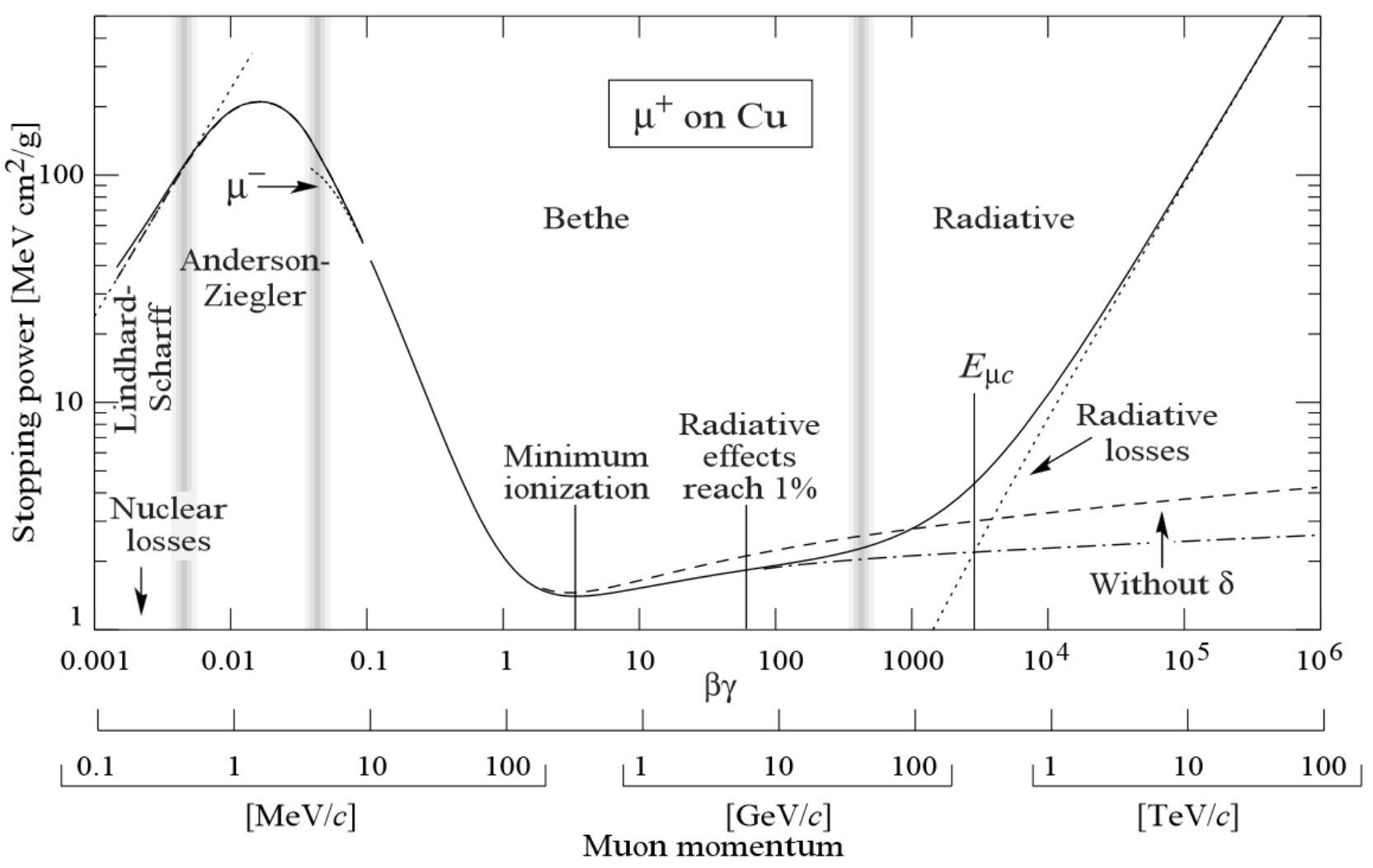

Figure 2.18: Bethe-Bloch Formula describing the differential energy loss of heavy charged particles. Note the region of nearly constant energy loss for minimum ionizing particles around a few GeV, as well as the sudden rise due to radiative losses at higher energies.
[M. Tanabashi, K. Hagiwara, et al. (Aug. 2018). "Review of Particle Physics"]

55 *typically* 10 m *with the length scaling logarithmic with energy*

56 $\sim 10^5$ *photons per GeV, in the sensitive wavelength range*

The amount of light emitted scales proportional with the cascade energy[56]. In case the cascade vertex is contained in the detector, the detector can be thought of as a calorimeter and the deposited energy resolution is dominated by statistical fluctuations in the collected light up to 100 TeV, where it reaches $\sim$ 10%. For higher energies, systematic uncertainties relating to the photon propagation dominate the energy resolution[57].

57 *Aartsen et al., "Energy Reconstruction Methods in the IceCube Neutrino Telescope"*

A directional reconstruction of cascades is possible through slight light intensity and timing asymmetries. On average more and earlier light is detected in the direction of the Cherenekov cones of the secondary particles. The angular cascade resolution approaches $\sim 15°$ above 100 TeV[58], where it again becomes systematics dominated.

58 *Aartsen et al., "Energy Reconstruction Methods in the IceCube Neutrino Telescope"*

### 2.3.2.2 *Muon neutrino CC interactions*

Muons resulting from $\nu_\mu$ CC interactions can propagate for many kilometers before stopping. The resulting detector signature is an extended column / track of light which surrounds the muon trajectory, as seen in Figure 2.17. In case the muon starts within the detector the event signature will be slightly altered with additional cascade contributions.

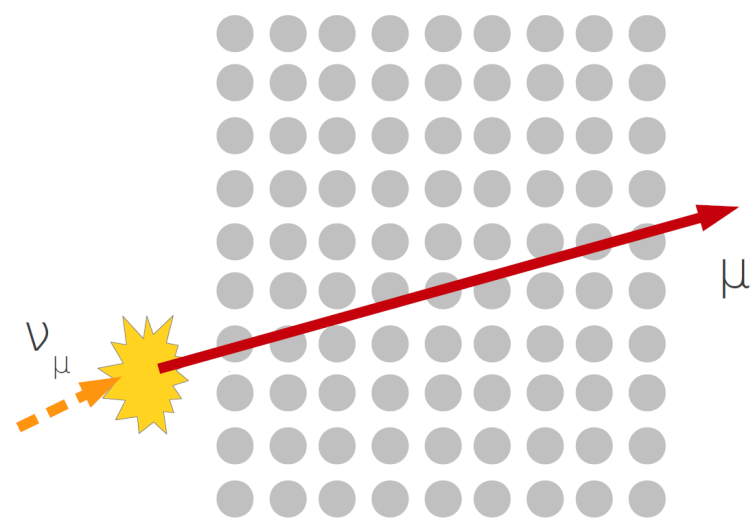

Figure 2.16: Principle of a muon neutrino charged-current interaction taking place outside the detection volume resulting in a muon traversing the entire length of the detector.
[Schukraft, "Search for a diffuse flux of extragalactic neutrinos with the IceCube neutrino observatory"]

The muon direction can be measured tracing the column of light. The kinematic opening angle between the neutrino and the muon is generally small and less than 0.5° above 1 TeV[59]. This is on the order of, but still smaller than the experimental resolution of 1° at the same energy.

As muons are usually not contained within the detector volume, their energy can not be measured calorimetrically. It is instead inferred from differential energy losses along the track. For GeV

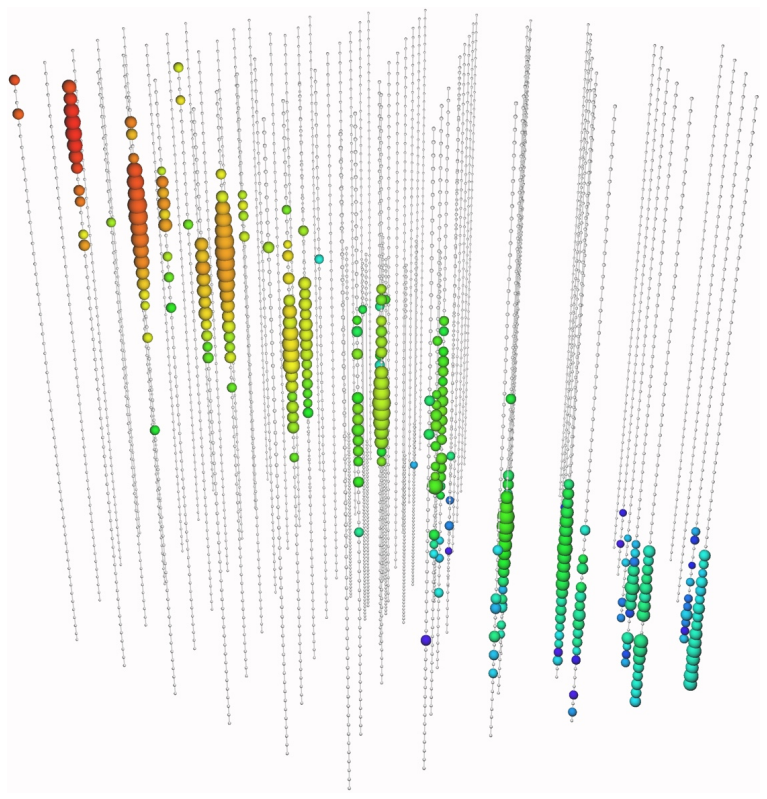

Figure 2.17: Signature of a high energy track event in IceCube. Same notations as in Figure 2.14 apply. [Rädel, "Measurement of High-Energy Muon Neutrinos with the IceCube Neutrino Observatory"]

muons the differential energy loss $\mathrm{d}E/\mathrm{d}x$, as given by the Bethe-Bloch mechanism and shown in Figure 2.18, is minimal and nearly independent of energy. While minimum ionizing muons provide a nice standard candle for calibration, their energy can essentially not be reconstructed. Above ~1 TeV the energy loss is dominated by stochastic losses. Reconstructing their strength and density along the track enables an energy resolution of $\sim 0.2$ in $\log_{10}(E_\mu)$[60].

While discussed here in terms of neutrino event signatures, it should be pointed out that cosmic ray muons, the dominant experimental background, produce essentially the same signature. In addition to single muons, air showers also produce closely confined muon bundles[61]. These can be distinguished from a high energy muon as their combined track is very homogeneous, with nearly no stochastic energy losses.

#### 2.3.2.3 *Tau neutrino CC interactions*

A CC $\nu_\tau$ interaction can be understood as a superposition of two cascades and a track event. The initial CC interaction creates a hadronic cascade and a tau lepton. This tau keeps propagating, producing a faint track signature, until it decays into a further cascade.

The two cascades can only be resolved when the separation length, given by the time delayed lifetime to be ~ 50 m per PeV, exceeds the experimental cascade vertex resolution. Otherwise, the tau is indistinguishable from a NC or $\nu_e$ CC signature.

59 *Aartsen et al., "All-sky Search for Time-integrated Neutrino Emission from Astrophysical Sources with 7 yr of IceCube Data"*

60 *Abbasi et al., "An improved method for measuring muon energy using the truncated mean of dE/dx"*

61 *Aartsen et al., "Characterization of the Atmospheric Muon Flux in IceCube"*

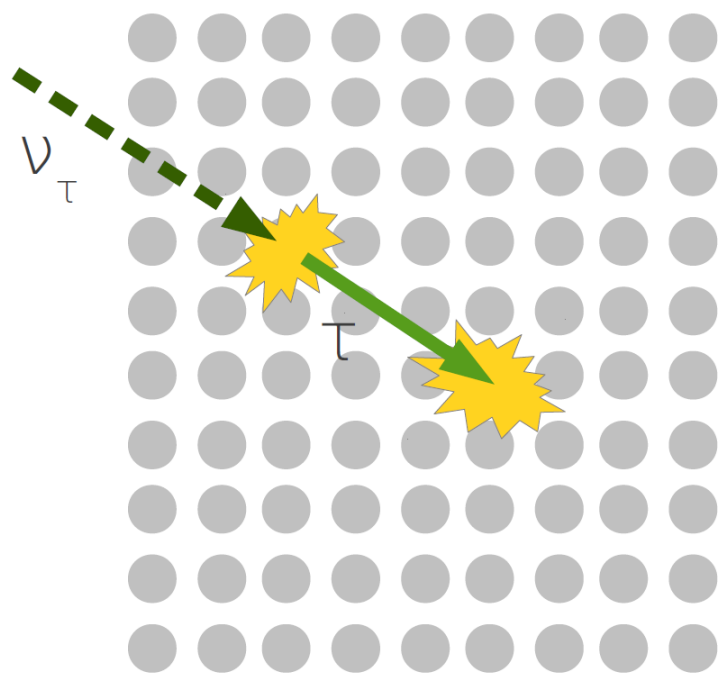

Figure 2.19: Tau neutrino signature. Both the primary interaction, as well as the decay of the resulting tau lepton result in a cascade-like signature. The vertex separation scales with energy. [Schukraft, "Search for a diffuse flux of extragalactic neutrinos with the IceCube neutrino observatory"]

### 2.3.3 *Experiments*

The idea to instrument natural, heavy, transparent media as Cherenkov telescopes was first conceived in 1960 by Moisei Markov[62] and started being actively pursued in 1975[63] with the first DUMAND[64] planning workshop. DUMAND was envisioned to be a massive cubic-kilometer detector comprising over 22'000 photo-detectors to be deployed at nearly 5 km depth of the coast of Hawaii.

While laying a lot of groundwork for the subsequent experiments, the reality of funding and technology eventually caught up. Prototype strings deployed in 1982 and 1984 failed during or shortly after deployment[65]. By 1988 the planned project scope had been reduced to 216 sensors in a 0.002 km$^3$ grid. The project was, after more deployment failures, finally terminated in 1995[66].

While initially Russian physicists participated in the DUMAND experiment, cold war politics forced them to pursue an independent project[67]. Lake Baikal, the world's largest freshwater lake by volume, located in southern Siberia, was chosen as site and the first string was deployed in 1984. The Baikal experiment was also the first to achieve three fully operational strings in 1993 and the first to measure atmospheric neutrinos in a deep underwater detector. The final detector configuration of the original experiment was installed in 1998, but Lake Baikal remains an active site with the new Baikal-GVD[68] experiment currently being deployed[69].

In 1988 Franzis Halzen among others, after learning about a Russian experiment at Vostok station to measure Askaryan radio emission from neutrino interactions, proposed what would become the first Cherenkov neutrino telescope in ice, AMANDA[70,71].

The first four string of the AMANDA(-A) telescope, located near the Amundson-Scott Research Station at the Geographic South Pole were installed during the 1993/94 season at a depth of 800 m to 1000 m. The data revealed that while the optical absorption length was better than expected, the effective scattering lengths had been dramatically over-predicted and were in-fact consistently below one meter. This was found to be caused by residual air bubbles[72] and prohibited event reconstructions with sufficient directional resolution.

[62] *M. A. Markov (1960). "On high energy neutrino physics"*

[63] *Ulrich F. Katz and Christian Spiering (July 2011). "High-Energy Neutrino Astrophysics: Status and Perspectives"*

[64] *acronym for: Deep Underwater Muon and Neutrino Detector*

[65] *Christian Spiering (July 2012). "Towards high-energy neutrino astronomy"*

[66] *Katz and Spiering, "High-Energy Neutrino Astrophysics: Status and Perspectives"*

[67] *Spiering, "Towards high-energy neutrino astronomy"*

[68] *acronym for: Gigaton Volume Detector*

[69] *A.D. Avrorin et al. (2018). "Baikal-GVD: status and prospects." Ed. by V.E. Volkova et al.*

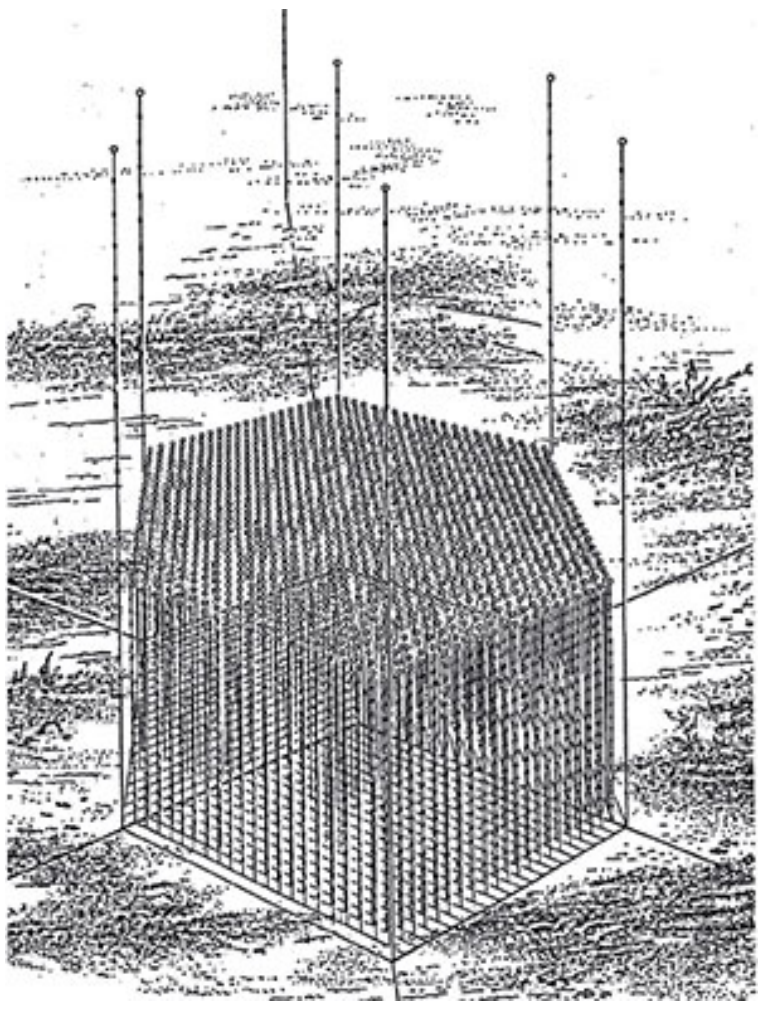

Figure 2.20: Original concept sketch of the DUMAND array [Katz and Spiering, "High-Energy Neutrino Astrophysics: Status and Perspectives"]

[70] *acronym for: Antarctic Muon And Neutrino Detector Array*

[71] *E Andres et al. (Mar. 2000). "The AMANDA neutrino telescope: principle of operation and first results"*

[72] *also see section 3.6*

Luckily ice core data from around Antarctica encouraged that below 1300 m even these bubbles should disappear and AMANDA(-B) was able to start deploying a second, deeper array in 1995/96 at depths between 1500 m and 2000 m. The optical properties were found to be significantly better. The detector was completed in 2000, comprising 19 strings with a total of 677 modules and kept taking data until 2009. At this point it was decommissioned as the successor experiment, IceCube, including its low energy extension DeepCore were nearing completion.

Following the termination of the DUMAND project, a fresh start for deep sea experiments was needed and the Mediterranean was identified as a suitable location with good infrastructure support. Of a number of projects, only ANTARES[73,74], which was proposed in 1999[75] and constructed between 2002 and 2008, reached completion. It consists of 12 cables with 25 levels of three photo-sensors each, spanning a volume of roughly $0.01\,\mathrm{km}^3$.

[73] *acronym for: Astronomy with a Neutrino Telescope and Abyss environmental RESearch*

[74] *Ageron et al., "ANTARES: The first undersea neutrino telescope"*

[75] *ANTARES Collaboration, "A Deep Sea Telescope for High Energy Neutrinos"*

In 1999, the same year ANTARES was proposed, the AMANDA collaboration started planning the successor experiment, now called IceCube. Deployment started in 2004 and finished in December 2010. As of today, IceCube is the largest Cherenkov neutrino telescope in the world with a detector volume of about one cubic kilometer. A full description of the IceCube instrumentation, deployment and calibration is given in the following chapters.

Future projects are envisioned both in Antarctica, the Mediterranean and Lake Baikal. In the Mediterranean, KM3Net[76,77] is planned to consist of independent detectors for the measurement of high-energy astrophysical neutrinos and low-energy oscillation physics. Construction started in 2016, but has been delayed, largely due to failures in the deep-sea infrastructure. More details about the planned IceCube expansions are discussed in section 7.1.

[76] *acronym for: Cubic Kilometre Neutrino Telescope*

[77] *Adrian-Martinez et al., "Letter of intent for KM3NeT 2.0"*

# 3

# GLACIAL ICE AND ITS OPTICAL PROPERTIES

**Being deployed into a natural glacier, IceCube does not have the luxury of being able to manipulate or characterize its detection medium in a laboratory setting. Instead, the experiment has to rely on the in-situ calibration of the optical properties. This chapter gives an introduction into ice and ice sheets, with a focus on optical properties.**

## 3.1 BULK PROPERTIES OF PURE ICE

Light propagation in an ideal block of ice, with no impurities, no enclosed gases and no stresses acting on it, can be described solely by its density and complex refractive index.

For reasons which will become obvious when discussing the crystal structure in section 3.6, the density of ice is lower than the density of water and ranges from $0.919\,\mathrm{g\,cm^{-3}}$ at $0\,°\mathrm{C}$ to $0.925\,\mathrm{g\,cm^{-3}}$ at $-50\,°\mathrm{C}$[78].

[78] *Petrenko and Whitworth,* Physics of Ice

The real part of the refractive index, as relevant for the Cherenkov yield and speed of the resulting photons, is nearly temperature independent and varies mildly between 1.35 and 1.30 in the considered wavelength range of 250 nm to 900 nm[79]. The complex part of the refractive index relates to the absorption length $\lambda_a$ and the absorption coefficient $a$ as:

[79] *Warren and Brandt, "Optical constants of ice from the ultraviolet to the microwave: A revised compilation"*

$$\frac{1}{\lambda_a} = a = \frac{4\pi \cdot Im(n)}{\lambda}. \tag{3.1}$$

The wavelength $\lambda$ dependence of the absorption coefficient $a$ as measured in different purity ice samples can be seen in Figure 3.1. It is empirically approximated as[80,81]:

[80] *Ackermann, Ahrens, et al., "Optical properties of deep glacial ice at the South Pole"*

[81] *The individual contributions are discussed in the following. $a_{dust}$, $A_U$, and $A_{IR}$ denote the impurity contribution, the Urbach tail and IR absorption respectively. $\delta\tau$ is the temperature difference with respect to a reference temperature.*

$$a(\lambda) = a_{dust}(\lambda) + A_U e^{-B_u \cdot \lambda} + A_{IR} e^{-\lambda_0/\lambda} \cdot (1 + 0.01 \cdot \delta\tau) \tag{3.2}$$

with

$$a_{dust}(\lambda) = a_{dust}(400) \cdot \left(\frac{\lambda}{400}\right)^{-\kappa} \tag{3.3}$$

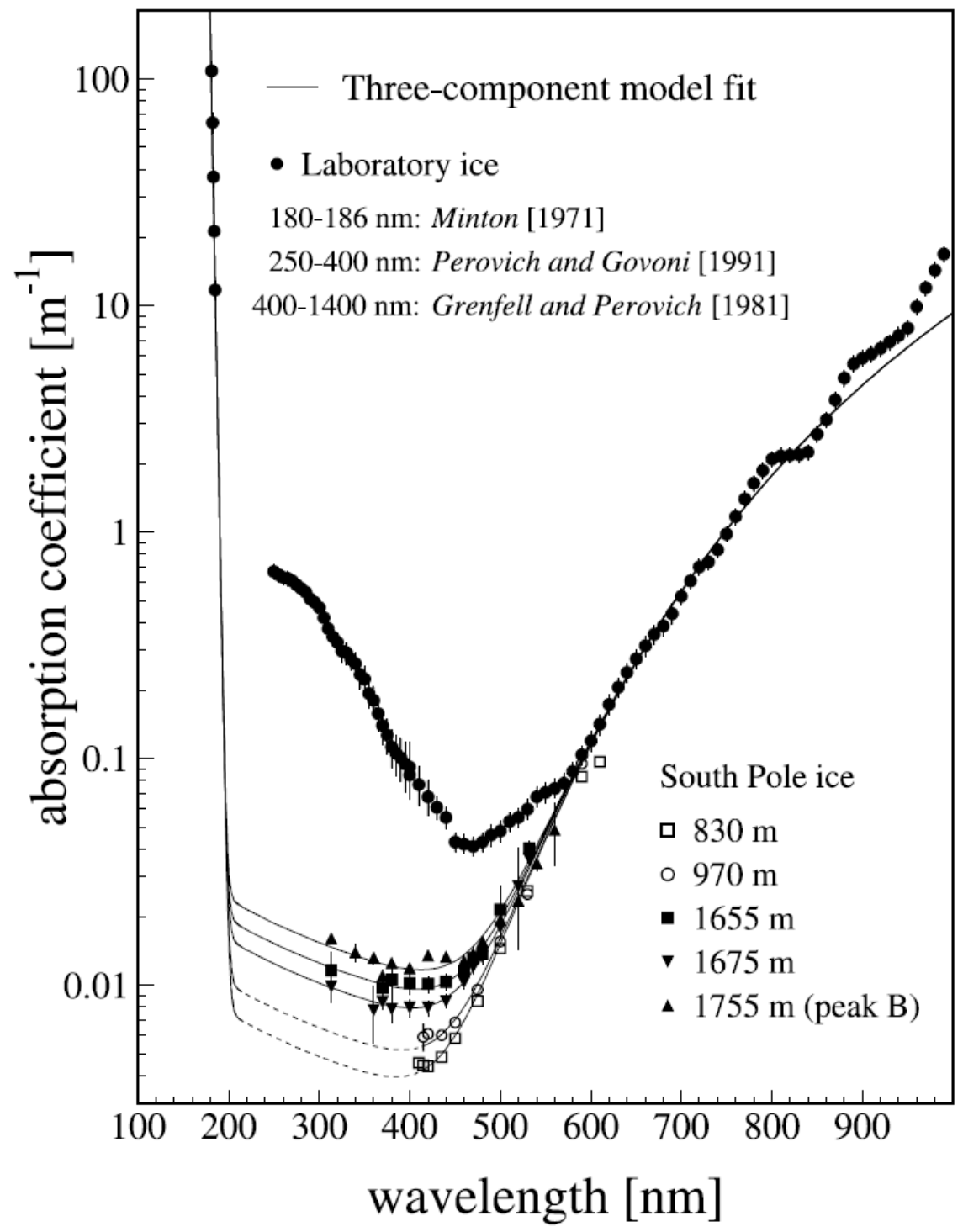

Figure 3.1: Absorption length as a function of wavelength for ice samples of varying quality. In the range from 200 nm to 500 nm the absorption length is limited by impurities, with the deep ice at the South Pole being the clearest known specimen.
[M. Ackermann, J. Ahrens, et al. (2006). "Optical properties of deep glacial ice at the South Pole"]

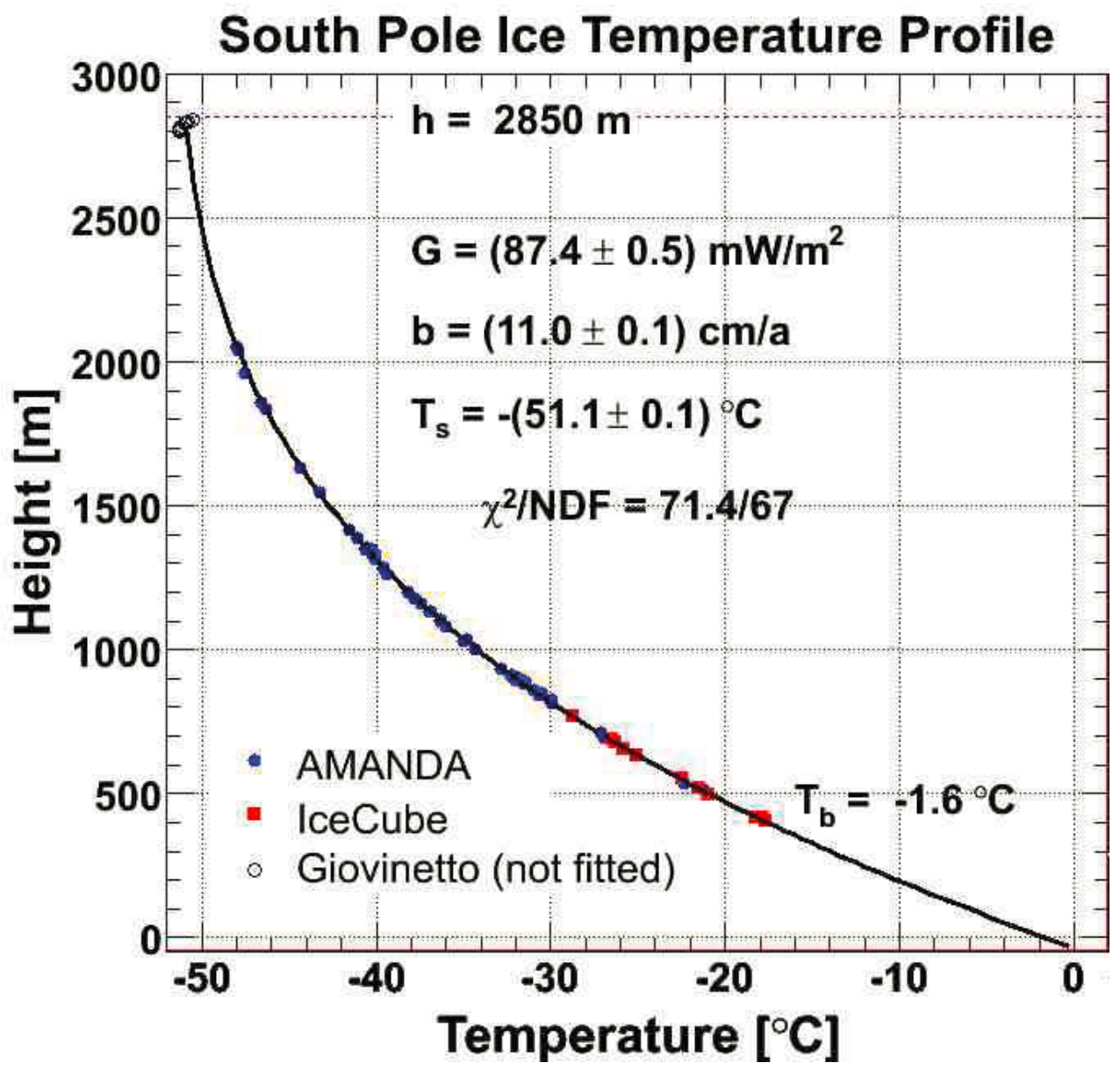

Figure 3.2: Temperature profile of the ice at the South Pole as measured in AMANDA and IceCube drill holes. Extrapolation indicates a wet bedrock interface.
[Ryan Bay (2018b). "Private communication"]
Updated version of:
[P. B. Price et al. (June 2002a). "Temperature profile for glacial ice at the South Pole: Implications for life in a nearby subglacial lake"]

Between ∼200 nm and ∼500 nm only upper limits on the complex refractive index are known and ice itself can be considered optically pure. In this range realistic optical properties are driven solely by impurities as will be discussed in the following sections. Some of these limits were obtained using IceCube and AMANDA measurements[82], as the instrumented ice has some of the lowest known impurity concentrations. This makes it one of the clearest known solids. In addition, the method of using a large timing array and nanosecond light pulses[83] enabled measurements of propagation distances not achievable with ice core segments in laboratories.

82 *Ackermann, Ahrens, et al., "Optical properties of deep glacial ice at the South Pole"*

83 *as explained in chapter 5*

Below ∼200 nm, at the so called "Urbach tail"[84], a steep exponential increase in absorptivity is observed as the photon energy exceeds the ice band-gap energy. The exact onset of the Urbach tail is unknown[85]. Above ∼500 nm the excitement of vibrational modes in the red and infrared leads to a gradual increase in absorptivity. The complex refractive index of the infrared region is believed to have a weak temperature dependence, with an increase of ∼ 1% per degree[86].

84 *Franz Urbach (Dec. 1953). "The Long-Wavelength Edge of Photographic Sensitivity and of the Electronic Absorption of Solids"*

85 *Warren and Brandt, "Optical constants of ice from the ultraviolet to the microwave: A revised compilation"*

86 *Ackermann, Ahrens, et al., "Optical properties of deep glacial ice at the South Pole"*

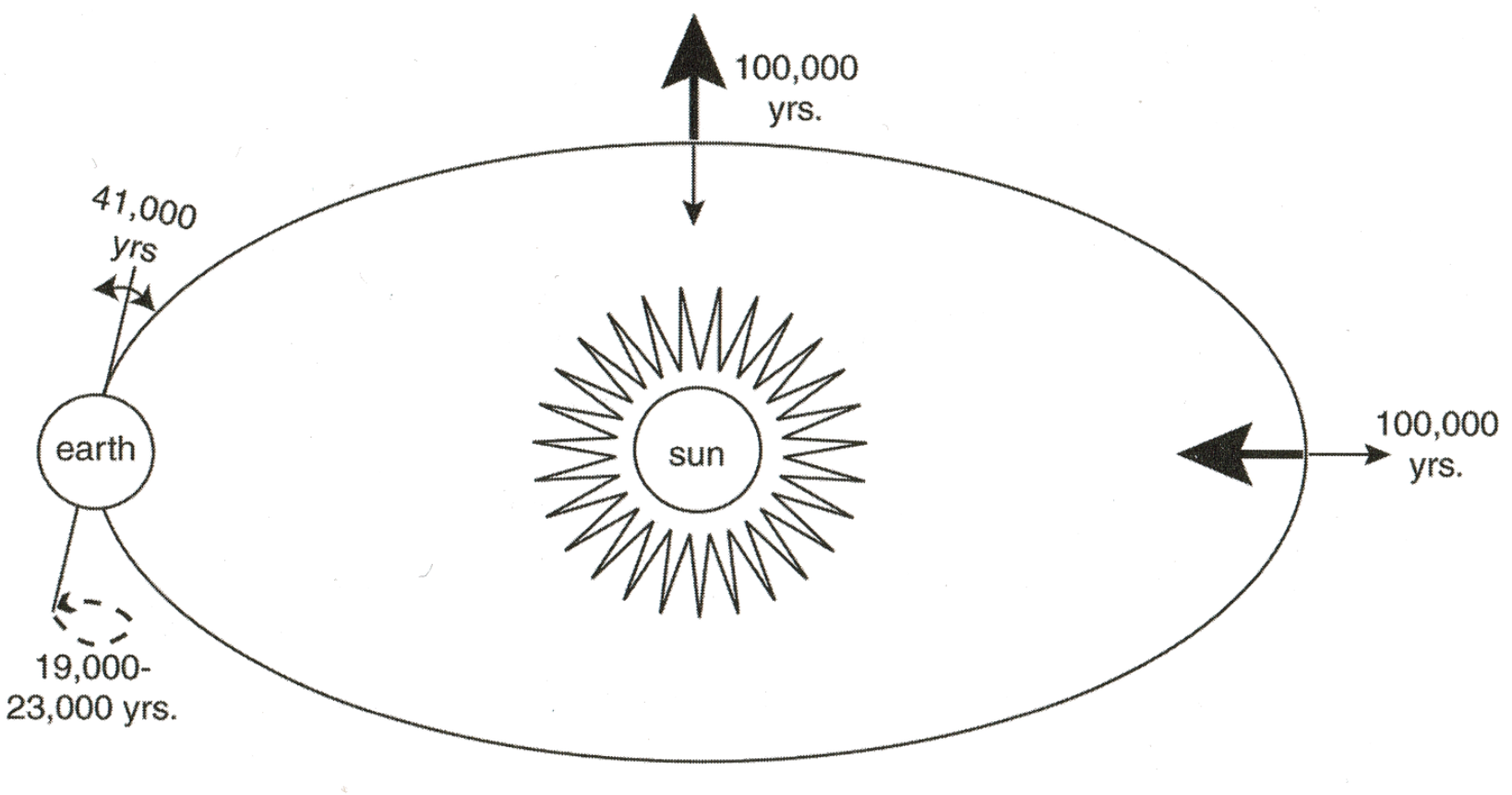

Figure 3.4: Periodic changes in Earths orbit as associated to the Milankovitch Cycles. Adapted from: [Richard B. Alley (2014). *The Two-Mile Time Machine: Ice Cores, Abrupt Climate Change, and Our Future - Updated Edition (Princeton Science Library)*. Princeton University Press]

The temperature profile of the ice at the South Pole is given in Figure 3.2. The local geothermal flux heats the ice from below. Extrapolating AMANDA and IceCube temperature measurements to the bedrock predicts a wet interface of $-1.6\,°C$, with a pressure-induced melting point of $-4\,°C$. The wet bedrock interface is also supported by seismic sounding measurements[87].

[87] *Peters et al., "Seismic detection of a subglacial lake near the South Pole, Antarctica"*

## 3.2 A SHORT CLIMATE HISTORY

The continent of Antarctica has been covered by ice for at least the past 15 million years[88]. Today it contains about 90% of the world's ice with an average ice thickness of $\sim 2\,km$ (see Figure 3.3).

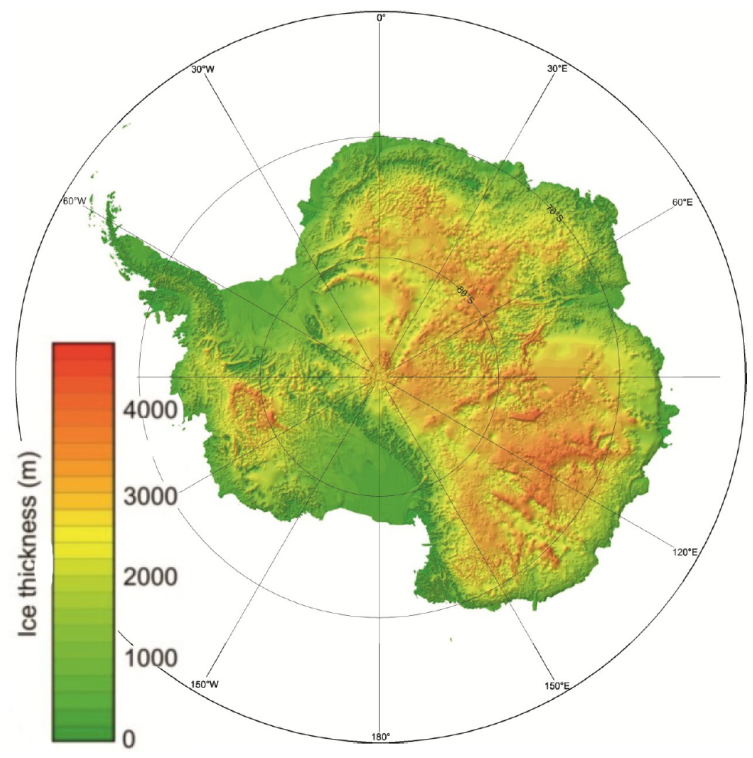

Figure 3.3: Map of ice thicknesses around Antarctica as measured in the Bedmap2 survey. [Fretwell, Pritchard, et al., "Bedmap2: improved ice bed, surface and thickness datasets for Antarctica"]

[88] *Levy et al., "Antarctic ice-sheet sensitivity to obliquity forcing enhanced through ocean connections"*

Neglecting climate forcing mechanisms[89,90], Earth's long term temperature changes are mainly driven by reoccurring patterns in Earth's orbit and rotation axis[91], as outlined in Figure 3.4. Due to orbit interactions with the outer planets, Earth's eccentricity varies between zero and five percent on a roughly 100'000 year cycle. In addition, the tilt of the Earth's rotation axis changes between 22.1° and 24.5° with a periodicity of 41'000 years and the axis precesses with a periodicity between 19'000 and 23'000 years.

While non of these three so called Milankovitch Cycles[92] significantly change the total yearly average solar energy reaching Earth, they do strongly impact the seasonality and distribution of solar radiation on the globe. Seasons arise due to the Earth's tilt and it follows that a larger tilt directly results in more pronounced seasonal variations. The precession dictates which hemisphere experiences which season at each point of the orbit. This becomes relevant at non-zero eccentricities, as the solar flux then changes between apogee and perigee[93].

[89] *Charlson, Schwartz, et al., "Climate Forcing by Anthropogenic Aerosols"*

[90] *That is difference between sunlight absorbed by Earth and energy radiated back to space.*

[91] *Alley,* The Two-Mile Time Machine: Ice Cores, Abrupt Climate Change, and Our Future - Updated Edition (Princeton Science Library)

[92] *named after Milutin Milankovitch*

[93] *Due to Kepler's second law the eccentricity also slightly changes the duration of the seasons.*

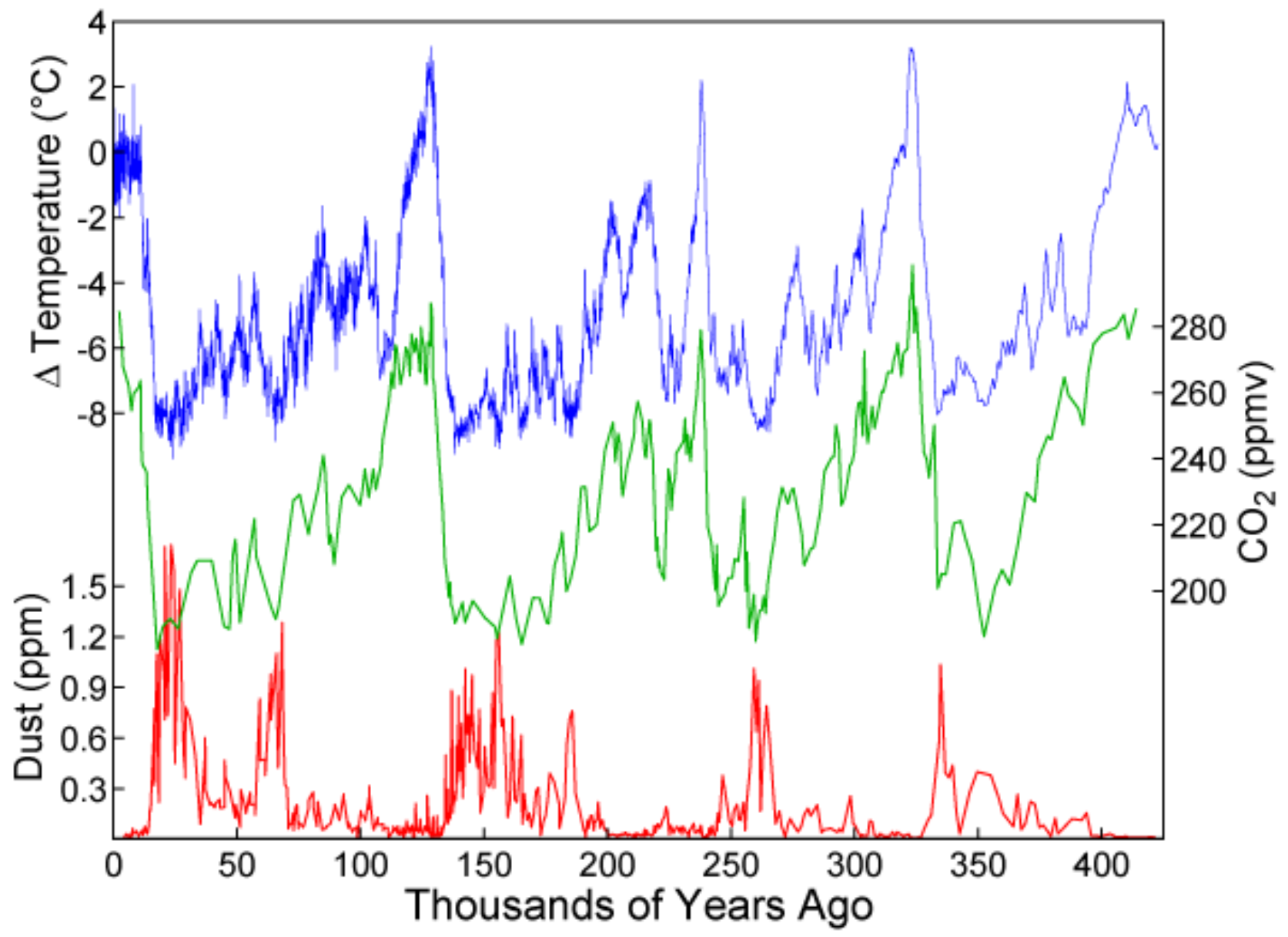

Figure 3.5: Ice core record of the past 500'000 years showing anti-correlation between dust concentration and temperature.
[Wikimedia Commons (2018e). *Vostok ice core dust and CO2*]

Currently the northern hemisphere experiences summer at the orbit apogee. Asymmetries between the hemisphere's seasons change the global heat balance as two-thirds of Earth's landmass can be found north of the equator.

As evident from Figure 3.5, Earth's climate, as traced in its mean temperature and $CO_2$ concentration, has been following the regular patterns of the Milankovitch cycles over at least the past 500'000 years. The strongest effect is caused by the eccentricity cycle, resulting in cold, glacial periods of roughly 90'000 years during which ice masses gradually grow, followed by a rapid 10'000 year collapse of the ice masses during warm interglacial periods.

This cycle is superimposed by additional modulations following the complex interaction between tilt, precession and eccentricity on the age scales of 20 to 40 ka. Periods of colder than average climate within a glacial period are called stadials[94]. Warmer than average periods are referred to as interstadials.

[94] *Alley*, The Two-Mile Time Machine: Ice Cores, Abrupt Climate Change, and Our Future - Updated Edition (Princeton Science Library)

As will be explained in the following, the cyclic change in climate conditions directly affects the deposited dust concentrations and thus the depth dependent optical properties observed in IceCube today.

It should be noted that the observed temperature swings require a larger change in heat balance than provided by the Milankovitch Cycles alone, hinting at correlated forcing effects[95]. The relative strengths and importance of the Milankovitch three cycles is also still debated as the predominant glacial cycle used to be $\sim 41$ ka as given by the axis inclination, up to 1 million years ago[96].

[95] *Alley*, The Two-Mile Time Machine: Ice Cores, Abrupt Climate Change, and Our Future - Updated Edition (Princeton Science Library)

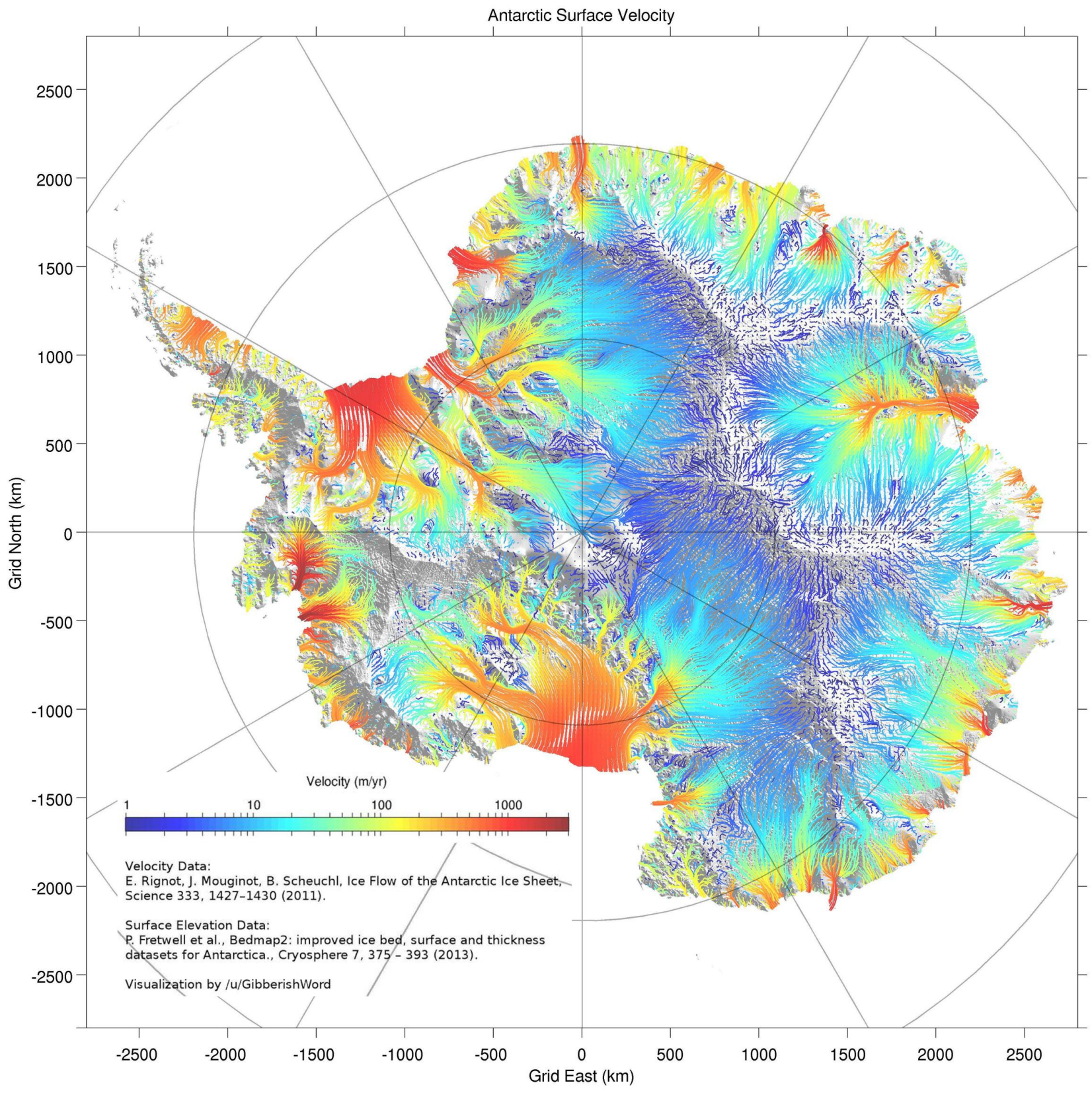

Figure 3.6: Map of Antarctic ice flow velocities, showing the same kind of river systems as found on any continent.
[E. Rignot et al. (Aug. 2011). "Ice Flow of the Antarctic Ice Sheet"][with adaptations as noted in the figure]

The last growth of the ice sheets reached its maximum about 26.5 ka ago. Earth has now been in a warm period, the so called Holocene interglacial, for the past $\sim$ 11'700 years, which has in turn allowed for the rise of human civilization.

## 3.3 Ice sheet dynamics

On shorter timescales the overall mass balance of the Antarctic ice sheet can be considered static. Yet averaged over the entire continent $\sim 170$ mm of liquid equivalent snow precipitates each year[97]. It follows that a roughly equivalent amount of ice is discharged into the oceans after being drained from the inland regions through a system of glacial rivers as mapped in Figure 3.6.

Ice flow can occur in two forms: Through basal sliding, where the entire ice sheet slips uniformly on top of a liquid interface to the bedrock, or through plastic deformation[98]. While generally perceived as a rigid body, naturally occurring ice is always within a few dozen degrees of its melting point. Therefore, it is classified as a "hot solid", which readily deforms under stress[99] through re-crystallization as described in section 3.6.1.

[96] *Shackleton et al., "An alternative astronomical calibration of the lower Pleistocene timescale based on ODP Site 677"*

[97] *Bromwich et al., "An Assessment of Precipitation Changes over Antarctica and the Southern Ocean since 1989 in Contemporary Global Reanalyses"*

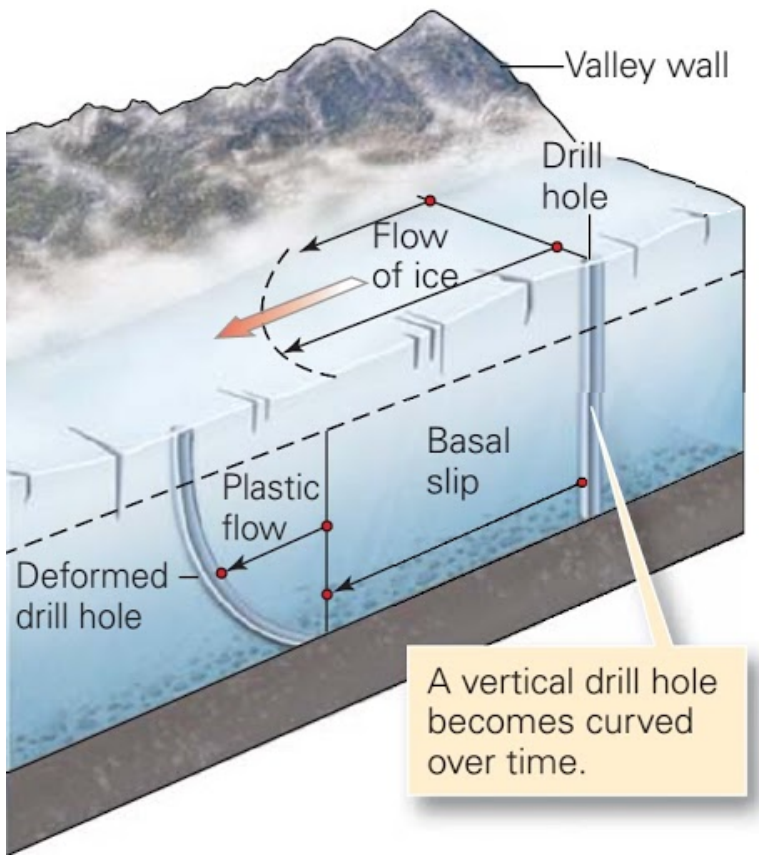

Figure 3.7: Sketch comparing basal (sliding along bedrock) and plastic (deformation of the solid under strain) ice flow.
[Marshak, *Essentials of Geology (Fourth Edition)*]

[98] *E. Rignot et al. (Aug. 2011). "Ice Flow of the Antarctic Ice Sheet"*

[99] *approachable equivalents might be a glowing hot iron, or flowing pancake batter*

Consider a glacial accumulation region as seen in Figure 3.8, where a new constant-thickness layer is added each year. The resulting overburden weighs on all layers below. As ice is incompressible, this causes a thinning of the deeper layers accompanied by a sideways spreading, inducing a flow outwards from the accumulation region. In a simplified model, each year's layer moves down just enough to make room for next years accumulation and a layer n% down the ice sheet in turn has thinned by n% from its original thickness. Shallow ice primarily moves downward, while deeper ice, containing more yearly layers per depth, primarily moves outward[100].

[100] *Alley,* The Two-Mile Time Machine: Ice Cores, Abrupt Climate Change, and Our Future - Updated Edition (Princeton Science Library)

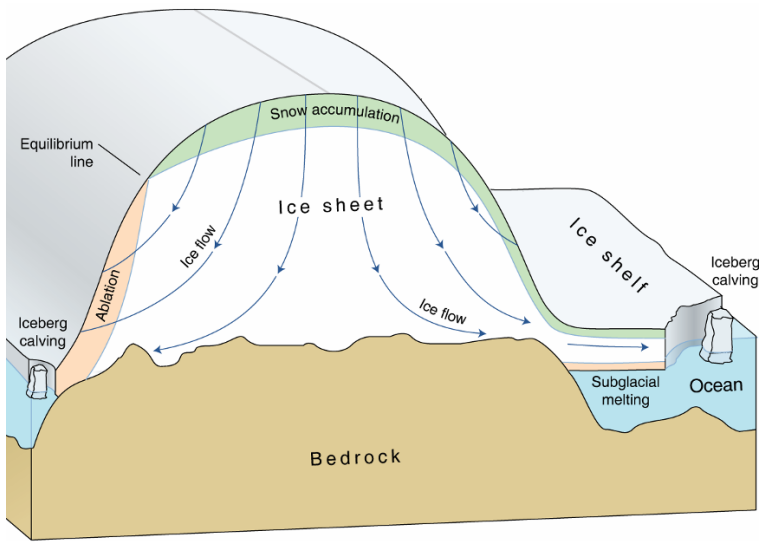

Figure 3.8: Sketch highlighting the principles of ice sheet dynamics. Under the pressure of new snow accumulation deeper ice layers thin and spread outwards resulting in a flow of the entire ice sheet away from the accumulation region. [NASA, *Ice sheet dynamics*]

At the IceCube site the glacier moves at 10.1 m per year. The movement is dominated by basal slip[101]. Nevertheless, a strong optical anisotropy in the flow direction as discussed in chapter 10 is observed. As neither the underlying mechanism causing the optical anisotropy, nor an adequate empiric parametrization have been found to date, it remains one of the largest ice systematics and its study is a major part of this thesis.

[101] *David A. Lilien et al. (July 2018). "Holocene Ice-Flow Speedup in the Vicinity of the South Pole"*

## 3.4 impurities

Glacial ice contains impurities carried by the wind which also brought in the snow precipitation. While the total concentration of impurities at the South Pole is remarkably low due to the remote location, these sub-ppm concentrations still drive the optical properties. Impurities are classified in the following categories:

- Mineral dust: Wind blown dust, believed to primarily originate from around Patagonia.[102]
- Marine salt crystals: Crystalline sea salt reaches the Antarctic high plateau after being picked up from the surrounding seas by winds and evaporation.
- Acid droplets: The primary acids found in ice are sulfuric acid and carbonic acid, created when carbon dioxide and sulfur dioxide react with snowflakes or water in the atmosphere. They are believed to persist in the ice as droplets which follow the grain boundary network.[103]
- Soot: Just like acids, soot originates from combustion processes or volcanic activity.

[102] *Alley,* The Two-Mile Time Machine: Ice Cores, Abrupt Climate Change, and Our Future - Updated Edition (Princeton Science Library)

[103] *Jérôme Weiss et al. (2002). "Dome Concordia ice microstructure: impurities effect on grain growth"*

Stadials, periods of colder-than-average climate within a glacial period, are generally dryer with larger dessert extents. They also tend to be windier. Both effects lead to larger impurity concentrations in the atmosphere[104]. Colder temperatures around Antarctica also locally result in less evaporation and larger sea ice extent leading to less precipitation. In conclusion, colder climates result in larger dust concentrations found in the ice, as can also be seen in Figure 3.5.

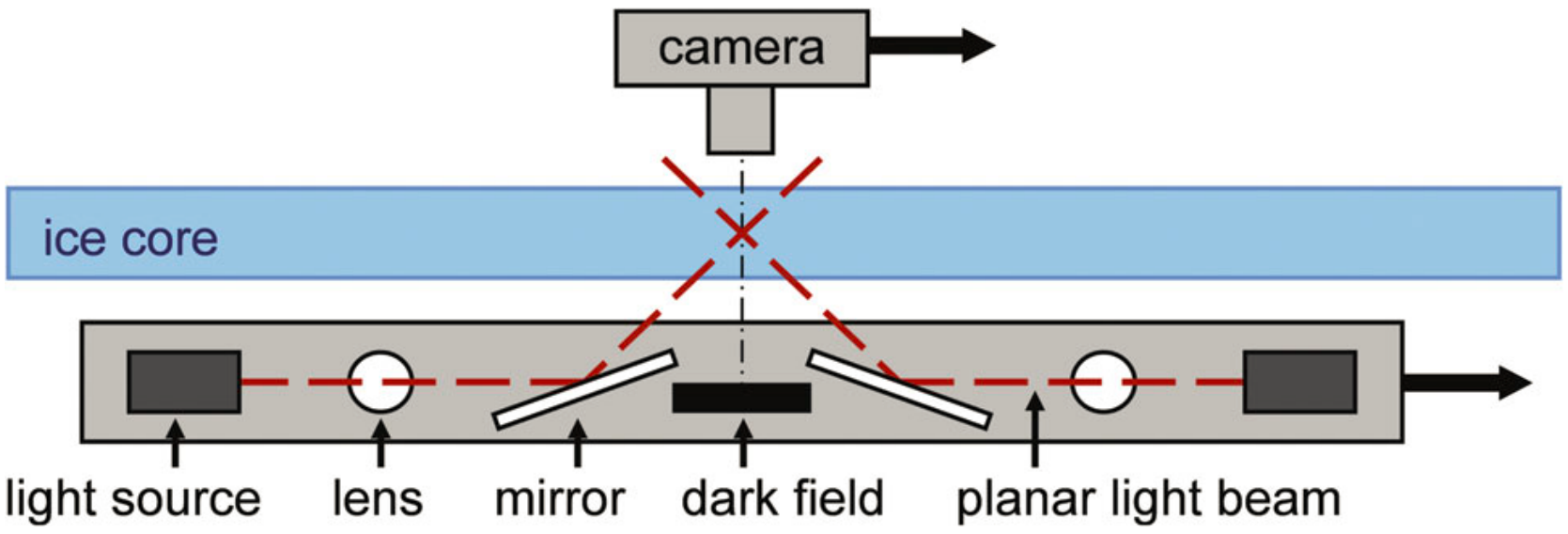

Figure 3.9: Working principle of a line scanner. Light can only reach the camera by scattering of impurities in the ice core.
[Sergio Henrique Faria et al. (Oct. 6, 2017). *The EPICA-DML Deep Ice Core*. Springer, Berlin]

[104] *Alley,* The Two-Mile Time Machine: Ice Cores, Abrupt Climate Change, and Our Future - Updated Edition (Princeton Science Library)

The same mechanisms also leads to seasonal variations, with most dust occurring in late winter. Seasonal variations also affect the impurity composition, with salts and acids showing the strongest variability[105]. The modeling of impurities in terms of optical scattering processes is discussed in section 5.2.1.

[105] *He and Price, "Remote sensing of dust in deep ice at the South Pole"*

## 3.5 STRATIGRAPHY

Except for rare volcanic lines, impurity concentrations are generally too low for the yearly layering / stratigraphy to be visible to the naked eye[106]. Instead, so called line scanners, see Figure 3.9, are utilized. Two opposing light sources are used to illuminate the ice at a 45° angle relative to the ice surface, with a camera observing the intersection against a black surface. Without any impurities the image appears black as only scattering can direct light onto the camera. An example image obtained from a line scanner can be seen in Figure 3.10.

[106] *Alley,* The Two-Mile Time Machine: Ice Cores, Abrupt Climate Change, and Our Future - Updated Edition (Princeton Science Library)

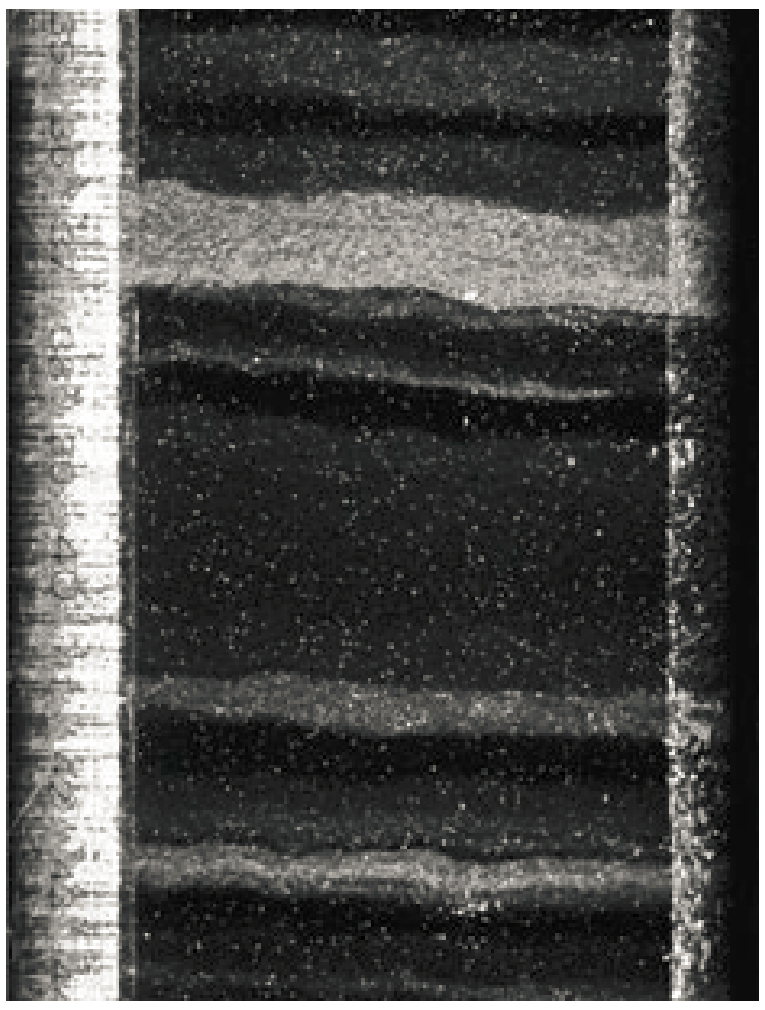

Figure 3.10: Ice stratigraphy image from the EDML ice core showing a ~10 cm segment at a depth of 1813 m. Bright bands are the result of higher dust concentrations in the later winter months and can also be used for dating of the core.
[Faria et al., *The EPICA-DML Deep Ice Core*]

The seasonal variation of impurity concentration and precipitation leads to a periodic layering as the glacier slowly builds up. The dusty late-winter months appear as brighter lines, so called cloudy bands.

A basic method for dating ice cores involves counting cloudy bands versus depth. This method is generally reliable unless differential flow distorts the layers or layers are too faint to be detected. Figure 3.11 shows the age versus depth relation for the South Pole. The annular layer thickness ranges between 1 cm and 2 cm[107] at the IceCube depths. The instrumented depths correspond to ages between 40 and 100 thousand years ago.

[107] *Aartsen et al., "South Pole glacial climate reconstruction from multi-borehole laser particulate stratigraphy"*

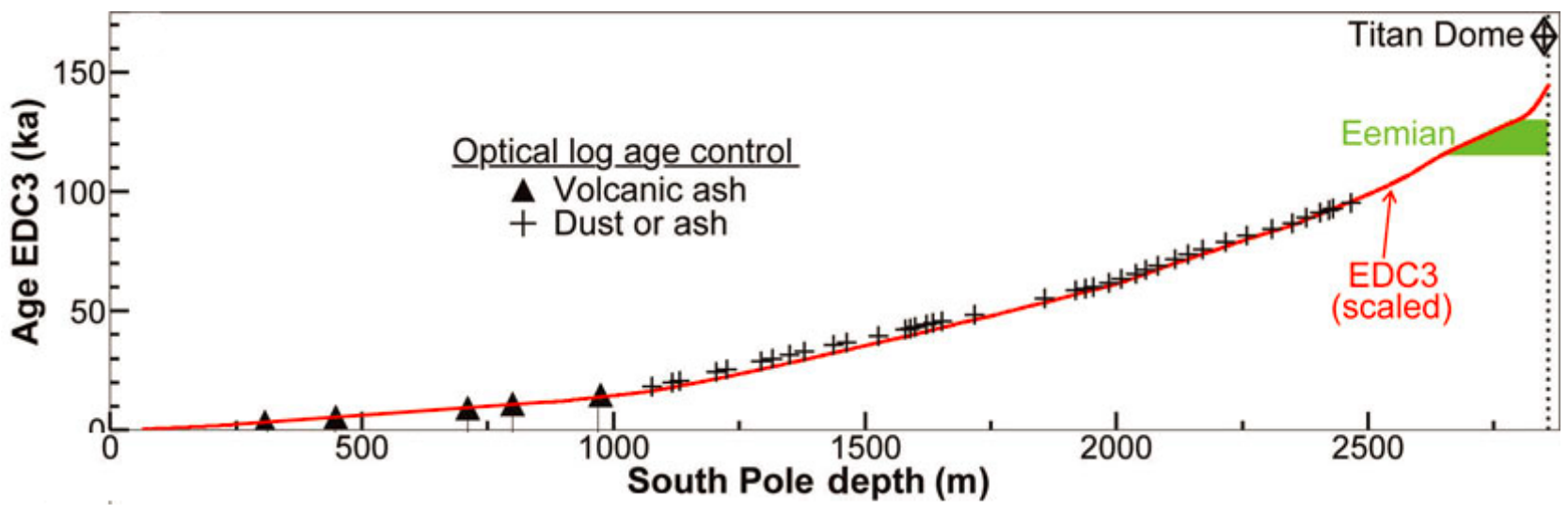

Figure 3.11: Relationship between depth and age of the South Pole ice. IceCube is deployed into 34′000 to 100′000 year old ice.
[M. G. Aartsen et al. (2013d). "South Pole glacial climate reconstruction from multi-borehole laser particulate stratigraphy"]

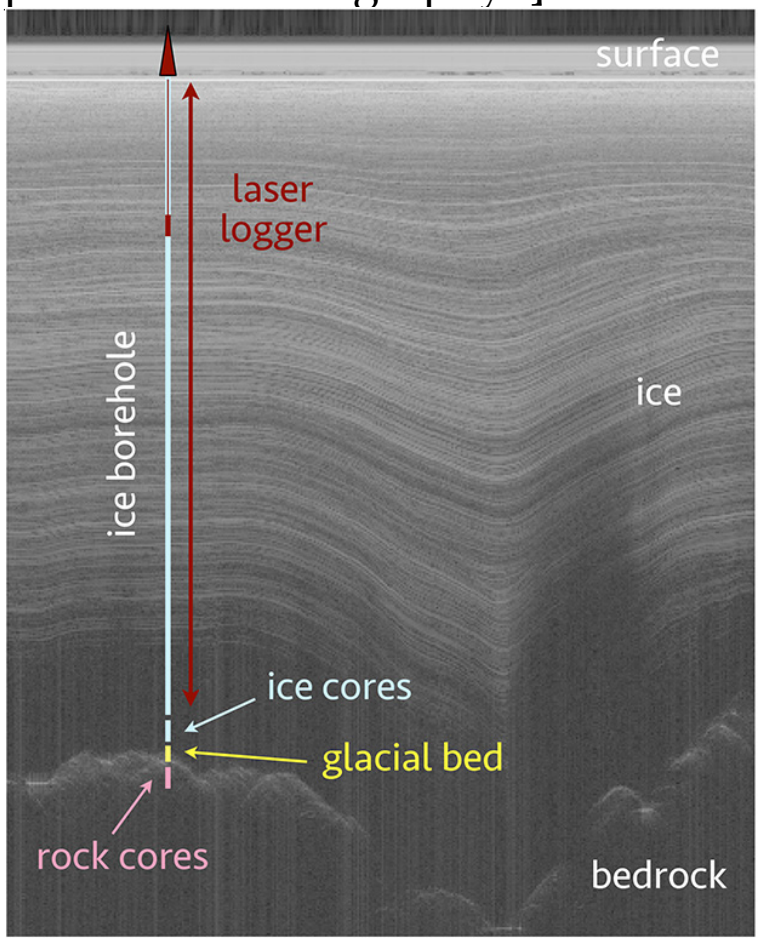

Figure 3.12: Example radar profile of an ice sheet. Changes in the dielectric properties result in reflections. The propagation delay and signal strength yield the depth contours. In the deep ice the radar profiles are seen to follow the contour of the underlying bedrock.
[RAID collaboration, *Ground penetrating radar*]

The stratigraphy of entire glaciers can be probed using radar sounding. A radar beam, emitted most commonly from a plane, propagates through the glacier and is partially reflected where there is a change in conductivity, usually acid content[108]. The propagation delay and intensity are recorded, yielding radar images as for example shown in Figure 3.12.

At depths close to the bedrock the contour lines are seen to follow the structure of the underlying bedrock. The features wash out as valleys are gradually filled in. In IceCube the depth variability of isochrons, layers of contemporaneous ice, is usually referred to as tilt.

## 3.6 CRYSTALLINE STRUCTURE

Ice is the comprehensive term for all 12 known solid phases of water[109]. For temperatures and pressures naturally occurring on Earth, only the hexagonal crystal form, $I_h$ occurs. Each monocrystal consists of a neat stack of layers and each layer can be thought of as a tessalation of hexagonal rings, with oxygen atoms at each node (see Figure 3.14). The plane defined by these layers is called the basal plane, with the normal vector referred to as the principal axis or c-axis, as it is the axis of highest rotational symmetry.

Due to the low atomic packing factor of the hexagonal crystal, ice is one of the few solids with a density ($0.919\,\mathrm{g\,cm^{-3}}$) smaller than its liquid phase. This increase in volume during freezing explains why ice exhibits a pressure-induced reduction of its melting point, as seen in Figure 3.13.

The ice found in glaciers such as on the Antarctic plateau has been created through compactification of snow layers into firn and at a later stage ice. Therefore, it consists of a large number of interlocked mono-crystals, also called grains. They are typically sub-millimeter to centimeter in size[110].

The surface where two grains meet is called a grain boundary. The general rule of thumb for grain size is that old and clean ice

[108] *Petrenko and Whitworth,* Physics of Ice

[109] *Sérgio H. Faria et al. (Apr. 2014b). "The microstructure of polar ice. Part II: State of the art"*

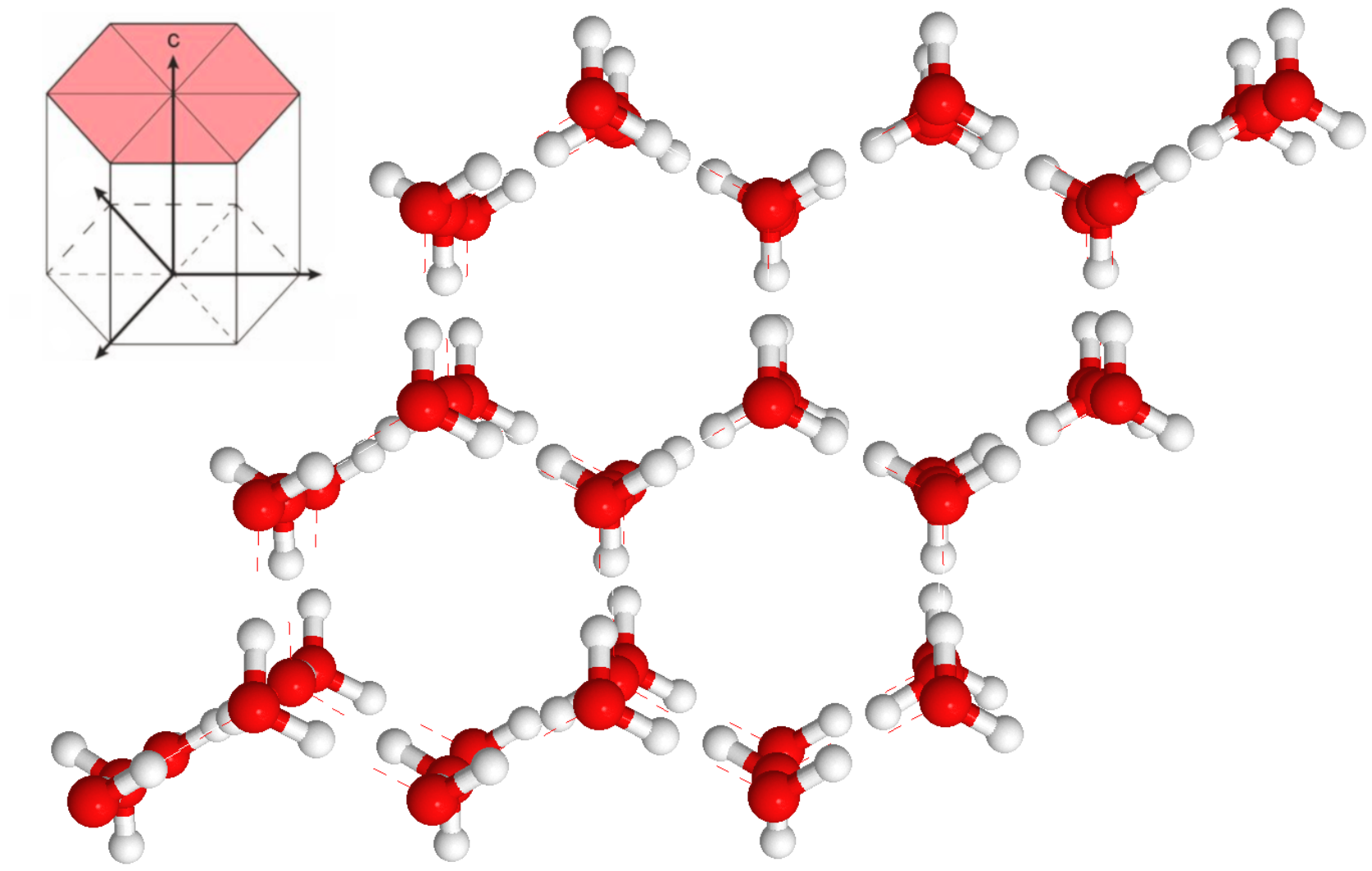

Figure 3.14: Crystal structure of $I_h$ ice (here in the proton ordered variant) viewed along the c-axis. Red dots denote oxygen atoms while white dots denote hydrogen. [Wikimedia Commons (2018b). *Ice XI View along c axis*]

develops large grains, while young or dirty ice is dominated by small grains.

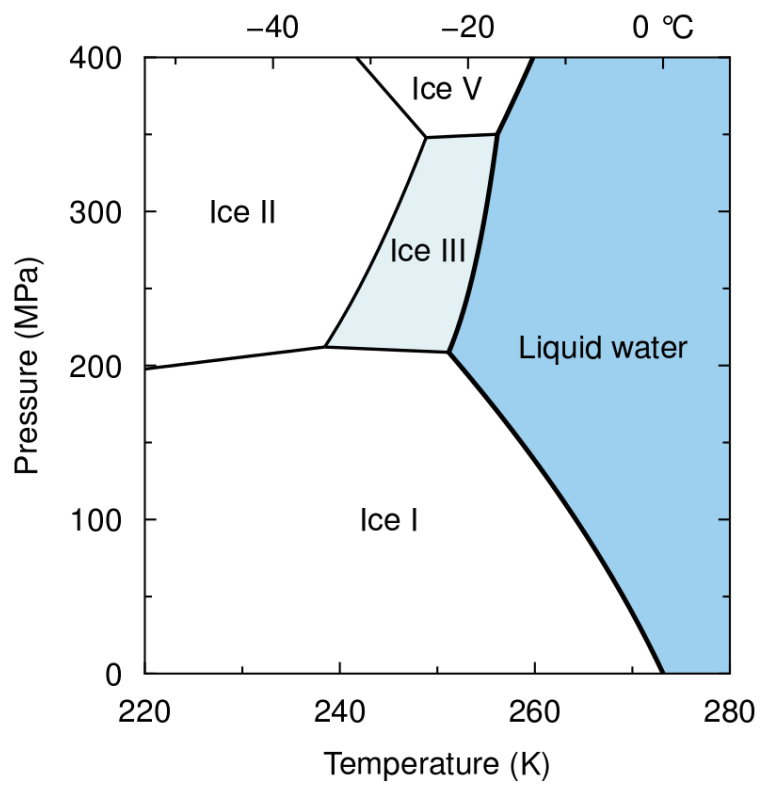

Figure 3.13: Phase diagram of water ice. Ice I naturally occurs on Earth. The pressure-induced reduction of the melting point is clearly seen. [Wikimedia Commons, *Ice III phase diagram*]

[110] *Faria et al., "The microstructure of polar ice. Part I: Highlights from ice core research"*

The density of fresh snow is generally smaller than $0.4\,\mathrm{g\,cm^{-3}}$ with about 65% of the volume being air in between snow grains[111]. As snow is compacted, most gases are released but about 10% by total volume is trapped between growing grains[112]. Figure 3.15 illustrates the firn-ice transition. Both images show microscopic views of thin sections of ice core samples. Gray areas are grains and black regions show air bubbles. The samples were left to sublimate for a while to achieve a smooth surface finish. The ice molecules at grain boundaries are less strongly bound to the lattice and sublimate faster. The resulting groves appear as gray lines in the chosen illumination.

Due to their large difference in refractive index compared to the surrounding ice grains, air bubbles act as optical scattering centers. As they have a negligible imaginary refractive index they do not absorb light. The high concentration of air bubbles in glacial ice generally results in a very short effective scattering length. As the pressure increases with overburden air bubbles shrink in size at larger depths.

Under even higher pressures (see Figure 3.17) air bubbles undergo a peculiar transition, where the individual gas molecules get incorporated into the crystal lattice of the ice. The resulting ice air clathrate hydrate[113], also called craigite, has a refractive index which is within five permille of that of ice and as a result appears nearly perfectly transparent[114]. Figure 3.16 shows a microscopic picture of a single craigite inclusion. The exact craigite formation mechanism remains elusive, but is believed to take place over at least several centuries, as extensive bubble-craigite transition regions, spanning up to several hundred meters[115], are observed in ice cores. Studies are further complicated because ice cores are usually left to relax at atmospheric pressure before being analyzed. This is required to be able to machine the cores but greatly alters craigite properties.

While AMANDA-A was still strongly impacted by light diffusion in air bubbles[116], IceCube is deployed below the craigite transition region.

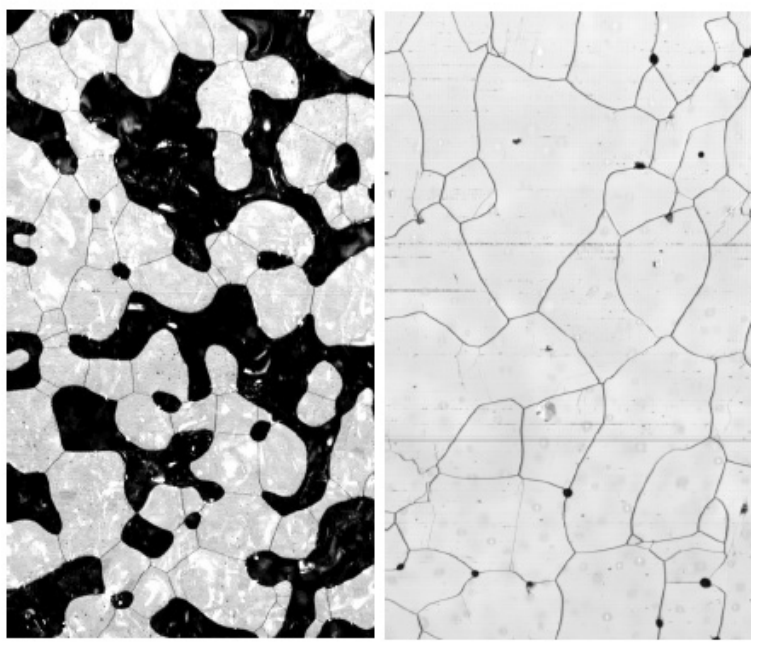

Figure 3.15: Microstructure mapping of thin sections of ice cores. Gas inclusions are black. Gray lines show grain boundaries. Left: Obtained in the firn at a depth of 11 m. Right: Obtained in the solid ice at 273 m.
[Eichler, "C-Axis Analysis of the NEEM Ice Core: An Approach based on Digital Image Processing"]

[111] *Michiel van den Broeke (May 2008). "Depth and Density of the Antarctic Firn Layer"*

[112] *Michael Bender et al. (1997). "Gases in ice cores"*

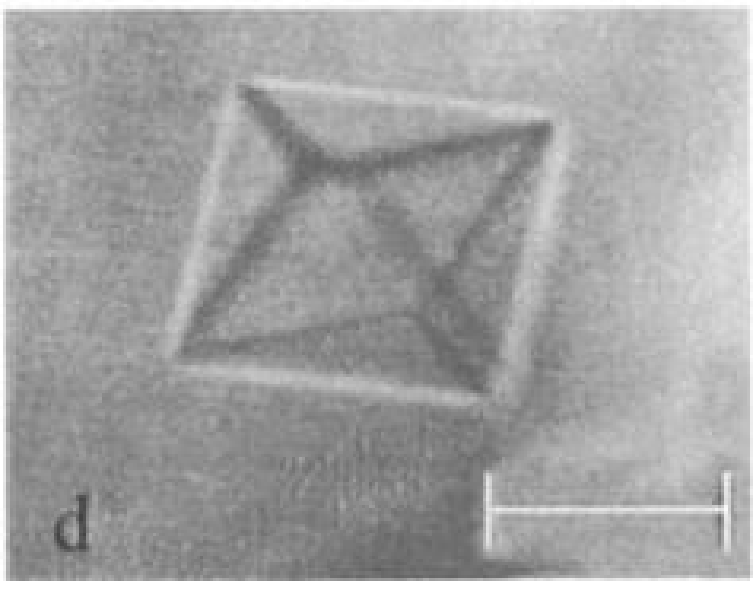

Figure 3.16: Fully developed craigite cage seen in the NGRIP ice core at a depth of 1272 m. The scale bar is roughly 100 µm across.
[Kipfstuhl et al., "Air bubbles and Clathrate hydrates in the transition zone of the NGRIP Deep Ice Core"]

## 3.6.1 *Crystal creep*

Plastic deformation of ice is mediated through deformation of individual grains as well as interactions between grains. A short overview as relevant for the discussion of the ice optical anisotropy in chapter 10 is given here. It should be appreciated that ice flow is a very complex and still quickly developing research topic. For more details please see for example: [117,118]

The viscosity of an individual crystal depends strongly on the direction of the applied strain. A hexagonal crystal will most readily deform as shear is applied orthogonal to the c-axis leading to slip of the individual basal planes. This is commonly visualized through a "deck of cards" metaphor. As a result individual grains elongate, with the major axis being perpendicular to the c-axis.

In a polycrystalline material such as ice, the deformation of individual grains is restricted by the presence of neighboring grains. Sustained strain will create a build-up of internal stresses. These are minimized through changes in the crystal structure, referred to as recrystallization processes.

One such process is called migration recrystallization, where grains realign atoms belonging to neighboring grains according to their lattice, extending their volume and moving the grain boundary. A competing process is rotation recrystallization, which rotates the c-axis of entire grains but in the process often also leads to a breakup of this grain. As grain boundaries are effectively lattice defects, recrystallization generally leads to a reduction in grain boundary area. Thus warm or old crystals, which do not experience repeated stresses, can grow large.

As grain boundaries move, small impurities are dragged with the boundary, referred to as impurity drag[119]. Larger impurities have a higher inertia and the boundary does not exert enough force to move them along. When a boundary hits an even larger impurity its movement can get arrested at that point, or the boundary may wrap around the particle. This is one possible explanation why ice samples with high impurity concentration generally tend to have small grains[120].

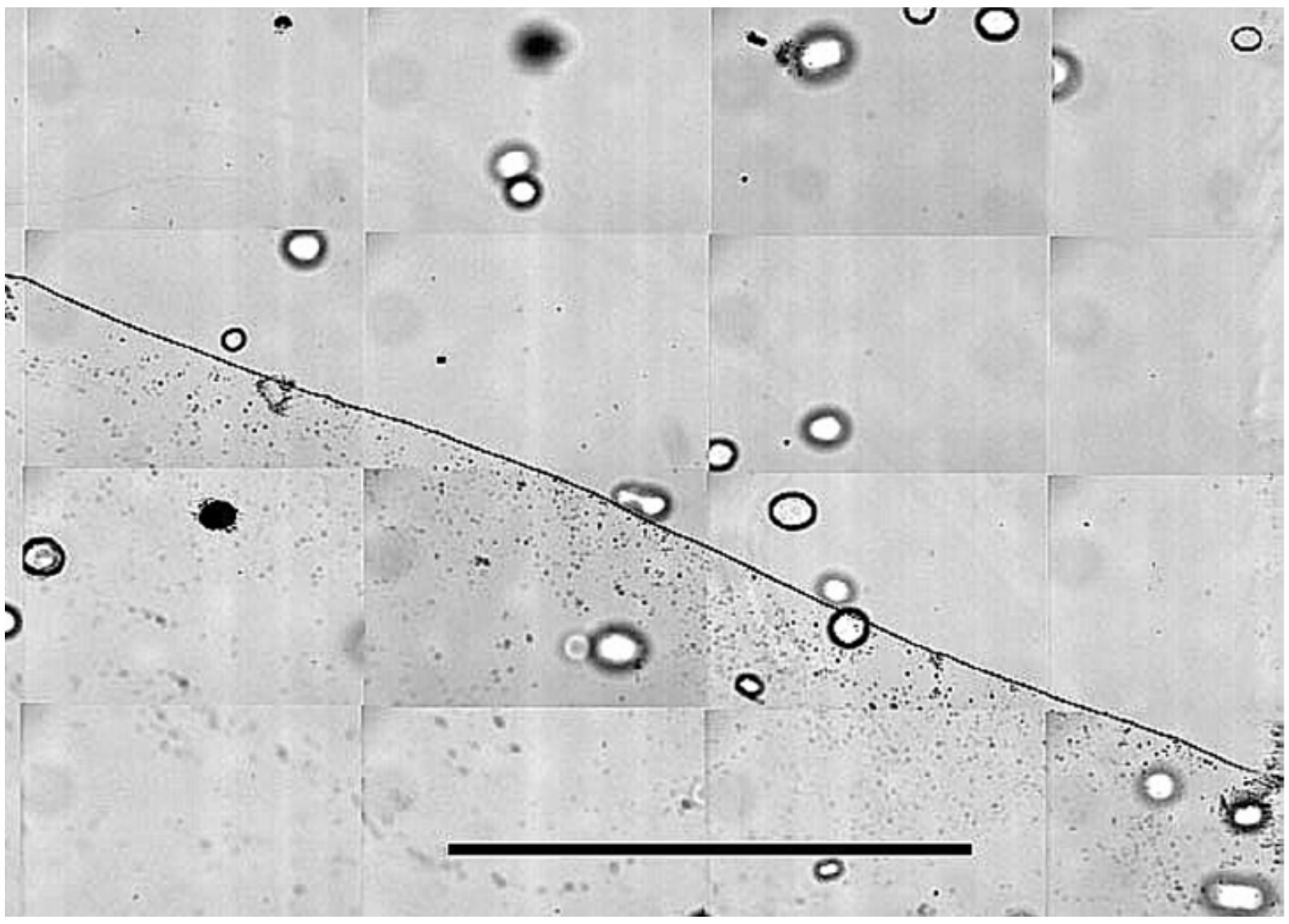

Figure 3.18: Microinclusions (tiny black dots) accumulated at a grain boundary of deep Antarctic ice (EDML core, 2656 m depth; scale bar: 3 mm).
[Sérgio H. Faria et al. (Apr. 2014b). "The microstructure of polar ice. Part II: State of the art"]

Both of these processes should lead to an aggregation of impurities on boundaries, so that the density of impurities in the grain is expected to be significantly smaller than on the boundaries. The extent to which this is realized at different depths in natural glacial ice is strongly debated. It has been established using conductivity measurements that acids move through the grain boundary network.

For other kinds of impurities, microscopic identification is difficult due to the small impurity concentrations and impurity sizes down to several dozen nanometers. Recent studies have tried to use Raman spectroscopy to identify the elemental composition at boundaries compared to the bulk[121]. Figure 3.18 shows an unusually strong example of grain boundary aggregation as found in the deep ice of the EDML (European Project for Ice Coring in Antarctica (EPICA) at Dronning Maud Land (DML), one of the best documented cores) core.

Grain boundary aggregation is discussed in this thesis as one possible cause of the optical ice anisotropy. It also plays an important role in ice sheet dynamic modeling (see section 3.3) and climate-change induced sea rise predictions, as impurities on grain boundaries effectively act as a lubricant reducing the plastic viscosity.

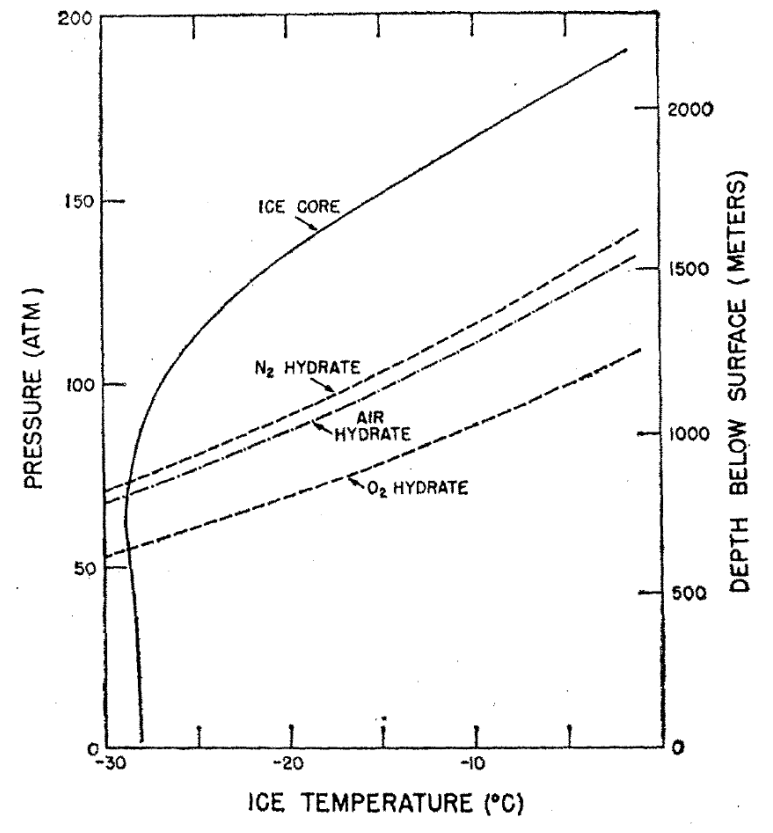

Figure 3.17: Craigite phase diagram compared to a typical ice core pressure-temperature relation. Craigite forms at large pressures and low temperatures.
[Miller, "Clathrate Hydrates of Air in Antarctic Ice"]

[113] *Craig et al., "Nonequilibrium air clathrate hydrates in Antarctic ice: a paleopiezomdter for polar ice caps."*

[114] *Uchida et al., "Refractive-index measurements of natural air–hydrate crystals in an Antarctic ice sheet"*

[115] *Kipfstuhl et al., "Air bubbles and Clathrate hydrates in the transition zone of the NGRIP Deep Ice Core"*

[116] *Price et al., "Optical properties of deep ice at the South Pole: scattering"*

[117] *Faria et al., "The microstructure of polar ice. Part II: State of the art"*

[118] *Durand et al., "Effect of impurities on grain growth in cold ice sheets"*

### 3.6.2 *C-axis measurements using birefringence*

The c-axis orientation of individual grains can be measured using polarized light microscopy. As seen in Figure 3.19, a thin section of an ice sample is placed between two orthogonal polarization filters. For any optically isotropic sample, no light would reach the detector behind the second polarizer.

But ice is a birefringent material. This means it possesses a distinct optical axis and light polarized along this axis (extraordinary ray) experiences a different index of refraction compared to light polarized orthogonal to this axis (ordinary ray). For more details on the theory of birefringence see the discussion on light diffusion in birefringent polycrystals in section 10.8.

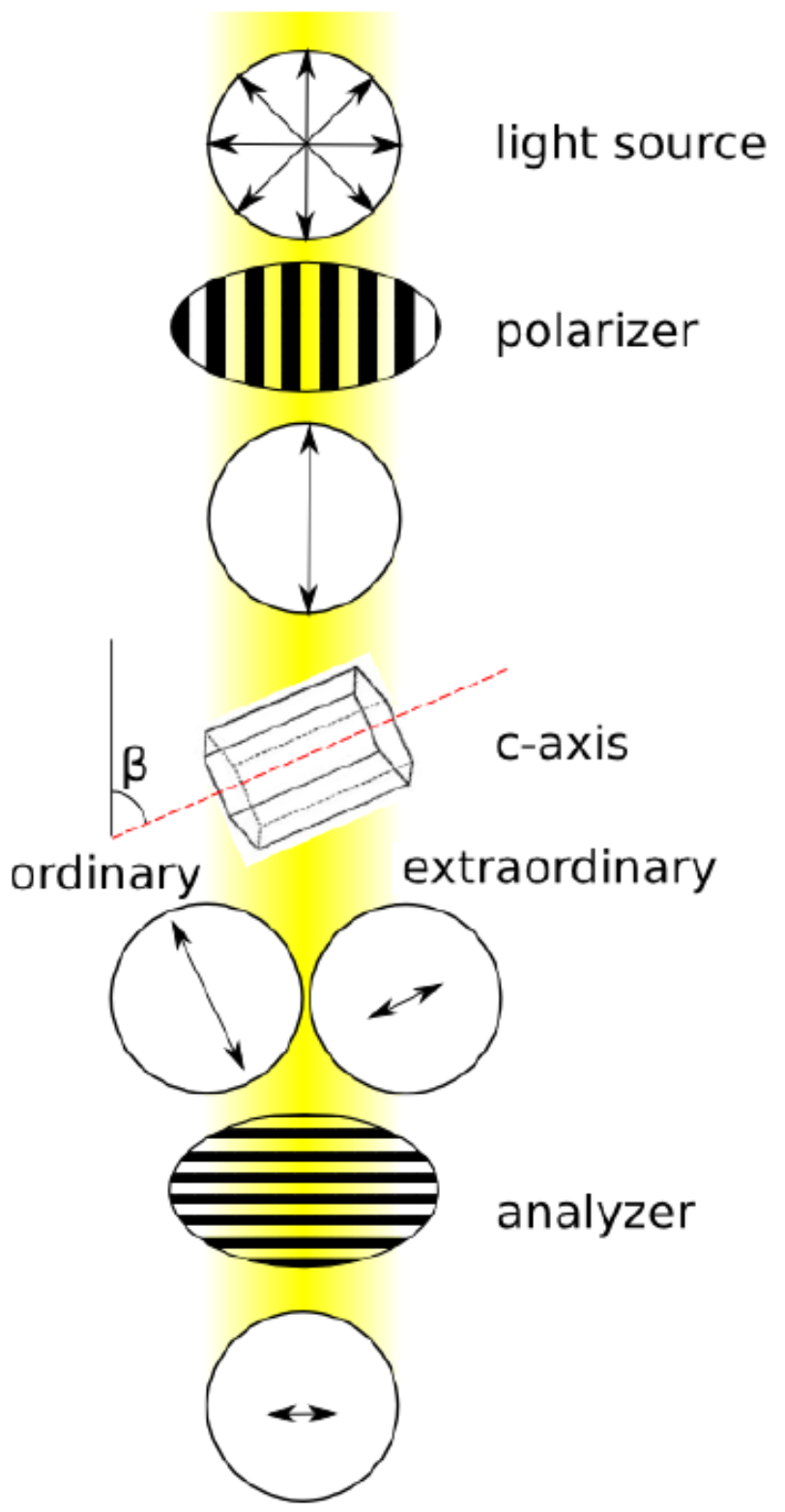

**Figure 3.19:** Sketch of a fabric analyzer used to measure the c-axis orientation of grains in a thin ice sample.
[Jan Eichler (Apr. 2013). "C-Axis Analysis of the NEEM Ice Core: An Approach based on Digital Image Processing." Freie Universität Berlin]

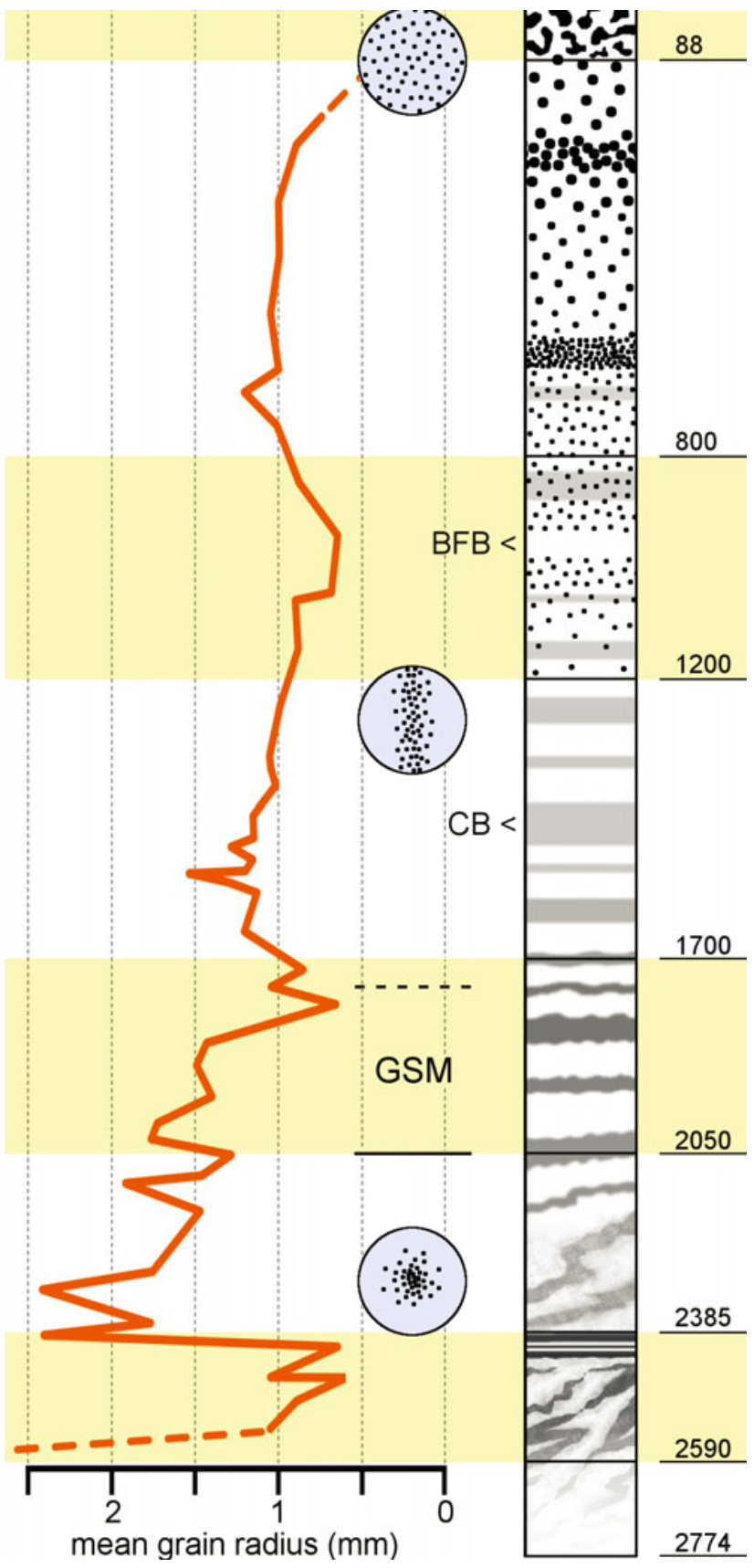

**Figure 3.20:** Summary of physical properties along the EDML ice core. Left: Grain size. Middle: C-axis distribution. Right: Visual stratigraphy.
[Sergio Henrique Faria et al. (Oct. 6, 2017). *The EPICA-DML Deep Ice Core*. Springer, Berlin]

119 *Durand et al., "Effect of impurities on grain growth in cold ice sheets"*

120 *Durand et al., "Effect of impurities on grain growth in cold ice sheets"*

121 *Jan Eichler et al. (May 2017). "Location and distribution of micro-inclusions in the EDML and NEEM ice cores using optical microscopy and in situ Raman spectroscopy"*

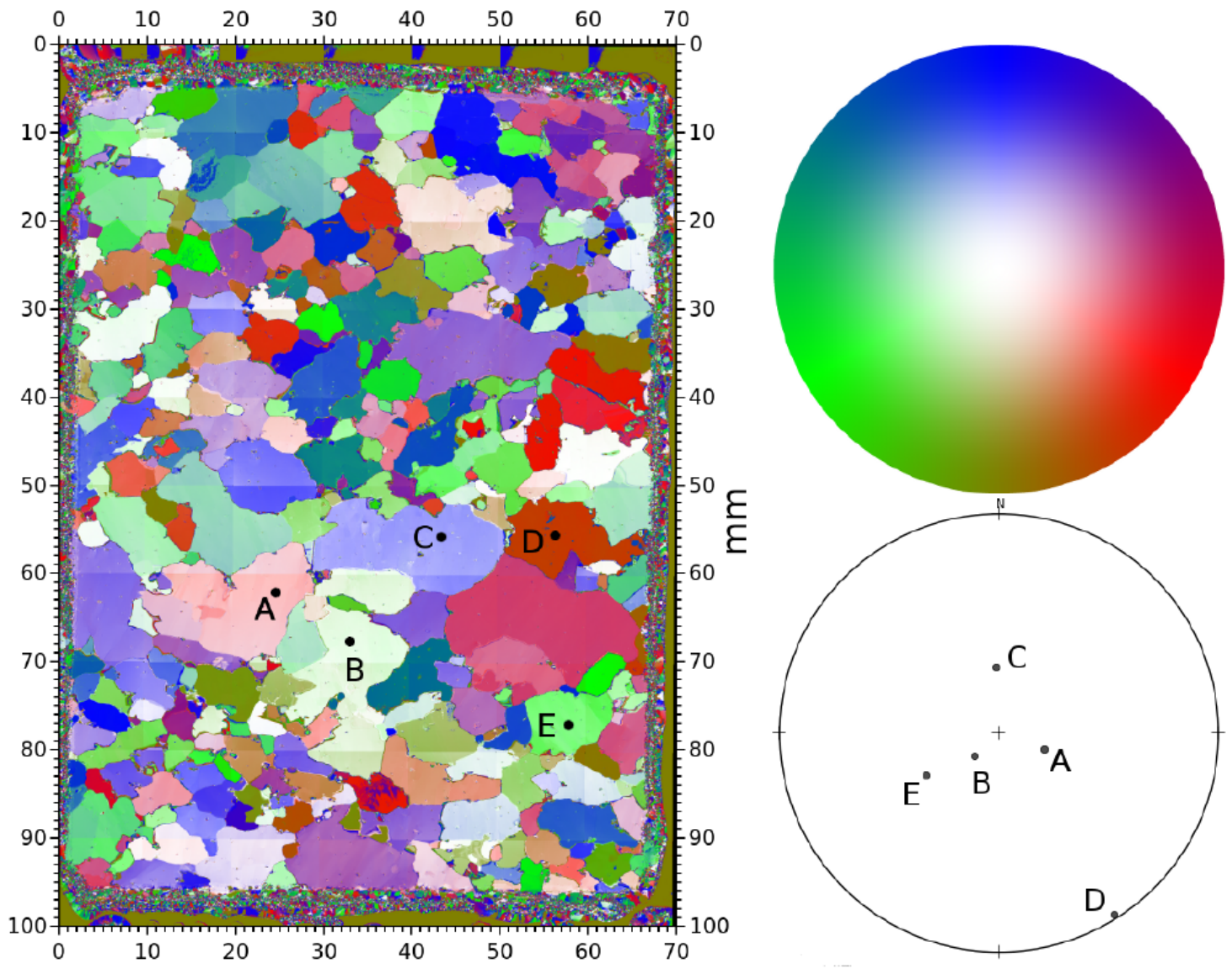

Figure 3.21: Fabric image obtained from the NEEM ice core at a depth of 784 m. Different c-axis orientations are color coded according to the sphere projection shown in the top right. The bottom right shows an example Schmidt diagram for the five selected grains A-E.
[Jan Eichler (Apr. 2013). "C-Axis Analysis of the NEEM Ice Core: An Approach based on Digital Image Processing." Freie Universität Berlin]

For ice the optical axis coincides with the crystal c-axis and the maximum relative difference in refractive index is roughly three permille over most of the optical spectrum[122]. The ordinary ray always propagates with the ordinary refractive index, while the refractive index of the extraordinary ray depends on the opening angle $\theta$ between the c-axis and the wave vector.

[122] *Petrenko and Whitworth*, Physics of Ice

Therefore the polarized light propagating through the ice sample gets split into two components in the base of the crystal system, which propagate at slightly different speeds and interfere behind the crystal. The resulting intensity measured behind the second polarizer is a function of the angle $\beta$ between the crystal c-axis and the initial polarization.

Just as in the analogous system of three polarizers, the intensity is zero whenever the c-axis is parallel to either of the two polarizer. This defines an extinction angle. The true three-dimensional orientation of the c-axis is thus obtained by repeatedly reorienting the crystal and rotating the polarizer system to find the resulting extinction angle. This process is nowadays being automated by so-called fabric analyzers[123], which test for all possible c-axis orientations.

[123] *Eichler, "C-Axis Analysis of the NEEM Ice Core: An Approach based on Digital Image Processing"*

An example of the resulting image can be seen in Figure 3.21, where each grain of the sample has been color-coded according to its c-axis orientation as defined in the color half-sphere in the top-right of the figure. Using the Lambert azimuthal equal-area projection, the orientation of an ensemble of grains can be visualized as shown in the lower-right of the figure.

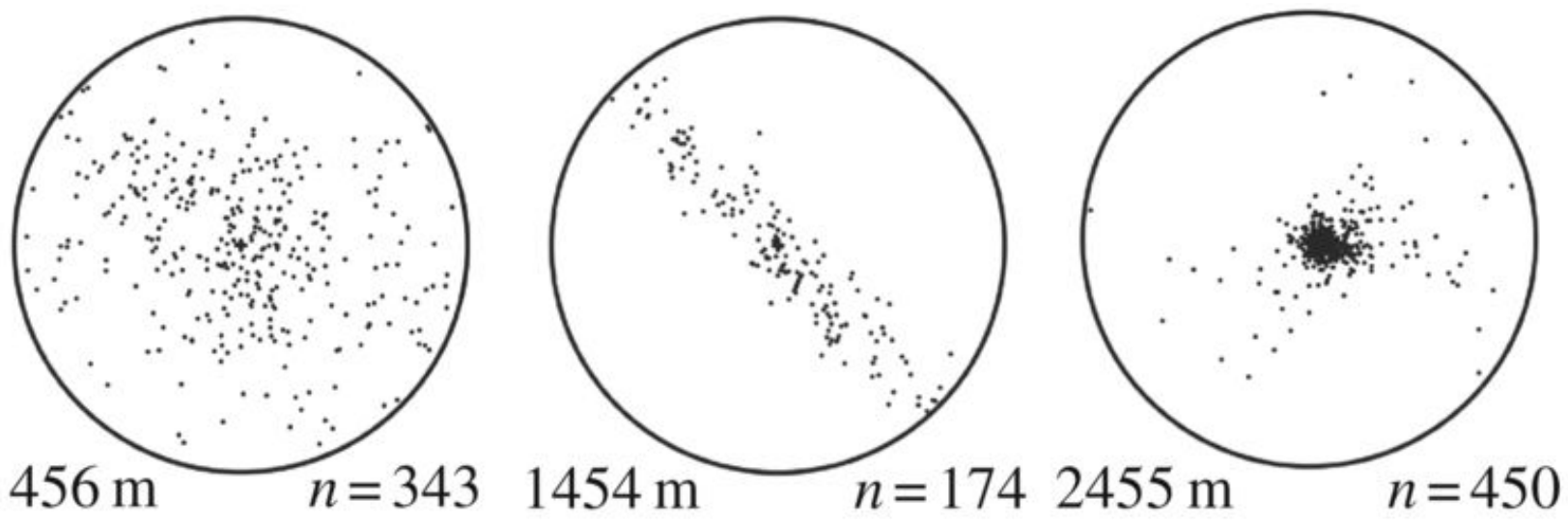

Figure 3.22: Typical LPO distributions along the EDML ice core. Random distribution in the shallow ice. Girdle in the deep ice and unimodal in the final 10% of the glacier.
[Ilka Weikusat et al. (2017). "Physical analysis of an Antarctic ice core—towards an integration of micro- and macrodynamics of polar ice"]

Such c-axis distribution plots are usually referred to as lattice-preferred orientation (LPO) diagrams. The orthogonal relation between elongation and c-axis orientation as expected from basal slip is generally found to hold (see Figure 3.23). C-axis measurements are more reliably compared to direct imaging of the grain shapes, which suffers from biases introduced through different viewing directions and irregular grain geometries. Thus, LPO diagrams are a commonly used tool to assess the ice strain and flow history.

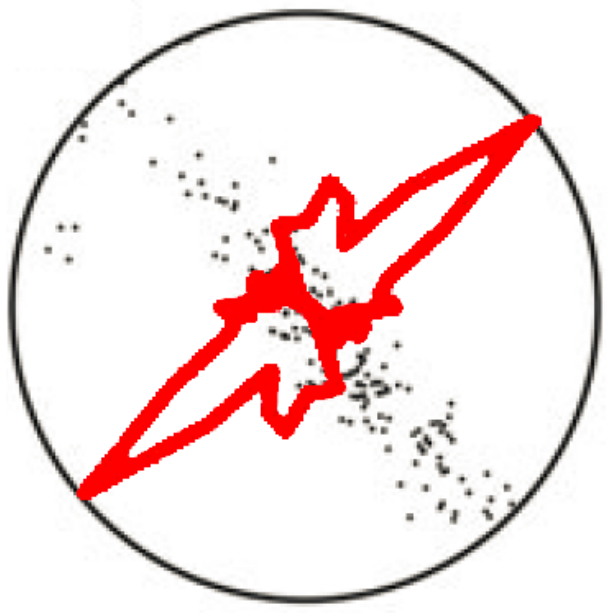

Figure 3.23: Correlation between grain elongations (red) and c-axis orientations (black dots) as measured in the EDML ice core at 1505 m depth. Adapted from: [Weikusat et al., "Physical analysis of an Antarctic ice core—towards an integration of micro- and macrodynamics of polar ice"]

LPO diagrams have a typical depth evolution as show in Figure 3.22. In the top third of the glacier, c-axes are distributed randomly. This slowly develops into a so called girdle fabric where most c-axes are found in a plane orthogonal to the flow. At 80% to 90% of the total depth of the glacier a second rapid transition to a single vertical cluster can be observed. This is commonly attributed to a transition from vertical compression with transverse extension to horizontal shear[124].

Figure 3.20 summarizes the described typical depth dependence of the grain size, stratigraphy and c-axis fabric using EDML data.

[124] *Faria et al., "The microstructure of polar ice. Part II: State of the art"*

4

# THE ICECUBE NEUTRINO OBSERVATORY

**This chapter describes the layout, instrumentation, and deployment of the IceCube detector. The characterization of each component will be discussed in the following chapter.**

The IceCube detector consists of deep, glacial ice as Cherenkov medium with supreme optical properties that has been instrumented with photo-sensors and electronics needed to digitize the signals.

## 4.1 PHOTOMULTIPLIERS AS PHOTODETECTORS

Photo-sensors to be used in a Cherenkov telescope need single photon detection capability with a large sensitive area and nanosecond timing resolution. Most commonly used devices are so called photomultipliers (PMTs).

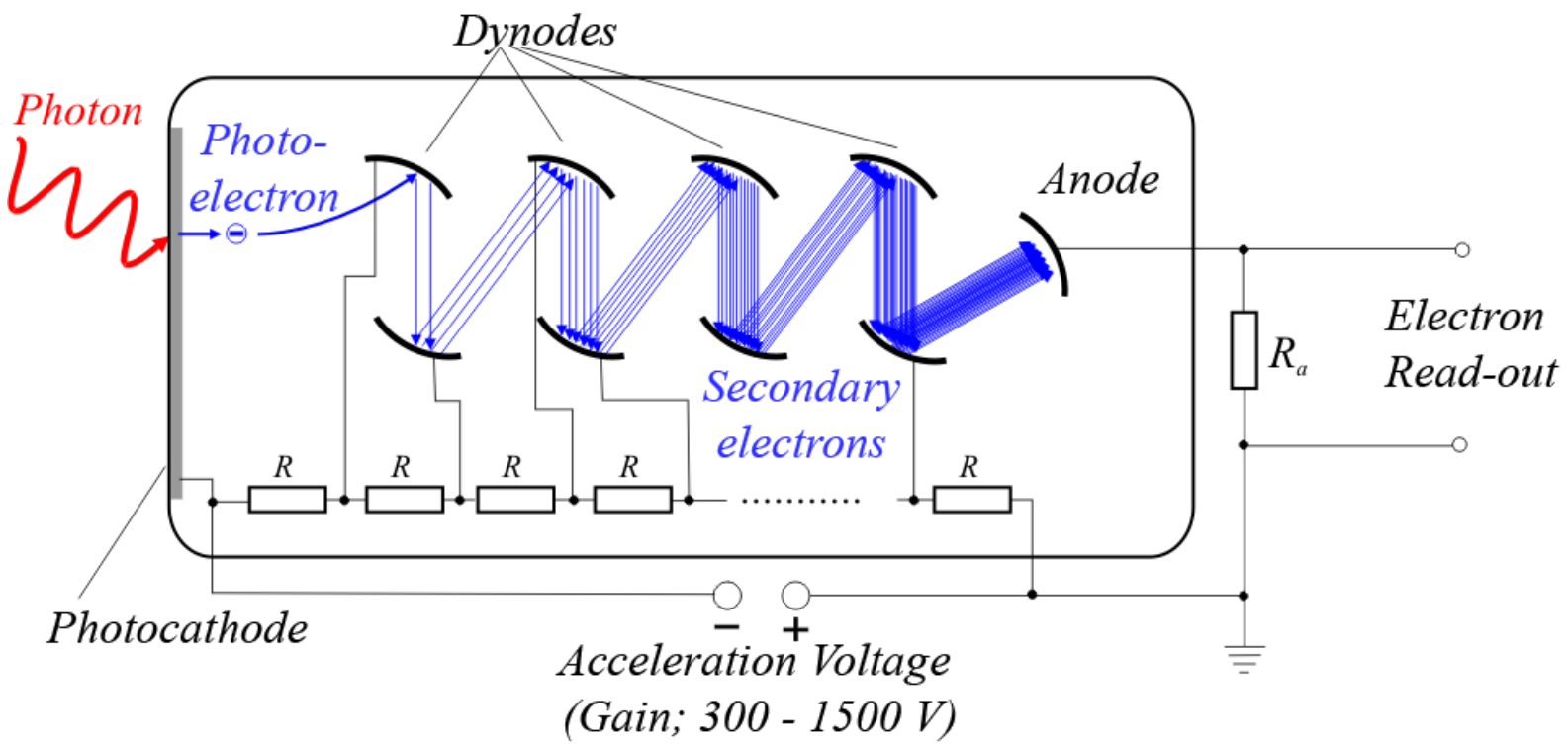

Figure 4.1: Typical assembly of a photomultiplier consisting of a photocathode and a series of metal dynodes on successively larger high voltage. A single detected photon results in a measurable current pulse at the last dynode.
[Wikimedia Commons (2018c). *Photomultiplier schema*]

A PMT consists of a photocathode and a series of metal plates on successively larger high voltage, called dynodes, inside an evacuated glass housing as sketched in Figure 4.1. A photon above a threshold energy may convert to an electron by the photoelectric effect in the photocathode. This so called photoelectron (PE) is accelerated towards the first dynode where it releases secondary electrons upon impact. This electron multiplication repeats at the following dynodes, resulting in a measurable current pulse at the final dynode, the PMT's anode[125].

In the following, the characteristics of the Hamamatsu R7081-02[126] PMT, as used in IceCube, are presented as a representative

[125] *Hamamatsu Photonics*, Photomultiplier Tubes: Basics and Applications

[126] *Hamamatsu Photonics*, Large Photocathode Area Photomultiplier Tubes

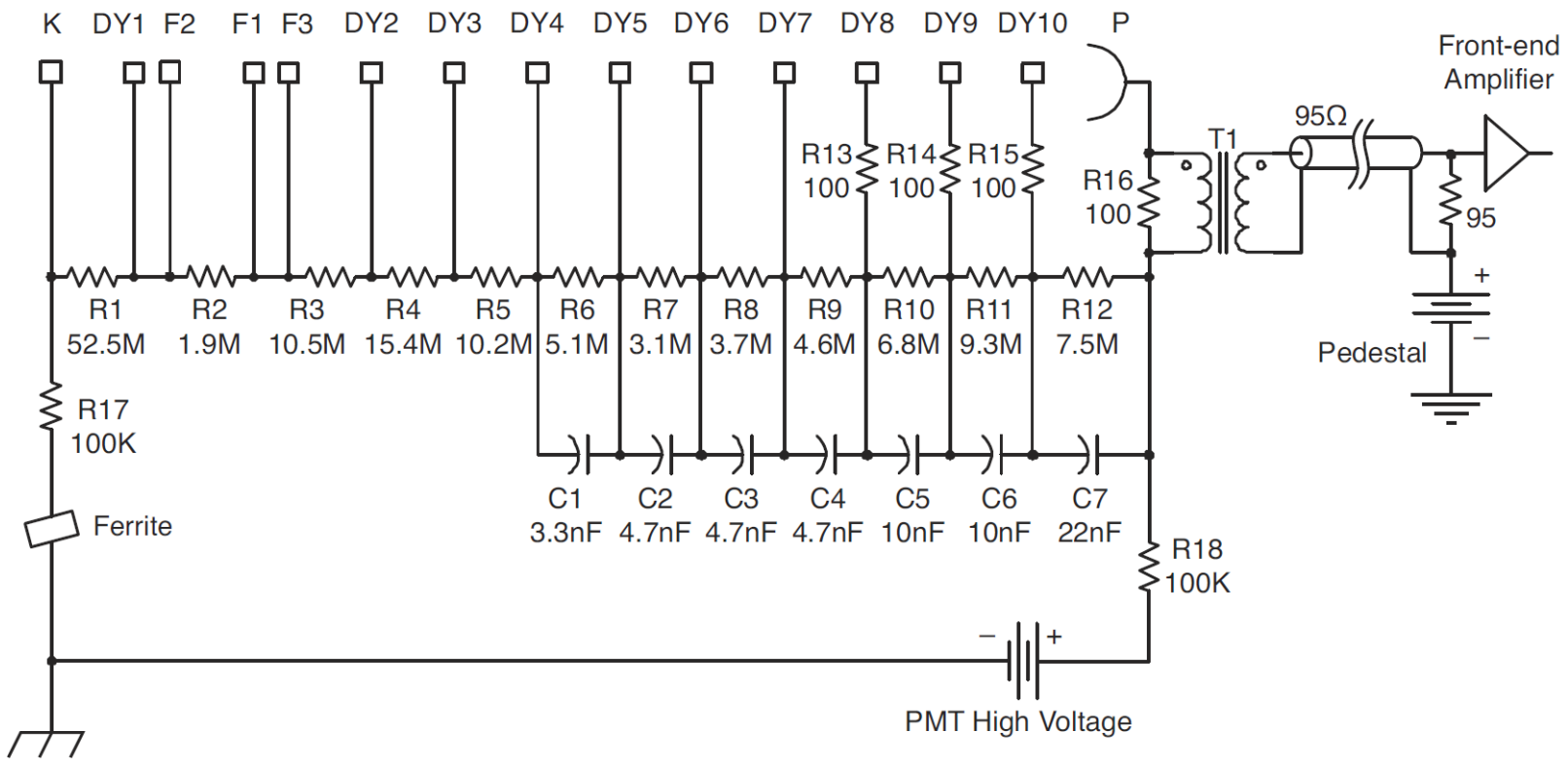

Figure 4.2: Schematic of the passive HV divider circuit and toroid coupling to the front-end amplifiers [R. Abbasi et al. (June 2010). "Calibration and characterization of the IceCube photomultiplier tube"]

example for large area PMTs. Some characteristics depend on the mode of operation. IceCube supplies the PMTs with a positive high voltage[127], using a resistive high voltage divider as shown in Figure 4.2. The RF PMT pulses are decoupled from the supply voltage using a bifilar-wound 1:1 toroidal transformer.

[127] *meaning that the photocathode is on ground potential*

As the trajectories of electrons in the PMT can be influenced by external magnetic fields, the IceCube PMT is shielded with a $\mu$-metal wire cage, which reduces magnetic fields to about a third of their strength[128].

[128] *R. Abbasi et al. (June 2010). "Calibration and characterization of the IceCube photomultiplier tube"*

### 4.1.1 *Detection efficiency*

The detection efficiency of PMTs is given by the quantum efficiency (QE) of the photoelectric effect at the photocathode times the collection efficiency (CE), which denotes the probability of the electron to be captured by the first dynode.

For the IceCube PMT at nominal gain, the collection efficiency is close to 100%[129], so that only the QE is considered in the following. The QE depends on the chosen photocathode alloy and thickness as well as the optical properties of the glass used for the PMT. Figure 4.3 depicts the wavelength dependence of the quantum efficiency of the IceCube PMT as measured by the manufacturer.

[129] *Abbasi et al., "Calibration and characterization of the IceCube photomultiplier tube"*

As the photon energy needs to be sufficient to release a photoelectron, the PMT is only sensitive for wavelengths smaller ~700 nm. It peaks around 400 nm before vanishing again around 300 nm due to strong absorption in the borosilicate PMT glass[130].

[130] *Hamamatsu Photonics,* Photomultiplier Tubes: Basics and Applications

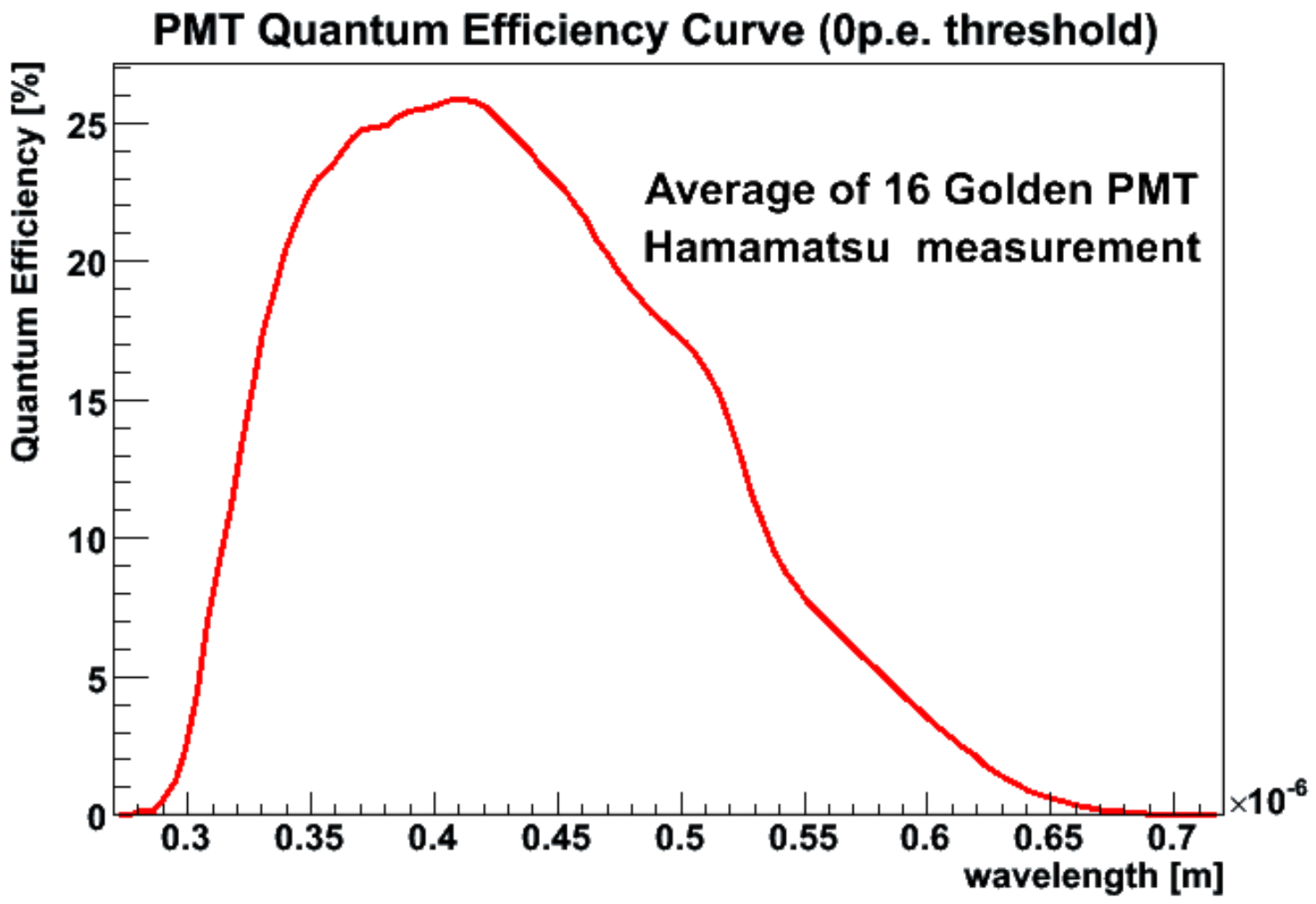

Figure 4.3: Average quantum efficiency of 16 IceCube PMTs as measured by Hamamatsu [Hamamatsu Photonics (2016). *Large Photocathode Area Photomultiplier Tubes*. Hamamatsu Photonics]

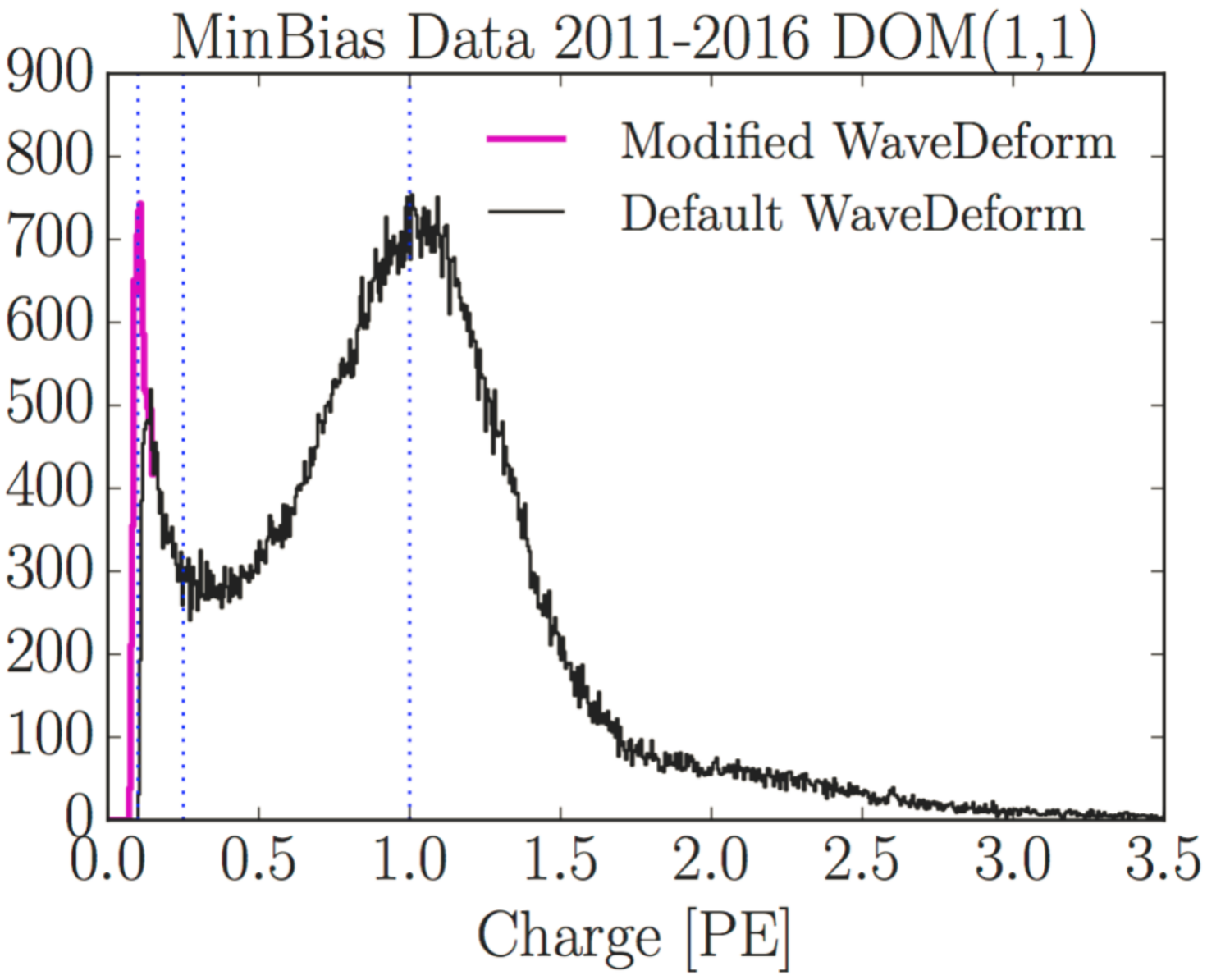

Figure 4.6: Example integrated charge distribution for a DOM illuminated with single photons. IceCube defines the unit photoelectron (PE) to be the charge of the mean of the Gaussian component. [M. G. Aartsen et al. (2019a). "In situ calibration of the single photoelectron charge response of the IceCube photomultipliers"]

## 4.1.2 *Gain characteristics*

The amplification at each dynode is empirically found to be proportional to $E^k$ with $E$ being the interstage voltage as set by the voltage divider[131] and $k$ typically ranging from 0.7 to 0.8. The total gain of the dynode system is the product of the amplifications at each dynode and as a result also follows a power law with an index proportional to the product of $k$ and the number of dynodes.

[131] *Hamamatsu Photonics*, Photomultiplier Tubes: Basics and Applications

IceCube nominally operates the PMT at a gain of $10^7$, which for single photons and a $100\,\Omega$ impedance results in a typical voltage waveform with an amplitude of roughly $8\,\mathrm{mV}$, as seen in Figure 4.5.

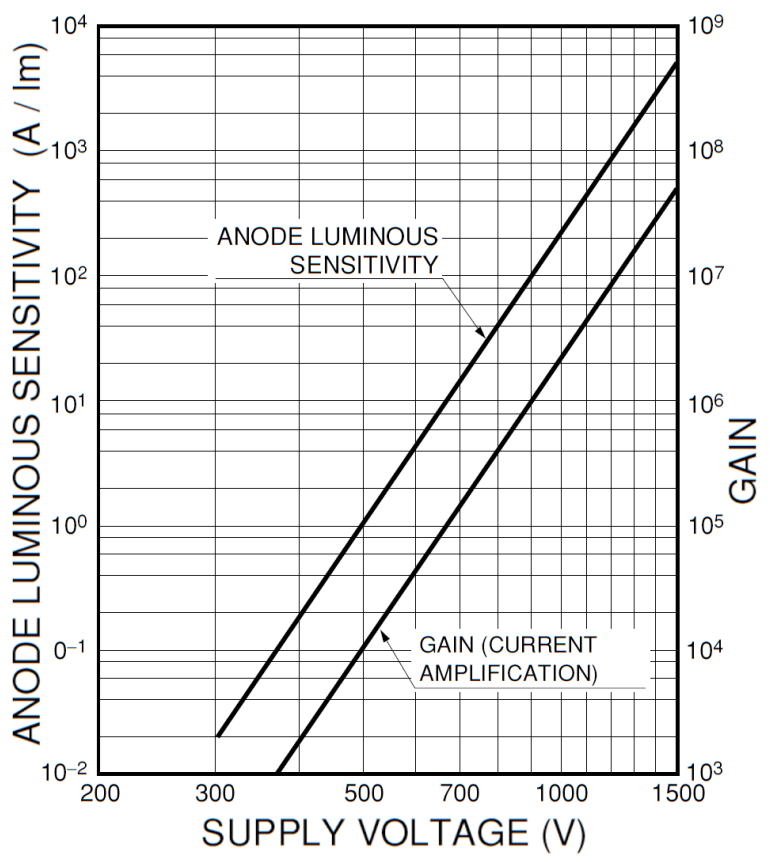

Figure 4.4: The general behavior of photomultiplier gain as a function of total applied high voltage follows a power law.
[Hamamatsu Photonics, *Photomultiplier Tubes: Basics and Applications*]

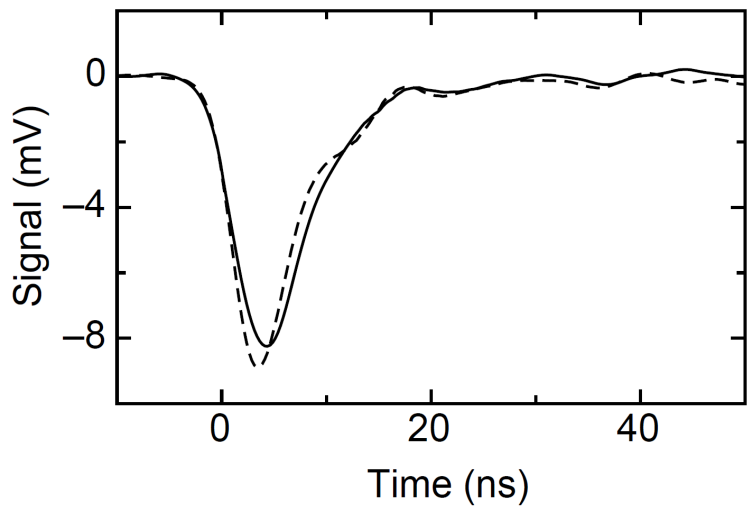

Figure 4.5: Average of 10,000 SPE waveforms at a gain of $10^7$, as seen at the secondary winding of the decoupling toroid. The solid and dashed curves correspond to new and old transformer designs.
[Abbasi et al., "Calibration and characterization of the IceCube photomultiplier tube"]

Since the electron multiplication at the dynodes is a statistic process, pulses from single photons follow a broad integrated charge distribution. The distribution for one of the in-ice sensors in shown in Figure 4.6. A Gaussian distribution with a roughly 30% standard deviation is seen for the nominally amplified pulses. In addition a steeply rising exponential at small charges is seen, which is caused by under-amplified pulses.

IceCube often classifies the received brightness in terms of the total charge delivered by a PMT. The unit photo-electron (PE) is defined to be the charge of the mean of the Gaussian component, although it is not the average charge of the single photon charge histogram.

The modeling and calibration of the per-PMT in-situ SPE charge distributions is discussed in further detail in appendix A.

### 4.1.3 *Timing characteristics*

The current pulse at the base is delayed with respect to the moment the photon is converted at the photocathode by the time the photoelectron requires to be accelerated to the first dynode and the resulting cascade requires to propagate through the subsequent dynode system. This so called transit-time scales as $1/\sqrt{\text{HV}}$ and is typically 55 ns for IceCube PMTs at nominal gain[132].

[132] *M. G. Aartsen et al. (2017c). "The IceCube Neutrino Observatory: instrumentation and online systems"*

The transit time is subject to a natural spread, the so called transit time spread, which limits the single photon timing resolution of PMTs. A histogram of recorded transit times with respect to a triggered light source is shown in Figure 4.7[133]. In addition to a $\sim$2 ns spread for photoelectrons following the nominal trajectory, a long tail terminating in a peak roughly 75 ns later is seen[134].

[133] *the offset is larger then the transit time by* $\sim$75 ns *due to the DOM delay board discussed in section 157*

[134] *The pedestal at around 10 counts results from random noise coincidences.*

These so called late-pulses are caused by electrons not undergoing full amplification at the first dynode, but rather being back-scattered towards the photocathode. The terminating peak corresponds to elastically scattered electrons and its time delay is approximately twice the travel time between the photocathode and the first dynode. The intermediate times correspond to inelastically scattered electrons.

The majority of all incident photons pass through the photocathode. For a small solid angle range, they may reach the first dynode and there convert to a photoelectron[135]. These so called pre-pulses arrive roughly 30 ns prior to the nominal pulses.

[135] *The photocathode is applied in an evaporation process after assembly, coating most of the PMT internal assemblies with photocathode material.*

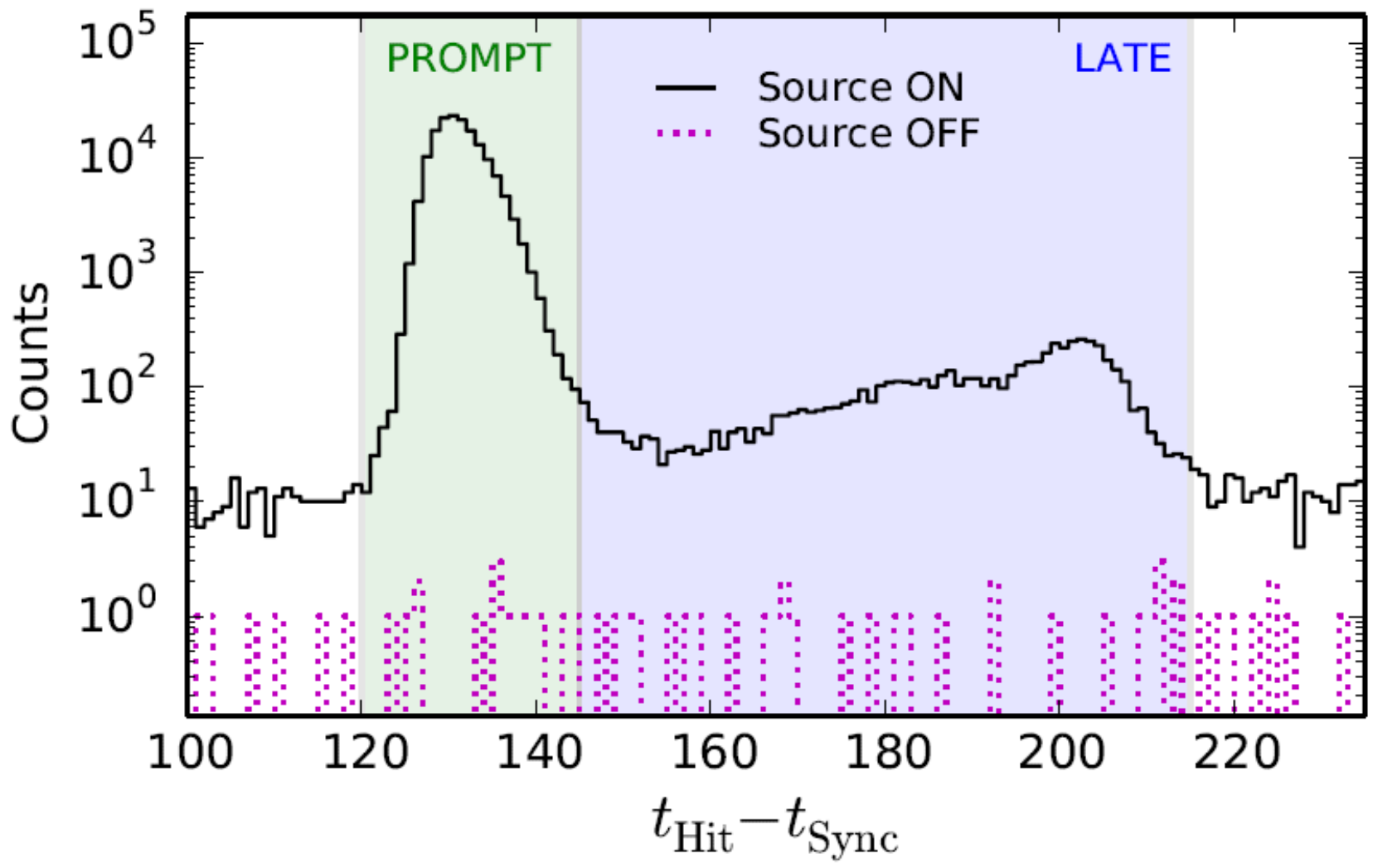

Figure 4.7: Timing distribution for nominal (prompt) pulses and late pulses, which have been elastically or inelastically scattered of the first dynode
[Chris Wendt (2018c). "private communication"]

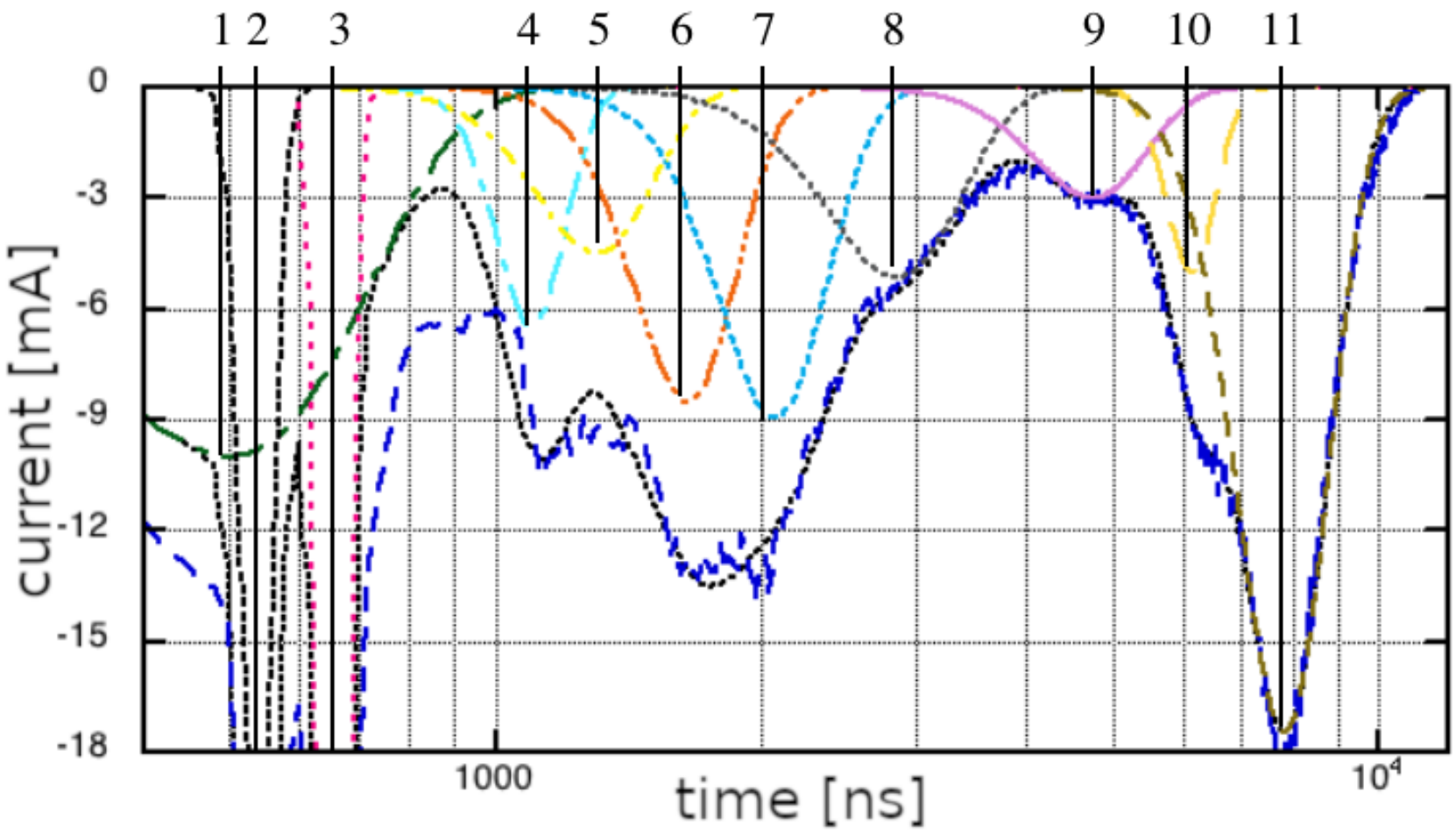

Figure 4.8: Afterpulse distribution arising from ions created close to the dynodes. The data is seen fitted by eleven Gaussians. The different contributions are attributed to different ion types or positions of origin.
[Chris Wendt et al. (2018). *Afterpulse data*]

Because they lack the amplification from the fist dynode, they are of very low amplitude. For the IceCube PMT the probability for a pulse to be a pre-pulse is estimated to be 0.3 % with an average charge of less than 0.05 PE[136].

[136] *Wendt, "PMT Charge Spectrum vs Time"*

Additional signals are also seen for several microseconds after illumination[137]. Electrons close to the first dynode may create ions from residual gas or dynode material. These slowly follow the inverse path of the electrons back to the photocathode, where they may create a number of electrons on impact[138]. The timing distribution of these so called afterpulses is seen in Figure 4.7.

[137] *Hamamatsu Photonics,* Photomultiplier Tubes: Basics and Applications

[138] *Ma et al., "Time and Amplitude of Afterpulse Measured with a Large Size Photomultiplier Tube"*

## 4.1.4 *Saturation*

The linearity of PMTs, that is the linearity of the output current relative to the rate of incident photons, is fundamentally limited by the resistivity of the photocathode and space charge effects close to the later dynodes[139]. For PMTs operated at gains

[139] *Hamamatsu Photonics,* Photomultiplier Tubes: Basics and Applications

required for single photon detection, the effects of the photocathode can be neglected.

Space charge effects are relevant but usually sub-dominant to the dynamic range of the employed PMT base[140]. As the electron cascade develops in the dynode chain, the dynodes may discharge faster than the voltage divider can supply charge. This effect can be mitigated by adding decoupling capacitors, in particular to the last dynodes where the largest current is required[141].

[140] *Hamamatsu Photonics*, Photomultiplier Tubes: Basics and Applications

[141] *Hamamatsu Photonics*, Photomultiplier Tubes: Basics and Applications

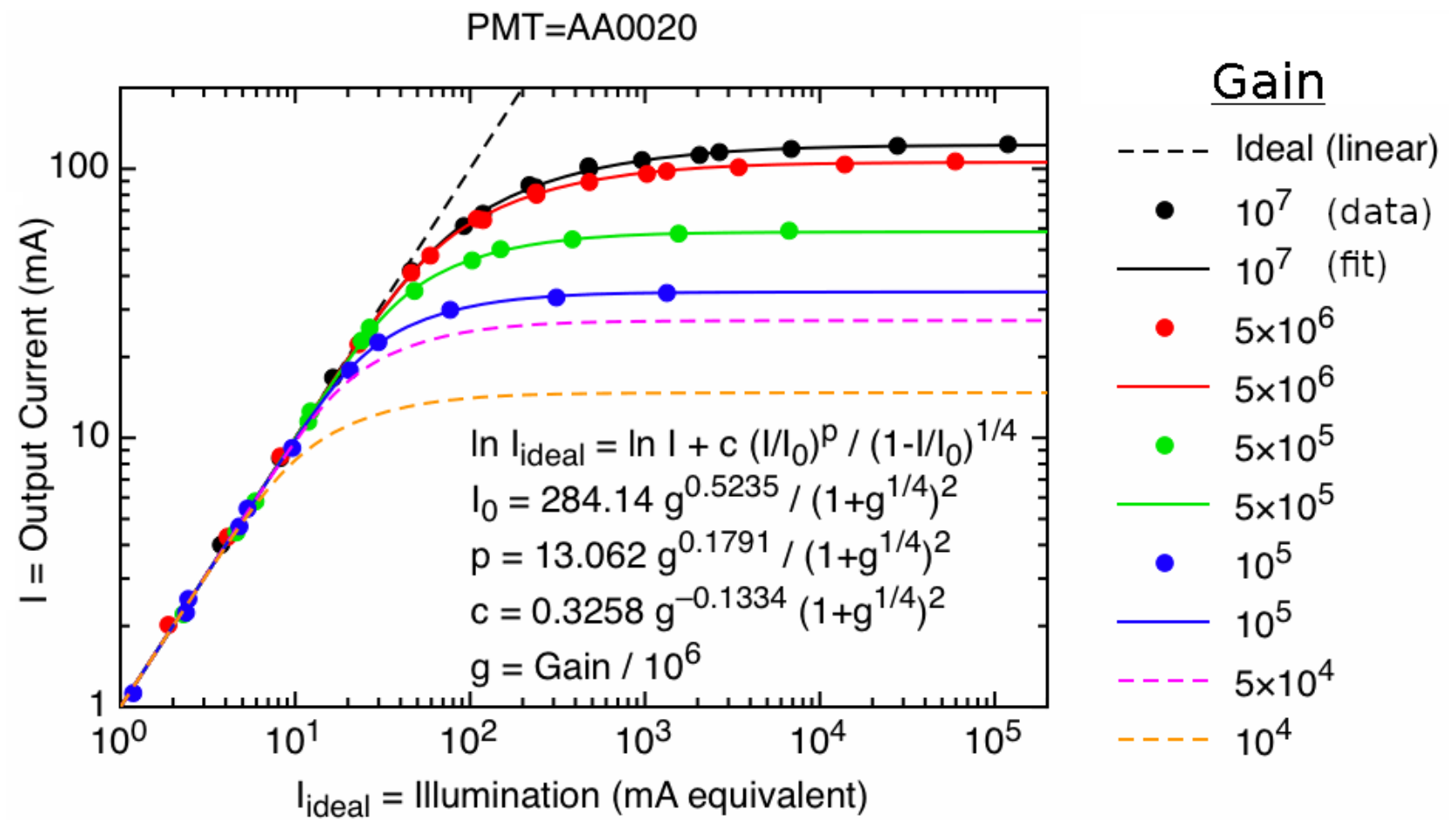

Figure 4.9: PMT saturation characteristics measured as the resulting instantaneous current versus the expected current at different gains.
Adapted from:
[Chris Wendt and Dahai Liu (2006). "PMT Saturation Lab, In-Ice, FAT"]

Figure 4.9 shows the linearity of an example IceCube PMT measured at gains between $10^5$ and $10^7$. The PMT was illuminated using a 200 ns, 410 nm LED pulse of variable intensity. The peak current at the base is calculated from the maximum of the resulting voltage pulse. The illumination where the resulting current deviates by 10% from the ideal expectation was found to be independent of the illumination duration as tested with pulse durations down to 3 ns and is also largely independent of the applied gain[142] .

[142] *Chris Wendt and Dahai Liu (2006). "PMT Saturation Lab, In-Ice, FAT"*

Therefore, PMTs operated at a reduced gain will saturate at proportionally larger light intensities. At a gain of $10^7$ the saturation level is equivalent to roughly 80 PE/ns.

## 4.2 DETECTOR LAYOUT

The IceCube Neutrino Observatory is located at the geographic South Pole, 1 km grid northwest of the Amundsen Scott South Pole Station. While other locations on the Antarctic plateau would also have been suitable in terms of ice quantity and quality, the South Pole offers the logistics required for this experiment[143].

*[143] Other permanently inhabited inland stations are the Russian Vostok station and the Italian Concordia station.*

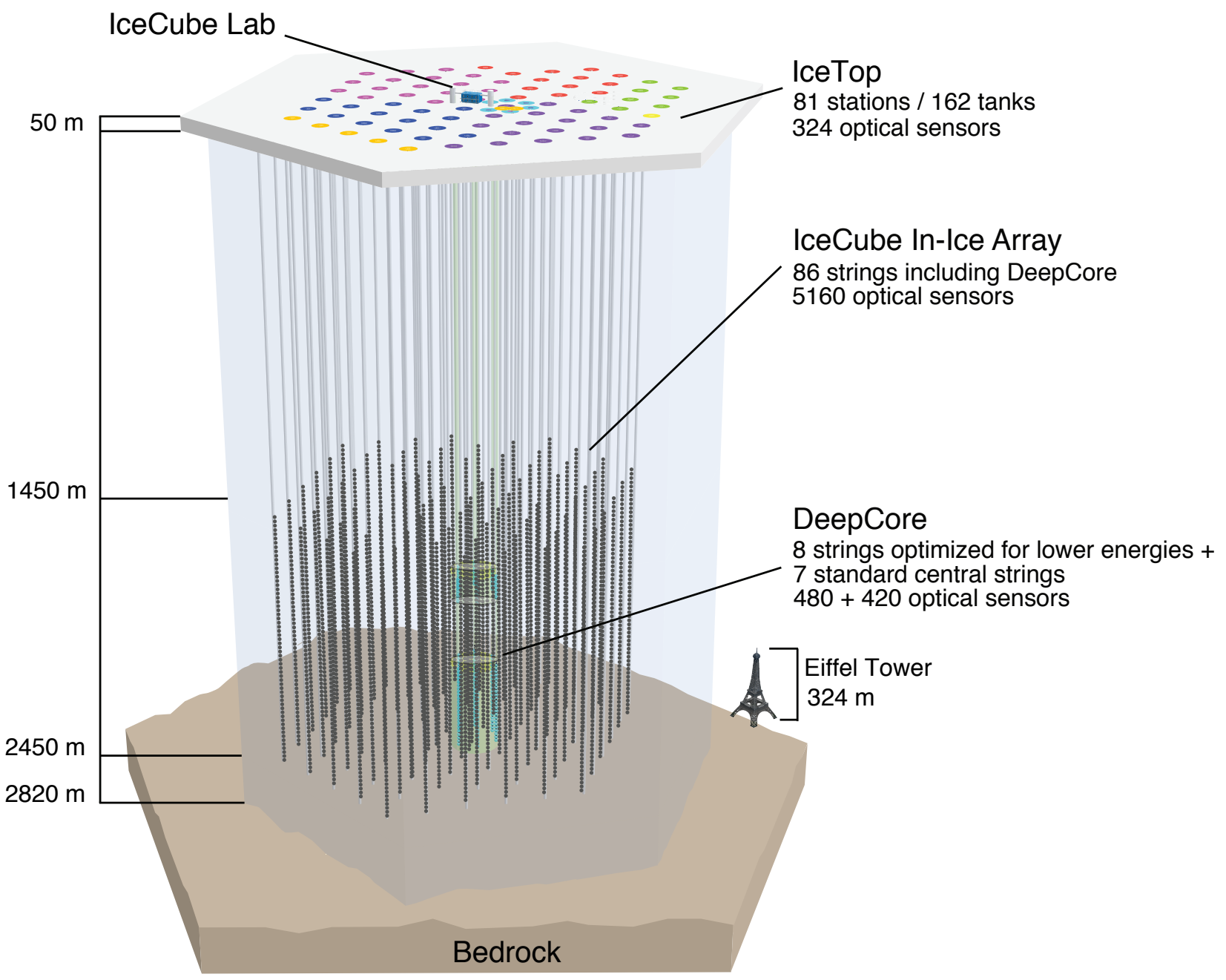

Figure 4.10: Overview of the IceCube detector array. Strings are indicated as gray lines, with black dots for DOMs. The different color of the IceTop stations indicate the year of deployment.
[M. G. Aartsen et al. (2017c). "The IceCube Neutrino Observatory: instrumentation and online systems"]

The in-ice detector consists of 86 cables[144] equipped with 60 Digital Optical Modules (DOMs) (see section 4.3.1) each. The strings reach a depth of ∼2500 m with DOMs instrumenting the last kilometer at an even spacing of 17 m. Strings are arranged on a hexagonal pattern spaced 125 m apart. This spacing results in an energy threshold of 100 GeV. The overall detector footprint covers ∼1 $km^2$.

*[144] Aartsen et al., "The IceCube Neutrino Observatory: instrumentation and online systems"*

A lower energy threshold of ∼5 GeV, as required for the measurement of atmospheric neutrino oscillations, is achieved with the DeepCore sub-detector[145]. It uses eight strings as a dense, central infill. Of those, six strings are equipped with PMTs of ∼ 35% higher quantum efficiency and the remaining two have mixed QE DOMs.
The average inter-string spacing between DeepCore strings is 72 m, about 1.5 times the effective scattering length at these depths. 50 DOMs are located in the cleanest ice below the dustlayer at a vertical spacing of 7 m. The other 10 DOMs form a veto cap above the dustlayer.

*[145] Abbasi et al., "The Design and Performance of IceCube DeepCore"*

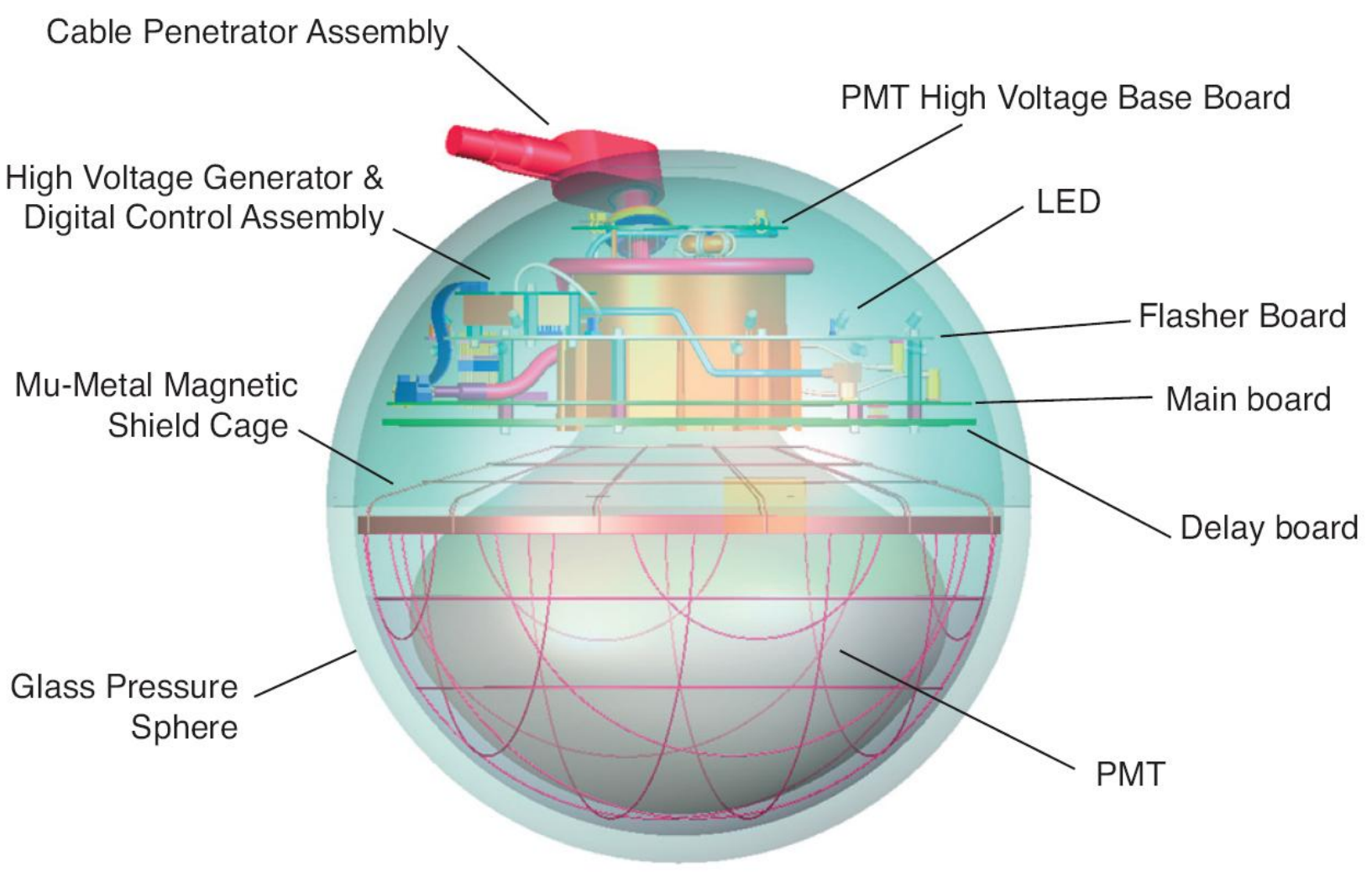

Figure 4.12: Sketch of a DOM showing all major sub-components.
[R. Abbasi et al. (2009). "The IceCube data acquisition system: Signal capture, digitization, and timestamping"]

In addition to the in-ice array, IceCube features an air shower array, IceTop[146], at the surface. It has 81 stations, each consisting of two tanks located near the location of a string, filled with $\sim$1.9 m$^3$ of artificial clear ice and equipped with two DOMs.

[146] *R. Abbasi et al. (Feb. 2013). "IceTop: The surface component of IceCube"*

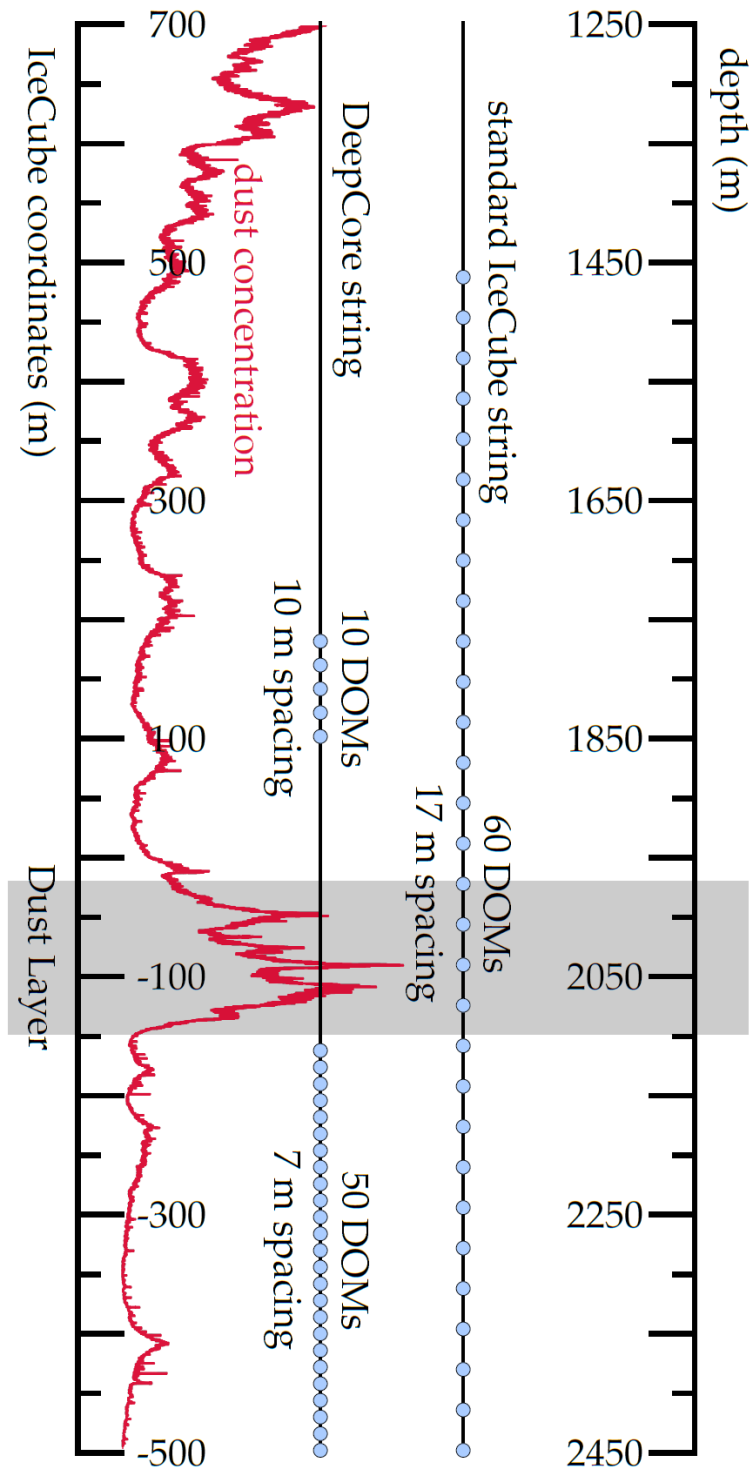

Figure 4.11: Vertical cross-section of the detector, showing both the sensor spacing, as well as the layered dust concentration of the instrumented ice.
[Euler, "Observation of oscillations of atmospheric neutrinos with the IceCube Neutrino Observatory"]

IceTop is sensitive to cosmic rays above 100 TeV, with a focus on the region around the knee. In conjunction with the in-ice array it has an excellent separation power between the muonic and non-muonic airshower components and can, within its limited solid angle coverage, also be utilized as a cosmic ray veto for IceCube. Due to snow accumulation IceTop continuously looses sensitivity to the electromagnetic shower component.

All cables merge at a central building called the IceCube laboratory (ICL), which primarily houses a server room. The ICL handles power and communication to the DOMs, local processing and storage and the satellite data transfer to data centers in the northern hemisphere.

## 4.3 INSTRUMENTATION

### 4.3.1 *The Digital Optical Module*

The IceCube DOMs are the fundamental building blocks of the detection array. Each DOM is an independent photon detection unit, containing a 10" Hamamatsu R7081 PMT, high voltage generation, full-waveform digitization, calibration LEDs, digital communication to the surface as well as neighboring DOMs, and all peripheral electronics required[147]. These are contained in a spherical glass pressure vessel.

[147] *R. Abbasi et al. (2009). "The IceCube data acquisition system: Signal capture, digitization, and timestamping"*

The DOM design was driven by the following considerations:

- Given that the drilling cost grows quickly with the hole diameter and the detector sensitivity scales with the photocathode area, a compromise between PMT and hole size has to be found[148].
- As the instrumentation is inaccessible after deployment, it has to be reliable over the planned operational lifespan of more than 25 years[149].
- Due to scattering in the ice complex arrival time distributions are expected. These require full waveform digitization to be able to achieve a good event reconstruction[150].
- Each DOM needs to cover a dynamic range from single photons to hundreds of instantaneous photons, with nanosecond timing resolution.
- Given the power limitations by the Amundsen-Scott South Pole station and the capabilities of the down-hole cables, each DOM can on average only consume ∼3.5 W.

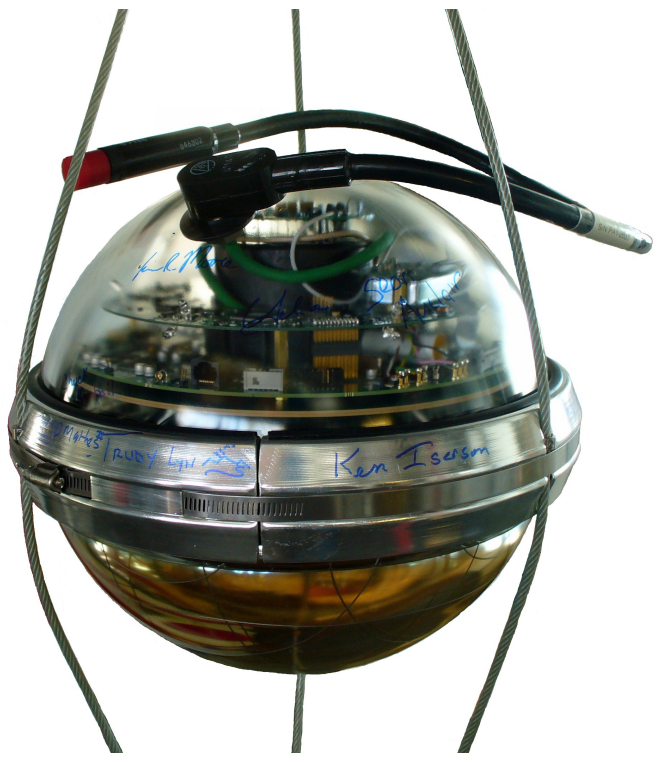

Figure 4.13: Photo of a fully assembled DOM.
[IceCube collaboration, "Internal graphics resource"]

The pressure vessel, photomultiplier, digitization and calibration LEDs are detailed in the following. For specific informations on other components please see [151].

#### 4.3.1.1 *The pressure vessel*

All components of a DOM are housed within a 33 cm diameter spherical glass pressure vessel composed of two symmetric half spheres[152]. The PMT is cemented into the lower hemisphere using an optical grade silicone gel. Both the gel and the glass sphere feature an optical transmittance larger than 90% above 400 nm. The PMT neck extends well into the upper hemisphere and is surrounded by the electronics boards.

Cables are fed through the upper hemisphere using a so called cable penetrator. After assembly, the sphere is sealed using a 0.5 bar nitrogen atmosphere. The penetrator and sphere are rated to be watertight at pressures of up to 690 bar, in order to withstand the hydrostatic and freeze-in pressure in the hole.

The DOMs are attached to the down-hole cable using an aluminum waistband, which runs around the equatorial joint of the two hemispheres. This waistband adds an additional 2 cm to the horizontal diameter.

148 *Ahrens et al.*, IceCube Preliminary Design Document Rev.1.24

149 *National-Science-Foundation*, IceCube Neutrino Observatory Funding FY2019

150 *Ahrens et al.*, IceCube Preliminary Design Document Rev.1.24

151 *Aartsen et al., "The IceCube Neutrino Observatory: instrumentation and online systems"*

152 *Aartsen et al., "The IceCube Neutrino Observatory: instrumentation and online systems"*

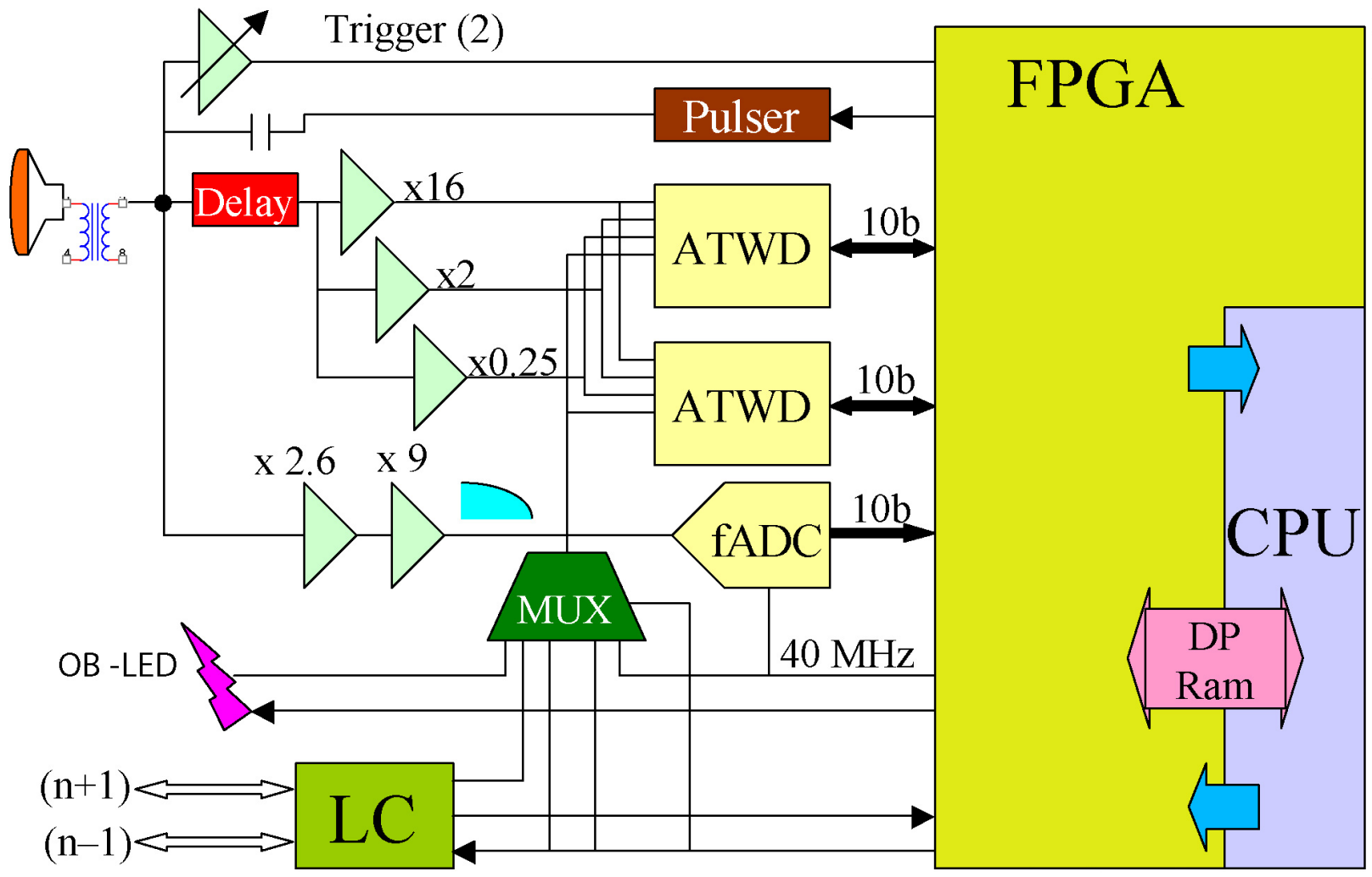

Figure 4.14: Block diagram of the DOM mainboard frontend showing the different ATWD and FADC amplification stages.
[R. Abbasi et al. (2009). "The IceCube data acquisition system: Signal capture, digitization, and timestamping"]

### 4.3.1.2 *The analog frontend and digitization*

The analog signal coming from the PMT base is digitized and processed in each DOM. Full waveform readout captures the entire extend of the event information, even after diffusion in the ice, and therefore allows for good event reconstruction.

As the power requirements for continuous ADCs at hundreds of mega-samples per second (Msps) exceeded the available power-budget when the mainboard was engineered[153], it was designed with triggered digitization. An overview of the system architecture is given in Figure 4.14.

[153] *Ahrens et al.,* IceCube Preliminary Design Document Rev.1.24

The primary ADCs are a pair of Analog Transient Waveform Digitizer (ATWD) chips[154]. These application specific integrated circuits, upon a trigger, sample the input voltage at 300 Msps into a 128 element deep switched capacitor array. Upon a later request, the sampled voltages can be sequentially digitized by a slow 10 bit ADC or they can be discarded.

[154] *S. Kleinfelder (Aug. 2003). "Gigahertz waveform sampling and digitization circuit design and implementation"*

To obtain precision information over the PMT's full dynamic range, from single photons to the saturation limit, each ATWD features three channels[155] . After amplification in the front-end these have least-significant-bit resolutions of 0.125 mV, 1 mV, and 8 mV, with saturation levels of 100 mV, 800 mV and 7.5 V.

[155] *and a fourth channel for auxiliary information from a multiplexer, such as LED current pulses*

Like any capacitive storage array the ATWDs show strong pedestals. These are non-zero fixed pattern waveforms measured with no signal present on the input and result from parasitic capacitances and other electronics artifacts. Figure 4.15 shows an example pedestal corrected baseline waveform, compared to a typical SPE waveform in the inset. The residual

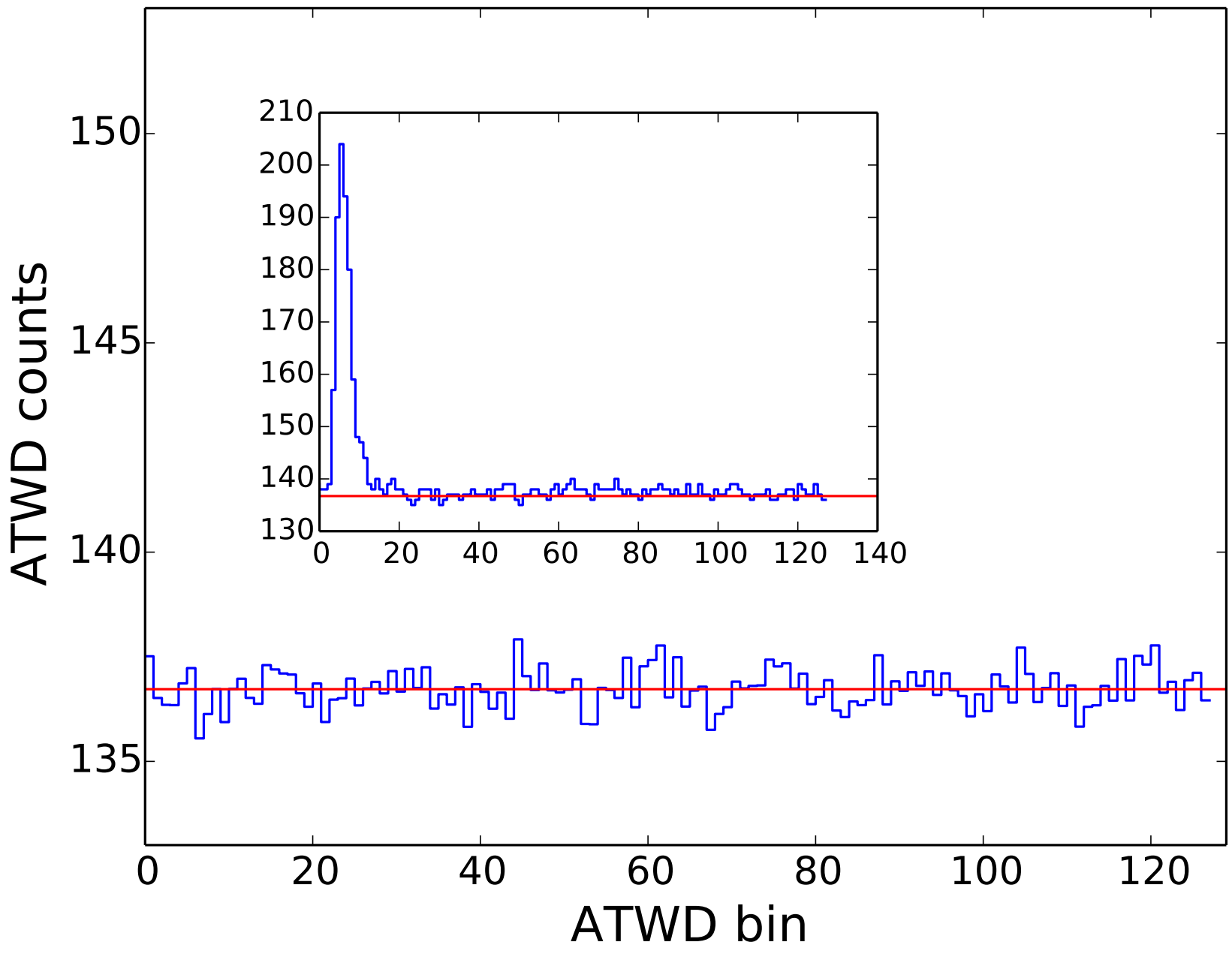

Figure 4.15: Averaged ATWD pedestal corrected baseline waveform. A typical SPE waveform is inset.
[M. G. Aartsen et al. (2017c). "The IceCube Neutrino Observatory: instrumentation and online systems"]

noise level is seen to be ~1 LSB.

The trigger for the waveform acquisition is provided by an analog comparator (see Figure 4.14), which for in-ice purposes is usually set to an amplitude corresponding to 0.25 PE[156]. As generating the trigger decision and communicating it to the ATWDs requires some time, the signal going to the ATWDs is delayed by ~ 75 ns using a dedicated analog delay board.

[156] *Aartsen et al., "The IceCube Neutrino Observatory: instrumentation and online systems"*

The trigger decision is also communicated to the neighboring DOMs up and down the same string, providing a so called local coincidence (LC) signal[157]. Only if a LC signal is received within ±1 µs of the own trigger, is the ATWD waveform digitized. This is a so called HLC (hard local coincidence) readout. As seen in Figure 4.16, the digitization of the stored values requires 29 µs per digitized amplification channel. Lower amplification channels are only digitized if the higher amplification channel saturates.

[157] *Aartsen et al., "The IceCube Neutrino Observatory: instrumentation and online systems"*

To extend the readout window beyond the 427 ns provided by the ATWDs, the mainboard also features a 40 MHz 10 bit FADC. The FADC samples continuously and can save waveforms at any time, with the length only being limited by memory availability.

In the current DOM firmware, the FADC is only launched in conjunction with an ATWD acquisition and saves samples for a fixed duration of 6.4 µs. To reduce dead-time the two ATWDs alternate and the idle ATWD can start a waveform acquisition

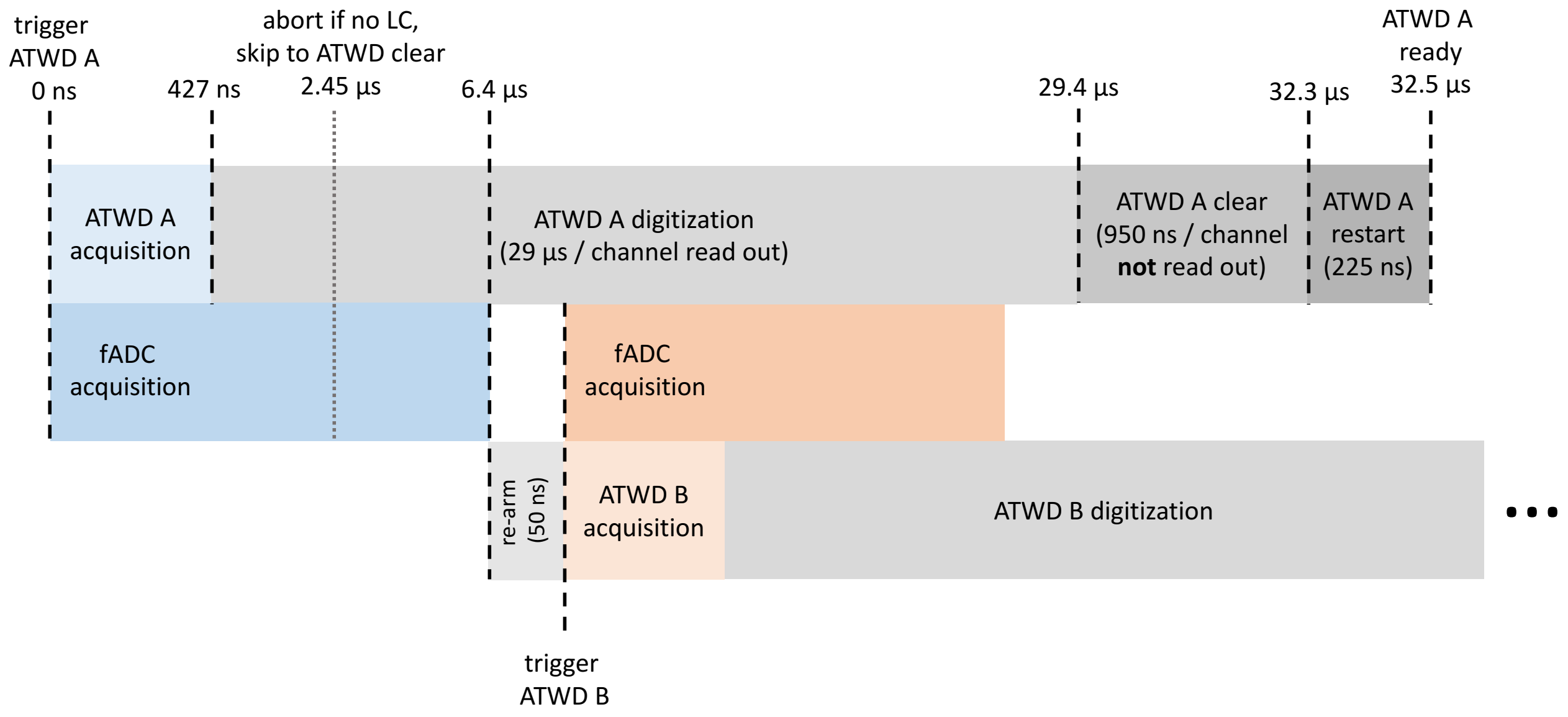

Figure 4.16: Timing of ATWD and fADC acquisition and associated deadtimes.
[M. G. Aartsen et al. (2017c). "The IceCube Neutrino Observatory: instrumentation and online systems"]

while the other ATWD still digitizes.

The FADC is also used when no coincidence signal is present. This is a so called SLC (soft local coincidence) readout. In this case no waveforms are sent to the surface. Instead, only a timestamp and three FADC samples around the maximum are recorded. For a full overview of the timing interplay between the ATWDs and the FADC see Figure 4.16.

Significant dead-time only occurs when both ATWDs are triggered in quick succession, with the maximum dead-time being 74.8 µs. The average fractional dead-time of in-ice DOMs[158] based on the fraction of comparator crossings when neither an ATWD nor the FADC was ready to digitize is $\sim 10^{-5}$.

158 *Aartsen et al., "The IceCube Neutrino Observatory: instrumentation and online systems"*

## 4.3.2 *Drilling and deployment*

### 4.3.2.1 *Drilling*

The IceCube deployment required the drilling of 86, 2500 m deep, 60 cm diameter holes within a few years. This was only possible through hot water drilling. 760 liters of water per minute were heated to $\sim 90\,°C$ and pumped down the drill hose to maintain a steady melting of at most 2.2 m per minute.

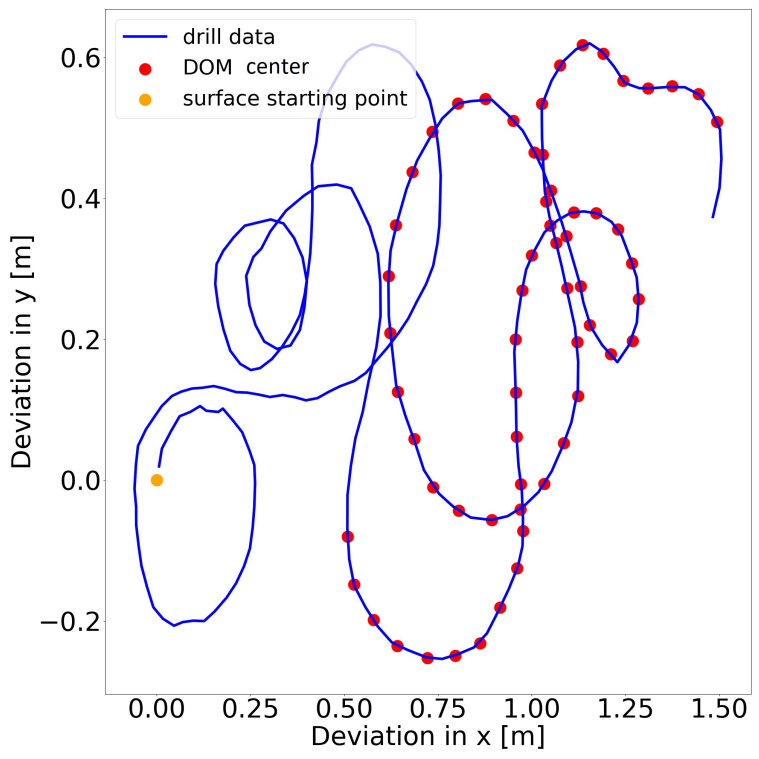

Figure 4.17: Example of the lateral drill head movement as reconstructed from the inertial measurement unit
[Peters, "Untersuchung des Einflusses der Unsicherheiten der Geometriekalibration auf die Richtungsrekonstruktion in IceCube"]

The drill head was equipped with an inertial measurement unit which enables reconstruction of the lateral displacement of the hole. The maximum deviation is typically within one meter, as exemplified in Figure 4.17.

The hole remained water filled and on retracting of the drill hose more hot water was added in a process called "reaming". This allows to obtain the desired hole radius, as monitored by a set of spring calipers on the end of the drill hose. More details on the drilling process can be found in [159].

[159] *Benson et al., "IceCube Enhanced Hot Water Drill functional description"*

#### 4.3.2.2 *Deployment and refreezing*

With a hole prepared, the drill moved to the next location and a separate deployment tower was moved in. As DOMs are buoyant, a 400 pound string weight was attached to the bottom of the down-hole cable before attaching a pair of DOMs to each of the 30 cable breakouts. To avoid the cable tilting or otherwise misplacing the DOMs, they were attached via a cable assembly, as seen in Figure 4.18.

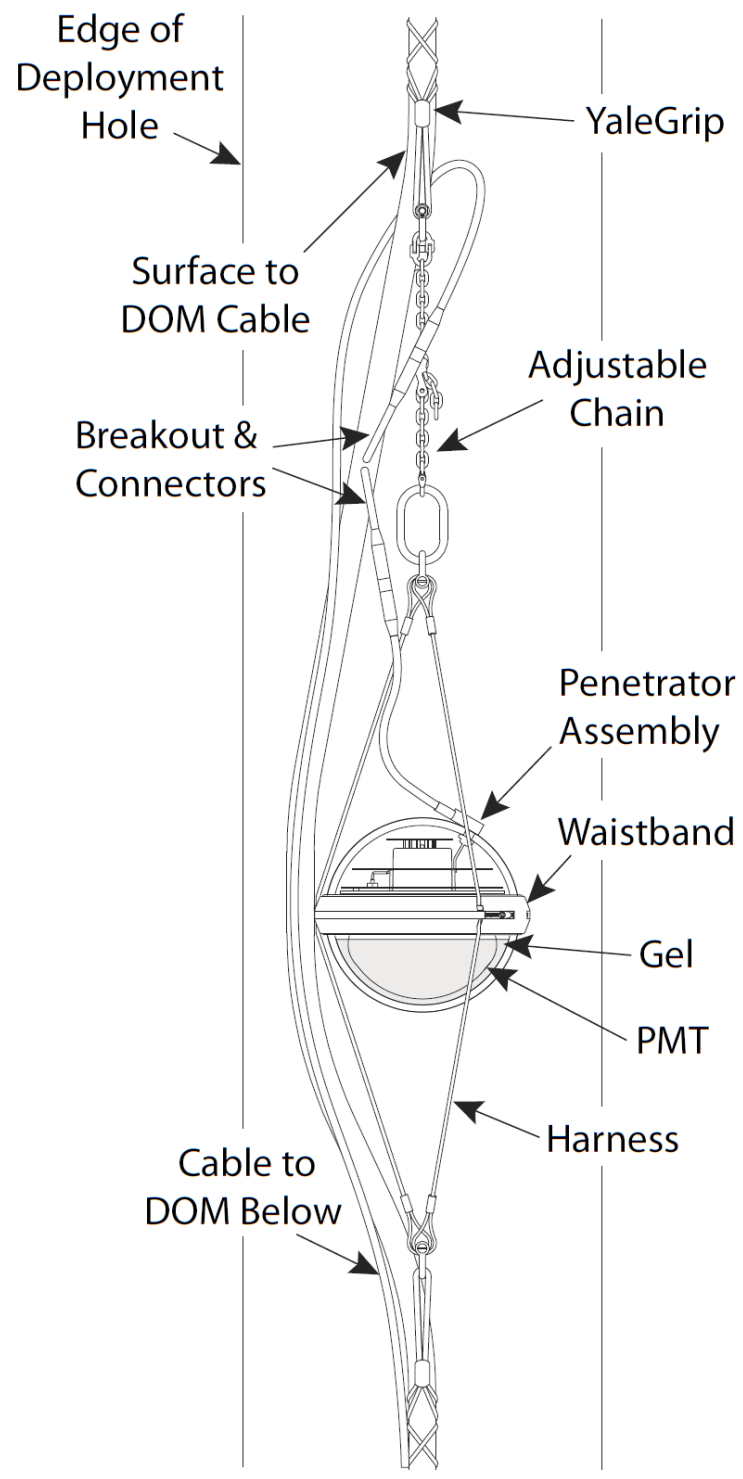

Figure 4.18: DOM cable assembly showing both the mechanical assembly as well as the cable breakout
[Aartsen et al., "The IceCube Neutrino Observatory: instrumentation and online systems"]

The tension is transferred via metal cables attached to the DOM harness, so the cable sits loose next to the DOM. In order to have a fixed azimuthal orientation of the cable at each DOM, it was zip-tied to a specific position of the harness. This translates to a specific orientation with respect to the flasher board.

In addition to the DOMs, two pressure sensors were installed on each string[160]. These were used to monitor the cable drop and to verify the final deployment depths. Since the pressure sensors only measure the extend of the water column, an additional measurement of the water level in each hole was needed to obtain the true depths under the snow surface.

[160] *Aartsen et al., "The IceCube Neutrino Observatory: instrumentation and online systems"*

After deployment the holes were left to refreeze. This typically took two to three weeks, with the hole freezing from the top down, as the ice closer to the surface is coldest.

Locally the freezing process is believed to be cylindrical, with the walls advancing towards the center. Freezing is an excellent purification process, with the freeze boundary pushing ahead impurities. These impurities, including vast amounts of air bubbles released from the glacial craigite, became trapped in the center of the hole, forming the so called bubble column. The characterization of the optical properties of this column and its impact on the DOMs acceptance is largely unknown and explored as in chapter 9 of this thesis.

In addition to the hydrostatic pressure of the water column, the freeze-in process results in over-pressures of up to 450 bar. Out of the 87 DOMs which have failed[161], 55 failed before the string was commissioned. It is suspected that in these cases water entered the pressure housing via the cable penetrator[162].

[161] *as of fall 2017*

[162] *Aartsen et al., "The IceCube Neutrino Observatory: instrumentation and online systems"*

## 4.4 DATA ACQUISITION AND PROCESSING

In the ICL, where the cables from all strings terminate, each string is associated with a dedicated server which handles all communication to these DOMs. These so called DOMhubs receive and cache all SLC and HLC hits generated by the DOMs for up to one week[163,164].

[163] *so called hitspooling*

[164] *Aartsen et al., "The IceCube Neutrino Observatory: instrumentation and online systems"*

The time-ordered series of HLC hits is sent to a trigger system, which identifies spatial and temporal clustering between the DOM, as expected from the different event topologies[165].

[165] *Aartsen et al., "The IceCube Neutrino Observatory: instrumentation and online systems"*

All hits in an at-least 10 μs window around identified trigger times are merged into events. This results in a raw-event rate of ~2.7 kHz, which is dominated by atmospheric muons. The total data rate of triggered events is ~1 TB per day. This is reduced to ~150 GB using a set of roughly 25 filters optimized to identify events of interest for different physics applications.

Events which pass the filters are sent to data centers in the northern hemisphere via satellite. In order to preserve bandwidth all calibration and reconstructions data is at that point discarded, to be recalculated in the North. At this processing stage the individual working groups and analyzers take over, applying their specific reconstructions and selections.

Events which are identified to have a large probability of being of astrophysical origin, generate so called alerts, where the event data generated at Pole is automatically sent to partner observatories around the world[166].

[166] *Aartsen et al., "The IceCube Neutrino Observatory: instrumentation and online systems"*

In addition to discrete hits, DOM noise rates are gathered and examined for temporal fluctuations as expected from the subthreshold interactions of many MeV neutrinos generated by a galactic supernova[167]. In case IceCube is informed of a supernova event by another observatory, the hitspool cache is saved and can be examined offline.

[167] *Lutz Köpke (Apr. 12, 2017). "Improved Detection of Supernovae with the IceCube Observatory"*

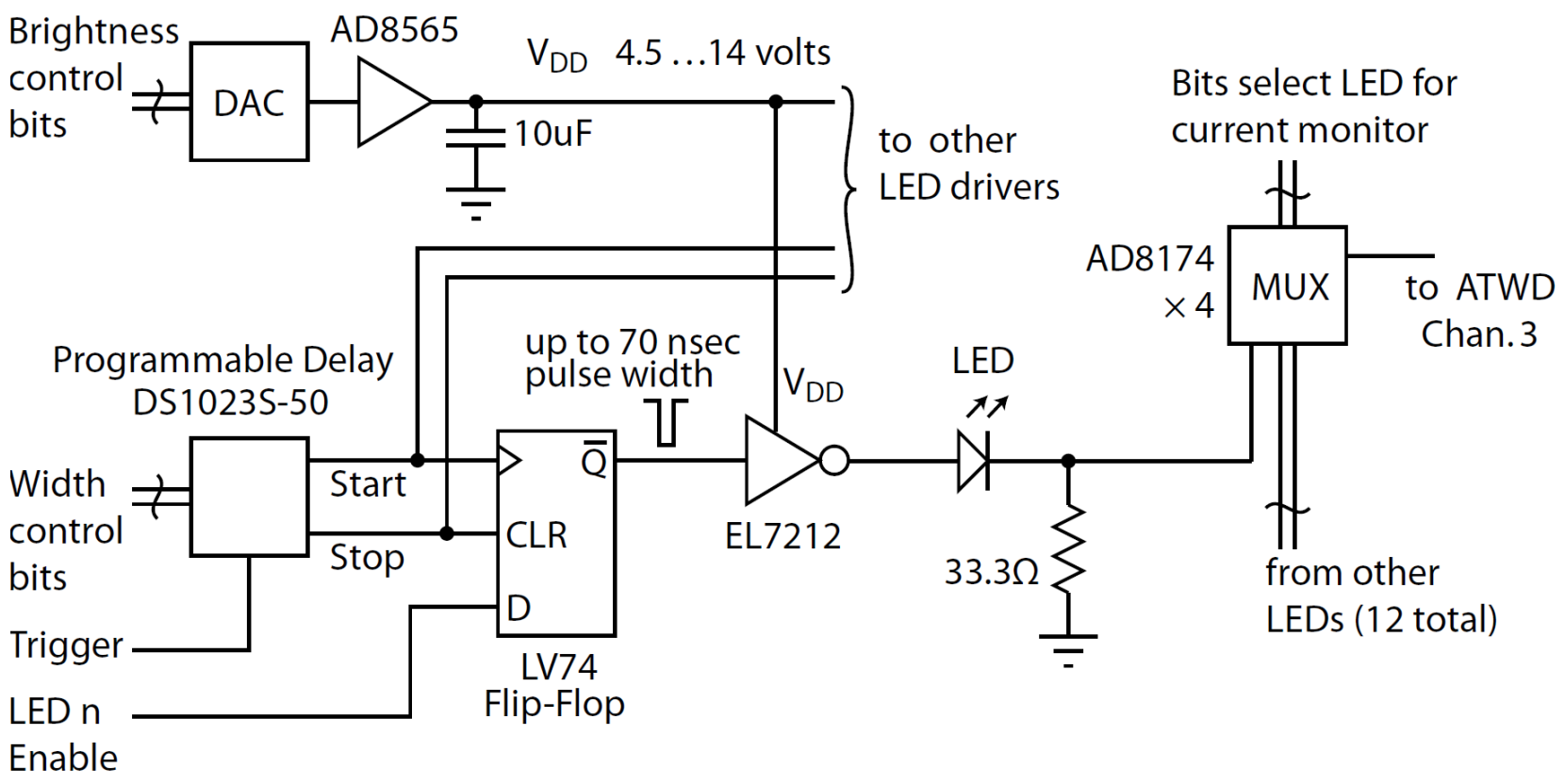

Figure 4.19: Simplified circuit diagram of the flasher board. The total light output can be varied through the pulse amplitude or pulse width. [M. G. Aartsen et al. (2017c). "The IceCube Neutrino Observatory: instrumentation and online systems"]

## 4.5 Calibration Devices

### 4.5.1 *The flasher board*

In addition to the electronics required to measure incoming light, each DOM can also emit controlled light flashes to be received by the rest of the array. This allows for the calibration of the DOMs positions, their timing response, the calibration of ice optical properties and can also be used to test the reconstruction performance of cascade-like events.

Every DOM has a dedicated flasher board, with twelve 405 nm LEDs[168], mounted above the mainboard (see Figure 4.12). The LEDs are arranged in outward pointing pairs, spaced 60° apart and located on the outer edge of the PCB[169]. The LEDs on the top of the PCB are bend to an angle of 50° above the plane, the lower LEDs are bend to an angle of ∼ 10° below the plane.

The angular emission profile of each LED has a Gaussian component with a standard deviation of approximately 13°. About 10% of the light is known to be emitted outside the Gaussian beam, with the exact profile still being under investigation and not included for the simulation[170]. At the air-glass boundary[171] the light is refracted, changing the emission angle and narrowing the angular profile in the ice to ∼10°. The light from the lower LEDs, called horizontal LEDs, is emitted into the ice nearly perfectly horizontally and the light from the top LEDs, called tilted LEDs, enters the ice at ∼48° above the horizontal[172].

During deployment, it was ensured that the down-hole cable was secured to the DOM at a specific azimuth orientation with respect to the flasher board[173]. This enabled a measurement of the absolute DOM orientations, as described on page 135.

[168] *except for 16 so called color-DOMs or cDOM with 340 nm, 370 nm, 450 nm and 505 nm LEDs*

[169] *Aartsen et al., "The IceCube Neutrino Observatory: instrumentation and online systems"*

[170] *Wendt and Tao, "Flasher LED Emission - Lab Studies"*

[171] *and to a lesser extend the glass-ice boundary*

[172] *Wendt,* LED angular emission profile

[173] *between LEDs 5 and 6*

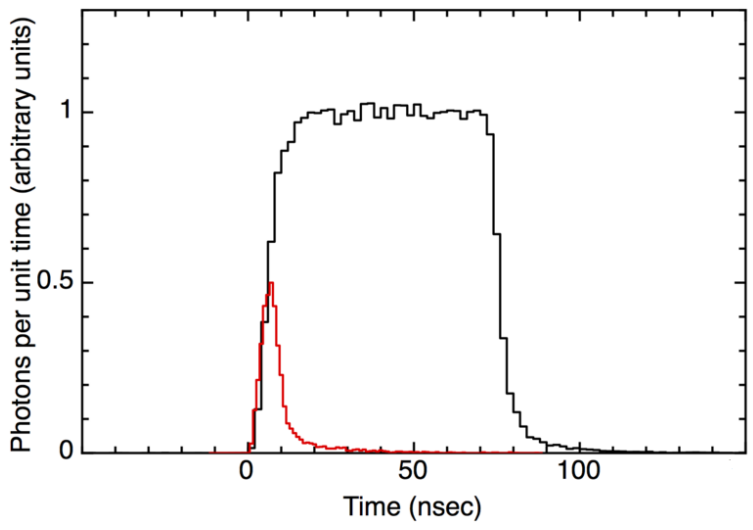

Figure 4.20: Flasher light curve at minimum and maximum width settings.
[Aartsen et al., "Measurement of South Pole ice transparency with the IceCube LED calibration system"]

While each LED has its own high-speed MOSFET driver and can be enabled individually, the pulse intensity and duration are set for all LEDs globally. Figure 4.19 shows the schematic diagram. The width of the light pulse can be adjusted between 5 ns and 70 ns (see Figure 4.20) using a programmable delay. The intensity is controlled via a variable voltage between 4.5 V and 14 V.

The maximum integral photon output is obtained at maximum brightness and width has been deduced from in-situ MC-to-flasher data comparisons to be $\sim 1.2 \cdot 10^{10}$ photons per LED. This is compatible with lab measurements performed on a few LEDs. The relative integral flasher output was found to depend on the brightness setting $B$ and width setting $W$ (both seven bit) as[174]

[174] *Chris Wendt (2018b).* LED light output

$$L = (0.0006753 + 0.000055593 \cdot B) \cdot (W + 13.9 - \frac{57.5}{1 + B/34.4}). \tag{4.1}$$

The current pulse through the LED is monitored using a spare channel on the ATWDs. It was found that there is a time offset of roughly 8 ns between the light flash and the current pulse[175]. This offset has to be fitted as a nuisance parameter in flasher fits and was found to depend slightly on the DOM production and deployment date.

[175] *Dawn Williams and othters (n.d.).* Flasher Timing Offset. *IceCube Wiki*

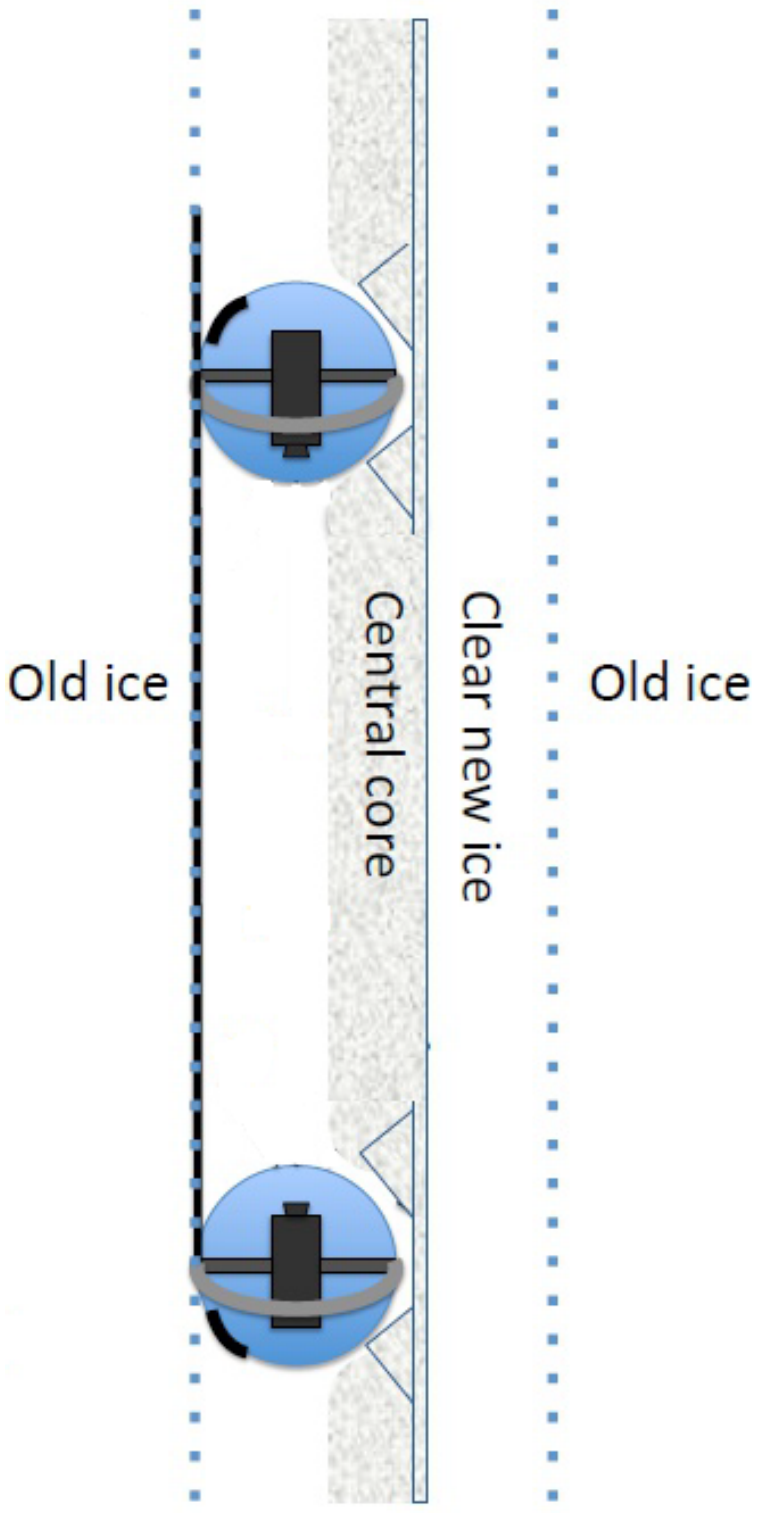

Figure 4.21: Illustration of the Sweden Camera assembly after deployment.
[Hulth, "Sweden Camera Special Device Review"]

The utility of the flashers in fitting the ice optical properties is discussed in the following chapter.

## 4.5.2 *The Sweden Cameras*

The optical quality of the refrozen drill holes has been a major source of uncertainty ever since AMANDA. To obtain a better visual understanding of the in-situ properties, cameras have been deployed into the holes.

The AMANDA camera[176], deployed to roughly 2270 m, observed the wall smoothness, triboluminescence sparks of freezing water and the cylindrical advance of the freezing hole walls. It failed after 9 days, prior to the hole being completely frozen.

[176] *footage is available at youtube.com/watch?v=hiJVajBA9TQ*

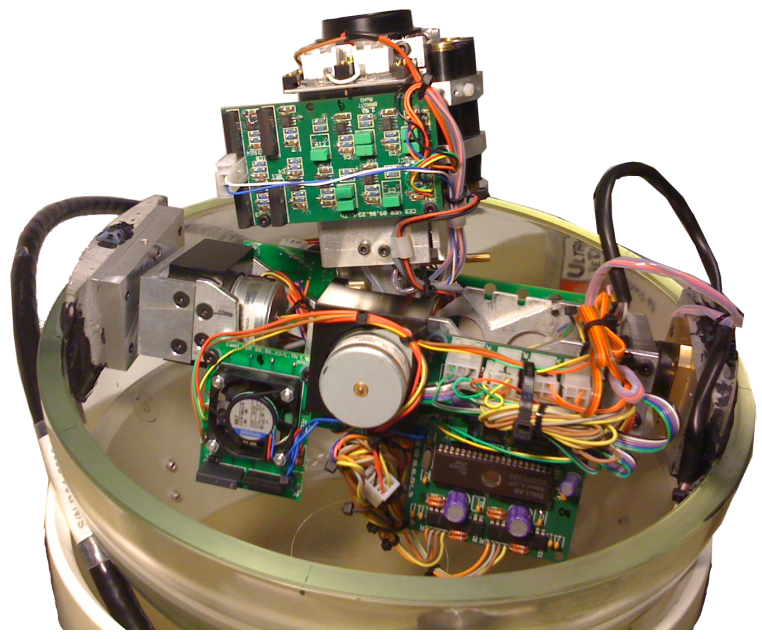

Figure 4.22: One of the Sweden Cameras during assembly.
[Hulth, "Sweden Camera Special Device Review"]

In IceCube, a first pair of stationary cameras, supposed to be looking straight up and down the hole, was deployed in 2007 but failed during deployment. A second attempt was made in 2010, during the deployment of string 80.

The system, developed by the University of Stockholm, Sweden, and called Sweden Camera, consists of two video cameras contained in two glass pressure vessels, identical to those used for the DOMs. They are situated ∼6 m apart, at a depth of roughly 2455 m, between the last DOM and the string weight[177].

[177] *Aartsen et al., "The IceCube Neutrino Observatory: instrumentation and online systems"*

The cameras are mounted on two independent rotation axes (see Figure 4.22), enabling a near $4\pi$ coverage. In addition to the camera, four white LEDs and a set of red, green and blue lasers is incorporated in each pressure vessel. Analog video is transmitted to the surface, where an independent control and digitization system is available.

Both cameras remain operational to date, but as of 2018 both have at least one failed rotation axis, severely limiting their utility. Results from the cameras are discussed in chapter 9.

### 4.5.3 *Dust loggers*

At deployment, no ice core was available to determine the stratigraphy at the depth of the detector. For this purpose, dust loggers[178] were deployed into IceCube holes. In essence, the flasher fits described later only add an absolute scale to the relative layering observed by these devices.

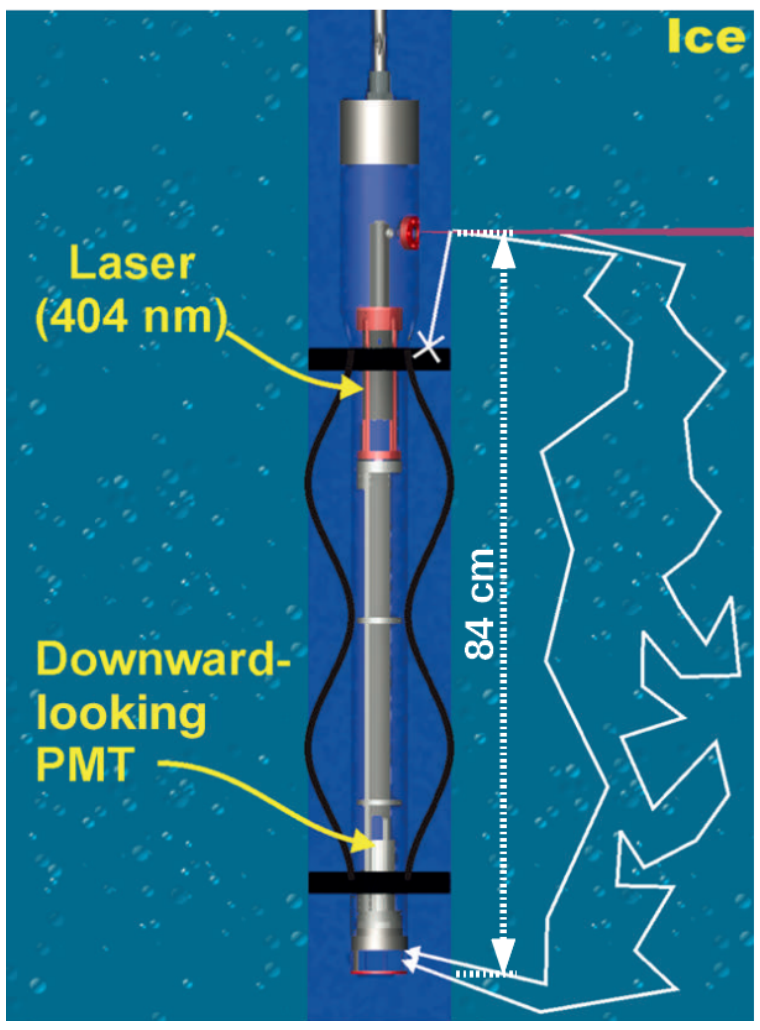

Figure 4.23: Schematic view of the dust logger concept. Light emitted in a horizontal fan can only reach the PMT below through scattering on nearby impurities.
[Aartsen et al., "South Pole glacial climate reconstruction from multi-borehole laser particulate stratigraphy"]

[178] *Aartsen et al., "South Pole glacial climate reconstruction from multi-borehole laser particulate stratigraphy"*

A dust logger consists of a 404 nm laser line source, emitting a 2 mm thin, horizontal fan of light about 60° across. A small fraction ($10^{-10}$ to $10^{-6}$) of photons is back-scattered on impurities and returns to the bottom section of the dust logger where a Hamamatsu 1" photon-counter module is located.

The intensity of the laser can be adjusted throughout the logging process. This allows to keep the count rate within the dynamic range of the photon counter, even in strongly changing conditions. To avoid stray light contamination from reflections on the water-ice interface, two sets of black nylon baffles are attached to the side of the pressure housing. They, in addition to a set of spring-loaded calipers, also help to keep the logger centered in the hole. The depth of the payload is monitored through the cable payout and on-board pressure sensors.

A total of eight IceCube holes were logged. In two cases, a dust logger was attached to the bottom of the string, deployed with it and left in the hole. For the other six holes, the dust logger was deployed and recovered with a dedicated winch, yielding logs on both the way down and up.

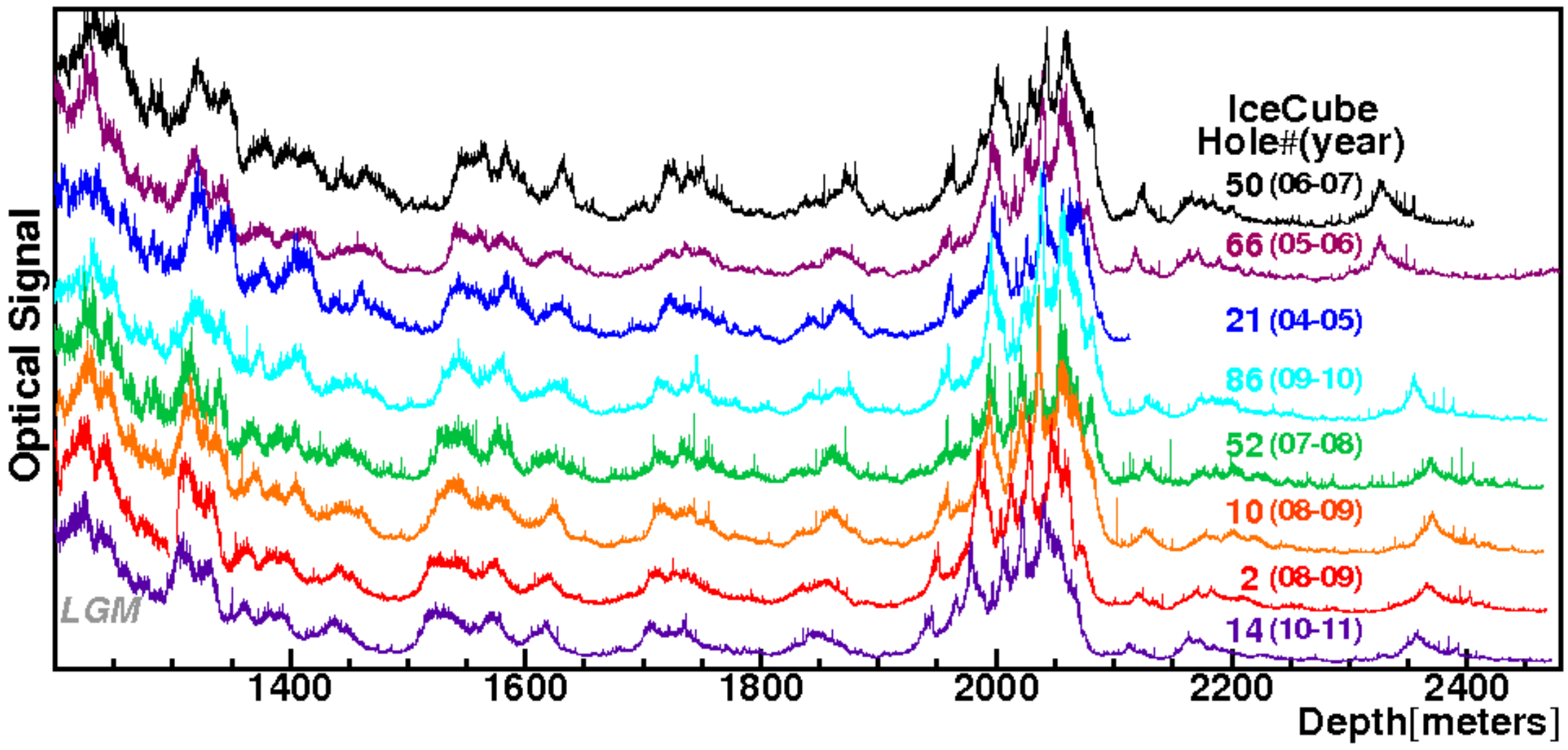

Figure 4.24: Particulate stratigraphy as obtained from the 8 dust logs. The dust layer created by one of the last stadials (compare to figure 3.5 at 60 ka) is seen at around 2100 m. Layer tilting can be mapped by matching volcanic lines.
[M. G. Aartsen et al. (2013d). "South Pole glacial climate reconstruction from multi-borehole laser particulate stratigraphy"]

Logging several holes over the surface footprint of the array has provided information on the layer tilting due to the underlying bedrock. These logs also provide the highest resolution particulate stratigraphy in Antarctica. A zoom-in into the IceCube depths is given in Figure 4.24.

Air bubbles start being significant for depths shallower than 1300 m. At greater depths, note the periodic increase and decrease of dust concentrations as associated with the Milankovitch Cycles of the axis precession[179]. An order of magnitude increase in impurities is seen in the so called dust layer at ~2000 m which results from one the last stadials.

[179] *see section 3.2*

While exactly aligned in the shallow ice, millimeter sized volcanic dust layers can be offset by several dozen meters between different logging locations in the deep ice. From these offsets a three-dimensional map of layer tilts as described in section 3.5 has been inter-/extrapolated. The resulting gradients, as presented in Figure 4.25, are mostly orthogonal to the measured surface ice flow direction, confirming the tie to the bedrock topology.

Since its application in IceCube, the dust logger has seen use in a number of coring holes around the world. It has recently been upgraded with an optional add-on magnetometer to obtain precision orientation information throughout the log and has been deployed down the SpiceCore[180] hole. This has enabled a view of the ice anisotropy complementary to the one provided by flashers, as discussed in section 10.9.

[180] *K.A. Casey et al. (2014). "The 1500 m South Pole ice core: recovering a 40 ka environmental record"*

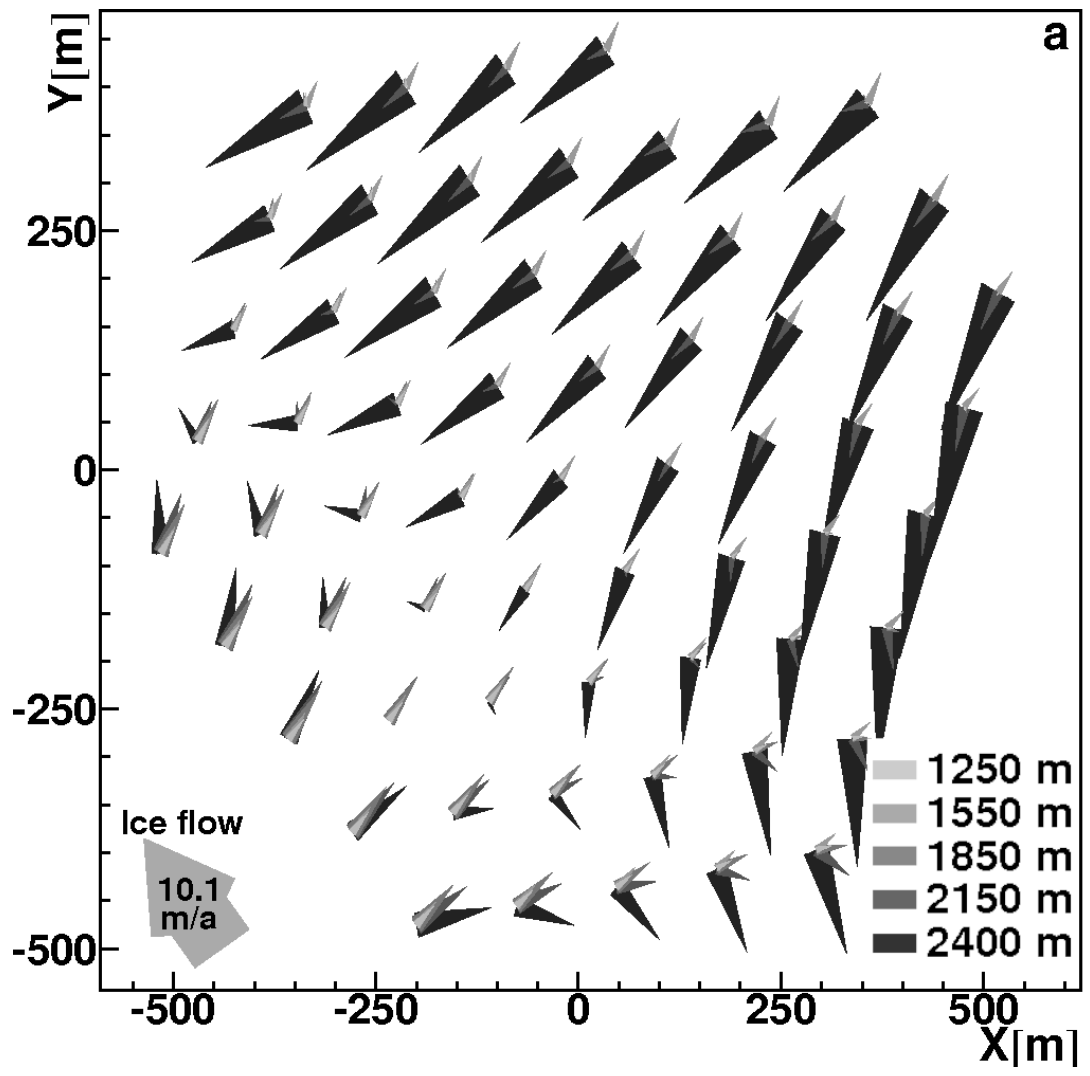

Figure 4.25: Layer tilt at five depths shows the influence of basal topography increasing toward the bedrock. Arrows depict downhill direction and relative tilt magnitude.
[M. G. Aartsen et al. (2013d). "South Pole glacial climate reconstruction from multi-borehole laser particulate stratigraphy"]

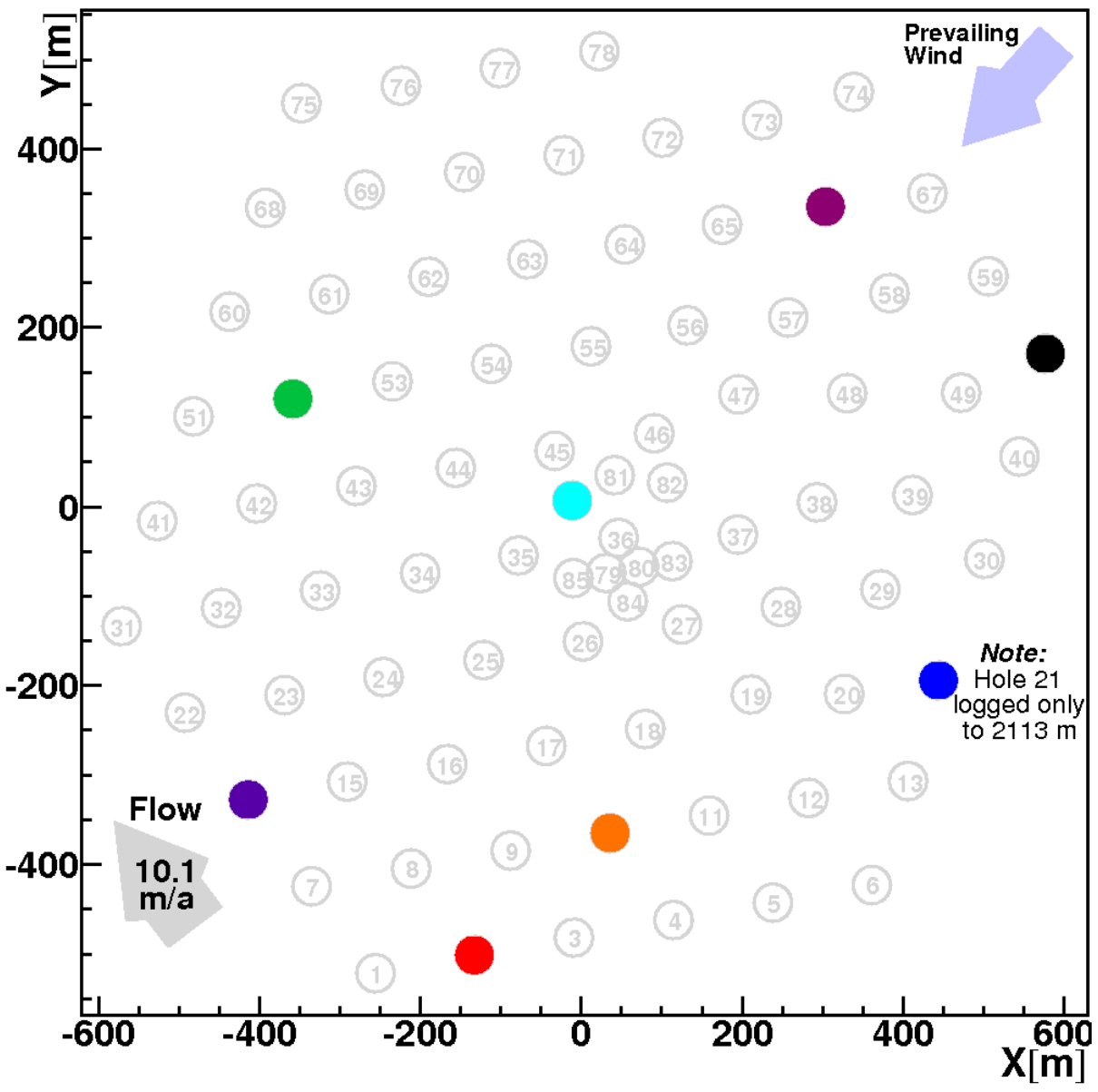

Figure 4.26: Locations where the dust logger has been deployed.
[M. G. Aartsen et al. (2013d). "South Pole glacial climate reconstruction from multi-borehole laser particulate stratigraphy"]

## 4.5.4 *Inclinometers*

Extrapolating temperature measurements obtained in AMANDA, it was expected that the ice is frozen to the bedrock[181] and thus stationary. Given the surface flow velocity of ∼10 m per year a substantial amount of shear was expected in the deep ice, which also informed the maximum deployment depth.

[181] *Price et al., "Temperature profile for glacial ice at the South Pole: Implications for life in a nearby subglacial lake"*

In order to verify the extrapolation and to measure the shear profile, 51 inclinometers have been installed throughout the IceCube array. Three are highly sensitive and stable, biaxial, electrolytic devices as commonly used in geophysics and are housed in dedicated aluminum pressure housings. These were deployed on strings 44, 45 and 68. In addition, 48 DOMs have been equipped with micro-electromechanical (MEMS) tilt sensors and distributed throughout the array.

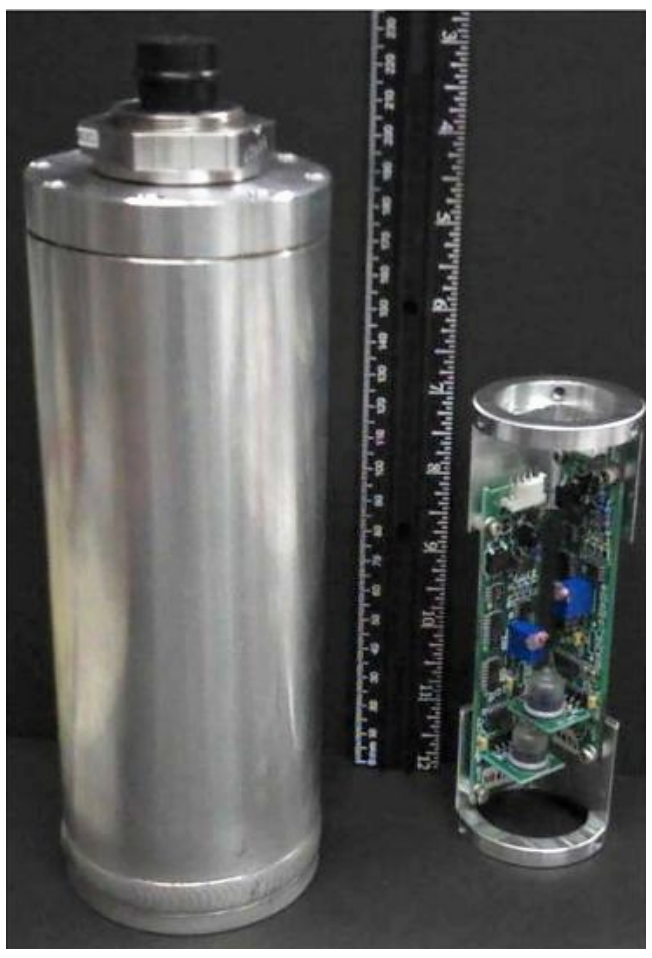

Figure 4.27: Electrolytic inclinometer in its pressure housing
[Bay, *Inclinometers wiki page*]

The average yearly rotation of all inclinometers is summarized in Figure 4.28. The measured shear is far less than predicted from a frozen bedrock interface. Only the deepest inclinometer shows a systematically significant trend. The temperature model has since been updated to predict a wet interface, resulting in larger contribution from basal sliding and far less shear in the deep ice[182].

[182] *Bay, "Private communication"*

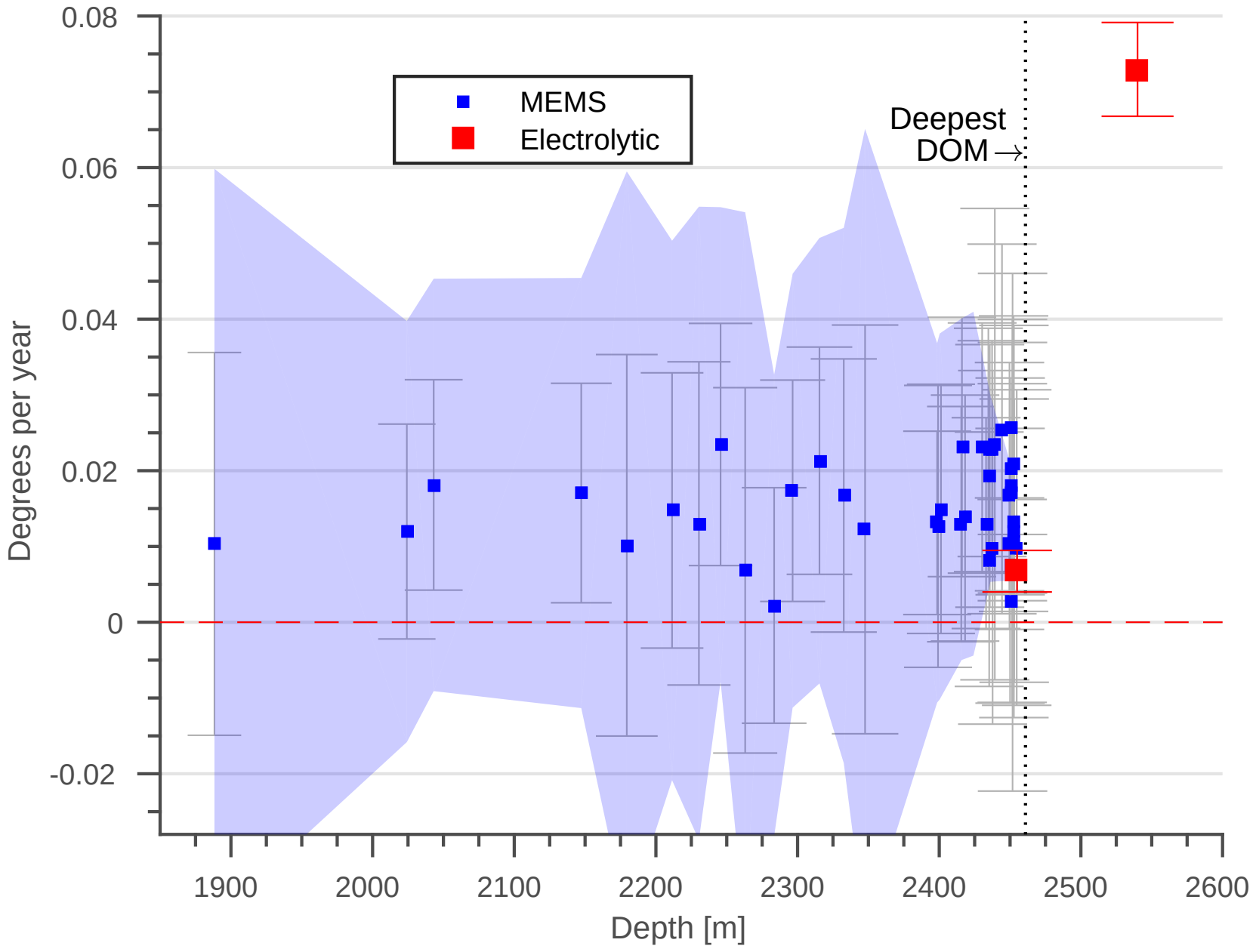

Figure 4.28: Average yearly rotation of all inclinometers versus deployment depths. The shaded area indicates the 95% confidence region for the MEMS sensor, neglecting possible drift due to aging.
[M. G. Aartsen et al. (2017c). "The IceCube Neutrino Observatory: instrumentation and online systems"]

## 4.5.5 *Standard candles*

The LED flashers provide light intensities at cascade equivalent energies up to 100 TeV, but with a rather unrealistic angular emission distribution. In order to better mimic the light deposition of real cascades and to extend the energy range, the standard candles were developed[183].

They consist of a 337 nm nitrogen laser, illuminating a reflective cone, which shapes the beam to a Cherenkov-like profile[184]. The intensity can be selected through neutral-density filters and is at most $\sim 10^{13}$ photons[185] in a short, 4 ns pulse. One upward and one downward facing standard candle were deployed on strings 40 and 55 respectively.

While the light intensity could never be calibrated satisfactorily, the standard candles still provide an interesting tool to test cascade vertex reconstructions or to investigate the DOM's saturation behaviors.

183 *J. Kiryluk et al. (Nov. 2, 2007). "IceCube Performance with Artificial Light Sources: the Road to Cascade Analyses"*

184 *Kurt Woschnagg et al. (n.d.).* Standard Candle. *IceCube Wiki*

185 *equivalent to a* 100 PeV *cascade*

# 5 DETECTOR CALIBRATION

**This chapter summarizes the state of the detector calibration, both in terms of hardware and ice knowledge, prior to this thesis. The reader is also introduced to the methods required for fitting ice properties using likelihood fits comparing LED flasher data to photon propagation simulation.**

The calibration of the IceCube Neutrino Observatory includes the characterization of the detection medium and the response of the sensors & electronics to at least the precision required to perform the set science goals.

## 5.1 HARDWARE CALIBRATION

One of the strengths of IceCube is the homogeneity and simplicity of the deployed hardware. There is only one kind of device, the DOM, which needs to be understood and characterized.

These calibration efforts can be classified as follows:

- Laboratory efforts performed during the development and production of the IceCube hardware, performed on both individual subsystems such as the detailed evaluation of over 400 PMTs[186], the Final Acceptance Testing (FAT) of all assembled DOMs prior to deployment, as well as continued laboratory efforts on existing surplus hardware[187].
- Continuous calibration performed as part of the standard data taking, in particular the time synchronization of all DOMs in the array to within 1 ns through RapCal as described in[188].
- The yearly re-calibration of the DOMs ADC calibration constants, PMT gain and transit time through DomCal[189].
- Calibration performed offline to access more (computationally) challenging topics such as the absolute detection efficiency, noise modeling or precision evaluation of PMT effects.

The constant low temperatures and low light levels provide ideal conditions for the operation of precision electronics. As an example of the detector stability, Figure 5.1 highlights the distribution of fractional changes in PMT gain from 2011 to 2016 for all in-ice DOMs, at the 2011 operating high voltage for each DOM.

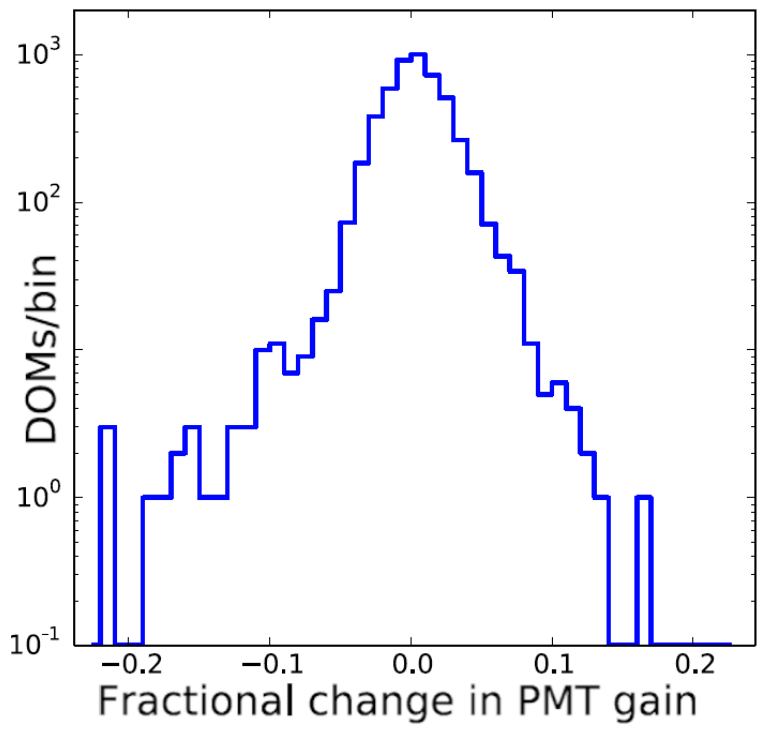

Figure 5.1: Fractional changes in PMT gain from 2011 to 2016 for all in-ice DOMs, at the 2011 operating high voltage for each DOM. [Aartsen et al., "The IceCube Neutrino Observatory: instrumentation and online systems"]

[186] *Abbasi et al., "Calibration and characterization of the IceCube photomultiplier tube"*

[187] *Tosi et al., "Calibrating photon detection efficiency in IceCube"*

[188] *Aartsen et al., "The IceCube Neutrino Observatory: instrumentation and online systems"*

[189] *except for Icetop, where the calibration is updated monthly due to the strong seasonal drifts*

In most aspects the hardware calibration has been well-established for many years and is precise enough not to be considered as an uncertainty at analysis level. The following sections exemplify two aspects where a substantial uncertainty remains and which are not further pursued in this thesis. The angular acceptance and the modeling of SPE charge distributions are discussed in chapter 9 and appendix A respectively.

### 5.1.1 *Detection efficiency*

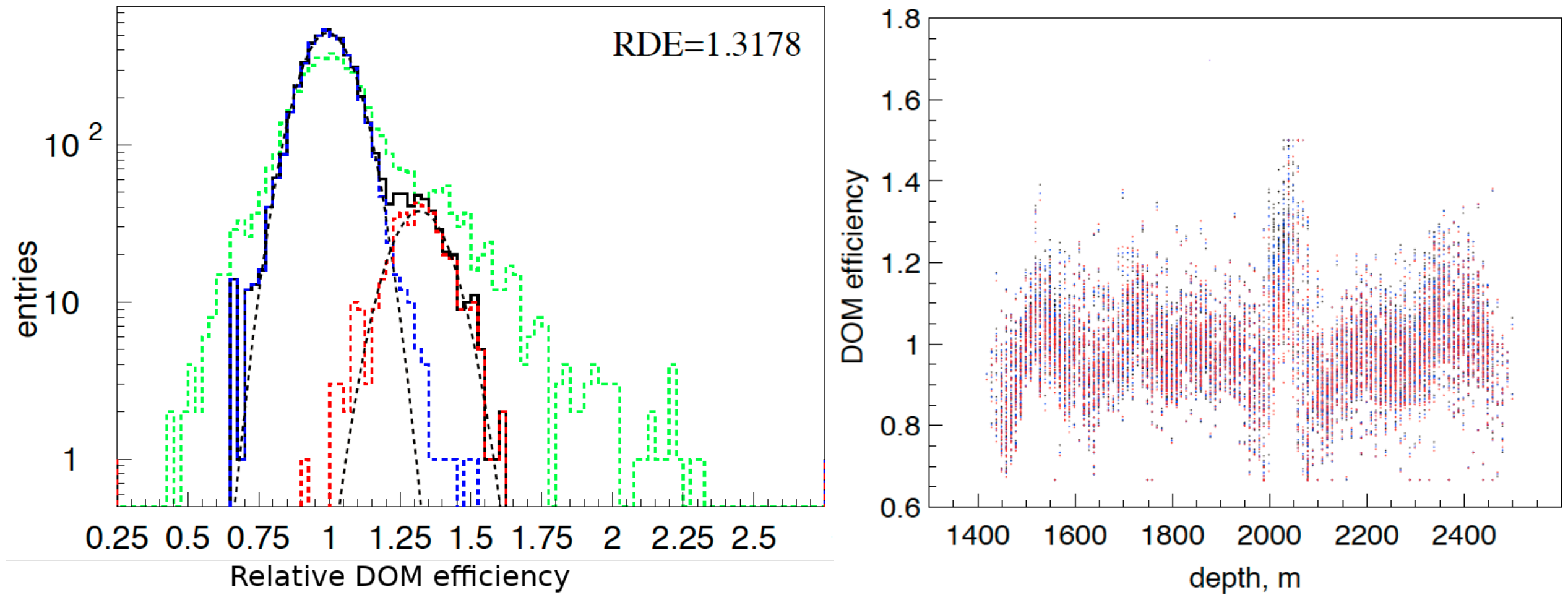

Figure 5.2: Left: Distribution of relative DOM efficiencies measured using flasher data. Blue denotes the histogram for standard DOMs, while red denotes the high quantum efficiency DOMs.
Right: Same relative DOM efficiencies plotted against the DOMs depths. The correlation to the ice properties is seen most clearly in the region of the dust layer around 2000 m.
[Dmitry Chirkin (2017). "Ice model with flasher unfolding"]

While the photon detection efficiency of bare PMTs and integrated DOMs has been extensively studied in the laboratory, the average in-situ absolute DOM efficiency is re-evaluated for every ice model iteration. Here the DOM efficiency denotes a multiplicative correction factor to the effective photon detection efficiency with respect to a laboratory reference.

The LEDs can not be used for this purpose as their absolute brightness is not sufficiently constrained. Instead, minimum ionizing muons and their experimentally and theoretically very well established photon yield are utilized as absolute reference light sources.

The primary challenge in applying this method is the selection of single, low-energy muons. Atmospheric muons usually arrive in bundles close to the shower core[190].

[190] *Tania Wood (2014). "Absolute DOM Efficiency with Muon Data in IceCube"*

More recently, relative DOM efficiencies, that is the relative difference in detection efficiency between individual DOMs, is being considered. While this can be achieved with both LEDs and muons, all measurement attempts to-date have yielded larger spreads than expected and strong correlations with the bulk ice properties.

### 5.1.2 *Geometry calibration*

While differential velocities in the ice movement could in principle disturb the detector shape over time, the deployed tilt sensors show measurable shear only below the instrumented volume[191]. Therefore, the detector geometry is assumed to be constant in a coordinate system relative to the snow surface, which slowly moves with the overall ice flow.

[191] see section 4.5.4

#### 5.1.2.1 *From deployment information*

Given the small lateral displacement observed throughout the drilling process[192], the lateral position for all DOMs on a string is taken to be the surface location of the drill hole as measured through GPS surveying.

As the spacing between modules is fixed, the depth of all DOMs is given by the depth offsets of any given string to a reference coordinate system. This was determined as the sum of the GPS height of the local snow position, the air column of the drill hole as measured by a laser range finder reflecting on the water surface and readings from pressure sensors along the string, combined with the known cable payout[193].

#### 5.1.2.2 *From flasher data*

The depth offset of each string was further corrected down to an accuracy of 0.2 m using inter-string flasher timing measurements. An example is shown in Figure 5.4.

Trilateration of the x-y positions originally proved difficult due to insufficient ice modeling but was later successfully tested on three strings in DeepCore. The trilateration matches the drill profile in shallow ice, with integral errors of the inertial measurement unit likely causing the discrepancy in the deep ice. An example profile is shown in Figure 5.3.[194]

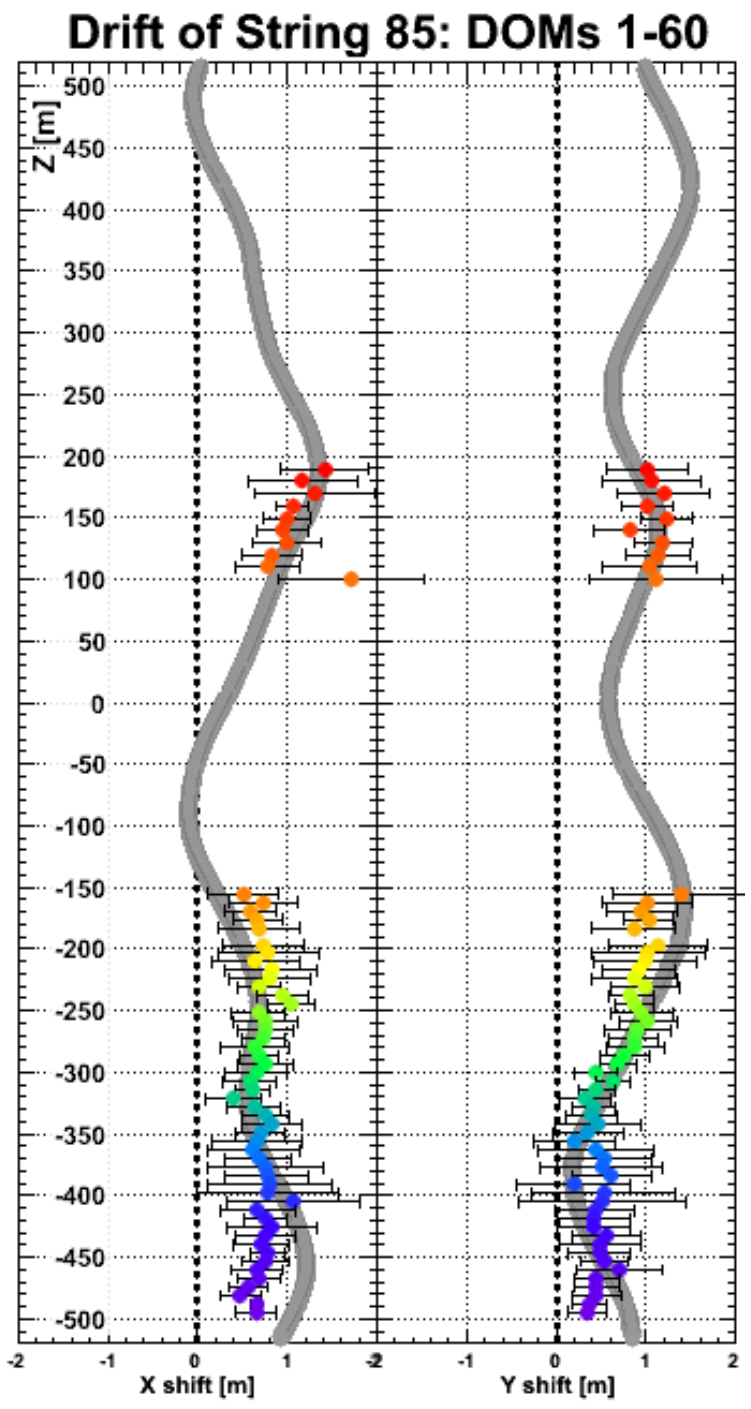

Figure 5.3: Trilateration result for DOMs on one DeepCore string compared to drill data (gray band). [Sheremata, "DeepCore DOM Profiles with the Trilateration Algorithm"]

[192] see section 4.3.2.1

[193] Aartsen et al., "The IceCube Neutrino Observatory: instrumentation and online systems"

[194] New attempts are currently underway, utilizing the now advanced understanding of the light propagation.

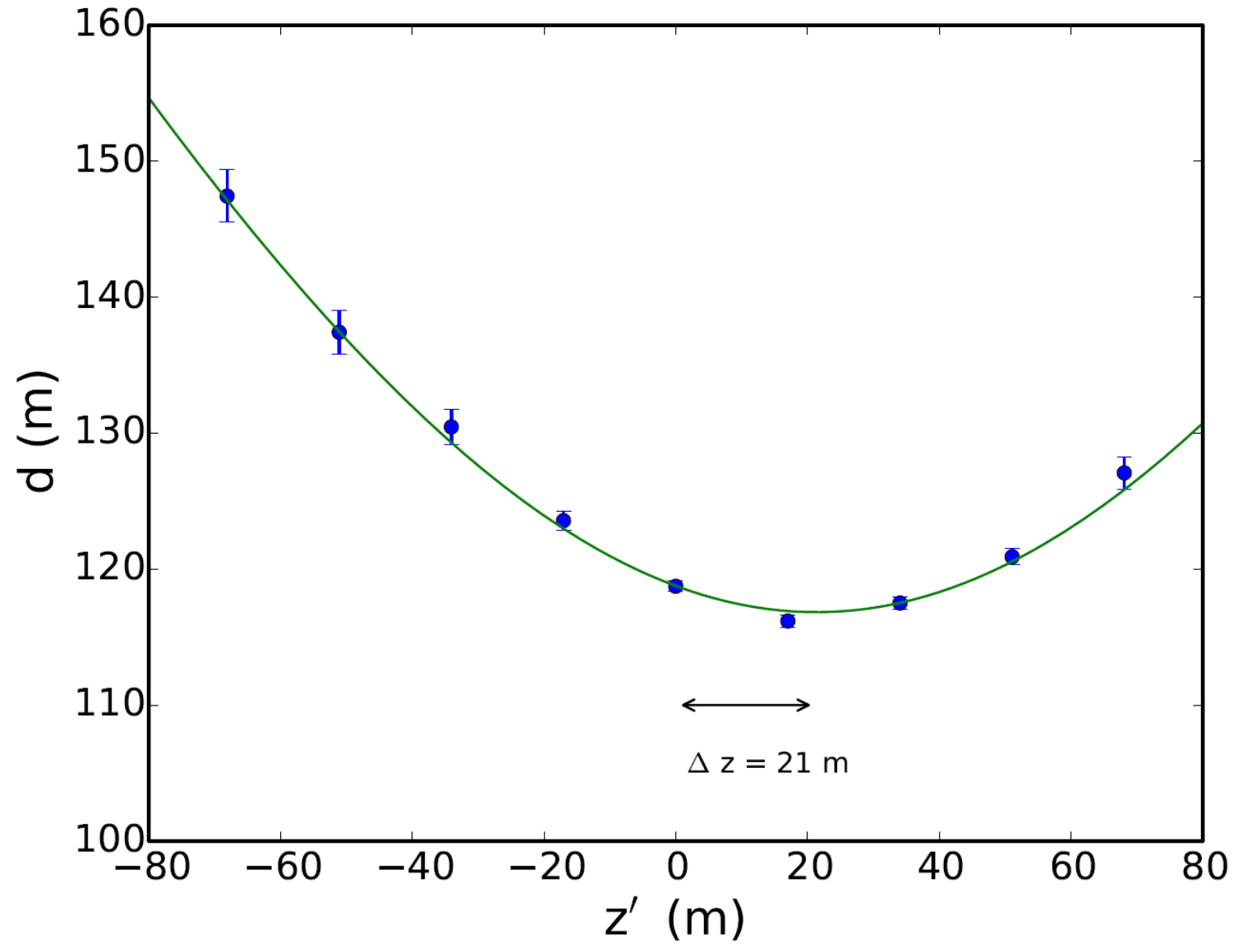

Figure 5.4: Distance $d$ vs. relative depth $z'$ of two string as measured using flasher data. In this example the pressure sensor on the flashing string failed during deployment. [M. G. Aartsen et al. (2017c). "The IceCube Neutrino Observatory: instrumentation and online systems"]

## 5.2 MODELING THE ANTARCTIC ICE

As reviewed in chapter 3, the optical properties of ice can be fully[195] described through absorption and scattering. In the deep glacial ice these are[196] given by scattering and absorption on impurities.

[195] *this paradigm has changed recently with regards to the anisotropy, see chapter 10*

[196] *except for intrinsic absorption at large (>500 nm) wavelengths*

In this section the theory of electromagnetic scattering and the analytic derivation of ice properties based on ice core information are summarized. Additionally, the methods applied to data obtained from the calibration devices, in particular the ice fitting from flasher data, and the resulting ice modeling employed in the IceCube detector simulation are described. The ice fitting methods lay the foundation for the later presented work on the hole ice properties and the discussion of the ice optical anisotropy.

### 5.2.1 *Mie scattering theory*

Electromagnetic scattering theory describes the passage of radiation through a heterogeneous material, where electromagnetic fields most conform both to the Maxwell equations and to the boundary conditions imposed by transitions in the propagation medium. Such is the case when a dust particle is illuminated by a beam of light and the amount and angular distribution of the light scattered by the particle, as well as the amount absorbed are to be deduced.

The probability density of scattering angles $\theta$ is usually called the scattering function. Neglecting its functional form[197], it is described through the asymmetry parameter $g = \langle \cos\theta \rangle$.

[197] *which is specified in section 5.2.2*

The geometric shapes of the Antartic dust constituents are not well constrained. Furthermore, for most particle shapes analytic solutions do not exist. As a result, one usually resorts to the simplest assumption of spherical particles, for which the Mie theory[198] yields analytic solutions.

[198] *Mie, "Beitraege zur Optik trueber Medien, speziell kolloidaler Metalloesungen"*

Assume a single spherical particle $j$ of radius $r$ and complex refractive index $n$ embedded in a medium with a refractive index $n_{ice}$ being illuminated by a beam of light with wavelength $\lambda$. From these parameters define the relative complex refractive index $m_j$ to be

$$m_j(\lambda) = \frac{n_j(\lambda)}{n_{ice}(\lambda)} \tag{5.1}$$

and the relative ratio of particle to wavelength size, the size parameter $x$, to be

$$x_j = \frac{2\pi \cdot r_j n_{ice}(\lambda)}{\lambda}. \tag{5.2}$$

From these two quantities[199] the so called Mie coefficients are given as

[199] *and assuming the permeability of the two media to be identical*

$$\begin{aligned} a_k &= \frac{m \cdot \Psi_k(mx)\Psi_k'(x) - \Psi_k(x)\Psi_k'(mx)}{m \cdot \Psi_k(mx)\xi_k'(x) - \xi_k(x)\Psi_k'(mx)} \\ b_k &= \frac{\Psi_k(mx)\Psi_k'(x) - m \cdot \Psi_k(x)\Psi_k'(mx)}{\Psi_k(mx)\xi_k'(x) - m \cdot \xi_k(x)\Psi_k'(mx)} \end{aligned} \tag{5.3}$$

where $\Psi$ and $\xi$ are $k$-th order Riccati–Bessel functions. For a detailed derivation see for example[200]. From these, the efficiency factors for absorption and scattering $Q_{abs}$ & $Q_{sca}$ as well as the asymmetry parameter $g$ (not shown here) can be calculated:

[200] *Bohren and Huffman,* Absorption and Scattering of Light by Small Particles

$$\begin{aligned} Q_{ext} &= \frac{2}{k^2}\sum_{k=1}^{\infty}(2k+1)\Re(a_k + b_k) \\ Q_{sca} &= \frac{2}{k^2}\sum_{k=1}^{\infty}(2k+1)(|a_k|^2 + |b_k|^2) \\ Q_{abs} &= Q_{ext} - Q_{sca} \end{aligned} \tag{5.4}$$

The efficiency factors denote the proportionality between the interaction cross sections and geometric area of the particle:

$$\begin{aligned} A(r,\lambda,m) &= Q_{ext}(r,\lambda,m) \cdot \pi r^2 \\ B(r,\lambda,m) &= Q_{sca}(r,\lambda,m) \cdot \pi r^2 \end{aligned} \tag{5.5}$$

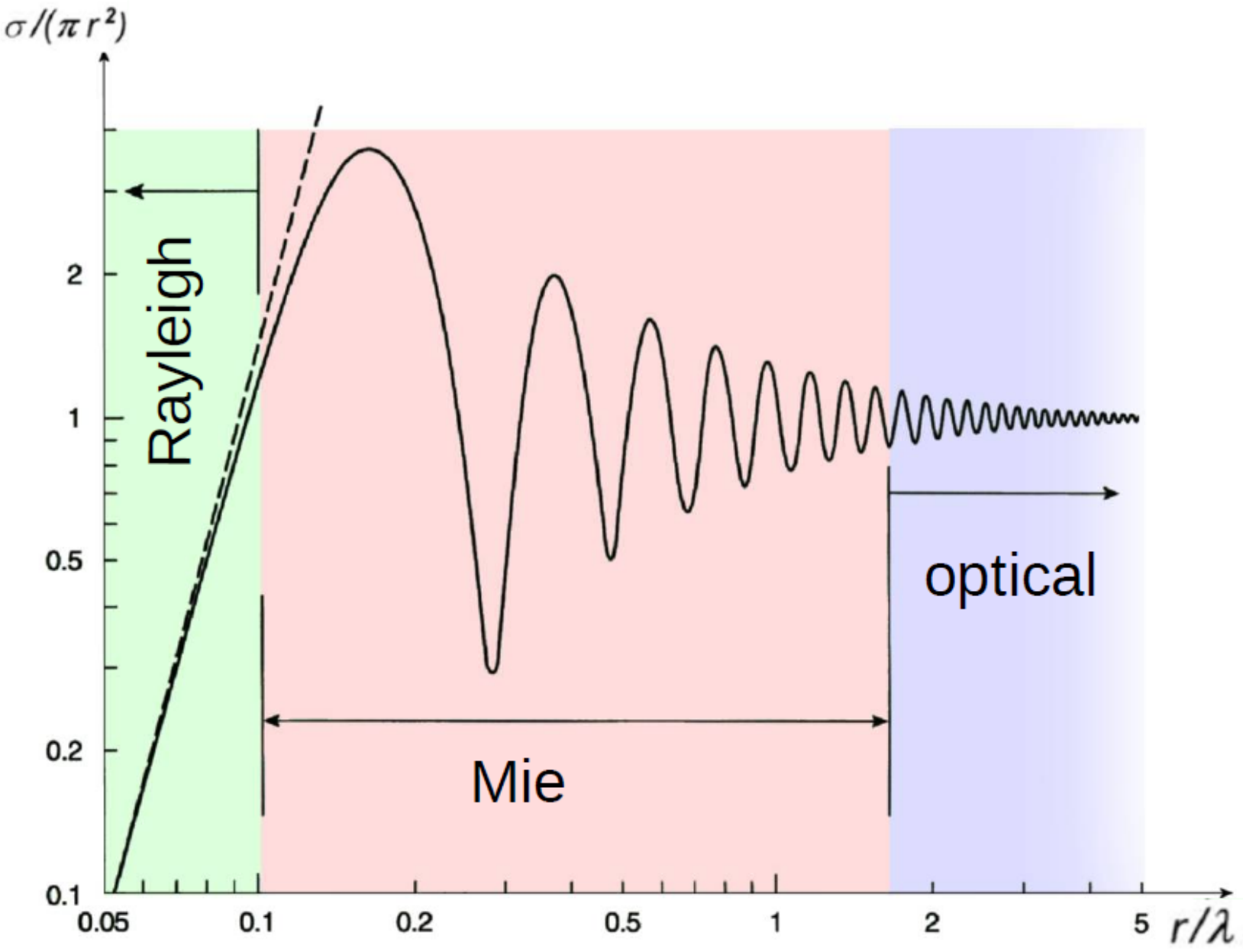

Figure 5.5: Generic behavior of the Mie efficiency (cross section over area) as a function of the size parameter of the scattering particle. [Christian Wolff (2018). *Radar Mie Efficiency*]

Figure 5.5 depicts the general behavior of Mie efficiencies as a function of the size parameter. The observed oscillations arise from Riccati-Bessel functions found in the Mie coefficients. As evident from these efficiencies, absorption is driven by the imaginary part of the particle's refractive index, while the absolute value drives the scattering. Note that as $m$ approaches unity, that is the particle exhibits the same index of refraction as the embedding medium[201], the amplitudes vanish.

[201] *as with Craigite in ice*

While the Mie Theory holds for any given ratio of particle to wavelength size, it is often used synonymously for the transition region between Rayleigh scattering ($x \ll 1$ , isotropic scattering $g \approx 0$) and geometric optics ($x \gg 1$, $g \approx 1$).

Given a medium containing a mixture of different impurities of known number densities and normalized fractional size distribution $\frac{df_j}{dr} = \frac{1}{N_j} \cdot \frac{dN_j}{dr}$,

$$N_j = N_j \int_{r_{min}}^{r_{max}} dr \left(\frac{df_j}{dr}\right) = \int_{r_{min}}^{r_{max}} dr \left(\frac{dN_j}{dr}\right), \tag{5.6}$$

the total absorption and scattering coefficients $a(\lambda)$ & $b(\lambda)$ as well as the average scattering asymmetry $g(\lambda)$ can be calculated by summing the individual contributions[202]:

[202] *He and Price, "Remote sensing of dust in deep ice at the South Pole"*

$$a(\lambda) = \sum_j N_j \int_{r_{min}}^{r_{max}} dr \left[A_j(r, \lambda, m_j)\frac{df_j}{dr}\right] \tag{5.7}$$

$$b(\lambda) = \sum_j N_j \int_{r_{min}}^{r_{max}} dr \left[B_j(r, \lambda, m_j)\frac{df_j}{dr}\right] \tag{5.8}$$

$$g(\lambda) = \langle \cos\theta \rangle = \sum_j N_j \left\{ \int_{r_{min}}^{r_{max}} dr \left[ Q_j(r,\lambda,m_j) \cdot g_j(r,\lambda,m_j) \frac{df_j}{dr} \right] \bigg/ \int_{r_{min}}^{r_{max}} dr \left[ B_j(r,\lambda,m_j) \frac{df_j}{dr} \right] \right\} \bigg/ \sum_j N_j \tag{5.9}$$

Larger absorption/scattering coefficients equate to a more strongly absorbing/scattering medium. Absorption and scattering lengths are given as the inverse of these coefficients. The absorption length describes the distance at which the survival probability of a photon drops to $1/e$. The scattering length, also referred to as the scattering mean free path, is the average distance between two scattering processes. A (mostly) scattering function independent effective scattering length $\lambda_{eff}$, denoting the distance at which a pencil beam becomes diffuse is given as[203]

$$\lambda_{eff}(\lambda) = \lambda_{sca}(\lambda) / \left[1 - g(\lambda)\right] = \frac{1}{b_e}. \tag{5.10}$$

#### 5.2.1.1 *Application to the South Pole ice*

Using the above relationships He and Price[204] in 1998 attempted to reproduce the optical properties reported by AMANDA a year earlier[205,206].

[203] *Aartsen et al., "Measurement of South Pole ice transparency with the IceCube LED calibration system"*

[204] *He and Price, "Remote sensing of dust in deep ice at the South Pole"*

[205] *Price et al., "Optical properties of deep ice at the South Pole: scattering"*

[206] *Askebjer and others, "Optical properties of deep ice at the South Pole: absorption"*

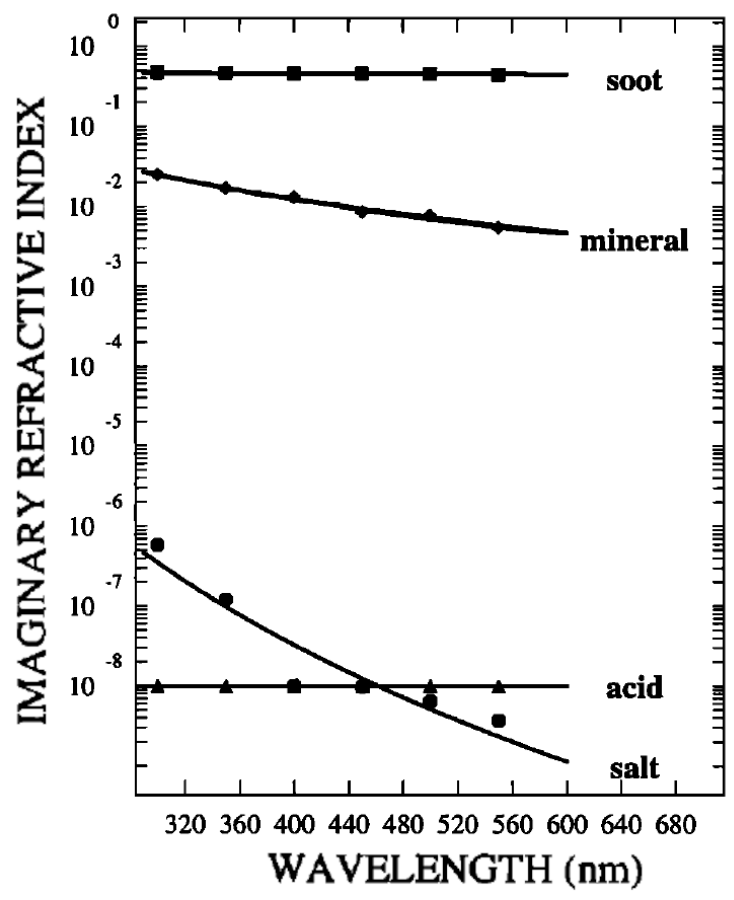

Figure 5.6: Imaginary part of the refractive index of various impurity types.
[He and Price, "Remote sensing of dust in deep ice at the South Pole"]

They identified the four major impurity components already discussed in section 3.4 to also drive the optical properties. Impurities are dozens of nm to few µm in size. The optical regime is never fully reached and detailed Mie scattering has to be considered.

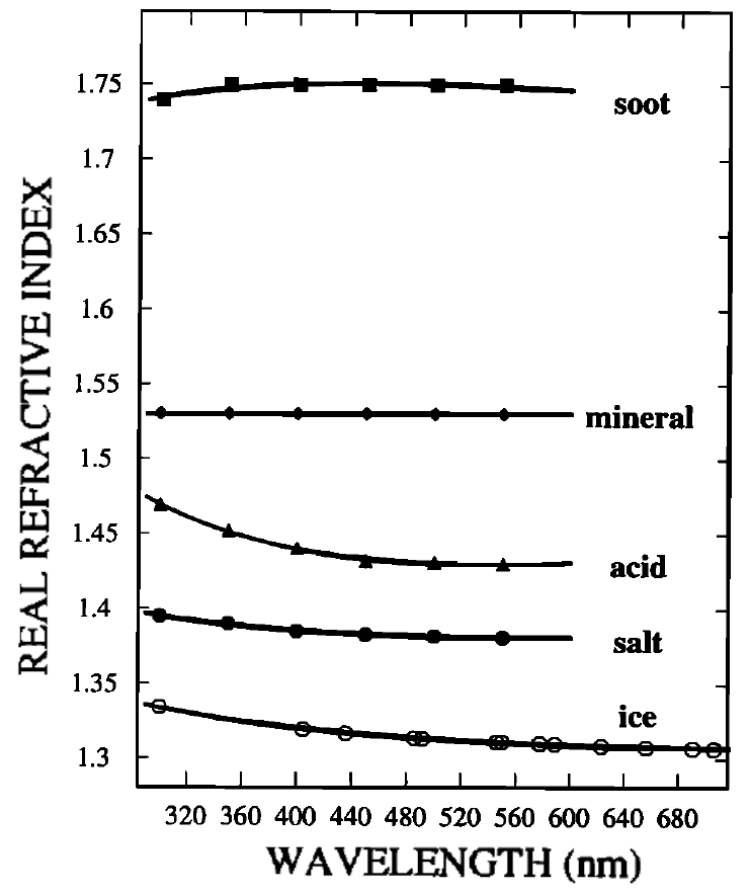

Figure 5.7: Real part of the refractive index of various impurity types.
[He and Price, "Remote sensing of dust in deep ice at the South Pole"]

For each of these impurities the refractive index, number density and size distributions need to be known. At that time no ice core from the South Pole was available. Therefore, these had to be deduced form ice cores from other regions of Antarctica and by comparing the more easily accessible contemporary surface deposition. Figures 5.6 and 5.7 show the used indices of refraction. Table 5.1 summarizes the assumed mass densities and size distributions.

The mass concentrations $M_j$ relate to the densities $\rho_j$ and size distributions as:

$$M_j = \frac{4\pi}{3} N_j \left( \frac{\rho_j}{\rho_{ice}} \right) \int_{r_{min}}^{r_{max}} dr \left( r^3 \frac{df_j}{dr} \right) \tag{5.11}$$

Table 5.1: Physical properties of main impurity contributions.
[Y. D. He and P. B. Price (July 1998). "Remote sensing of dust in deep ice at the South Pole"]

| Component | Minerals | Sea salt | Acids | Soot |
|---|---|---|---|---|
| Modal radius, $\hat{r}_j$ (µm) | 0.27 | 0.4 | 0.07 | 0.0118 |
| Geometric std. dev., $\hat{\sigma}_j$ | 2.67 | 2.03 | 1.86 | 2.00 |
| Mean density, $\rho_j$ ($\mathrm{g\,cm^{-1}}$) | 2.6 | 2.2 | 1.68 | 2.3 |
| $M_j$ in snow ($\mathrm{ng\,g^{-1}}$) | 12 | 33 | 171 | 0.1-0.6 |
| $M_j$ at 0.83 km ($\mathrm{ng\,g^{-1}}$) | 12 | 36 | 171 | 0.1-0.6 |
| $M_j$ at 1.7 km ($\mathrm{ng\,g^{-1}}$) | 126 | 139 | 247 | 0.1-6.0[1] |
| $M_j$ at 2.5 km ($\mathrm{ng\,g^{-1}}$) | 12 | 15 | 172 | 0.1-0.6 |

Information on size distributions is sparse and often limited by the detection threshold of the used instrumentation or biased by non-spherical particles. Given the information available at the time, the normalized fractional size distribution were assumed to be log-normal[207]:

[207] *He and Price, "Remote sensing of dust in deep ice at the South Pole"*

$$\frac{df_j}{dr} = \frac{1}{r}\frac{df_j}{d\log r} = \frac{1}{r\sqrt{2\pi}\log\hat{\sigma}_j}\exp\left\{-\frac{1}{2\left(\log\hat{\sigma}_j\right)^2}\left[\log\left(\frac{r}{\hat{r}_j}\right)\right]^2\right\} \tag{5.12}$$

with modal radia $\hat{r}_j$ and geometric standard deviations $\hat{\sigma}_j$.

From the refractive indices (see Figure 5.7) alone it is evident that salts and acids contribute negligibly to the absorption. Calculating the per-component absorption and scattering coefficients reveals that sea salt is the most strongly scattering component. Soot is highly absorbing, but it contributes the least to scattering. Yet, as it has the smallest particles, in the tenth of nm range, it is the only constituent acting as Rayleigh scatterer, making it distinct for back-scattering experiments such as the dust logger.

Summing the individual contributions yields absorption and scattering coefficients quantitatively compatible with the AMANDA measurements. It also confirms a power law wavelengths dependency for both scattering and absorption as were already assumed for the AMANDA analyses (see Figure 5.8).

The exponents are found to depend only weakly on dust size and composition. This motivates a depth independent wavelength scaling as used in all subsequent ice models. He and Price also correctly predicted significantly cleaner ice at the lowest IceCube depths, as measured and confirmed some years later[208].

[208] *Ackermann, Ahrens, et al., "Optical properties of deep glacial ice at the South Pole"*

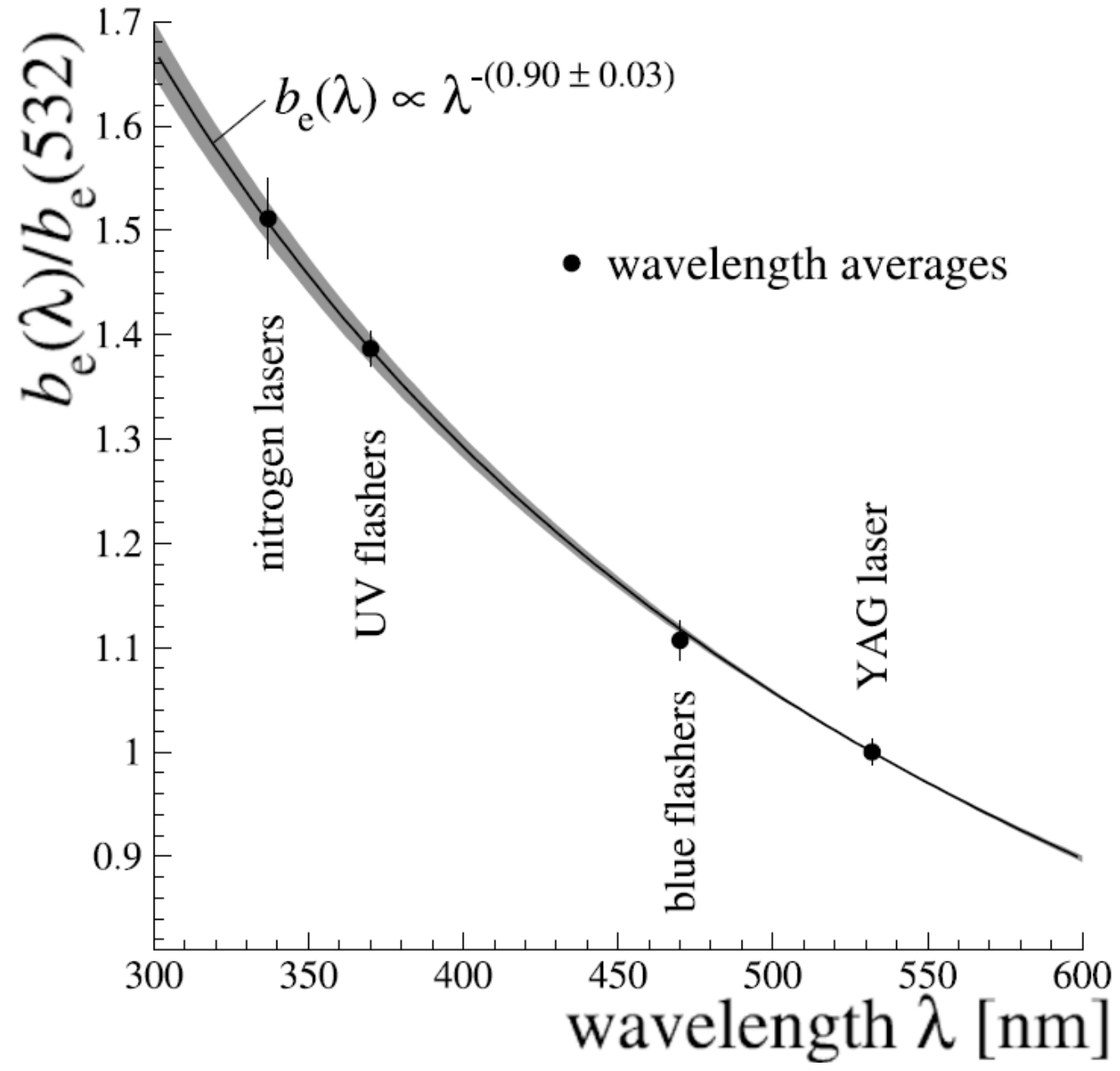

Figure 5.8: Wavelength dependence of the measured scattering coefficients compared to a fitted power law as expected from Mie theory [M. Ackermann, J. Ahrens, et al. (2006). "Optical properties of deep glacial ice at the South Pole"]

The assumption of spherical particles is generally wrong and the error induced by this assumption is hard to assess. The ice parameters, *a*, *b* and the scattering function, result from contributions from all types of impurities and their size distributions. This luckily washes out most of the detailed features in the size parameters. In addition, none of the direct Mie calculations are actually used for the in-situ calibration measurements as described in section 5.2.6.

The parametrization of the scattering function is the only model input derived from theory, with the mean scattering angle *g* left as a free parameter. Still, when discussing subtle concepts, such as the rotation of impurities in the context of anisotropy, care has to be taken not to generalize the scalings deduced from Mie calculations.

## 5.2.2 *Photon propagation as a simulation approach*

IceCube detects individual photons as produced through Cherenkov radiation or as emitted by the calibration LEDs. On their way from their source to a potential detector these photons are subject to absorption and scattering in the layered ice, shaping both the intensity pattern in the detector as well the arrival time distributions on every module.

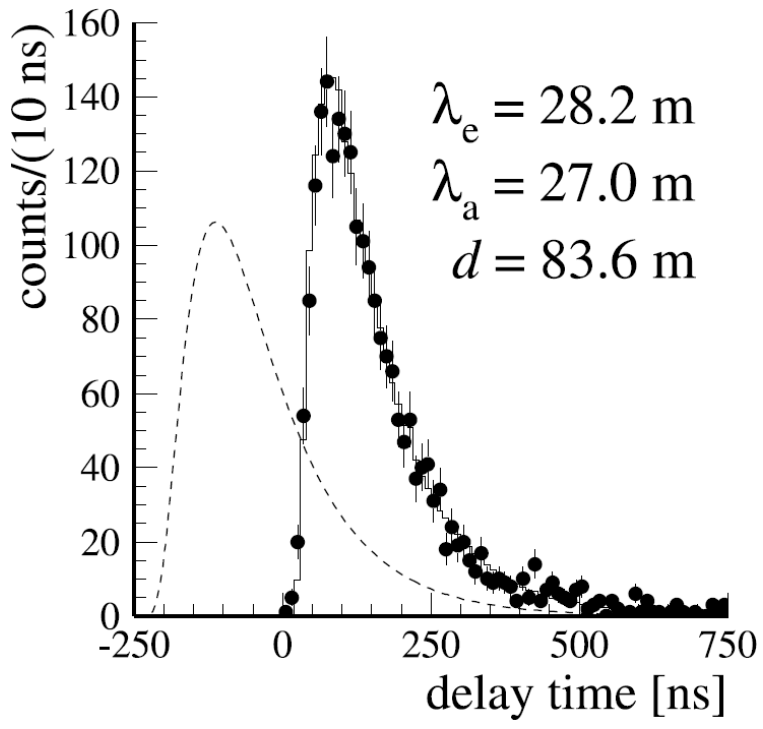

Figure 5.9: Green's function (dashed line) compared to photon propagation simulation (solid line) and data given the same optical properties and at the AMANDA-B string spacing of 83 m.
[Ackermann, Ahrens, et al., "Optical properties of deep glacial ice at the South Pole"]

For an emitter in the far field ($d \gg \lambda_e$) and given a weak absorption coefficient compared to the scattering coefficient, the arrival time distribution $u(t)$ at a distance $d$ from an isotropic source is described by Green's function[209] as

[209] *Ackermann, Ahrens, et al., "Optical properties of deep glacial ice at the South Pole"*

$$u(d,t) = \frac{1}{(4\pi \cdot Dt)^{3/2}} \cdot \exp -\frac{d^2}{4Dt} \cdot \exp -\frac{tc_{ice}}{\lambda_a}, \tag{5.13}$$

where $D = c_{ice}\lambda_{eff}/3$ is the diffusion constant. As evident from this equation, the position of the rising edge is sensitive to the scattering coefficient, while the slope of the tail is determined by the absorption coefficient. While this is approximately true even outside the far field, Green's function is inapplicable in the clean ice and at the sensor spacings used in IceCube, as can be seen in Figure 5.9.

For IceCube, the photon propagation needs to be fully modeled in the simulation. This is achieved through the use of photon propagation software, namely the "photon propagation code" (PPC)[210] and its derivative clSim. Their concept is a full first principle simulation by tracking each photon individually, as described in the following.

[210] *Dmitry Chirkin (Oct. 2013c). "Photon tracking with GPUs in IceCube"*

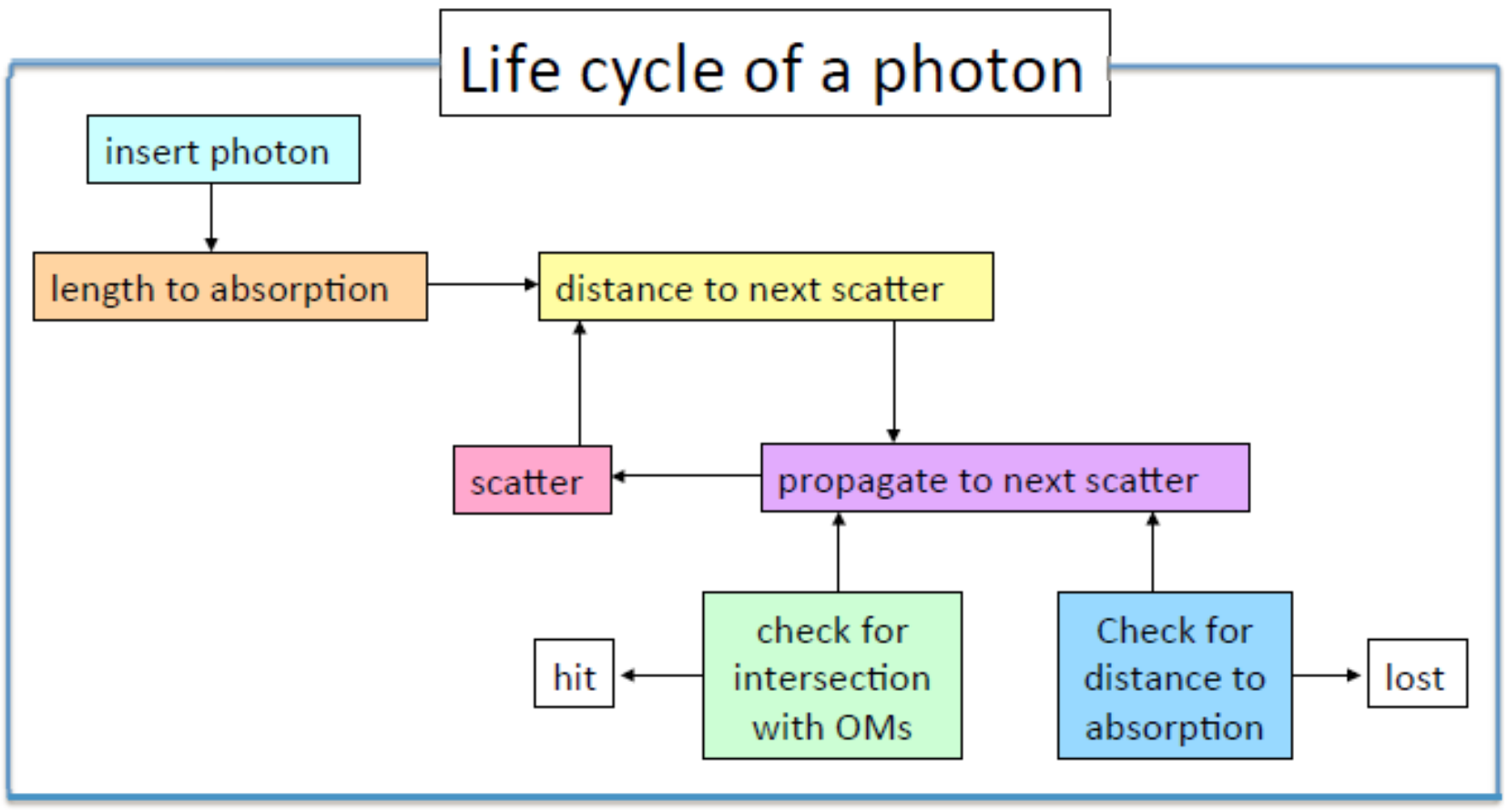

Figure 5.10: Photon life cycle as implemented for the photon propagation
[Dmitry Chirkin (2015b). "Photon Propagation with GPUs in IceCube." en]

For every created photon the total lifetime / absorption weight in multiples of absorption lengths is sampled from an exponential distribution with unity scale. Next the distance to the next scattering process is determined in the same fashion and the photon is moved through a depth layered ice model along its current propagation direction towards the next scattering center.

For each layer traversed, the length times the local[211] absorption/scattering coefficient is subtracted from the current absorption / scattering weight. As the scattering weight reaches zero, the scattering site has been reached and the photon is deflected according to the chosen scattering function. The scattering transport process is repeated until the photon is either absorbed, as the absorption weight reach an epsilon cut-off value, or the photon is incident on a DOM and stored for later processing.

[211] *and potentially direction dependent*

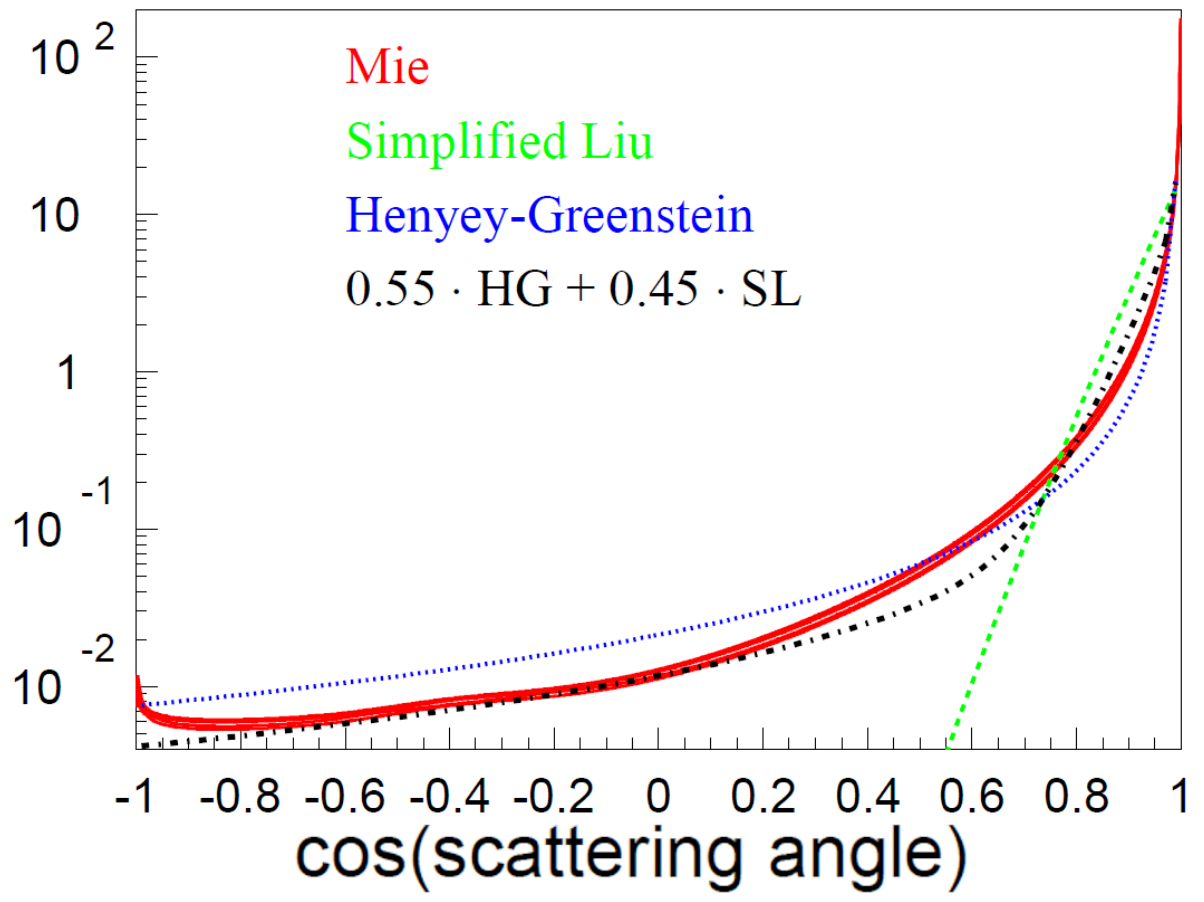

Figure 5.11: Parametrization of the Mie scattering function as calculated for the South Pole impurity composition as a superposition of the analytic Simplified Liu and Henyey Greenstein models.
[M. G. Aartsen et al. (May 2013c). "Measurement of South Pole ice transparency with the IceCube LED calibration system"]

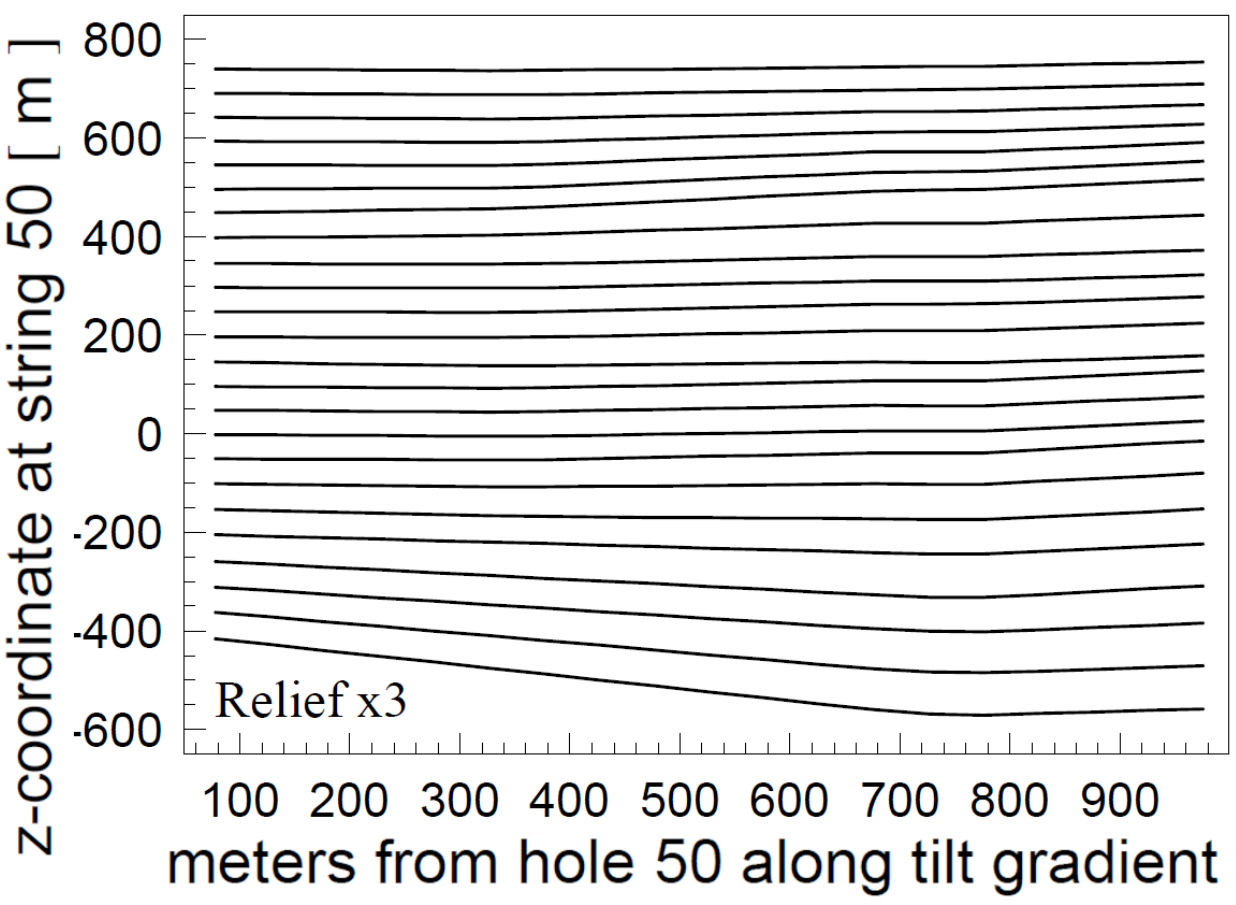

Figure 5.12: PPC ice tilt parametrization. Compare this one dimensional approximation to the true profile shown in figure 4.25.
[M. G. Aartsen et al. (May 2013c). "Measurement of South Pole ice transparency with the IceCube LED calibration system"]

PPC aims to mimic the Mie scattering distributions deduced from the analytic calculations summarized above through the superposition of two analytic approximations of the Mie scattering function, which can be inverted and thus sampled efficiently via random numbers. These are the Henyey-Greenstein[212]

[212] *Henyey and Greenstein, "Diffuse radiation in the Galaxy"*

$$p(\cos\theta) = \frac{1}{2}\frac{1-g^2}{[1+g^2-2g\cdot\cos\theta]^{3/2}} \tag{5.14}$$

and Simplified-Liu[213]

[213] *Liu, "A new phase function approximating to Mie scattering for radiative transport equations"*

$$p(\cos\theta) \propto (1+\cos\theta)^{\alpha}\text{, with } \alpha = \frac{2g}{1-g} \tag{5.15}$$

parametrizations. Both parametrizations require the average scattering angle $g = \langle\cos\theta\rangle$ as only and shared parameter. The superposition is given by a mixing parameter $f_{SL}$, which is fitted as part of the ice modeling, to be

$$p(\cos\theta) = (1-f_{SL})\cdot HG(\cos\theta) + f_{SL}\cdot SL(\cos\theta). \tag{5.16}$$

The photon propagation is additionally complicated by the optical ice anisotropy, which is discussed in details in chapter 10 and the tilting layers in the deep ice[214]. Tilt is currently implemented approximately by offsetting the depth at each scattering site based on a one dimensional tilt relief which only takes into account the depth and distance along the tilt gradient[215] as seen in Figure 5.12.

[214] *see section 3.5*

[215] *see Figure 10.4*

As the trajectories of individual photons are independent, each photon can be tracked independently. This enables significant computation speedups of up to a factor 250 when using parallelized implementations on GPUs instead of CPU implementations.

Another significant increase in computational speed can be achieved through an approximation called DOM oversizing, where the DOMs are simulated to be up to 16 times as large as in reality.

Oversizing, however, creates a small timing bias. In addition, when using oversizing, photons are not deleted upon detection by a DOM in order to avoid unphysical shadowing effects. Whenever oversizing is used in this thesis it is stated explicitly and has been tested to have a negligible or sub-dominant effect.

PPC also includes a limited simulation of the detector hardware. This is done by down-sampling the incident photons based on their impact point or angle on the DOM and the DOM's wavelength acceptance. This makes it usable as a standalone software without any dependencies on the main IceCube software framework. All ice fits presented in this thesis or used to derive the SPICE family of ice models[216] use this software chain.

[216] *see section 5.2.8*

### 5.2.3 *Likelihood method for ice fitting*

The photon propagation described in the previous section enables reproducing (flasher) events in simulation, given a set of model parameters including a realization of the ice properties. Most ice calibration studies perform an optimization of the ice assumptions in order to minimize the discrepancies between simulated and measured flasher events.

In practice, best estimators for the ice properties are obtained through a log-likelihood minimization, where a single likelihood value is computed for every pair of emitter and receiver DOMs.

For this purpose, the experimental and the simulated events are averaged over the number of repetitions in this flasher configuration[217]. The light curve of each receiving DOM[218] is then binned using a Baysian Blocking[219] algorithm, where each bin is multiples of 25 ns long and contains roughly the same statistics as any other bin.

[217] *usually 10 in simulation and several hundred in data*

[218] *an example is shown in Figure 5.13*

[219] *Scargle, "Studies in Astronomical Time Series Analysis. V. Bayesian Blocks, a New Method to Analyze Structure in Photon Counting Data"*

The Baysian Blocking preserves detail during rapid changes of the light curves, such as during the rising edge, but avoids empty bins during the long tail of the falling edge. The binning is fixed based on the experimental data and then applied to the simulated data.

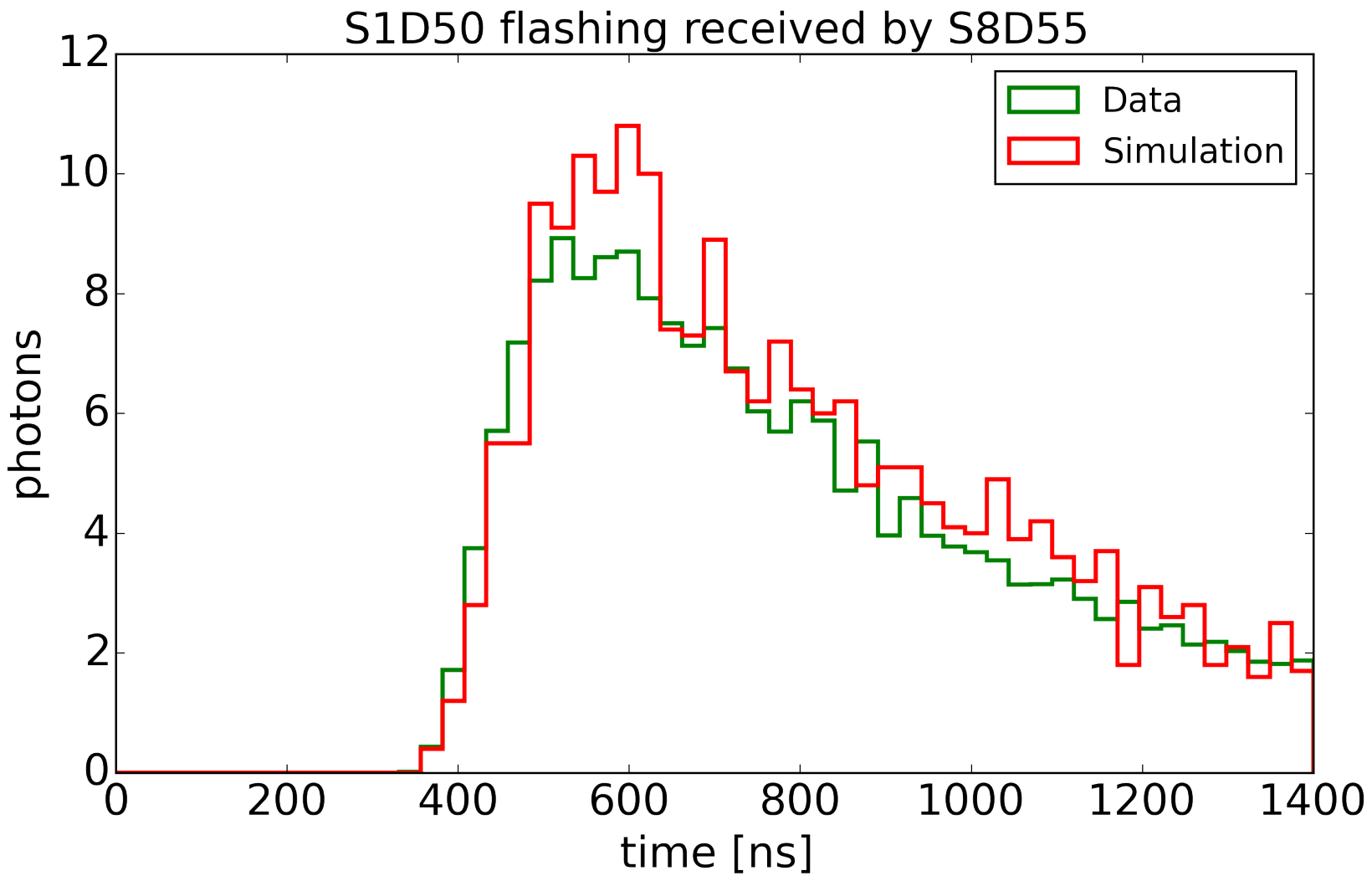

Figure 5.13: Example flasher light curve in 25 ns analysis binning. DOM 50 on string 1 emits light which is detected by DOM 55 on string 6. The data is averaged over 240 repetitions. The simulation is averaged over 10 repetitions.

The per-event average expectation in each bin is a function of the sampled ice properties and nuisance parameters. The nuisance parameters are further discussed in section 5.2.5.1. The likelihood function used for this purpose[220] is derived in[221] and is given as

[220] *commonly called the Dima likelihood*

[221] *Chirkin, "Likelihood description for comparing data with simulation of limited statistics"*

$$-\ln \mathcal{L} = \sum_i \left[ s_i \ln \frac{s_i/n_s}{\mu_s^i} + d_i \ln \frac{d_i/n_d}{\mu_d^i} + \frac{1}{2\sigma^2} \ln^2 \frac{\mu_d^i}{\mu_s^i} \right] \quad (5.17)$$

where $i$ denotes the current bin of the light curve, $s_i$ & $d_i$ the photon count in simulation and data for this bin, $n_s$ & $n_d$ the simulation repetitions and number of data events, $\sigma$ the so called model error and $\mu_s$ & $\mu_d$ the simulation and data expectation values.

The model error takes into account potential discrepancies in reproducing data with a simulation, which is potentially incomplete or uses non-ideal parametrizations[222].

[222] *It is currently set to 10%, but smaller values are being considered as current data-simulation discrepancies are already smaller.*

Using the model error, it is assumed that a difference between the expectation values of simulation and data can exist even at the best fit point, $\mu_s \neq \mu_d \neq (s_i + d_i)/(n_s + n_d)$. This is modeled through an additional penalty term in the likelihood[223]. This extension also requires an approximation of the now in-principle independent expectation values within the likelihood calculation and is performed as described in[224].

[223] *Chirkin, "Likelihood description for comparing data with simulation of limited statistics"*

[224] *Chirkin, "Likelihood description for comparing data with simulation of limited statistics"*

Due to the limited accuracy of the simulated detector response as implemented in PPC, DOM noise rates are not explicitly modeled but are added to the likelihood as an additional constant rate, assumed to be 500 Hz, summed to $s_i$. DOMs with a total charge exceeding either 500 PE or 1000 PE are excluded from the likelihood calculation, avoiding biases by saturated PMTs.

In addition, only bins within $-500$ ns to 1000 ns around the peak of each light curve are included in the likelihood calculation, thus minimizing the effect of noise, muons and other physics events accidentally overlapping with the flasher events. It also minimizes the impact of afterpulses which are not modeled as part of the flasher simulation.

This likelihood[225] is beneficial over a standard Poisson likelihood[226] as it takes into account the uncertainty of the expectation caused by the small statistics of the simulated data compared to the experimental data. Therefore, the expectation is optimized including the knowledge of the limited statistics of the simulated and experimental data. In the limit of infinite statistics of simulated data this likelihood converges to a saturated Poisson likelihood[227].

[225] *Chirkin, "Likelihood description for comparing data with simulation of limited statistics"*

[226] *Tanabashi, Hagiwara, et al., "Review of Particle Physics"*

[227] *Sebastian Schoenen (2017). "Discovery and characterization of a diffuse astrophysical muon neutrino flux with the iceCube neutrino observatory." en. RWTH Aachen University*

Two regularization terms are sometimes added to the likelihood function described in equation 5.17.

The first one is used to control the fluctuations of scattering and absorption coefficients in under-constrained ice layers[228] as a function of depth. It is formed of terms that are numerical expressions for second derivatives of scattering and absorption with respect to the position of the ice layer.

[228] *as for example given in the dust layer*

The second term is used to smooth the fluctuations in the correlation between scattering and absorption coefficients[229,230].

[229] *Chirkin, "Likelihood description for comparing data with simulation of limited statistics"*

[230] *enforcing the notion that both are proportional to the dust concentration*

### 5.2.4 *All-purpose flasher data set*

The currently most commonly used data set to fit the ice properties, and which is also employed for this work, is the so called all-purpose flasher data. It was collected in 2013 and includes all DOMs with reliably working flasher boards (4746) as emitters. For each emitter DOM several hundred events with all horizontal LEDs flashing at once and at maximum brightness and width were recorded.

In spring 2018, another data set containing events for all individual LEDs, both horizontal and tilted, was recorded. It was originally intended to measure the cable position with respect to every DOM[231]. In addition, this data set should in the future prove very useful to investigate potential biases introduced by only fitting to the horizontal LEDs and to further investigate directional dependencies.

[231] *see section 9.5.1*

### 5.2.5 *The ice model to be fitted*

The "ice model" is the comprehensive term for a set of parameters used to describe the ice in photon propagation. The detailed cm-sized stratigraphy associated with the yearly layering[232] can not be constrained through the flasher data, nor is it necessary to accurately describe the photon propagation over dozens of meters.

[232] *see section 3.5*

Instead, as a reasonable compromise, the ice layer width was early on chosen to be 10 m, in order to maintain at least one receiving DOM in each layer, and has not been changed since. The ice table, tabulating the absorption and scattering coefficients in all layers, extends from a depth of 1100 m to 2800 m (just above the bedrock).

As not all depths are within the instrumented volume, properties above the detector are taken from AMANDA measurements, or are extrapolated from dust logger data. Properties below are extrapolated using the stratigraphy as obtained from the EDML ice core.

Following the description in section 5.2.1 for scattering and section 3.1 for absorption[233], each layer is described by its dust induced absorption and scattering coefficients at a wavelength of 400 nm.

[233] *neglecting the Urbach tail, as the DOM is not sensitive below* 250 nm

These are scaled to other wavelengths and the total absorption coefficients including the ice intrinsic absorption as[234]:

[234] Aartsen et al., "Measurement of South Pole ice transparency with the IceCube LED calibration system"

$$a(\lambda) = a_{dust}(\lambda) + A_{IR}e^{-\lambda_0/\lambda} \cdot (1 + 0.01 \cdot \delta\tau) \tag{5.18}$$

with

$$a_{dust}(\lambda) = a_{dust}(400\,\text{nm}) \cdot \left(\frac{\lambda}{400\,\text{nm}}\right)^{-\kappa} \tag{5.19}$$

for the absorption coefficients, with tabulated temperature differences $\delta\tau$ compared to the temperature at 1730 m, and

$$b_e(\lambda) = b_e(400\,\text{nm}) \cdot \left(\frac{\lambda}{400\,\text{nm}}\right)^{-\alpha} \tag{5.20}$$

for the effective scattering coefficients.

While in principle all parameters are depth dependent, e.g. due to changes in the dust composition, some are deemed constant enough to be described by a single global value or functional parametrization. These are:

- The coefficients $\alpha$ and $\kappa$ describing the wavelength dependence of scattering and absorption.
- $A$ and $\lambda_0$ in the infrared absorption term.
- The scattering function parameters $g$ and $f_{SL}$.
- The parametrization for scattering on bubbles at shallow depths, which is not fitted but taken from AMANDA.
- The parametrization of tilt, which is depth dependent and is derived from the dust logger.
- Traditionally the anisotropy is described by one direction and two strengths coefficients. As part of this thesis a depth dependent parametrization is explored in chapter 10.

This leaves 9 global parameters (4 for the wavelength dependency and intrinsic ice absorption, 2 for the scattering function, and 3 for the anisotropy) and about 100 layers within the instrumented volume, with 2 parameters each, to be constrained by the fit.

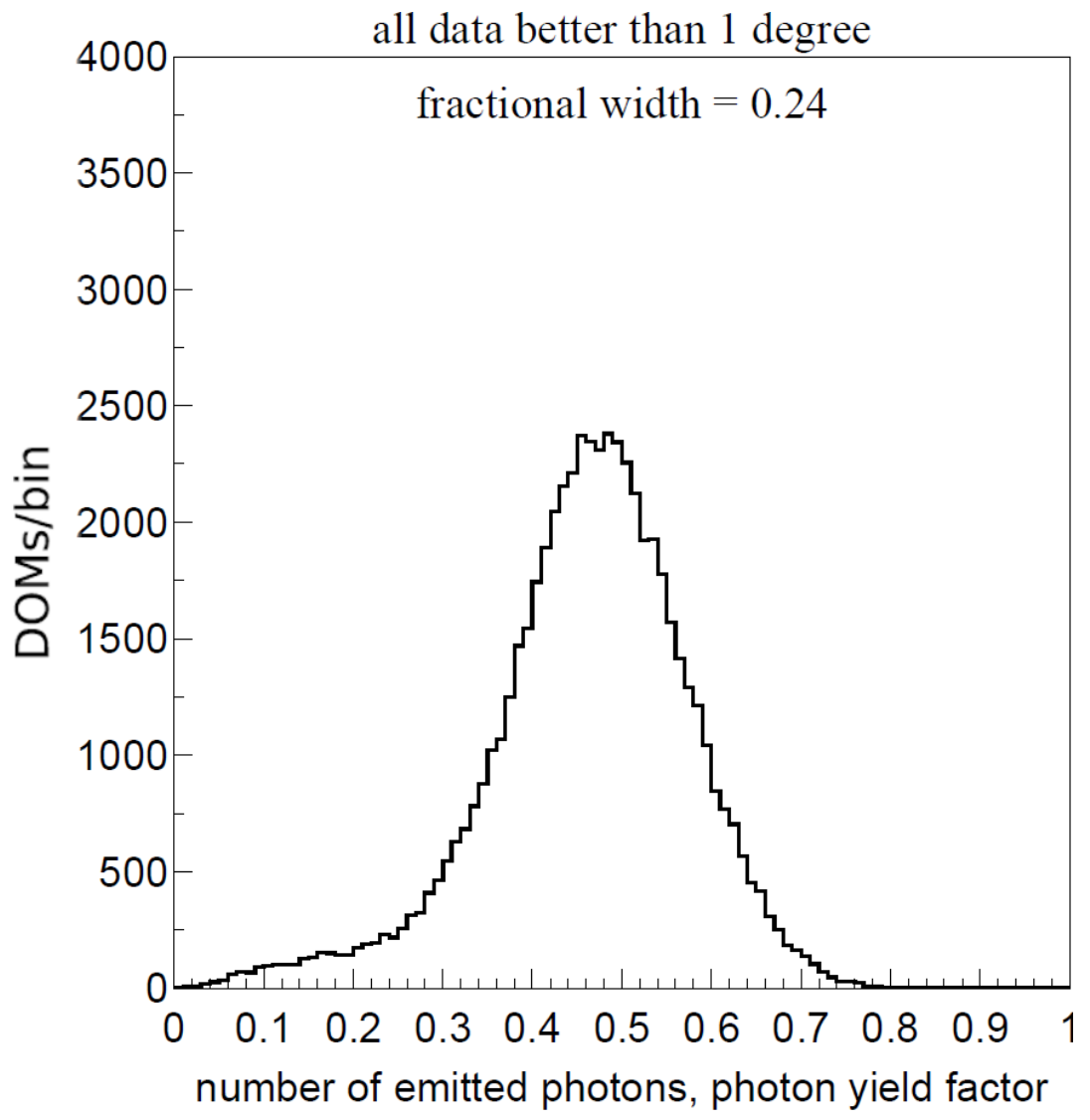

Figure 5.14: Relative light yield of all individual LEDs as fitted during the nuisance parameter evaluation. [Dmitry Chirkin (2019). "Private Communication"]

#### 5.2.5.1 *Nuisance parameters*

In addition to the parameters as defined by the ice model, the fit also constrains a number of per-DOM nuisance parameters required to accurately match the data and to reduce biases. These nuisance parameters can be categorized in being flasher specific properties or DOM specific properties. The later can, if deduced reliably, also be applicable to all other analyses. The LED specific nuisance parameters are the global time offset between the current pulse and the instance of light emission, the total photon yield, and the angular LED emission profile.

To make the fit independent of the badly constrained total LED intensities, the simulation simply keeps injecting photons until the total number of photons as measured in data is reached.

The fit is thus only sensitive to the relative registered intensity between DOMs and their timing profiles. The spread in flasher intensity required to match the data is seen in Figure 5.14 and found to be 24%. This value agrees with measurements performed during the final acceptance testing[235].

[235] *Wendt, "private communication"*

While the flashers appear as nearly isotropic point sources at large distances, at smaller distances their angular emission profile remains visible. In simulation it is modeled through a number of analytic models or deduced from the data itself. These LED emission profiles are based on lab measurements, with a simple Gaussian intensity profile being the default.

Prior to 2018 the individual azimuthal LED orientations[236] were not known. Instead, the summed emission of the six horizontal LEDs in the all-purpose flasher data was approximated as a rotationally symmetric "donut" around the DOM. Alternatively the flasher profile can be unfolded from the data itself.

[236] *see section 9.5.1*

The process is sketched in Figure 5.15 and works as follows:

1. For a given emitter DOM, simulate 74 flasher events assuming a single horizontal LED[237] positioned in 5 degree steps in azimuth, as well as an LED pointing straight up or down.

2. Use these simulated events to create a transfer matrix ($A$) between the flasher orientation bins ($\vec{x}$) and simulated flasher events. Where the simulation data is represented by a vector ($\vec{y}$) containing a concatenation of all light curves[238] on all receiver DOMs. This vector typically has around twenty million entries.

3. Apply the inverse of the transfer matrix ($A^{-1}$) to the vector representing real events of this DOM, yielding the real azimuthal intensity profile.

[237] *two-dimensional Gaussian (9.7°) intensity profile*

[238] *see Figure 5.13 for an example*

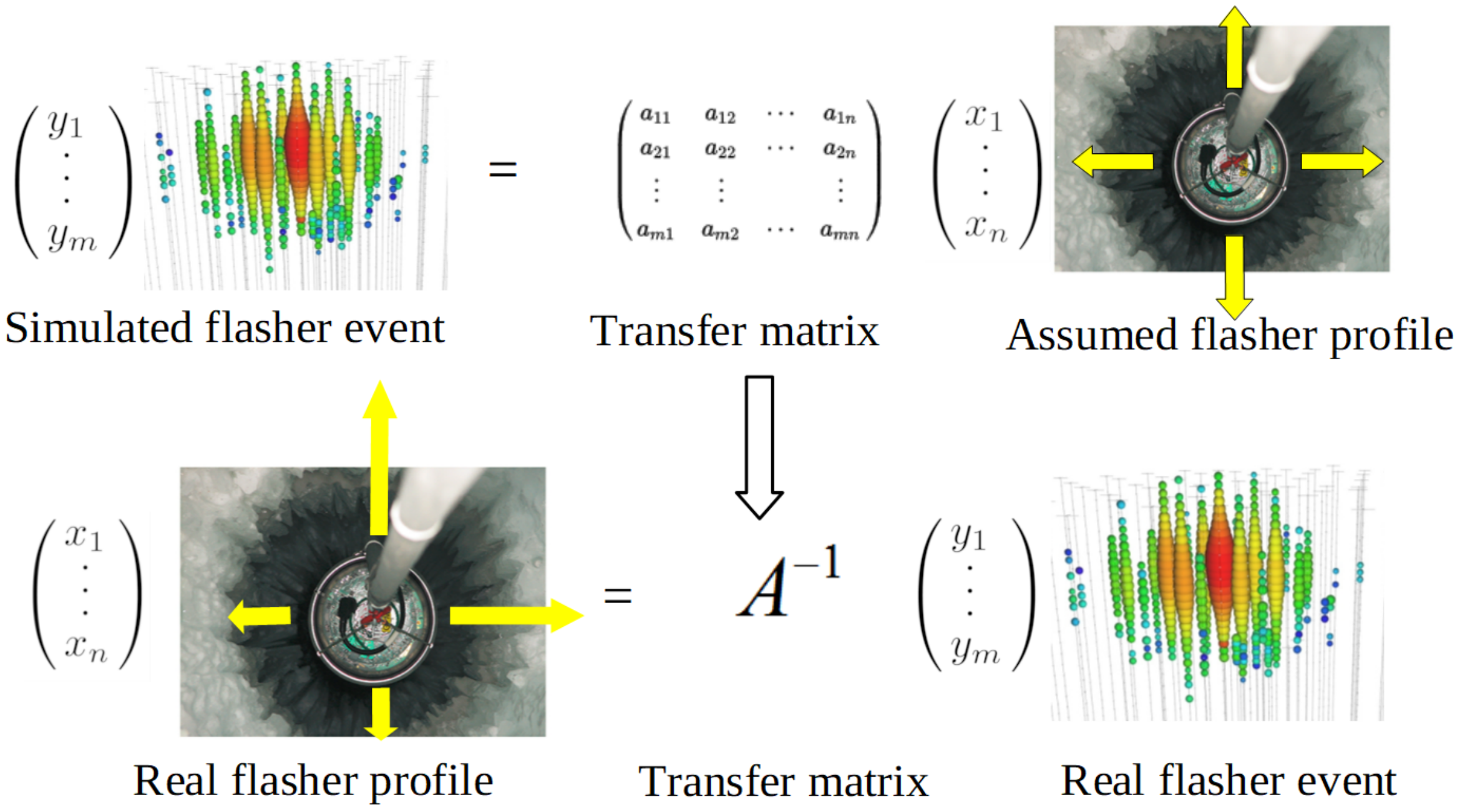

Figure 5.15: Principle description of the flasher unfolding technique. The transfer matrix between assumed flasher profiles and the resulting event characteristics is deduced from simulation and the inverse matrix applied to the data to yield the true flasher profile.

The flasher unfolding was tested to reliably work when reconstructing single LEDs and was subsequently also applied to the all-purpose flasher data. As described in the context of hole ice studies in section 9.4.2.1, the flasher unfolding breaks down when applied to the all-purpose flasher data and biases in particular the anisotropy strength.

The DOM specific nuisance parameters are the relative detection efficiencies and the hole ice modified angular acceptance curves. The relative DOM efficiencies were already presented as part of the hardware uncertainties in section 5.1.1. Angular acceptance curves are discussed in section 9.2.

### 5.2.6 *Fitting procedures*

The parameters of the ice model are generally obtained through likelihood scans, where each scan point is one realization of the ice model parameters tested against (the all-purpose) flasher data. The timing offset and LED intensity nuisance parameters are optimized for each realization through a number of low statistics iterations.

While automated grid scans are possible and have been used in the past when only considering data from one emitter string, the time required per likelihood evaluation[239] generally makes automated minimization / scanning, in particular of many parameters at once, prohibitive.

[239] $\sim$ 400 *GPU hours per emitter string at oversizing 1*

Instead, the likelihood space is evaluated manually with (at most two dimensional) slices through the parameter space where the next realization is guessed from the already covered likelihood profile or chosen to sufficiently enclose the found minimum.

The Dima likelihood does generally not fulfill Wilks Theorem[240,241]. As such, the log-likelihood (LLH, sometimes also denoted goodness of fit or GOF) contour of a one dimensional likelihood scan enclosing the minimum by a $\Delta LLH$ of 1 does not represent a $1\sigma$ statistical uncertainty.

[240] *Wilks, "The Large-Sample Distribution of the Likelihood Ratio for Testing Composite Hypotheses"*

[241] *Chirkin, "Private Communication"*

Instead, the spread in LLH values equivalent to the $1\sigma$ uncertainty is obtained by re-simulating a realization close to the optimum a number of times and computing the standard-deviation of the resulting $LLH$ values. Alternatively the uncertainty can be accessed through the spread of $LLH$ values of the grid scan close to the minimum with respect to a fitted paraboloid.

Using the all-purpose flasher data set, the statistical errors on the bulk ice properties, in particular the layered absorption and scattering coefficients, are generally below 1%.

At this level, systematic uncertainties and in particular also the choice of parametrization for currently not well understood effects[242] largely dominate the total uncertainty.

[242] *such as the hole ice or the anisotropy*

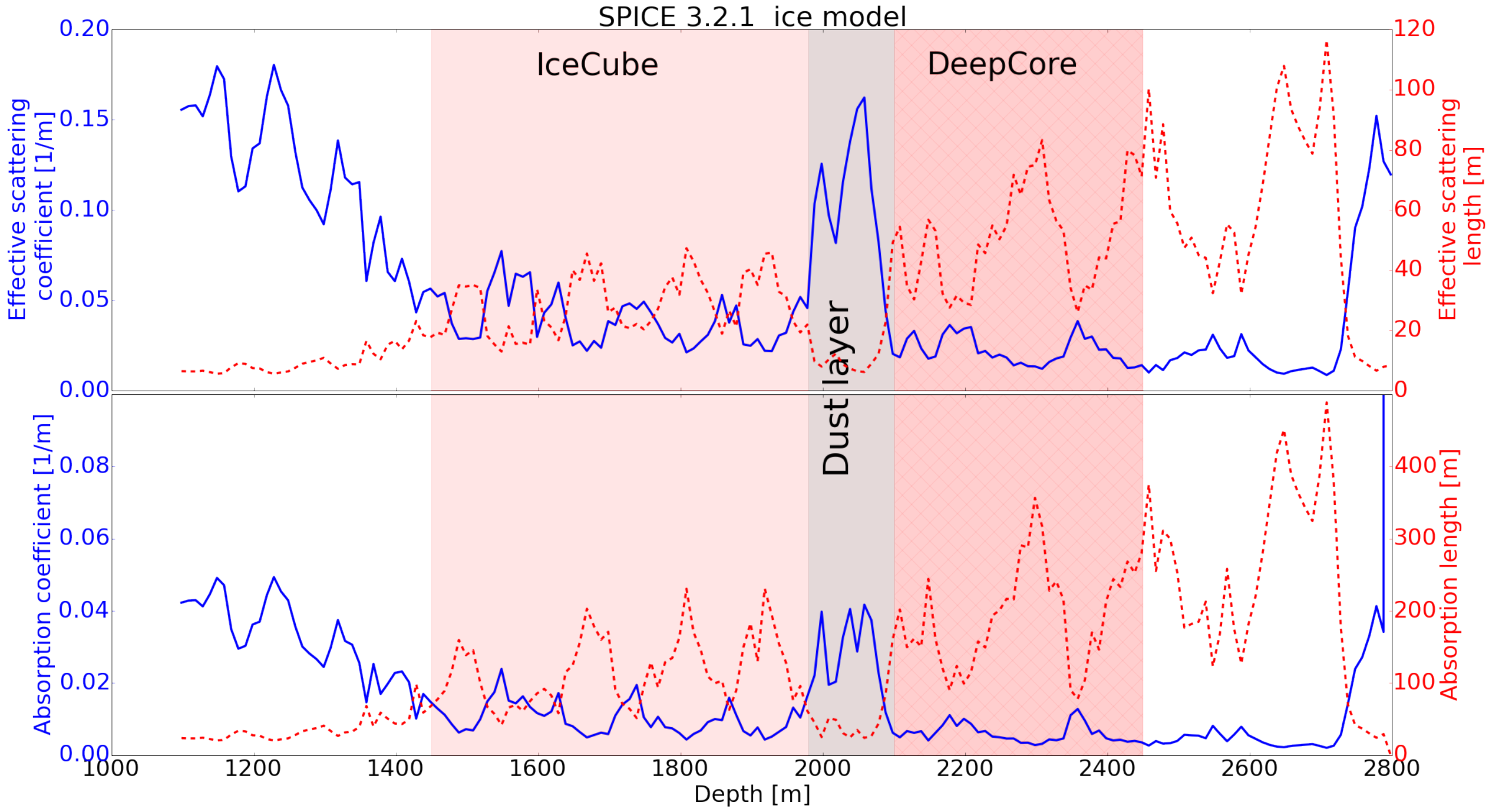

Figure 5.16: Layering of the scattering and absorption coefficients as realized in Spice3.2.1, the current default ice model.
[Summer Blot et al. (2018). "Ice model wiki"]

Evaluating the total uncertainty that is introduced by systematics is both computationally and conceptually challenging as the entire fit has to be repeated under the varied uncertainties. In addition, some systematics do not have well understood parameter ranges.

The total uncertainty introduced by a set of well-established systematics was last evaluated for the Spice MIE[243] ice model in 2013. The fit was repeated assuming either an instantaneous LED emission or a realistic rectangular timing profile of 62 ns width. This resulted in a $\sim$ 6% bias on the absorption and scattering coefficients. The fit was also repeated for different pulse extraction and saturation correction methods, yielding a bias of $\sim$ 4%.

[243] *Aartsen et al., "Measurement of South Pole ice transparency with the IceCube LED calibration system"*

As the saturation exclusion limit has since been reduced from 1000 PE to 500 PE and the true uncertainty on the flasher timing profile is far smaller, the contributions from these two unknowns are overestimated. A total systematic uncertainty of 10% is still used as a generic approximation for most analyses.

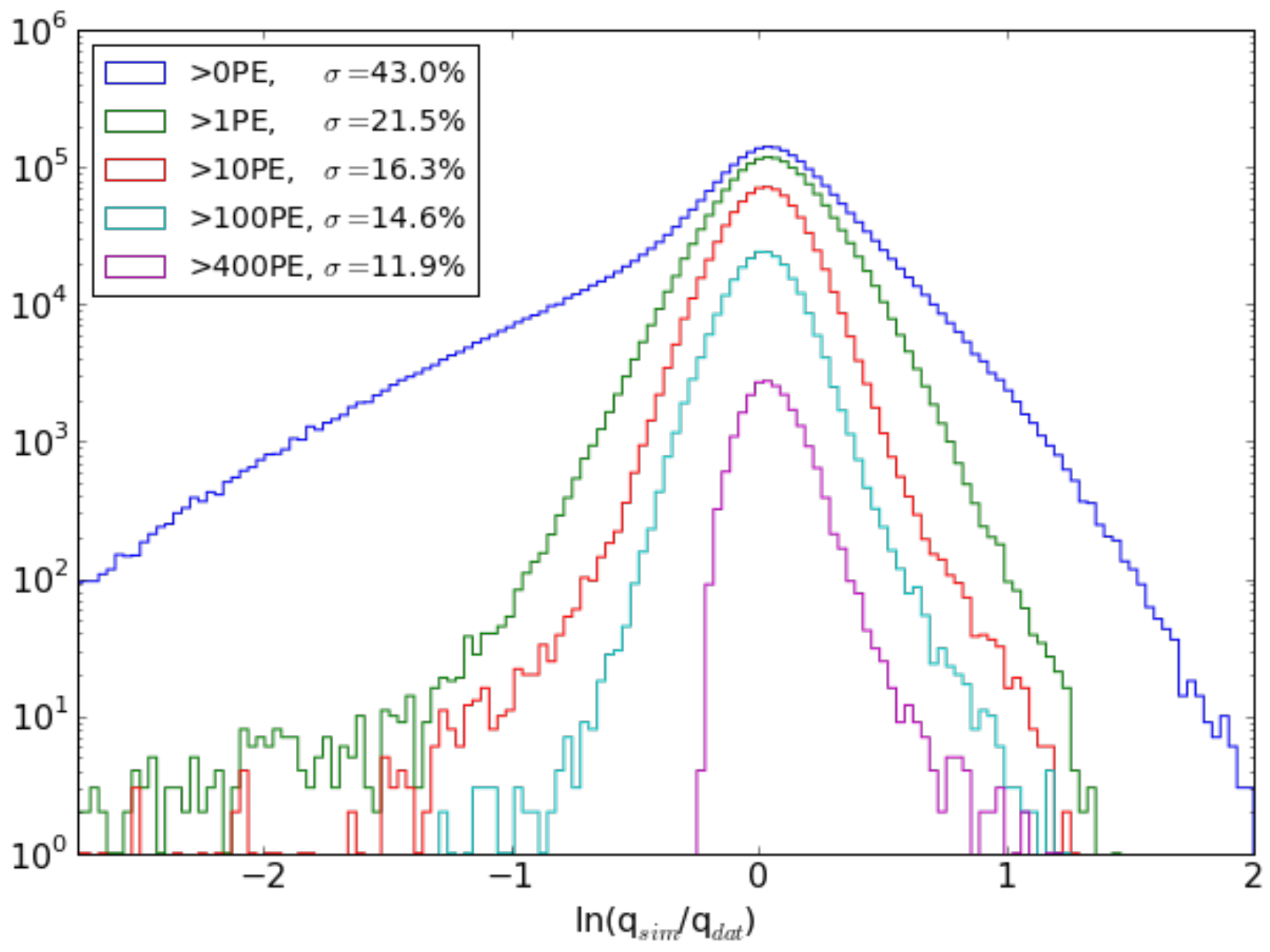

Figure 5.17: Example model error histogram. The different colors denote different required minimum integrated average charges in both data and simulation.

## 5.2.7 *Model error*

The improvement of a new ice model is usually characterized in terms of the absolute improvement in the flasher likelihood value. While this might seem unconventional, it is applicable as different models are always fitted to, or at least tested against, the same data and the number of parameters does not change significantly between models[244] compared to the number of flasher data bins, leaving the number of degrees of freedom essentially unchanged.

[244] *even when introducing new depth dependent quantities*

Still, in order to provide a more physically interpretable quality parameter, which can also be compared between different data sets and experiments, the so called model error[245] has been introduced[246].

[245] *which does share many properties with the model error employed in the likelihood description in section 5.2.3*

[246] *Aartsen et al., "Measurement of South Pole ice transparency with the IceCube LED calibration system"*

It is defined as the standard deviation of natural logarithms of the ratios of total average charge per emitter-receiver pair in data ($q_{dat}$) and simulation ($q_{sim}$). See Figure 5.17 for an example from a recent ice model.

As charges are subject to Poisson fluctuations, the model error is only a robust measure when requiring a minimum number of PE in both data and simulation and when simulating with enough repetitions. Usually the model error for emitter-receiver pairs with more than 10 PE and using 100 simulation repetitions is quoted.

### 5.2.8 *History of ice models*

The effects described above are implemented in the most recent ice models. Naturally the ice modeling has gradually evolved starting in AMANDA and is still being refined today. This section aims to give a short overview of ice model development. An extensive list can be found at [247].

[247] *Summer Blot et al. (2018). "Ice model wiki"*

Starting from a "bulk" model without any layering, at least ten ice models were released during AMANDA times between 1995 and 2005 using both muon data and artificial light sources. The most thorough and influential AMANDA ice model was the millenium/Y2k model published 2006[248]. An update called AHA was released in 2007. (Both yielded model errors of ∼55%.)[249]

[248] *Ackermann, Ahrens, et al., "Optical properties of deep glacial ice at the South Pole"*

[249] *An interesting footnote in ice model history is also the Lordi model released several months after Y2k. It tried to scale the high resolution dust logs to the absolute coefficients of Y2K.*

In 2011 WHAM[250] , the last model based on AMANDA analysis methods and software tools but based on IceCube data was released (42% model error).

[250] *acronym for: Water (Hardened) Antarctic Measurement*

Following the development of PPC and the Dima likelihood the first generation Spice[251] model was released in 2009. In addition to the new tools it introduced angular acceptance modeling in ice fitting and was one of the first models to include EDML ice core information for extrapolation to greater depths. To reduce the computational complexity Spice1 was based on data from string 63[252] (29% model error).

[251] *acronym for: South Pole Ice*

[252] *located in the center of IC40*

Spice2 in 2010 added tilt and abandoned AMANDA models as a seed for fitting, instead forcing a correlation between the absorption and scattering coefficients.

2010 also saw the release of Spice Mie[253] which for the first time allowed fitting to the scattering function. The ice anisotropy was introduced with Spice Lea in 2012, resulting in a large improvement in model error down to 20%.

[253] *Aartsen et al., "Measurement of South Pole ice transparency with the IceCube LED calibration system"*

Spice Munich released in 2013 introduced LED unfolding. Angular acceptance unfolding followed in 2015 with Spice3. At the same time the amount of flasher data utilized was extended to 7 and 85 strings respectively.

Spice3 was updated in spring 2016 to Spice 3.2, refining the layer fitting and resulting in a model error of 10%. This model is currently considered the default for most analyses in IceCube.

While Spice3.2 is still considered the default model today, several faults in its parameter values, such as an overestimated

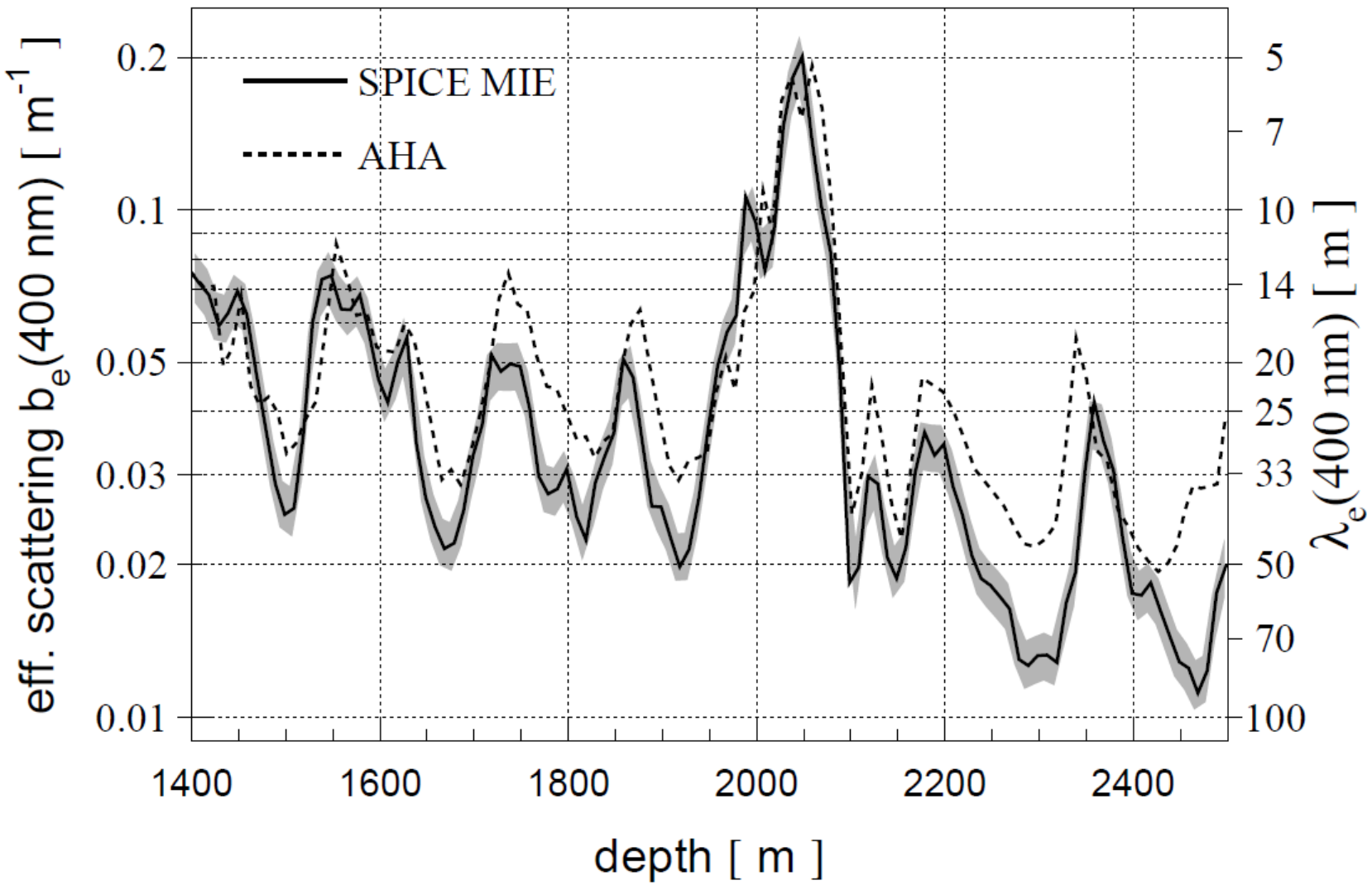

Figure 5.18: Comparison of the scattering coefficients as stated in the AHA (late AMANDA) and Spice Mie (early IceCube) ice models. [M. G. Aartsen et al. (May 2013c). "Measurement of South Pole ice transparency with the IceCube LED calibration system"]

anisotropy, as well as modeling insufficiencies have been identified. In fall 2016 SpiceHD[254] was released as a modification to Spice3.2. It introduces a more physically motivated modeling of the hole ice. The final iteration of DOM orientations fitted to single-LED data[255], enabling a precise simulation of the downhole cable, was presented in spring of 2018.

254 *see section 9.4*

255 *see section 9.5.1*

While several improvements with respect to the anisotropy description[256], have been achieved and released as updates to Spice3.2, a fully new ice model iteration will only be released following the completion of the fit of the birefringence based anisotropy model[257].

256 *see section 10.4*

257 *see section 10.8*

# 6

# DETECTOR SYSTEMATICS AND SELECTED ICECUBE RESULTS

**This chapter shall summarize how the detector systematics, some of which are explored in this thesis, impact the sensitivity of high-level physics analyses.**

## 6.1 SYSTEMATICS

The accuracy of any experiment is fundamentally limited by the statistics it can accumulate and the quality of understanding of the detector response.

During operation, the statistical gain in sensitivity approximately scales with the square-root of lifetime. As a result, the systematic uncertainties will at some point always become dominant to the overall error and future improvements can only be achieved by improving the understanding of the detector.

For most applications the relevant systematics in IceCube can be classified as originating from hardware effects, uncertainties in the ice properties or through the physics assumptions.

### 6.1.1 *Hardware*

A large number of hardware parameters such as the PMT transit times, ADC gains or DOM clock speeds are addressed by the calibration. Most of these are known precisely enough to not require further consideration at analysis level.

A hardware property universally considered at analysis level is the absolute photon detection efficiency averaged over all DOMs. Its impact can be understood as a shift in reconstructed energy[258] and assumed trigger threshold. As described in section 5.1.1, the determination of the absolute DOM efficiency is challenging and often correlates with ice properties. Usually a 10% uncertainty on the absolute photon detection efficiency is assumed at analysis level.

[258] *Cascade energy reconstructions directly and muon energy reconstructions indirectly scale with the total observed charge, as explained in section 2.3.2.*

### 6.1.2 *Ice*

Uncertainties in the ice parametrization can be classified as resulting from the bulk ice or the hole ice modeling. While the bulk ice has many uncertain parameters, at analysis level these are usually tested by either switching between different discrete iterations of the ice model or by globally scaling all absorption and/or scattering coefficients.

For the hole ice different angular acceptance curves as well as the more modern SpiceHD[259] model are tested. While the bulk ice primarily changes the energy scale and only introduces subtle changes to the event signatures, the hole ice parametrizations are constructed such that they do not change the total recorded charge but its distribution in the detector.

[259] *see section 9.4*

### 6.1.3 *Physics assumptions*

In addition to detector uncertainties, many analyses are sensitive to uncertainties of other physics parameters not targeted by the analysis.

For most analyses these include the atmospheric neutrino fluxes, cross sections or neutrino oscillation parameters. While often just as important as detector systematics, this source of uncertainty is not further discussed in this thesis.

## 6.2 SELECTED RESULTS

Research topics covered by the IceCube Neutrino Observatory reach beyond the already discussed search for astrophysical neutrinos and their sources or the measurement of neutrino oscillation parameters. These are topics such as dark matter searches, supernova physics, cosmic rays and solar physics.

In the following three kinds of analyses are highlighted, in order to point out the different detector systematics effecting each.

### 6.2.1 *Astrophysical neutrinos*

#### 6.2.1.1 *Diffuse spectrum*

After having been discovered by the HESE analysis[260], the astrophysical neutrino flux has also been measured using through-going, upwards pointing muons[261].

[260] *M. G. Aartsen et al. (Sept. 2014d). "Observation of High-Energy Astrophysical Neutrinos in Three Years of IceCube Data"*

[261] *M. G. Aartsen et al. (Dec. 2016c). "Observation and Characterization of a cosmic muon neutrino flux from the northern hemisphere using six years of IceCube data"*

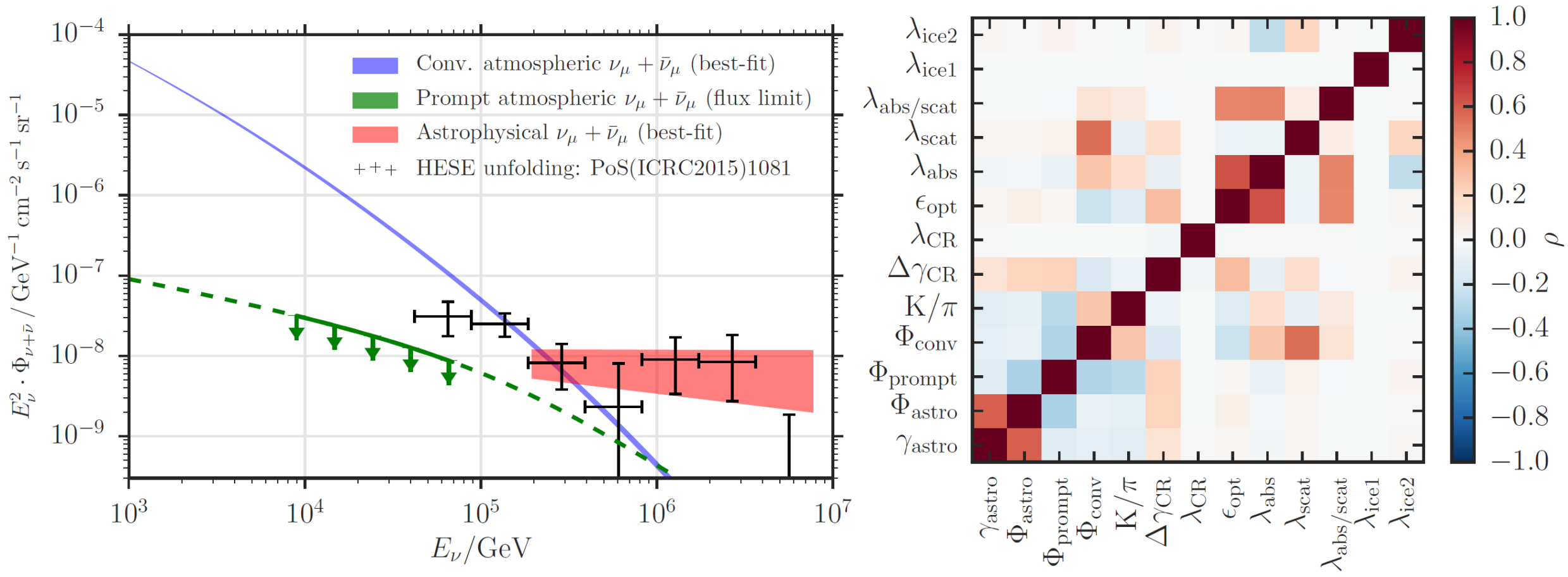

Figure 6.1: Left: Energy spectrum of atmospheric and astrophysical neutrinos as measured using upgoing muon neutrinos. Right: Correlation matrix of signal and nuisance parameters. $\lambda_{ice}$ describes the linear combination between different ice models
[M. G. Aartsen et al. (Dec. 2016c). "Observation and Characterization of a cosmic muon neutrino flux from the northern hemisphere using six years of IceCube data"]

This analysis provides a larger statistics and allows for more detailed systematics studies due to the inclusion of a large, low-energy background region.

Figure 6.1 features both the final atmospheric and astrophysical spectra measured by this analysis as well as the correlation between the physics and nuisance parameters considered.

While the choice of ice-model is seen to have a marginal effect, the overall absorption and scattering strengths as well as the DOM efficiency are seen to impact the conventional (atmospheric) flux normalizations as they alter the energy scale.

#### 6.2.1.2 *Point source searches*

Given a sample enriched with astrophysical neutrinos, their distribution on the sky can be analyzed for clustering as expected from point sources which emitted several registered neutrinos.

Even for point source analyses the DOM detection efficiency can be a relevant systematic as a change in energy scale changes the assumed relative contribution of background and signal events.

For a given assumed spectrum and large enough statistics the sensitivity at any point in the sky scales as[262]

$$\frac{S}{\sqrt{B}} \propto \frac{1}{\sqrt{\pi \cdot \sigma^2}} \propto \frac{1}{\sigma}, \tag{6.1}$$

with $S$ being the number of signal and $B$ the number of background events. It is thus directly inversely proportional to the experimental angular resolution $\sigma$.

262 *Reimann, "Private communication"*

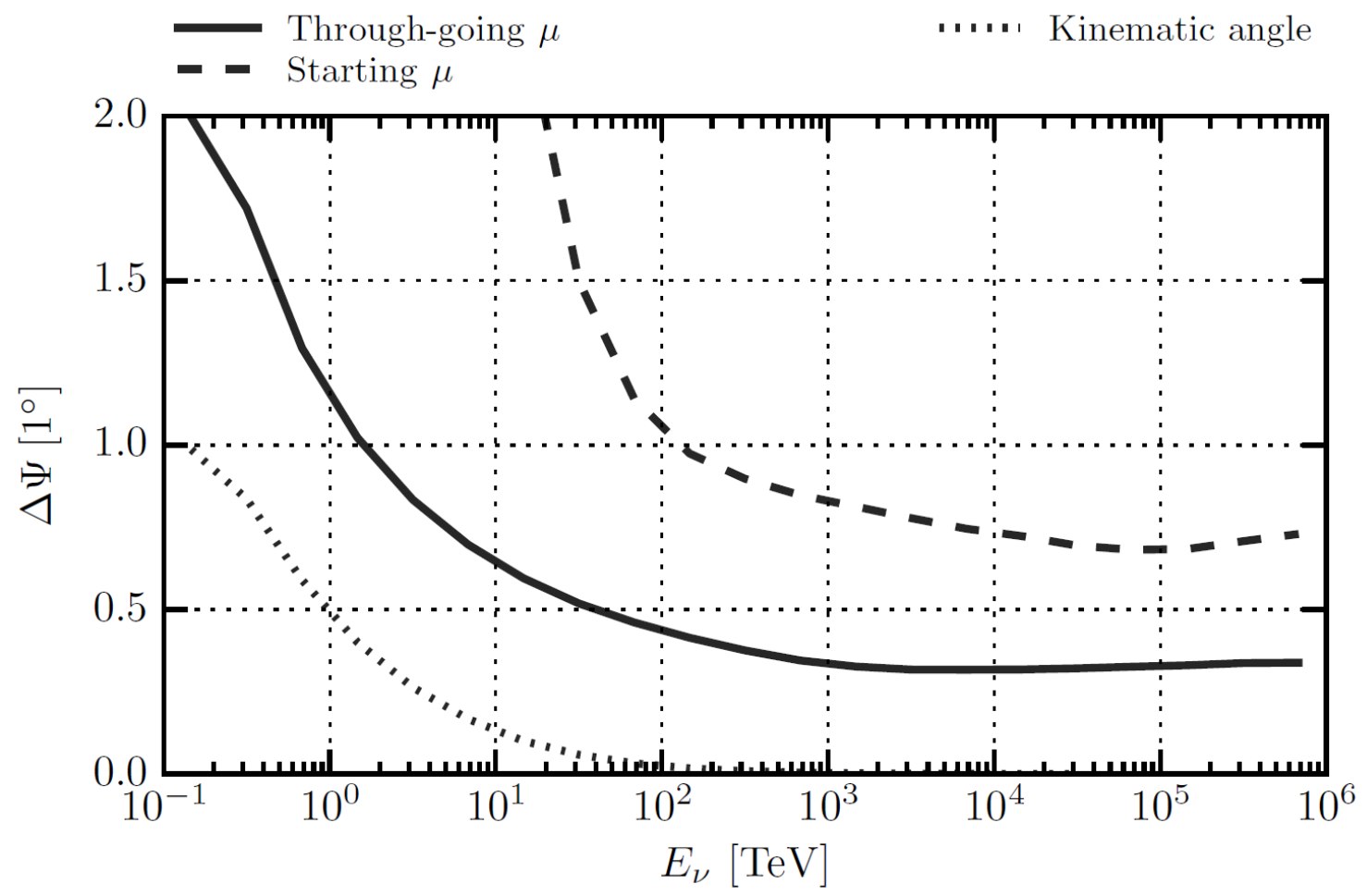

Figure 6.2: Angular resolution for muon tracks compared to the kinematic opening angle. This plot considers the statistical error only. [M. G. Aartsen et al. (Jan. 2017a). "All-sky Search for Time-integrated Neutrino Emission from Astrophysical Sources with 7 yr of IceCube Data"]

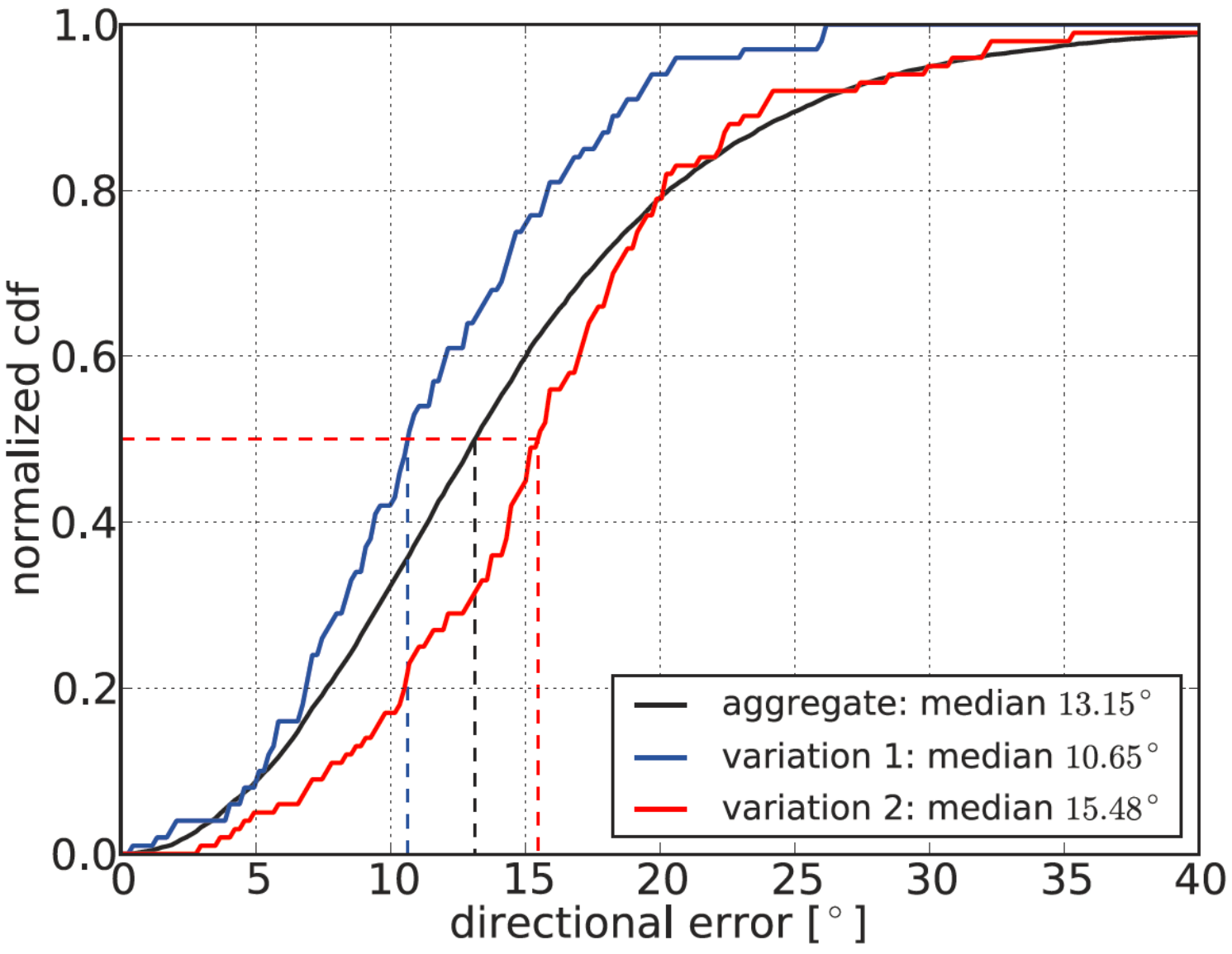

Figure 6.3: Angular resolution for an individual example cascade event, which was re-simulated using two different ice model variants while not changing the ice assumptions during reconstruction. [Claudio Kopper and other (2013). "Analysis of the High-Energy Starting Events in IceCube"]

The angular resolution is fundamentally limited by the kinematic opening angle between the charged lepton and the neutrino. The difference between the kinematic limit and the experimental resolution results from a limited number of DOMs sampling each event and uncertainties induced through the limited knowledge of the ice properties.

Figure 6.2 presents the statistical angular uncertainty of track events compared to the kinematic limit. Figure 6.3 shows the uncertainty introduced in cascade reconstructions due to the ice uncertainty.

#### 6.2.1.3 *Flavor composition*

Apart from the energy spectrum and angular distribution, the astrophysical events can also be studied with respect to their flavor composition.

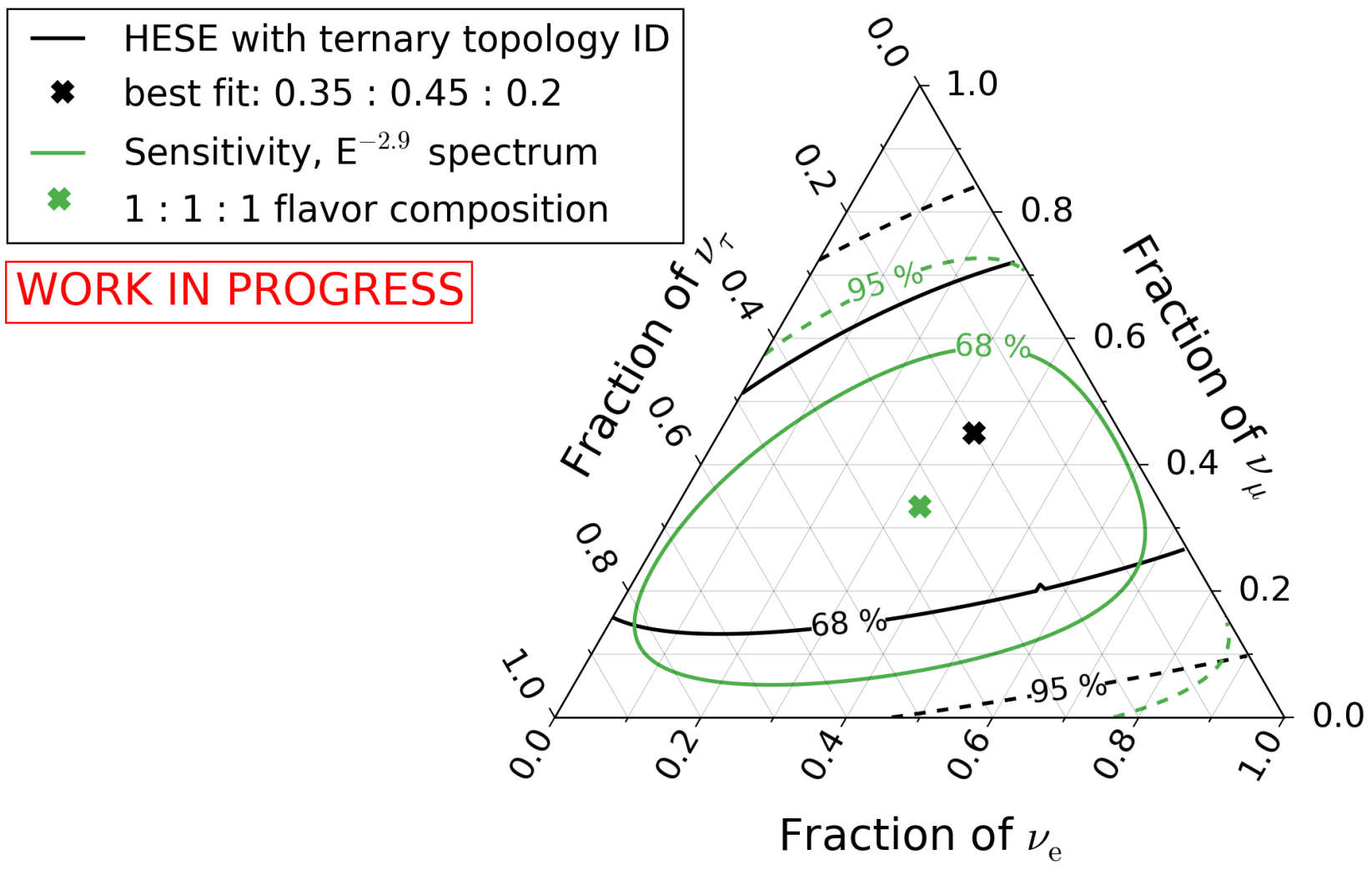

Figure 6.4: Measured flavor composition of astrophysical neutrinos in the HESE analysis. [Juliana Stachurska et al. (2018). "New Measurement of the Flavor Compositionof High-Energy Neutrino Events with ContainedVertices in IceCube"]

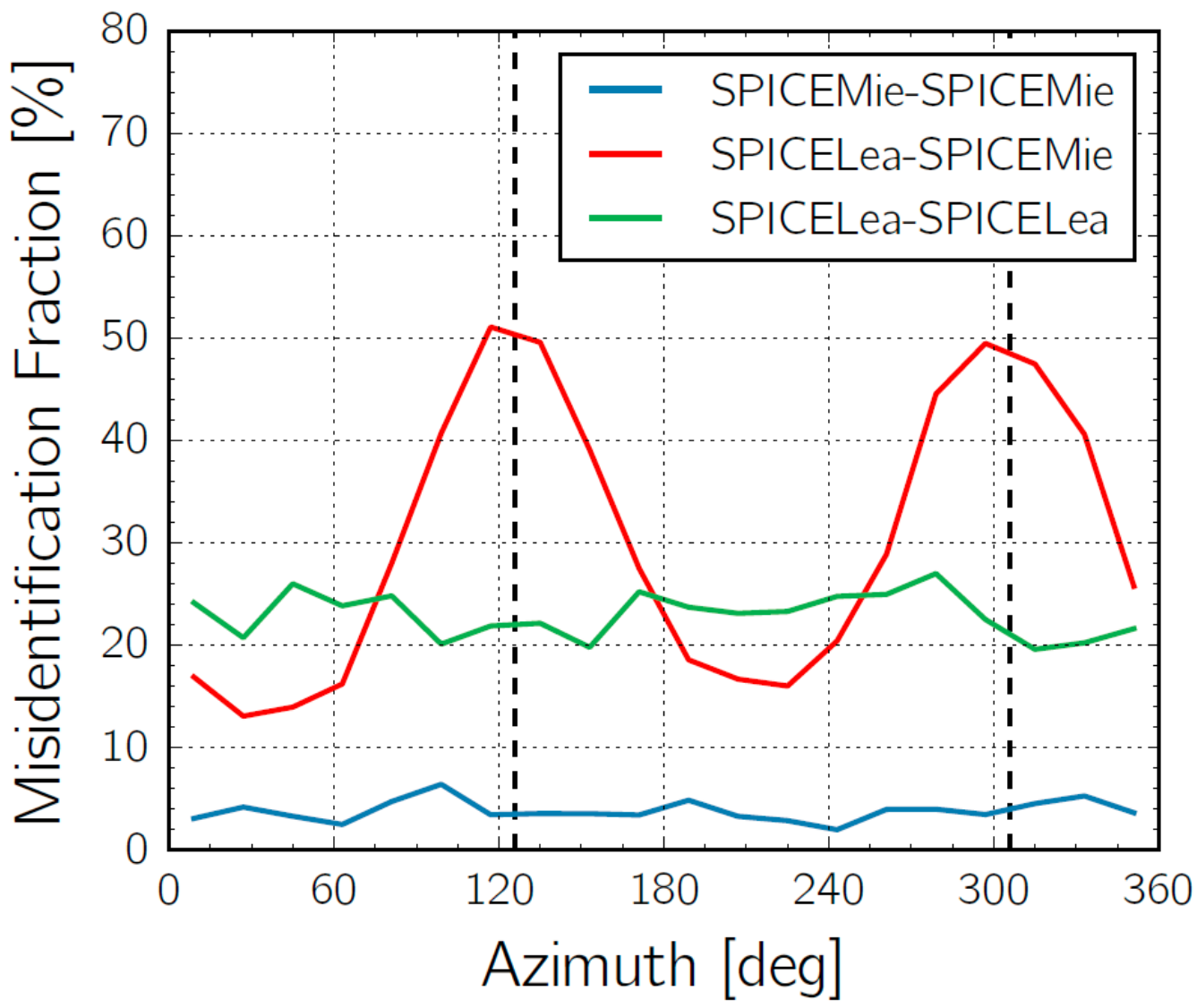

Figure 6.5: Monte carlo study on double cascades being misidentified as single cascades when reconstructing with a different ice model compared to the simulation. [Marcel Usner (2018). "Search for Astrophysical Tau-Neutrinos in Six Years of High-Energy Starting Events in the IceCube Detector." en. Humboldt-Universität zu Berlin]

The primary flavor distinction is given by the event signature, as electron neutrino interactions do not[263] result in a muon track. In order to identify tau neutrinos, the vertex separation that is expected for double cascade interactions (see section 2.3.2.3) is searched for in terms of a spatial separation[264] of the light signature.

In addition to the already discussed ice uncertainties, these searches are highly sensitive to mis-modeling of the anisotropy, as it can fake the characteristic spatial separations.

Figure 6.4 highlights the current knowledge of the flavor mixing of astrophysical neutrinos. The misidentification fraction of reconstructed double cascade events as a function of the reconstructed azimuth angle for different ice model combinations in simulation and reconstruction is shown in Figure 6.5.

263 *except at very high energies through the Glashow resonance*

264 *Wille et al., "Search for Astrophysical Tau Neutrinos with the IceCube Waveforms"*

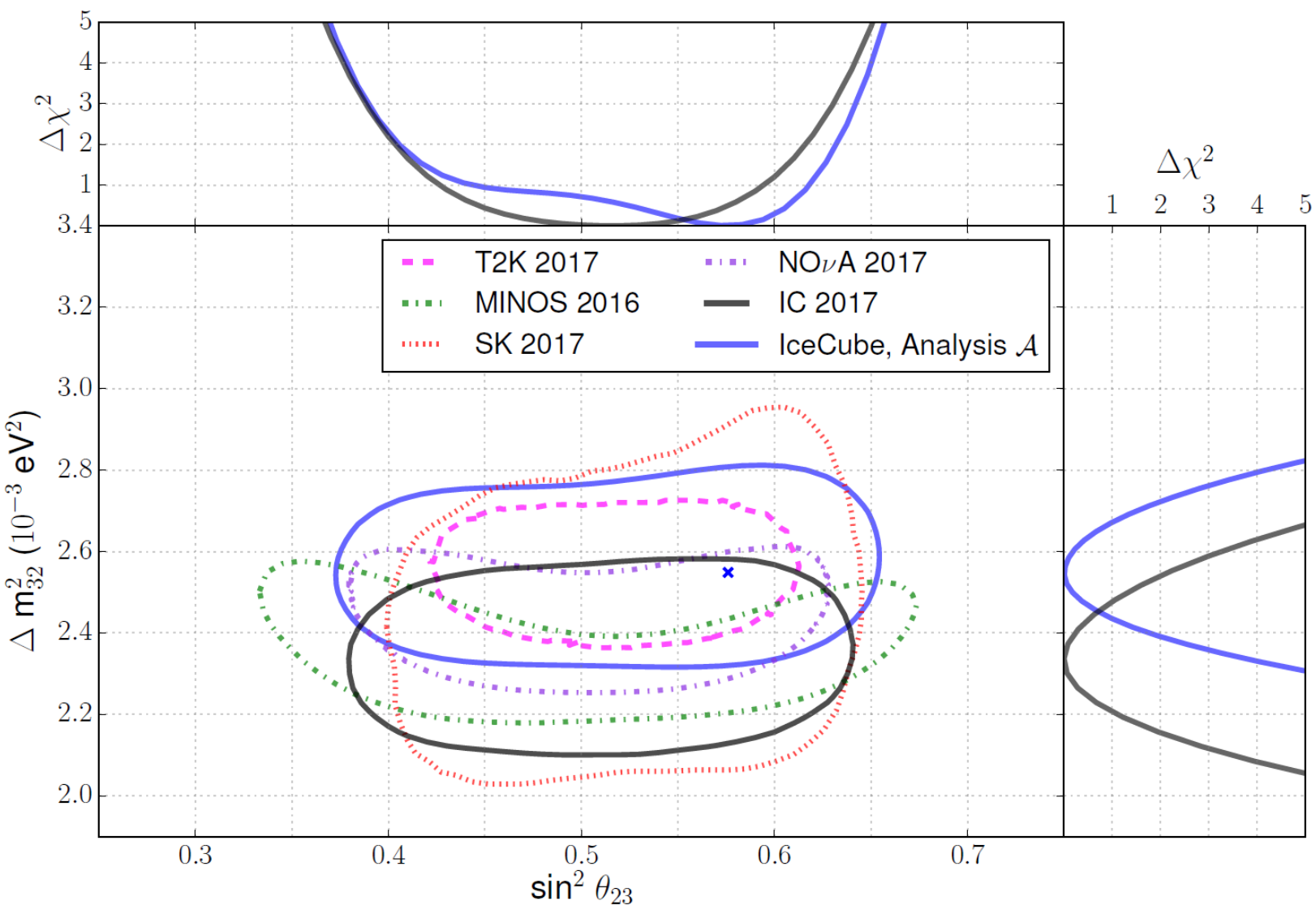

Figure 6.7: Recent IceCube neutrino oscillation result. Associated systematics are shown in Figure 6.6. [M. G. Aartsen et al. (Jan. 16, 2019c). "Measurement of Atmospheric Tau Neutrino Appearance with IceCube DeepCore"]

As the available anisotropy models are known to not fully describe the data, a currently unaccounted for and unknown misidentification fraction remains regardless of the employed ice model. Chapter 10 discusses a re-evaluation of the anisotropy modeling.

## 6.2.2 *Neutrino oscillations*

Given the oscillation parameters as described in section 2.1.1 and the Earth's diameter as the length scale for propagation baselines, neutrino oscillations can be resolved at GeV energies[265]. For this purpose IceCube features the infill detector DeepCore. At the short propagation distances probed in DeepCore, errors in the bulk ice modeling are usually considered sub-dominant.

[265] *M. G. Aartsen et al. (Jan. 16, 2019c). "Measurement of Atmospheric Tau Neutrino Appearance with IceCube DeepCore"*

Instead, systematics studies focus on local effects, mostly through different parameterizations of the hole ice using angular acceptance curves and the resulting changes on the zenith distribution.

Figure 6.7 depicts the result of a neutrino oscillation measurement as described in [266]. The uncertainty contour contains statistical and systematic uncertainties, with the relative contribution of the systematic uncertainties to the one sigma contour being broken down in Figure 6.6. The detector systematics contribute 41% to the total uncertainty. The hole ice modeling has the largest single contribution.

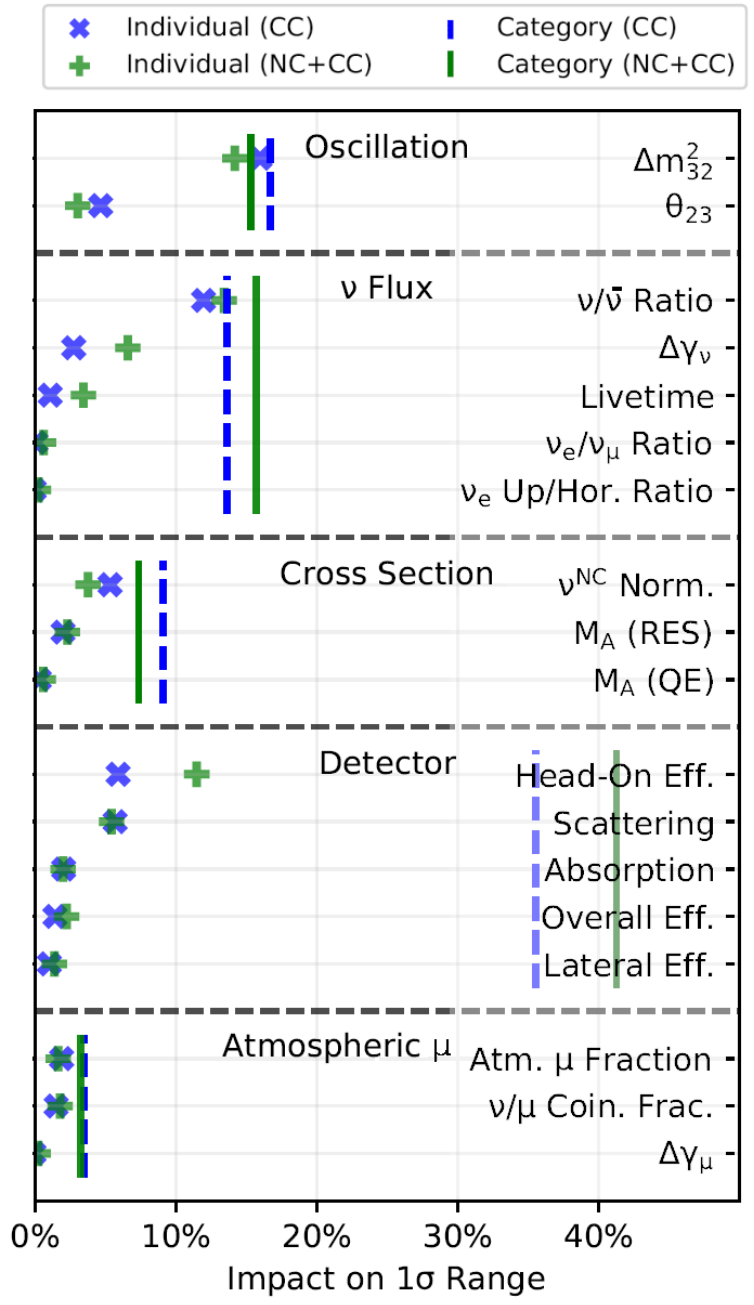

Figure 6.6: Overview of systematics effecting the DRAGON tau appearance analysis. The detector systematics introduce the largest single effect on the total uncertainty. [Aartsen et al., "Measurement of Atmospheric Tau Neutrino Appearance with IceCube DeepCore"]

[266] *Aartsen et al., "Measurement of Atmospheric Tau Neutrino Appearance with IceCube DeepCore"*

# 7

# THE ICECUBE UGRADE AND ICECUBE GEN2

**Given the success of IceCube, the collaboration is currently considering upgrades to extend and improve the detectors capabilities. The details surrounding these detector extensions are rapidly evolving. This chapter outlines the status as of Spring 2019.**

The idea to significantly extend the existing detector for both high and low energy applications has first been broadly presented in 2014[267,268]. As of now, the goals of IceCube-Gen2 and the IceCube Upgrade[269] can be summarized as follows:

267 *Aartsen et al., "IceCube-Gen2: A Vision for the Future of Neutrino Astronomy in Antarctica"*

268 *Aartsen et al., "Letter of Intent: The Precision IceCube Next Generation Upgrade (PINGU)"*

269 *National-Science-Foundation,* IceCube Gen2-Phase1 (Upgrade) Award

- Re-calibration of the in-situ optical properties. This will help to constrain in particular low-energy systematics and enable a better angular resolution for cascade reconstructions, thus enabling better point source search sensitivities.
- Decreasing the atmospheric neutrino energy threshold to enable competitive measurements of fundamental neutrino properties, including the neutrino-mass ordering and the unitarity in the tau sector of the PMNS matrix.
- Increasing the rate of detected astrophysical neutrinos to increase the range of the energy spectrum. This will enable the resolution of spectral features, potentially even towards the GZK region, and allows for a quicker identification of astrophysical sources.

As these goals are quite varied, a large suite of different detector components as depicted in Figure 7.1 has been suggested.

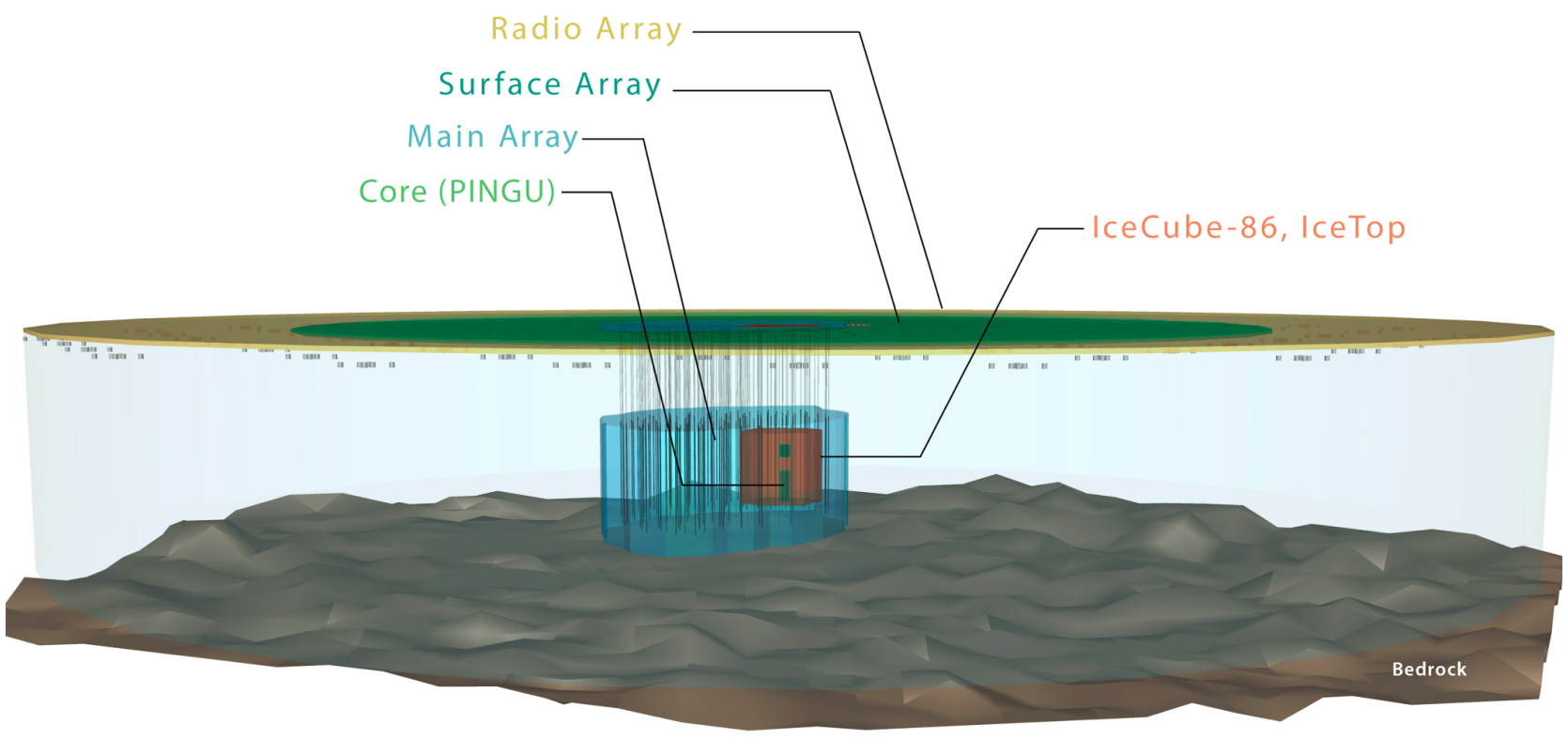

Figure 7.1: Envisioned extend of the IceCube Gen2 facility, comprising both in-ice and surface extensions [IceCube collaboration (2019). "Internal graphics resource"]

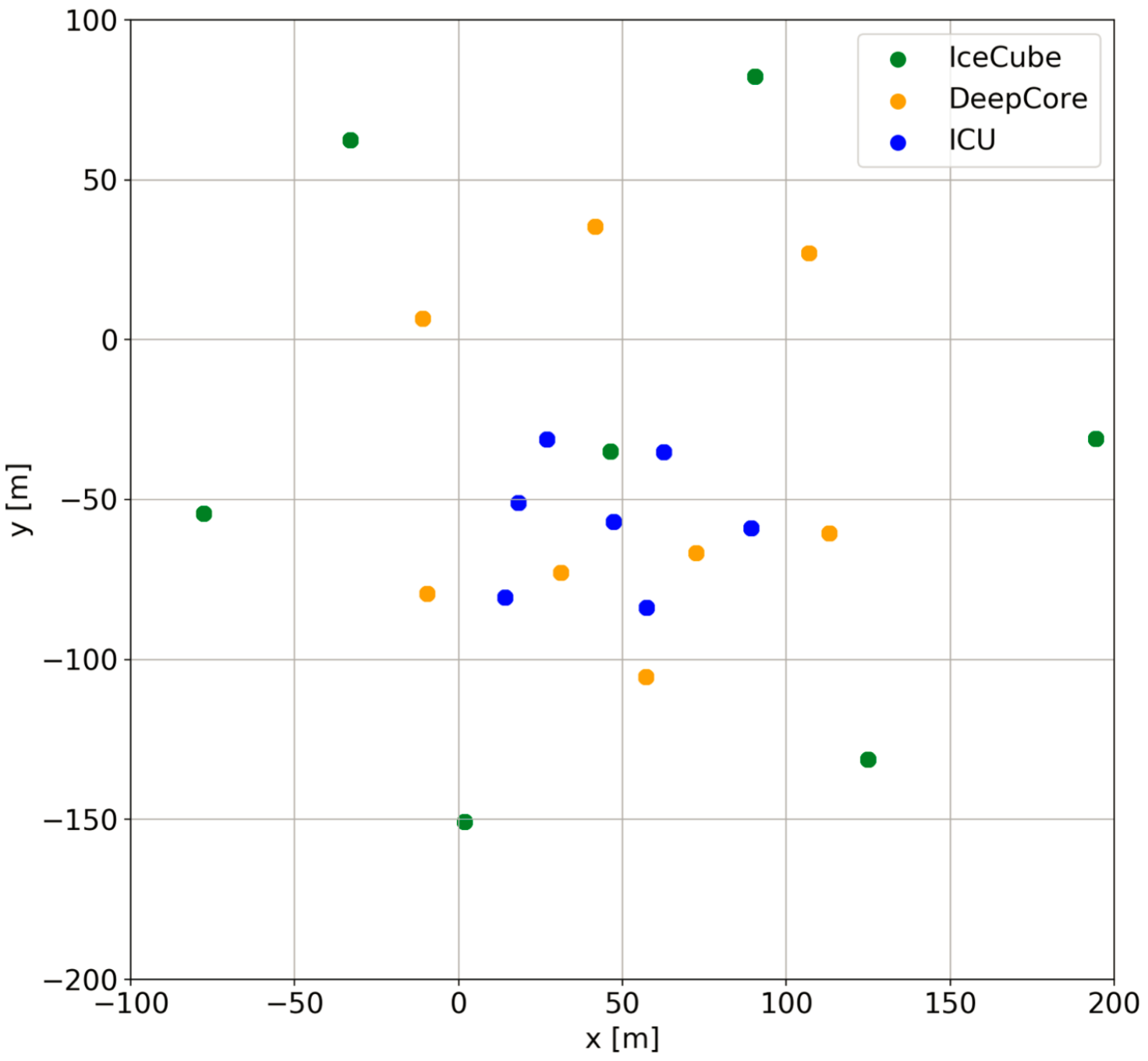

Figure 7.2: Currently proposed geometry and sensor arrangement for the IceCube Upgrade. For calibration purposes several existing strings are to be surrounded. [Tom Stuttard (Nov. 2018). *ICU simulation upgrade*. Extension call slides. 28]

## 7.1 IN-ICE EXTENSIONS

Low and high energy extensions require different in-ice arrays, with low-energy extensions aiming at more densely instrumenting the DeepCore volume and high-energy extensions hoping to increase the sensitive volume to up-to 10 $km^3$.

As of now, IceCube has been approved to prepare the so called IceCube Upgrade[270]. It is to extend the array with seven high density infill strings, as seen in figure 7.2. Each string will be equipped with a mixture of DOM variants at a 3 m vertical spacing.

[270] *National-Science-Foundation*, IceCube Gen2-Phase1 (Upgrade) Award

While immediately advantageous for low energy studies, an improved ice understanding will be achieved through the new sensor spacing as well as dedicated light sources and will improve the calibrating the already existing and future high energy data.

### 7.1.1 *Optical modules*

The collaboration is currently considering a number of designs for the future DOMs. A selection is shown in Figure 7.3.

For the low-energy extension the so called mDOM[271] is the baseline design. It features 24 three-inch PMTs distributed over the entire surface of the pressure sphere, yielding a nearly homogeneous photon detection efficiency in all directions, as well as directional resolution. This is particularly advantageous for small sensor spacings where a large portion of detected photons has not been scattered and can thus been traced to the Cherenkov cone.

[271] *Lew Classen (2017). "The mDOM - a multi-PMT digital optical module for the IceCube-Gen2 neutrino telescope." Friedrich-Alexander-Universität Erlangen-Nürnberg*

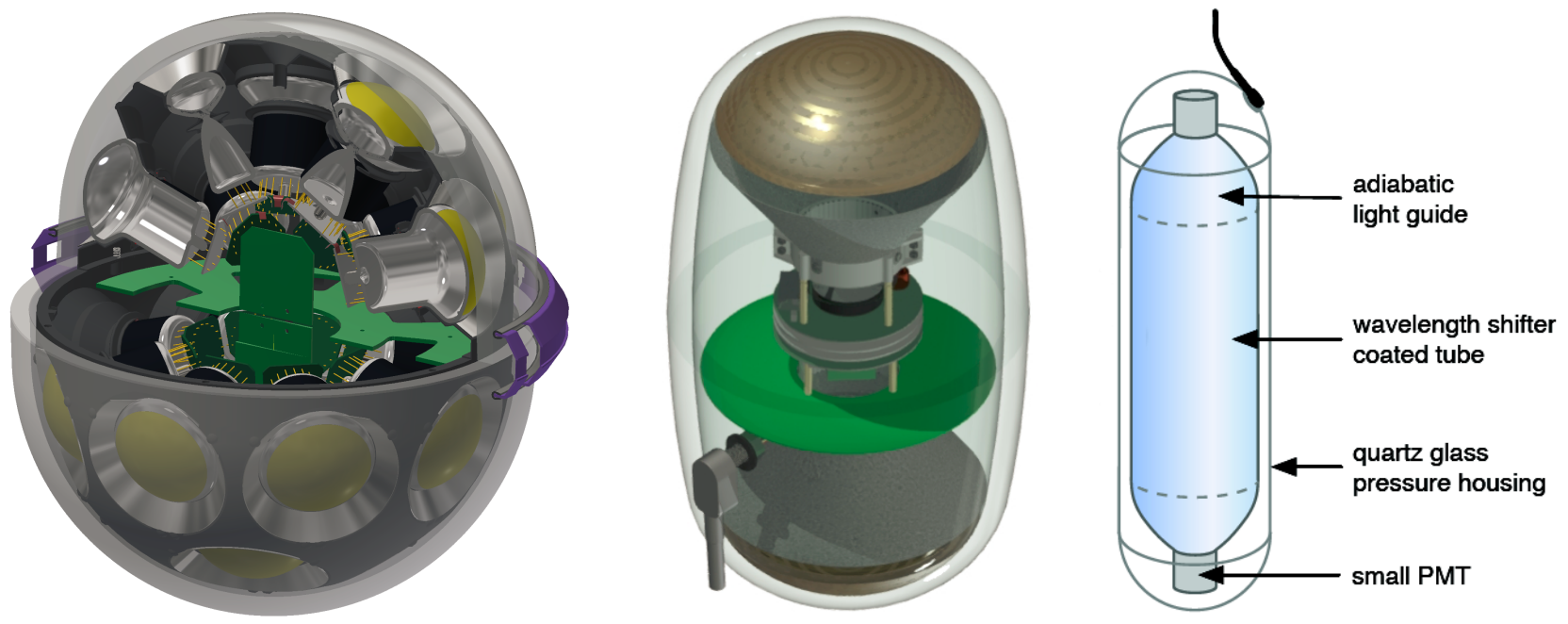

Figure 7.3: Selection of proposed optical modules for the IceCube in-ice extensions. From left to right: mDOM, D-EGG and WOM.
[M. G. Aartsen et al. (2017b). "IceCube-Gen2, the Next Generation Neutrino Observatory"]

Other designs under consideration for both the low- and high-energy extension are an updated version of the original IceCube DOM, called pDOM, and a so called D-EGG[272,273]. The D-EGG utilizes the electronics developed for the pDOM but integrates two slightly smaller PMTs inside an elongated pressure vessel.

272 *Dual optical sensor in an Ellipsoid Glass for Gen2*

273 *Ishihara et al., "Overview and performance of the D-Egg optical sensor for IceCube-Gen2"*

More experimental designs try to utilize wavelength shifting to achieve a large photo-collection area while using small photosensors. This can either be realized using plastic fibers[274] or a glass tube coated with a wavelength shifting paint (WOM[275]). At the expense of timing resolution, these designs can be elongated to instrument a large fraction of the hole. Both design are narrow compared to a standard DOM, potentially reducing drilling cost.

274 *Sandstrom, "Fiber Optical Module (FOM)"*

275 *Peiffer, Hebecker, et al., "Overview and Performance of the Wavelength-shifting Optical Module (WOM) for IceCube-Gen2"*

## 7.1.2 *Proposed calibration devices*

In addition to increasing the instrumented photocathode area, new in-ice calibration devices are planned.

### 7.1.2.1 *POCAM*

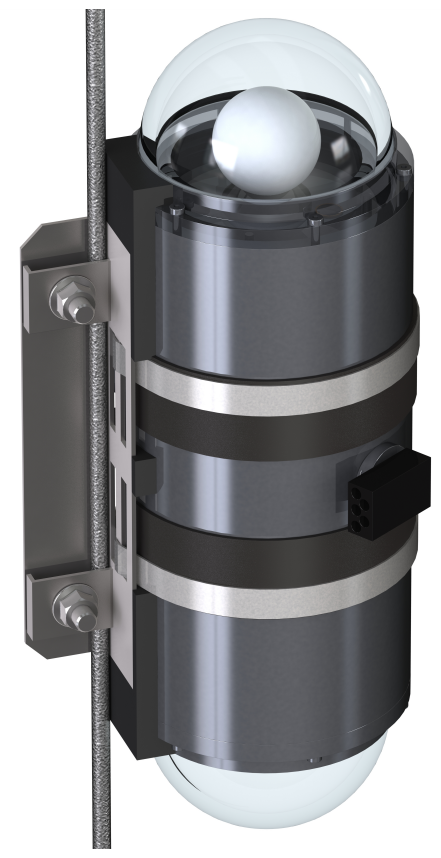

Figure 7.4: CAD rendering of a POCAM. One of the two homogenizing teflon spheres is seen in the upper glass hemisphere.
[Resconi, Rongen, et al., "The Precision Optical CAlibration Module for IceCube-Gen2: First Prototype"]

The current DOMs feature LEDs emitting light either horizontally or at a $\sim 45°$ angle into the ice. This means that studies requiring illumination at different angles, such as to determine hole ice properties, relative DOM efficiencies, or the ice anisotropy, currently need to rely on a correct understanding of the light diffusion through scattering. For these applications a fully isotropic light source would proof very valuable.

Such a device, called the Precision Optical CAlibration Module for IceCube-Gen2 (POCAM)[276], is currently under development. It features two semi-transparent integrating PTFE spheres,

276 *Resconi, Rongen, et al., "The Precision Optical CAlibration Module for IceCube-Gen2: First Prototype"*

which homogenize the light emitted by an array of LEDs situated below each sphere. In addition to the light sources, the POCAM includes SiPMs and photodiodes to be able to self-calibrate the per-flash photon count and timing profile.

A simulation study suggesting the application of the POCAM to study hole ice properties will be discussed in Section 9.7.

#### 7.1.2.2 *Laser system*

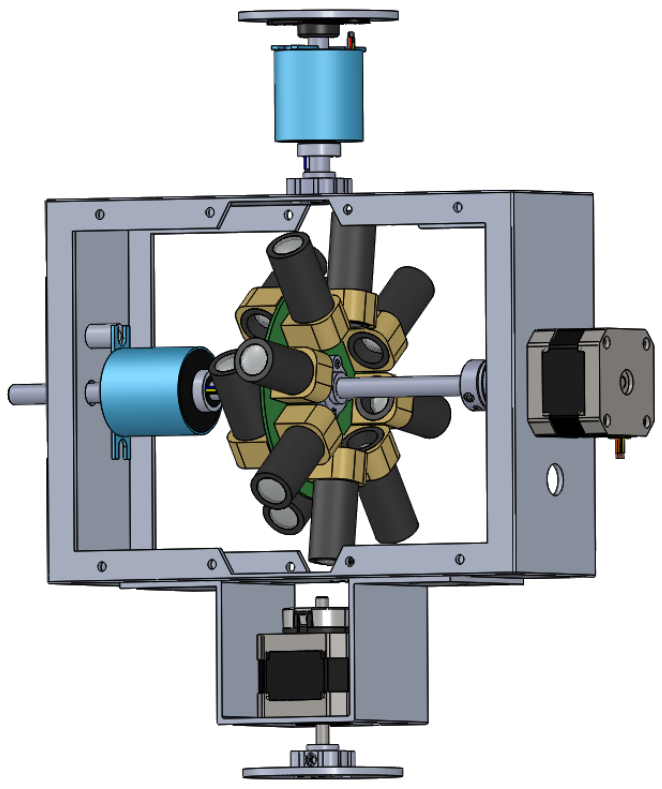

Figure 7.5: CAD drawing of the envisioned mechanical holding structure of a pencil beam with $4\pi$ coverage
[Nuckles and Wendt, *Pencil Beam Development Status*]

Complementary to the POCAM a stand-alone module providing narrow laser based pencil beams is under development[277]. Through a mechanical assembly as seen in Figure 7.5 and similar to the Sweden Camera, the beam is supposed to be freely rotatable with sub-degree accuracy.

[277] *Nuckles and Wendt,* Pencil Beam Development Status

Given a precise understanding of its orientation and relative position with respect to neighboring modules, the device could prove very useful in measuring the scattering function or exploring the hole ice. The scattering function would for example be accessible by measuring the detection rate as a function of the opening angle between the beam and a receiving module.

#### 7.1.2.3 *Camera system*

The Sweden camera revealed that the local ice around each DOM is a very complex environment. It is still unclear if the observations from the Sweden Camera are generally applicable to the rest of the detector, as it is situated in a non-representative position at the very bottom of a hole.

This motivates the idea to integrate at least one camera into every future DOM. Such a system is currently under development [278].

[278] *JongHyun Kim et al. (Aug. 2017). "A camera system for IceCube-Gen2"*

#### 7.1.2.4 *Acoustic positioning system*

In the current IceCube array the absolute depth of each string is triangulated using measurements of light delay between the DOMs[279]. At the envisioned extended string spacings of up-to 300 m, as proposed for the high energy extension, this method will no longer be feasible as the distance corresponds to several extinction lengths. Instead, it has been suggested to use acoustic triangulation, as the acoustic extinction length in the deep ice can reach up to 300 m[280].

[279] *see section 5.1.2.2*

[280] *R. Abbasi et al. (Jan. 2011). "Measurement of acoustic attenuation in South Pole ice"*

Based on experience from the SPATS[281] engineering array and further sensor development within the Enex-Range collaboration[282], acoustic sensors to be integrated into the DOMs

[281] *Y. Abdou et al. (Aug. 2012). "Design and performance of the South Pole Acoustic Test Setup"*

[282] *Dirk Heinen et al. (2017). "EnEx-RANGE - Robust autonomous Acoustic Navigation in Glacial icE." Ed. by S. Buitink et al.*

and stand-alone acoustic emitters are currently under development[283].

[283] Wickmann et al., *"Development of an acoustic sensor for the future IceCube-Gen2 detector for neutrino detection and position calibration"*

In addition, the acoustic sensors can be used to record transient signals as expected from ultra-high energy neutrino interactions in the ice[284].

[284] Askariyan et al., *"Acoustic detection of high energy particle showers in water"*

## 7.2 SURFACE EXTENSIONS

Several surface extensions are also planned, in addition to extending the in-ice instrumentation. The primary motivation is to provide an efficient veto for atmospheric muons over a larger solid angle than currently provided by IceTop. This would increase the effective area for example for the HESE study, which relies on starting events in the detector, as ideally no veto region is required any longer.

The detector can either be realized by IceTop-like stations which register the passage of electrons and muons on the surface, via surface-radio detectors or via imaging air-cherenkov telescopes which image the air-shower development in the atmosphere.

### 7.2.1 *Scintillator Array*

It is planned to replace the water Cherenkov tanks by plastic scintillator panels[285], with the light being digitized by SiPM detectors.

[285] Haungs, *"A Scintillator and Radio Enhancement of the IceCube Surface Detector Array"*

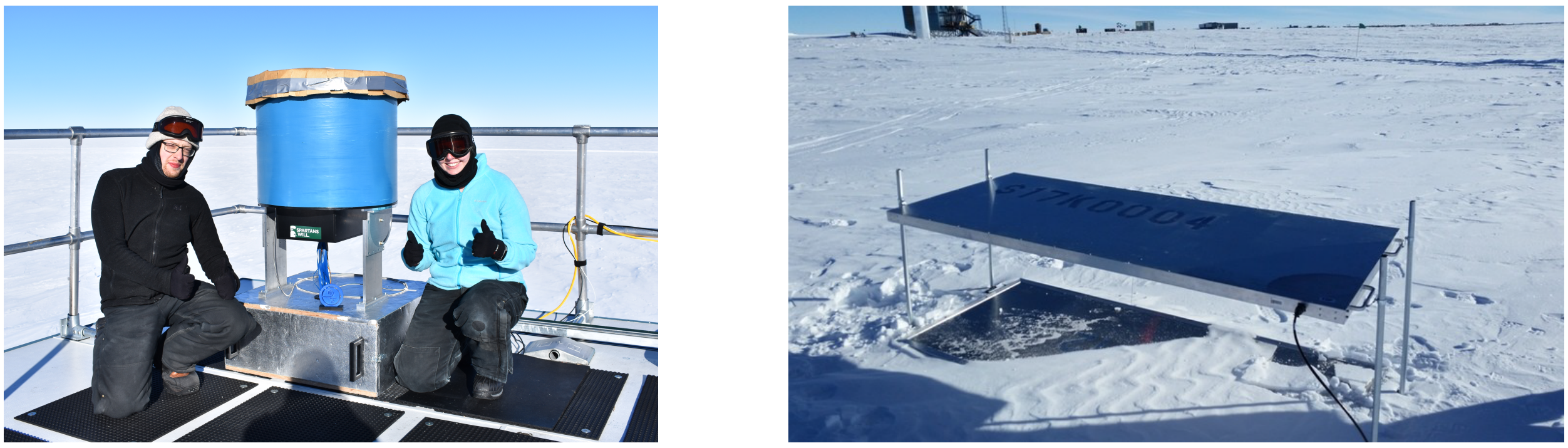

Figure 7.6: Demonstrator experiments of potential IceCube Gen2 surface extensions. Left: Second iteration IceAct telescope being deployed on the roof of the ICL; Right: Two second iteration scintillator panels as part of an IceTop infill array
[Matt Kauer (2018). *Private communication*]

The IceTop stations slowly loose their ability to register the electromagnetic air shower components, as they are buried by an increasing overburden of snow. Scintillators placed on top of existing stations could temporarily recover the detector sensitivity to the electromagnetic component.

A scintillator test-array consisting of seven pairs of scintillators has been deployed during the 16/17 and 17/18 South Pole seasons[286].

[286] S. Kunwar et al. (Aug. 2017). *"The IceTop Scintillator Upgrade." Sissa Medialab*

The scintillator stations may in the future also be equipped with radio air-shower detectors to boost the sensitivity for very inclined showers. A first prototype antenna was deployed during the 18/19 season[287].

[287] Haungs, *"A Scintillator and Radio Enhancement of the IceCube Surface Detector Array"*

### 7.2.2 *IceAct*

IceAct[288] features small scale, imaging Air-Cherenkov telescopes based on Famous Telescope designed for the Pierre Auger Observatory[289]. Prototype telescopes have been deployed during the 15/16, 17/18 and 18/19 South Pole seasons.

[288] T. Bretz et al. (July 2018). *"A compact and light-weight refractive telescope for the observation of extensive air showers"*

[289] Thomas Bretz et al. (Aug. 2016). *"FAMOUS - A fluorescence telescope using SiPMs." Sissa Medialab*

Through imaging of the air shower development in the atmosphere, it also increases the sensitivity to the cosmic ray composition in the IceTop energy range, in addition to enabling an improved cosmic ray veto.

## 7.3 RADIO EXTENSIONS

In the scope of the high energy extension there may also be a radio neutrino detector operated in conjunction with the future optical Cherenkov detector.

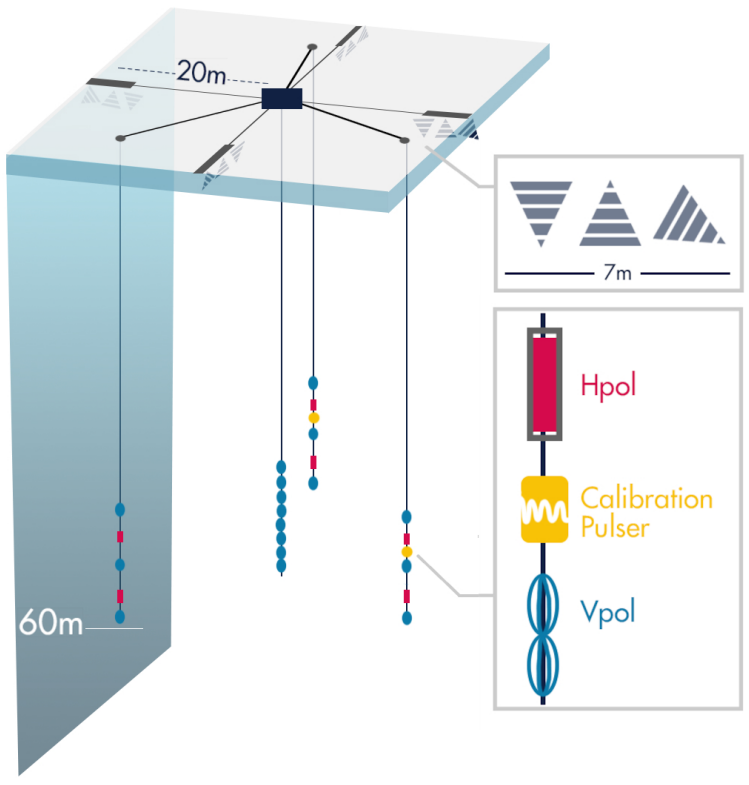

Figure 7.7: Schematic drawing of a single RNO detector station, consisting of both shallow and deep antennas.
[Deaconu, "RNO - The Radio Neutrino Observatory"]

This can either be realized with antennas just below the surface (as in ARIANA[290]) and/or with antennas in shallow holes (as in ARA[291]).

[290] Lisa Gerhardt et al. (Dec. 2010). *"A prototype station for ARIANNA: A detector for cosmic neutrinos"*

[291] P. Allison et al. (Feb. 2012). *"Design and initial performance of the Askaryan Radio Array prototype EeV neutrino detector at the South Pole"*

This project is currently being pursued by the independent "Radio Neutrino Observatory"[292] experiment, but might in the future cooperate with the IceCube Gen2 efforts.

[292] Deaconu, *"RNO - The Radio Neutrino Observatory"*

# 8

# DEVELOPMENT OF A SUB-NS LIGHT SOURCE FOR ICECUBE GEN2

**This chapter describes the design and performance of a low-cost and easy to build light source that provides sub-ns light pulses of variable intensity at a nearly arbitrary wavelengths.**

---

**Declaration of Pre-released Publications**
The study presented in this chapter has in-part already been published[293]. The author of this thesis has written this publication together with Merlin Schaufel. The content is, where not otherwise cited, fully based on own work.

---

In order to characterize photo-detectors with single photon timing resolutions (SPTR) in the picosecond to nanosecond range, such as PMTs and SiPMs[294], light sources with light curves faster than the studied sensor are required.

Such sources are commercially available, but expensive. In contrast, simple circuits commonly used in the community, such as the Kapustinsky pulser[295], are limited to light pulses of few nanosecond duration.

## 8.1 THE ELECTRIC PULSE DRIVER

The central component of any fast light source is a circuit that can supply a sufficiently narrow electric drive pulse. For this purpose an avalanche transistor based circuit as described in [296] was chosen. The schematic is depicted in Figure 8.1.

The *LT1082* switching regulator[297] provides an adjustable bias voltage which is applied to a *2N2369* avalanche transistor[298]. The bias voltage (typically 70 V) is chosen just below the breakdown point of the transistor so that no random pulses occur. An input pulse to the base then triggers an avalanche, discharging the discharge capacitance $C_D$ through the emitter.

[293] *Rongen and Schaufel, "Design and evaluation of a versatile picosecond light pulser"*

[294] *Ronzhin et al., "Tests of timing properties of silicon photomultipliers"*

[295] *Kapustinsky et al., "A fast timing light pulser for scintillation detectors"*

[296] *Williams,* Triggered 250 Picosecond Rise Time Pulse Generator

[297] *Linear Technology,* LT1082 1A High Voltage, Efficiency Switching Voltage Regulator Datasheet

[298] *Motorola,* 2N2369 Switching Transistors

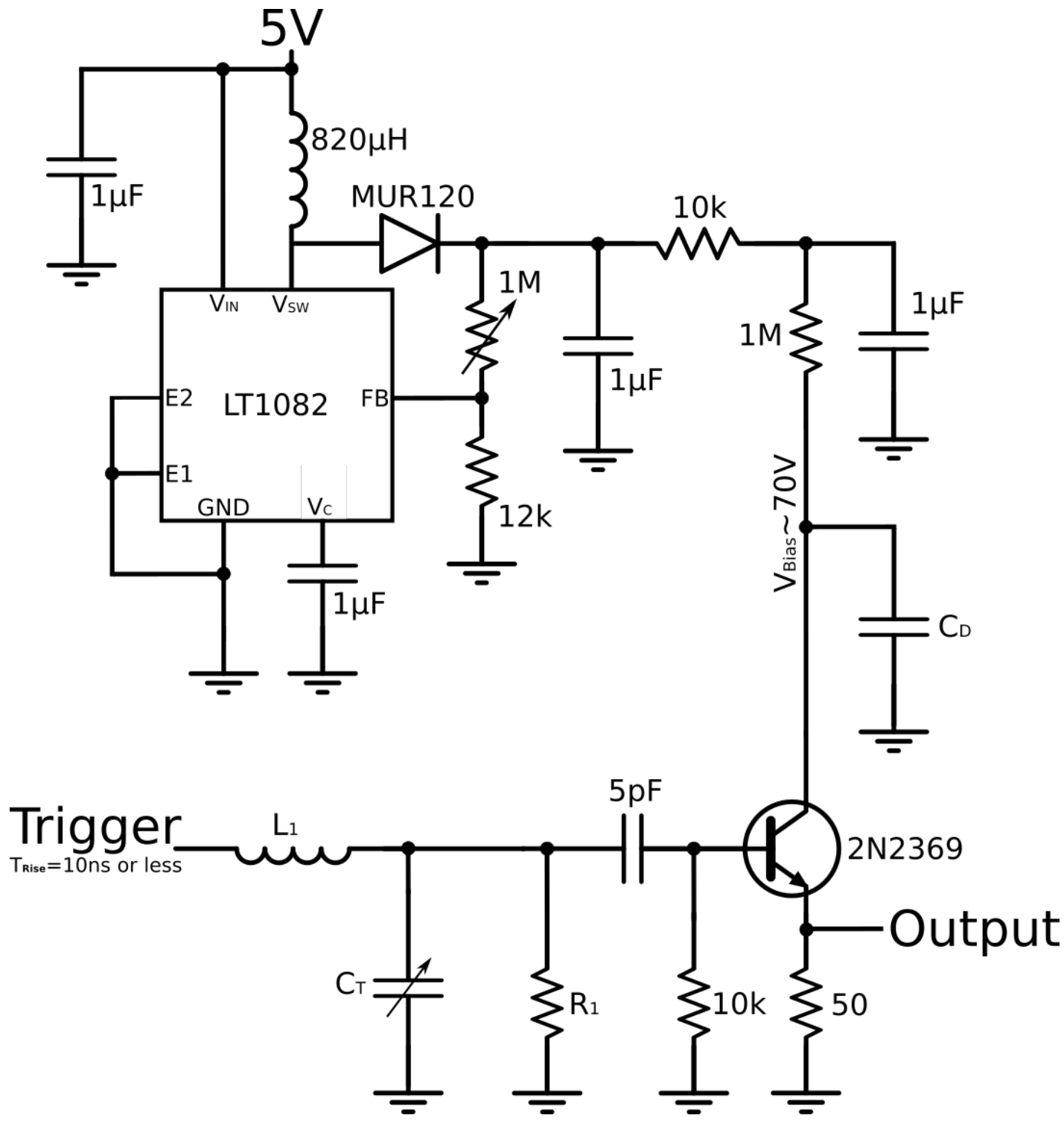

Figure 8.1: Triggered pulse generator based on a reverse biased avalanche transistor as proposed in [Jim Williams (1994). *Triggered 250 Picosecond Rise Time Pulse Generator*. Linear Technology]

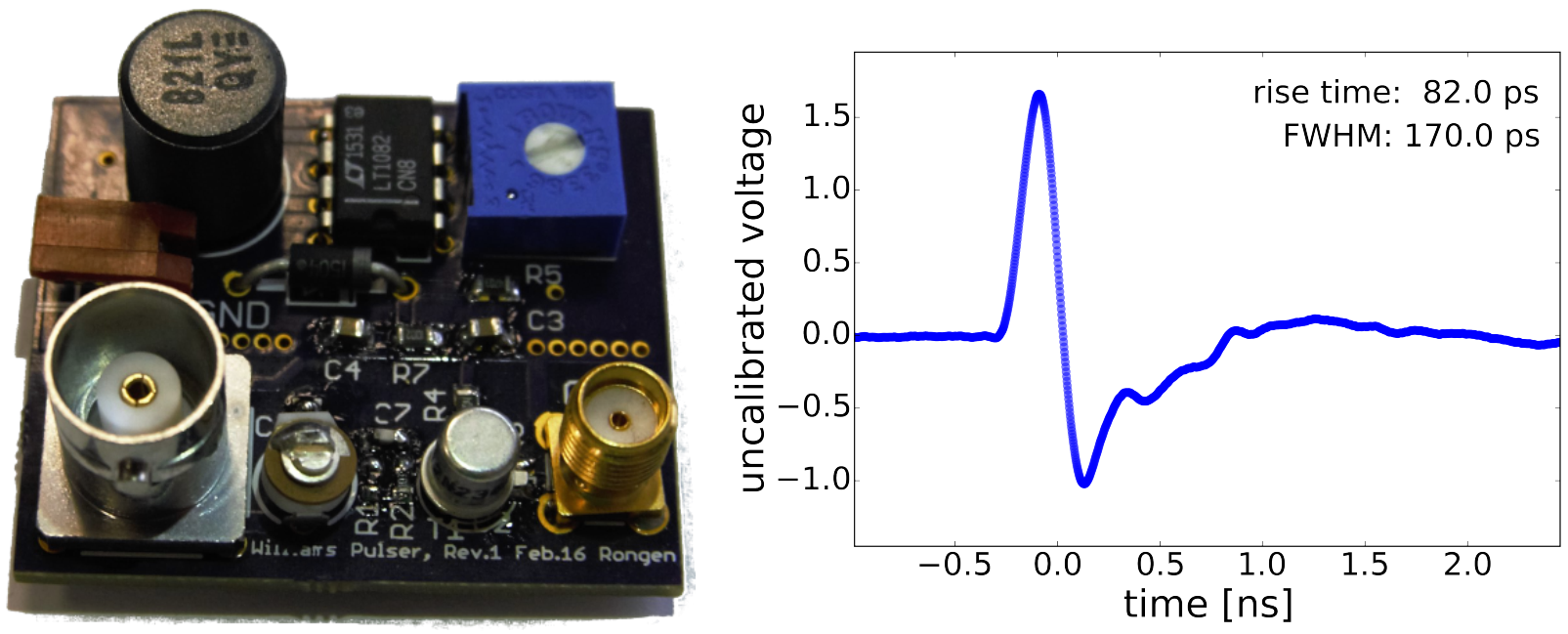

Figure 8.2: Prototype PCB (left) and voltage output (right) of the electric pulse driver. The undershoot ensures a clean turn-off of the LED

As the collector voltage drops the transistor self-resets and can be triggered again. The pulse height and width scale with the discharge capacitance. For this study values of $C_D$ between 0 pF (parasitics only) and 12 pF were chosen.

The pulse shape of the drive circuit (as prototyped on a dedicated board, see Figure 8.2) was evaluated using a *Tektronix MSO71254C* 100 GS/s 12.5 GHz *Mixed Signal Oscilloscope*[299] with a *P7380A 8GHz* differential probe[300] and single use solder contacts. As the probe only has a ±2.5 V dynamic range, a passive voltage divider was needed. Figure 8.2 shows a measured drive pulse, with only parasitic discharge capacitance in the circuit.

The pulse is bipolar, with a distinct, clean primary pulse of 170 ps FWHM. While the bipolar pulse shape is non-intentional, it does offer an advantage when applied to LEDs and solid state lasers, as the subsequent negative pulse helps to remove charges

299 *Tektronix (2018).* MSO/DPO70000 Series Datasheet. *Tektronix*

300 *Textronix (2006).* P7380A User Manual. *Textronix*

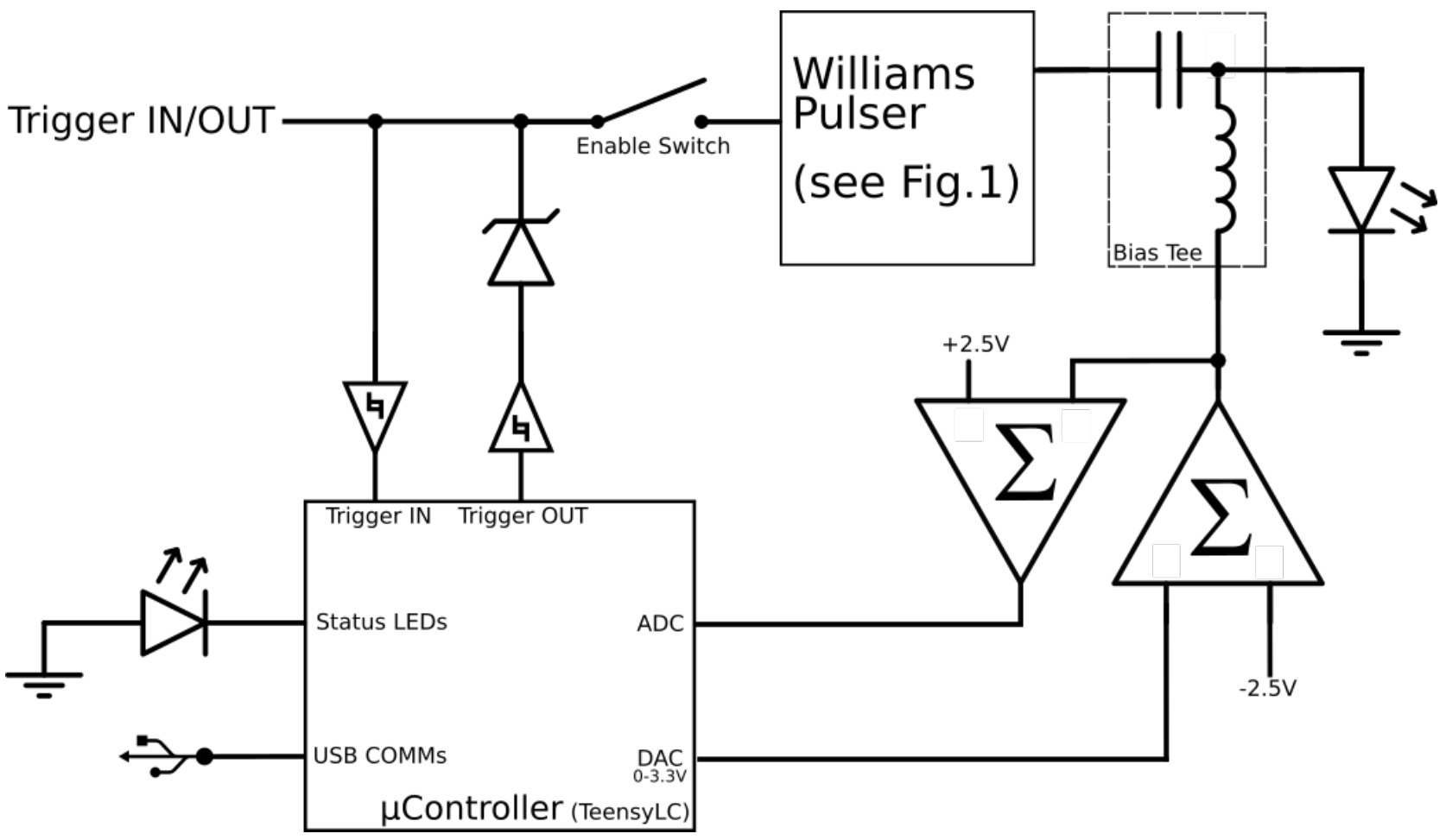

Figure 8.3: Schematic layout of a fully integrated picosecond light pulser module, including biasing and the micro controller for triggering and control.

from the depletion layer, thus reducing the turn-off time. This so called active sweep-out technique was for example suggested in [301].

[301] Lee, *"Effect of Junction Capacitance on the Rise Time of LEDs and on the Turn-on Delay of Injection Lasers"*

While the overall bipolar shape has been observed reproducibly, the additional ringing in the tail changed significantly between repeated measurement with different voltage dividers and contact points. Therefore, it results at least partially from the measurement setup.

## 8.2 full light pulser module

In order to provide a simple to use light source, the electric pulse driver as described in the previous section is incorporated into a stand-alone light pulser module (see Figure 8.4). It provides an additional micro-controller [302] for trigger control and a *Mini-Circuits TCBT-14+* 10 MHz-10 GHz bias tee[303] to shift the DC level of the drive pulse.

[302] *a Teensy LC [PJRC,* Teensy LC*]*

[303] *Mini-Circuits,* TCBT-14+ Surface Mount Bias Tee

[304] *TT Electronics,* Vertical Cavity Surface Emitting Laser in T-1 Package

The light source is attached via a SMA breakout board, so that different sources can be easily used with the same module. Tested light sources are an OPV332 850 nm vertical cavity surface emitting laser (VCSEL) from TT Electronics[304] (expected rise/fall time ∼100 ps) and several 3 mm through-hole LEDs purchased from Roithner Lasertechnik.

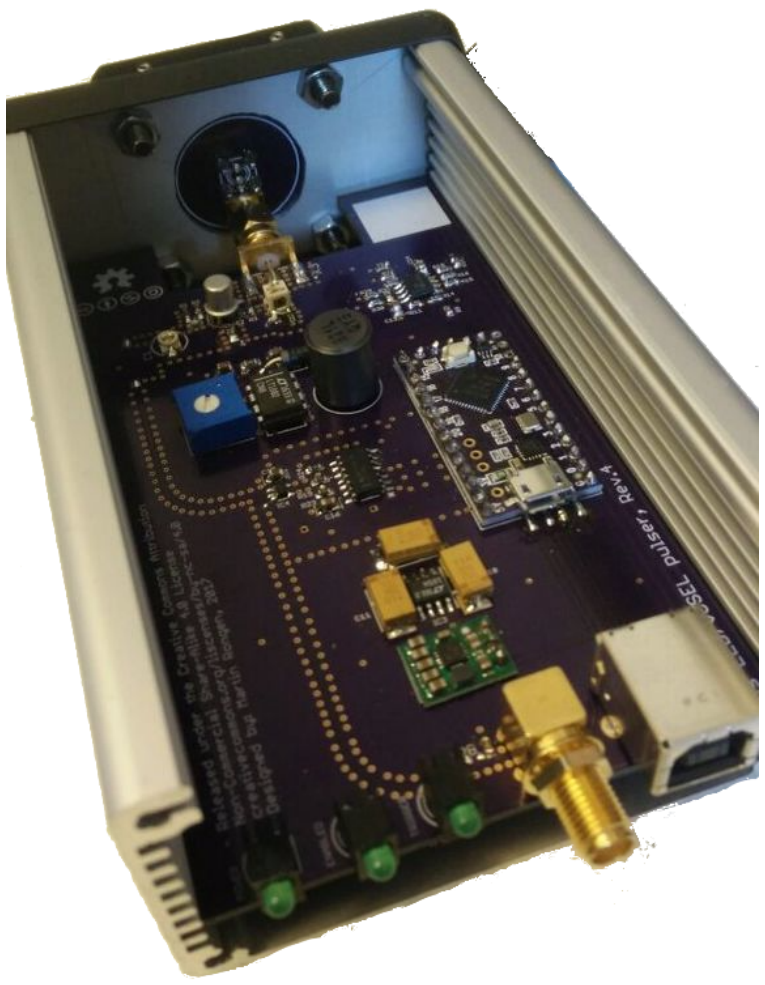

Figure 8.4: Photo of a fully assembled module.

A trigger pulse can either be provided externally, with the micro-controller monitoring the frequency, or can be generated by the micro-controller directly. In both cases the light output can be inhibited by an analog switch to the trigger input off the electric pulse driver.

The bias-tee in conjunction with a DAC on the micro-controller can shift the DC pulse level between −2.5 V and 0.8 V. This allows for an effective intensity control, without changing the shape of the drive pulse. As the integral of the drive pulse over the forward voltage, as well as the sweep-out efficiency, changes with biasing, the width of the light curve inevitable also changes. Negative bias voltages result in a narrower but dimmer light curve.

To achieve the highest possible intensity, no additional current limiting resistor is placed in series with the LED/VCSEL. This is safe as the total energy of the pulse is limited by the discharge capacitor. In fact, depending on the biasing, the voltage of the LEDs is barely over the forward voltage and no aging has been observed for either LEDs or VCSELs over many days of continuous operation at high repetition frequencies.

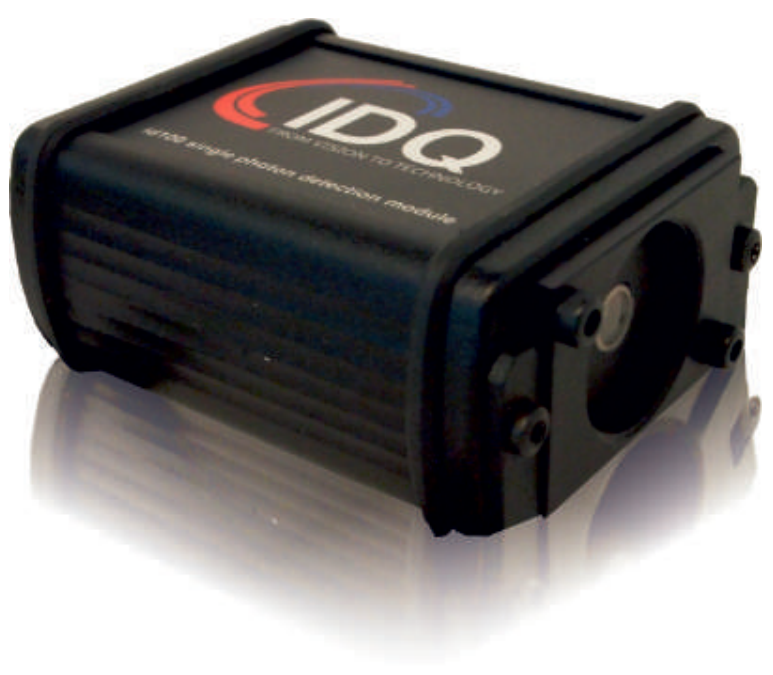

Figure 8.5: ID100 detector featuring a 50 µm APD with 60 ps FWHM timing resolution.
[SA, *IDQ ID100 Visible Single-Photon Detector*]

The total material cost for the module including an extruded aluminum enclosure and a one-inch optics adapter is ∼ 150€.

## 8.3 CHARACTERIZATION

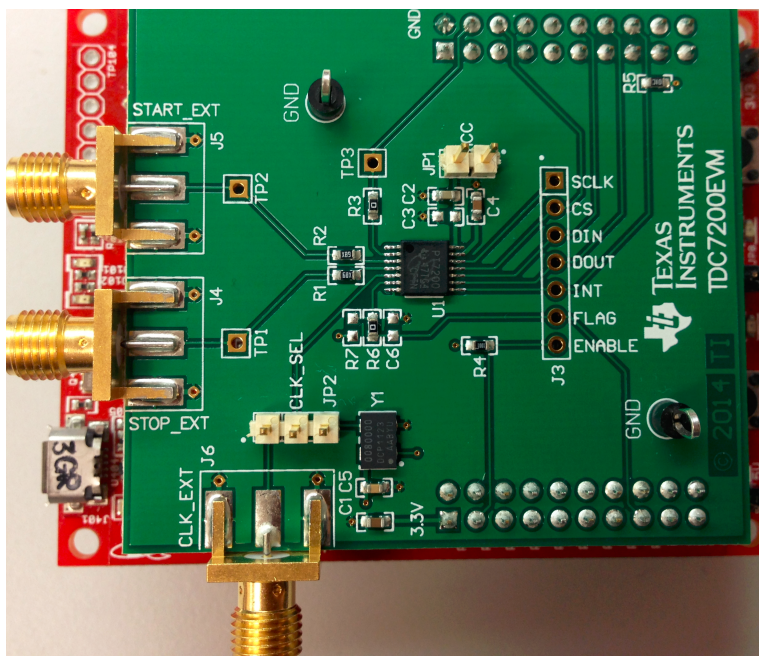

Figure 8.6: TDC7200 evaluation board to digitize the delay between the trigger pulse and the photon detection by the ID100.
[Texas Instruments, *TDC7200 Evaluation Module Users Guide*]

The light source is being characterized in terms of the timing profile of the light curve as well as the integral photon output.

### 8.3.1 *Timing*

#### 8.3.1.1 *APD setup*

The light curve is obtained by using a 50 µm avalanche photodiode based *IDQ ID-100* APD detector[305]. The device outputs a 10 ns, 2 V rectangular pulse when one or more incident photons are detected. The single photon time resolution (SPTR) is guaranteed to be smaller than 60 ps FWHM and is usually below 40 ps[306].

In this setup, the light curve is measured as the distribution of timing delays between the internal trigger pulse of the light source and the arrival time of single photons at the APD. For this purpose a *TDC7200 evaluation board*[307] is being used as a Time-to-Digital Converter.

As the APD can not distinguish the number of registered photons, the occupancy, that is the increase in APD detection rate relative to the trigger rate, needs to be small (∼10 %) to guarantee mostly single photons. In order to measure the occupancy, the APD rate is monitored with a custom frequency counter,

[305] *Becker & Hickl GmbH (2005).* TCSPC Performance of the ID100-50 Detector. *Becker & Hickl GmbH*

[306] *P Jani et al. (Nov. 2006). "Timing resolution (FWHM) of some photon counting detectors and electronic circuitry"*

[307] *Texas Instruments (Mar. 2016).* TDC7200 Time-to-Digital Converter for Time-of-Flight Applications. *Texas Instruments*

consisting of a *SN74LVC*[308] monostable multi-vibrator and a micro-controller.

The TDC7200 measures delays, by counting the number of clock cycles of a multiple-GHz ring oscillator, between a rising/falling edge of a start and a stop signal. As the frequency of ring oscillators is hard to determine and can drift over time, it is calibrated against an external, precision 8 MHz clock after each trigger. The TDC7200 datasheet[309] claims a resolution of 55 ps (one cycle of the ring oscillator) and a standard deviation of 35 ps (calibration uncertainty).

The timing accuracy of the TDC7200 has been verified by measuring two TTL pulses at 100 ns delay as provided by a *Quantum Composer 9518+* pulse generator[310]. The standard deviation of the timing delay as seen in Figure 8.7 agrees with the specification.

[308] *Texas Instruments,* SN74LVC2G07 Dual Buffer and Driver With Open-Drain Outputs

[309] *Texas Instruments,* TDC7200 Time-to-Digital Converter for Time-of-Flight Applications

[310] *Quantum Composer,* 9510+ Series Digital Delay Pulse Generators

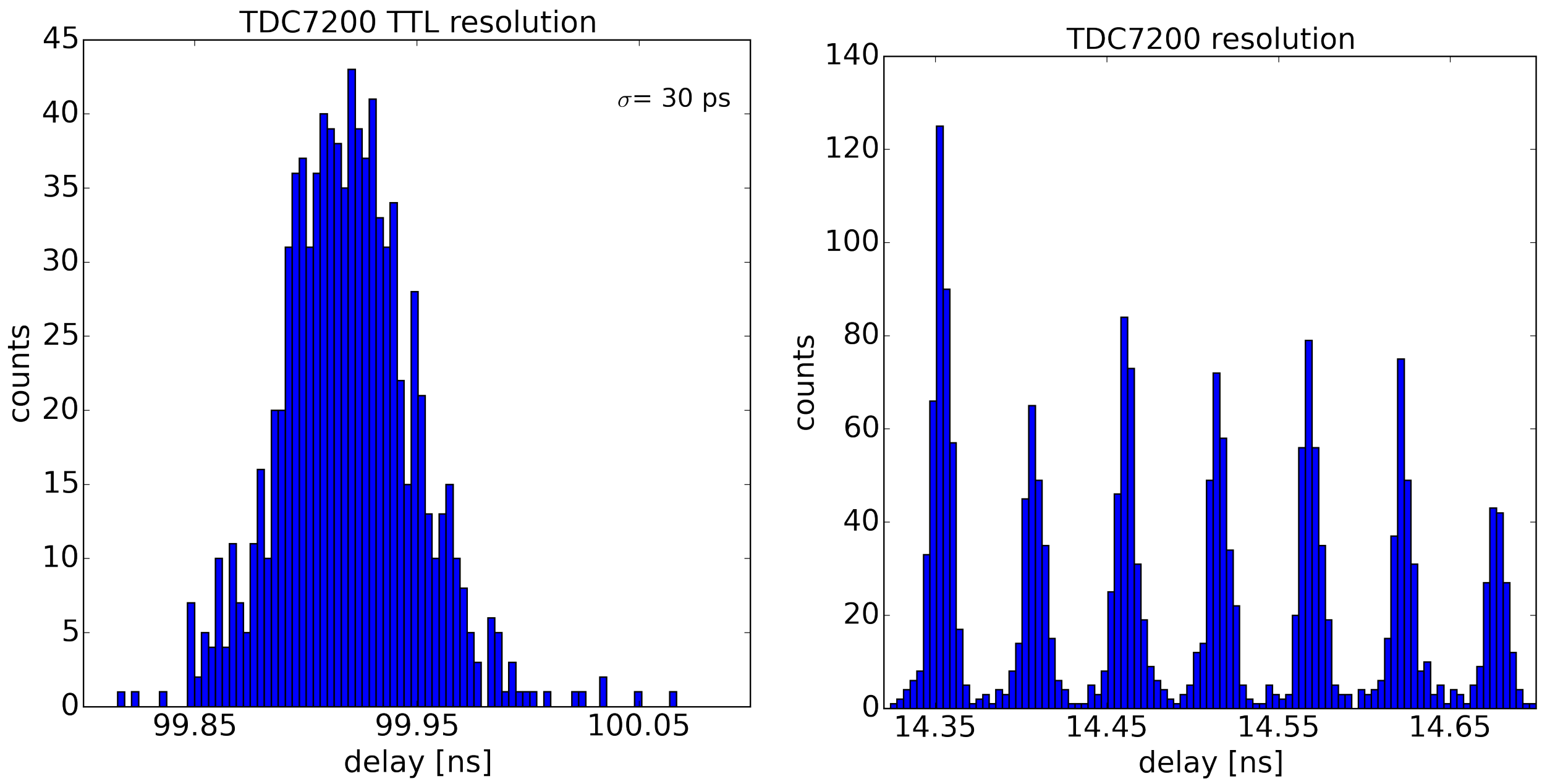

Figure 8.7: Performance of the TDC7200 evaluation board.
Left: Timing precision when measuring a precise delay.
Right: TDC7200 binning behavior for broad time spreads.

When measuring the delay of a less well determined signal (see Figure 8.7), the fact that the resolution is larger than the standard deviation becomes apparent as "fingers" in the histogram. In the following, the binning of all delay histograms has been chosen to match the TDC resolution.

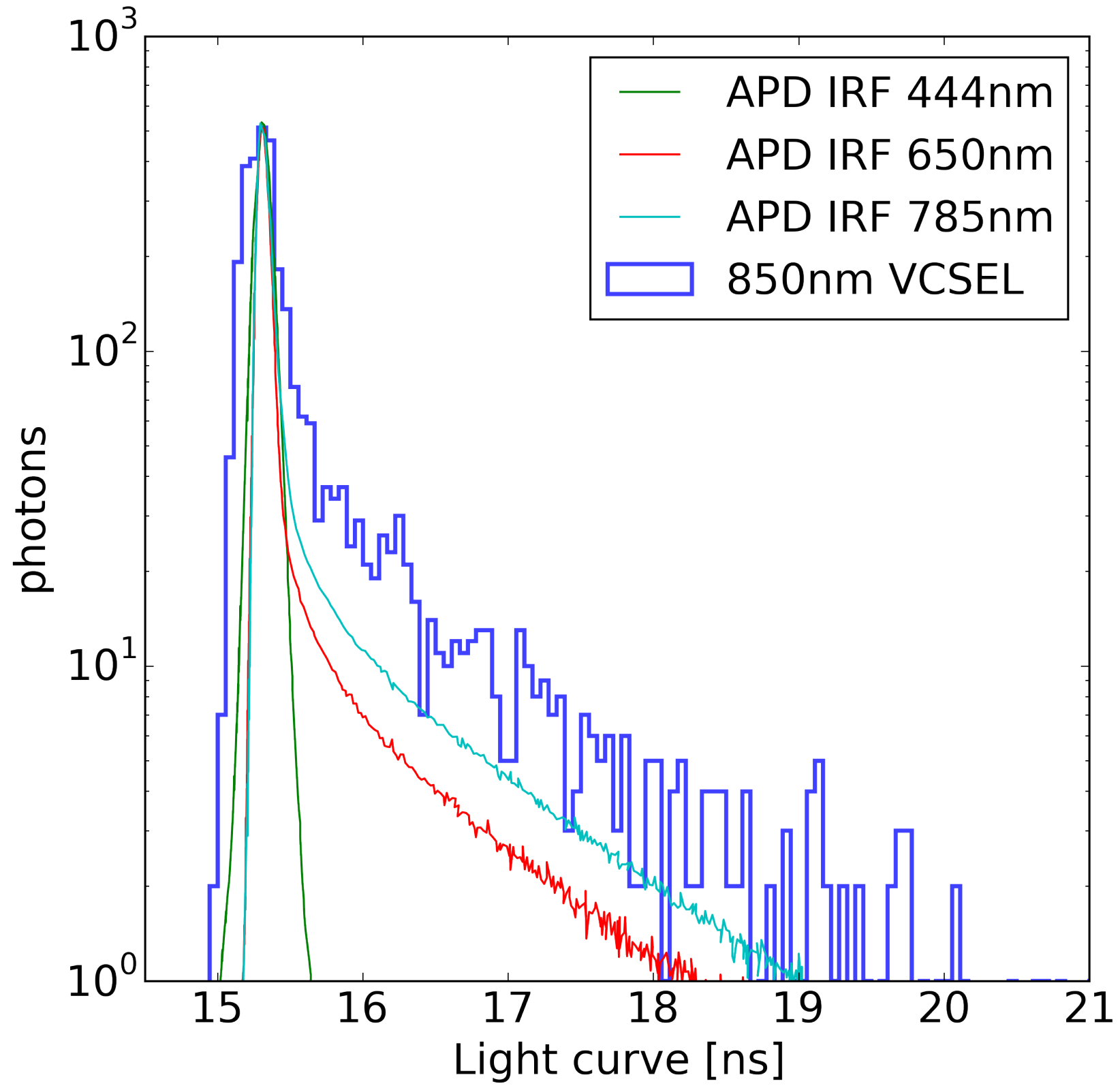

Figure 8.8: VCSEL light curve at −2.5 V biasing with no discharge capacitance, compared to the APD IRFs at different wavelength. Sadly no IRF is available at 850 nm. The IR diffusion tails are seen to be well described by exponentials, as is the tail seen for the VCSEL.

#### 8.3.1.2 *IR-VCSEL timing performance*

Out of the tested light sources the VCSEL achieves the best timing performance. Figure 8.8 shows the measured light curve compared to the expected APD IRF at different wavelengths. In addition to a primary, short duration peak a long duration tail is observed. This is most likely due to an APD detector effect, as APDs suffer from significant infrared diffusion tails, limiting the ability to quantify potential, low brightness, long duration turn-off tails of the light curves at long wavelengths[311].

[311] *Becker & Hickl GmbH*, TCSPC Performance of the ID100-50 Detector

As no exact APD IRF is available at the VCSEL wavelength of 850 nm, the IR diffusion tail is fitted to the data assuming a simple exponential behavior. Figure 8.9 shows the cleaned light curve after subtracting the IR diffusion component. The light pulser output shows a clean Gaussian profile with a standard deviation of ∼100 ps.

#### 8.3.1.3 *LEDs timing performance*

As the wavelength availability of VCSELs is very limited, a selection of 26 LEDs ranging from 365 nm to 810 nm was tested. All LEDs are 3 mm trough-hole variants ordered from *Roithner Lasertechnik*. These LEDs are not designed for pulsed applica-tions and no rise-times are stated in the datasheets.

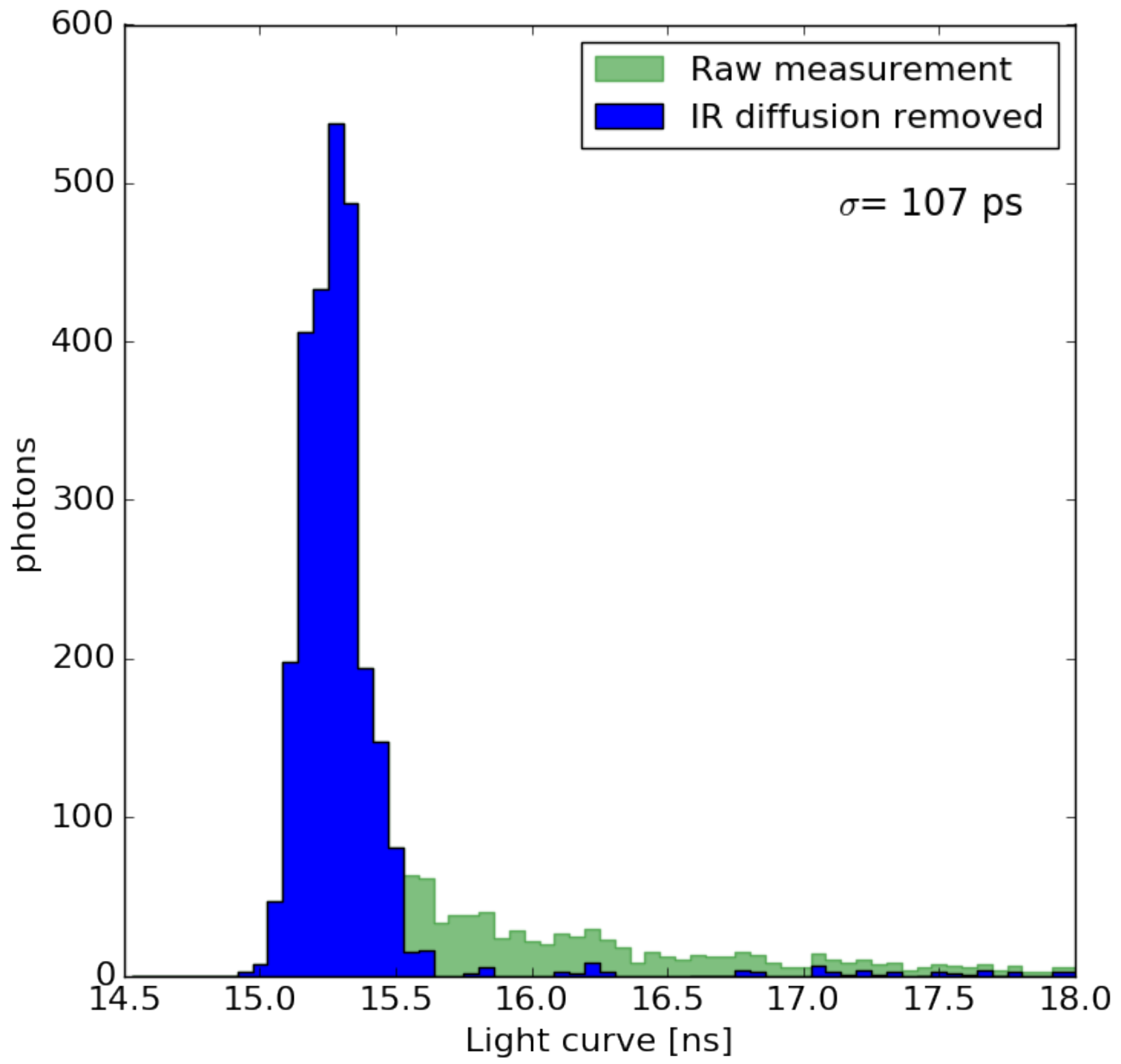

Figure 8.9: Same measurement as in Figure 8.8, but after correcting for the APD IR diffusion tail with an exponential fit to the tail.

As the emitted intensity has been observed to be significantly lower compared to VCSEL measurements, the APD setup for timing measurements had to be adapted. Instead of starting the TDC on the rising edge of the trigger pulse, it is started on an APD signal and stopped on the falling edge of the trigger pulse. The occupancy is still checked with the frequency counter but is always below 1%. As the width of the trigger pulse is not perfectly controlled ($\sigma \sim$60 ps) the SPTR of this configuration is slightly worse at $\sim$80 ps.

Only 20 LEDs emitted a detectable amount of light. Of these eight (370 nm, 375 nm, 385 nm, 590 nm, 605 nm, 680 nm, 700 nm and 770 nm) exhibit light curves shorter than 1 ns standard deviation. In general, high wavelength LEDs are easier to pulse due to the smaller band gaps, yet the best light curve was obtained with a 385 nm LED as shown in figure 8.10. For the non-IR LEDs no long-duration tail is observed, confirming that the effect seen for the VCSEL is indeed a detector artifact caused by the APD.

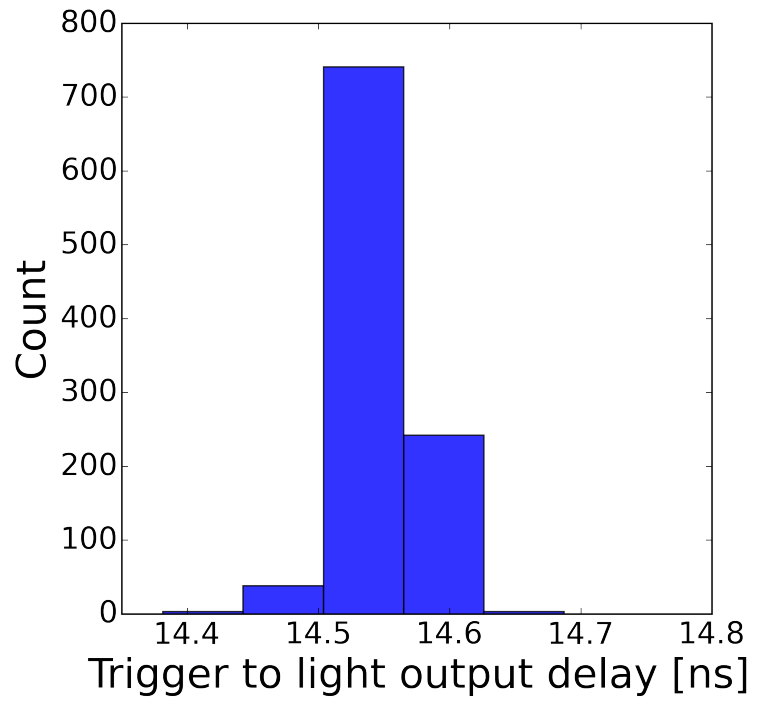

Figure 8.11: Trigger stability. Measured as the timing delay at the largest possible intensity.

### 8.3.2 *Trigger stability*

The trigger stability is defined as the standard deviation of the timing delay between trigger pulses and the beginning of the light curve. Thus it includes both: the jitter between the trigger and the drive pulse as well as possible turn-on variations of the LED/VCSEL. It is measured by illuminating the APD at the maximum possible brightness.

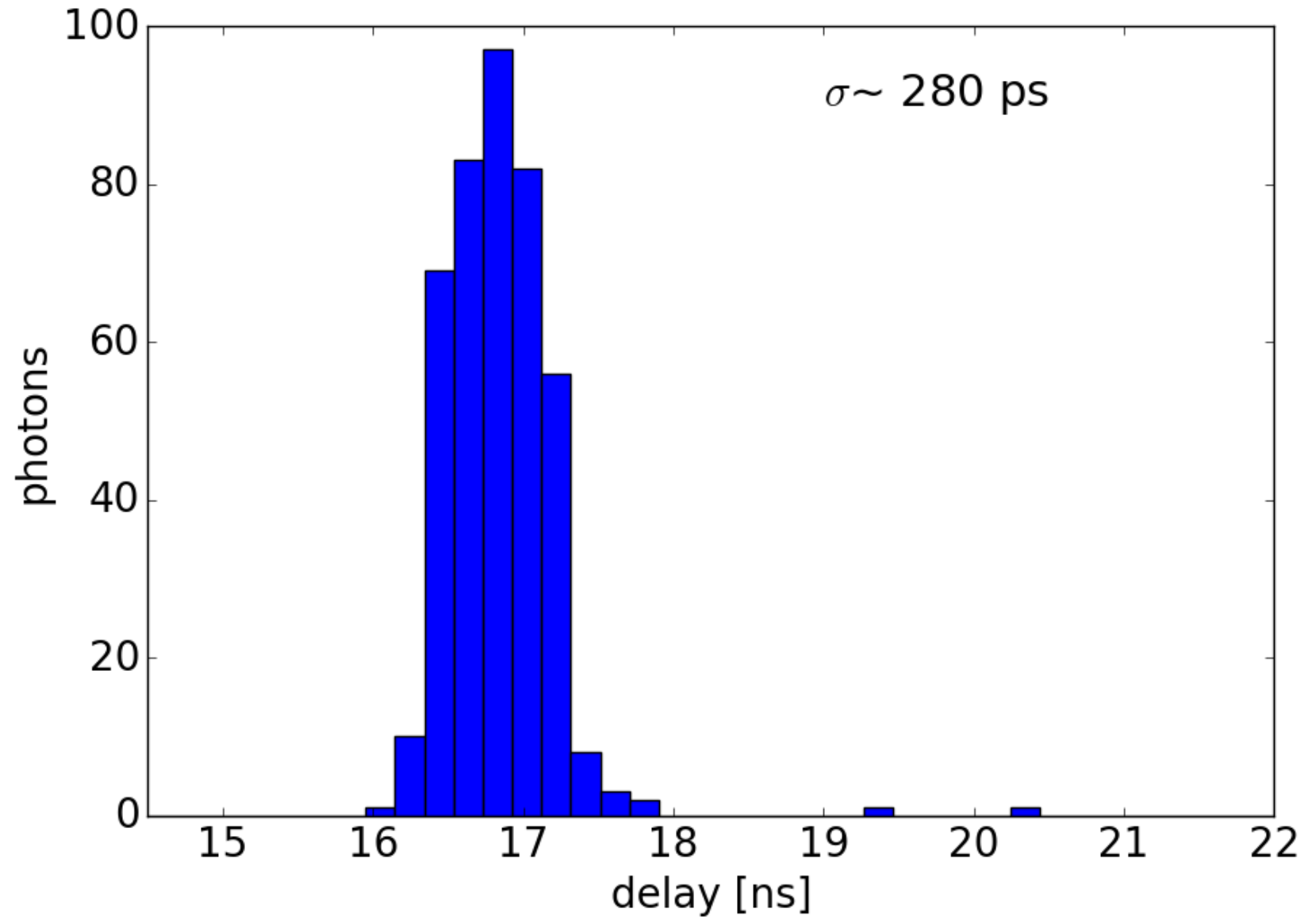

Figure 8.10: 385 nm LED light curve at $-2.0$ V biasing and 10 pF discharge capacitance, measured with the APD setup.

As several photons reach the APD nearly simultaneously, this measurement is not limited by the APD SPTR. No dependence of the trigger stability on the used light source was found. A histogram of the registered timing delays is shown in Figure 8.11. The spread is barely detectable at $\sim 40$ ps and matches the spread seen when measuring the jitter of the electric drive pulse.

## 8.3.3 *Intensity*

In order to quantify the pulse intensity (photons per pulse) a *Hamamatsu S2281* photodiode has been used in conjunction with a custom preamplifier as described in [312]. The intensity of a single pulse is not sufficient to be distinguished from the noise of the photodiode system. Instead, the average pulse intensity can be measured as the slope of the dark noise corrected intensity versus the light source repetition frequency.

[312] Tosi et al., *"Calibrating photon detection efficiency in IceCube"*

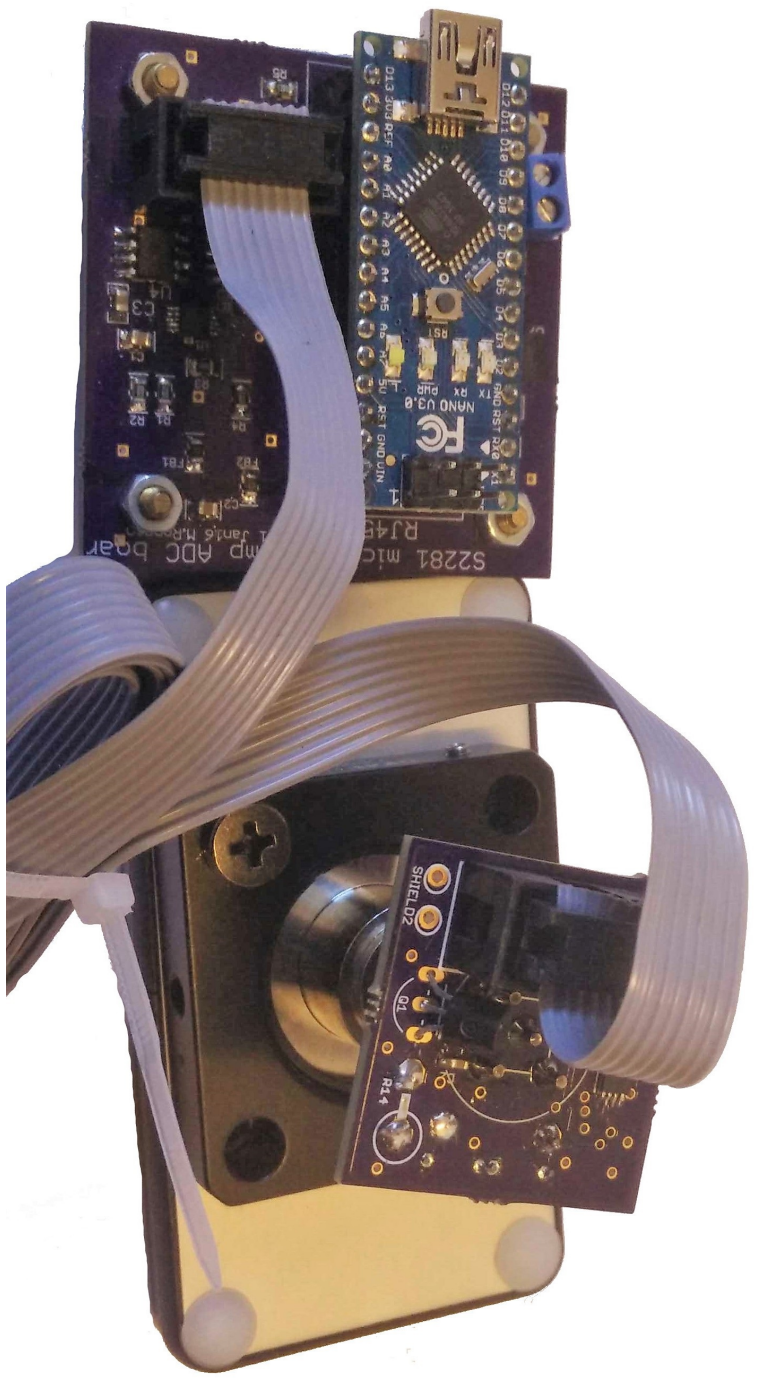

Figure 8.12: Photodiode system. The preamplifier is seen attached to a photodiode in the bottom. The ADC board is seen in the top.

Figure 8.13 shows the measured photo-current as a function of repetition frequency for a VCSEL with 0 pF discharge capacitance at $-2.5$ V biasing. The slope and the quantum efficiency from [313] yield $\sim 4 \cdot 10^7$ photons per pulse. An additional 10 % systematic error on the quantum efficiency has to be considered.

[313] *Hamamatsu (2015).* S2281 series Si photodiodes. *Hamamatsu*

Increasing the bias voltage to 0.7 V results in a threefold increase of the photon number at the expense of a slightly longer light

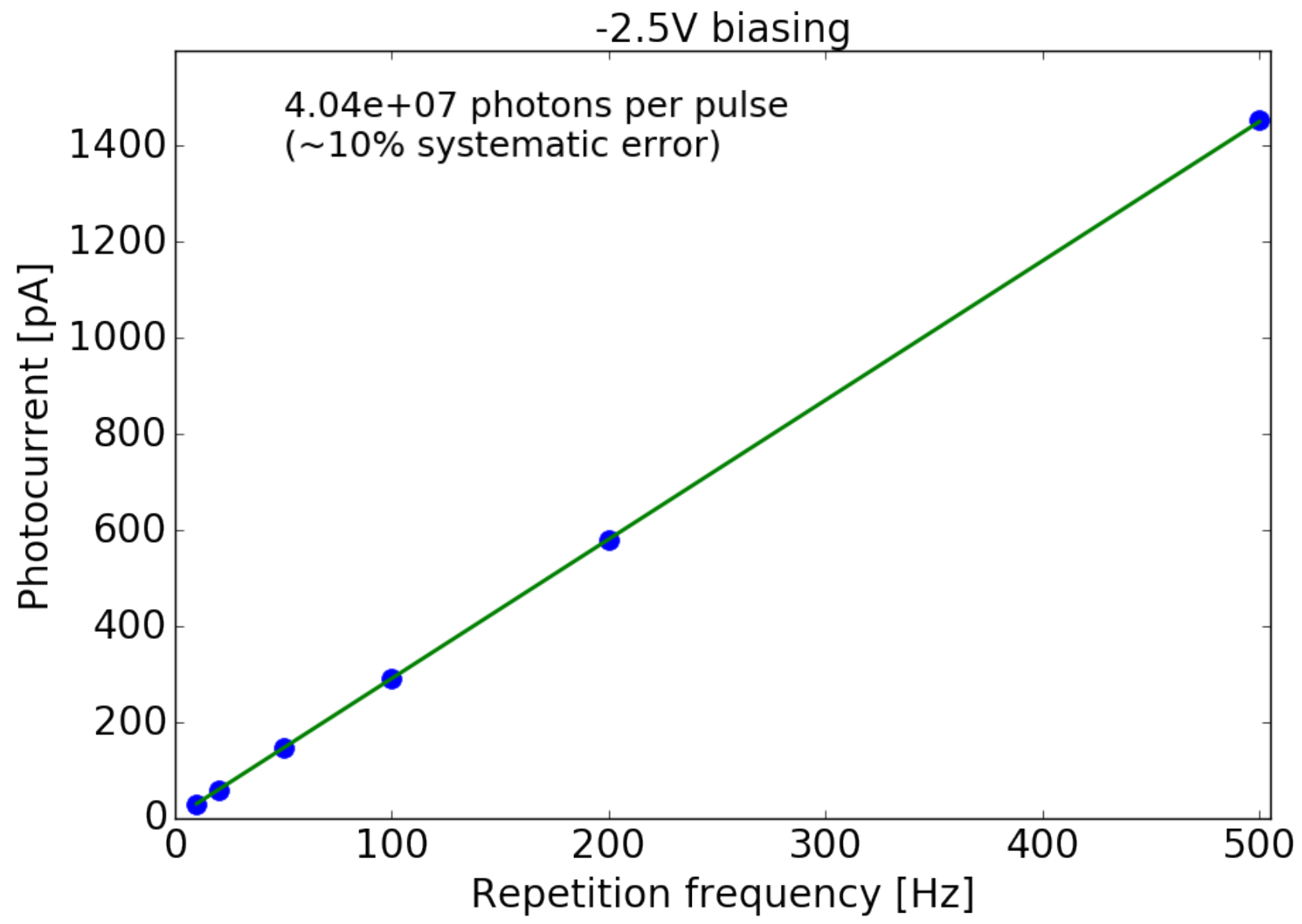

Figure 8.13: Example measurement of the VCSEL per-pulse light output at −2.5 V biasing. The slope, corrected for the quantum efficiency of the photodiode, yields the number of photons per pulse.

curve.

For the LEDs, pulse intensities between $10^4$ photons per pulse (no discharge capacitance, low biasing) and $10^8$ photons per pulse (10 pF and positive biasing) are seen. For the earlier mentioned 385 nm LED at 10 pF the photon number goes from $\sim 5 \cdot 10^4$ photons per pulse at −2.0 V biasing to $\sim 10^7$ photons per pulse at 0.5 V biasing. This leads to an increase in the light curve standard deviation from ∼280 ps to ∼800 ps.

## 8.4 Summary

Using a well-established avalanche transistor based pulse driving circuit, combined with biasing and trigger control, a versatile, low-cost and easy to use light source has been built and characterized. Using LEDs sub-ns light pulses at nearly arbitrary wavelengths can be realized. In conjunction with a 850 nm VCSEL light curves as short as 100 ps have been achieved.

Within the IceCube collaboration the device is now being used as a calibration pulser for IceAct, in lab setups at RWTH Aachen, University of Wisconsin-Madison, Technical University of Munich, Chiba University and University of Münster, and an integration into the POCAM is also being considered. Outside of IceCube the device is known to be in use by the Pierre-Auger Observatory (RWTH Aachen), Super-Kamiokande (The University of Tokyo), CMS (INFN Turin and the University of Maryland) and SABRE (University of Canberra).

# 9 THE HOLE ICE PROBLEM

**Measuring the optical properties of the refrozen drill holes has been an unresolved problem ever since the installation of the AMANDA detectors and thus for over 20 years. As it is one of the dominating detector systematics, in particular for low-energy oscillation analyses, several new measurement concepts have been explored.**

After the propagation of Cherenkov photons through the bulk glacial ice, each photon detected by a DOM has to also propagate through the refrozen water of the drill holes, called "hole ice".

A quantitative evaluation of the hole ice was first conducted with IceCube's predecessor AMANDA. Since then, multiple AMANDA and IceCube studies using inter-string LED data[314,315] or muon data[316,317] have qualitatively broadened our understanding of the hole ice. However, the obtained results are still partially conflicting.

## 9.1 FROM THE SWEDEN CAMERA

The hole ice was directly imaged by a pair of cameras installed at the bottom of drill hole 80[318], in the center of IceCube[319]. It suggests two hole ice components: A clear outer region and a central column of $\sim 8\,\mathrm{cm}$ diameter with a very small scattering length.

This observation is consistent with cylindrical freezing, where impurities or air bubbles are pushed along the freezing boundaries until they merge in the center. For this reason the small inner column is also referred to as the "bubble column". As the clear outer region is believed to have no impact on the detector sensitivity, the terms hole ice and bubble column are used interchangeably in the following.

## 9.2 MODELING VIA ANGULAR ACCEPTANCE CURVES

The impact of the hole ice is usually modeled as a modification of the DOM's angular acceptance curve with respect to the

314 Niessen, *"Simulated bubbes and their effect on some observables in AMANDA B4"*

315 Chirkin, *"Fitting the hole ice angular sensitivity model with the all-string flasher data"*

316 Olbrechts et al., *"On the Angular Sensitivity of Optical Modules in the Ice"*

317 Gerhardt, Effects of Hole Ice on Hit Time

318 *see section 4.5.2*

319 Hulth, *"Results from the IceCube video camera system at 2455 meters ice depth"*

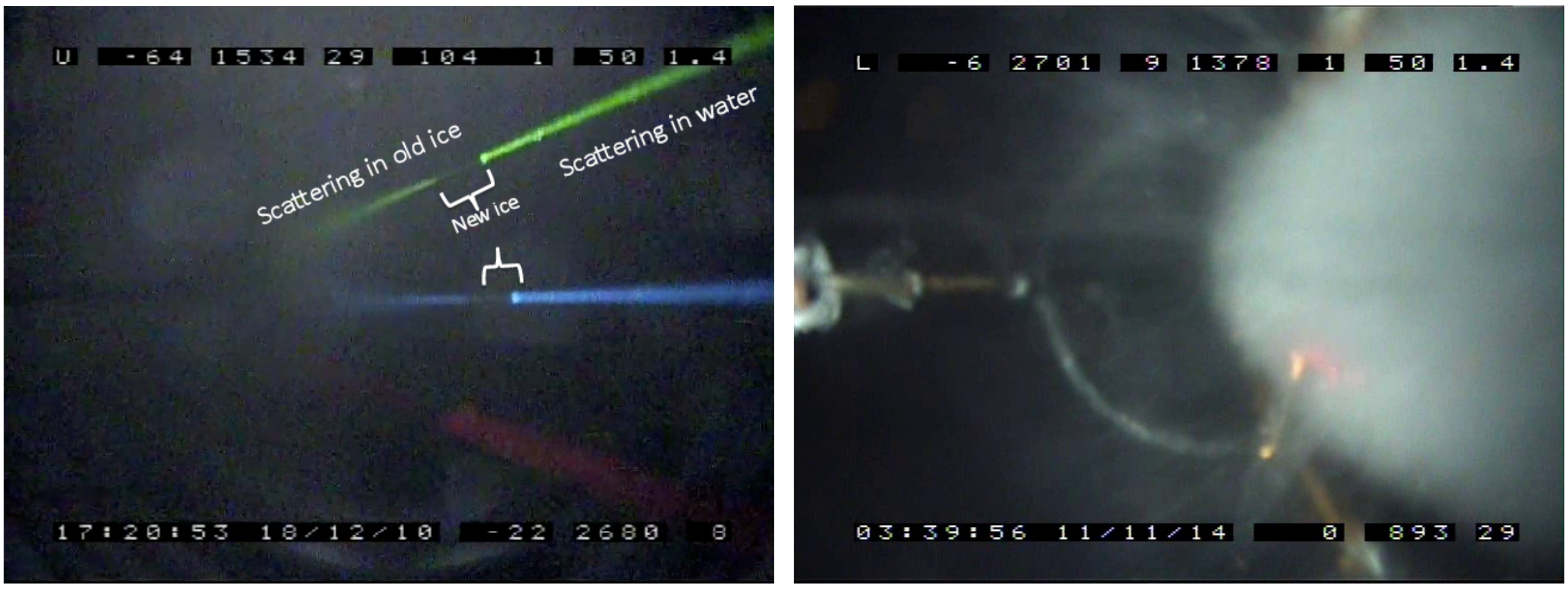

Figure 9.1: Sweden Camera images. Left: Camera facing sideways shortly after deployment. The light is seen to scatter least in the newly refrozen outer hole ice. Right: Lower camera looking straight up at the upper camera, with the hole ice fully developed in the right half of the image. Filament like outgrowth from the column indicate that the scattering centers are indeed air bubbles.
[Per Olof Hulth (2013). "Results from the IceCube video camera system at 2455 meters ice depth"]

laboratory measurement (see Figure 9.27). The angular acceptance curve abstracts the DOM as a point-like sensor, whose acceptance depends only on the incident angle with respect to the PMT axis. A re-evaluation of the DOM's intrinsic angular acceptance is presented in section 9.2.4.

In this model the acceptance in the forward direction of the PMT ($\cos(\theta) = 1$) is reduced with respect to the lab measurements as the hole ice scatters away incident photons. The acceptance in the backward direction ($\cos(\theta) = -1$) is increased as photons have a chance to be scattered to the front of the DOM.

Measurements are difficult as the forward region can not be directly probed with the installed LEDs or with Cherenkov light from atmospheric muons.

Different measurements and models are currently being used and their range reflects the current uncertainty with respect to the angular acceptance.

### 9.2.1 *The AMANDA H2-model*

The default angular acceptance model considered in IceCube until 2015 was derived from AMANDA measurements[320].

A frequency doubled, pulsed YAG-laser at a wavelength of 530 nm sent light to a spherical diffuser in the ice. From this data, timing distributions of close-by but non-saturated[321] OMs were extracted.

MC simulations assuming a bubble column of varying scattering length but with a diameter equivalent to the drill hole, indicated

[320] *A. Karle and K. Woschnagg (Sept. 1998). "Hole Ice Studies with YAG data." AMANDA internal report*

[321] *In AMANDA meaning SPE dominated, due to a vastly larger gain.*

a dependence of the widths of these time distributions on the hole ice scattering length.

To reduce the influence of bulk ice properties, the ratios of widths of OMs observing at the same altitude below and above the emitter were constructed. Figure 9.2 shows the data ratios superimposed to simulation predictions. This YAG laser analysis suggests a hole ice that is best described by a geometric scattering length in the range between 50 cm and 100 cm.

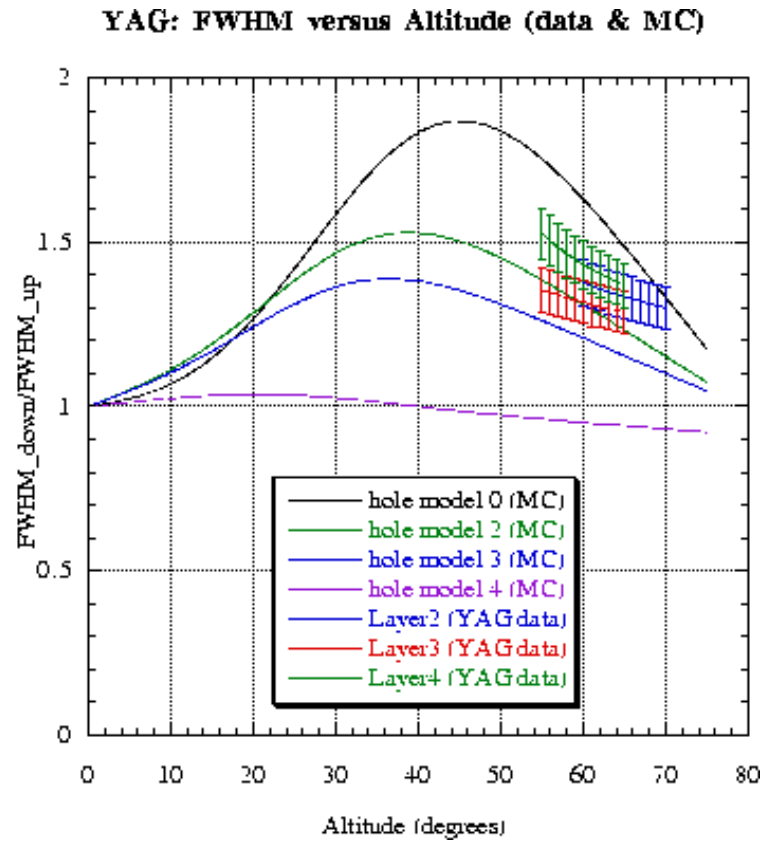

Figure 9.2: AMANDA YAG hole ice analysis comparing timing distributions for OMs below and above the emitter to simulations assuming a large bubble column. [AMANDA internal, credit: Albrecht Karle]

From this estimate the angular acceptance function was constructed by moving a plane wave of light around an OM in a simulation with the given hole ice properties. The resulting function, dubbed H2-model is shown in comparison to newer models in Figure 9.4.

As described later, the AMANDA model is nowadays understood to be conceptually flawed, in that the assumed column diameter drives the angular acceptance rather than the deduced scattering length[322].

[322] *see section 9.3.8*

## 9.2.2 *Ice cavern study*

In 2009 the hole ice timing impact was studied in IceCube[323]. The chosen method tried to identify the hole ice impact by comparing the timing distributions of DOMs in varying amounts of hole ice.

[323] *Gerhardt*, Effects of Hole Ice on Hit Time

Nominally, the drilled hole diameter is accurate to within centimeters. But on some occasions the drill did get stuck with no means to shut-off the water circulation. This resulted in drill caverns with estimated diameters of up to 10 m.

The study identified three DOMs in vast drill caverns and compared the time residuals of close by muon tracks to the average distribution of all DOMs. No timing differences were observed.

## 9.2.3 *The Dima model*

In 2015 the first widely used angular acceptance model based on IceCube (flasher) data was constructed by Dmitry Chirkin[324].

[324] *Chirkin, "Fitting the hole ice angular sensitivity model with the all-string flasher data"*

Similar to the flasher unfolding[325], it utilizes a matrix inversion technique to constrain the angular acceptance curve, binned into 40 equally distributed points in $\cos(\theta)$. To ensure an optimal impact on the flasher likelihood, the transfer matrix is constructed between the angular acceptance bins and all bayesian-blocked time bins for all emitter-receiver DOM pairs in the all-purpose flasher data. This roughly $40 \cdot 20,000,000$ matrix is then constructed from simulation assuming a seed angular acceptance, inverted and applied to data to yield the optimal angular acceptance. The process is repeated for a couple of iterations, such that the final result converges and is ideally independent of the seed.

[325] *see section 5.2.5.1*

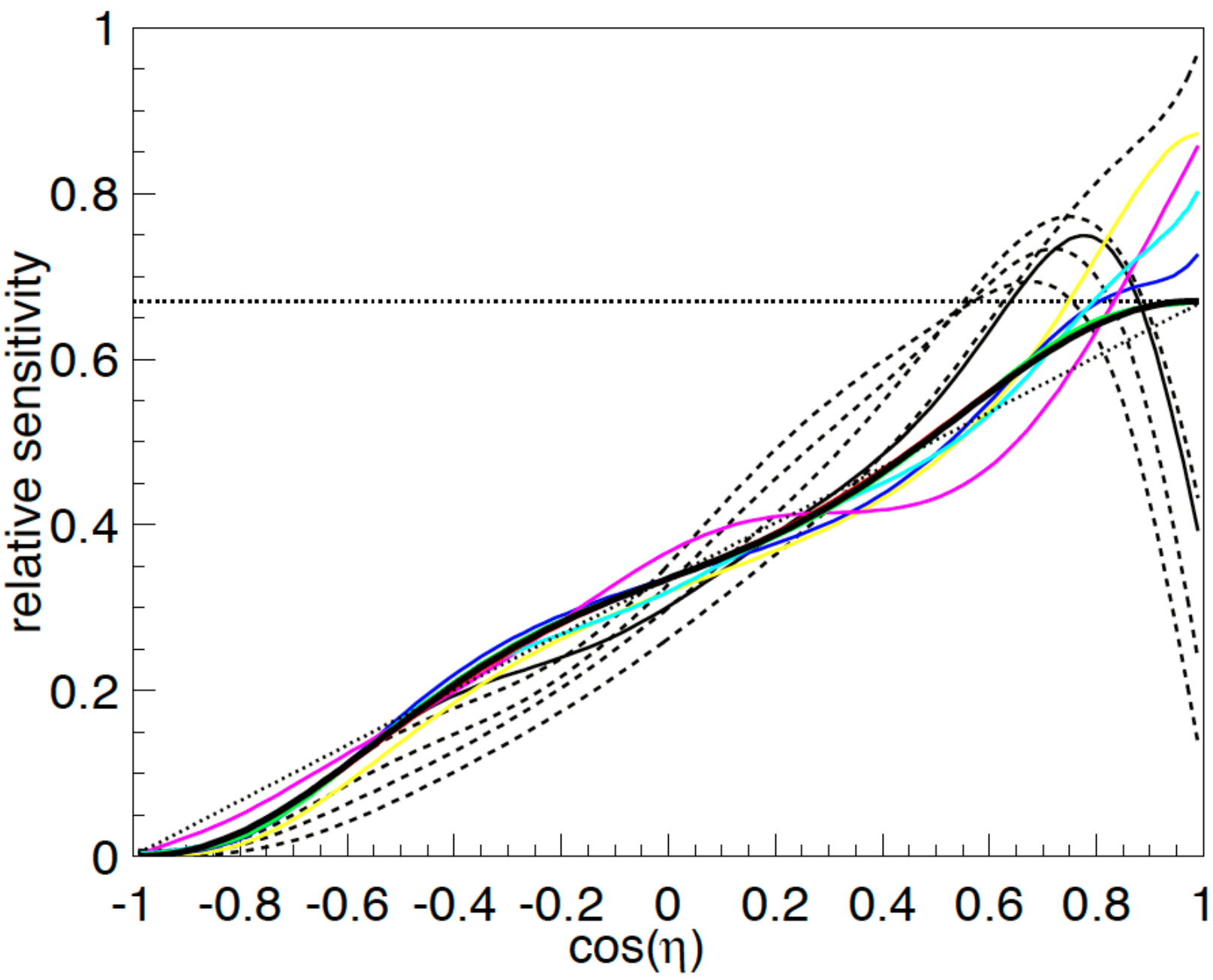

Figure 9.3: Unfolded angular acceptance curves (colored lines) resulting from different seeds (dotted lines). Even unfoldings seeded with a small forward acceptance converge towards large fractional forward acceptances.
[Dmitry Chirkin (2015a). "Fitting the hole ice angular sensitivity model with the all-string flasher data"]

A laboratory parametrization, several variants of the AMANDA model, as well as simply a constant value, have been tested as seeds. The resulting angular acceptance curves can be seen in Figure 9.3. While there is still substantial spread between the final iterations, generally a far larger forward-acceptance is found compared to the AMANDA model.

The overall improvement to the all-purpose flasher data is modest at $\sim 20$ LLH counts.

To enable an interpolation between the final realizations, a single parameter analytic parametrization was approximated as:

$$A_{\text{Dima}}(p) = 0.35 \cdot (1 + 1.5 \cdot \cos(\theta) - 0.5 \cdot \cos^3(\theta)) + p \cdot \cos(\theta) \cdot (\cos^2(\theta) - 1)^3 \tag{9.1}$$

Angular acceptance curves within the claimed uncertainty of $p$ are given in Figure 9.4.

Following the studies which will be presented in section 9.3.8, it is now understood that the forward acceptance relates to the size of the bubble column, while the scattering length alters the shape of the maximum. As the chosen parametrization fixes the forward acceptance, it does not allow for any variation of the assumed bubble column size.

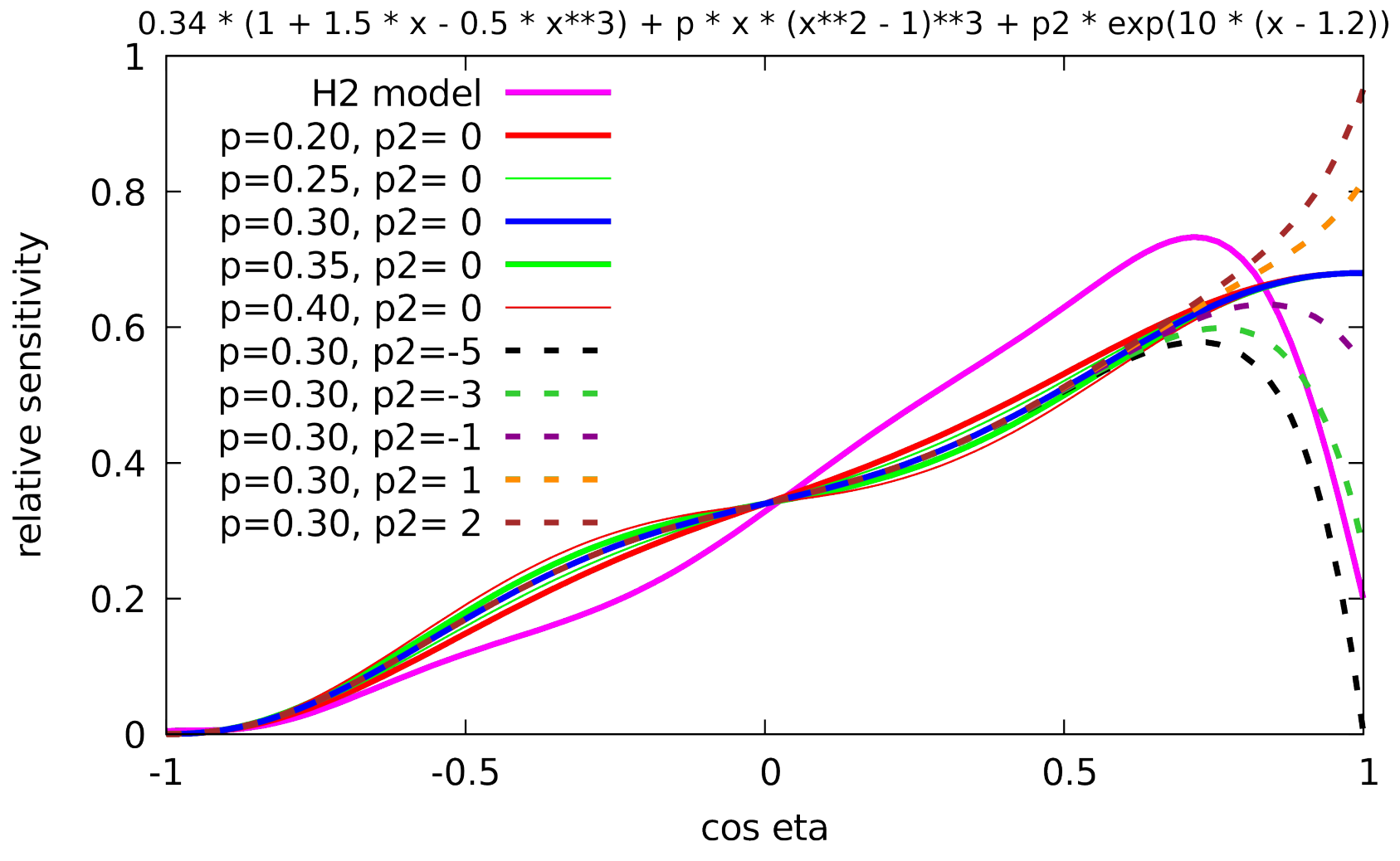

Figure 9.4: $p$, $p2$ parametrization of the angular acceptance curve, allowing interpolation between the AMANDA models and the best fit unfolding.
[Juan Pablo Yanez (2018). *MSU Forward Hole Ice*]

To overcome this limitation the IceCube low-energy group introduced an ad-hoc extension of the parametrization which introduces a variable damping of the forward-acceptance:

$$A_{\text{MSU}}(p, p2) = A_{\text{Dima}}(p) + p2 \cdot \exp(10 \cdot (\cos\theta - 1.2)) \quad (9.2)$$

Allowed $p2$ parameter values range between 2, giving a lab-like acceptance, and -5, yielding zero forward acceptance as expected from a strongly scattering bubble column covering the entire DOM.

### 9.2.4 *Re-evaluation of the DOMs intrinsic angular acceptance*

Prior to investigating the effect of the hole ice on the DOM's angular acceptance function, the angular acceptance intrinsic to the DOM shall be re-investigated.

This study utilizes the laboratory for low-volume, absolute DOM calibration, constructed by Christopher Wendt et al.[326] at WIPAC, Madison. For a general overview of the setup see Figure 9.5.

[326] *Tosi et al., "Calibrating photon detection efficiency in IceCube"*

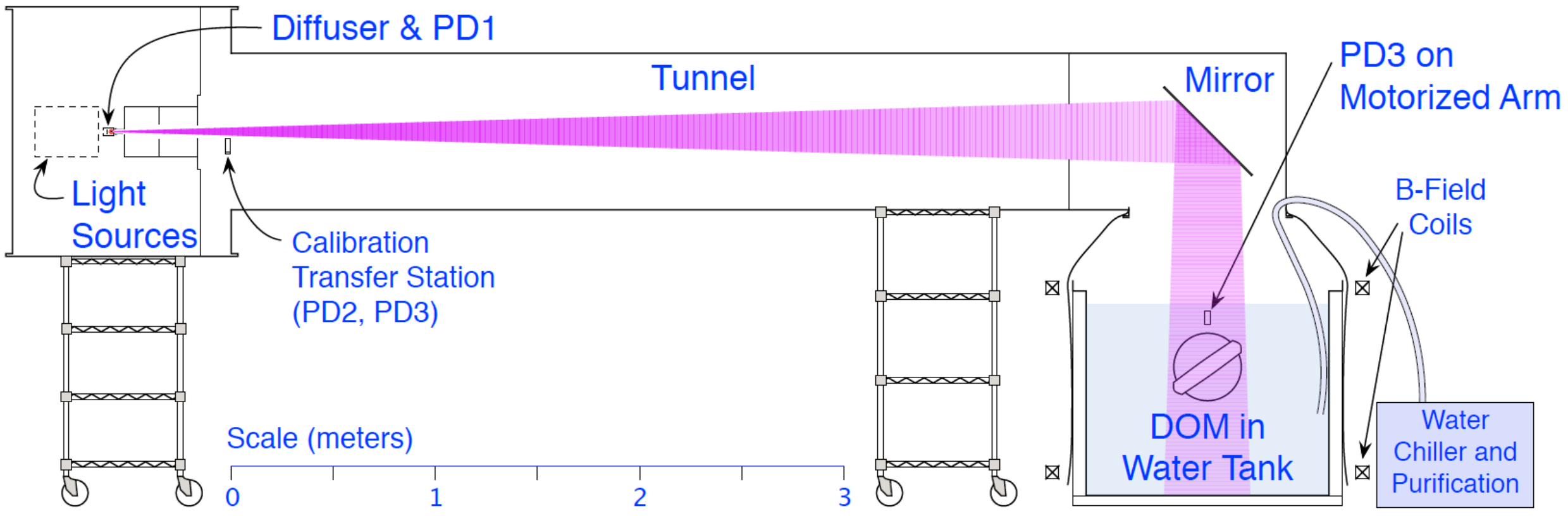

Figure 9.5: Overview of the Madison lab for DOM testing. A DOM is positioned in a water tank. Light sources are located in a source box ~4 m away.
[Delia Tosi et al. (2014). "Calibrating photon detection efficiency in IceCube"]

Light coming from a source box expands under its natural divergence in a 5 m tunnel. It is then restricted to the desired size, slightly larger than a DOM, by an aperture before being reflected down onto a DOM, positioned in a water tank. This results in a near planar wave of light.

The water, which is de-ionized during filling, is filtered continuously and is kept at a stable temperature around 3 °C. These precautions guarantee a stable and low PMT dark noise rate.

As the electron optics inside the PMT can be influenced by magnetic fields, a set of three Helmholtz coil pairs on the outside of the water tank is used to set any desired field vector and are usually configured to negate the Earth's magnetic field.

#### 9.2.4.1 *Light sources*

Of the light sources available (see Figure 9.6), the multi-color, programmable intensity LEDs as well as the Hamamatsu PLP-10 405 nm pulsed laser[327] are used. The LEDs can only be operated continuously. In this case, the SPE scaler (counting the trigger rate above 0.25 PE) is used as data acquisition.

[327] *Hamamatsu Photonics (2017).* PLP-10Laser diode head Series

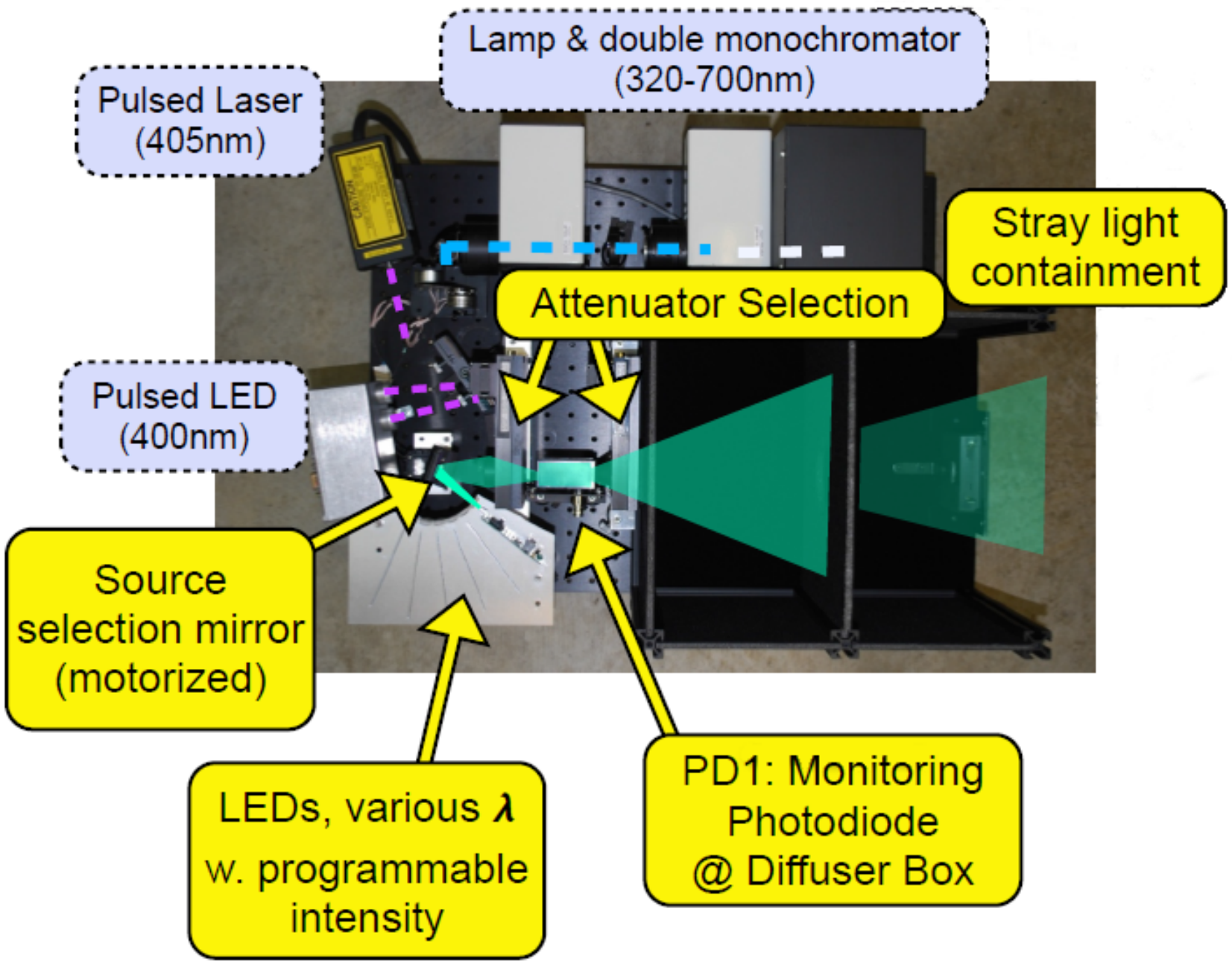

Figure 9.6: Overview of available light sources. For this study the pulsed laser and intensity programmable LEDs in conjunction with a variety of neutral density filters have been used.
[Wouter Van De Pontseele (Nov. 7, 2015). "Studies Supporting PMT Characterization for the IceCube Collaboration: Precise Photodiode Calibration"]

The laser features a synchronization pulse, which triggers a second DOM mainboard. A local coincidence setup between this mainboard and the DOM in the tank allow for a full waveform acquisition. At $\sim 70\,\mathrm{ps}$ FWHM the laser pulse appears instantaneous to the DOM. The output power of the laser is fixed, but neutral density filters can be used to attenuate the light. Without any filters the DOM typically registers 70 PE per pulse. In a second configuration with 99.9% attenuation the SPE occupancy is around 3%.

#### 9.2.4.2 *Rotation mechanism*

The harness of the DOM is attached to a rotational axis, which in turn gets rotated by a multi-turn servo. In addition, an inclinometer is attached to the shaft, which can be used to read back the set angle. The angle as read by the inclinometer has been tested to agree with the desired angle to within 1°.

The inclinometer has been positioned to agree with the orientation of the equator band. As the equator band does not necessarily match the true equator of the DOM exactly, a systematic offset has to be considered.

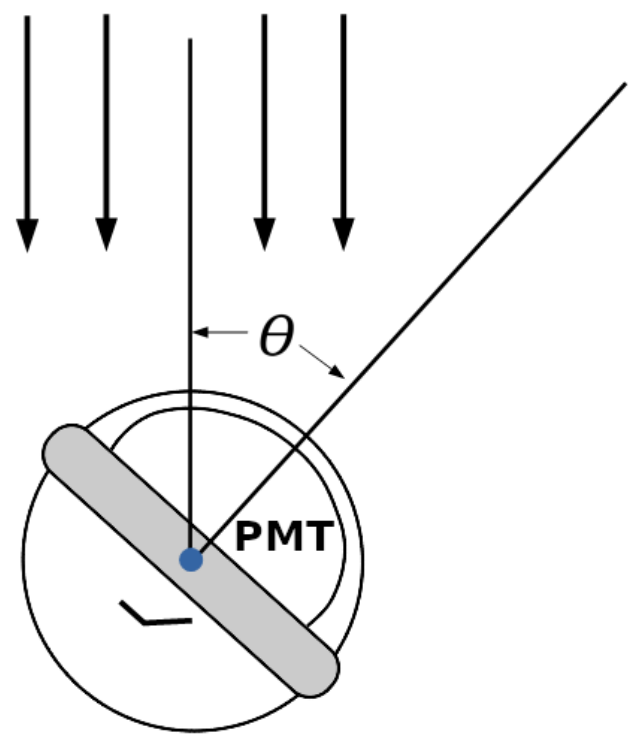

Figure 9.7: Definition of the DOM rotation angle. This is consistent with the zenith angle used for the angular acceptance curves.

The DOM angle is defined to be zero as the DOM is pointing upwards. Positive angles denote the DOM being rotated to the right in the perspective defined in Figure 9.7.

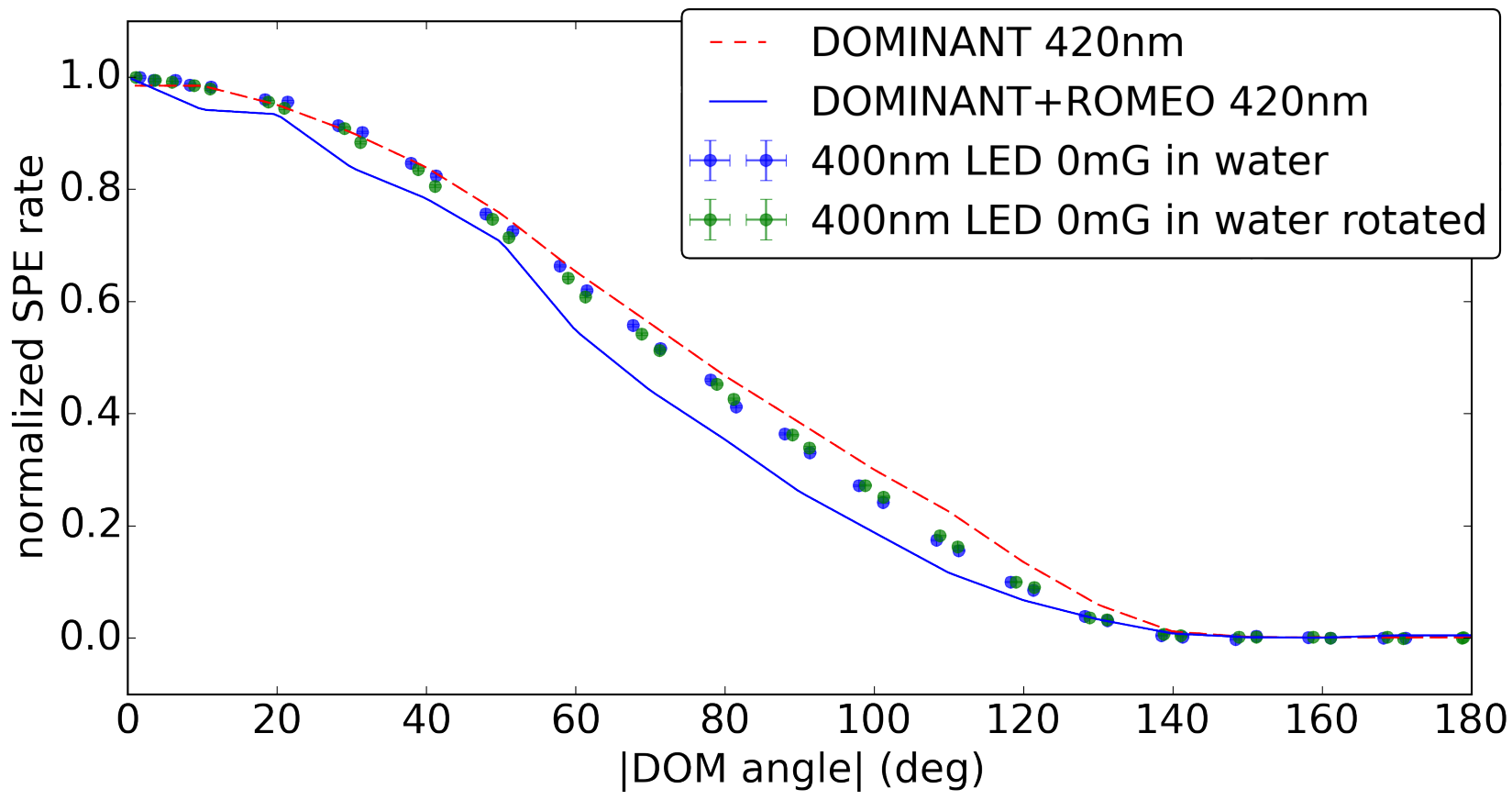

Figure 9.8: Angular photon counting acceptance. The DOM angle was corrected for a small misalignment of the waist band. The blue and green data points denote different orientations of the first dynode. The new measurement is compared to the DOMINANT Geant simulation and the IceCube default ROMEO model.

### 9.2.4.3 *Photon counting acceptance*

Using the LEDs or the laser at low occupancy and high repetition frequencies the photon counting acceptance has been measured as the SPE discriminator rate versus the DOM angle. The laser and an LED of equivalent wavelengths agree to within less than 1%. LEDs between 360 nm and 520 nm show at most a 10% variation over the entire angular range.

The measured photon counting angular acceptance at 400 nm can be seen in Figure 9.8. It is compared to the default IceCube DOM-intrinsic angular acceptance models DOMINANT and ROMEO. DOMINANT is the result of a Geant simulation of the DOM being illuminated by a plane wave of light[328]. It gets folded with PMT characteristics, such as the average position dependent photocathode acceptance, to result in the ROMEO model[329].

All curves are normalized to 1 at vertical incidence. The inclinometer angles were corrected by 2° to account for a slightly tilted waistband and to match the acceptance of positive and negative angles[330]. The new lab measurement falls right between the purely geometric expectation by DOMINANT and the adjusted ROMEO model.

### 9.2.4.4 *Gain variations*

While the photon counting acceptance is the quantity usually considered for the angular acceptance, it should be noted that the average gain was found to also show a dependence on the DOM orientation.

Using the pulsed laser at low occupancy SPE charge spectra were recorded at different angles. The mean SPE charge in units of DOMCal calibrated SPE versus angle is seen in Figure 9.10. The data has been split into several time ranges with respect to the laser trigger according to the definition in Figure 4.7.

[328] *Kotoyo Hoshina et al. (2007). "DOMINANT : Dom Optical-photon to Material INteraction ANd Tracking simulator"*

[329] *Kotoyo Hoshina et al. (2004). "ROMEO: the Root-based Optical Module EmulatOr"*

[330] *This was done after verifying that the asymmetry of the first dynode does not induce an asymmetry for positive and negative rotation angles, by rotating the Benthos sphere in the waistband.*

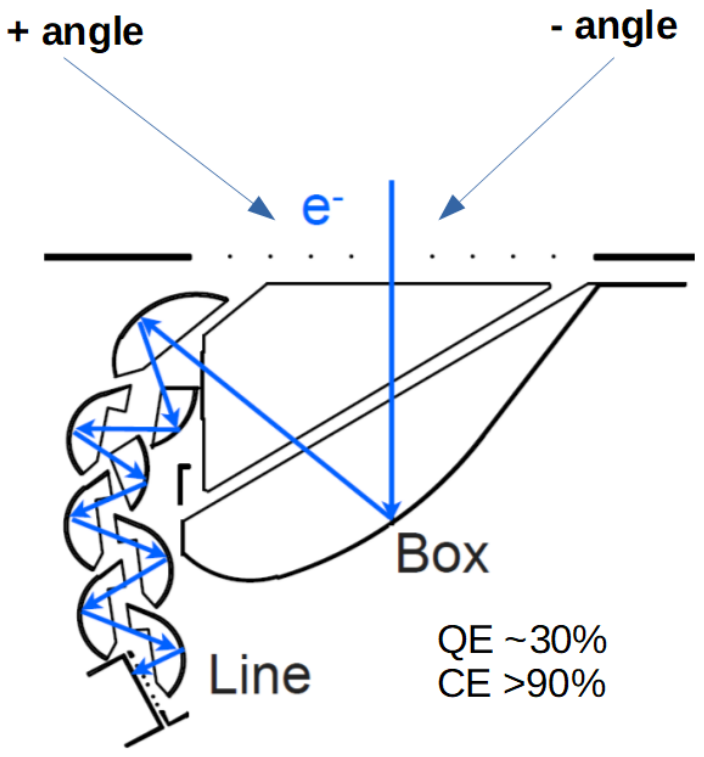

Figure 9.9: Sketch of a box-line dynode chain. The initial photoelectron impacts the first dynode at different angles depending on the emission position on the photocathode, resulting in different gains. Adapted from:
[Okajima et al., "Detailed performance evaluation of a new 20-inch photomultiplier tube with a Box and Line dynode"]

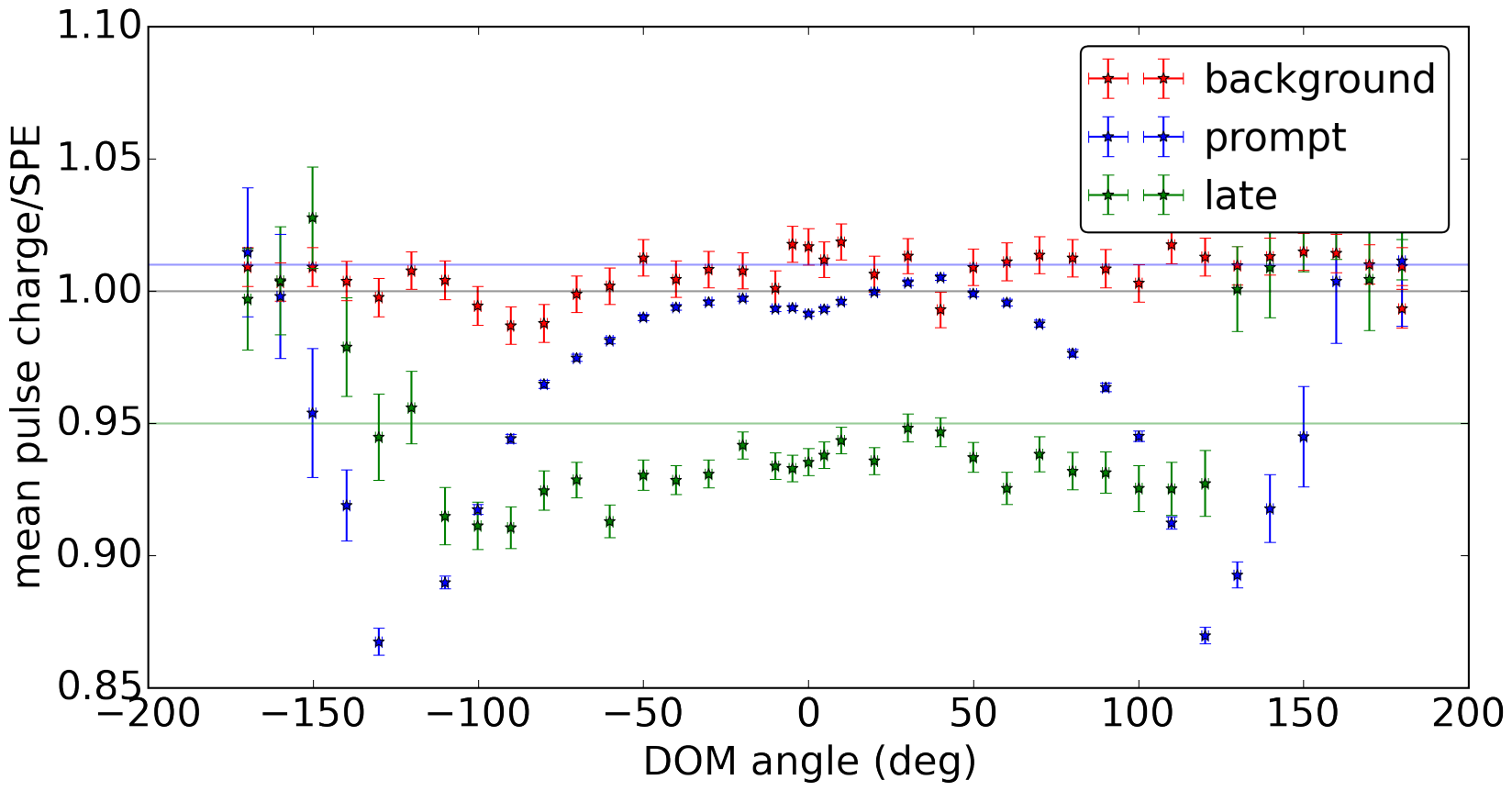

Figure 9.10: Variations of the PMT gain as a function of photon incidence angle for various pulse types. Nominal (prompt) pulses are amplified less at high angles. The effective charge angular acceptance function is the product of the gain and the photon detection efficiency.

For background hits[331], which are understood to originate from random locations on the photocathode, no dependence is observed, establishing a stable measurement. Prompt signals, for which the associated electrons followed the nominal electron trajectory, show a roughly 15% decrease in gain as the DOM is turned away from the light source. For late pulses, where the electron initially gets deflected off the first dynode and then returns from a randomized direction, the gain is flat.

[331] *selected to be uncorrelated with the light source*

This behavior can be understood considering the dynode structure of the employed PMT. The Hamamatsu R7081-02 features a box-line dynode chain as sketched in Figure 9.9. To achieve the best possible collection efficiency, the first dynode is large and shaped like a tilted, parabolic mirror. For prompt electrons the incident angle on the first dynode and therefore the amplification depends directly on the illuminated photocathode location.

Event reconstructions often use measured charge instead of hit probability.It follows that the relevant DOM acceptance is the product of the photon counting acceptance and the angular gain variation. This was also confirmed in a measurement of the pulsed laser without attenuation by constructing the angular acceptance curve as the average charge per pulse as a function of rotation angle.

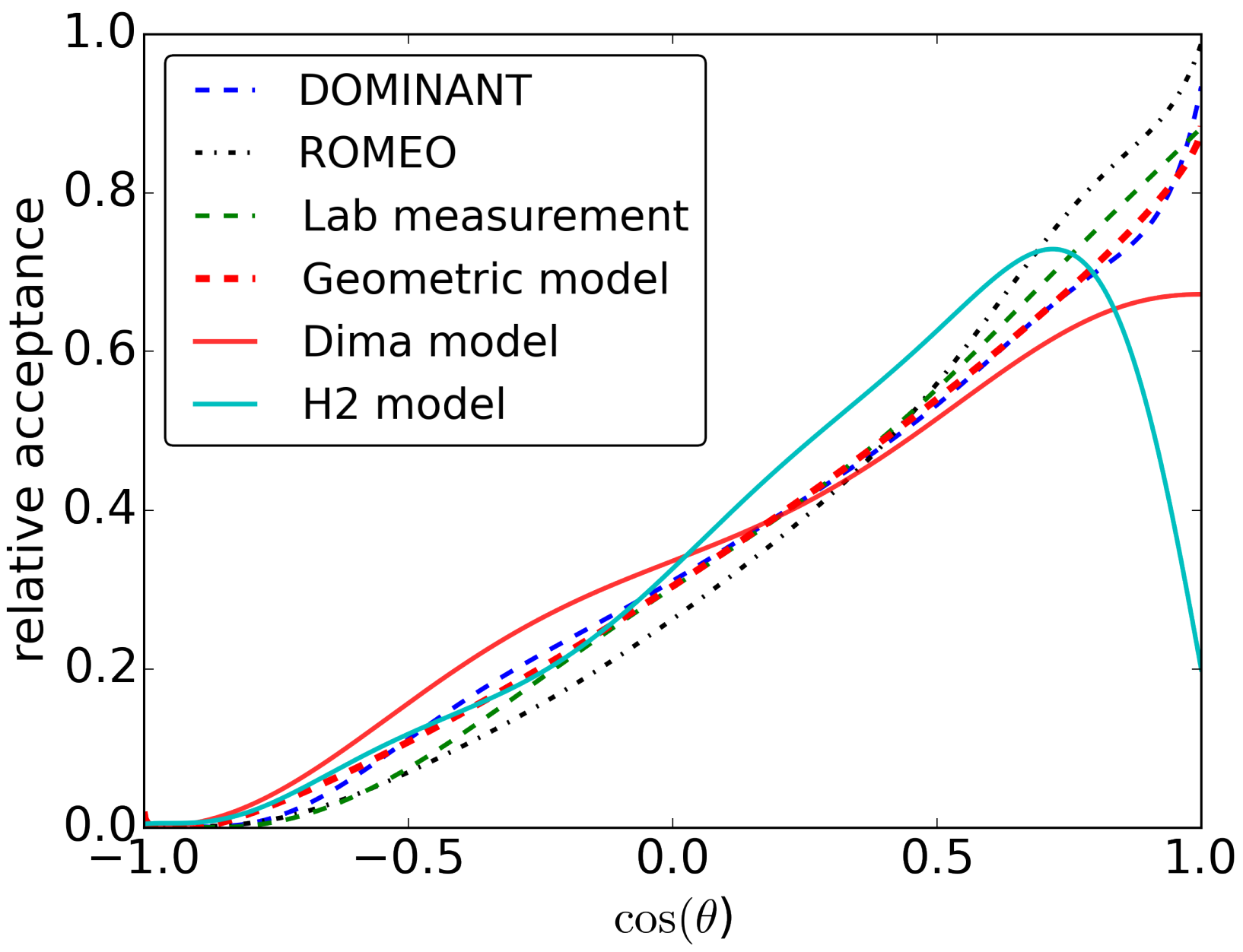

Figure 9.12: Angular acceptance as calculated from a simplified geometric model of the DOM, compared to various angular acceptance curves. The simple assumption that the sensitive area extends over the entire lower hemisphere of the Benthos sphere up to the waistband is sufficient for most use cases. The applied modifications due to the hole ice are far larger. (Curves are area normalized.)

#### 9.2.4.5 *Comparison to the geometric expectation*

While the angular acceptance function might appear complex at first, it can actually be described by an extremely simplified geometric model of the DOM.

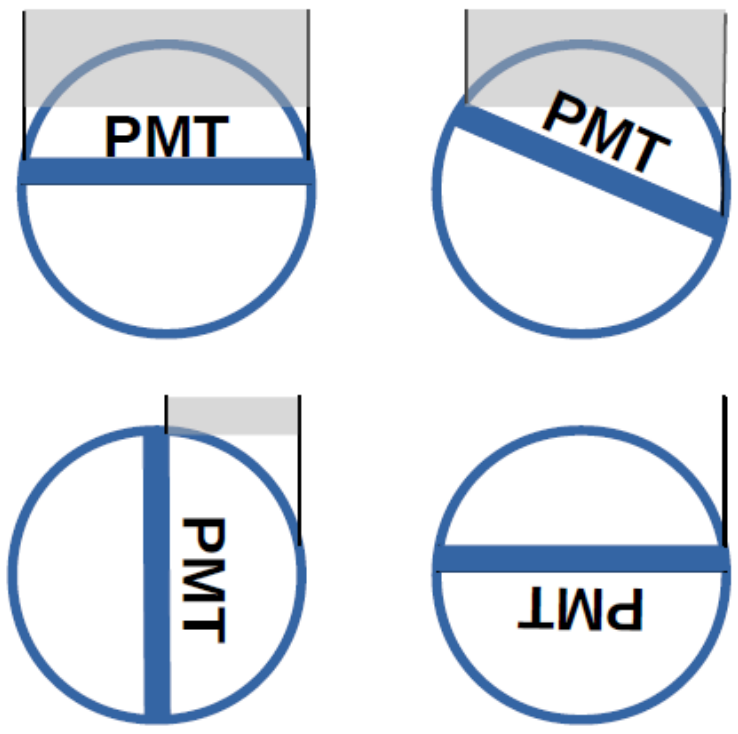

Figure 9.11: Construction used to convert simple geometric DOM model into an angular acceptance curve.

Assuming the sensitive area of the PMT to be well described by the surface of the lower hemisphere of the pressure sphere, where unobstructed by the waistband, the acceptance can be constructed as seen in Figure 9.11. Assuming that the effective area scales as the illuminated, horizontal distance[332] it is proportional to:

$$A \propto 16.51[\mathrm{cm}] \cdot (1 + \cos(\theta)) - 4[\mathrm{cm}] \cdot \sin(\theta) \tag{9.3}$$

[332] *as the waistband (incorporated through the sin-term) only obscures one dimension under rotation of the DOM*

Figure 9.12 compares this approximation to the above measured angular acceptance curve as well as the IceCube default DOMINANT and ROMEO models. The geometric model describes the Geant based DOMINANT simulation at a good accuracy. The additional PMT effects as modeled in ROMEO and measured in the laboratory setup are generally small compared to common hole ice modifications. Therefore, it is justified to describe the DOM using this simplification, as will be done in section 9.4.1.

## 9.3 Same-DOM flasher measurements

Except for the Sweden Camera images, no direct observation of the hole ice has been performed to date. The following study started out with the aim to perform a direct imaging of the bubble column extent by utilizing the different relative orientations of flasher LEDs with respect to the assumed column.

---

**Declaration of Pre-released Publications**
The study presented in this section has in-part already been published by the IceCube collaboration[333]. The author of this thesis has contributed all presented work.

---

333 *Rongen, "Measuring the optical properties of IceCube drill holes"*

### 9.3.1 *The measurement idea*

For the determination of hole ice properties it is unfavorable to measure the light emitted by one DOM with another DOM, as the bulk ice properties will dominate the propagation.

By sequentially turning on the LEDs of one DOM and measuring the intensity of light returning to the PMT of the same DOM a measurement similar to the concept of the dust logger can be performed, where the measured intensity is proportional to the scattering coefficient in the direction of the LED.

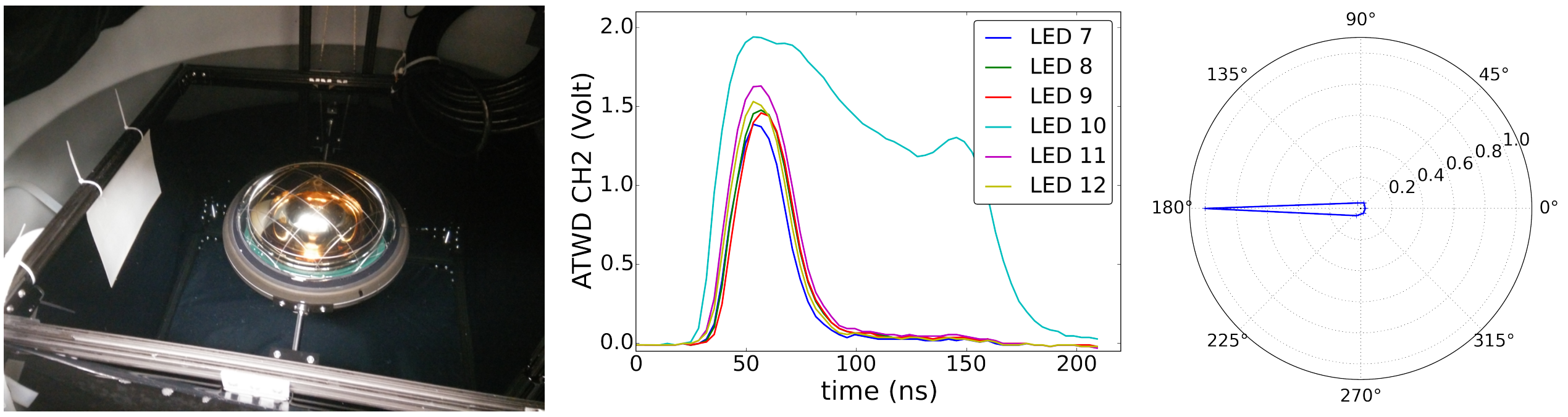

Figure 9.13: Lab demonstration of the measurement principle. Left: DOM in water tank with a reflector to one side. Middle: The PMT voltage waveform shows the reflected intensities of the different LEDs. Right: Azimuthal pattern of the relative detected brightness as derived from the late pulse peak (see section 9.3.3) to account for PMT saturation. A characteristic tear-drop pattern is seen.

This idea was demonstrated by conducting the described procedure with a DOM in a water tank. The tank was lined with an absorptive black textile, which contained a reflective sheet on one side. Figure 9.13 shows the average PMT waveforms for each tilted LED. The LED pointing in the direction of the reflective sheet (LED 10) yields the strongest PMT signal. The orientation and azimuthal size of the reflecting object with respect to the DOM, can be gauged from the so called tear-drop pattern shown in the same Figure.

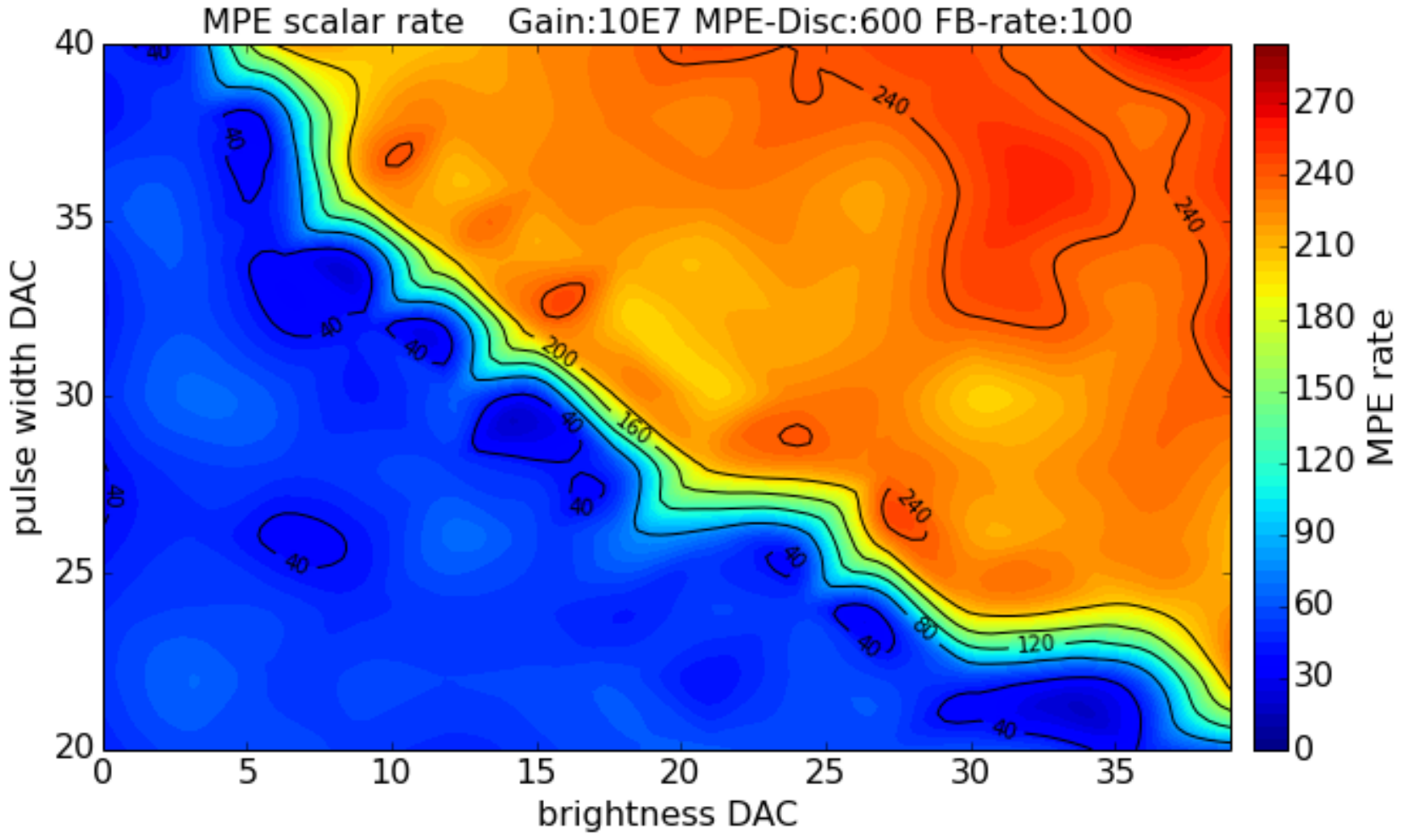

Figure 9.14: Trigger efficiency of the flasher LED measured as the MPE scaler rate as a function of LED DAC settings. In the blue region the LEDs do not reliably emit light.

## 9.3.2 *The data acquisition*

The DOM firmware used for standard data acquisition disables the high voltage to the PMT whenever the flasher board is enabled. Thus, a custom DAQ called DARD[334] based on the debugging firmware IceBoot had to be developed. The resulting system controls a single IceCube string one DOM at a time, while the rest of the detector remains in normal operation.

*334 Data Acquisition foR a flashing Dom*

As the receiving PMT is very close to the emitting LED, the smallest possible LED intensity and PMT gain are desirable. To test the dimmest possible LED operating condition, resulting in a stable light emission, the above mentioned lab setup was utilized. With the PMT at nominal gain, a single LED was flashed at 100 Hz and using a variety of width and brightness settings.

The resulting MPE scalar[335] rates are depicted in Figure 9.14. For low values of both the width and the brightness DACs the trigger rate is below the set rate of 100 Hz. This indicates that the LED did not flash reliably.

*335 a discriminator triggering on PMT signals above several PE*

For in-situ operation the width has been chosen at a fixed DAC value of 30, with brightness settings between 25 and 35.

Even at the lowest possible pulse setting each LED is expected to emit $10^8$ photons[336]. At the nominal gain of $10^7$ this saturates the PMT of the flashing DOM. Therefore, a reduced high voltage of 750 V was chosen. This results in a gain of roughly $10^5$ while retaining most timing and efficiency[337] characteristics.

*336 see section 4.5.1 for the scaling*

*337 see section 9.3.6.1 for the collection efficiency*

### 9.3.3 *In-situ data collection and processing*

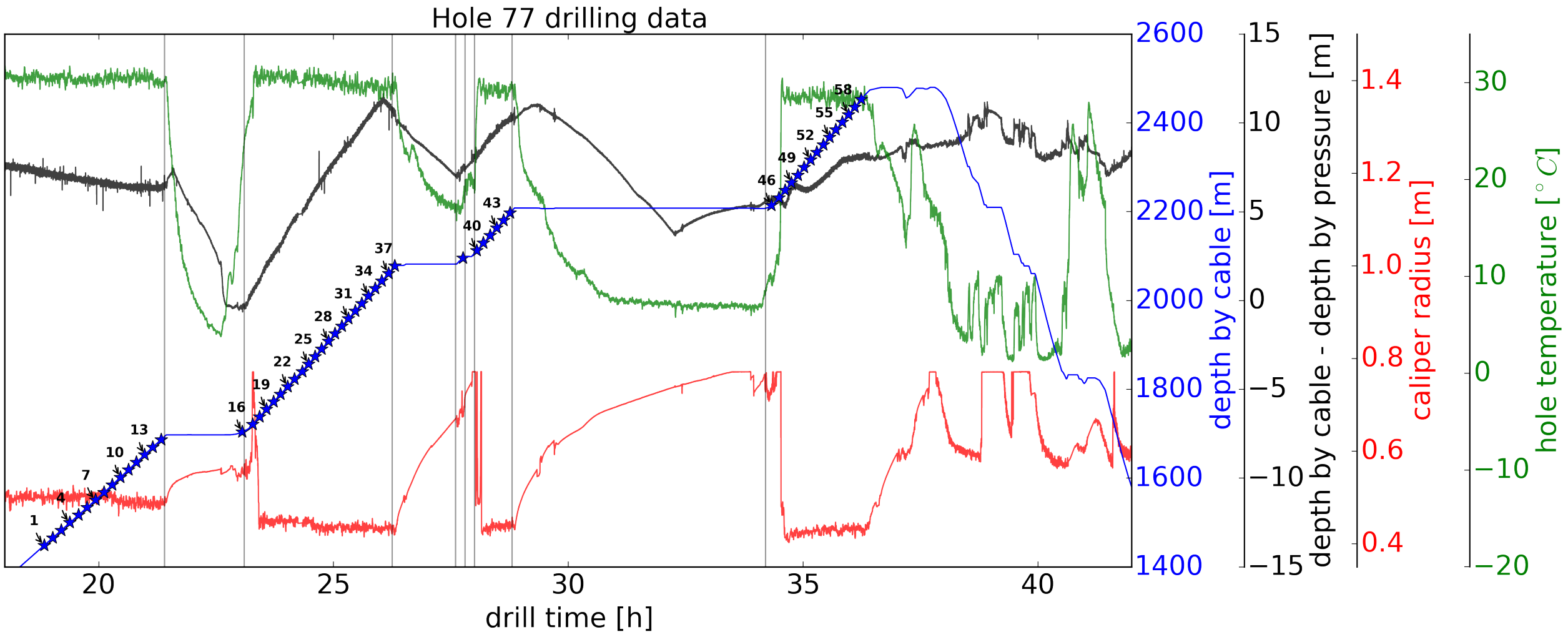

Figure 9.15: Drill data for String 77. The drill got interrupted at three points during the drilling process. With water still circulating, large caverns were melted into the ice, potentially resulting in more significant hole ice at these locations. This string was picked for the ice cavern run as it exhibits the largest number of significant drill caverns.

The DAQ has been applied to a total of 81 DOMs in four runs:

- In a pilot run, 5 DOMs above the Sweden Camera, which is the only location of known properties, and 3 DOMs in shallow ice on the same string were measured.
- Lacking obvious tear-drop patterns, a run was performed for DOMs suspected to be in drill caverns. String 77, for which the drill profile can be seen in Figure 9.15, was selected for being the string with the largest number of significant drill problems and resulting caverns. 12 DOMs surrounding 3 caverns were chosen.
- Still lacking concrete evidence the statistics was increased by measuring all 60 DOMs on string 80. Eight of these had already been part of the pilot run.
- In 2017 an independent analysis of flasher data identified nine strongly tilted (up to 50°) DOMs. DARD was applied to these, on the premise that some LEDs on these DOMs might be closer to the hole center and might illuminate the bubble column, even if it is too small to be otherwise seen.

After setting up the high voltage, 100 waveforms are recorded for each LED configuration and once without any light.

During post-processing the raw waveforms are averaged, corrected for the digitizer pedestals as obtained from the dark events[338] and then calibrated to absolute times and voltages using up-to-date DOMCal conversion values[339].

[338] *ATWD pedestals are very sensitive to the operational conditions and the default pedestals were found not to be sufficient*

[339] *Achterberg et al., "First year performance of the IceCube neutrino telescope"*

As the presented method probes very small distances, no resolvable timing features are expected. The only observable for each LED is the total returning charge. The integral over the resulting waveform voltage bins ($V_i$) is directly proportional to the number of recorded photoelectrons (PEs), where $Z$ is the frontend impedance and $f$ is the sampling frequency of the digitizers:

$$\text{PE}_{\text{exp}} = \frac{1}{Z(\Omega)} \cdot \frac{1}{e} \cdot \frac{1}{\text{f(Hz)}} \cdot \left( \sum_i V_i \right) \cdot \frac{1}{\text{gain}} \tag{9.4}$$

The frontend impedance and the PMT gain are calculated/extrapolated from the DOMCal calibration constants. The voltages are samples from the highest gain, non-clipped ATWD channel.

As discussed in section 4.1.4, the PMT behaves non-linearly in its output current for instantaneous currents above ~20 mA. To extend the photon counting capability into the saturation onset region the currents are corrected using the parametrization described in section 4.1.4.

For waveform amplitudes beyond 1.5 V, this correction becomes unreliable as the saturation levels off. To retain some photon counting capability the late pulse region is considered.

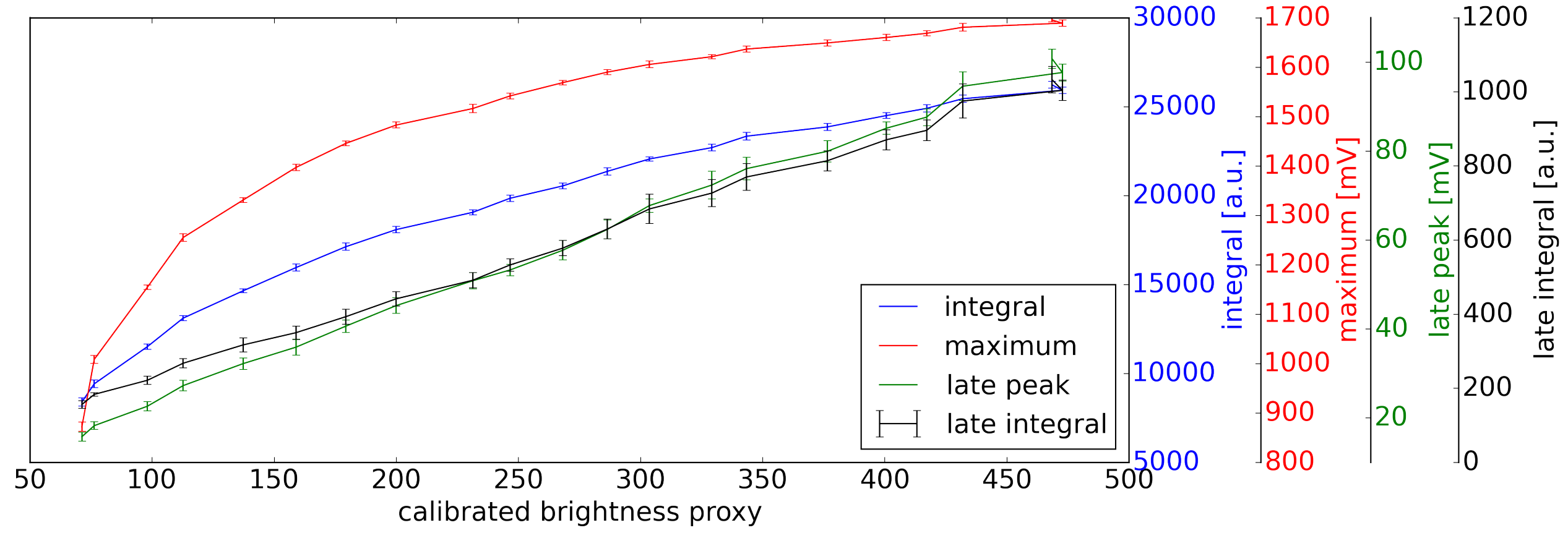

Figure 9.16: Saturation characteristics of different brightness proxies used to calculate the number of detected photons from the ATWD voltage waveforms.

In the lab setup the waveform integral of the least amplified channel was compared to the integral of the most amplified channel in the late pulse region. This region was defined to be 22 to 35 ATWD bins after the waveform maximum, which is consistent with the definition in Figure 4.7. For the tested DOM the late pulse region was found to contain ~ 2.75% of the full waveform charge. As the spread of this fraction over an ensemble of PMTs is not known, the found value is applied as a rough approximation.

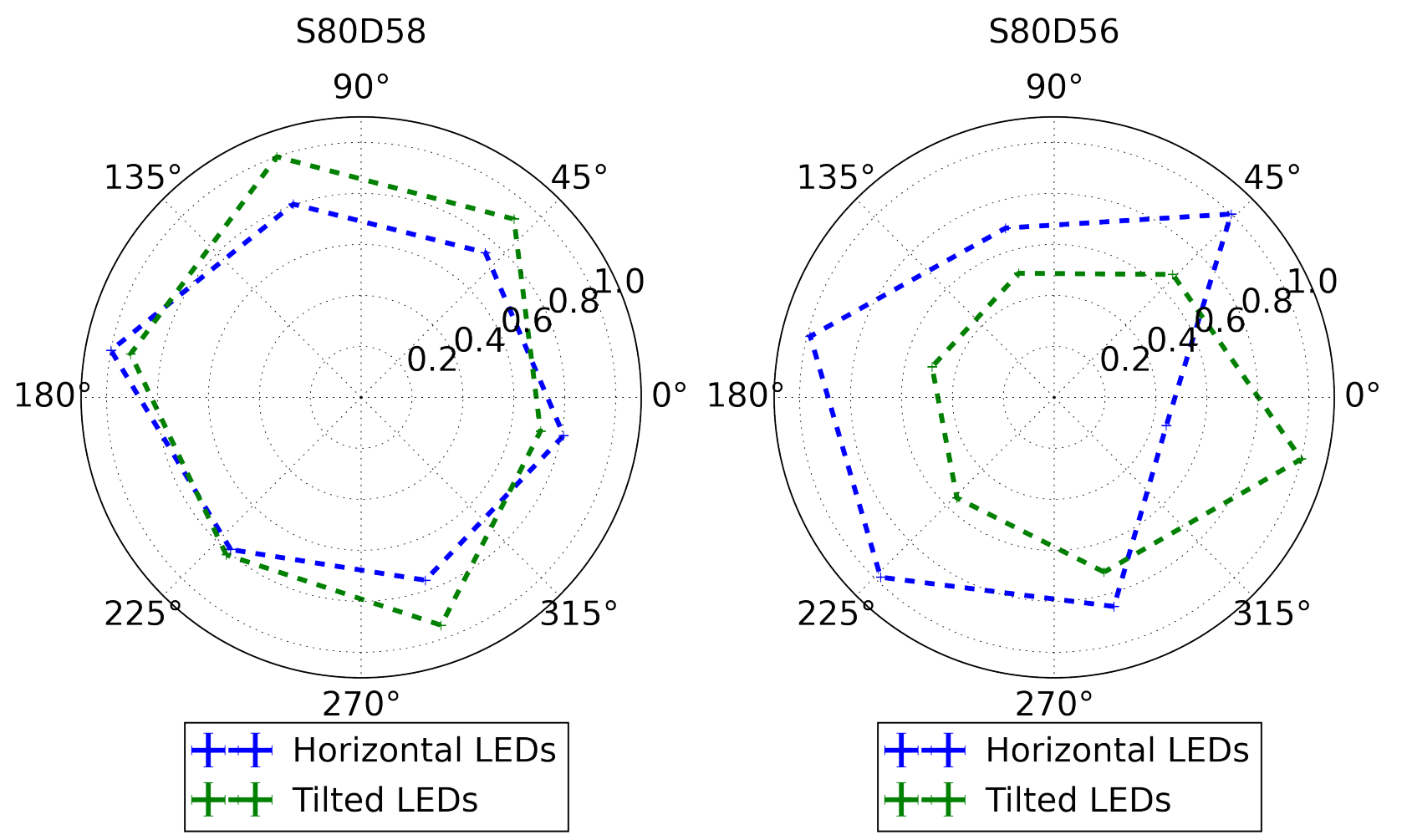

Figure 9.17: Typical azimuthal patterns of detected PEs for two DOMs, ranging from ∼2000 to ∼7000 PE. No strong tear-drop patterns are seen. Note that the photon counts are normalized on the brightest LED of each configuration.

The linearity of the different brightness proxies as a function of an arbitrary but linear brightness can be seen in Figure 9.16. The linear brightness proxy is obtained by operating the same light source behind a neutral density filter and scanning the average waveform maximum as a function of the light source brightness setting.

Two examples of the number of measured PE as a function of the azimuthal LED direction can be seen in Figure 9.17. Note, that light output is known to vary by about 20% between different LEDs given bright operation conditions, as discussed in section 5.2.5.1. The spread is expected to increase at dimmer settings.

For this study, data for both horizontal and tilted LEDs has been gathered. As the horizontal LEDs show far stronger saturation effects and are possibly subject to internal light contamination[340], only the tilted LEDs are considered for quantitative interpretations, while the horizontal LEDs provide a cross-check for potential tear-drop shapes.

[340] *see section 9.3.6*

## 9.3.4 *Teardrop patterns*

For the majority of DOMs the measured number of returning photons does not show a strong dependence on the azimuthal flasher direction . Out of the 81 DOMs tested only three show tear-drop patterns, where the maximum exceeds the median photon count by more than a factor three in the horizontal or tilted LEDs.

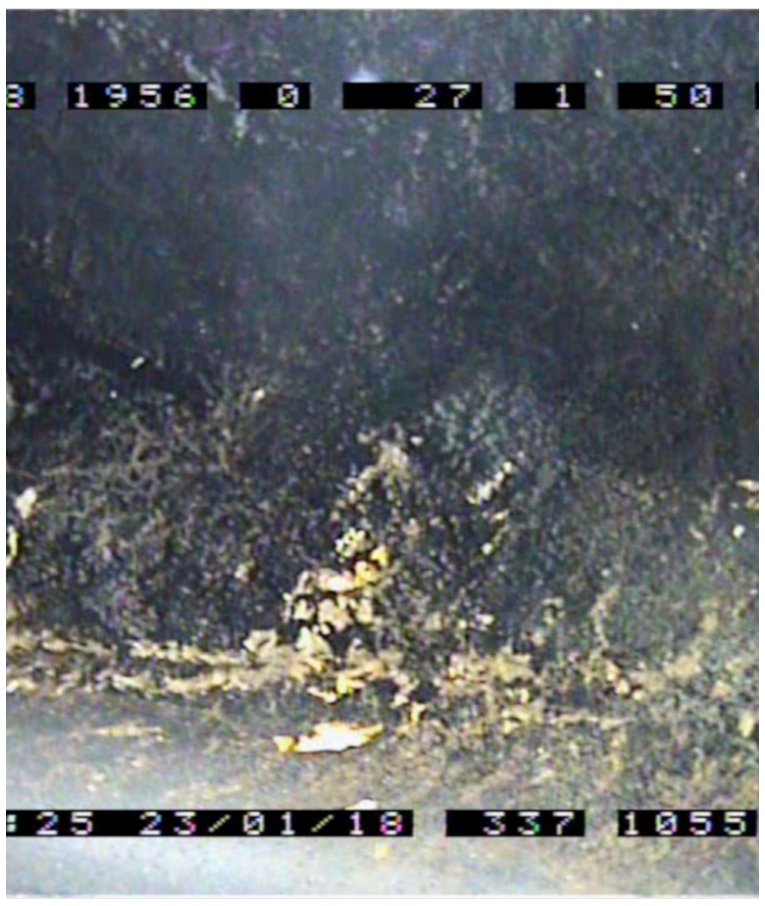

Figure 9.19: Sweden Camera view on yellow dirt patches found on the upper glass hemispheres. These are suspected to be glycol introduced by the drill and could act as strong scattering centers leading to isolated teardrop patterns.

While the number of returning photons between tilted and horizontal LEDs at the same DOM location is generally well correlated (Figure 9.18), these three DOMs do not show a con-sistent tear-drop shape in the other set of LEDs as evident in Figure 9.20.

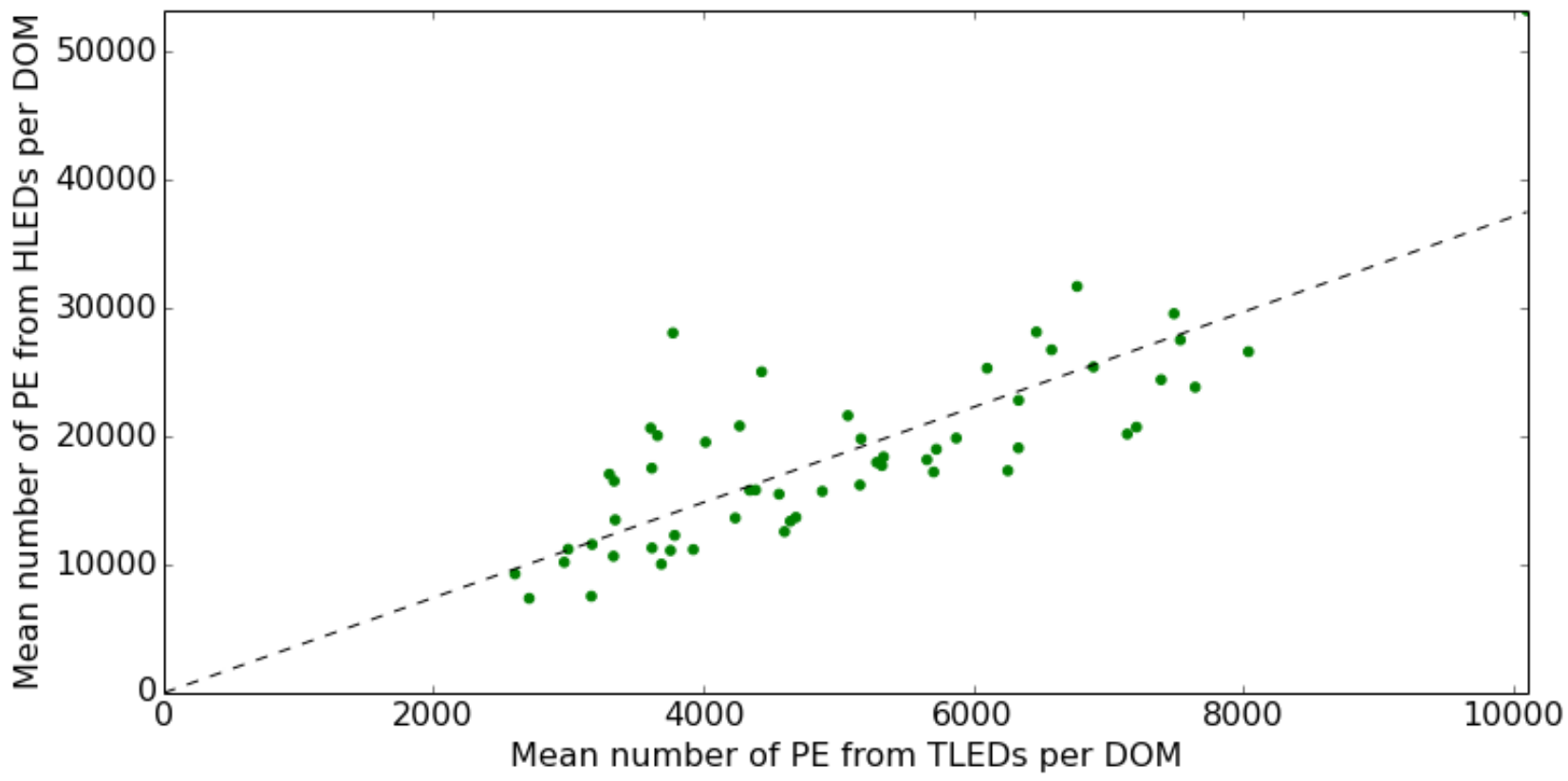

Figure 9.18: Correlation between the mean number of PE registered on the horizontal and tilted LEDs per DOM. The strong correlation indicates that the horizontal LEDs are in fact not dominated by internal light contamination (see below), but by the properties of the surrounding ice.

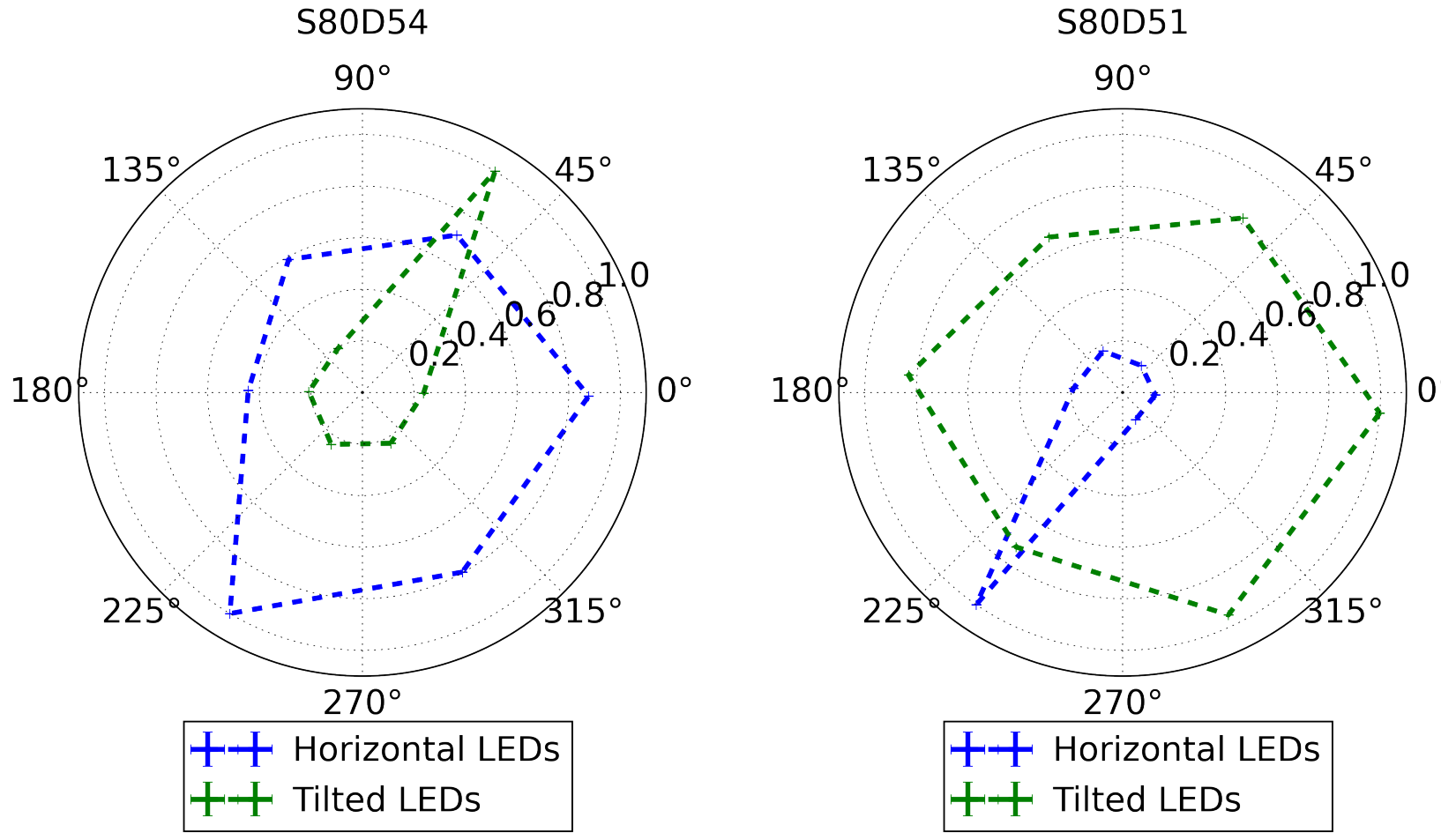

Figure 9.20: Two of the three DOMs showing significant teardrop patterns. Note that the pattern is not consistent between horizontal and tilted LEDs.

These outliers are probably the result of local anomalies such as defects in the PMT aluminium coating[341] or an LED directly illuminating an impurity deposition on the upper glass hemisphere, as observed in Figure 9.19.

[341] *see section 9.3.6.2*

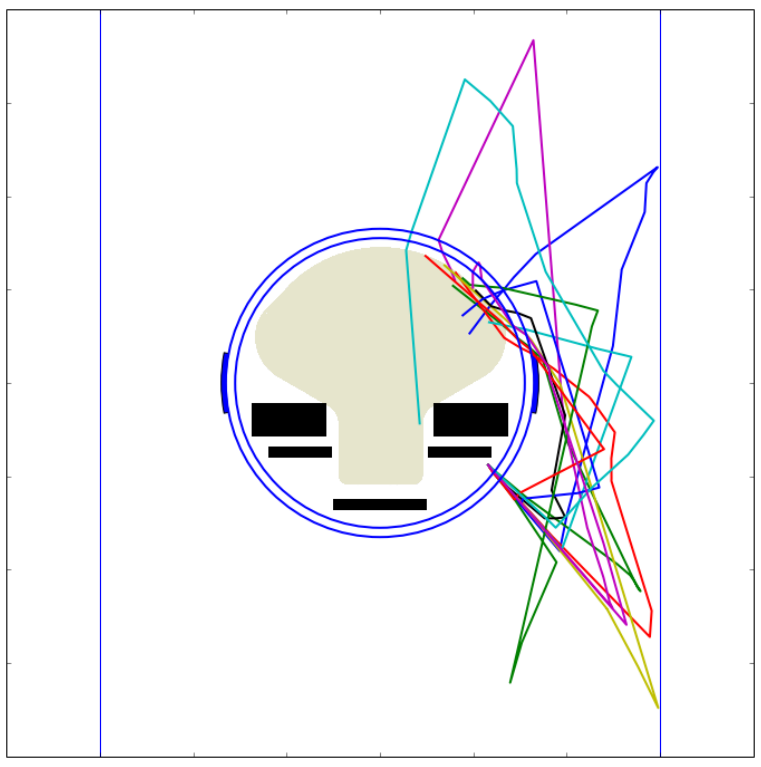

Figure 9.21: Geant simulation of photons being scattered to the PMT by a hole ice column with 100 cm effective scattering length. (Photons not reaching the PMT are not shown.)

A strong inhomogeneity of the hole ice, such as small column that encompasses some but not all of the flashers, is generally not supported by the current data.

## 9.3.5 *Simulation*

Though not originally intended as a quantitative study, the average absolute photon counts obtained per DOM can be interpreted in terms of the scattering properties of a homogeneous hole ice column when compared to simulation.

A detailed Geant4 model of a DOM based on DOMINANT[342] has been placed in the center of a 60 cm diameter hole ice col-

[342] *Hoshina et al., "DOMINANT : Dom Optical-photon to Material INteraction ANd Tracking simulator"*

umn surrounded by a 600 m bulk ice volume. For both media the Henyey-Greenstein scattering function[343], which is a good approximation of Mie scattering and which is also used in the analysis of the bulk ice, is assumed with an average deflection angle $\langle\cos(\theta)\rangle = 0.9$ [344]. The number of PEs detected by the DOM ($PE_{sim}$) is given by the amount of photons simulated to arrive at the photocathode ($N_{hits}$), the quantum efficiency ($QE$), mean charge per detected photon in PE ($\langle SPE\rangle$)[345] and the collection efficiency ($CE$):

[343] Henyey and Greenstein, "Diffuse radiation in the Galaxy"

[344] This was chosen in order to achieve consistence with the bulk ice parametrization. Assuming that the bubble column actually consists of spherical air bubbles, g should be rather ~0.7. Results can still be compared in terms of effective scattering lengths.

[345] according to the TA004 SPE template

$$PE_{sim} = N_{hits} \cdot QE \cdot CE \cdot \langle SPE\rangle = N_{hits} \cdot 0.25 \cdot 0.6 \cdot 0.85\,[\mathrm{PE}] \quad (9.5)$$

The simulation has been performed for two extreme values of the geometric scattering length of the bulk ice of 20 m and 2 m and a wide range of effective scattering coefficients of the hole ice column as shown in Figure 9.26. The results stated below are for a bulk ice of 20 m scattering length. This applies to most of the selected DOMs.

### 9.3.6 *Systematics*

The primary systematic uncertainty for this simulation is a 20% uncertainty on the absolute photon output of the flasher LEDs, both due to uncertainties in the flasher parametrization used[346], and LED-to-LED fluctuations[347].

[346] see section 4.5.1

[347] see Figure 5.14

Extrapolation of the collection efficiency to low voltages is discussed in the following.

The strong discrepancy in back-scattered intensities between the horizontal and tilted LEDs was originally not understood and from Geant simulations as exemplified in Figure 9.22 a suspicion arose that the aluminum coating used as grounding on the back of the PMT bulbs might transmit by as much as 4%. This internal light contamination be sufficient to explain the entire intensity seen in the horizontal LEDs.[348] To exclude this scenario the transmittance of the aluminium coating was tested and is presented in section 9.3.6.2.

[348] The tilted LEDs are not prone to this problem as they are on the other side of the flasher board which shadows them from the mainboard.

Other systematics, including absorption in the hole ice, shadowing by the DOM cable as well as positional uncertainties of the LEDs and the quantum efficiency of individual PMTs, were not considered.

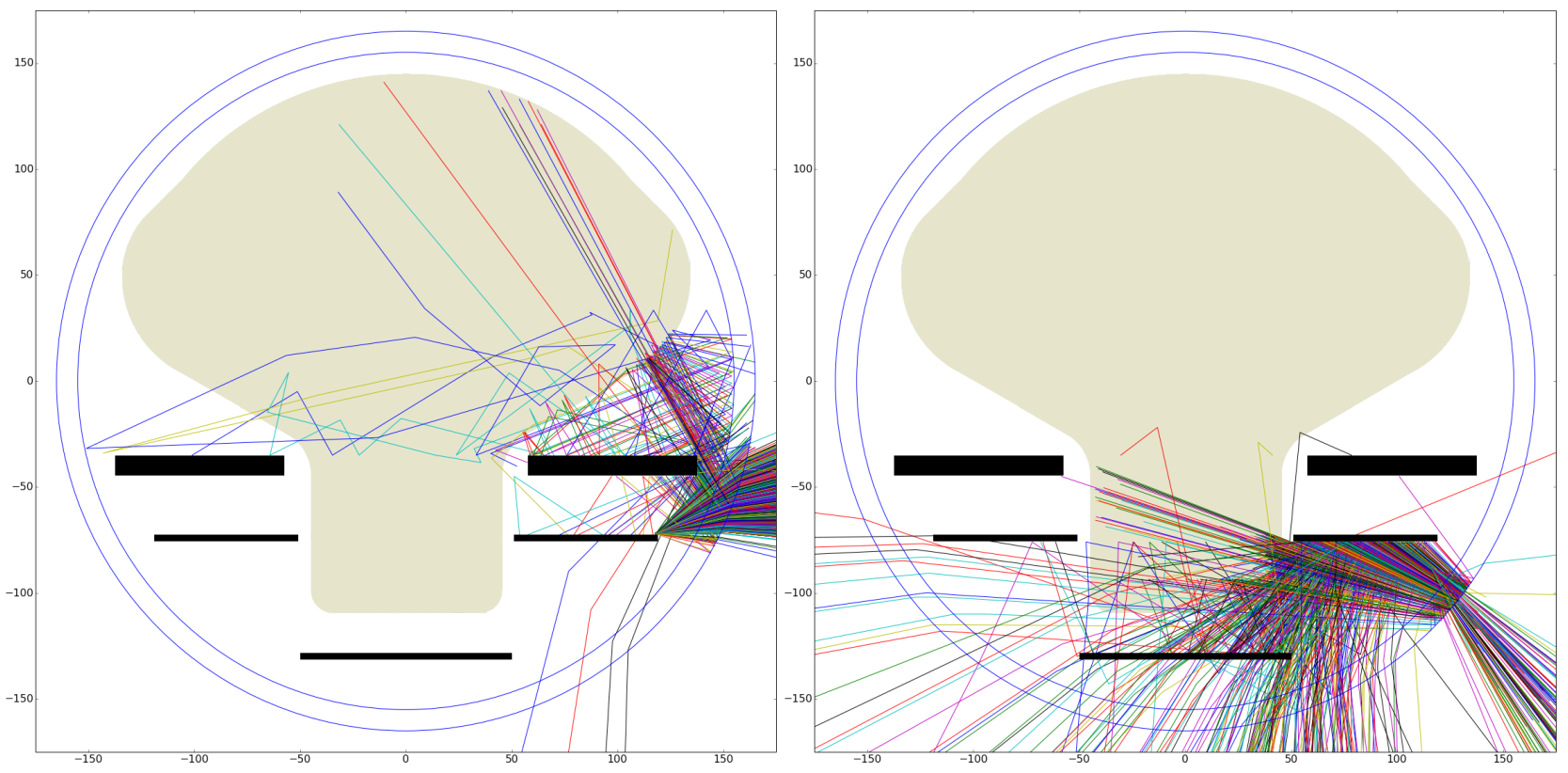

Figure 9.22: Simulation of internal light contamination. Left: About 4% of the photons emitted by the horizontal LEDs are reflected of the air/glass interface and can reach the aluminium coated back of the PMT. Right: The flasher board shadows reflected photons in the case of the tilted LEDs. Only the neck of the PMT can be reached. This should not result in significant light contamination.

### 9.3.6.1 *Collection efficiency and gain*

While the collection efficiency (CE) is near unity at the nominal gain of $10^7$ [349], it degrades at smaller voltages where the photoelectrons are on less well defined trajectories to the first dynode.

349 *Hamamatsu Photonics,* Large Photocathode Area Photomultiplier Tubes

The best available measurement of the collection efficiency was performed on 62 DOMs during the final acceptance testing. The result is shown in Figure 9.23. A linear extrapolation of the resulting curve indicates a collection efficiency of ∼70% at 750 V.

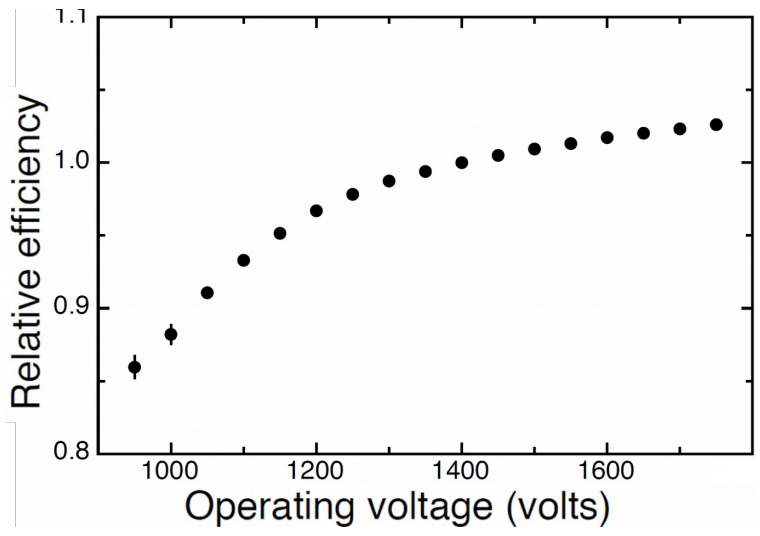

Figure 9.23: FAT measurement of the collection efficiency as a function of high voltage (Normalized to 1300 V.)
[IceCube internal, Chris Wendt]

As this measurement does not actually extend to 750 V, it was cross-checked with a DOM in the Madison lab[350].

350 *Tosi et al., "Calibrating photon detection efficiency in IceCube"*

Illuminating the DOM with a constant intensity, the high voltage was reduced and the resulting average charge measured. After correcting for the change in gain, the remaining reduction in charge is attributed to the reduced collection efficiency.

This calculation for this individual DOM yields a 60% CE at 750 V. Although the difference to the FAT value has been traced to a potentially wrong gain description at low voltages, where the SPE discriminator threshold cuts into the Gaussian peak of the SPE charge distribution, both values are deemed reasonable.

As a result the uncertainty on the collection efficiency introduces a ∼10% uncertainty on the total collected charge. This is still subdominant to the ∼20% uncertainty on the LED intensity.

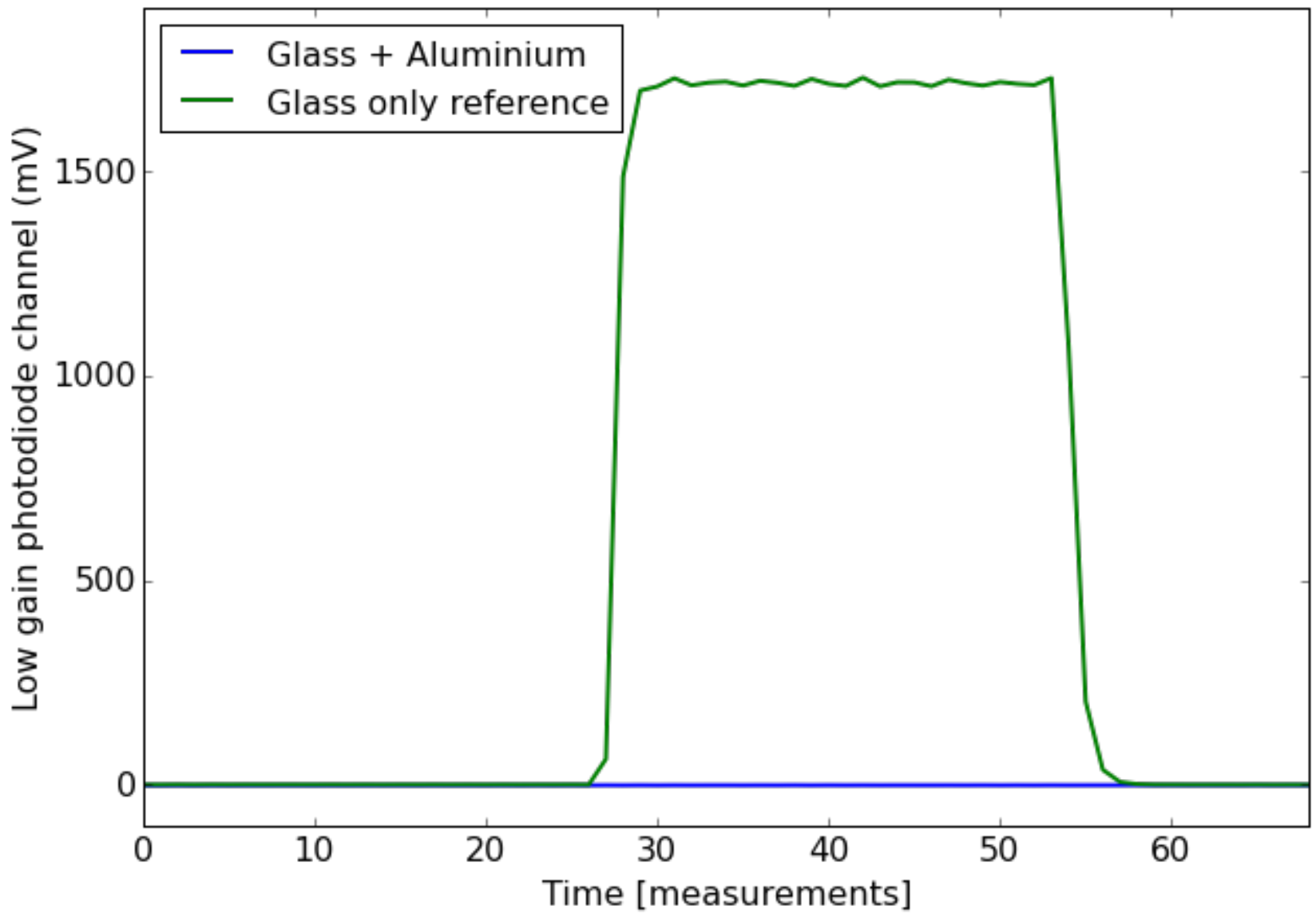

Figure 9.25: On-Off-On cycles of transmittance measurements through the PMT glass and the PMT glass as well as the aluminum coating. No light from the opposing LED is seen through the coating.

#### 9.3.6.2 *PMT grounding transparency*

To investigate the possibility of light from the horizontal LEDs entering the PMT directly through the metal grounding, a small test was setup.

Hamamatsu supplied an empty PMT glass bulb with only the aluminium grounding already applied. Inspired by magnetic aquarium glass cleaners, an LED driver and a photodiode system as described in[351] were fitted to a pair of toroidal neodymium magnet and felt gliders. This pairing, as seen in Figure 9.24, can be attached to the glass bulb, such that the LED can be moved on the outside with the photodiode following on the inside while staying perfectly aligned.

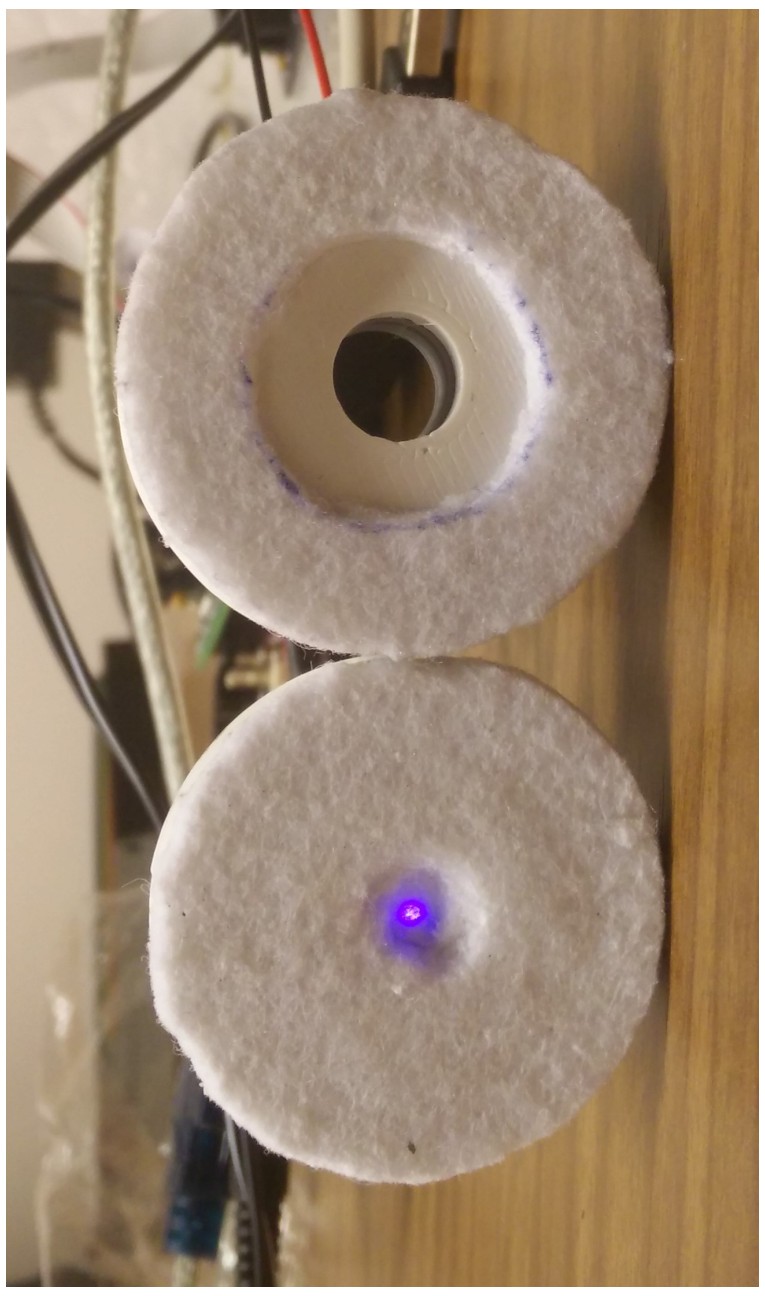

Figure 9.24: Photodiode (top) and LED (bottom) in magnetic housings with felt gliders. This pairing can be slid along the PMT glass like a magnetic aquarium cleaner.

The aluminum transparency has been measured by comparing the photocurrent at several locations with and without coating. An example measurement cycle can be seen in Figure 9.25.

For locations with coating no photocurrent above the residual noise level could be detected. This results in an upper limit on the transmittance of 1‰and makes it practically negligible for DARD.

It should be noted that Hamamatsu does not individually control the aluminium coating, except for a visual inspection[352]. Therefore regions of increased transmission still remain plausible, although none could be found on the supplied bulb.

351 *Tosi et al., "Calibrating photon detection efficiency in IceCube"*

352 *Dille, "private communcitation (Hamamatsu)"*

### 9.3.7 *Homogeneous Hole Ice Interpretation*

#### 9.3.7.1 *Scattering length*

In order to quantify the average hole ice properties, the number of PEs observed from the different tilted LEDs have been averaged for each DOM on String 80. The distribution of measured PEs compared to the simulated expectation for different scattering coefficients is shown in Figure 9.26.

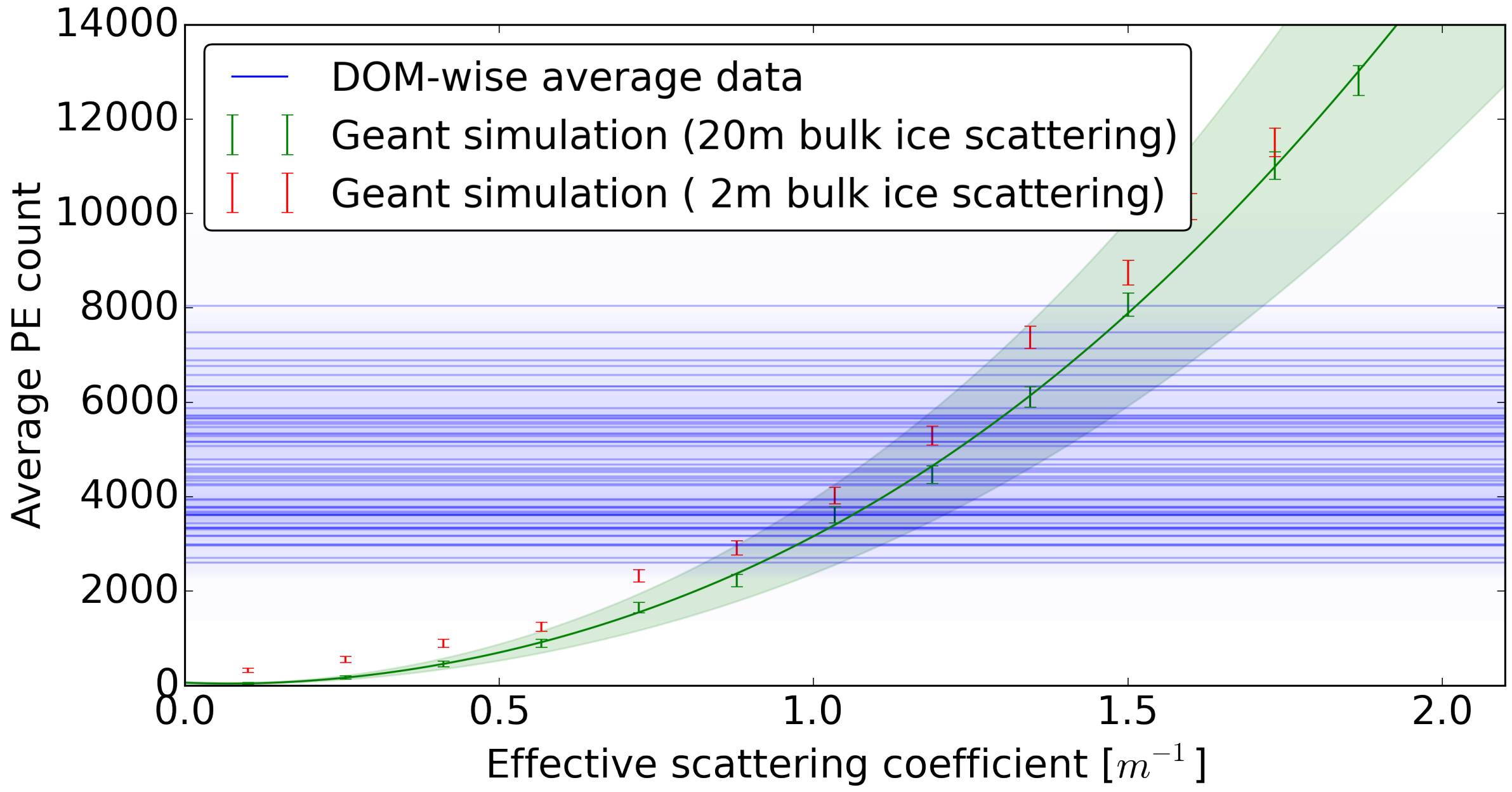

Figure 9.26: Comparison of the experimental (blue lines) and simulated (green function) light intensities arriving at the PMT of the emitting DOM. From the intersections the most fitting scattering length of the local ice can be extracted. The simulation assumes the hole ice to occupy the entire drill hole.

The scattering coefficient at which the simulated number of PEs crosses the measured PE count describes the mean properties of the hole ice around this specific DOM. The 20% systematic uncertainty of the simulation is denoted by the green shaded area.

The average effective scattering length for all DOMs is $84 \pm 15$ cm, while all DOMs are covered by effective scattering lengths ranging between ∼60 cm and ∼125 cm. This result indicates a slightly stronger scattering than found by the original AMANDA analysis[353] (100 cm to 350 cm effective).

353 *see section 9.2.1*

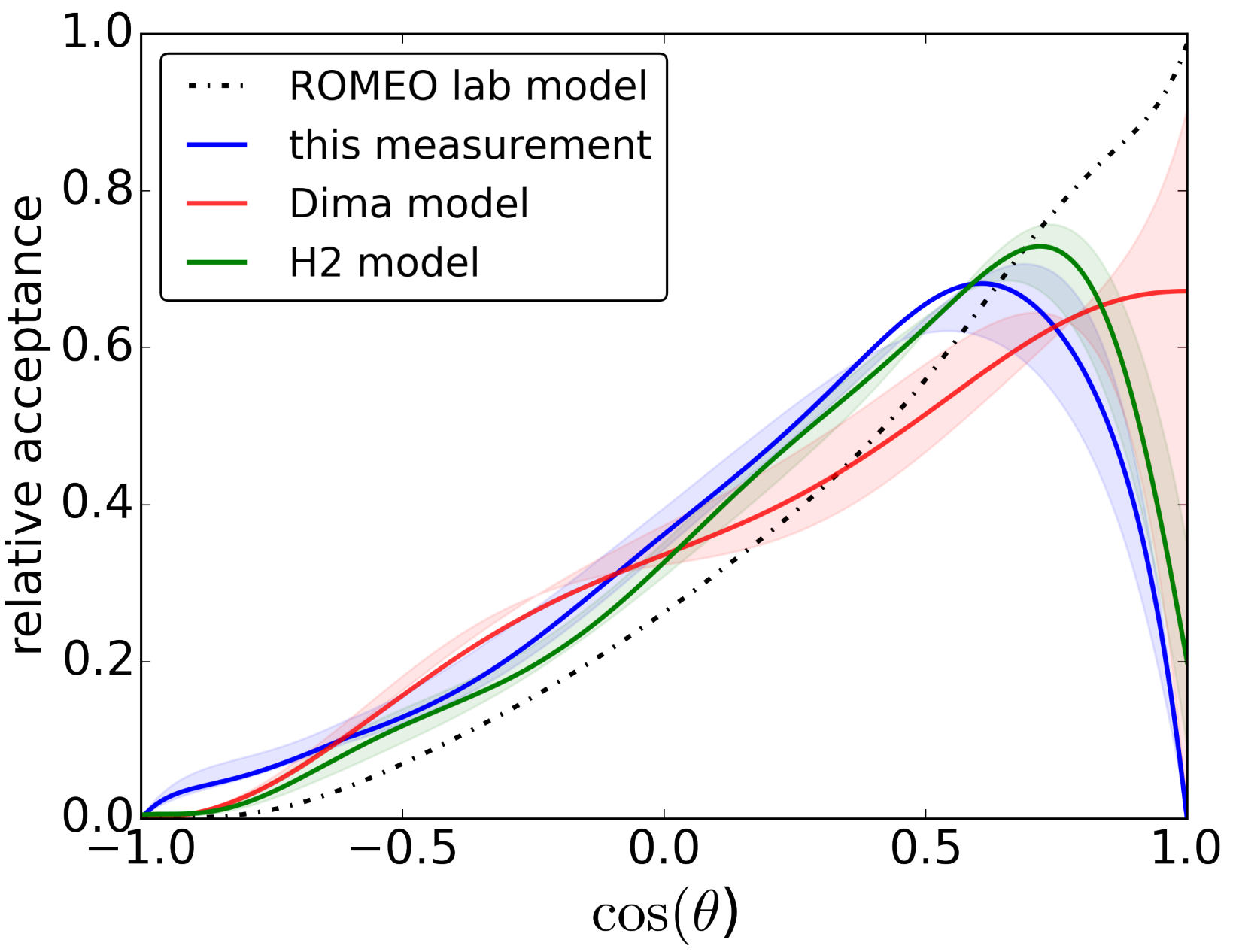

Figure 9.27: Comparison of angular acceptance models derived from laboratory measurements without scattering and modified for the hole ice properties. The new DARD curve as well as the default IceCube model H2 assume the bubble column to occupy the entire drill hole.

#### 9.3.7.2 *Angular acceptance curve*

One can employ the same Geant4 simulation described in section 9.3.5 to calculate an effective modification to the angular acceptance due to the hole ice. This is done by simulating uniform beam some distance away from the DOM.

As the beam is rotated around the DOM, the relative photon counts between simulations with no hole ice and with a given hole ice column yield a relative correction factor for each simulated angle. Applying this correction factor to the angular acceptance measured in the lab yields the average DOM angular acceptance in the ice, assuming a homogeneous, 60 cm diameter hole ice column with the above deduced range of scattering lengths.

The simulation has been performed for a variety of sizes and distances of the uniform beam. The resulting correction curve was found to be stable for beams larger than 1 m in radius placed more than ∼2.5 m away from the DOM.

Figure 9.27 depicts the obtained angular acceptance for the best fit effective scattering length, compared to the laboratory measurement and currently used default IceCube hole ice models[354]. The default IceCube hole ice model and the acceptance obtained from the presented method agree well, especially when considering the very different experimental approaches and simulation assumptions.

[354] *Aartsen et al., "Measurement of South Pole ice transparency with the IceCube LED calibration system"*

### 9.3.8 *Angular acceptance models for small hole ice columns*

As the obtained angular acceptance curve seemed "suspiciously close" to the default model, the dependence of the angular acceptance curve on the hole ice properties was further investigated.

For that purpose, the above described procedure to generate angular acceptance curves was repeated for a variety of scattering length and bubble column sizes. Each time assuming that the bubble column is centered on the DOM.

As evident from Figure 9.28, for a bubble column filling the entire diameter of the drill hole, the scattering length only has a small effect on the angular acceptance function by shifting the position of the maximum. The size, on the other hand, dramatically changes the forward acceptance, which is generally associated to be the primary effect of the hole ice. For hole ice radii larger than a DOM the forward acceptance depletes fully. Smaller bubble columns result in shading proportional to the covered photocathode area.

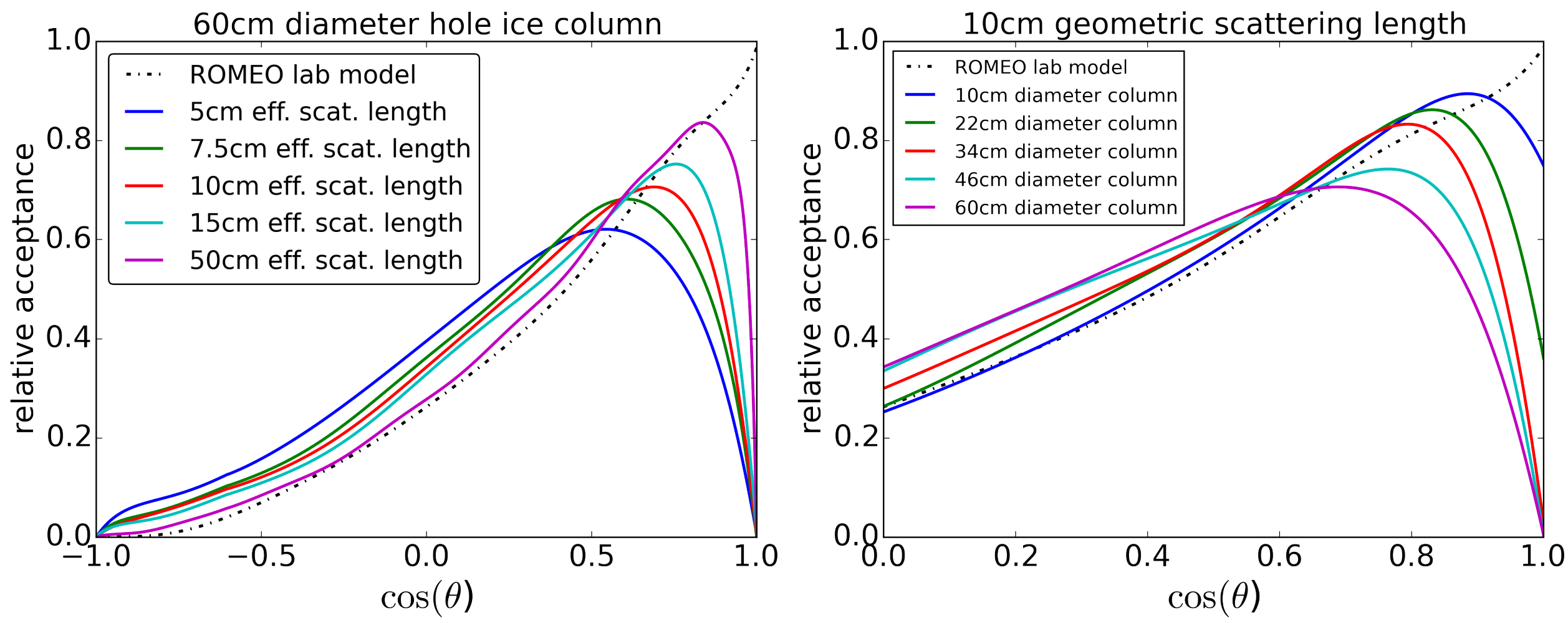

Figure 9.28: Dependence of the angular acceptance curve on the size and scattering length of the bubble column. The size is the most relevant quantity as the forward acceptance simply scales with the shadowed of photocathode area.

The AMANDA H2 model and its variants derive the hole ice scattering strength and angular acceptance curve assuming the entire drill hole to be filled with a homogeneous scattering medium. Given the above explored dependencies of the bubble column diameter and scattering length on the angular acceptance curve, it follows that these models are not driven by their underlying measurement but by the assumed size and that these models thus do not cover the true uncertainty.

## 9.4 SPICEHD

DARD was intended as a straight-forward tool to measure the geometry of the bubble column on a DOM-to-DOM basis. With DARD not being able to convincingly pinpoint the bubble column, a new attempt in describing a detector average bubble column based on inter-string flasher data was tried.

It is based on a novel parametrization of the DOM's geometric acceptance, combined with discrete photon propagation through the ice of the bubble column.

### 9.4.1 *Direct detection as an alternative DOM modeling approach*

Historically, when photon propagation was not executed as part of the physics simulation and only previously parametrized acceptance tables were used, angular acceptance curves were the only possible means to include the OM's geometric acceptance.

While the angular acceptance function is an appropriate parametrization in the far-field, where DOMs can be considered point-like, it exhibits problems for light sources or light diffusion in the vicinity of the DOM. In the far-field the incident angle of a photon with respect to the PMT axis[355] can be thought of as a representative, bulk ice diffused, incident angle sampled from all photons belonging to a plane wave of light much larger than the DOM.

[355] *as evaluated in the angular acceptance function*

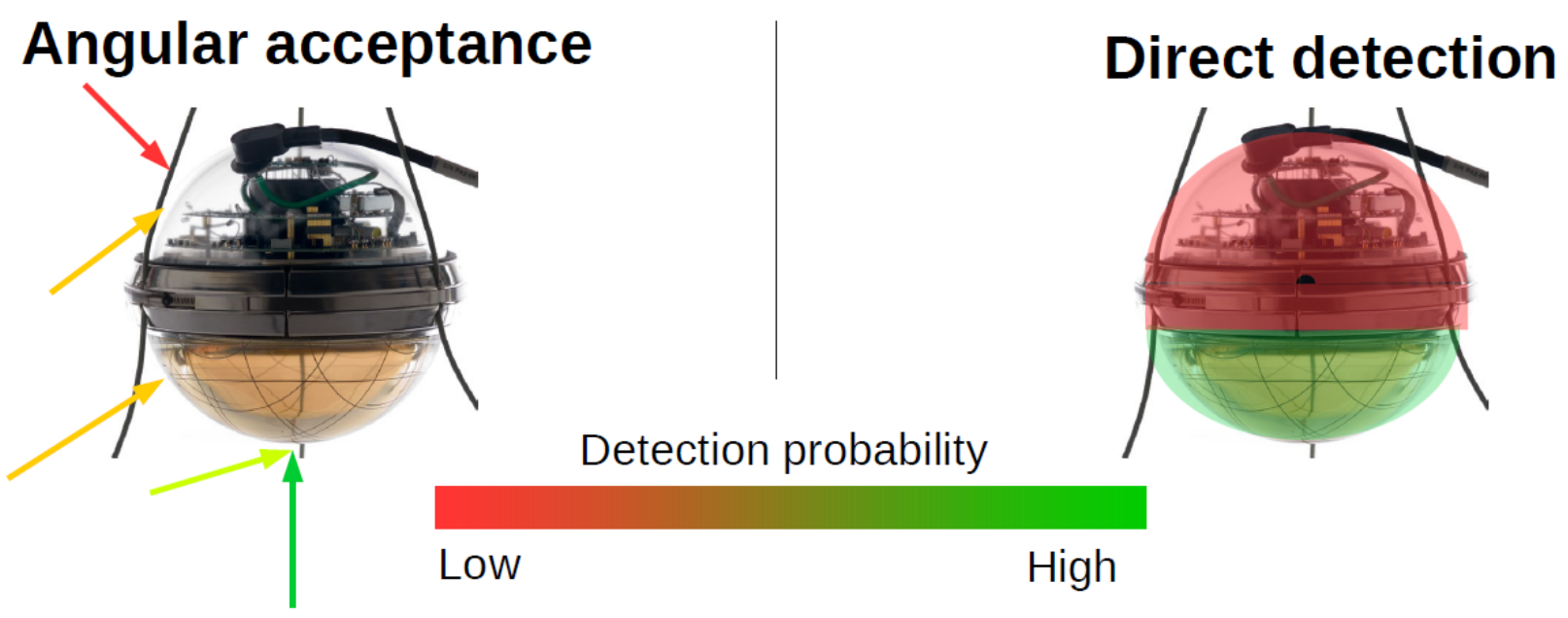

Figure 9.29: Conceptional difference between simulations using the angular acceptance function or direct detection. The angular acceptance only considers the impact angle and neglects the impact position.

Yet, this concept breaks down as photons originate in the vicinity of a DOM or are locally strongly scattered. Here individual photons can no longer be though to originate from a common plane wave. As can be seen in the sketch in Figure 9.29, this can lead to situations where photons reach the DOM's upper hemisphere at an angle considered sensitive in the angular acceptance model although no PMT is present. Alternatively, photons reaching the PMT may be discarded because of a shallow incidence.

For a number of DeepCore neutrino oscillation analyses it has been shown that, even when using angular acceptance models, non-unity oversizing factors introduce a significant data-MC mismatch. In these cases it is self-evident that the angular acceptance functions can neither be applicable nor correctly represent hole ice effects.

Given that most low-energy physics simulations are nowadays based on photon propagation with DOMs having a true physical extend and, as we have previously shown in section 9.2.4.5, that the DOM's intrinsic angular acceptance is well described by its geometry alone, the angular acceptance is no longer required to describe the DOM's acceptance. Instead, one can employ a method coined "direct-detection" where photons in the simulation are accepted or rejected based on their impact point on the glass sphere.

To preserve the same total detection efficiency in simulation and since the exact extent of the photocathode is not precisely known, a height cut[356] is placed such that the total area fraction of the spherical cap representing the PMT is identical to the integral below the angular acceptance curves.

[356] *see Figure 9.29*

In this concept the bubble column is not a part of the DOM parametrization but is instead introduced surrounding the DOM with an assumed hole ice column of varied optical properties and by continuing the photon propagation through this additional medium.

Assuming the bubble column to have constant properties over the entire detector, this new hole ice model has four free, global parameters: the scattering length, the scattering function as parametrized by $g$[357], the absorption length and the diameter of the bubble column cylinder. In addition, the relative position of every DOM is a free geometrical parameter. It is expressed by a radial offset and an azimuth angle, yielding another $2 \cdot 5160 = 10320$ free parameters.

[357] *see section 214*

The presented concept is only applicable with unity oversizing simulation. This requires $\sim 7$ GPU hours of photon propagation per emitter and considered realization.

### 9.4.2 *Fitting procedure*

Given the large number of free parameters and long simulation time for each LLH evaluation, the fit has to be broken down into a sequence of manageable, low dimensional steps, each exploring one particular set of parameters at a time.

Strictly speaking, this is only applicable when the parameters are uncorrelated. As this is not the case, step-wise fitting is not guaranteed to yield the true set of optimum and unbiased parameters. This can and is in the ice-fitting often mitigated through iterative fitting[358] and is, to the extend possible, also attempted in this case.

[358] *Aartsen et al., "Measurement of South Pole ice transparency with the IceCube LED calibration system"*

The purpose of the first step is to identify the most relevant optical parameters, to be studied in more detail later. It assumes all optical properties to be orthogonal and minimizes them independently using one dimensional likelihood scans.

With a rough idea of the optical parameters, the next step attempts to constrain the relative position of the bubble column with respect to each DOM.

Finally, the approximated geometry is applied to perform a detailed two-dimensional likelihood scan of the two most relevant optical properties, which are the bubble column size and effective scattering strength.

#### 9.4.2.1 *Frist step: Pre-evaluation of optical parameters*

With the bubble column assumed to be centered on all DOMs, a rough idea of the optical parameters is obtained through one dimensional fits.

As the bubble column is assumed to be constant over the entire detector and to reduce computation times, not all DOMs of the all-purpose flasher dataset are included as emitters. Instead, 7 strings (19, 24, 57, 62, 63, 80 & 81) are selected. This results in a good overall coverage of the array, being somewhat centered on DeepCore.

As a first guess, mostly based on Sweden Camera information, the diameter is assumed to be 0.5 DOMradii or 16 cm in diameter, with an effective scattering length of 5 cm, a $g$-parameter of 0.9 and an absorption length of 100 m.

The likelihood scans for the absorption length and the scattering parameters show no significant trends. The presented method is not sensitive to them. The absorption length is in the following always assumed to be 100 m. This is consistent with the assumption that the bubble column indeed contains mostly air bubbles and no other impurities.

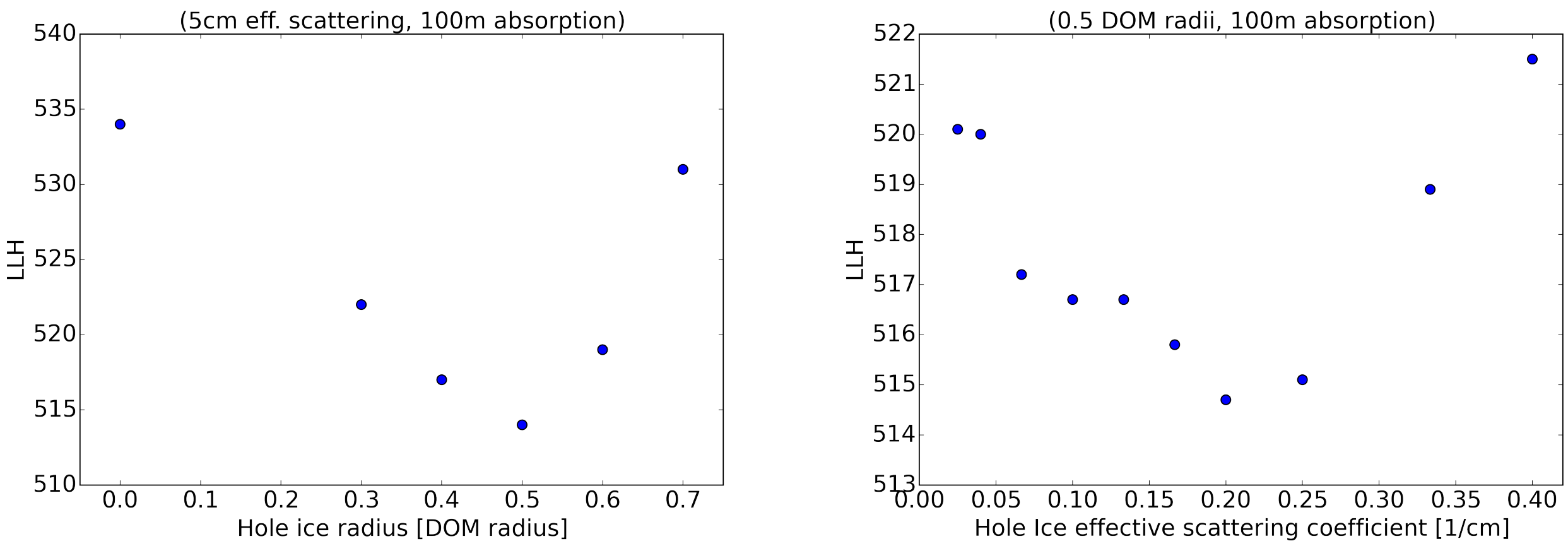

Figure 9.30: Spice HD pre-fits. Left: Fit of the bubble column diameter, assuming a 5 cm effective scattering length. Right: Fit of the effective scattering length assuming a 0.5 DOM diameter bubble column.

Likelihood scans for the effective scattering coefficient and the size are shown in Figure 9.30. Both parameters can be fitted successfully with minima close to the original guesses. The likelihood spaces are obviously more complicated than a simple paraboloid. The reason will become apparent when performing the 2-D fit in step three.

#### 9.4.2.2 *Second step: The position fitting*

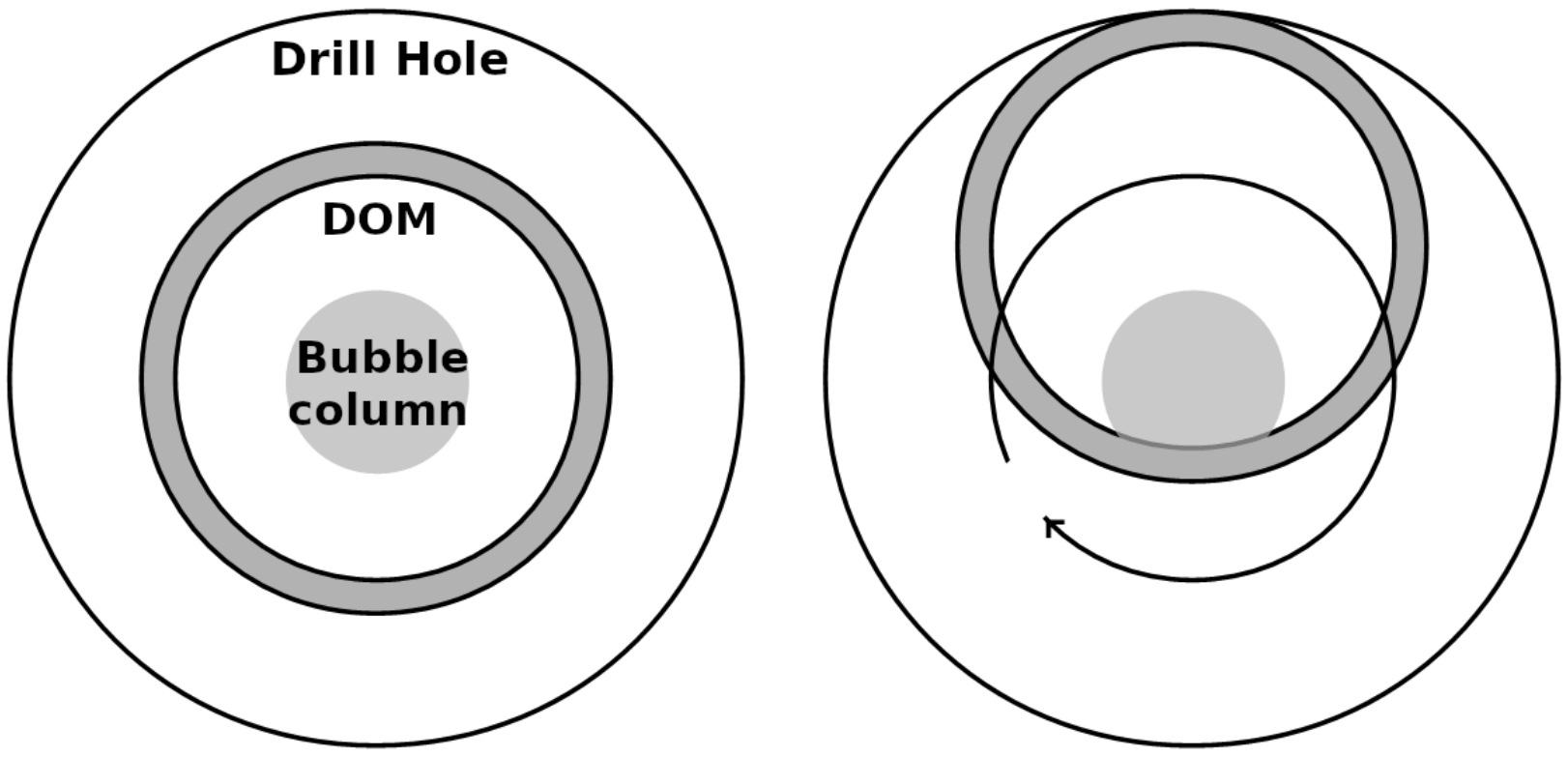

Figure 9.31: Relative orientations considered in the geometry fit. Due to the refreezing mechanism the bubble column is always assumed to be centered in the hole. Left: The DOM is center in the hole as well. Right: The DOM rests against the drill wall. The azimuth is rotated in discrete steps of 60°.

The effect of the bubble column depends strongly on the relative orientation to each DOM, as it effectively shadows light from different directions. Given a rough idea of the optical properties, the next step is to constrain these geometries.

Assuming that the DOMs move in the hole with the bubble column always being at the hole center, the maximum possible displacement between the two is 10 cm. From this constrain seven discrete possible orientations were picked to be tested. One has the DOM centered on the bubble column. The other six have the DOM against the wall, at a 10 cm offset from the hole center, and are spaced 60° apart in azimuth.

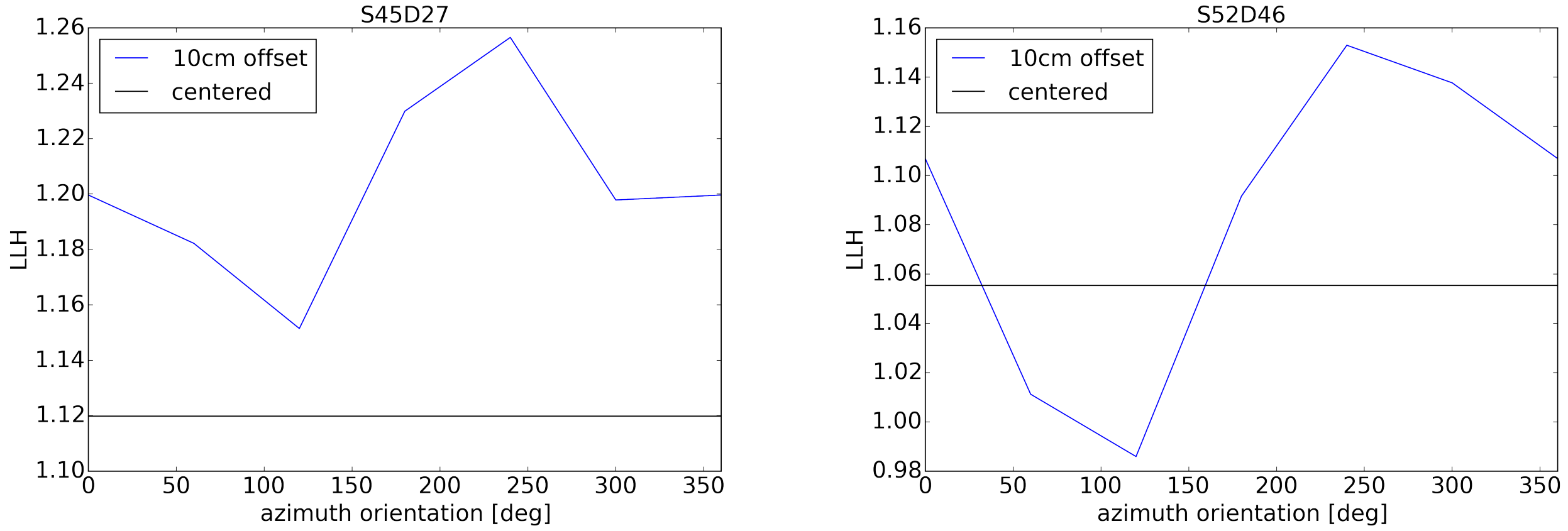

Figure 9.32: Orientation fits for two example DOMs. The likelihood for the offset orientations follows a sinusoidal profile as expected. The statistical fluctuation of individual LLH points is known to be ∼0.02. Left: The best fit orientation is the DOM being centered on the bubble column. Right: The likelihood for offset DOM orientations is seen to oscillate around the value for a centered DOM. The best fit is obtained for an offset position at 120°.

Assuming the likelihood value evaluation of a receiving DOM to be independent of the position of the flashing DOMs, when averaging over a large enough number of flashers, the best positions for all receiving DOM can be fitted from a single 85 string MC set for each of the seven assumed geometries.

In the available implementation, PPC places the hole ice cylinder in the lateral center of gravity of each string. To obtain the correct hole ice effect, the tested geometry of each set therefore contains 180° offset orientations for alternating DOMs.

The likelihood space for individual DOMs appears mostly smooth, with the expected sinusoidal behavior as a function of azimuth. Two examples are given in Figure 9.32. DOM S45 D27 is fitted to be centered on the bubble column, while DOM S52 D46 prefers an offset position at an angle of 130°.

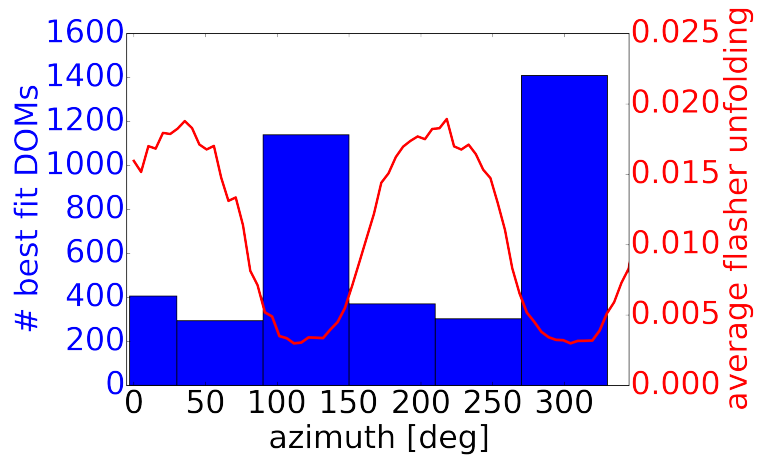

Figure 9.33: Azimuth distribution for offset DOMs as fitted with a 11% anisotropy bulk ice simulation. The distribution is expected to be flat but shows a strong bias towards the anisotropy axis, indicating an overestimated anisotropy strength.

In a MC round-trip test ∼ 1500 DOMs are recovered at the true position. Given the seven possible positions, this equates to twice the random expectation. The position fit, although yielding an average improvement, is evidently of low sensitivity and has a large potential for future improvement.

Applying the fit to data, two-thirds of all DOMs are found to be offset / resting against the wall. One expects these DOMs to be randomly distributed in azimuth. Yet, as evident from Figure 9.33, about half of these DOMs are found on an axis which correlates with the 130° anisotropy axis.

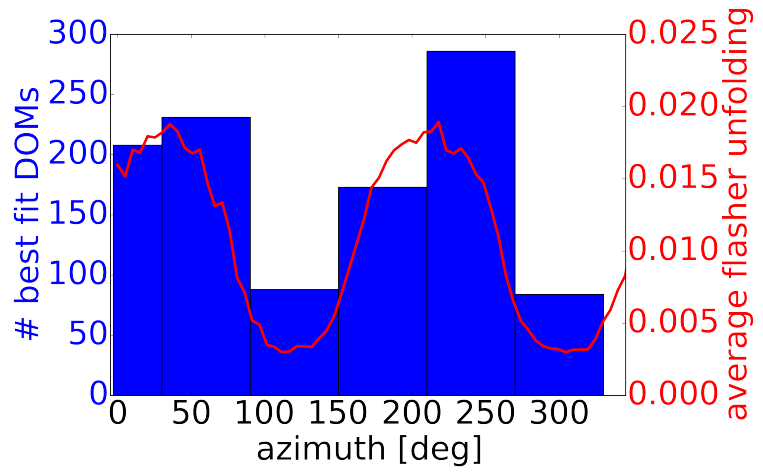

Figure 9.34: Azimuth distribution for offset DOMs as fitted with a 0% anisotropy bulk ice simulation. The inversion of the bias confirms that it is caused by the anisotropy.

While initially puzzling, this effect can be understood assuming an overestimated anisotropy strength in the MC compared to the true ice. For an isotropic light source, such as a flashing DOM at several effective scattering lengths, this results in too much light being received on the anisotropy axis in the MC. Introducing a partially shading bubble column into the photon propagation, allows the fit to move this bubble column onto the anisotropy axis so to reduce the additional light.

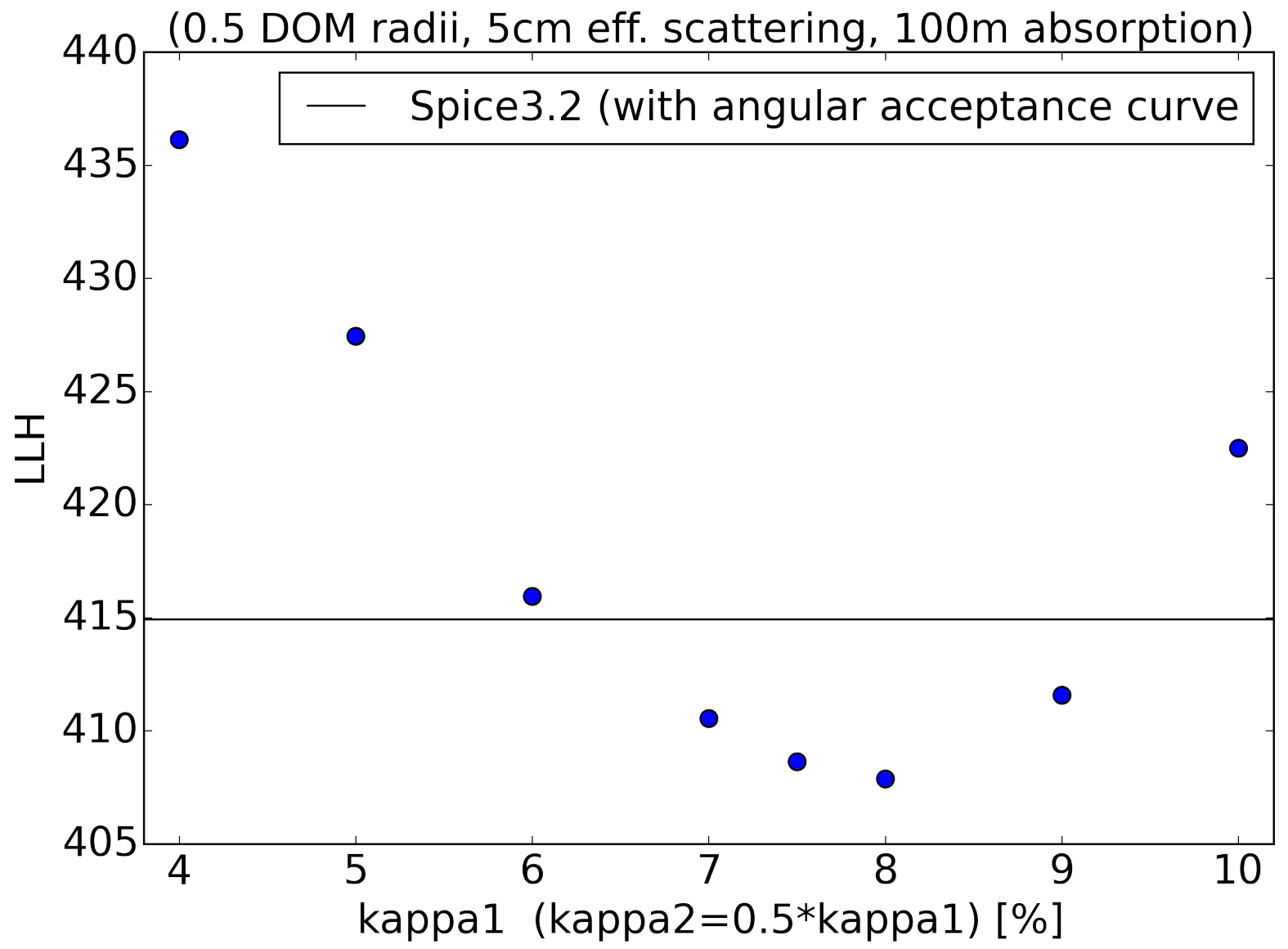

Figure 9.38: Fit to the $\kappa_1$ anisotropy strength assuming $\kappa_2 = 0.5 \cdot \kappa_1$. Without the bias introduced through flasher unfolding the best fit is at the Spice Lea value of 8%.

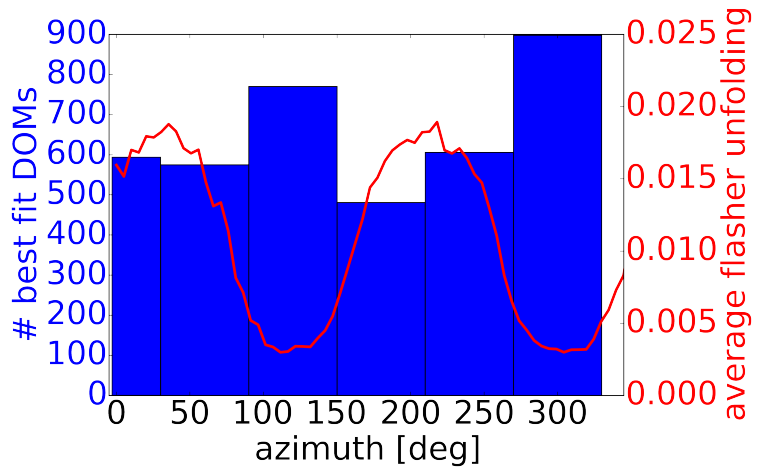

Figure 9.35: Azimuth distribution for offset DOMs as fitted with the best fit 8% anisotropy bulk ice simulation. A small bias towards the anisotropy axis remains.

This idea is tested by repeating the position fitting on a subset of emitters with the anisotropy disabled in photon propagation. As expected and apparent in Figure 9.34, the bias inverts.

The anisotropy being incorrectly fitted in Spice3.2 has been traced to a bias introduced by the flasher unfolding, which is new in Spice3[359]. While successfully tested to recover the profile and orientation of individual flashing LEDs, the unfolding was never tested for the all-purpose flasher data where all six horizontal LEDs are used simultaneously.

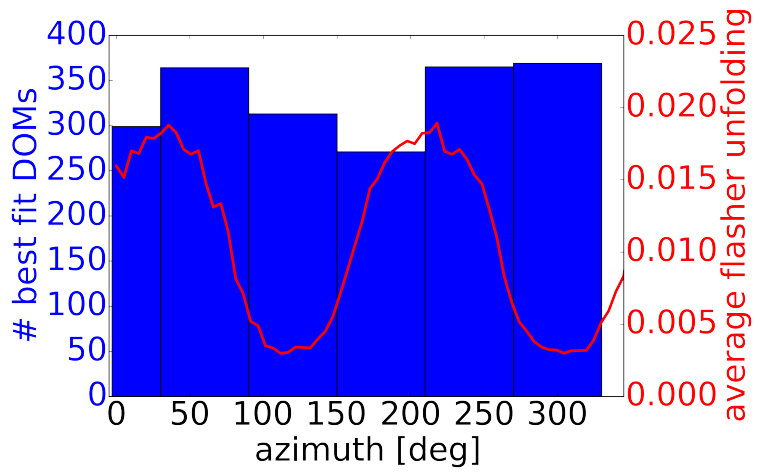

Figure 9.36: Azimuth distribution for offset DOMs above the dust layer as fitted with the best fit 8% anisotropy bulk ice simulation. While not perfectly uniform, no bias towards the anisotropy axis is seen.

The unfolded flasher profile has to be uniform over all emitting DOMs, as we know the DOMs to be randomly oriented[360]. Yet the average flasher profile unfolded during the Spice3.2 ice fit shows a strong sinusoidal modulation with a factor three amplitude and aligned with the anisotropy axis.

It should be noted that while the average flasher modulation is unexpected, the flasher unfolding is intentionally not used for the SpiceHD fit as it was anticipated to potentially absorb local ice effects.

Acknowledging that the flasher unfolding biased the fitted anisotropy strength in Spice3.2, the anisotropy strength is refitted without flasher unfolding and based on the 7 strings of data also used previously. The likelihood scan can be seen in Figure

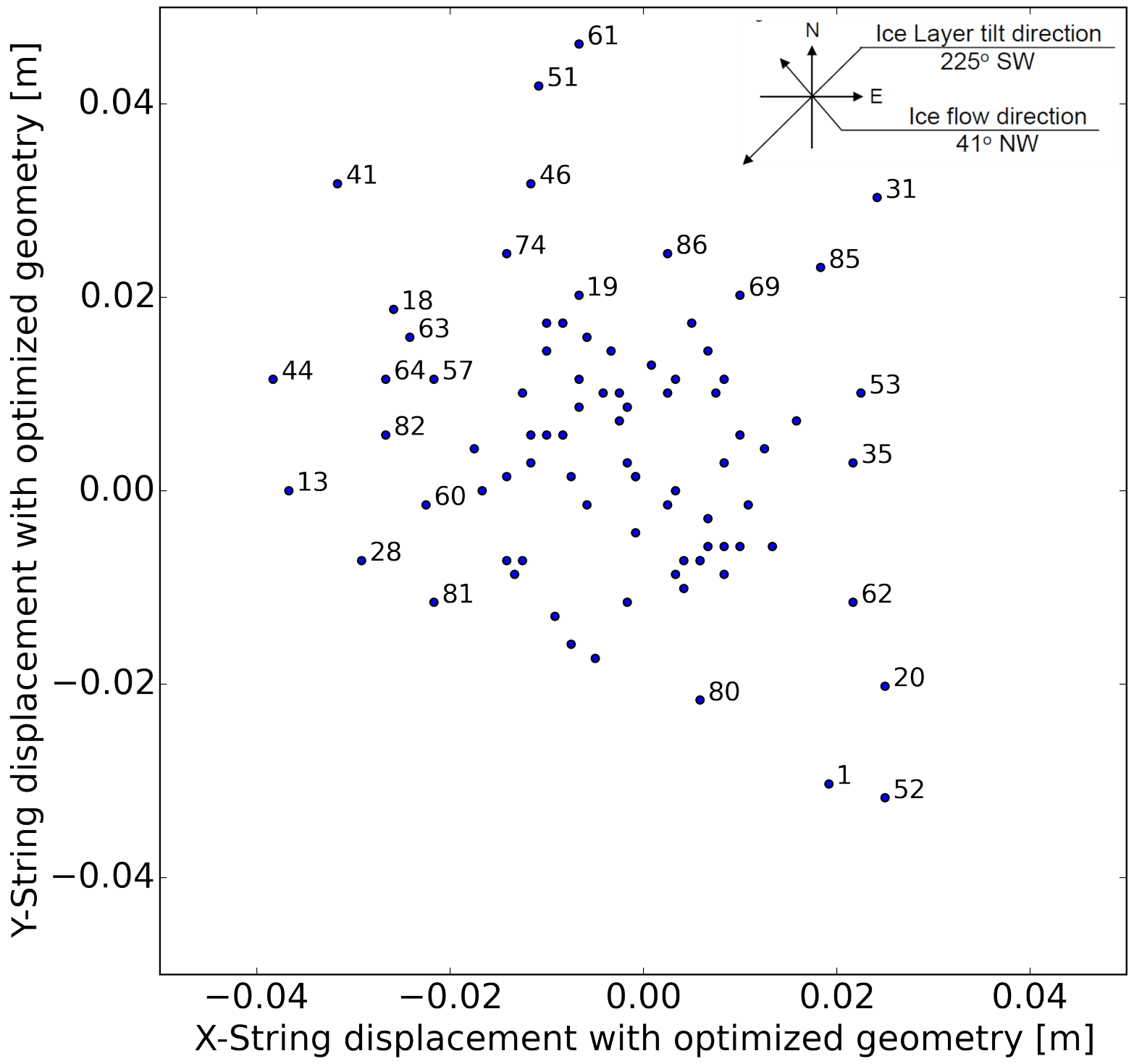

Figure 9.39: PPC positions the bubble column in the center of gravity of all DOMs along each string. The plot shows the per-string displacement between the original COG and the COG after applying the per-DOM best fit orientations. Large outliers are denoted with their string number. This displacement smears the fitted orientations in the new simulations. Notice the small bias towards the anisotropy axis.

9.38. It yields a best fit anisotropy strength of 8%, as was also originally fitted in SpiceLea.

Repeating the position fitting using an 8% anisotropy results in a nearly uniform azimuthal distribution of bubble column orientations above the dust layer (Figure 9.36).

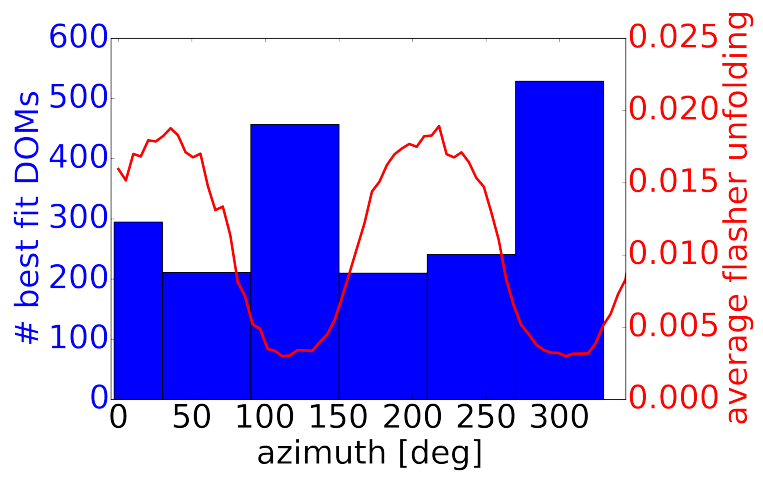

Figure 9.37: Azimuth distribution for offset DOMs below the dust layeras fitted with the best fit 8% anisotropy bulk ice simulation. The bias towards the anisotropy axis indicated that the anisotropy strength is not constant throughout the detector and significantly weaker at depth.

Below the dust layer a residual anisotropy bias remains, hinting at a strongly reduced anisotropy strength at these depths. This observation sparked an interest in the depth dependence of the anisotropy and the underlying process governing the anisotropy. These topics will be further elaborated on in chapter 10.

In absence of a better anisotropy model, the positions fitted using the 8% anisotropy are taken to be the final result of this fitting step. The fitted geometry is only retained in the later applied geometry if DOMs on a string are randomly oriented, so that the center of gravity is not moved. Given the observed anisotropy bias, this is not exactly true.

Figure 9.39 shows the displacement of the COG for all strings. Offsets are generally smaller than 2 cm, with outliers labeled in the plot. A small correlation of the displacements with the anisotropy axis is indeed observed.

359 *see section 5.2.5.1 for an explanation of the unfolding*

360 *see section 9.5.1*

### 9.4.3 *Step three: Refined 2-D optical properties*

With the optimized geometry in place, the two most relevant optical properties, the size and the effective scattering lengths are refined in a 2-D LLH scan using the seven string flasher data that was already employed in step one.

Starting from the estimated parameters, new realizations are hand-picked trying to follow the slopes of the likelihood space to a new minimum, or alternatively trying to extend the covered space to be able to fully constrain the contour. Overall LLH values for roughly 200 realizations were calculated in that fashion.

The resulting likelihood space, plotted once as the size vs. the effective scattering length, and once against the scattering coefficient is seen in Figure 9.40. From repeated re-simulation of a selected realization the statistical likelihood spread is known to be $\sim 1.3$ LLH units. This spread is used to construct a statistical confidence region.

While the overall constrain on the bubble column properties is weak, the following conclusions can be drawn:

- The fit essentially excludes a no-bubble column hypothesis. LLH points below 0.2 DOM-radii are disfavored at any scattering strengths.
- For bubble column diameters which do not extend outside the DOM's perimeter even for offset DOMs (<0.45 DOM-radia), the effective scattering length is essentially unconstrained between 3 cm and 20 cm.
- For larger bubble column sizes the fit can not constrain a small but strongly scattering bubble column versus a weakly scattering bubble column which fills the entire drill hole.

The stability of the fit has been tested against potential influences of a non-optimal geometry, by only including central or offset DOMs. In a different test only DOMs above or below the dust layer were included in the likelihood construction. The resulting contours agree in all cases.

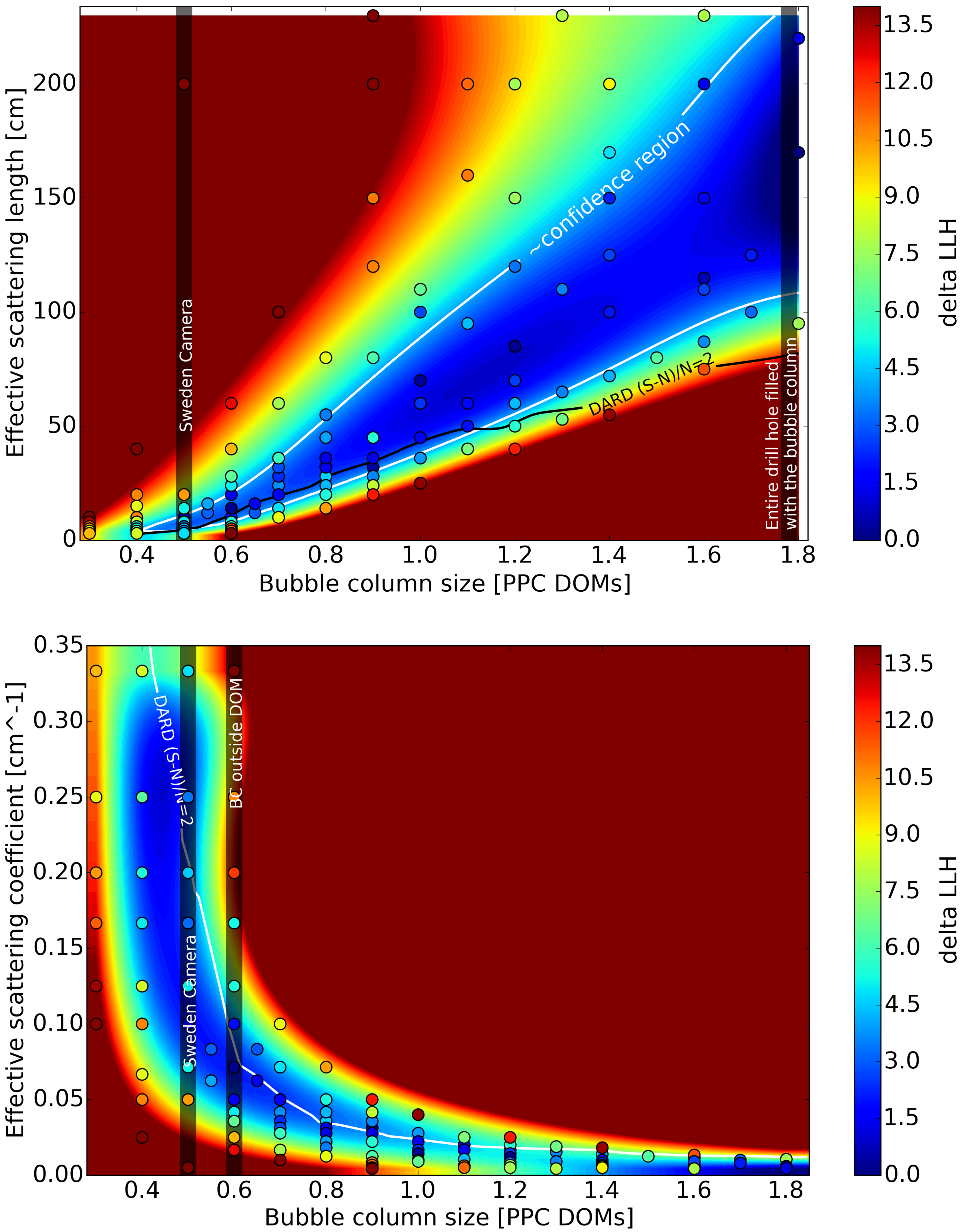

Figure 9.40: SpiceHD likelihood contours as a function of the bubble column size in units of DOM extents and the bubble column scattering strength. Top: Size versus effective scattering length. Colored circles denote the delta LLH of tested realizations compared to the best fit. The background color is a smoothing spline used to construct the statistical confidence contour. Bottom: Plotted against the scattering coefficient instead of length. The optical properties are largely unconstrained when the bubble column does not extend to the LED perimeter.

#### 9.4.3.1 *Full circle MC challenge*

The sensitivity of the measurement has been verified by trying to recover an injected MC truth. For that purpose one of the simulation sets used to span the likelihood space in the SpiceHD fit was declared the MC truth and the likelihood of other simulation sets was calculated assuming that truth to represent the data[361].

[361] *Nuisance parameters are thus not re-optimized against the MC truth.*

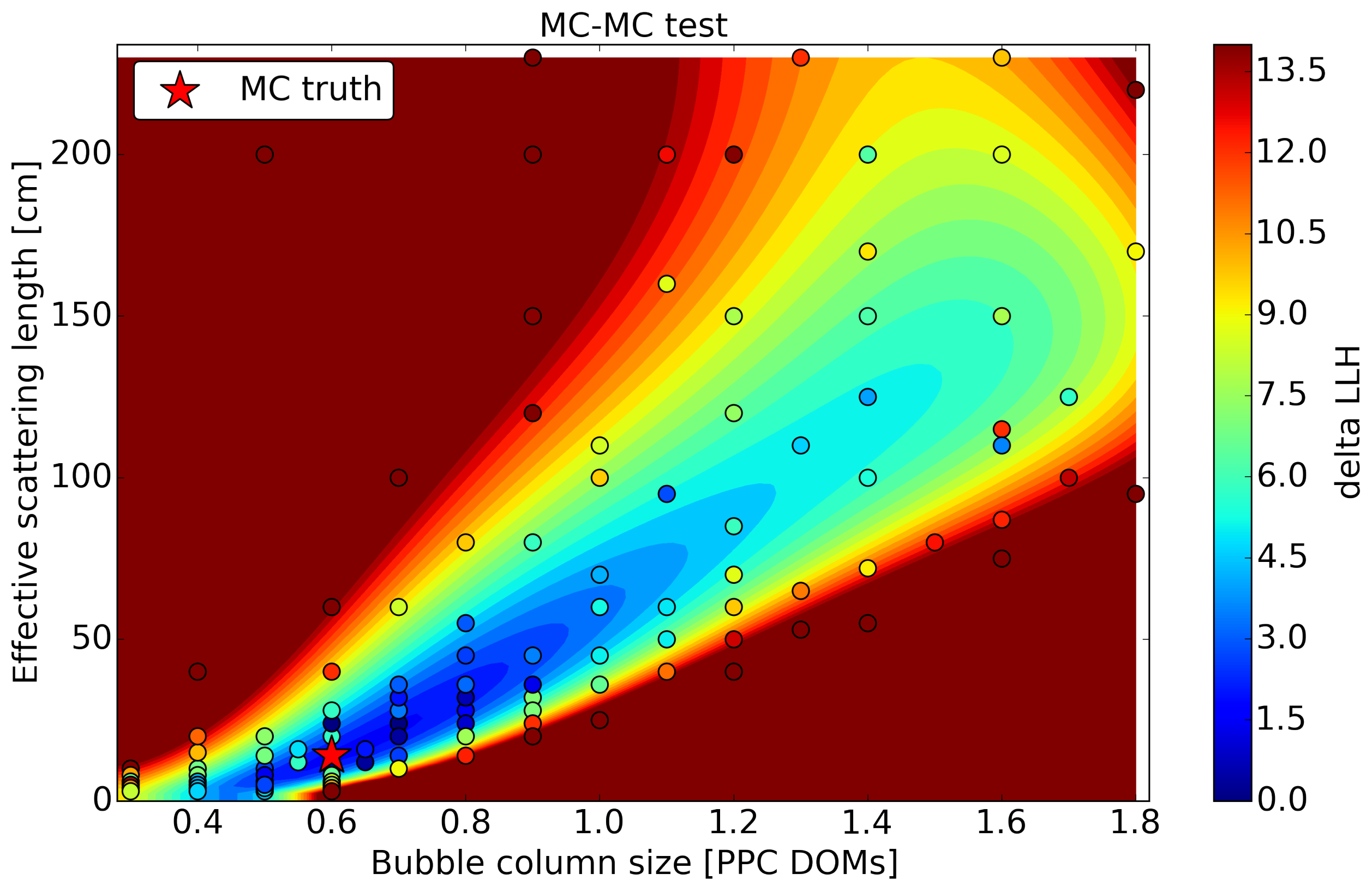

Figure 9.41: SpiceHD sensitivy test. One realization used for the fitting was picked as MC truth as denoted by the red star and the likelihood for all other realizations calculated with respect to this pseudo-data. The likelihood contour largely resembles the contour found when fitting the experimental data, indicating that the uncertainty is largely driven by the limited sensitivity of the analysis.

The resulting LLH landscape is depicted in Figure 9.41. While the truth is generally recovered, the MC test shows a similar degeneracy between a small, strongly scattering and a large, weakly scattering bubble column as seen in data.

The overall uncertainty of the MC fit is slightly reduced compared to the data fit, indicating that mis-modeling, like a depth dependent hole ice, or errors in the fitted bubble column positions introduce additional smearing.

#### 9.4.3.2 *DARD and Sweden Camera compatibility*

While the Sweden camera does not allow for a quantitative evaluation of the bubble column scattering length, it can be used to precisely infer its diameter. Figure 9.42 shows the camera view in the water filled hole, directly after deployment, and later with the bubble column fully developed.

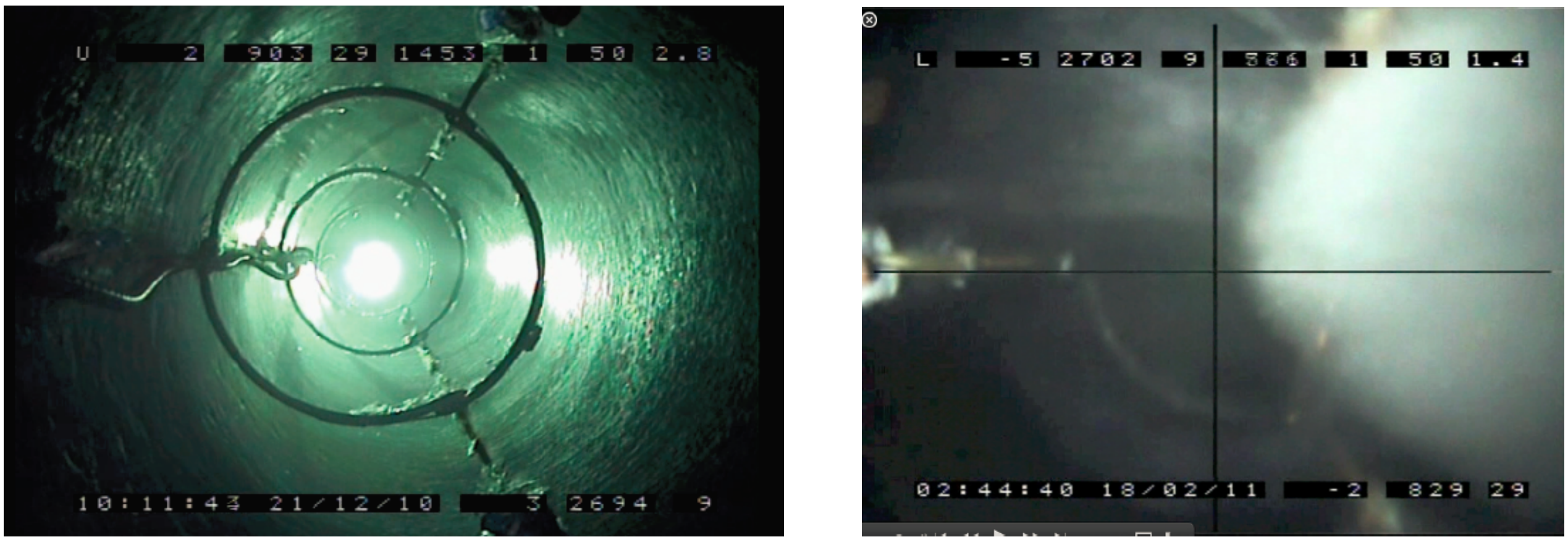

Figure 9.42: Sweden Camera images used to deduce the orientation and size of the bubble column. Left: Upper camera Camera facing downwards directly after deployment. The spacers are seen resting against the drill wall. Right: Lower camera looking straight up at the upper camera, with the hole ice fully developed and taking up half the field of view.
[Per Olof Hulth (2013). "Results from the IceCube video camera system at 2455 meters ice depth"]

In the inital image the camera can be seen resting against the wall. The spacer rings between the two cameras are made from DOM waist bands and have a known diameter of 39 cm. From this the diameter of the hole can be estimated to be roughly 55 cm, which is consistent with expectation.

With the bubble column fully developed and the camera looking straight up, the clear outer hole ice extends to the center of the image. Given that the camera rests against the wall, it is known to be 19.5 cm from the wall. Assuming the column to be cylindrical and centered in the hole it follows that its diameter is roughly 140 to 180 mm, or 0.4-0.55 DOM radii given the DOM diameter in PPC[362].

[362] *which is defined by the Benthos glass, excluding the waist band*

The original DARD result assumed the bubble column to be centered on the DOMs. Given the SpiceHD position fit, where the majority of DOMs are rather offset from the bubble column, the DARD measurement can be reinterpreted as an upper limit on the scattering length given an assumed size.

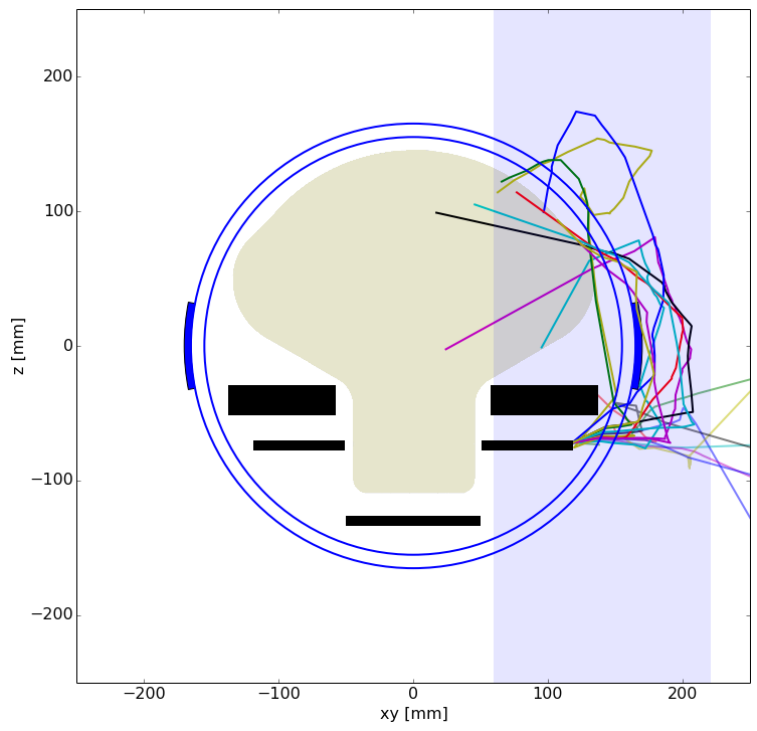

Figure 9.43: Geant4 view of ray traces for a DARD scenario with an offset bubble column.

Using the DARD tools, described in section 9.3, the number of photons returning to the PMT of the emitting DOM is simulated given an LED pointing directly into a 10 cm offset bubble column. If this scenario results in a substantial flux, a tear-drop pattern would have been observed in DARD, as the other LEDs do not point into the bubble column. As no significant tear-drop patterns were observed, size and scattering length combinations resulting in a distinguishable rate can be excluded.

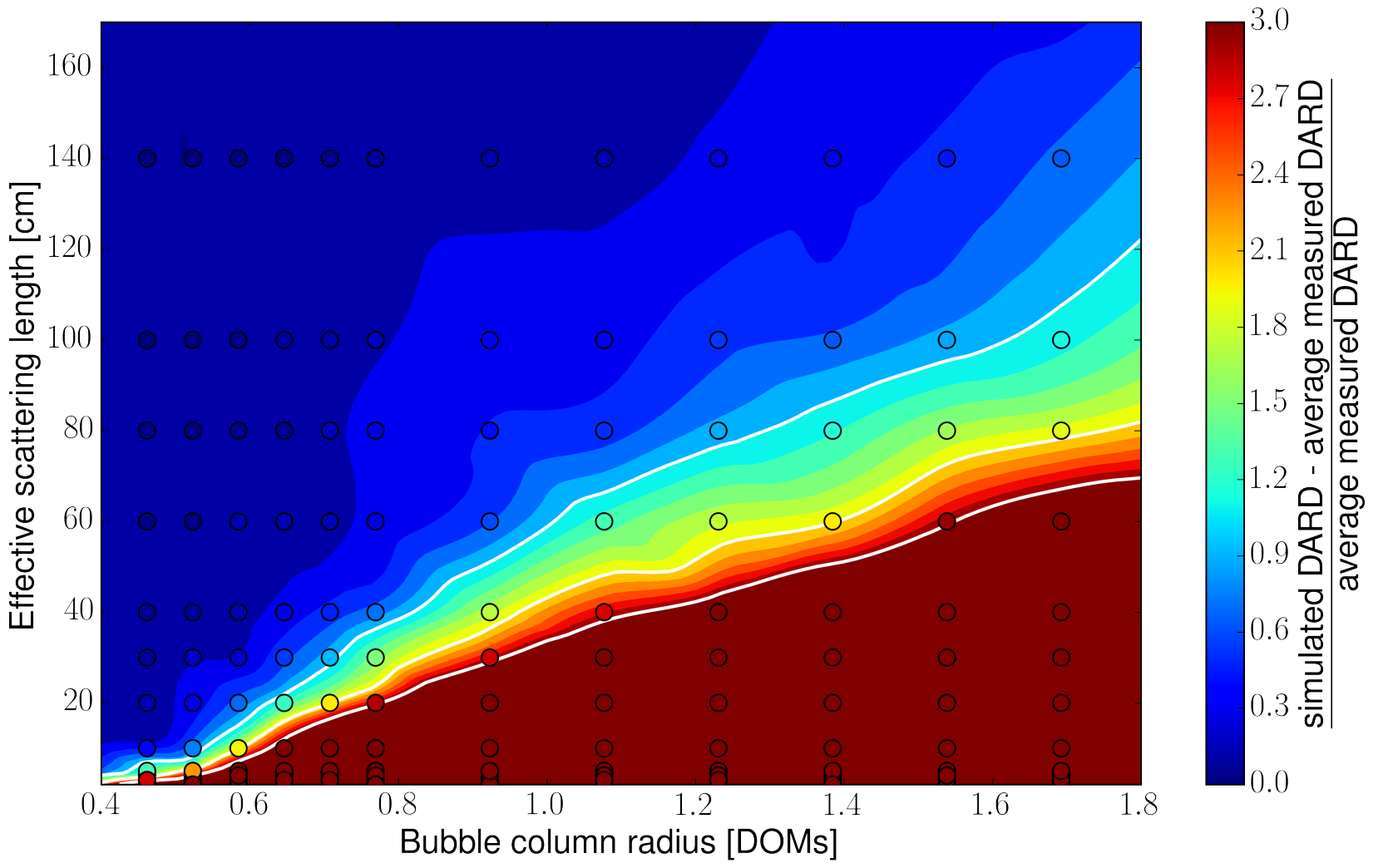

Figure 9.44: DARD limits obtained for the SpiceHD parameter space assuming an offset bubble column and a S/N detection threshold of three for teardrop patterns.

Marginalizing over all DARD DOMs and their LEDs, the average number of measured photons was assumed to represent the background. A tear-drop pattern, as expected from the simulation, is defined to be significant if the number photons to be detected exceeds the background by a factor 3. At this level no tear-drop pattern were consistently observed between horizontal and tilted LEDs in all tested DOMs.

This signal over noise development as a function of size and scattering length as well as the defined exclusion contour is plotted in Figure 9.44 and has also already been shown in the SpiceHD likelihood spaces in Figure 9.40. It traces the upper limit scattering strength also excluded by SpiceHD.

## 9.5 SPICEHD AND THE DOWN-HOLE CABLE

After the SpiceHD model had been developed, the interpretation of its results had to be adjusted following the findings from two new analyses performed by Dmitry Chirkin and summarized below. These analyses were designed to determine the azimuthal position of the down-hole cable with respect to each DOM.

### 9.5.1 *Emitter orientation analysis*

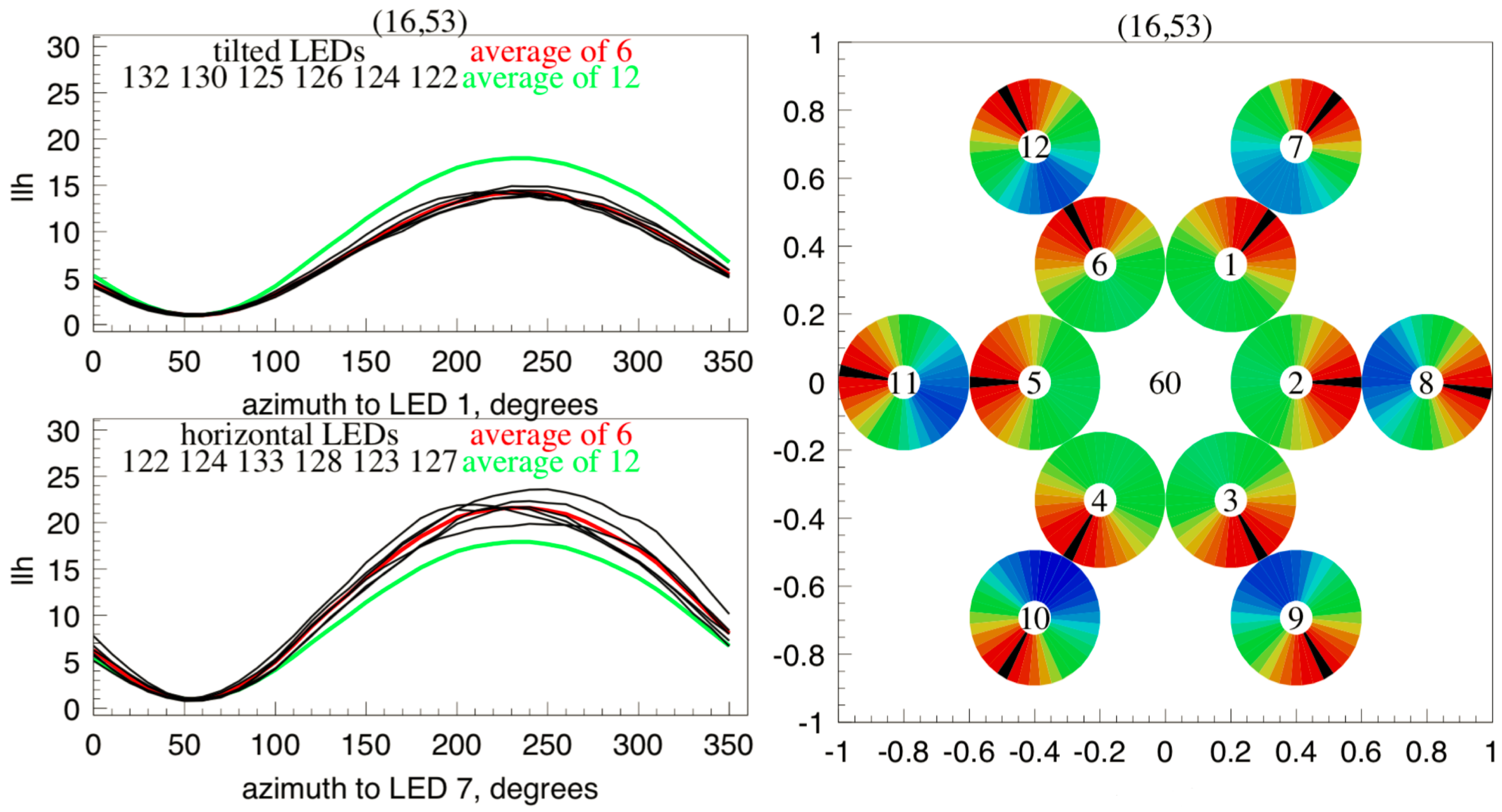

Figure 9.45: Flasher board orientation measurement for DOM S16D53 using single-LED flasher data. The orientation of all 12 LEDs on the flasher board is fitted by comparing simulation at a 1° spacing to the data. Left: Likelihood profiles for all six horizontal and tilted LEDs. Right: Different representation of the same data. Each colored circle denotes the position of one LED on the flasher board. The color gradient decodes the likelihood profile, with the best fit orientation indicated as a black bar. When all LEDs fit the same overall flasher board orientation, the black bars should point straight outwards.
[Dmitry Chirkin (2018b). "Single LED data-taking campaign: summary and results"]

The most accurate method takes advantage of the deployment procedure detailed in section 4.5.1, where the DOMs are attached to the down-hole cable such that the cable is always positioned between LEDs 5&6 and 11&12. Therefore, the cable positions can be deduced by knowing the absolute orientation of each flasher board.

This is achieved by performing per emitter DOM LLH-scans of single-LED data to PPC simulations of this DOM with a horizontal or tilted LED in arbitrary azimuthal orientations. Example likelihood profiles are shown on the left side of Figure 9.45 for the 12 LEDs of DOM 53 on String 16. The absolute orientation of the flasher board is given by the mutual minimum of the likelihood profiles.

The flasher board orientations obtained by this method were verified against a previous analysis which utilized the rising edge timing on adjacent strings and an analysis applying flasher unfolding (see section 5.2.5.1) to single LED data. All methods agree within their uncertainties, with the presented method having the smallest average uncertainty of below one degree.

The distribution of orientations is homogeneous[363] making a potential bias by anisotropy effects unlikely and no patterns of correlated orientations along strings could be identified.

[363] *Dmitry Chirkin (2018b). "Single LED data-taking campaign: summary and results"*

### 9.5.2 *Cable shadow analysis*

To verify that the DOMs were deployed in the specified orientation and that the cable did not break loose from the DOMs, a cross-check analysis tries to directly identify the cable shadow on receiving DOMs.

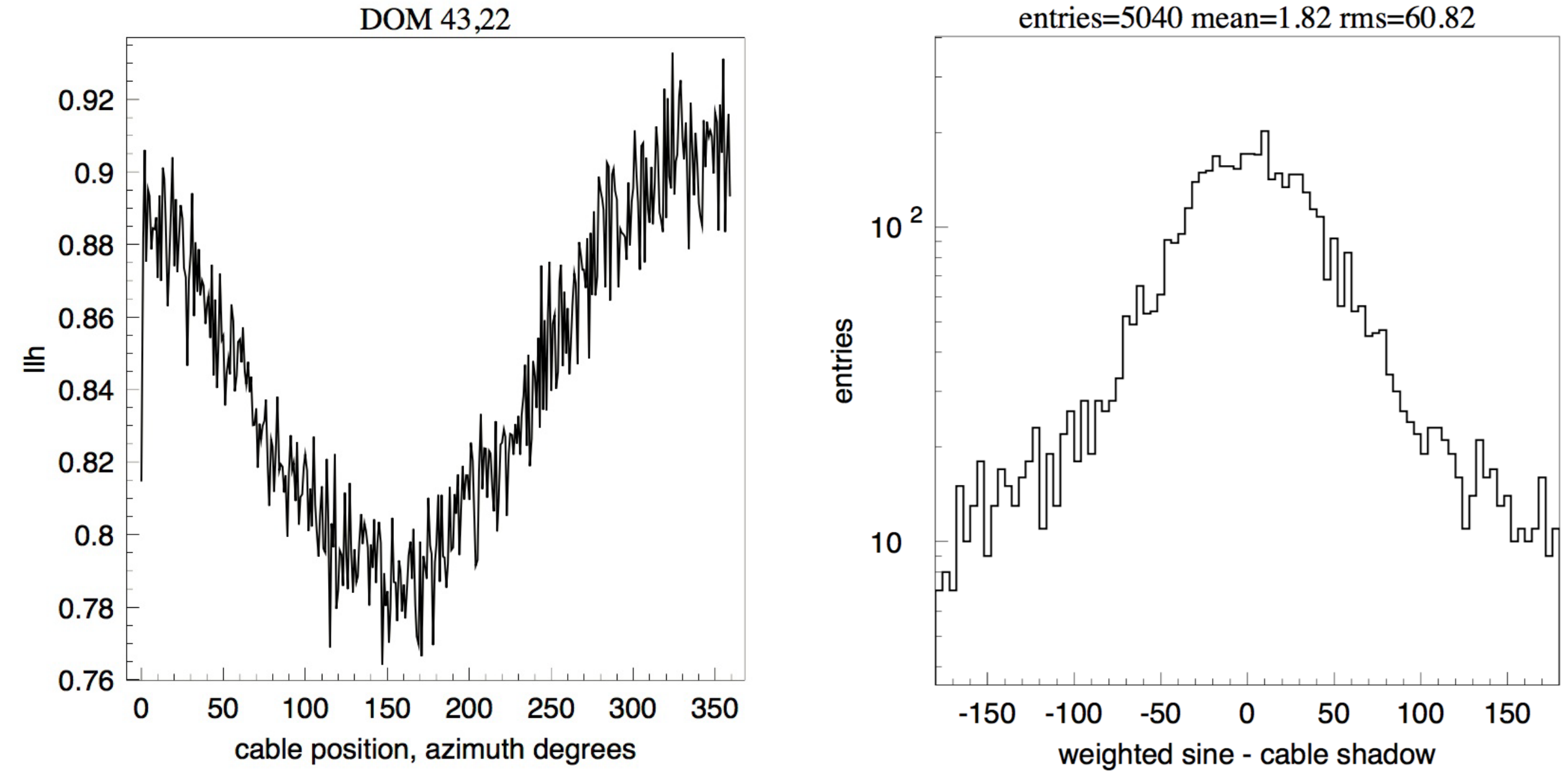

Figure 9.46: Left: LLH scan for DOM S43D22 comparing flasher data to a PPC simulation with a cable at different orientations. Right: Angular difference between the cable shadow and the flasher board orientation measurements [Dmitry Chirkin (2018b). "Single LED data-taking campaign: summary and results"]

The all-purpose flasher data LLH is computed for each receiving DOM is and using a PPC simulation with an approximate cable implementation. Photons emitted from another, arbitrary DOM arriving at the receiver DOM sphere are back-traced to test for an intersection with a proposed (curved) cable cylinder. The proposed cable is moved around the DOM and photons intersecting the cable are in each case excluded from the LLH calculation.

The left side of Figure 9.46 shows an example LLH profile for DOM S43 D22 as the simulated cable is moved around the DOM. The minimum yields the cable orientation, while the spread of LLH values near the minimum indicates a resolution of $\sim 50°$. As shown on the right side of the same Figure, the cable orientations found in this manner agree with the cable positions as deduced from the flasher board orientations, with a typical deviation on the order of the resolution of the cable shadow analysis.

### 9.5.3 *Correlation with SpiceHD*

Conceptually the cable shadow analysis is very similar to the SpiceHD analysis. Both use the all-purpose flasher data and evaluate per-receiver LLH values. In the case of the flasher shadow analysis the local perturbation of the ice is a perfectly absorbing cylinder of known size and radial distance from the DOM, with the only free parameter being the orientation. For SpiceHD also the size, distance and optical properties are free variables.

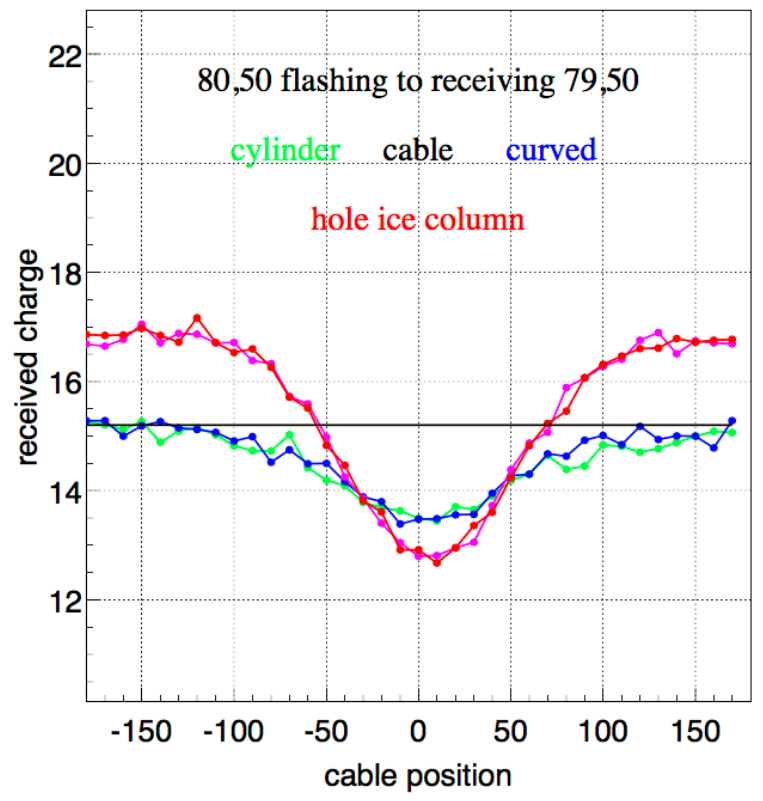

Figure 9.47: Relative azimuthal shadowing by different cable implementations (green, black and blue) compared to a strongly scattering bubble column simulation (red). [Chirkin, "Single LED data-taking campaign: summary and results"]

Neglecting timing effects, which were previously argued to be unresolvable, both the cable and the bubble column lead to an azimuth and zenith dependent detection efficiency.

A direct comparison for a typical SpiceHD configuration and horizontal illumination is shown in Figure 9.47. Both scenarios lead to a deficit as the bubble column / cable shadows the DOM. As a diffuse scatterer, the bubble column partially reflects light when it is situated behind the DOM, leading to a small intensity enhancement.

In the case of the cable shadow analysis, a correlation with the flasher board orientation analysis has confirmed that the employed method is sensitive to and predominantly identifies the cable position.

The availability of reliable cable orientations allows testing if SpiceHD has fitted something other than the cable, which could in turn be the bubble column.

Figure 9.48 shows the histogram of opening angles between the cable position and the bubble column position as fitted by SpiceHD. For a sizable fraction of DOMs, the fitted bubble column position agrees with the cable position within the $60°$ binning used in SpiceHD. Therefore the dominant effect picked up by the SpiceHD analysis seems to be the cable shadowing. The validity of the presented constraints on the size and scattering length are put into question.

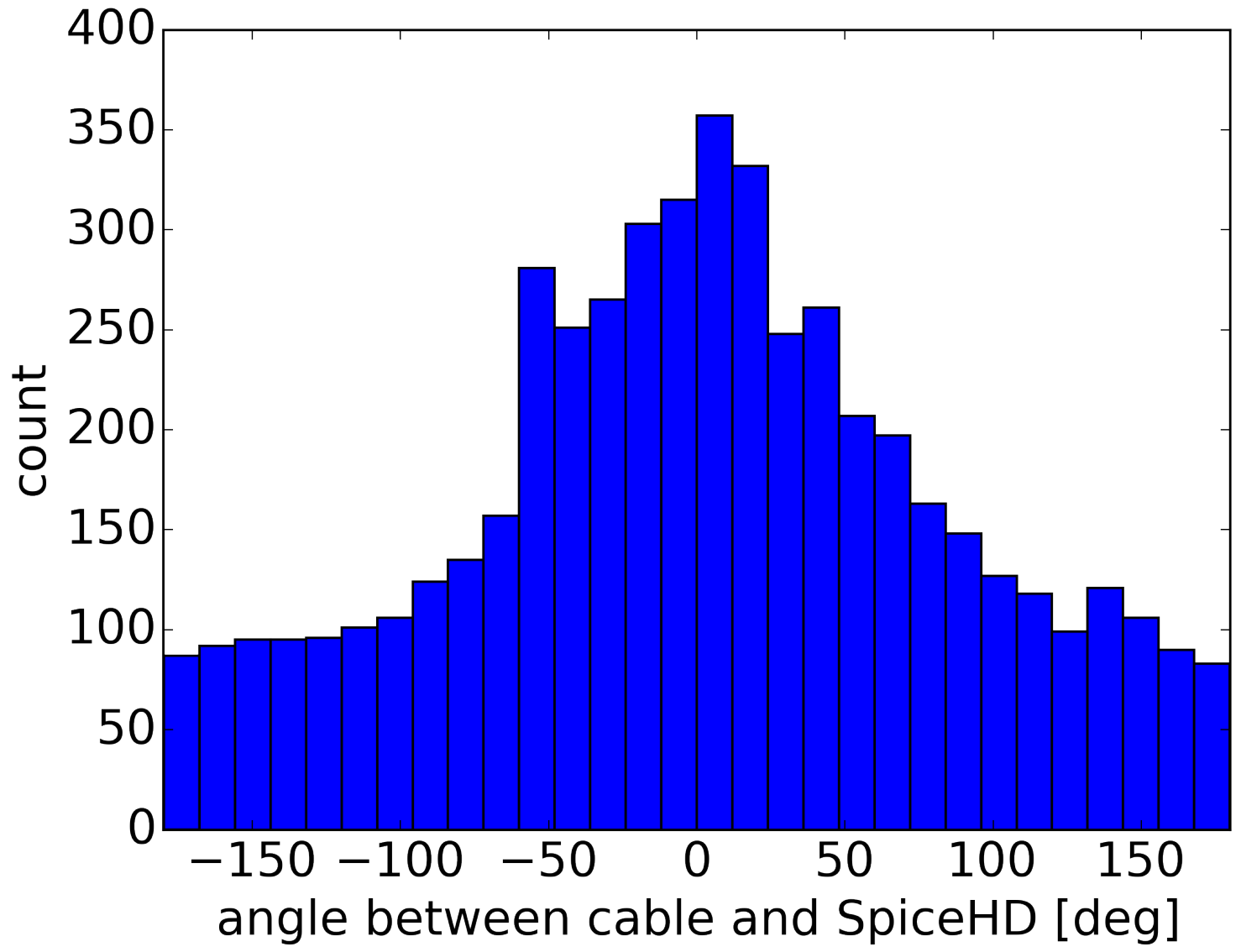

Figure 9.48: Correlation between the measured cable position and the bubble column orientation as fitted in SpiceHD. In most cases SpiceHD has picked up on the cable position.

## 9.5.4 *Local ice effects in the flasher board analysis*

The measurement of the flasher board orientation is highly overconstrained as twelve emitters are used to fit a single direction. Therefore the per-LED likelihood profiles as seen in Figure 9.45 have also been tested for possible hole ice effects[364]. Indications for local diffusion would be isolated LEDs which point to a different DOM orientation, or LEDs with a much reduced likelihood amplitude compared to the average on the DOM.

[364] *Dmitry Chirkin (2018a). "Characterizing anomalies of local ice"*

Similar to DARD, this search[365] is primarily sensitive to a bubble column outside the LED perimeter. For DOMs where a beam broadening or orientation deviation has been observed, it is located at the position of the fitted cable. Just as with DARD no clear hole ice signature is found.

[365] *Chirkin, "Characterizing anomalies of local ice"*

## 9.5.5 *Impact of SpiceHD*

Since its release in September 2016, the SpiceHD model has been generally accepted as the most realistic modeling approach. In particular low energy oscillation analyses, which already previously used non-oversized MC, have adopted SpiceHD as one hole ice realization to be incorporated into final uncertainties[366].

[366] *Aartsen et al., "Measurement of Atmospheric Tau Neutrino Appearance with IceCube DeepCore"*

## 9.6 SUMMARY OF THE CURRENT KNOWLEDGE

While the hole ice has been known and a nuisance ever since the times of AMANDA, progress in understanding its true nature has been very slow.

The two presented studies, DARD and SpiceHD, introduce new measurement and modeling concepts, but still fail to produce conclusive results.

Reviewing the existing body of work, the following observations can be made:

- A central strongly scattering column of roughly 16 cm diameter is visually observed at the location of the Sweden Camera.
- Flasher data as well as physics data is better described using a variety of hole ice parametrizations when compared to the angular acceptance intrinsic to the DOM itself.
- Emitter-side analyses like DARD or the measurement of the flasher board orientations are primarily sensitive to a bubble column outside the flasher perimeter and find no evidence for a bubble column.
- Receiver-side analyses (SpiceHD and the cable shadow analysis) find a shadowing object at the position of the cable. SpiceHD, assuming a non-absorbing but scattering object, sets constrains on the size of the fitted object, which is significantly larger than the cable.

From these observation it seems reasonable to conclude that either of the following statements must be true:

- In most of the detector there is no significant bubble column and the shadowing of the cable has been confused with a hole ice effect.
- The bubble column is small in diameter and centered around the DOM, so that it can not be detected by an emitter-side analysis.
- The bubble column is small and usually at the cable position. Given that the DOMs are randomly oriented in azimuth and that the drill holes can locally be considered straight[367], it seems plausible for the cable to be located in the center of the hole and thus at the same location as the bubble column.

[367] *see section 4.3.2.1 for the drill performance*

## 9.7 FUTURE MEASUREMENTS WITH THE POCAM

Given the remaining uncertainty on the hole ice, several new measurements are anticipated as part of the IceCube upgrade.

---

**Declaration of Pre-released Publications**
The study presented in this section has already been published by the IceCube collaboration[368]. The author of this thesis has written this publication as corresponding author together with Elisa Resconi and Kai Krings. All material shown in the following is based on own work.

---

[368] *Resconi, Rongen, et al., "The Precision Optical CAlibration Module for IceCube-Gen2: First Prototype"*

To investigate the possibility to improve the in-situ angular acceptance curve measurement using the POCAM, photon tracking simulation equivalent to those performed for SpiceHD have been performed. A single POCAM is simulated as a point-like and perfectly isotropic emitter situated in the center of the IceCube infill array, DeepCore.

Due to expected improvements in the drilling technology, it is assumed that the hole ice surrounding the POCAM has the same properties as the bulk ice. As an initial simplification the bulk glacial ice is assumed to be free of scattering. The same study can be performed in realistic bulk ice, by applying stringent timing cuts on the photon propagation delay from the POCAM to each receiving DOM, ensuring a straight, unscattered propagation. The light emitted from the POCAM can be detected by any IceCube DOM.

Two kinds of datasets have been simulated. The reference datasets account for bulk ice propagation, but do no propagation through hole ice and assume the DOMs to be able to measure photons over their entire spherical surface. All other datasets assume a given realization of the hole ice and only allow photon detection at the PMT surface, representing the data from a potential measurement.

The DOM-wise ratio of detected photons in the data-like set and the reference dataset is equivalent to the DOMs in-situ angular acceptance function at the zenith angle of this DOM relative to the POCAM. By plotting the relative acceptance of many DOMs the overall angular acceptance of an average DOM is obtained.

Figure 9.49a shows an example where the entire drill hole is assumed to be filled with a weakly scattering medium and all

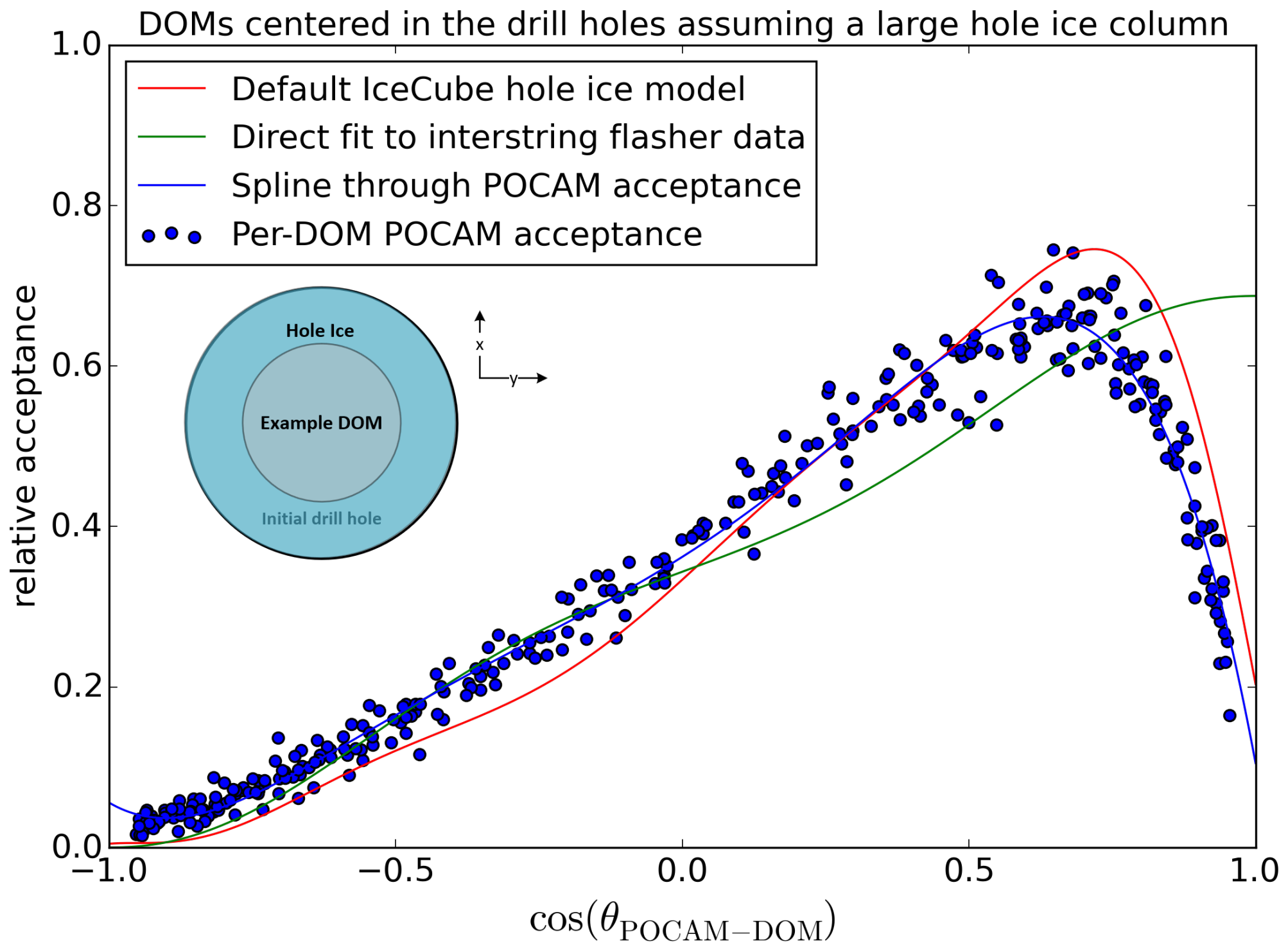

(a) Assuming the entire drill hole to be filled with a weakly scattering medium and the DOMs to be centered inside the holes (IceCube default).

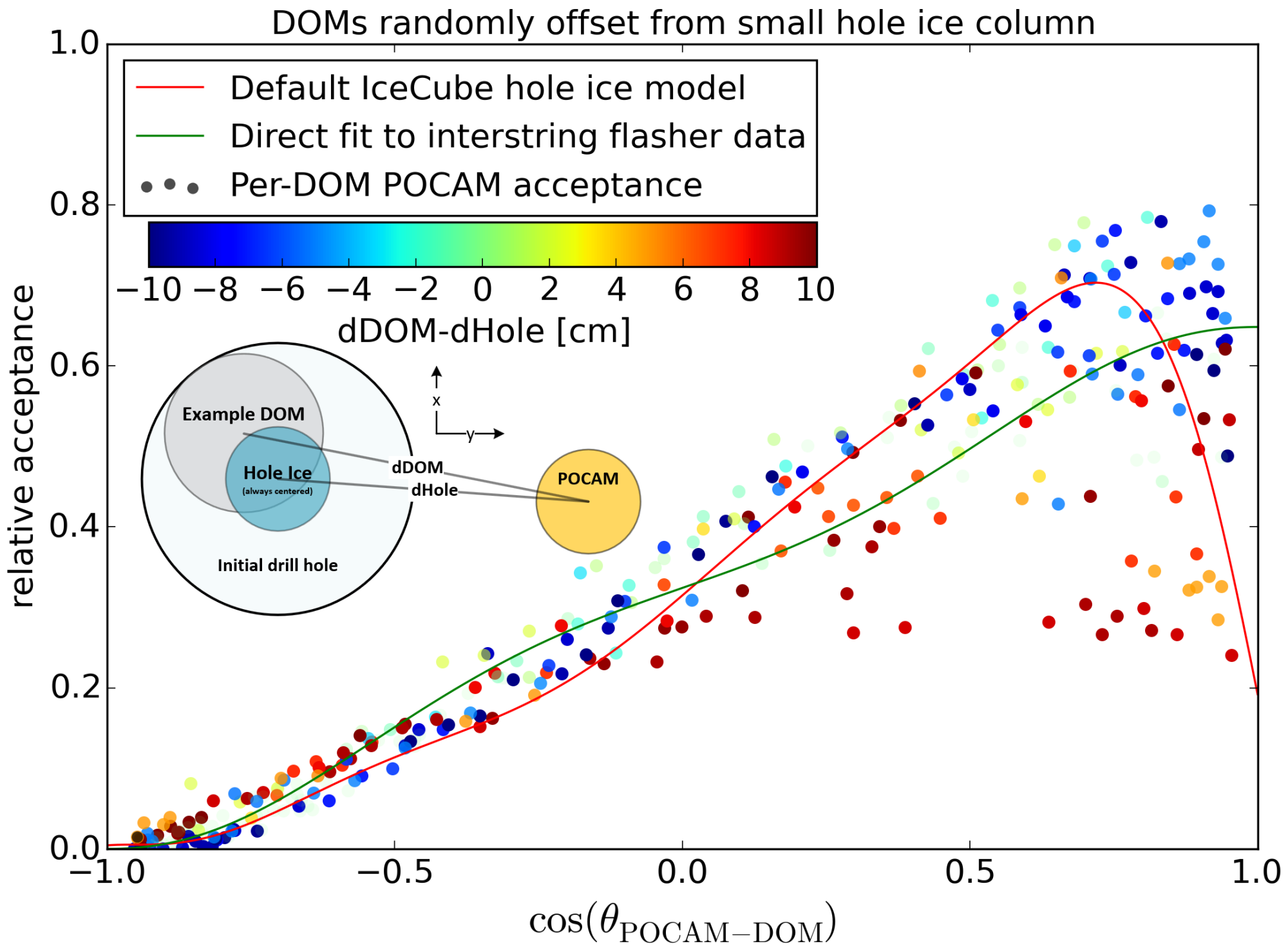

(b) Assuming a central, small and strongly scattering column with the DOMs randomly positioned inside the holes.

Figure 9.49: POCAM capability for measuring the IceCube drill hole properties. The points represent different DOMs measuring light from a single POCAM at different angles.

DOMs are located in the center of their respective hole. The obtained angular acceptance curve agrees well with the H2-model ("Default IceCube hole ice model") which is based on the same assumptions.

Figure 9.49b shows an example for a small, but strongly scattering hole ice column. In addition, the DOMs are now randomly positioned inside the drill hole. Obviously no consistent average angular acceptance curve can be obtained, as different parts of the DOM surface are being shaded off relative to the emitter. Using Monte Carlo knowledge about the geometry, we can identify causally connected curves which belong to similar relative geometries. In the experiment the geometric ambiguity can be resolved by multiple POCAMs illuminating each DOM from different azimuthal angles.

Given a satisfactory measurement in Gen2 Phase-1, the individual angular acceptance curves can also be interpreted in terms of the underlying parameters, namely the size and scattering length of the hole ice and position of the DOMs. This will allow treating the hole ice via direct photon propagation instead of the effective description via the angular acceptance curve, which neglects azimuthal effects.

# 10

# THE OPTICAL ICE ANISOTROPY

**The bias observed while fitting the orientation of the bubble column motivated a depth dependent evaluation of the anisotropy strength. This in turn revealed that our current parametrization can only poorly describe the data and triggered a series of investigations of the true underlying cause of the anisotropy.**

## 10.1 THE OBSERVED EFFECT

One does not expect the amount of received light to depend on the orientation of the receiver to the emitter, when observing an isotropic light source, as for example given by averaging over many emitting flasher DOMs.

Figure 10.1 shows the total charge observed per emitter-receiver pair in the all-purpose flasher data compared to simulation without any anisotropy modifications. A strong modulation with a period of 180° in the azimuth of the IceCube coordinate system can be seen.

Along the direction of the ice flow axis about twice as much light as expected is received. The inverse is found along the orthogonal axis which is the direction of the ice tilt. The impact of the anisotropy on the arrival time distribution is discussed in section 10.5.

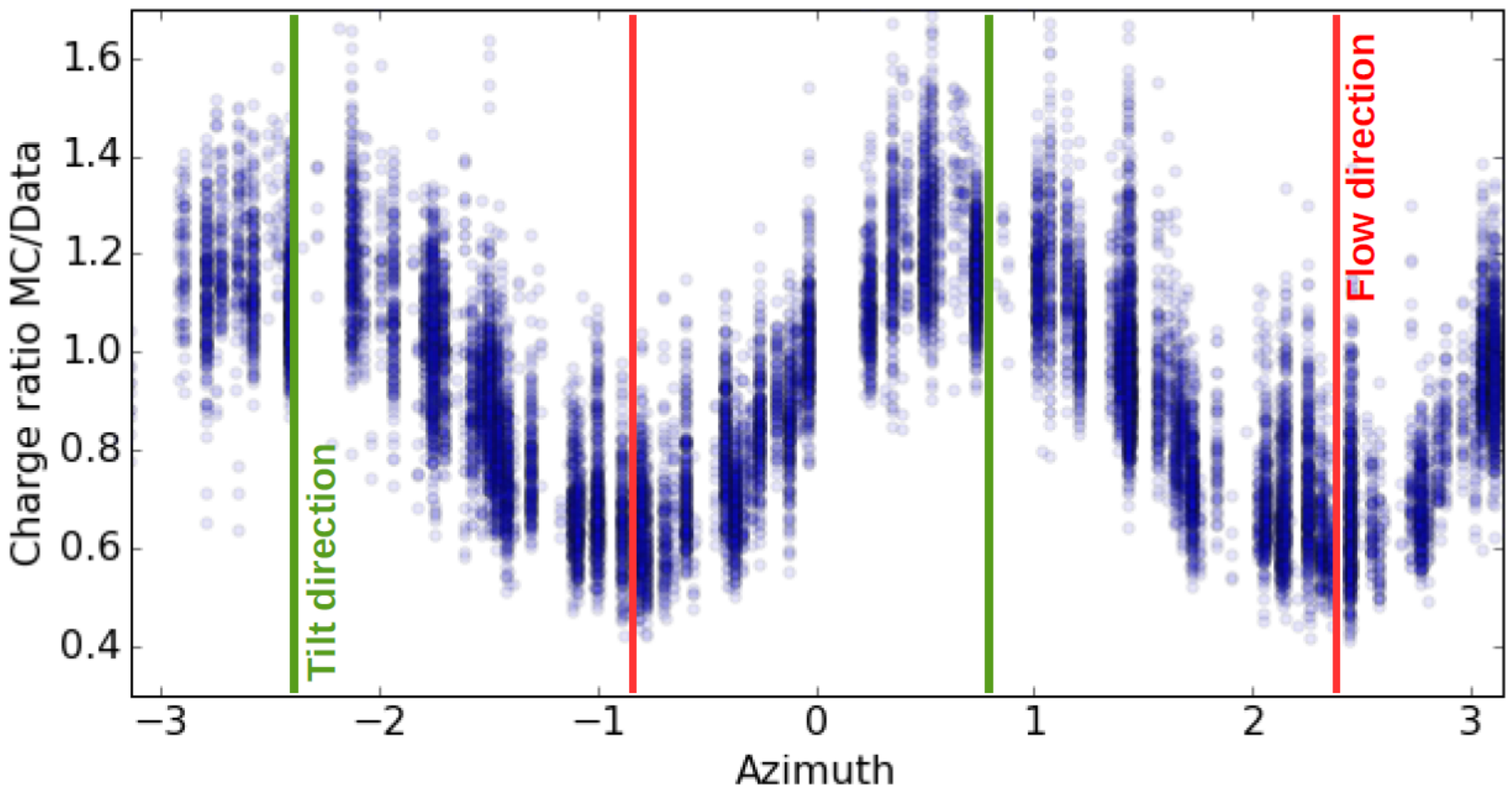

Figure 10.1: Ratio of total charges received in each DOM emitter-receiver pair in the all-purpose flasher data as a function of azimuth. Comparing experimental events to simulation with no anisotropy.

## 10.2 THE ORIGINAL PARAMETRIZATION

The original parametrization and fit[369] argued that due to time and space reversal symmetries the absorption length and geometric scattering length can not be direction dependent. Therefore the anisotropy was implemented as a modification to the scattering function, the only remaining Mie scattering parameter.

[369] *Dmitry Chirkin (2013a). "Evidence of optical anisotropy of the South Pole ice"*

Instead of evaluating the scattering function $f(\cos(\theta))$, that is the probability distribution to scatter by a given angle, with respect to the true orientation of the ingoing and outgoing photon directions $\vec{n}_{in}$ and $\vec{n}_{out}$, the scattering function is evaluate in terms of a stretched coordinate system:

$$f(\vec{n}_{in} \cdot \vec{n}_{out}) \rightarrow f(\vec{k}_{in} \cdot \vec{k}_{out}), \tag{10.1}$$

where the transformation is introduced through a matrix $A$:

$$\vec{k}_{in,out} = \frac{A\vec{n}_{in,out}}{|A\vec{n}_{in,out}|}. \tag{10.2}$$

When written in terms of a basis with the flow direction along the x-axis and the z-axis along the true zenith, the matrix is diagonal and can be expressed as:

$$A = \begin{pmatrix} \alpha & 0 & 0 \\ 0 & \beta & 0 \\ 0 & 0 & \gamma \end{pmatrix} = \exp \begin{pmatrix} \kappa_1 & 0 & 0 \\ 0 & \kappa_2 & 0 \\ 0 & 0 & \kappa_3 \end{pmatrix}. \tag{10.3}$$

[370] *M. A. C. Potenza et al. (June 2016). "Shape and size constraints on dust optical properties from the Dome C ice core, Antarctica"*

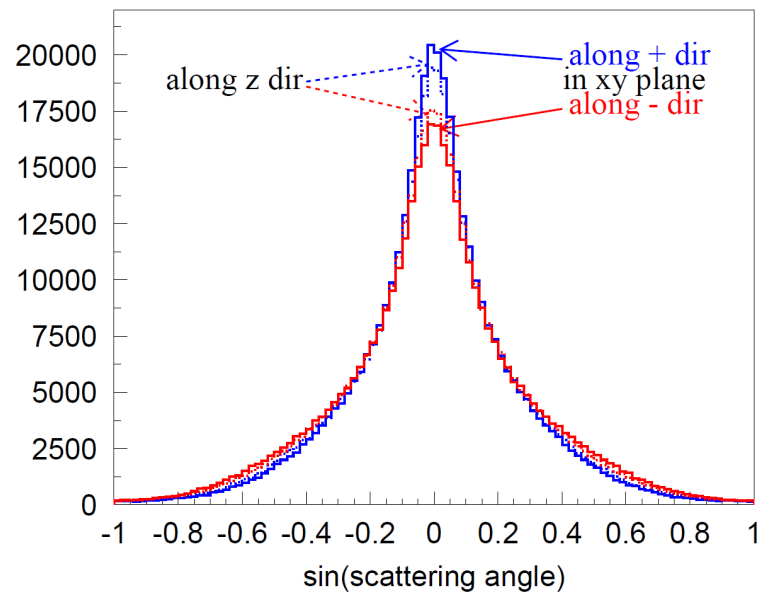

Figure 10.2: Comparison of scattering functions for a number of selected propagation directions as given by the original anisotropy parametrization. (Positive and negative scattering angles are functionally identical.)
[Chirkin, "Evidence of optical anisotropy of the South Pole ice"]

In order to conserve the direction-averaged effective scattering coefficients, which have been fitted to a high accuracy prior to the discovery of the anisotropy, it is required that:

$$\alpha \cdot \beta \cdot \gamma = 1 \quad \text{or equally} \quad \kappa_1 + \kappa_2 + \kappa_3 = 0 \tag{10.4}$$

While not derived from first-principle Mie calculations, the parametrization was justified to be a plausible result of elongated impurities becoming preferentially aligned by the flow and thus introducing a direction dependence to the scattering function. While several glaciological studies[370] explore the shapes of impurities, elongations for different impurities are not well established, nor is there any evidence for impurities becoming oriented with the flow.

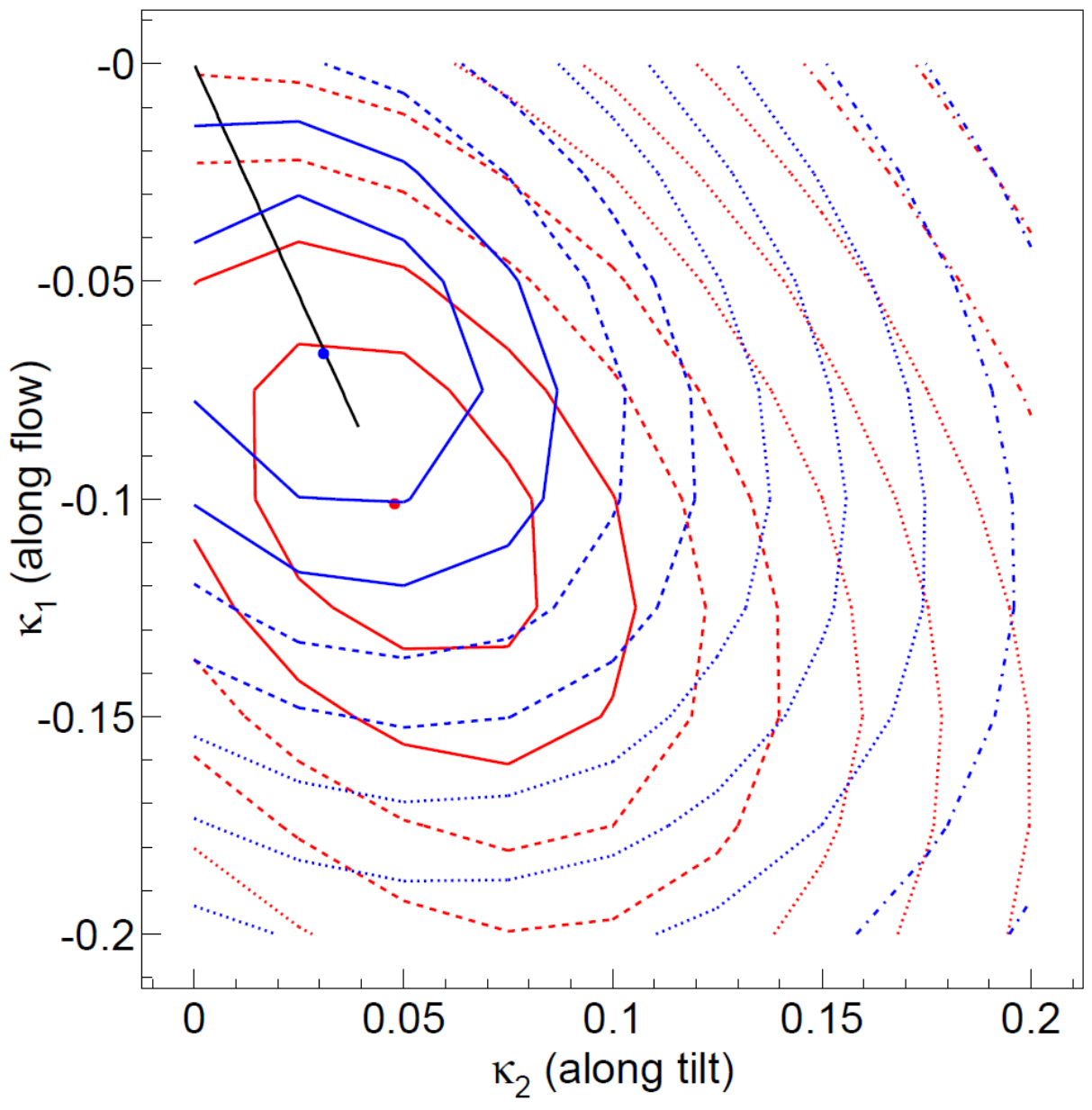

Figure 10.3: Spice Lea anisotropy strength fit. The red contour uses only charge information in the likelihood. The blue contour uses both charge and timing.
[Dmitry Chirkin (2013a). "Evidence of optical anisotropy of the South Pole ice"]

The parametrization was implemented into the photon propagators[371]. The resulting scattering functions for different example propagation directions and assuming an 8% anisotropy strength ($\kappa_1 = -0.08$) are shown in Figure 10.2. In order to get a better understanding of the effect, it is helpful to derive a small-angle approximation of the modification to the average scattering angle[372]:

[372] *Chirkin, "Evidence of optical anisotropy of the South Pole ice"*

$$1 - \langle \cos\theta \rangle = (1-g) \cdot \frac{1}{2} \cdot (B_{in}B_{in} - n^i B_{in} n^j B_{jn}) \cdot |A\vec{n}|^2 \quad (10.5)$$

with $B = A^{-1}$ and g being the default $\langle \cos\theta \rangle \sim 0.9$. It is evident that the parametrization changes the effective scattering coefficient as a function of the propagation direction. Photons propagating along the flow axis experience less scattering than photons propagating along the tilt axis or inclined from the horizontal.

Using this parametrization, the anisotropy axis and strength averaged over the entire detector were fitted. The axis is 130° and the LLH-contours of a grid-scan in $\kappa_1$ & $\kappa_2$ are shown in Figure 10.3. The red contour only considers charge, while the blue contour uses both charge and timing information. The best fit anisotropy strength for the combined fit are $\kappa_1 = -8.2\%$ & $\kappa_2 = 4.2\%$, while the charge-only fit prefers a stronger anisotropy.

With the introduction of flasher unfolding in Spice3[373], the anisotropy strength has been fitted to 10.6%. The effect that charge and timing information do not fit the same anisotropy strength will be of major relevance for the following discussions.

[373] *see section 5.2.8*

## 10.3 FITTING THE ANISOTROPY AXIS

In departing from an anisotropy which is assumed constant throughout the detector, the first step is a re-evaluation of the anisotropy axis. This can be achieved without a likelihood fit and is thus independent of any model assumption. To obtain spacial resolution the data is binned in emitting DOMs, either within a tilt-corrected depth range or by string number.

The axis can simply be fitted as the phase of the sinusoidal charge modulation as originally shown in Figure 10.1. As the number of emitters is limited after binning, the very strong, correlated modulation seen in the flasher unfolding profiles[374] from a simulation without anisotropy is used instead. Figure 10.4 shows the resulting anisotropy axes, either as function of depth or as function of lateral position.

[374] *see section 5.2.5.1*

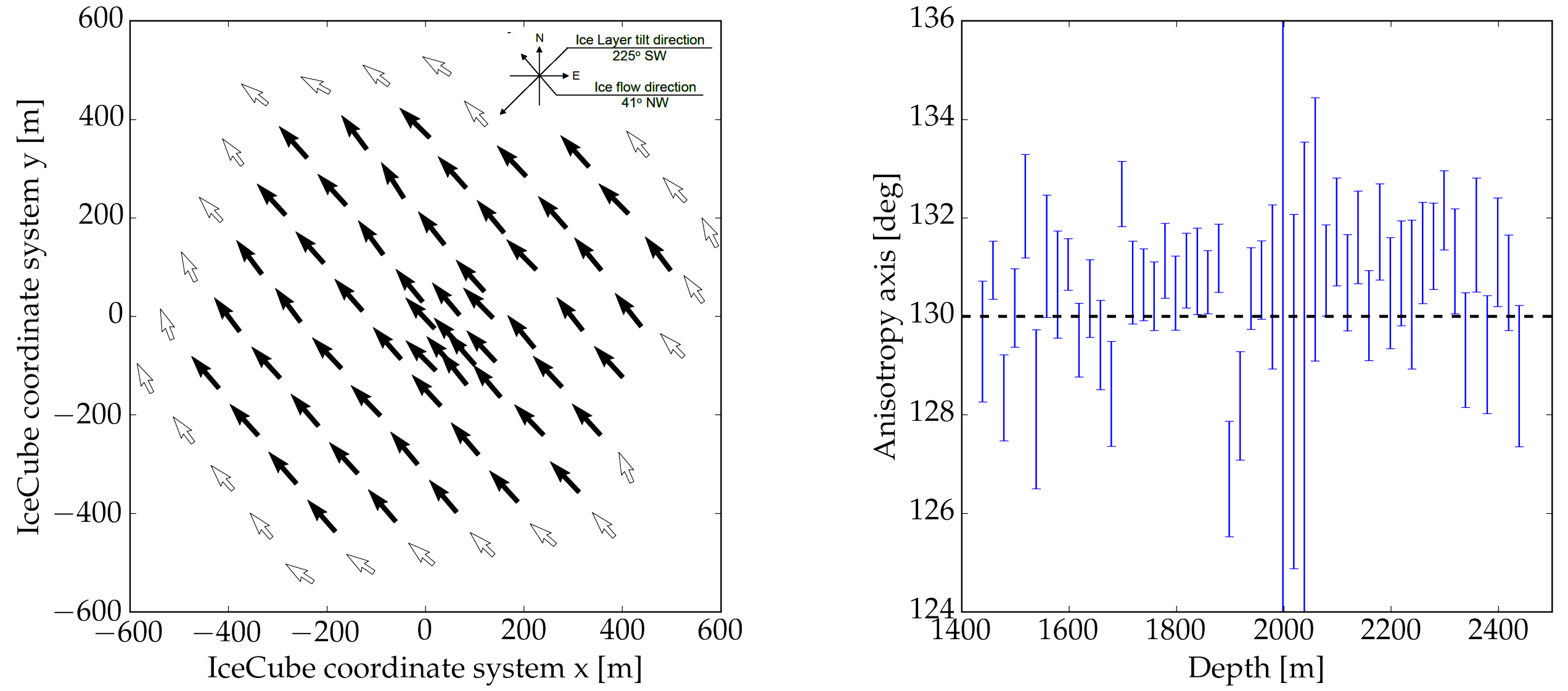

Figure 10.4: The anisotropy direction is found to be constant throughout the detector. Left: Average anisotropy axis orientations along each string. Edge strings are unreliable as no neighboring strings exist for some directions, potentially biasing the fit. Right: Average anisotropy direction of all strings binned by emitter depth.

The anisotropy axis is seen to be constant throughout the entire detector and is considered constant for all following investigations. The resolution is generally below 1°. Edge strings should be disregarded as light can only be sampled from a limited angular range, in addition the strong attenuation in the dust layer strongly reduces the sensitivity in that depth range.

The anisotropy is an effect which only develops throughout the photon propagation. As receiving DOMs might not be in the same depth bin / on the same string as the emitter, there is a correlation in the found anisotropy axis of neighboring bins. The depth resolution of this approach is discussed in the following section.

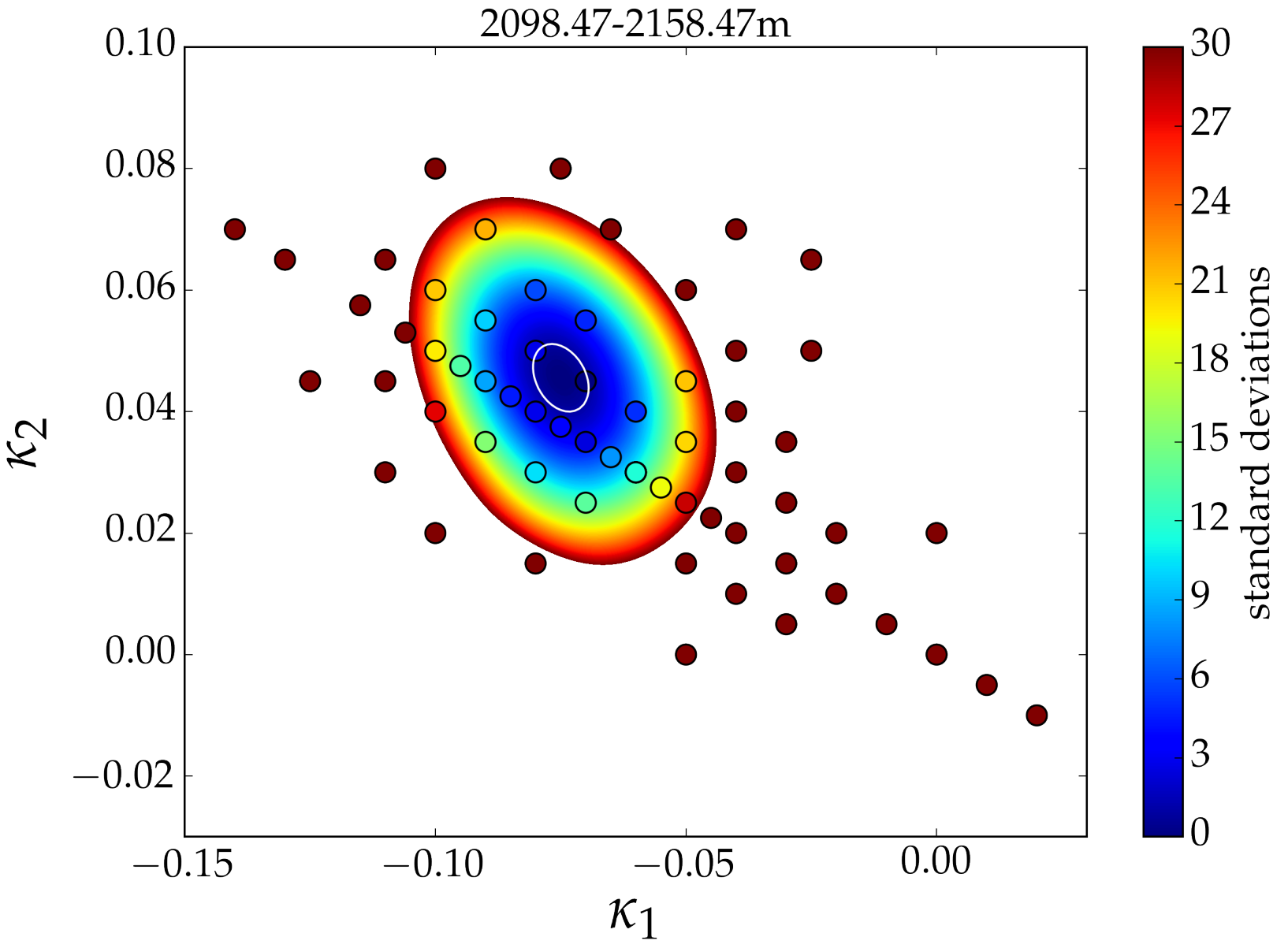

Figure 10.5: Example 2D likelihood scan of the anisotropy strength in the depth range of 2098 m to 2158 m. Colored circles denote the delta LLH of tested realizations compared to the best fit in units of standard deviations of the statistical fluctuations of the LLH near the best fit. For the likelihood evaluation only emitter-receiver pairs within the depth range are considered. The background color is a smoothing spline used to construct the statistical confidence contour.

## 10.4 DEPTH DEPENDENT LLH FITTING

As the anisotropy axis has been established to be constant over the detector, the anisotropy strength can be evaluated simply as a function of depth. It is fitted using an likelihood grid-scan. A DOM oversizing of 16 is used to speed up the simulations. To achieve depth resolution the likelihood calculation is split into depth bins according to the tilt corrected emitter locations.

To be able to choose appropriately sized depth range, the resolution has been tested in a MC round-trip fit. Pseudo-data, resembling flasher events has been generated assuming a step function in the anisotropy strength. The anisotropy strength was then fitted in an equivalently binned likelihood scan.

The resulting depth profile was found to be well fitted by an error function

$$\kappa(z) = \kappa_1 + (\kappa_2 - \kappa_1) \cdot \frac{1}{2} \left[ 1 + erf\left( \frac{z - z_0}{\sqrt{2}\sigma} \right) \right] \quad (10.6)$$

assuming a depth resolution $\sigma$ of 55 m, when all receivers are included in the likelihood calculation. The depth resolution is 30 m, when only receivers within a 10 m depth range to the emitter are included.

As the trade-off in statistics is large compared to the gain in resolution, the fit was performed with all receivers included in the likelihood calculations. Two different depth binnings for the emitters of 60 m, according to the resolution, and 30 m, as a cross-check, were selected.

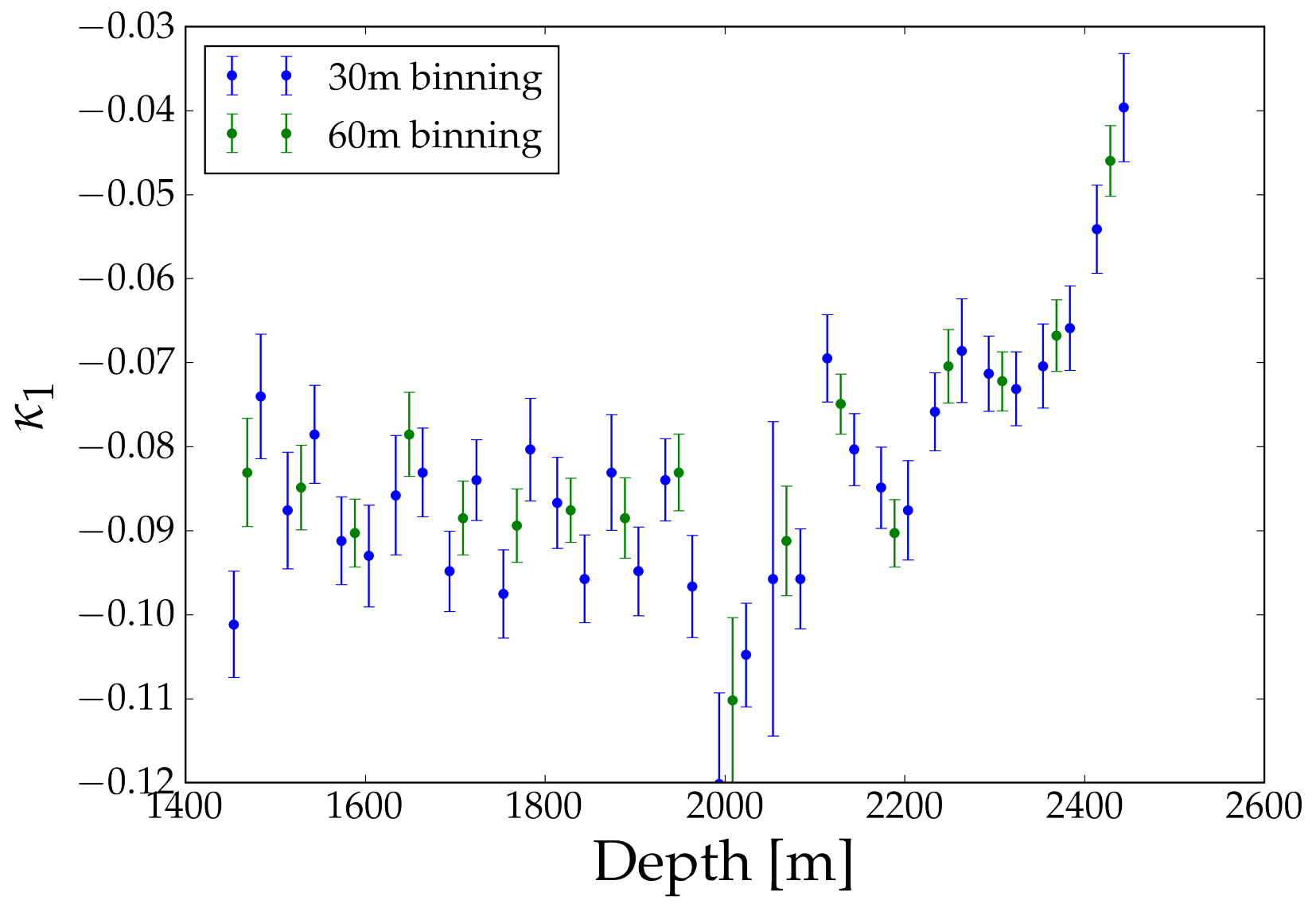

Figure 10.6: Depth dependence of the anisotropy strength $\kappa_1$ using the default scattering function based parametrization. Each point and uncertainty is derived from a likelihood scan as presented in figure 10.5. The anisotropy is seen to be of constant strength, with no correlation to other optical properties, above the dust layer. It is slightly weaker below and seems to vanish quickly at the lower edge of the detector.

Figure 10.5 shows an example likelihood space as a function of the anisotropy coefficients $\kappa_1$ and $\kappa_2$ for a depth of ~2130 m. Each point represents a tested anisotropy configuration with the color denoting the $\Delta LLH$ from the fitted minimum in units of standard deviations. As the Dima likelihood does not fulfill Wilks Theorem the standard deviation is obtained by re-simulating a point close to the minimum several times, as explained in section 5.2.6.

The best-fit anisotropy strengths and uncertainties obtained from all depth ranges are summarized in the depth profile for the $\kappa_1$ anisotropy strength coefficient shown in Figure 10.6.

The anisotropy is seen to have a constant strength of ~ 9% above the dust-layer, with no-correlation to the absorption and scattering coefficients. Below the dust-layer the strength is reduced to a fairly constant 7%, before undergoing a sudden weakening below 2400 m.

The slope of the weakening is consistent with the expected smearing by the depth resolution and therefore consistent with an abrupt change. As the instrumented depth ends at ~2450 m it can not be concluded if the anisotropy vanishes completely.

No sudden change in the ice properties was previously known in the clean ice at 2400 m. The property most likely correlated with this change is the c-axis distribution. While c-axis distributions at the South Pole are only known to the depth of the SpiceCore hole[375,376], the fabric is expected to turn from girdle to a single maximum in the bottom 10% of the glacier. This transition is known, for example from EDML[377], to take place within several dozen meters[378].

[375] *see Figure 10.23*

[376] ~1850 m *in IceCube coordinates*

[377] *see section 3.6.2*

[378] *Weikusat et al., "Physical analysis of an Antarctic ice core—towards an integration of micro- and macrodynamics of polar ice"*

## 10.5 DISCREPANCY BETWEEN TIMING AND CHARGE OBSERVABLES

Instead of fitting the anisotropy strength using the standard likelihood method, a variety of other observables dedicated to certain timing aspects has also been tested. This is motivated by the observed bias introduced through the (charge sensitive) flasher unfolding as seen in section 9.4.2.2.

The average photon arrival time, the direct photon arrival time[379] and the standard deviation of the photon arrival time for each emitter-receiver pair were chosen as observables. Because absolute timing variables are difficult to match precisely, even without considering anisotropy, their absolute data-MC deviations are not considered.

[379] *defined as the* $10^{th}$ *percentile*

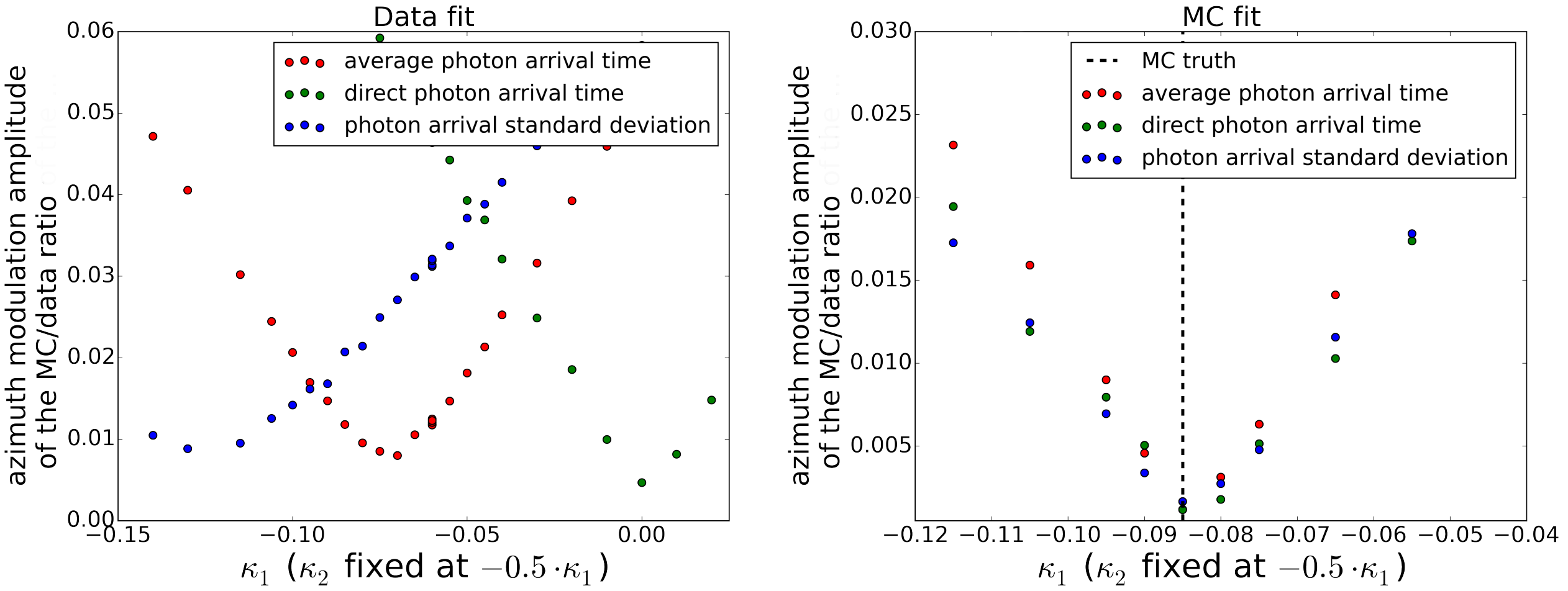

Figure 10.7: Fit of the anisotropy strength using timing variables. Left: The different timing variables prefer different anisotropy strengths when applied to data. Right: Applied to simulation, all observables recover the injected truth. This indicates that the current anisotropy implementation can not parametrize the true effect.

Instead, the data-MC ratio of each observable is calculated for all emitter-receiver pairs above the dust layer. The anisotropy strength is then fitted by minimizing the amplitude of the azimuth modulation of these ratios. This is analogous to minimizing the amplitude of the charge ratio as shown in Figure 10.1.

Figure 10.7 shows the fit results when fitting against a simulation truth of 8.5% on the right and against data on the left. In the MC case all observables fit the injected truth, ensuring that the selected observables are indeed sensitive to the timing variations introduced by the simulated scattering coefficient based anisotropy.

Applied to the data, the fits return vastly different anisotropy strengths $\kappa 1$ for the different observables. The direct photon arrival time, which is most sensitive to scattering, is best described without any anisotropy. The average photon arrival time yields a value similar to the $\kappa_1 = -8.5\%$ anisotropy strength in Spice Lea and the photon arrival time standard deviation prefers a significantly larger anisotropy strength.

The fit has been re-run for various realization of oversizing factors, scattering function parameterizations, hole ice models and scaled bulk ice properties. None of these systematics are found to have an impact on the observed behavior.

To be able to better grasp this behavior, Figure 10.8 shows simulated light curves of different anisotropy strengths, as well as data, after averaging over DOMs aligned with or orthogonal to the flow axis[380].

[380] *The IceCube detector hexagon is aligned such that this is exactly possible on the flow axis. On the tilt axis only strings off by* $30°$ *can be selected.*

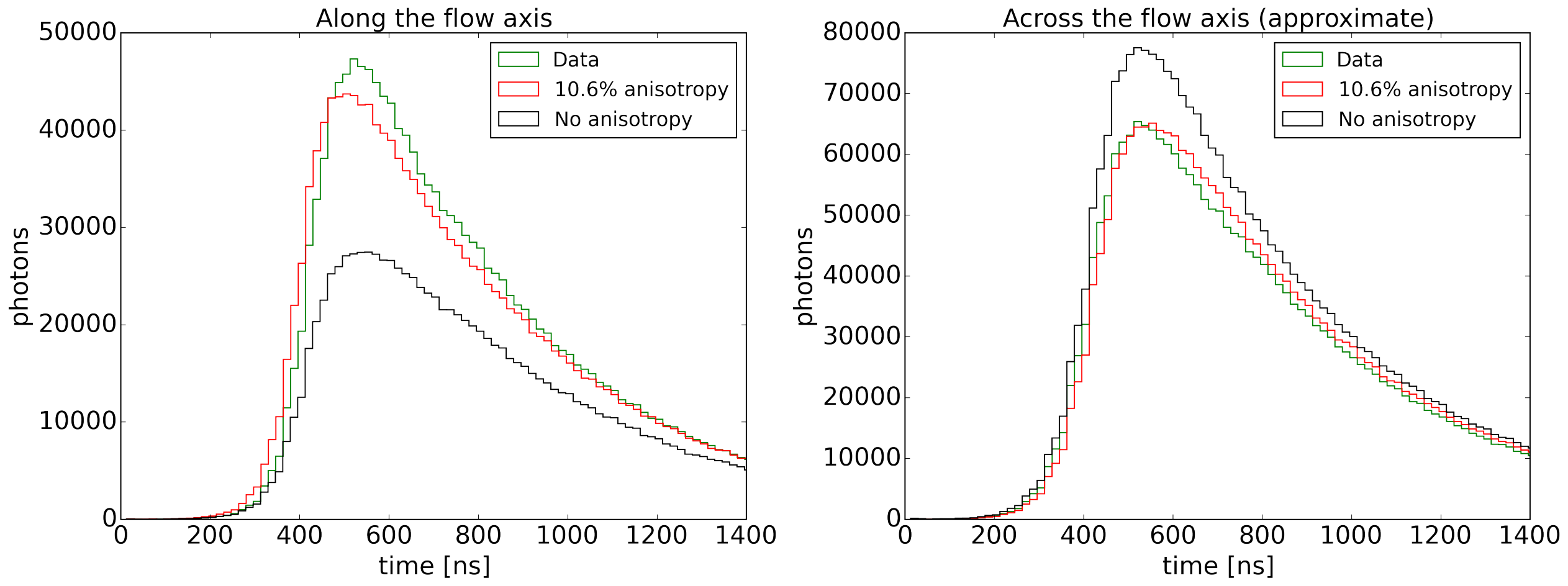

Figure 10.8: Average light curves along and across the flow axis, simulated with different anisotropy strengths $\kappa_1$ and compared against the measured data.

Evidently, the position of the rising edge is already well matched without any anisotropy. The scattering function based anisotropy reduces the effective scattering length in the flow direction and increases it in the tilt direction. As a result, more light reaches DOMs along the flow and less light reaches DOMs along the tilt axis. For the same reason, however, the previously well matched arrival time is now delayed on the tilt axis, and the first photons reach the receivers too early on the flow axis.

Judging from the light curves a scattering based anisotropy parametrization seems to not match the data.

## 10.6 Absorption anisotropy

The scattering function based anisotropy parametrization can not describe timing and charge observables simultaneously. It is therefore evidently not a parametrization which reflects the true underlying process.

As the optical properties are assumed to be governed by impurities, several other modifications acting on the absorption and scattering coefficients and lengths have been explored. The physical plausibility of such models is discussed in section 10.7.

### 10.6.1 *Ellipsoid parametrization*

A direction dependent modulation of the absorption or scattering can be implemented as either a modification of the coefficients or as a modification of the absorption / scattering lengths. The correct choice is not obvious.

In either case the direction dependence is introduced through the local radius of a three-dimensional body:

$$a/b = a_0/b_0 \cdot \frac{r(\phi,\theta)}{r_{average}} \qquad l = l_0 \cdot \frac{r(\phi,\theta)}{r_{average}} \tag{10.7}$$

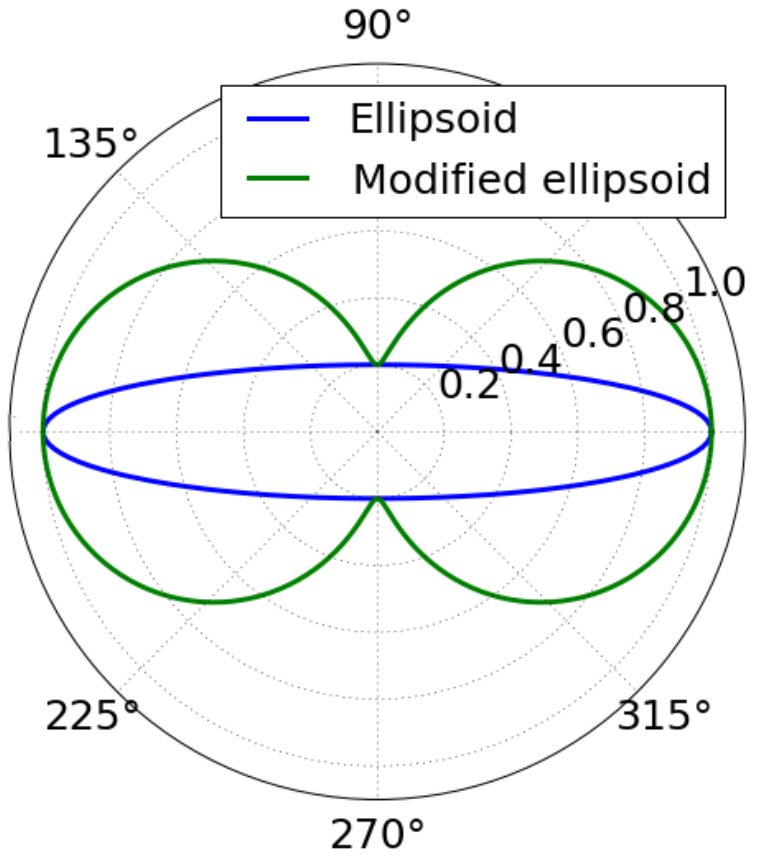

Figure 10.9: Sketch of the local radius of a standard ellipsoid and the modified parametrization, which was chosen to invert the curvature.

As motivated in the following section, an ellipsoid is a natural choice for the absorption length. For an ellipsoid with the axis $a$ aligned with the flow and in the following chosen to be 1, the axis $b$ along the tilt and the axis $c$ along the vertical, the radius in a given direction $(\theta, \phi)$ from the origin is given as:

$$r(\phi,\theta) = \frac{a \cdot b \cdot c}{\sqrt{b^2 c^2 \cdot \cos^2(\theta) \cdot \cos^2(\phi) + a^2 (c^2 \cdot \cos^2(\phi) \cdot \sin^2(\theta) + b^2 \sin^2(\phi))}}. \tag{10.8}$$

To test the choice of the geometric body, lets define a rather arbitrary, alternative body via its angle dependent radius as:

$$r(\phi,\theta) = \sqrt{(\sin(\theta) \cdot \cos(\phi))^2 + ((1-\alpha_\phi) \cdot \sin(\theta) \cdot \sin(\phi))^2 + ((1-\alpha_\theta) \cdot \cos(\theta))^2}. \tag{10.9}$$

This body is from now on called "modified-ellipsoid". It follows the same symmetry around the flow axis, but inverts the curvature of the body, as seen in Figure 10.9. $\alpha_\phi$ and $\alpha_\theta$ parametrize the azimuth and zenith anisotropy strength.

### 10.6.1.1 *In-time renormalization*

Just as with the scattering function based parametrization, it has to be assured that the anisotropy does not change the well-established, direction averaged bulk ice properties. Otherwise, larger anisotropy strength may be penalized in the likelihood fit as they change the detector average properties, although they may give a good description of the directionality. As a result, the above introduced $r_{average}$ has to be chosen appropriately. If the flashers would result in photons on average propagating uniformly in all directions[381] $r_{average}$, could simply be calculated as

[381] *as usually assumed*

$$r_{average} = \int_0^{2\pi} \int_0^{\pi} r(\theta, \phi) \cdot \sin(\theta) d\theta d\phi / 4\pi. \tag{10.10}$$

But, as the bulk ice was fitted to data from the horizontal LEDs and the absorption length is always only a small multiple of the effective scattering length, the overall directions sampled by the photons are not uniform, as evident from Figure 10.10. The averaging thus has to be done over the actual photon directions as sampled and averaged over in the bulk ice fit.

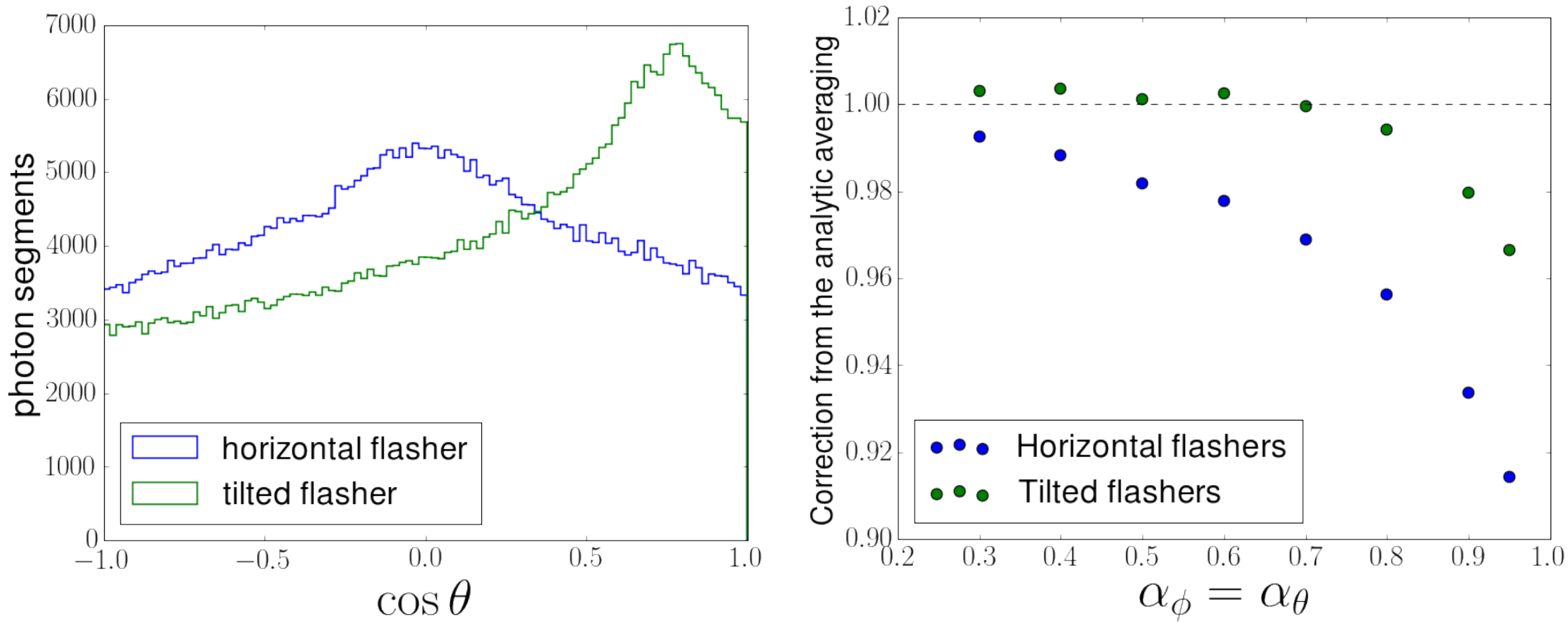

Figure 10.10: Left: Histogram of zenith directions of all photon propagation segments encountered during flasher simulations. As the fully diffuse limit is never reached, the initial propagation direction remains evident.
Right: Correction factor when performing in time renormalization compared to the analytic expectation.

The correct average can not easily be predicted and is instead obtained through an initialization phase in the photon propagation. In this phase the average radius is calculated as a running average of the ellipsoid radii along the propagated photon directions.

As the averaging process breaks the GPU parallelization and results in a massive performance decrease, it is only performed for the first couple of hundred thousand photons and then the sufficiently $r_{average}$ is applied for the rest of the simulation.

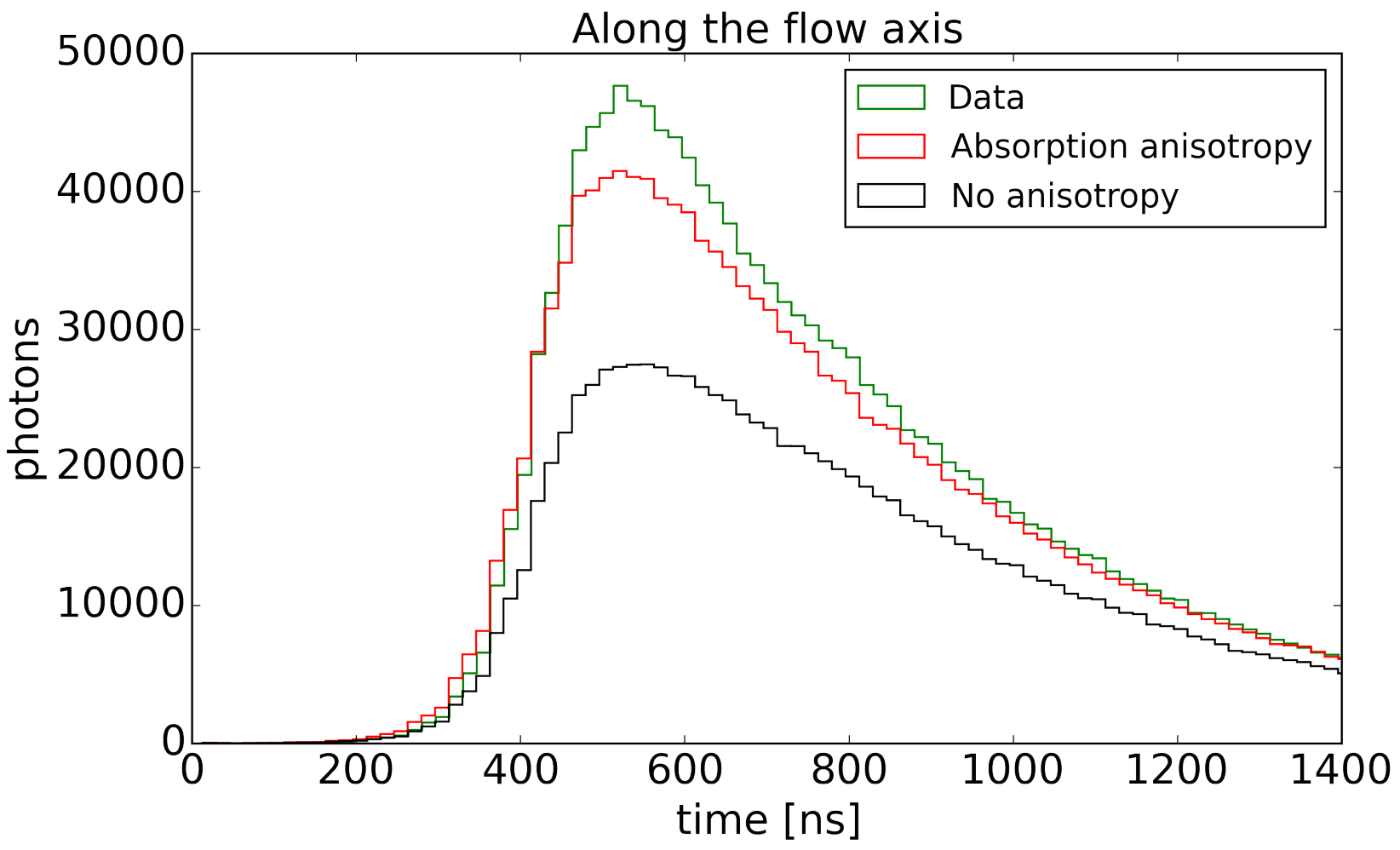

Figure 10.11: Average light curve along the flow axis for the absorption only anisotropy model.

Fitting with this so called in-time renormalization avoids an artificial penalty on larger anisotropy strengths.

## 10.6.2 *Fit results*

Assuming an absorption-only anisotropy, as suggested in section 10.5, the ellipsoid and the modified-ellipsoid have been both fitted as a modification to both the absorption lengths and the absorption coefficients.

The modified-ellipsoid applied to the absorption coefficients does not converge with the major axis trending towards infinity. Equally, the standard ellipsoid applied to the absorption lengths does not converge with the minor axes trending negative.

However, the modified-ellipsoid applied to the absorption length as well as the standard ellipsoid applied to the coefficients converge. The best fit points for the modified-ellipsoid are $\alpha_\phi \approx \alpha_\theta \approx 0.75$ and for the standard ellipsoid $b \approx 4$ and $c \approx 3.5$. Both models indicate a factor four difference in the absorption strength between the flow and the tilt axes, where the least absorption takes place in the flow direction. This modulation amplitude is too large to be consistent with glaciological findings, as investigated in section 10.7.

The all-purpose flasher data likelihood of both models is $\sim 3380$ at the best fit position. This is a significant improvement of about 215 likelihood units compared to the best scattering function based anisotropy model and one of the largest bulk ice improvements in recent years.

The anisotropy strength has also been re-fitted under globally varied bulk ice properties, which did not result in an improved fit.

The flow-axis selected light curve as seen in Figure 10.11 still has the same deficiencies as observed previously. Namely, the total charge is still under-predicted and the rising edge occurs a little too early.

### 10.6.3 *Mixture of absorption and scattering based anisotropy*

By applying the scaling to the scattering parameters as well as the absorption parameters, the new parametrizations allow for a continuous mixture between a pure absorption and an absorption and scattering based anisotropy.

The best fit anisotropy strength for the modified-ellipsoid applied to the lengths has been refitted for a number of mixture strengths $\zeta$. $\zeta = 0$ indicates that no scattering anisotropy is introduced, which is identical to the previous section. $\zeta = 1$ means that the same scaling is applied to both the absorption and scattering lengths, while a negative $\zeta$ means that the scattering lengths behave inversely to the absorption lengths, with the largest absorption lengths but smallest scattering lengths along the flow axis.

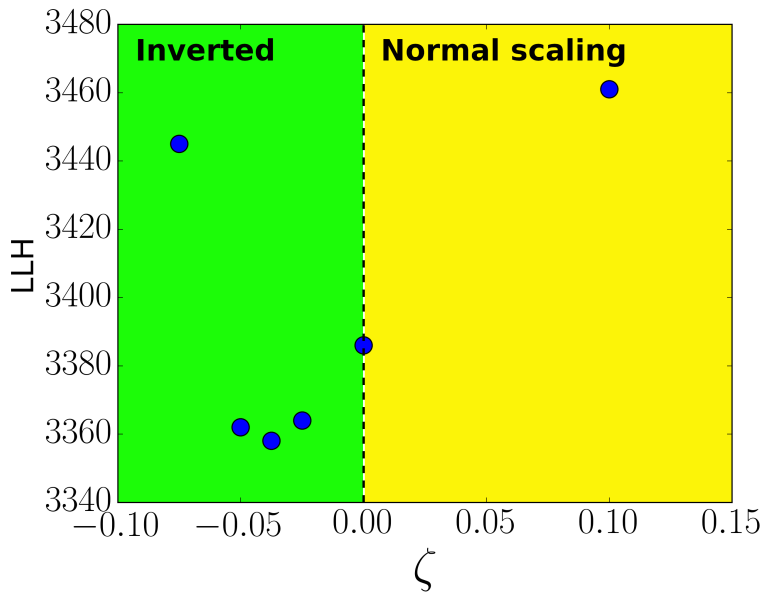

Figure 10.12: Best fit absorption and scattering anisotropy. The preference for a negative $\zeta$ indicates an inverse scaling of the absorption and scattering coefficients but is unphysical given the dust composition.

Figure 10.12 shows the best fit LLH values for a number of $\zeta$ values around a pure absorption anisotropy. Surprisingly the pure absorption anisotropy is not the best fit point. Instead, a scenario with less absorption but slightly more scattering along the flow axis is preferred. More scattering in the flow direction slightly delays the light curve, but also results in less light reaching the sensors. As a result, the required absorption modulation is even larger at $\alpha_{\phi/\theta} \approx 0.825$.

For very specific impurity shapes, Mie theory allows scattering and absorption to be inversely anisotropic[382]. Yet, averaging over several impurity types, shapes and size distributions one always expects a direct correlation between the absorption and scattering strength.

[382] *Shoji Asano (Mar. 1979). "Light scattering properties of spheroidal particles"*

Therefore and due to the large modifications required, while resulting in an overall improved data description, the newly proposed absorption and / or scattering length / coefficient based anisotropy parametrization appear likely unphysical.

## 10.7 PHYSICAL PLAUSIBILITY OF THE COEFFICIENT SCALING

In the Spice Lea paper[383], the authors argued that due to symmetry considerations no anisotropy could be introduced to the coefficients and the anisotropy was applied to the scattering function instead. While not rigorously investigated at that point it was argued that elongated impurities being aligned with the flow could lead to this effect.

[383] *Chirkin, "Evidence of optical anisotropy of the South Pole ice"*

Yet, as laid out in the previous sections, the data is not well described by a scattering anisotropy and is better described by direction dependent absorption coefficients.

This kind of anisotropy can be motivated assuming that impurities are preferentially found on the ice crystal / grain boundaries as introduced in section 3.6.1. Averaging over many grains yields a tri-axial ellipsoid as the average grain shape with the major axis aligned with the flow.

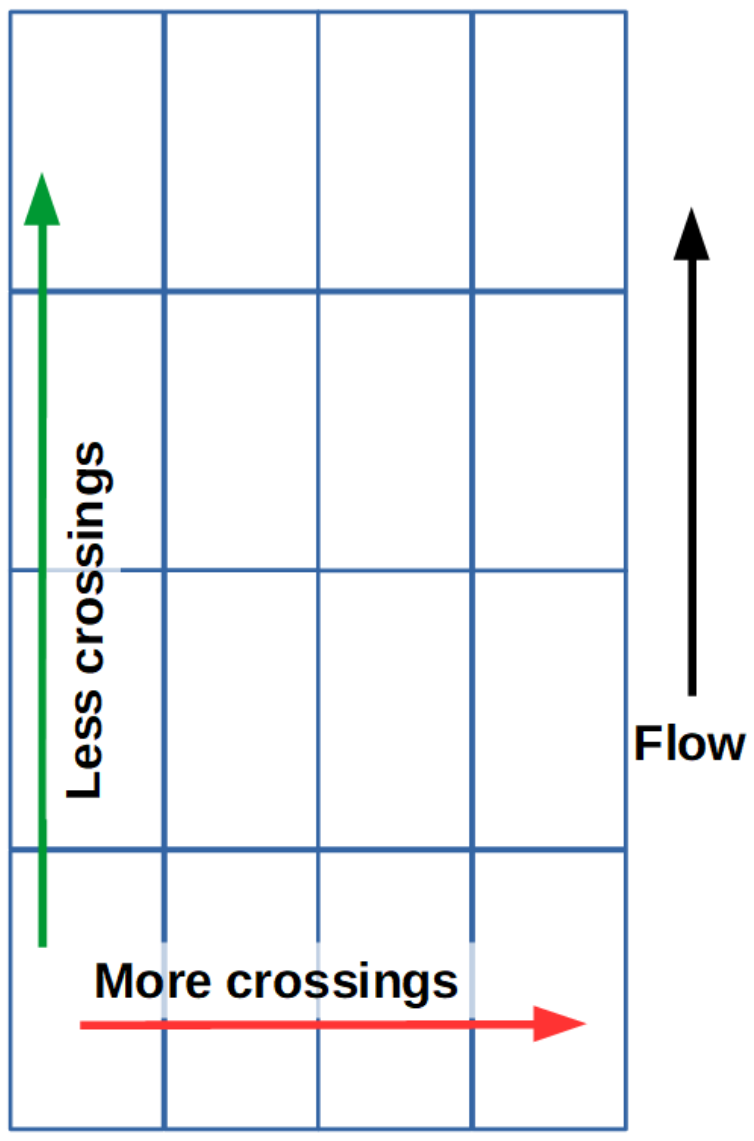

Figure 10.13: Assuming that impurities aggregate on grain boundaries and that the crystal structure is elongated, the number of impurity aggregation regions crossed by a photon depends on the propagation direction.

As sketched in Figure 10.13, this means that a photon propagating along the flow direction sees less grain boundaries and therefore also less impurity regions per unit length compared to a photon propagating along the tilt. As a result, it seems plausible that the absorption length of a photon traveling in a given direction should depend on the local radius of the ellipsoid in that direction.

Yet, the probability $p$ for any scattering or absorption process in an infinite slab $dx$ of material ($p = n \cdot \sigma \cdot dx$) only depends on the direction independent cross section for a single particle interaction $\sigma$ and the particle density $n$. As the density is a scalar quantity, no directional dependence should be introduced even if the density strongly varies in the medium.

This raises the general question, whether an irregular impurity density is at all able to exhibit anisotropic optical properties.

In the following a two-dimensional toy simulation is discussed to study this question. To simplify the modeling the following assumptions are made:

- All impurities are found on a grain boundary.
- All impurities have the same absorption /scattering cross section. (Same material, same size, no elongation.)
- Scattering/ absorption is considered in a geometric optics approximation. An impurity has a given size. When a pho-

ton encounters the impurity a discrete interaction takes place.

- Considering absorption, a photon is deleted with a constant probability on encountering an impurity.

### 10.7.1 *Simulation approach*

#### 10.7.1.1 *Defining the impurity distribution*

The simulation requires a black-and-white input image. It then generates coordinates uniformly distributed over the area of the image. If the image color at the coordinates of each point is black, this point is accepted as the location of an impurity. All other points are rejected. The simulation generates random coordinates until a set number of impurities is reached.

#### 10.7.1.2 *Evaluating the optical properties*

To evaluate the optical properties of the configured medium we consider only propagation along the vertical[384] and horizontal [385] direction.

[384] *picture columns*

[385] *picture rows*

Whenever a photon (which is also simulated as a single pixel) hits an impurity it is assumed to be deflected. For each horizontal row of pixels and vertical column of pixels count the number of impurities. The average distance between impurities, that is the count divided by the pixel size of the column/row, is the average scattering distance for photons traveling along that column or row. Averaging over all columns or rows gives the mean scattering length for horizontally and vertically propagating photons.

Consider one photon per column or row to be originating from the lower edge of the image for vertically propagating photons and from the left edge for horizontally propagating photons. Also assume that the photon is absorbed as soon as it encounters an impurity.

The absorption length for each photon is now given by the distance from the edge to the first impurity in the column/row. In order to obtain an unbiased result, the impurity density in the simulation has to be sufficiently large for each photon to get absorbed along each column and row of the picture.

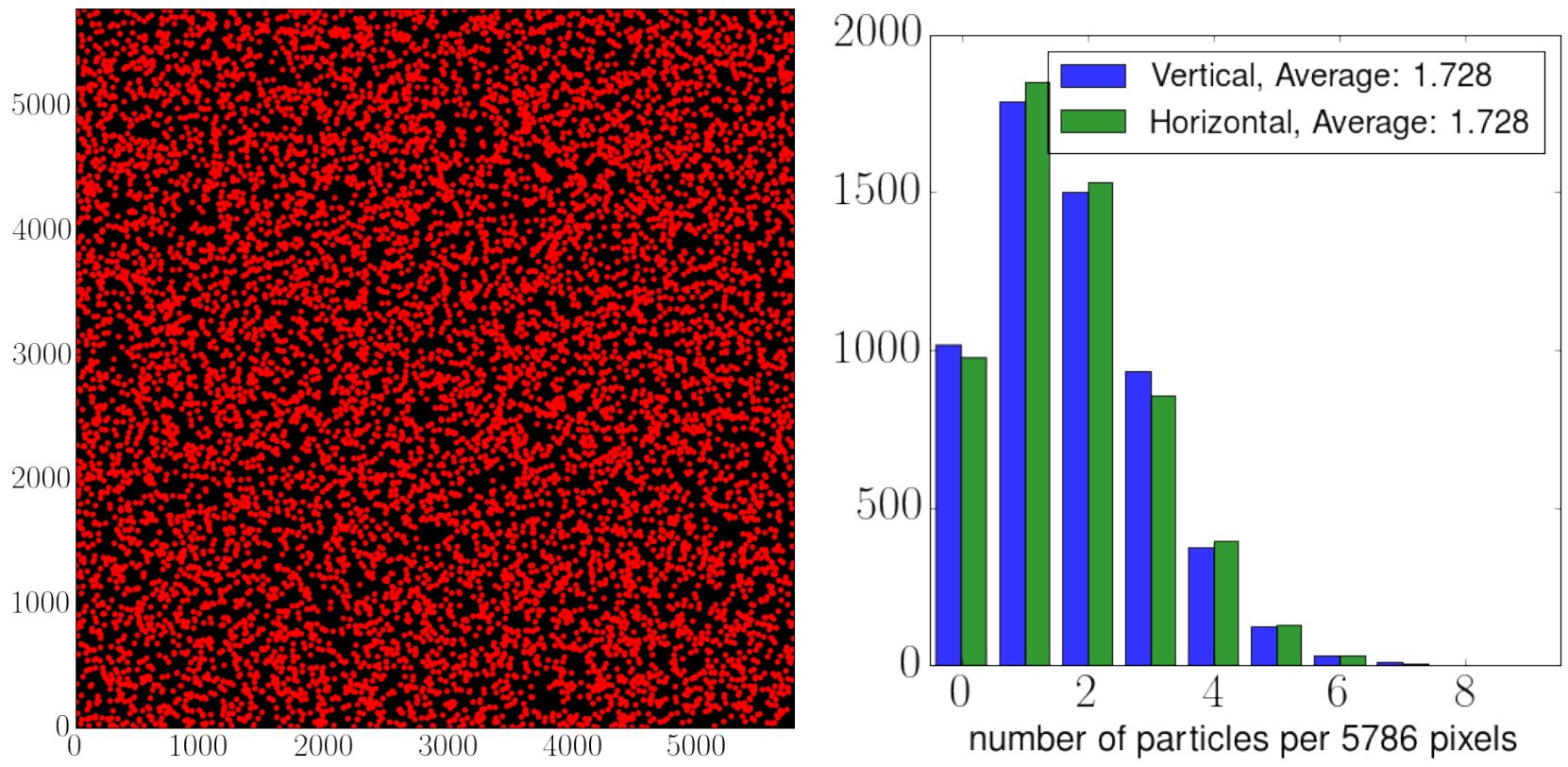

Figure 10.16: Uniform impurity distribution. Left: Impurity map. Each red dot denotes one scattering / absorption centrum. Right: Impurity number distribution for columns and rows.

## 10.7.2 *Isotropic medium*

In order to validate the toy simulation, first consider a medium in which impurities are entirely randomly distributed. A black square is generated as input image for the simulation. The resulting impurity density when generating 10,000 impurities is shown in Figure 10.16.

Counting the number of impurities in each column/row of the picture we obtain distributions as shown in Figure 10.16. Obviously the mean number of impurities is exactly identical. This makes sense as the averages are analytically identical to the surface density of impurities.

The absorption lengths are now evaluated as described above. The resulting distributions for two different impurity densities [386] are shown in Figures 10.14 and 10.15. As expected from the Beer-Lambert law[387] both distributions follow an exponential function. The scenario of higher impurity density also yields a shorter average absorption length, as can be seen from the steeper slope.

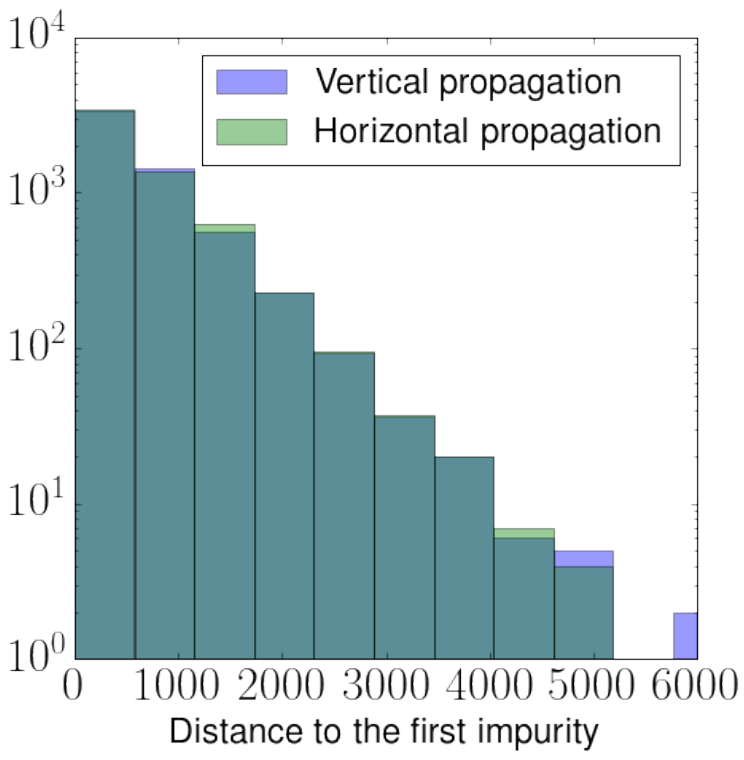

Figure 10.14: Distance to the first impurity, which yields the absorption length, in the case of a low impurity concentration.

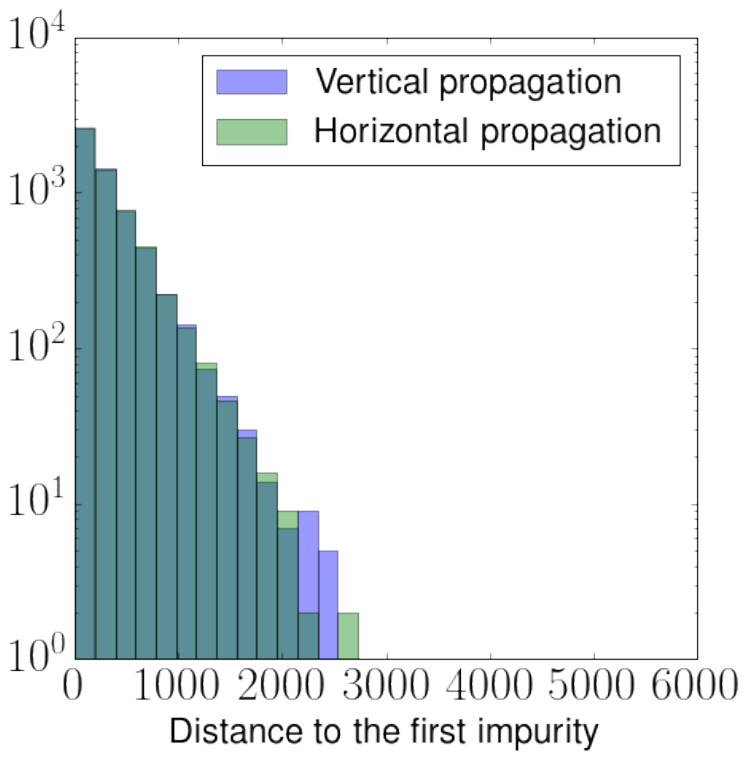

Figure 10.15: Distance to the first impurity, which yields the absorption length, in the case of a high impurity concentration.

## 10.7.3 *Anisotropic medium*

Now consider the originally discussed scenario of a girdle crystal fabric. An example of a measured grain boundary network is shown in Figure 10.17. In principle this picture could be used for the simulation. But as it only contains a few grains and the true elongation is unknown, it will likely generate biased results.

Instead, we generate a close approximation by partitioning a square surface into random polygons. The used Voronoi tessellation algorithm[388] takes a set of input points[389] and for each seed constructs the corresponding region consisting of all points

386 *which both satisfy the requirement that each photon is absorbed*

387 *Beer, "Bestimmung der Absorption des rothen Lichts in farbigen Flüssigkeiten"*

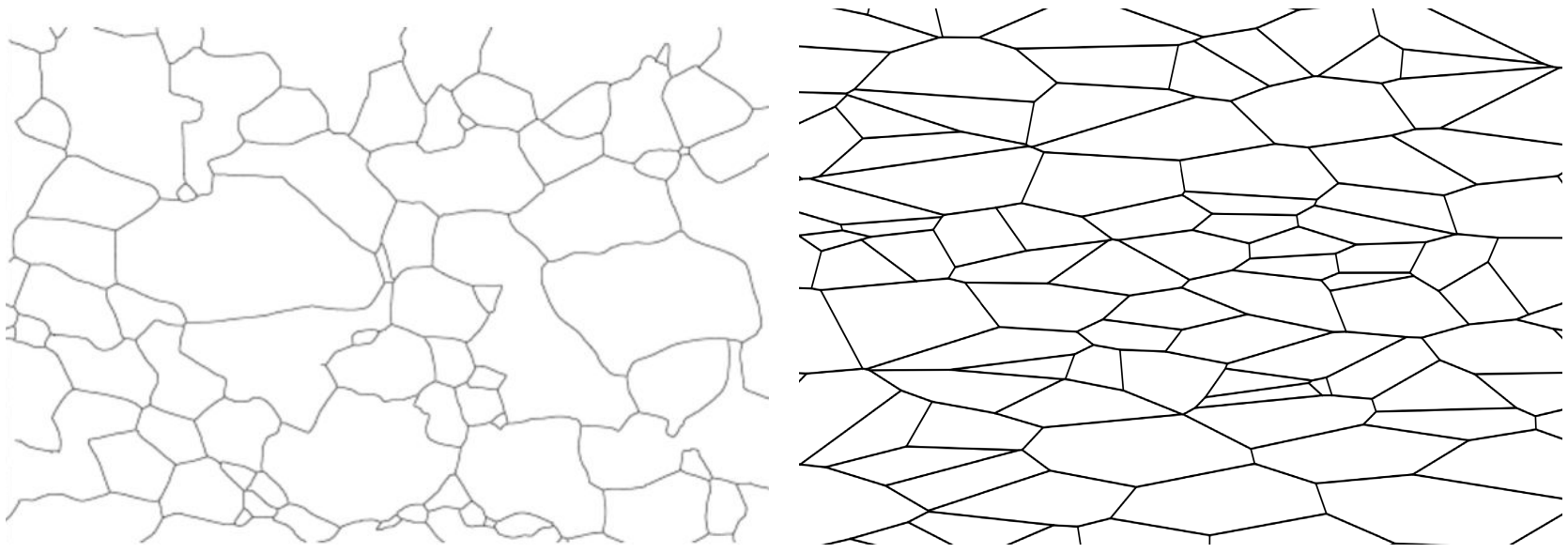

Figure 10.17: Left: Realistic grain boundary network as imaged from an ice core.
[Ilka Weikusat et al. (2017). "Physical analysis of an Antarctic ice core—towards an integration of micro- and macrodynamics of polar ice"]
Right: Simulated grain boundary network using Voronoi tessellation.

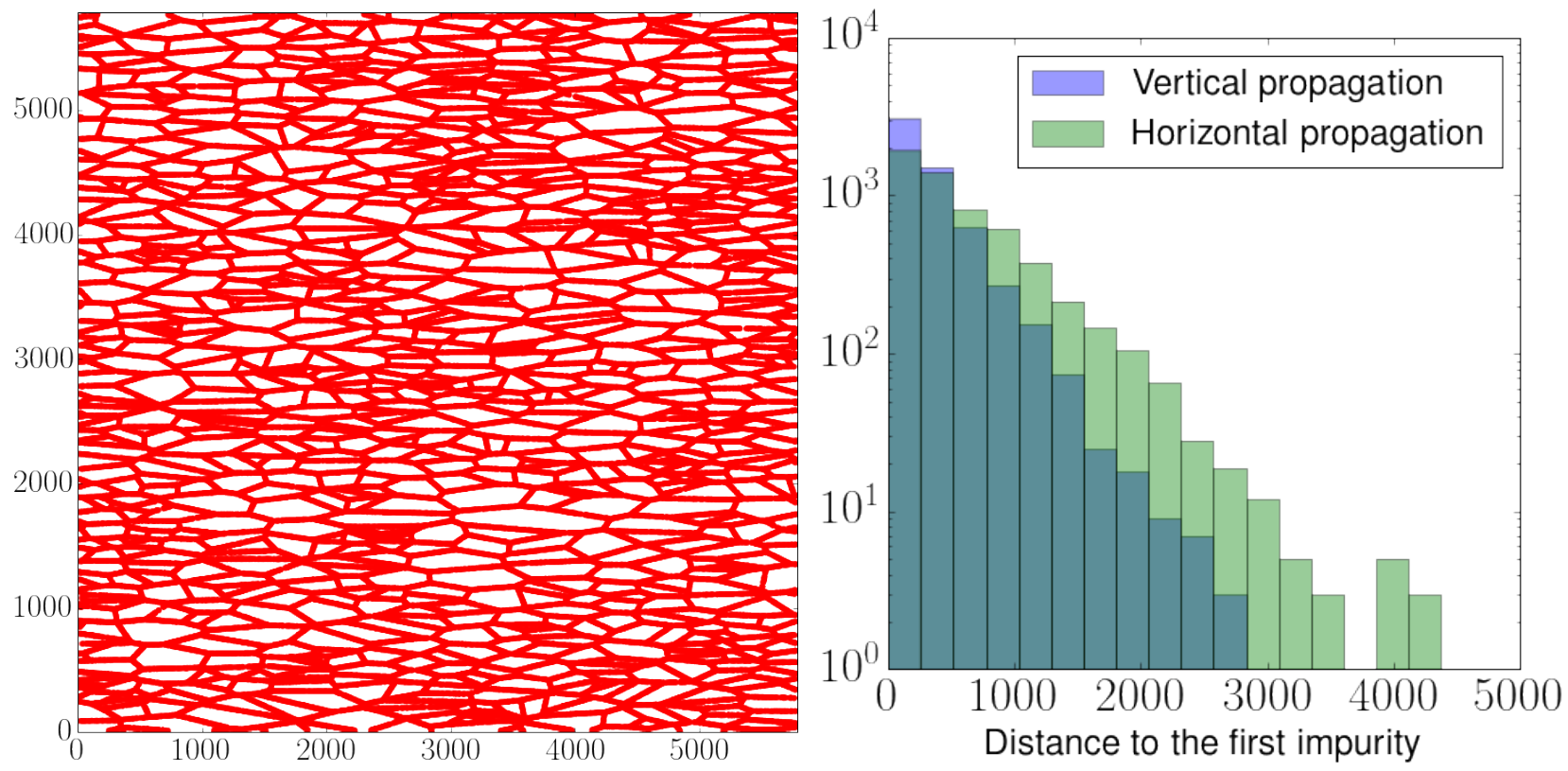

Figure 10.18: Left: Simulation example in a sparse girdle fabric.
Right: Evaluation of the absorption length for a sufficiently dense impurity distribution.

closer to this point than to any other point. This yields an image of polygons with on average equal aspect ratio. In order to approximate elongated grains, this image is stretched along the x-axis and then cropped back to a square.

Figure 10.18 shows the random distribution of 10,000 impurities in the simulated crystal fabric. The mean number of impurities when counted horizontally or vertically is again exactly identical, as it is simply the surface density. The scattering length can therefore never be anisotropic.

To properly evaluate the absorption length in the horizontal and vertical direction we again consider a more densely filled map where each column and row contains at least one impurity. The resulting distributions of the distance to the absorption still show an exponential form for both propagation directions. But the horizontal propagation has a longer absorption length than the vertical propagation. The absorption length depends on the direction of propagation.

### 10.7.4 *Scale considerations*

While the scenario above shows the validity of the general concept of an absorption anisotropy, it does not properly reflect the real length scales. The absorption length in IceCube is on the

388 *Steven Fortune (1997).* Handbook of Discrete and Computational Geometry. *CRC-Press*

389 *which for our application of generated randomly over the surface*

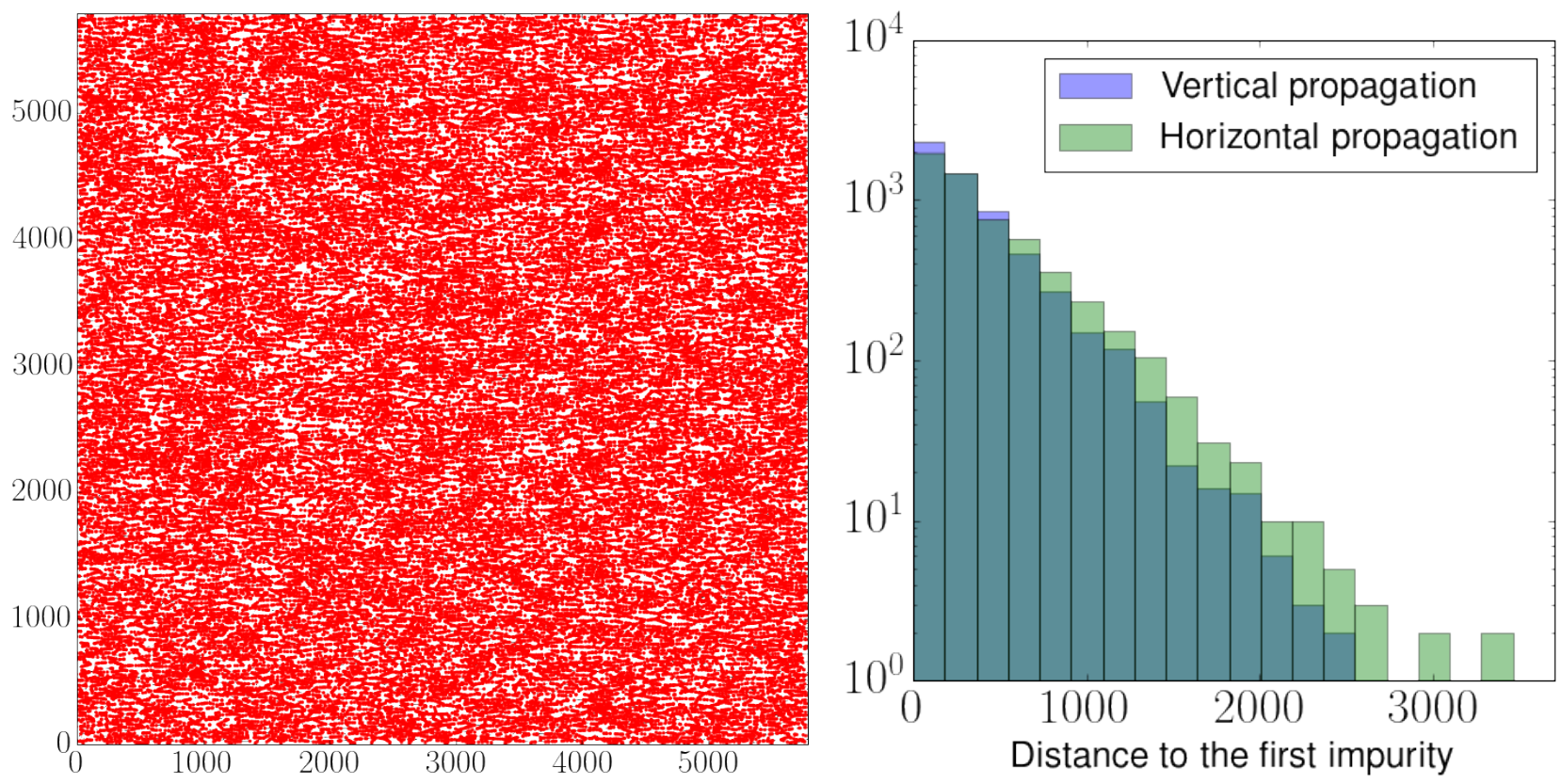

Figure 10.19: Evaluation of the absorption length in a more granular fabrics.

order of 100 m, while the size of impurities is about 1 µm[390].

[390] see section 5.2.1.1

The true grain sizes and elongations are currently unknown, but comparing to other ice cores the diameter is unlikely to exceed 1 cm[391]. Therefore a photon traverses at least $10^4$ grains within one absorption length.

[391] see section 3.6

In the example above all photons are absorbed after about 40 subsequent grains. It stands to argue that the anisotropy diminishes as the grain boundaries become less densely populated and the overall random relative positioning of the grains outweighs the average grain elongation. While a simulation at the true length scales is not feasible using the presented image based framework, the scaling can be tested by increasing the number of grains tenfold while keeping the particle number constant.

Figure 10.19 shows a simulated fabric, as well as the resulting absorption lengths. The anisotropy effect is significantly reduced.

To test the robustness of the toy model two alternative scenarios were tested. The first scenario increase the particle number tenfold, while having only a 10% absorption probability at each particle. The second reduces the image resolution, while keeping the impurity number per column/row identical, such scaling the size of the impurity relative to the grains. Both scenarios did not alter the resulting anisotropy.

### 10.7.5 *Conclusion*

In the considered scenario of impurities being located on the surface of random but on average preferentially elongated polygons, we obtain a direction independent scattering length and a direction dependent absorption length.

While the general concept does seem to be valid, the effect vanishes when applying realistic length scales. Absorption or scattering coefficient based anisotropies are thus deemed unphysical and are not well motivated for use in IceCube.

## 10.8 BIREFRINGENCE ANISOTROPY

So far we have considered impurities to be the only relevant cause for light diffusion in the ice. The presented complications in modeling the optical ice anisotropy through modifications of the Mie absorption or scattering properties have prompted a re-evaluation of this assumption.

Aside impurities, only the ice refractive index can be considered. Luckily glacial ice as a birefringent polycrystal with a girdle c-axis distribution is a complex medium, which might offer the possibility to deflect light.

In the following the progress in understanding the impact of the micro-structure of ice on the light propagation is presented.

---

**Declaration of Pre-released Publications**
The study summarized in this section has already been published as an IceCube internal report[392] and as a proceeding to ICRC 2019[393]. The author of this thesis is a major contributor alongside Dmitry Chirkin and has in particular contributed the original idea, the Zemax simulation and most of the crystallography considerations.

---

392 *Chirkin and Rongen,* Photon propagation through birefringent polycrystals

393 *Chirkin and Rongen, "Light diffusion in birefringent polycrystals and the IceCube ice anisotropy"*

### 10.8.1 *Crystallography context*

Birefringent crystals, such as calcite, are fascinating objects and light propagation through birefringent, polycrystalline materials has been discussed as early as 1955[394].

394 *Raman and Viswanathan, "The theory of the propagation of light in polycrystalline media"*

While the literature agrees that the combined effect of ray splitting on many crystal interfaces will lead to a continuous beam diffusion, the resulting diffusion patterns remain largely unexplored.

### 10.8.2 *Birefringence*

In a homogeneous, transparent, non-magnetic medium the relation between the electric field and the displacement field as well as the magnetic fields is given as[395]:

395 *Bohren and Huffman,* Absorption and Scattering of Light by Small Particles

$$\vec{B} = \vec{H} \tag{10.11}$$

$$\vec{D} = \epsilon \vec{E} \tag{10.12}$$

As the dielectric tensor $\epsilon$ is symmetric, one can always find a coordinate system where it is is orthogonal

$$\epsilon = \begin{pmatrix} n_x^2 & 0 & 0 \\ 0 & n_y^2 & 0 \\ 0 & 0 & n_z^2 \end{pmatrix}, \tag{10.13}$$

with $n_i$ being the refractive indices along the given axes. Uniaxial crystals, such as ice, have two distinct refractive indices: $n_x = n_y \equiv n_o \neq n_z \equiv n_e$. The axis with the unique refractive index defines the optical axis / c-axis.

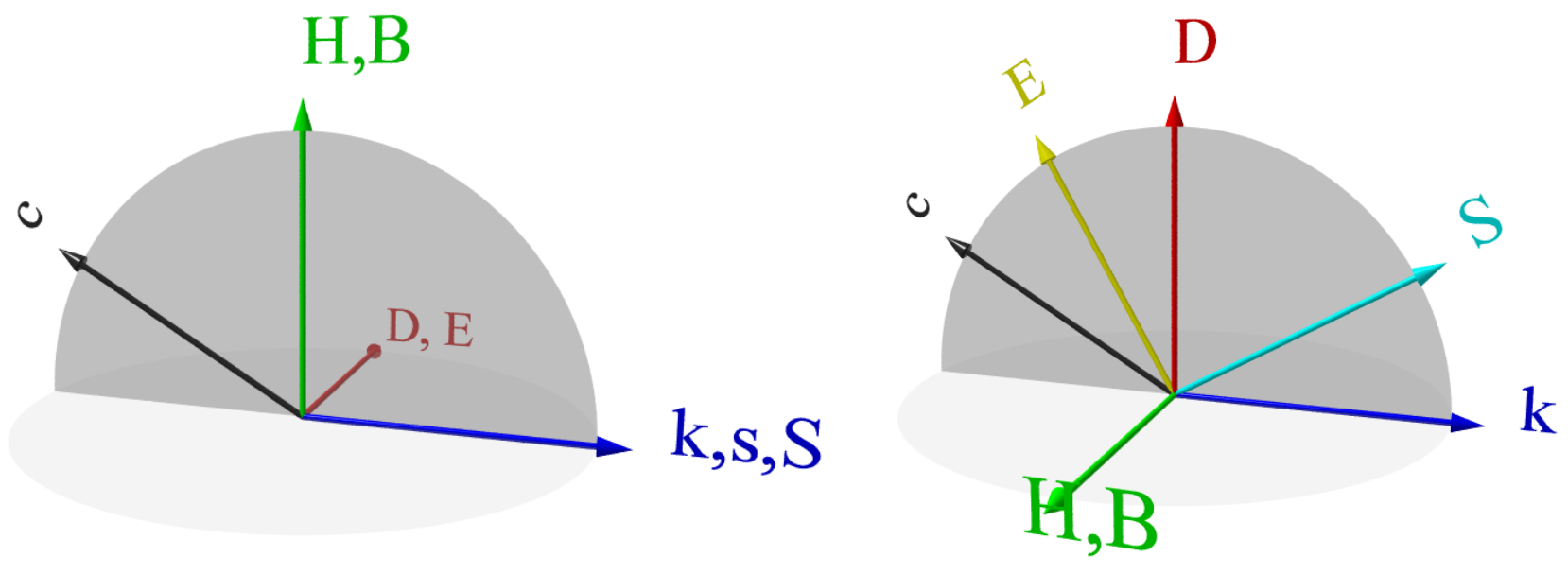

Figure 10.20: Orientation of all electromagnetic-vectors for the ordinary (left) and extraordinary (right) ray with respect to the c-axis [Christoph U. Keller (n.d.). "Crystal Optics." Lecture Notes, Leiden University]

Light propagating in a uniaxial crystal is split into an ordinary wave and an extraordinary wave of orthogonal polarizations. For the ordinary wave, the electric field vector $\vec{E}$ and the displacement vector $\vec{D}$ are always perpendicular to both the optical axis of the crystal and the propagation vectors $\vec{k}$ and $\vec{S}$, which are parallel.

For the extraordinary wave however, the electric field $\vec{E}$ is not, in general, perpendicular to the propagation vector $\vec{k}$. It lies in the plane formed by the propagation vector and the displacement vector. The electric field vectors of extraordinary waves are mutually orthogonal[396]. As the Poynting vector / the energy flow is given by $\vec{S} = \frac{c}{4\pi}\vec{E} \times \vec{H}$, it is for the extraordinary ray not parallel to $\vec{k}$.

[396] *Z. Zhang and H.J. Caulfield (1996). "Reflection and refraction by interfaces of uniaxial crystals"*

While the ordinary ray always propagates with the ordinary refractive index $n_o$, the refractive index of the extraordinary ray depends on the opening angle $\theta$ between the optical axis $\vec{c}$ and the wave vector $\vec{k}$ as given in equation 10.16.

The birefringence strength can be expressed as:

$$\beta = \left(\frac{n_e}{n_o}\right)^2 - 1 \tag{10.14}$$

which for ice is $\beta \approx 2 \cdot 10^{-3}$.

| wavelength $\lambda$ (nm) | $n_o$ | $n_e$ |
|---|---|---|
| 405 | 1.3185 | 1.3200 |
| 436 | 1.3161 | 1.3176 |
| 492 | 1.3128 | 1.3143 |
| 546 | 1.3105 | 1.3119 |
| 624 | 1.3091 | 1.3105 |
| 691 | 1.3067 | 1.3081 |

Table 10.1: Refractive indices of ice taken from [Victor F. Petrenko and Robert W. Whitworth (Jan. 2002). *Physics of Ice*. Oxford University Press]

### 10.8.3 *Zemax study*

As a first simulation attempt, a polycrystal was realized in the industry-standard ray tracing software Zemax OpticStudio[397].

[397] *Zemax, LLC,* Zemax OpticStudio

This was achieved by tessellating a $1\,\text{cm}^3$ cubical volume into 1000 grains using the open-source crystallography software Neper[398], exporting the individual grains as CAD objects and reassembling the object in Zemax.

[398] *Quey et al., "Large-scale 3D random polycrystals for the finite element method: Generation, meshing and remeshing"*

Tracing 1000 rays through 4 independent crystal realizations, each with c-axes aligned in arbitrarily oriented planes, revealed a dependence of the diffusion on the propagation direction. Light propagating in the plane of c-axes, i.e. orthogonal to the flow direction, undergoes the least diffusion. In addition, first hints for a deflection were seen when increasing $\beta$ to $\sim 10^{-2}$.

While qualitatively encouraging, the Zemax simulation does not technically scale to larger crystal sizes, nor does it allow the flexibility required for a thorough evaluation.

### 10.8.4 *Ad-hoc PPC implementation*

Given the idea of a photon deflection, an ad-hoc PPC implementation was realized in order to be able to quantify the typical deflection strength required to account for the observed anisotropy.

In this approach photons are deflected by a constant azimuth angle per meter towards the flow axis and a constant zenith angle per meter towards the horizontal.

It should be noted that this description has been chosen solely due to its simplicity both conceptional and in terms of implementation. Even without a detailed study of the birefringence diffusion, it is known to be un-physical, as the deflection has to vanish on the flow axis, where the optimum is reached, and on the tilt axis, where upstream or downstream deflection are ambiguous. Using this approach, the photon instead ends up oscillating around the flow axis.

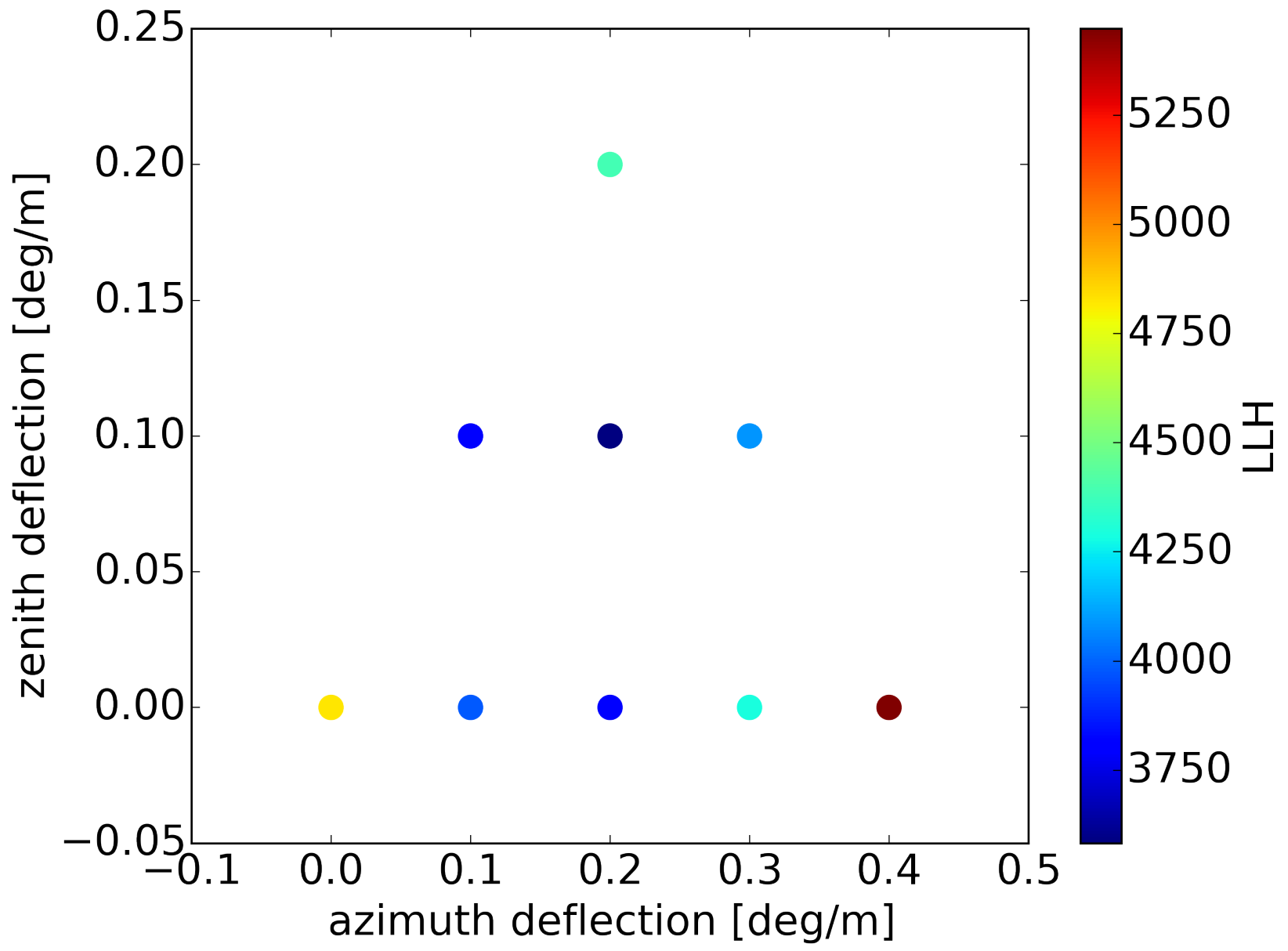

Figure 10.21: Sparse likelihood grid scan for constant azimuth and zenith angular deflection rates towards the flow axis and horizon. The best fit deflections are small enough to have gone unnoticed in laboratory measurements.

Figure 10.21 shows a likelihood grid-scan in the two angular deflection rates. The best description of the all-purpose flasher data is reached for an azimuth deflection of $0.2\,^\circ\,\mathrm{m}^{-1}$ and a zenith deflection of $0.1\,^\circ\,\mathrm{m}^{-1}$. This extremely simple model already describes the all-purpose flasher data just as well as the original scattering function anisotropy.

### 10.8.5 *Analytic calculation*

In order to be able to write ray-tracing software for birefringent polycrystals, first the electrodynamics taking place as a ray undergoes a single grain boundary crossing has to be understood.

Assuming an arbitrary ray incident on a plane interface, we calculate the four possible wave vectors, the ordinary and extraordinary refracted rays and the ordinary and extraordinary reflected rays.

Given the wave vectors, the four associated Poynting vectors, yielding the energy flow and as such probable photon directions, are calculated from the boundary conditions[399].

[399] *L.D. Landau and E.M. Lifshitz (1960).* Electrodynamics of Continuous Media (Course of Theoretical Physics, Vol. 8). *Addison-Wesley*

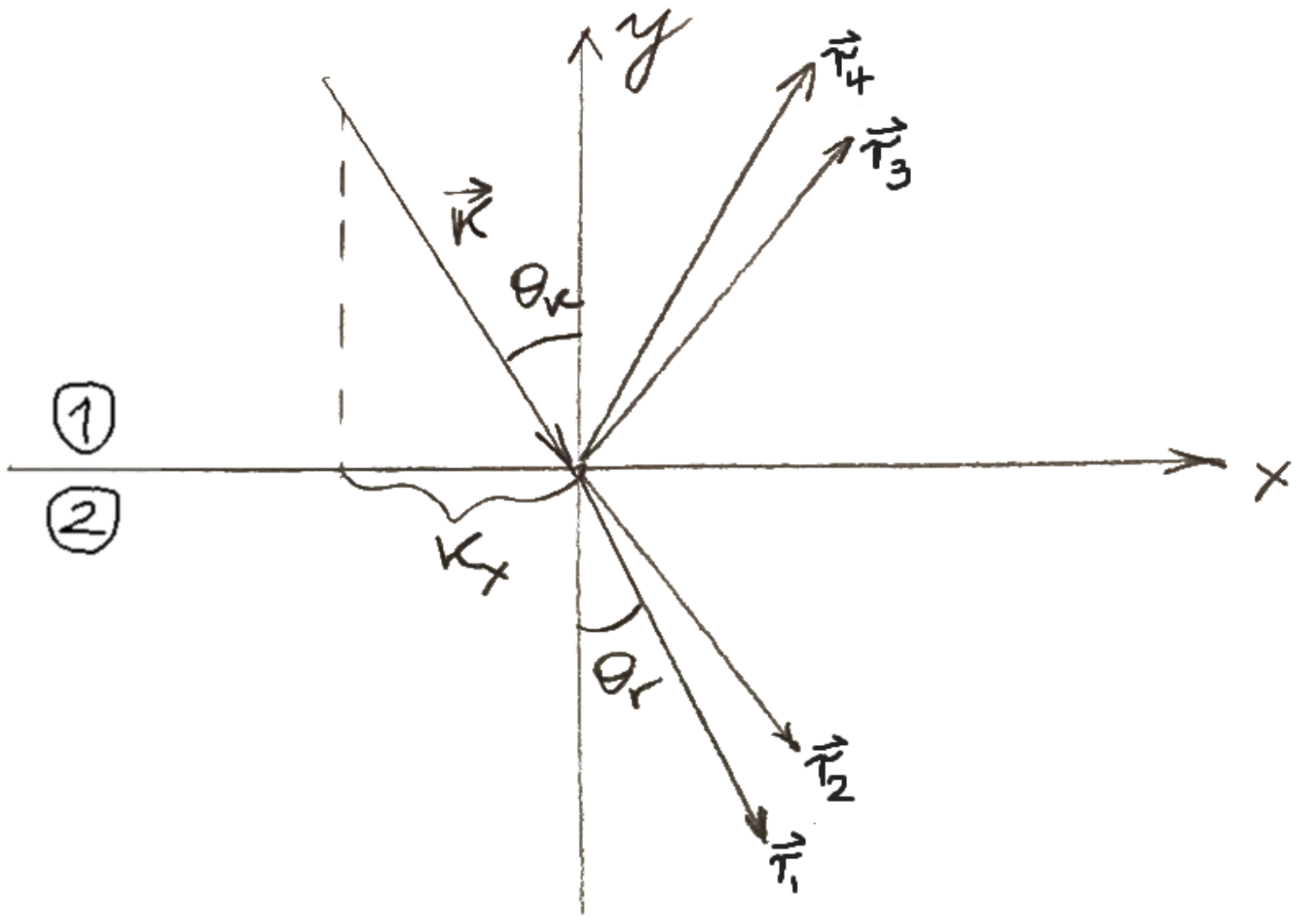

Figure 10.22: Sketch of wave vectors for the incident, reflected and refracted rays. The surface component is always conserved.

### 10.8.5.1 *Wave vectors*

Figure 10.22 shows the situation at hand. An incoming wave vector $\vec{k}$ intersects the interface and is split into four outgoing wave vectors $\vec{r}_i$. The coordinate system can always be chosen such that the surface normal $\vec{n}$ is along the y-axis and that the surface components of $\vec{k}$ and as such $\vec{r}$ are along the x-axis.

Because of translational symmetry of the interface surface, the surface components of all wave vectors are identical: $k_x = r_x$

As the wave number is given by $k = \frac{2\pi}{\lambda}$, we can define a vector $\vec{n}$ such that $\vec{k} = \omega\vec{n}/c$, whose magnitude n is the direction dependent refractive index $n = \sqrt{\epsilon(\theta)}$. Therefore, the magnitude of the wave vector is proportional to the refractive index and we shall simplify $|\vec{k}| = n$ in the following.

#### Outgoing ordinary rays

Given the magnitude $n_o$ and surface component $k_x$ of any ordinary wave vector the y-component is simply calculated as:

$$r_y = \pm\sqrt{n_o^2 - k_x^2} \tag{10.15}$$

Given two media with different refractive indices, one obtains Snell's law for refraction and the usual law for reflection ($k_y = r_y$). The outgoing ordinary ray of an inbound ordinary ray is obviously not deflected, as it does not see a change in refractive index.

### OUTGOING EXTRAORDINARY RAY

Determining $r_y$ for the extraordinary rays follows the same logic, however, using a refractive index which depends on the opening angles $\theta$ between the refracted wave vector $\vec{r} = (r_x, r_y)$ and the optical axis $\vec{a} = (a_x, a_y, a_z)$:

$$\frac{1}{n^2} = \frac{1}{n_e^2} + \left(\frac{1}{n_0^2} - \frac{1}{n_e^2}\right) \cdot \cos(\theta)^2 \tag{10.16}$$

The optical axis is given by the optical axis of medium 1 for the reflected and of medium 2 for the refracted ray. Rewriting the $\cos(\theta)$ as the scalar product between the wave vector and the optical axis gives:

$$\frac{1}{n_e^2} + \left(\frac{1}{n_o^2} - \frac{1}{n_e^2}\right) \cdot \frac{(a_x r_x + a_y r_y)^2}{n^2} - \frac{1}{n^2} = 0 \tag{10.17}$$

Multiplying with $n^2$, $n_e^2$ and using $r_x^2 + r_y^2 = n^2$ yields:

$$r_x^2 + r_y^2 + \beta \cdot (a_x r_y + a_y r_y)^2 = n_e^2 \tag{10.18}$$

This expands into the following quadratic equation in $r_y$:

$$r_y^2 \cdot (1 + \beta a_y^2) + 2\beta a_x a_y r_x r_y + r_x^2 \cdot (1 + \beta a_x^2) - n_e^2 = 0 \tag{10.19}$$

The solution of is given as:

$$r_y = \frac{-\beta a_x a_y r_x \pm \sqrt{D}}{1 + \beta a_y^2} \tag{10.20}$$

with:

$$\begin{aligned} D &= (\beta a_x a_y r_x)^2 - (1 + \beta a_y^2)(r_x^2(1 + \beta a_x^2) - n_e^2) \\ &= n_e^2 \cdot (1 + \beta a_y^2) - r_x^2 \cdot (1 + \beta \cdot (a_x^2 + a_y^2)) \end{aligned} \tag{10.21}$$

From these two solutions the direction appropriate for the reflected or refracted ray is chosen and the other is discarded. In the case of no birefringence ($\beta = 0$) we again obtain the solution for the ordinary ray.

#### 10.8.5.2 *Poynting vectors*

Once the directions of the wave vector are determined, the continuity conditions at the boundary can be written for the normal components of $\vec{D}$ and $\vec{B}$, and for the tangential components of

$\vec{E}$ and $\vec{H}$. If $\vec{n}$ is a normal vector perpendicular to the interface surface, we have:

$$\begin{aligned} \vec{n} \cdot \vec{D}_1 &= \vec{n} \cdot \vec{D}_2 \\ \vec{n} \cdot \vec{B}_1 &= \vec{n} \cdot \vec{B}_2 \\ \vec{n} \times \vec{E}_1 &= \vec{n} \times \vec{E}_2 \\ \vec{n} \times \vec{H}_1 &= \vec{n} \times \vec{H}_2 \end{aligned} \tag{10.22}$$

Here 1 indicates the sum of fields for incident and reflected waves, and 2 indicates the fields of the refracted waves propagating away from the boundary surface into the second medium. Since $\vec{B} = \vec{H}$, two of the eight equations above imply that $\vec{B}_1 = \vec{B}_2$ and $\vec{H}_1 = \vec{H}_2$. Together with the boundary conditions for $\vec{D}$ and $\vec{E}$, we have a system of 6 linear equations. These equations are sufficient to determine the amplitudes of the 4 outgoing waves: two reflected (ordinary and extraordinary), and two refracted (also, ordinary and extraordinary).

Since we only have 4 unknowns, 2 of these equations are necessarily co-linear to the rest, if the wave vectors are determined correctly.

Solving the linear equation system yields the Poynting vectors and as such the energy flux directions / photon directions of the four outgoing rays:

$$\vec{S}_i = \vec{E}_i \times \vec{H}_i \tag{10.23}$$

The relative intensity of these rays as usually denoted in Fresnel coefficients is derived from the Poynting theorem[400], which for our case (no moving charges, no temporal change in total energy) is given as

$$\oiint_{\partial V} \vec{S} \cdot d\vec{A} = 0 \tag{10.24}$$

where $\partial V$ is the boundary of a volume V surrounding the interface. The choice of volume is arbitrary. A simple choice is a box around the interface. In the limit of an infinitely thin but wide box, it is evident that the sum of Poynting vector components normal to the interface plane is conserved.

The outgoing photon is now chosen randomly, with a probability proportional to the four surface-normal components of the Poynting vectors.

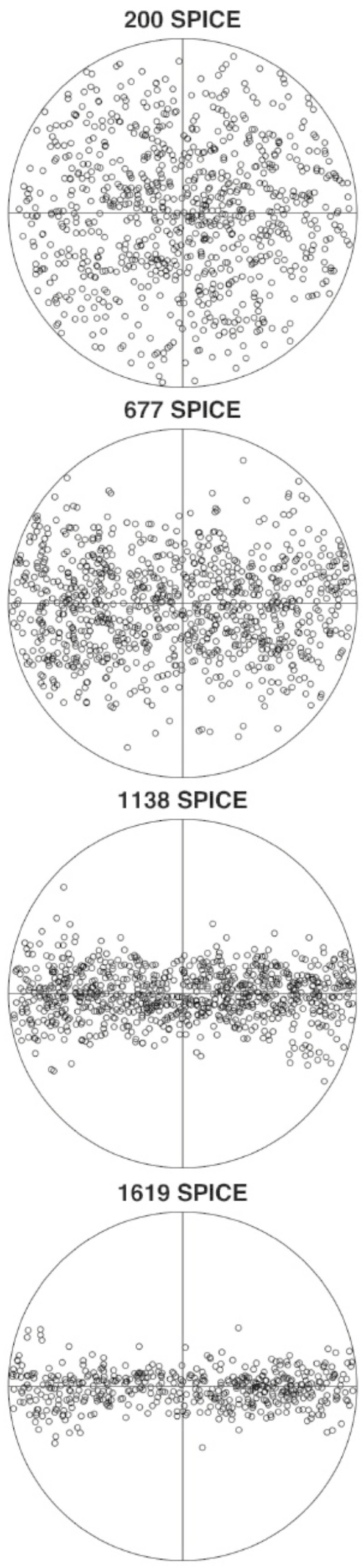

Figure 10.23: SpiceCore c-axis distributions at various depths [Voigt et al., "c-Axis Fabric of the South Pole Ice Core, SPC14"]

[400] *Poynting, "On the transfer of energy in the electromagnetic field"*

Up to three outgoing rays may exhibit imaginary $r_y$, as either $D$ or $n_o^2 - k_x^2$ become negative. This case is conceptually equivalent to total-internal reflection. The resulting waves are evanescent, do not significantly penetrate into the medium and instead have Poynting vectors parallel to the boundary.

After implementing the solution presented here, we came across this paper[401] and found the approach described there to be very similar.

[401] Zhang and Caulfield, "Reflection and refraction by interfaces of uniaxial crystals"

## 10.8.6 *Photon propagator software*

Based on the calculations above, a stand-alone photon propagator for birefringent polycrystals "bfr" was implemented in C++.

The distribution of planes representing the grain boundaries is sampled from the assumed ellipsoid, representing the average grain, as explained in the following section.

The average distance between grain boundaries is given by the average chord length through the assumed ellipsoid given the initial propagation direction. As this direction dependent chord length can in general not be calculated analytically[402], it is approximated through MC sampling. The c-axis fabric is assumed to be a perfect girdle, with all c-axes on the surface of a plane. This is a good approximation of the available SpiceCore data, as seen in Figure 10.23.

[402] W.J.M. de Kruijf and J.L. Kloosterman (Mar. 2003). "On the average chord length in reactor physics"

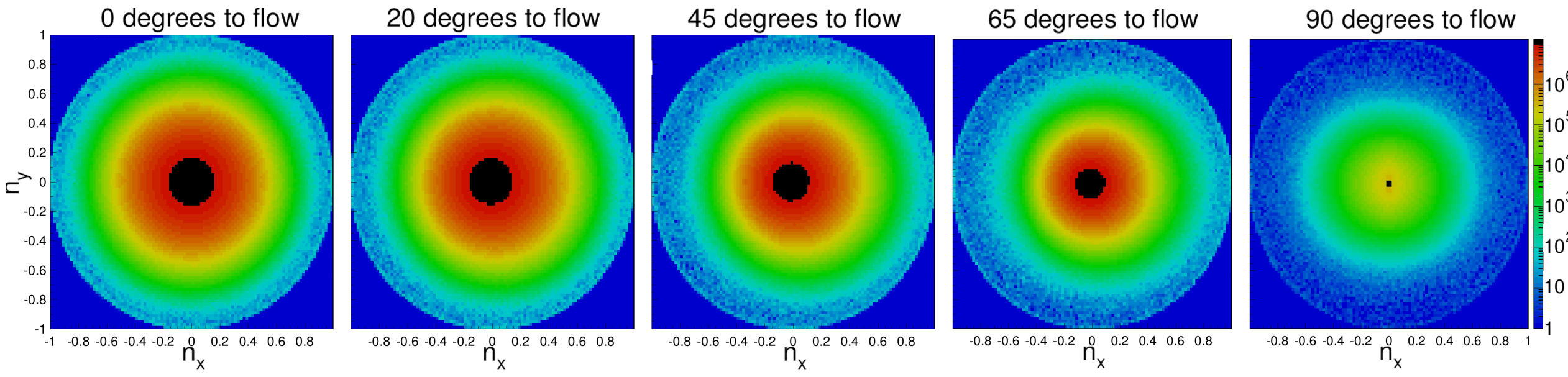

Figure 10.24: Example diffusion patterns after 1000 boundary crossings, assuming spherical grains and a perfect girdle. Photons are always injected along the z-axis. The normal vector of a perfect girdle rotates from being aligned with the photon direction in the left most plot to being orthogonal to the photon direction in the right most plot. The diffusion size is seen to reduce as photons approach the tilt direction. For intermediate angles an asymmetric distribution with a non-zero mean is observed.

The resulting diffusion patterns after 1000 boundary crossings for a variety of initial propagation directions relative to the flow axis and assuming a spherical average grain are shown in Figure 10.24. The overall diffusion is largest when propagating along the flow direction and gets continuously smaller towards the tilt axis. For intermediate angles the distribution is slightly asymmetric, resulting in a mean deflection as originally postulated.

The averages and standard deviations of the propagation direction components for the scenario in Figure 10.24 are shown in Figure 10.25. The maximum deflection is $\sim 0.1°$. This is on the order of magnitude which is required to describe the in-situ data.

For a generalized ellipsoid the diffusion patterns are not only a function of the opening angle between the initial photon direction and the flow, but depend on the absolute zenith and azimuth orientation of the propagation direction with respect to the flow.

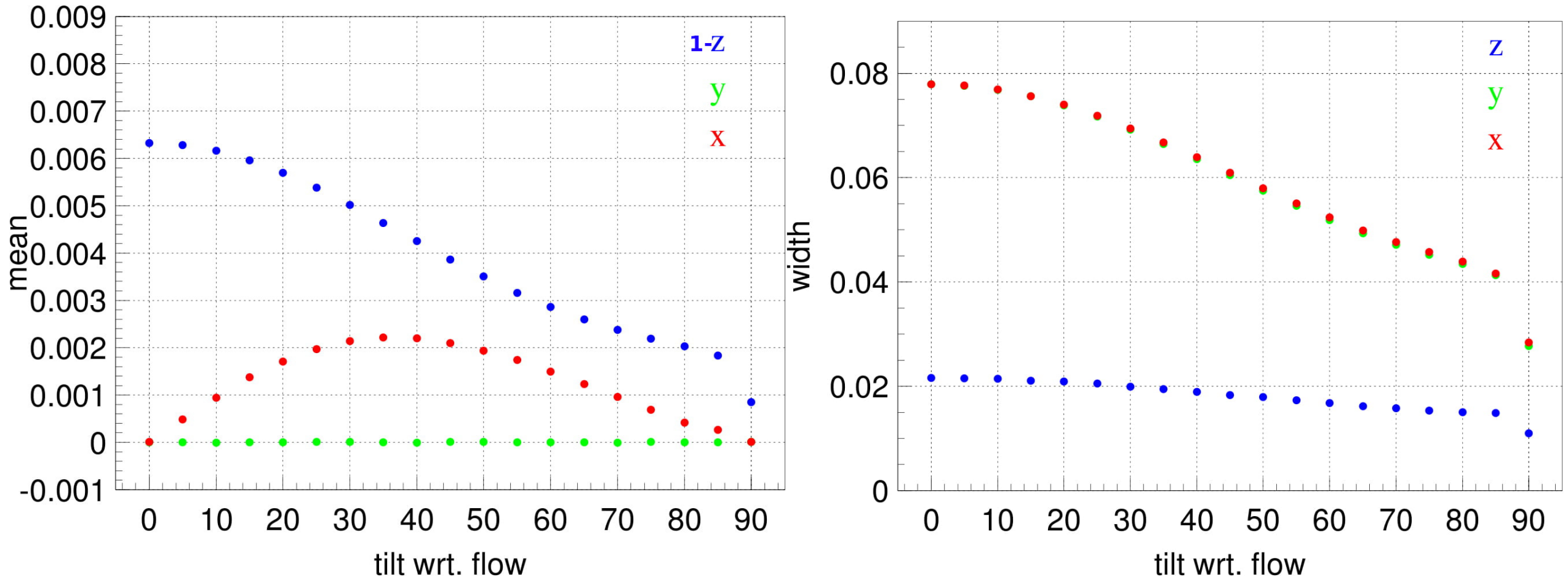

Figure 10.25: Diffusion as a function of the opening angle between the initial photon direction and the flow given on average spherical grains.
Left: Average propagation vector components for photons initially launched along the z-direction after 1000 grain boundary crossing.
Right: Standard deviation of photon vector components.

## 10.8.7 *Sampling surface orientations from an ellipsoid*

As the average grain shape deviates from a sphere, the encountered distribution of face orientations depends on the photon direction.

Assuming that the face orientation of a solid, tessellated into elongated polyhedra, to be described by the surface orientation density of an ellipsoid describing the average grain shape, one can sample the distribution as follows.

The surface of an ellipsoid is defined by the equation,

$$f(x', y', z') = \frac{x'^2}{a^2} + \frac{y'^2}{b^2} + \frac{z'^2}{c^2} = 1 \tag{10.25}$$

where $a$, $b$ and $c$ are the dimensions of the major and minor axes. The normal vector on any point of the surface is given by the gradient

$$\nabla f = [2 \cdot \frac{x'}{a^2}, 2 \cdot \frac{y'}{b^2}, 2 \cdot \frac{z'}{c^2}]. \tag{10.26}$$

For a given set of azimuth and zenith angles, the coordinates on a unit sphere $(x, y, z)$ and on the ellipsoid $(x', y', z')$ are given as:

$$\begin{aligned} x = \sin\theta \cdot \cos\phi \quad &\text{and} \quad x' = a \cdot x \\ y = \sin\theta \cdot \sin\phi \quad &\text{and} \quad y' = b \cdot y \\ z = \cos\theta \quad &\text{and} \quad z' = c \cdot z \end{aligned} \tag{10.27}$$

Substituting the ellipsoid surface position into equation 10.26 the surface normal at this position is then:

$$\vec{n} = [\frac{2}{a} \cdot \sin\theta \cdot \cos\phi, \frac{2}{b} \cdot \sin\theta \cdot \sin\phi, \frac{2}{c} \cdot \cos\theta] \tag{10.28}$$

One can now sample these gradients with angles chosen to be uniform on a sphere. As the surface density per solid angle of an ellipsoid is different from a sphere, the relative surface density

$$\mu_{(x,y,z)} = ||dS'||/||dS|| = \sqrt{(ac \cdot y)^2 + (ab \cdot z)^2 + (bc \cdot x)^2} \tag{10.29}$$

has to be applied as a weighting factor. Where the maximum weighting factor is given as

$$\mu_{\text{max}} = \max(ac, ab, bc). \tag{10.30}$$

Instead of weighting, one can also employ a rejection sampling with an acceptance probability of $\mu/\mu_{\text{max}}$.

In addition to the distribution of face orientations, the distribution of face orientations actually encountered by a photon can be obtained by weighting the distribution of face orientations with the scalar product of the photon's propagation vector with each face normal vector. The probability to encounter a given plane is therefore simply the projected area relative to the incident light.

Figure 10.26 shows the $\cos(\theta)$ distribution of (encountered) face normal vectors for an ellipsoid with unity minor axes and a major axes of two along the z-axis. The distribution is compared to a crystal-like Voronoi tessellation generated with Neper and assuming the same mean elongation. Lines have been traced through the tessellation, identifying grain boundary encounters and computing their incidence angles. The distributions are found to be indistinguishable, confirming that the ensemble of polyhedra faces follows the average ellipsoid.

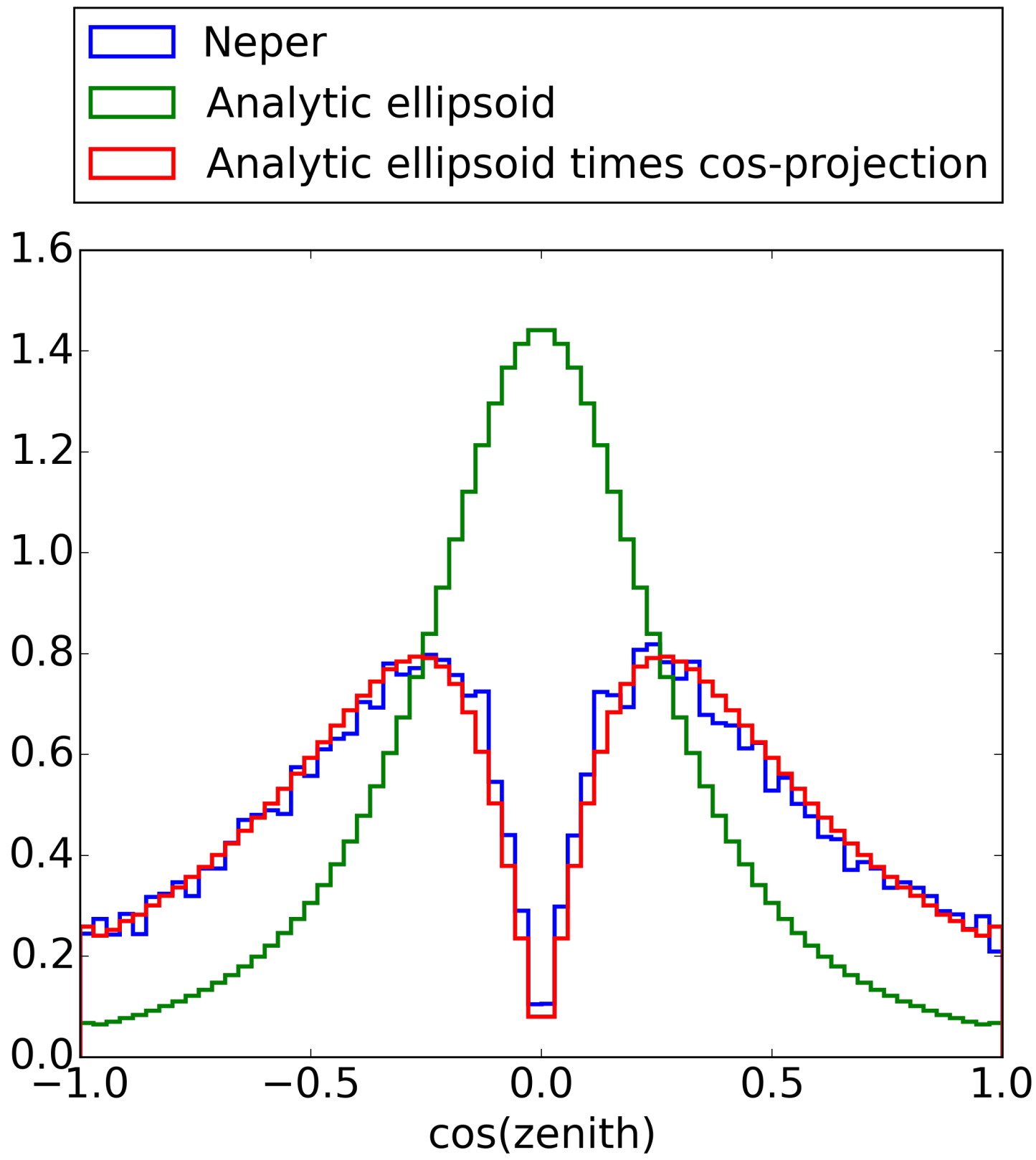

Figure 10.26: Ellipsoid surface sampling for an ellipsoid with unity minor axes and a major axes of two along the z-axis.
Green: Analytic cos($\theta$) distribution of face normal vectors.
Red: Analytic cos($\theta$) distribution weighted by the encounter probability, given by the scalar product with a photon propagating along z.
Blue: Encounter probability as found in a Neper crystal tessellation simulation when tracing photons along vertical lines.

### 10.8.8 *Future ice fit*

The bfr code could be integrated into the PPC photon propagation. This would result in a roughly 3000 times reduced simulation speed and render flasher fits and physics simulations impossible. As a result the birefringence diffusion is instead planned to be implemented as an effective parametrization, where diffusion templates are pre-computed for each photon direction with regard to the flow, given assumed c-axis distributions and ellipsoid shapes.

These templates are then applied at regular intervals, for example at each Mie scattering position, after scaling to the number of grains traversed for that distance. As the broadening is a statistical process, it scales as the square-root of distance, while the deflection has been shown to scale linearly with distance.

Given a perfect knowledge of the ice, the model has no free parameters. While the c-axis distribution is known, and can probably be simplified to be a perfect girdle, the average grain shape and size are currently not known from SpiceCore. These parameters will most likely have to be fitted from the flasher data.

## 10.9 DUST LOGGER ANISOTROPY

The optical ice anisotropy is not only observed in IceCube detector data but has recently also been studied using the dust logger[403]. For this purpose an optional extension consisting of an Applied Physics Systems Model 547 Directional Sensor[404], with an azimuth resolution of about 1°, has been fitted to the top of the logger.

[403] *see section 4.5.3 for details on the device*

[404] *Applied Physics Systems (2018).* Model 547 Directional Sensor

This configuration has to date been deployed down the SpiceCore hole twice during the 16/17 season using the Intermediate Depth Winch. Due to a limited cable length, it was only able to reach a depth of ∼1575 m of the 1751 m cored.

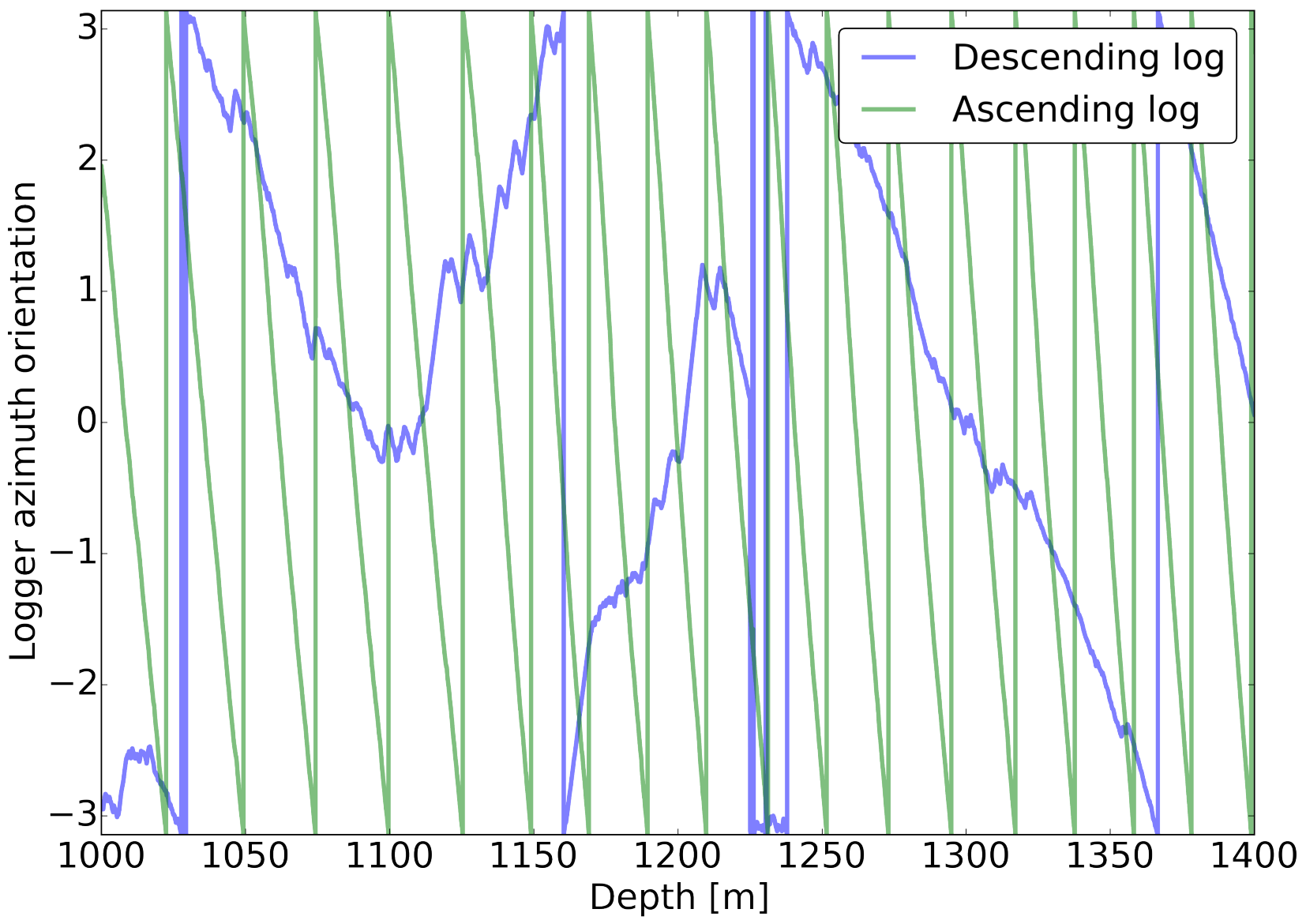

Figure 10.27: Azimuthal orientation during an upwards going and a downwards going log. A smooth and continuous rotation of slightly varying angular velocity is observed when ascending. During descent the rotational movement is rather erratic. This indicates the tool getting stuck and slipping along the wall of drill hole.

A total of four logs (two going down, two going up) are available for analysis. Figure 10.27 shows the rotation as the tool descents and ascents the hole. On ascent, as the cable is pulling the tool up, it undergoes a smooth rotation of slightly varying angular velocity.

On descent the logger sinks under its own weight and the rotation is not continuous, with the logger likely repeatedly getting stuck on the wall and then slipping.

After the logs have been depth aligned to within a couple of centimeters by Ryan Bay, using characteristic features such as volcanic lines, the anisotropy signature can be obtained as the ratio of two logs. The ratios of raw data are usually on average non-unity and show slow continuous variations. These global offsets, most likely induced by differences in the set laser intensity and drift, are corrected using a low order polynomial fitted

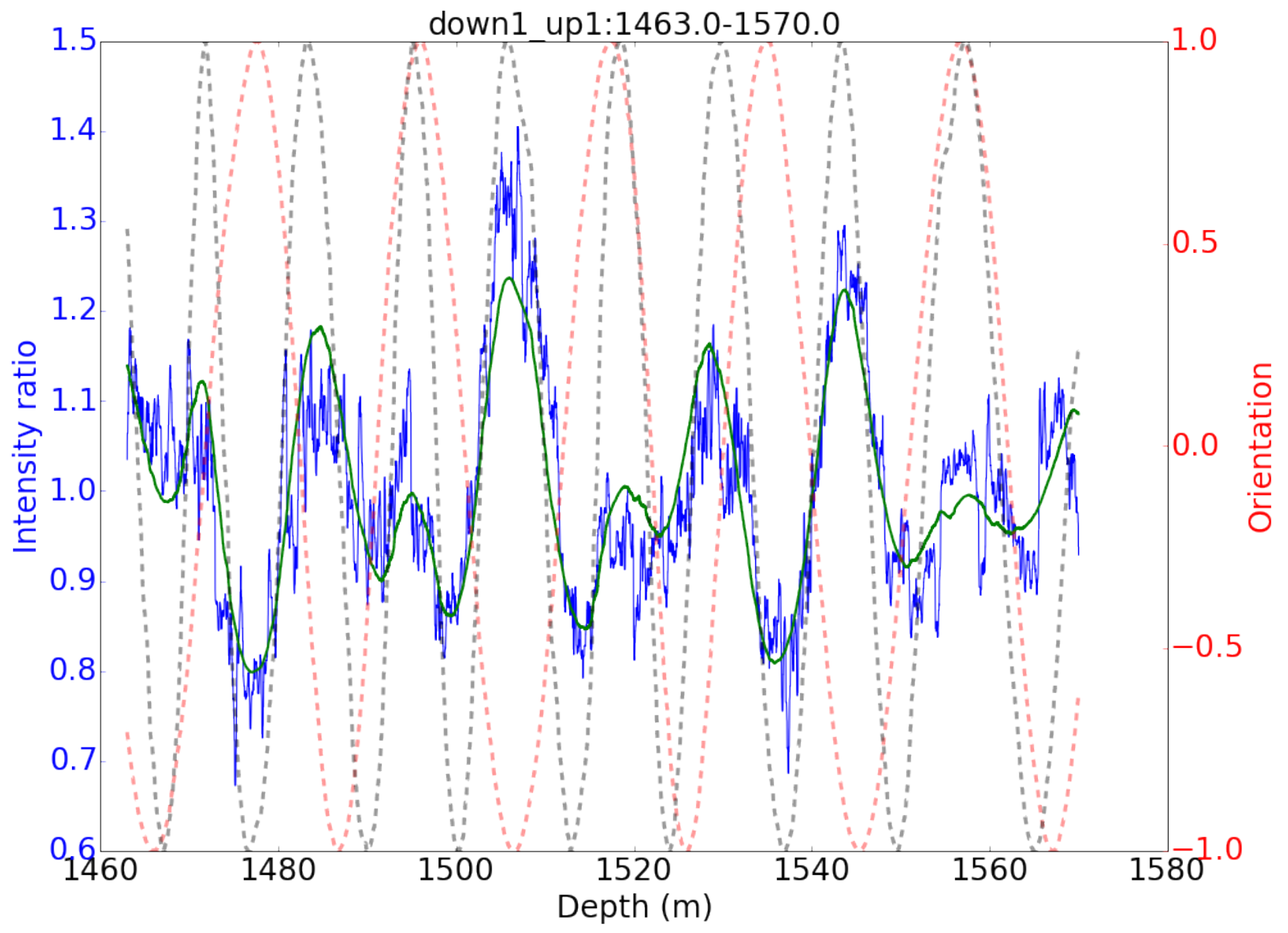

Figure 10.28: Example intensity ratio fit in a ∼100 m depth slice. The red and black dotted lines denote the orientations of the two used logs. (The orientation is defined as the cosine of the azimuth angle.) Blue is the intensity ratio. Green is the fitted intensity ratio. The shape is purely given by the relative orientation. Therefore the amplitude is the only free parameter.

to the ratio.

An example ratio for a 130 m depth slice and after correcting the global offset is given in Figure 10.28. When both logs are in phase, pointing in the same direction at the same depth, the intensities are observed to be equal and thus the ratio is unity. When they are out of phase the ratio becomes as large as $\sim 1.5$. Given this behavior and the observed intensity modulation in IceCube the anisotropy strength $a$ can be fitted using an empirical model given as,

$$ratio = \frac{1 + a \cdot \cos\left(2 \cdot \arccos(orientation1)\right)}{1 + a \cdot \cos\left(2 \cdot \arccos(orientation2)\right)} \tag{10.31}$$

with $orientation1$ and $orientation2$ being the cosine of the azimuth angles of the respective logs. It should be noted that the strength parameter $a$ represents the strength of the anisotropy signal in the data, but is not directly comparable to anisotropy parameters in any IceCube parametrization and is in particular also correlated with other bulk ice properties.

To obtain depth resolution, fits have been performed using ∼100 m depth slices. This retains at least several full modulations per slice, so not to pick up on local fluctuations. For each depth slice six unique ratios of the four logs are available. Though, as the downwards logs are not necessarily trustworthy, only one high quality ratio is available. In addition, one descending and one ascending log were operated above saturation above ∼1050 m, leaving no pair of logs without known problems above that depth.

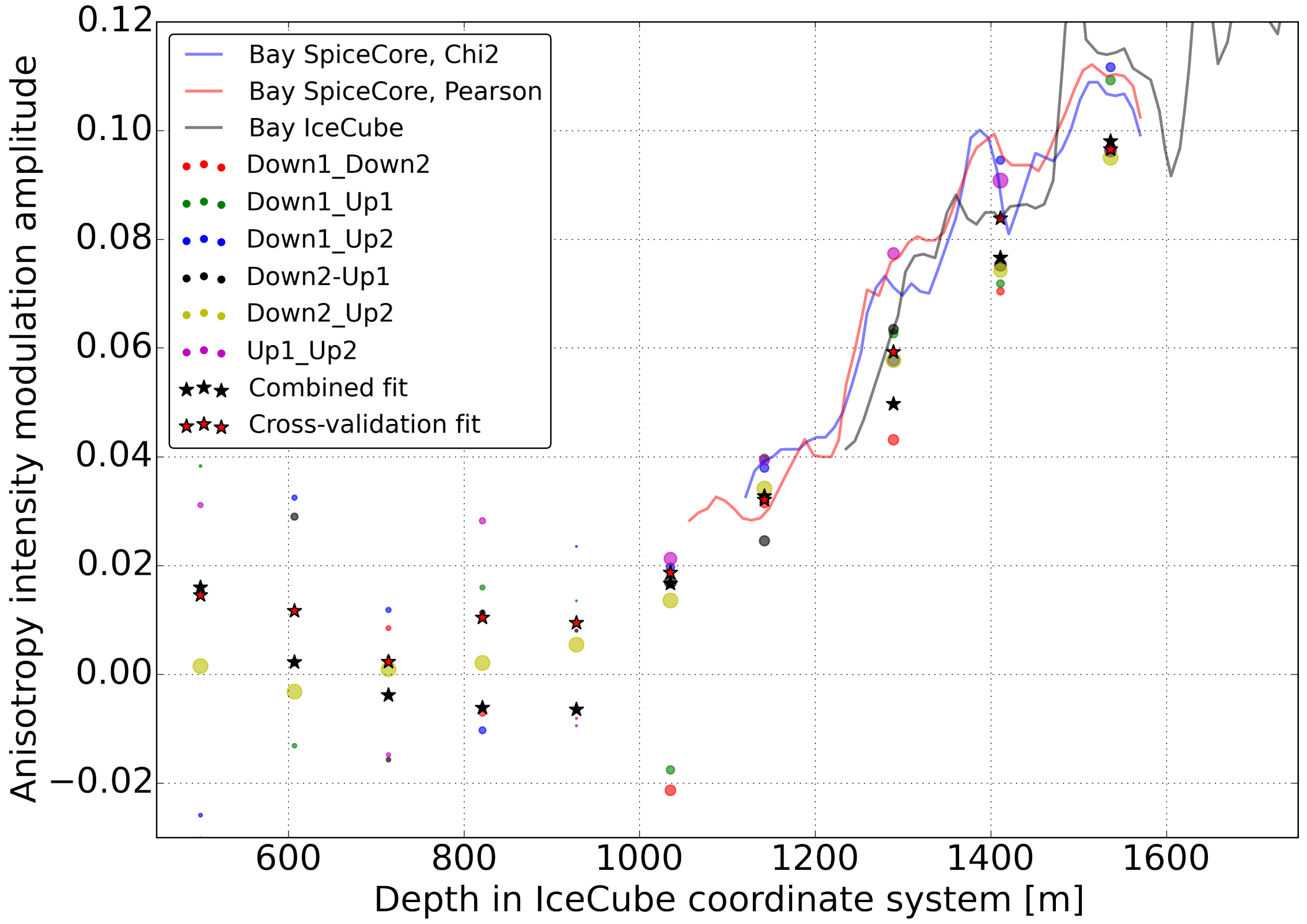

Figure 10.29: Anisotropy signal strength versus depths as fitted from all dust logger ratios and combined fitting methods. Points are scaled such that larger points indicate an overall better data description. The data points derived here are shown in comparison to sliding windows fits performed by Ryan Bay on the same data, as well as by comparing IceCube logs.
[Ryan Bay (2018b). "Private communication"]

While the fits trace the experimental ratios well at large depths, problems are encountered in the shallow ice. For instance anisotropy like oscillatory modulations of the intensity ratio are observed, which do not consistently align with relative orientation of the logs.

In order to obtain a robust and unbiased result, two methods have been suggested:

- Combined fit:
  Minimize the sum of $\chi^2$s of all six ratios at the same time
- Cross-validation fit:
  Use three ratios to fit the anisotropy parameter as described for the combined fit. Then use the fitted anisotropy strength to predict the three remaining ratios. Loop over all 20 combinations of three distinct ratios for fitting. Select the anisotropy strength that has the best predictive power as given by the smallest total $\chi^2$ of the predicted ratios.

A combined overview of the depth dependent anisotropy strength as fitted from dust logger data is seen in Figure 10.29. Below ~1000 m, where the fabric is not yet girdle, no anisotropy signature is seen in the dust logger data. The strength of the signature gradually increases below as the fabric gets stronger (see Figure 10.23). An interpretation of the absolute values in terms of an absolute ice anisotropy strength is not possible as discussed previously.

During the 2018/19 season a number of logs going down the entire extend of the hole will be sampled. The additional data should help distinguish the anisotropy signature from potential artifacts.

The dust logger information is complimentary to flasher data as it is only sensitive to locally back-scattered / reflected light. Therefore, a successful IceCube anisotropy parametrization which can also quantitatively describe the dust logger, likely describes the true underlying process. For such a quantitative description a dust logger detector model still needs to be implemented in the photon propagator software.

The birefringence based concept described in the previous section has the potential to also explain the dust logger data, as the induced diffusion as for example shown in Figure 10.24, introduces an effective and direction dependent probability for a photon to back-scatter.

## 10.10 SUMMARY AND OUTLOOK

The ice anisotropy is currently the most significant bulk ice effect known to be improperly modeled.

The presented studies suggest that the optical ice anisotropy is not caused by impurities and associated Mie scattering but rather a result of the ice micro-structure as a birefringent polycrystal. This introduction of an ice intrinsic diffusion process is a significant departure from the previous understanding that the ice optical properties are essentially only driven by impurities.

A full implementation and ice fitting using the birefringence concepts still requires a significant amount of effort. Given a successful modeling using the birefringence anisotropy, the results are also of interest to glaciology and crystallography as they offer an indirect measurement of the ice fabric and average grain shape and size at scales not accessible from ice cores.

# 11 CONCLUSION AND OUTLOOK

This thesis has explored a variety of topics related to the calibration of the IceCube detector.

The first two chapters have introduced the research field of neutrino astronomy. The third chapter has elaborated on the optical properties of ice. The forth chapter has introduced the IceCube Neutrino observatory. A neutrino telescope that instruments deep glacial ice.

Chapters 5, 6 & 7 explain the detector calibration, related systematics, their impact on physics analyses, as well as the planned and ongoing IceCube detector extensions.

Chapter 8 has presented the design and characterization of an innovative light source combining a very quick drive pulse, delivered by a Williams circuit, with the advantage of the sweep-out technique. It achieves light curves as short as 100 ps, is intensity controllable and can drive LEDs or solid state lasers of nearly arbitrary wavelengths. The design has since found wide use within and outside the IceCube collaboration.

Chapter 9 has presented attempts to resolve longstanding uncertainties surrounding the optical quality of the refrozen IceCube drill holes. While the properties could not be resolved conclusively, the presented studies have significantly narrowed the plausible parameter space. Additionally, and probably more importantly, they have introduced a number of new modeling concepts such as discarding angular acceptance curves in favor of a geometric model of the DOM and have introduced new measurement techniques such as the same-DOM flasher measurements. The resulting SpiceHD model has found application in systematic studies for low-energy neutrino oscillation analysis.

Chapter 10 has presented the first volumetric analysis of the strength of the ice optical anisotropy. It traces the depth profile also expected for the c-axis distribution. In doing so it has also been shown that the current parametrization, which is based on the scattering function, can not fit charge and timing observables of the data simultaneously. In fact no impurity driven absorption or scattering parametrization seems to be fully capable of explaining the observed data.

In departing from the paradigm that the optical properties are necessarily linked to impurities, the ice micro-structure as a birefringent polycrystal with a preferential c-axis distribution has now been identified as a plausible cause of the anisotropy. In photon propagation through individual crystals the idea of an effective deflection has been confirmed and a diffusion as a function of the propagating direction has been found. This concept will hopefully be refined to a point where it can be fitted to flasher data in the near future.

In addition, the new instrumentation and data that will be provided by the IceCube Upgrade as well as new dustlogger deployments will hopefully provide new insides regarding the nature of the ice and will in turn expand the physics capabilities of the detector and our appreciation for its unique properties.

*Where the glacier meets the sky, the land ceases to be earthly,*
*and the earth becomes one with the heavens;*
*no sorrows live there anymore, and therefore joy is not necessary;*
*beauty alone reigns there, beyond all demands.*

— *Halldór Laxness*

# A

# SPE CHARGE TEMPLATES

**To match the data, the IceCube MC needs to simulate all aspects of the detector including especially the PMT characteristics. The distribution of measured charges per detected photon, the so called single photo-electron (SPE) distribution, describes the PMTs gain characteristics.**

**Declaration of Pre-released Publications**
The study summarized in this section is currently being prepared for publication by the IceCube collaboration[405]. The author of this thesis is a mayor contributor alongside Spencer Axani and in particular contributed the fitting algorithm.

405 *Aartsen et al., "In situ calibration of the single photoelectron charge response of the IceCube photomultipliers"*

The SPE distribution, used in simulations up to 2018, was obtained by averaging over 118 PMTs in a 2004 lab measurement performed in Chiba[406]. The PMTs were at that point not integrated in DOMs, but operated at a higher than nominal gain of $3 \cdot 10^7$ using stand-along high voltage supplies. A triggered light source delivered low occupancy light pulses with the charge being digitized using a LeCroy 2249A charge ADC[407].

406 *Abbasi et al., "Calibration and characterization of the IceCube photomultiplier tube"*

407 *LeCroy*, 2249A Charge ADC

The resulting distributions were fitted using an empiric model consisting of a Gaussian for the electronic noise contribution, when no photon was detected, and an SPE distribution consisting of a Gaussian, centered by definition around 1 PE, to describe the nominally amplified pulses and a low-charge exponential to describe under-amplified charges[408]. The averaged distribution is seen in red in the left of Figure A.2.

408 *Abbasi et al., "Calibration and characterization of the IceCube photomultiplier tube"*

While this distribution has been used for all IceCube simulations, its accuracy has been put into question by other lab measurements. In a 2013 measurement performed on fully integrated DOMs at the Madison dark-freezer labs, originally designed for the DOM final acceptance testing, the relative contribution of low charge pulses was found to be significantly higher[409].

409 *Wendt*, DOM SPE Charge Spectrum

An appropriate description of the data could only be reached by introducing a second steeply falling exponential. Electric noise contribution to this low charge region was excluded by identical measurements with a shutter in front of the light source. In addition, the low amplitude pulses were found to follow the same arrival time distribution as the nominal pulses.

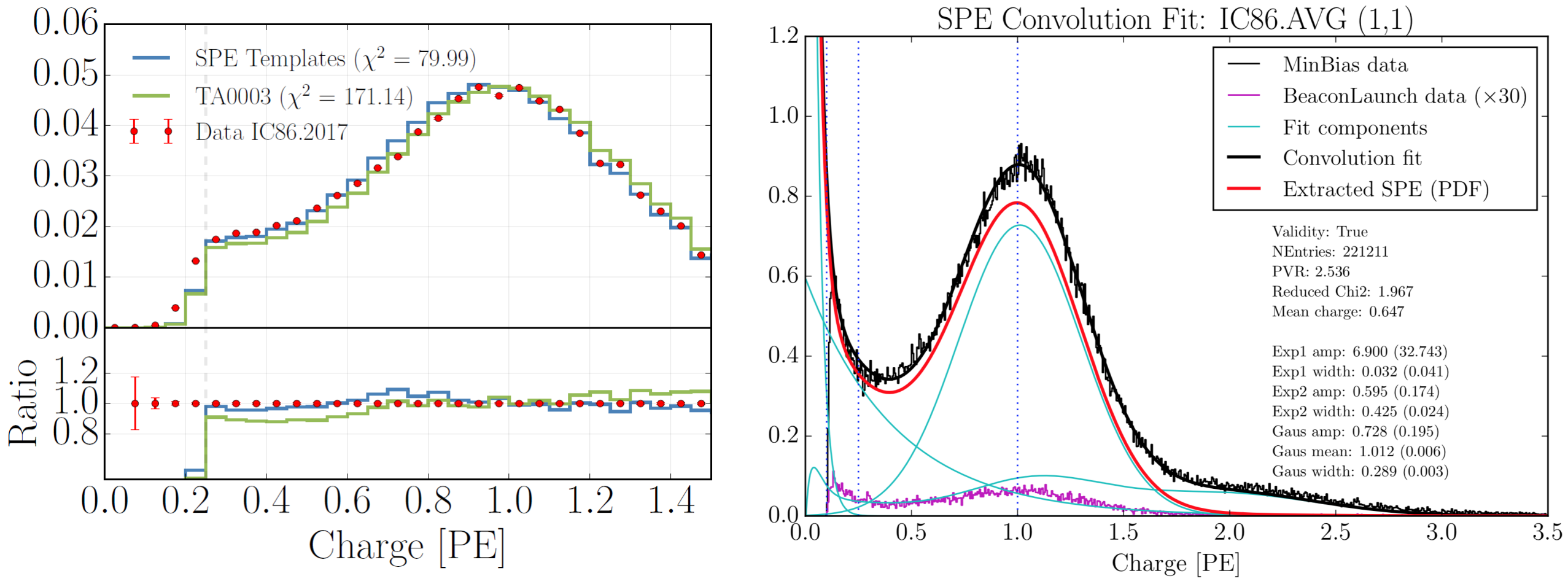

Figure A.1: Left: An analysis level comparison for the TA0003 distribution compared to the SPE Templates and data from IC86.2017; Right: An example fit result for DOM S1D1. The result from the convolutional fitter is shown in black and the components of the fit are shown in cyan. The extracted SPE template is shown in red.

Recently, several high level physics analyses confirmed a data-MC mismatch in the low-charge region around the valley. In the case of the GRECO analysis, this resulted in having to refrain from using charge information and instead using only hit probabilities. These discrepancies triggered an extensive re-calibration campaign, trying to extract per-DOM SPE charge distributions from in-situ data.

As no triggered, low-occupancy light sources are available in-situ, the distributions have to be obtained from noise or low amplitude muon data. Measurements in the disputed region of small charges are further complicated by the hardware trigger requiring at least 0.25 PE[410].

[410] *see section 4.3.1.2*

To overcome the later complication, a special pulse section has been developed, which searches all ATWD waveforms for a clearly separated but completely sampled second PMT pulse, which is of caused not biased by the trigger requirement. Special care about the timing of subsequent pulses has to be taken, not to be biased by the potentially different charge distributions of late pulses.

The pulse selection results in per-DOM charge histograms as for example seen in figure A.1, with charges down to 0.1 PE and sampled with negligible electronic noise contribution.

Also evident is a significant population of high charge pulses, indicating cases where at least two instantaneous photons con-tribute to the measurement. In order to extract an unbiased SPE template the entire distribution must be fitted at once, including the high charge tail.

For this purpose a so called convolutional fitter has been developed. At each likelihood call, with a given set of parameter values for the SPE template, the charge distributions for two or three instantaneous photons is convoluted from the assumed SPE template and the relative contributions of single, double and triple photons to the histogram fitted as nuisance parameters.

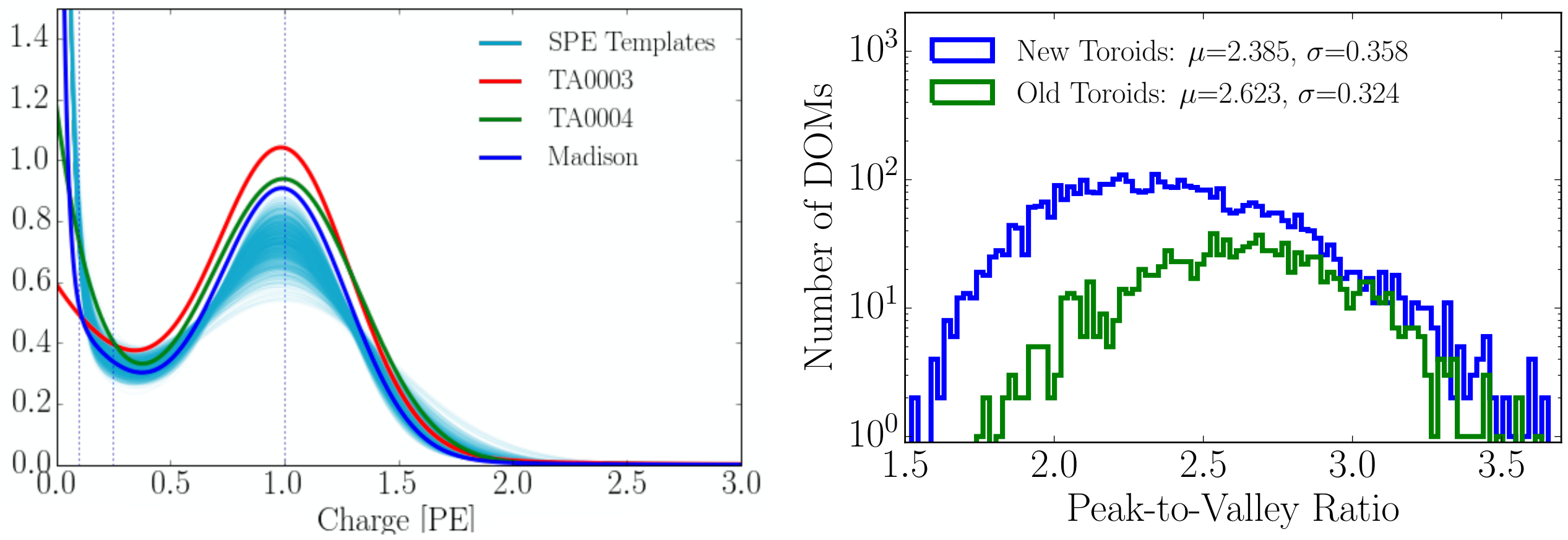

Figure A.2: Left: The normalized charge distributions. The TA0003 distribution is shown in red. The 2013 Madison measurement in blue and the the cumulative SPE templates in cyan; Right: The Peak-to-Valley ratio split between DOMs with new and old AC coupling toroids.

This fit has been applied to all in-ice DOMs and in all but a handful of cases achieves a fit quality similar to the example depicted in Figure A.1.

The spread and overall shape of extracted SPE templates is seen in comparison to the previous measurements in Figure A.2. The high statistics in-situ measurement has also enabled a number of studies investigating the detector stability and homogeneity. No temporal drift of individual DOMs or the detector was found over six years of available data. Yet, splitting the detector by minor hardware differences such as the PMT type or toroid version statistically significant differences in the average SPE template shape are observed.

The new SPE templates will be used in all IceCube simulation starting in 2019.

# B

# EVALUATION OF HIGH VOLTAGE MODULES

**As part of the design process for an updated version of the IceCube DOM, so called pDOM, to be used for the IceCube upgrade, two potential high voltage modules were evaluated.**

The HVM SMHV 0520[411] can provide a positive high voltage of up to 2 kV at 500 µA. It needs to be supplied with 5 V. The output voltage and current limits are set using two analog voltage inputs. It also features analog output monitors for both the achieved high voltage and the supplied current.

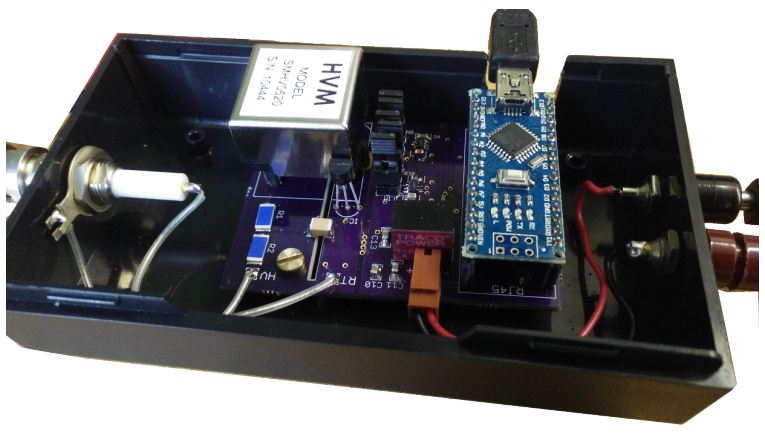

Figure B.1: Custom evaluation board for SHMHV modules

As the module was provided without any of the required peripherals, a custom evaluation board as depicted in figure B.1 has been developed. Two MCP4725 12 bit DACs[412] provide the control voltages required to set the voltage and current limit. An Arduino Nano acts as interface between the DACs and a control PC. It also provides a rough digitization of the monitor outputs and reads temperature data from a DS18S20[413] sensor glued to the case of the HV module.

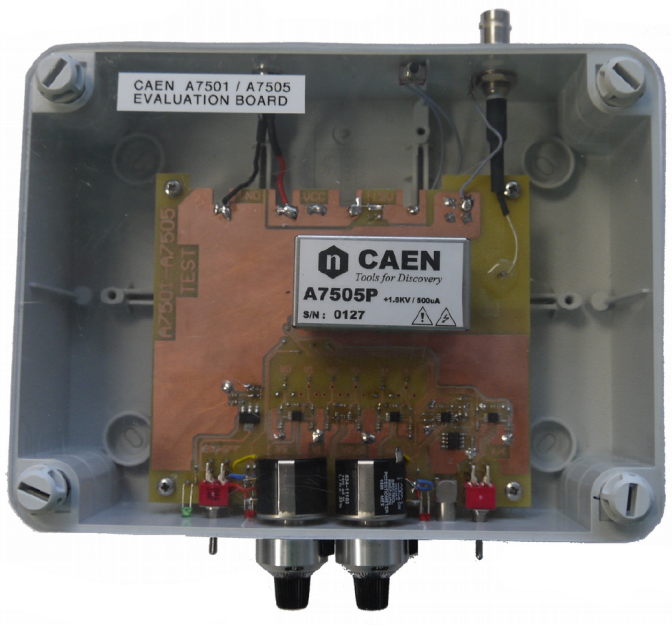

Figure B.2: Evaluation unit as provided by CAEN

The CAEN A7505P[414] is functionally identical, but only delivers a maximum of 1.6 kV and requires a 12 V input voltage. It is cheaper than the SMHV module but only available with a larger form factor. CAEN provided an all-analog evaluation unit as seen in figure B.2. The output voltage and current limit are set manually using precision potentiometers.

To be able to measure the characteristics of the high voltage output at different loads and without having to source dedicated testing equipment, a precision high voltage divider as seen in figure B.3 has been designed. It uses 1% tolerance, 100 ppm/°C

[411] *HVM Tech*, Precision Regulated HV DC/DC Converter

[412] *Microchip*, MCP4725 precision 12bit ADC

[413] *Maxim Integrated*, SD18S20 Digital Temperature Sensor

[414] *CAEN Electronics*, A7505 High Voltage Module

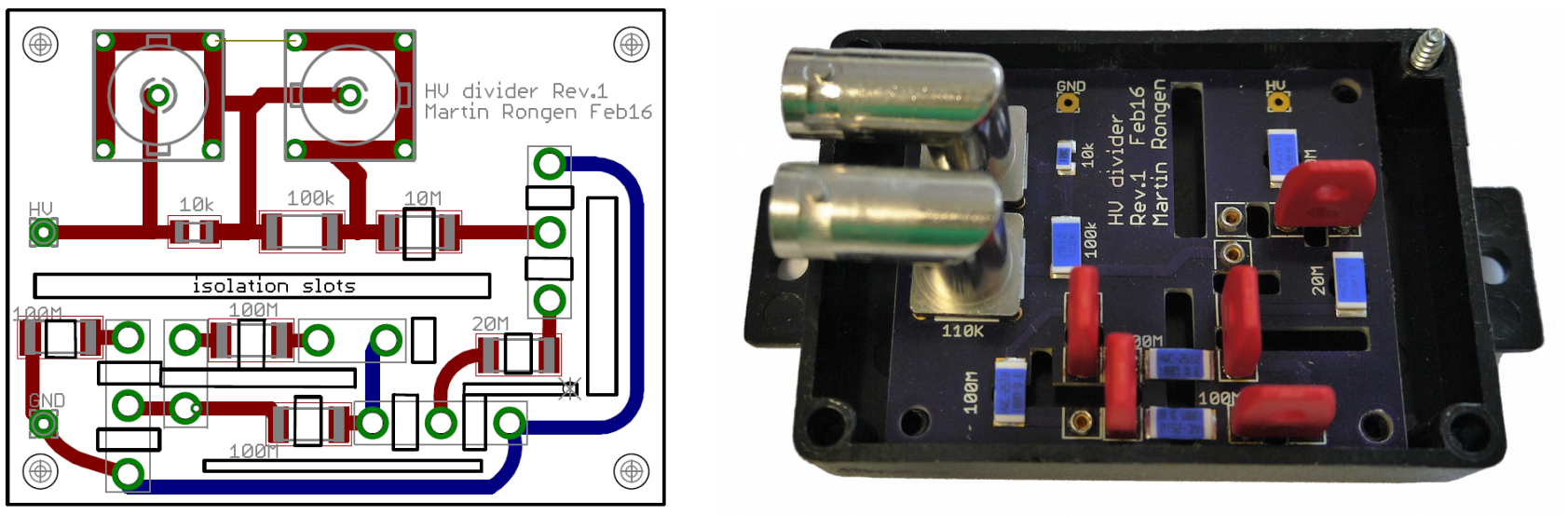

Figure B.3: Custom high voltage divider with jumper selectable load.
Left: PCB layout
Right: Assembled unit

415 *Nicrom Electronic (2017).* HVC High Voltage Resistors. *Nicrom Electronic*

precision high voltage resistors from Nicrom Electronics[415] to provide jumper selectable impedances of 10, 30, 80, 130 (likely PMT base impedance), 180 or 230 MΩ. Two outputs sample the voltage over 10 kΩ and 110 kΩ.

Both the CAEN and the HVM were tested for their long term voltage stability, ripple, stand-by power consumption and conversion efficiency. The CAEN module performed excellently in terms of noise performance and conversion efficiency but requires an additional 96 mW stand-by power. The HVM module has a similar efficiency but no measurable stand-by consumption.

Operated at loads of 30 MΩ and below no measurable ripple is present on the HVM output. At higher impedances the output oscillates as seen in figure B.4. This problem was confirmed by the manufacturer and is supposed to be resolved in an updated version by correcting an error in the feedback loop.

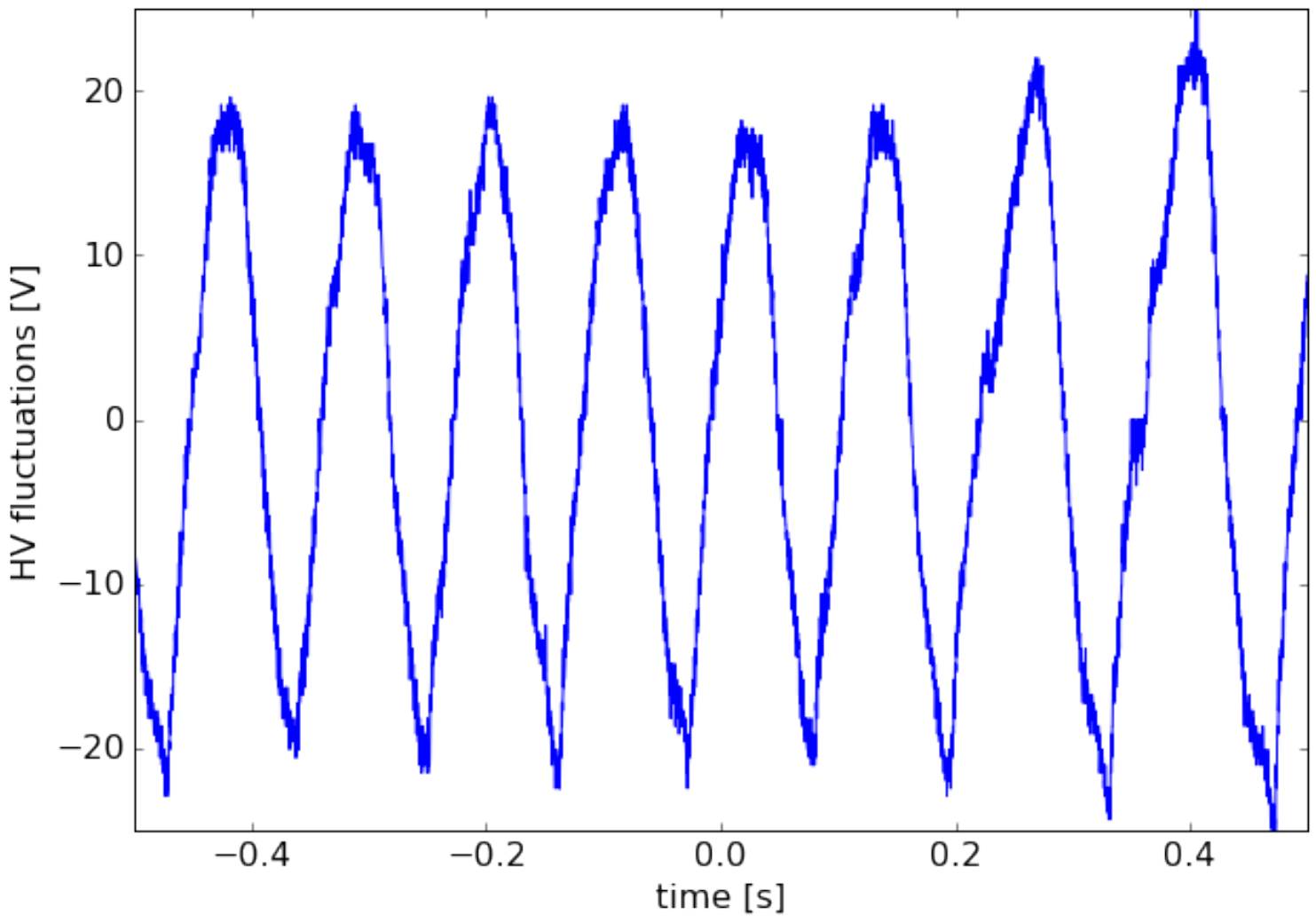

Figure B.4: Ripple voltage as observed with SHMHV 0520 the at high impedances. These were caused by an error in the feedback loop later corrected by the manufacturer.

# C

# FLARING DOMS

**During validation of the GRECO sample, targeted at low-energy neutrino oscillations, two DOMs sporadically emitting light and as such contaminating the sample were identified.**

The problem was raised when Martin Leuermann found two hot-spots in the vertex distribution of up-going events. The hot-spots are centered arounds DOMs "Pictured Rocks" S83D33 and "Acadia" S83D58. The events are few-GeV equivalent with over 90% of all charge being recorded in the lowest DOM. An example event is seen in figure C.1.

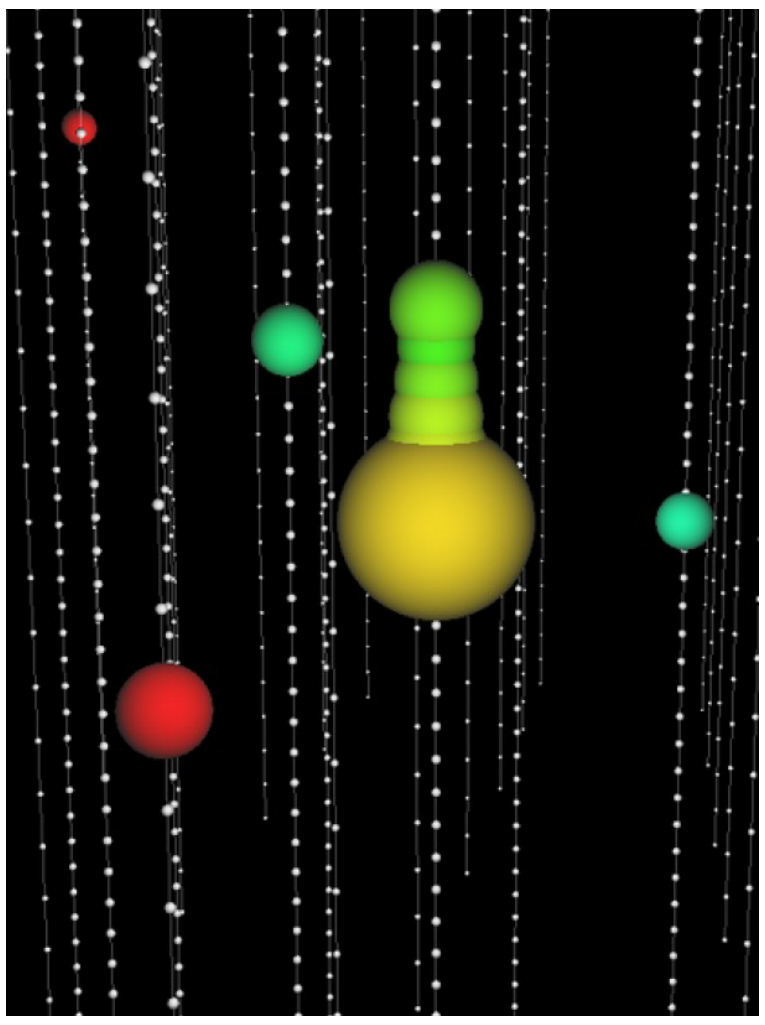

Figure C.1: Example event display of a flaring DOM. Light is seen propagating straight up, indicating light emission at the base.

As hits are only seen on the same string, accidental light emission of the flasher LEDs can be excluded and instead the high voltage system in particular the PMT base is suspected.

This is also supported by electronic ringing seen ∼60 ns prior to the rising edge of the PMT pulse on DOM S83D58. The delay is consistent with the PMT transit time, suggesting that something like a spark at the base creates electronic noise and light.

As a follow-up to these findings Jim Braun implemented an automated search for events compatible with flaring DOMs[416]. A total of 21 DOMs were identified to exhibit the problem, with flare rates ranging between ∼3300 and few flares per year. Except for one DOM the flare rate seems to be constant over

[416] *Braun, "Update on Light-Emitting DOMs"*

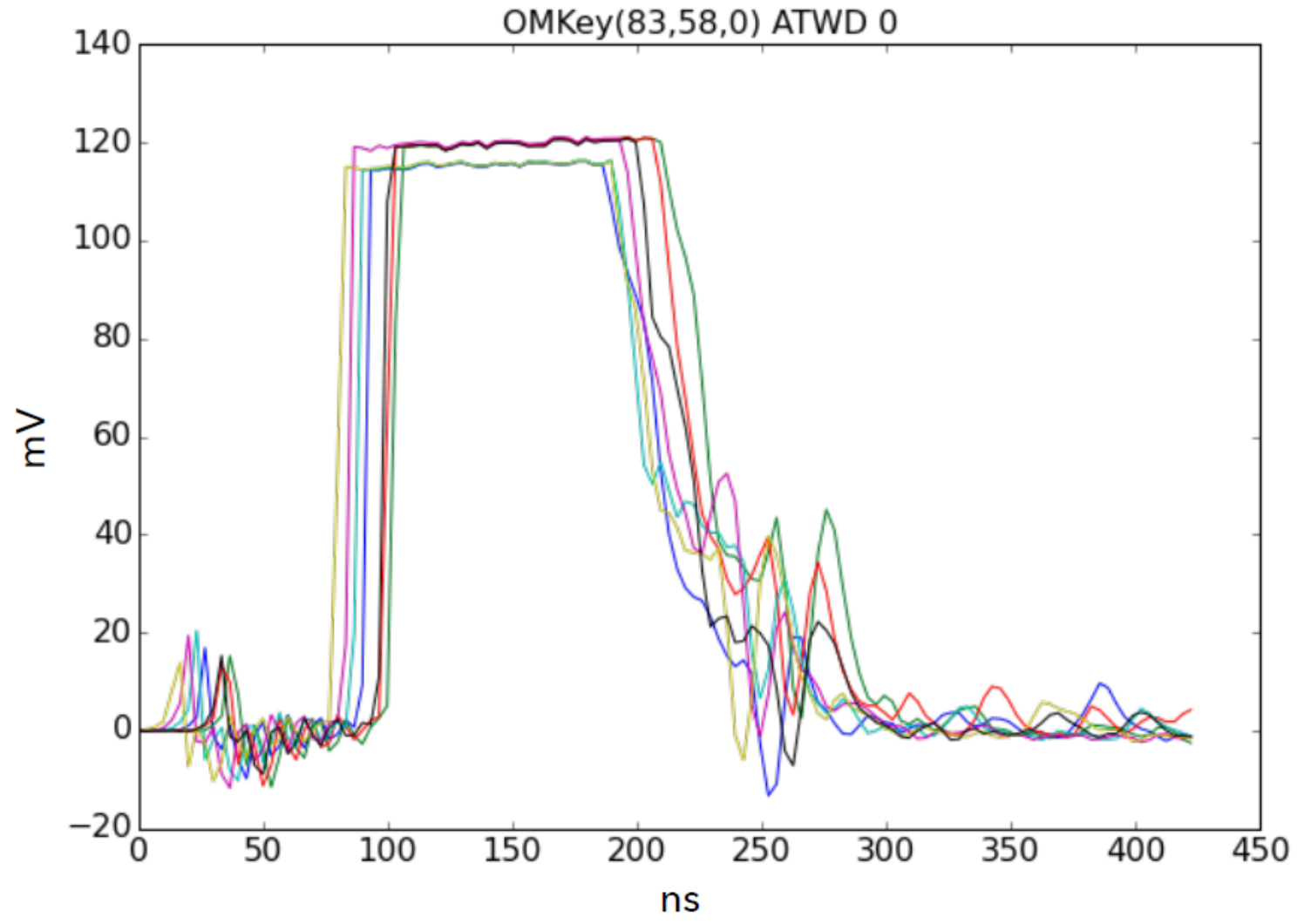

Figure C.2: ATWD0 waveforms of the flaring DOM S83D58. The brightnesses are remarkably similar. In addition electronic ringing is seen ∼60 ns prior to the rising edge of the PMT pulse. This delay is consistent with the PMT transit time, indicating that the ringing might be at the time of light emission

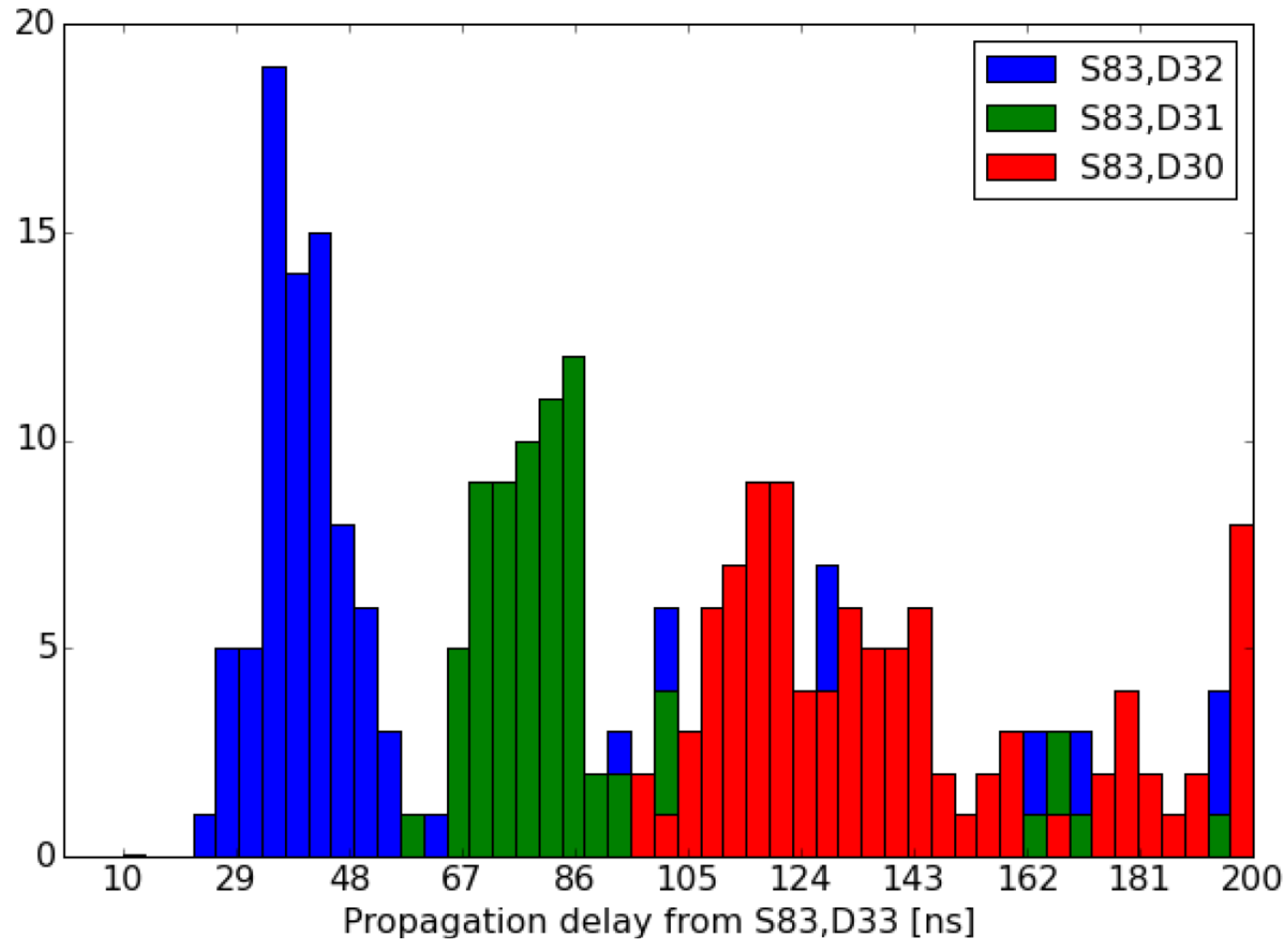

Figure C.3: The propagation delay between the suspected emitting DOM and subsequent DOMs on the same string is consistent with the geometric expectation of ∼44 ns per 10 m

time.

The high voltage of the three worst offenders was lowered to mitigate the problem and the flare search has been automated for the DeepCore data collection.

# D

# MUONS CROSSING DOMS

**The IceCube angular reconstruction can be tested and potentially significantly improved given a set of muons with precisely known directions. Such muons can be obtained in case they produce a distinct signature when crossing a DOM or any other detector coincident with IceCube, yielding precisely know pivot positions.**

This concept was first explored using muons measured in coincidence with the dark matter experiment DM-Ice[417] of which two detectors are deployed below strings 7 and 79. As muons are only ever observed at one reference point, no improvement to the angular reconstruction could be shown.

417 *Cherwinka, Grant, et al., "Measurement of muon annual modulation and muon-induced phosphorescence in NaI(Tl) crystals with DM-Ice17"*

This raised the question if muons may also induce a characteristic optical signal when crossing a DOM compared to just passing by at a close distance. To test this hypothesis a muon telescope, as seen in figure D.1, consisting of three scintillator panels as trigger and a central DOM in a dark box was constructed.

The scintillator panels were borrowed from a CosMO[418] kit. The analog PMT outputs were fed into NIM discriminators and a coincidence logic to generate a muon trigger signal. This provided as an analog input for a spare DOM mainboard which is operated as a local coincidence partner[419] for the DOM in the telescope.

418 *Franke et al., "CosMO - A Cosmic Muon Observer Experiment for Students"*

419 *see section 4.3.1.2*

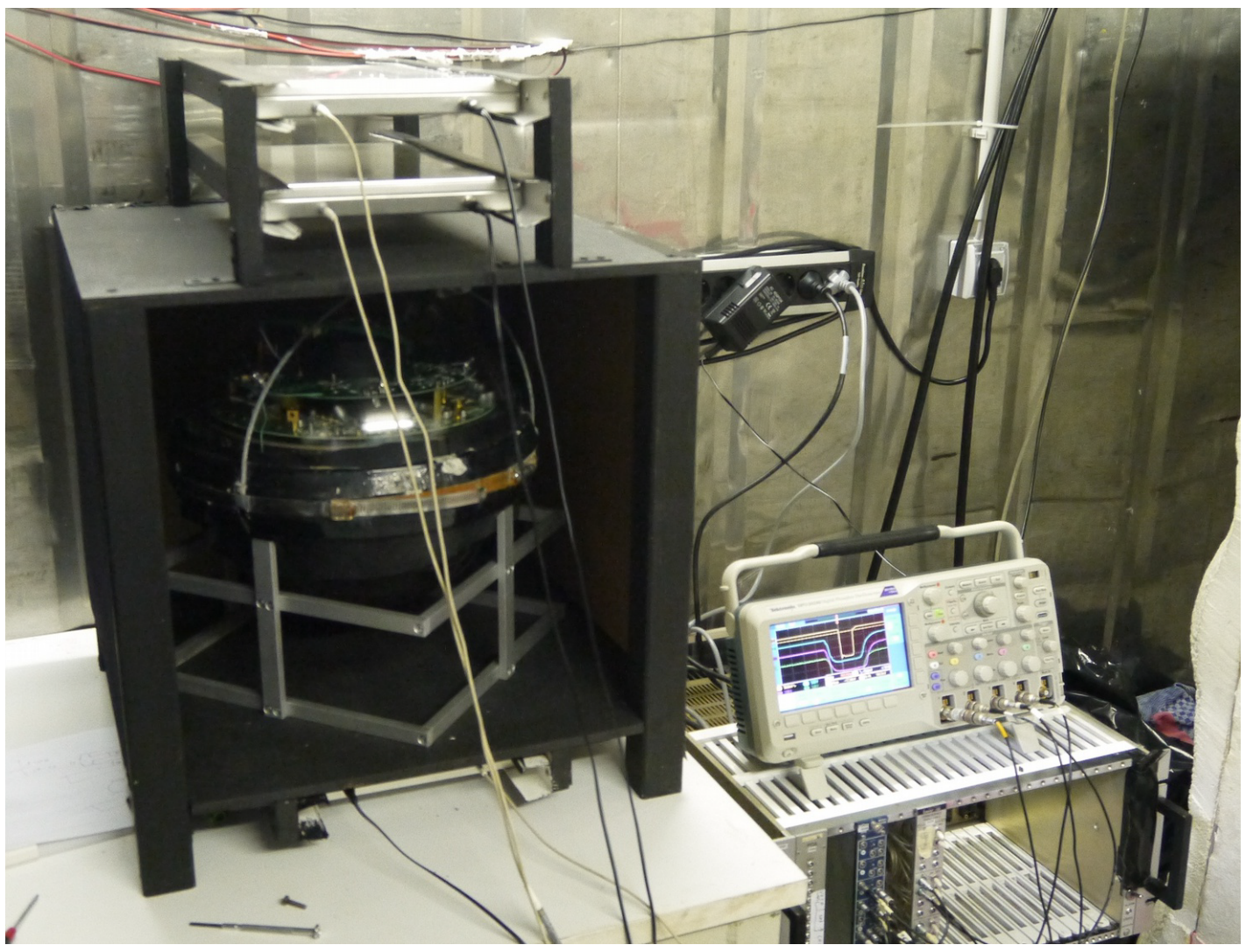

Figure D.1: Muon telescope composed of three CosMO scintillator panels (two above, one below) acting a coincidence trigger for the DOM in the center of the assembly.

Figure D.2 shows the charge spectra recorded in the DOM. Without any coincidence the data is dominated by single photo electrons as expected. Recording only muon events, identified by being coincident with either two or three scintillators, a characteristic charge peak of roughly 70 PE is observed.

The signature was confirmed by Matt Kauer using as similar setup[420]. Studies performed at the University of Alabama, targeted at muons which are reconstructed to pass close to DOMs, were not able to identify a similar signal in in-situ data.

[420] *Matt Kauer et al. (Sept. 2015).* Quick cross-check of the integrated charge from a muon passing through a DOM. *Calibration call*

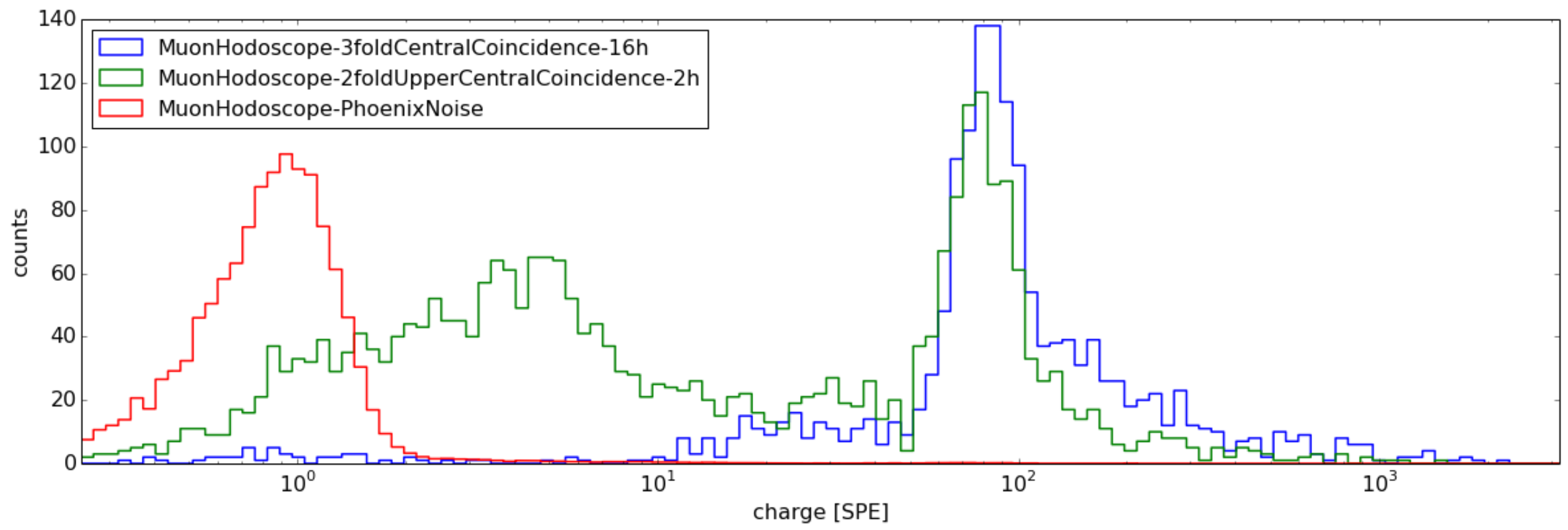

Figure D.2: DOM charge spectra for different coincidence settings as well as the DOM SPE spectrum. Muons crossing the DOM, as triggered on by the scintillators, result in a roughly 70 PE peak.

This is likely the case, because the refractive index of the glass pressure housing is so closely matched to the surrounding ice, that the DOM essentially appears transparent in the ice. As such the amount of Cherenkov light produced close to or inside of the DOM are indistinguishable and light created through any other precesses within the DOM will largely leak into the ice, further smearing out the signature.

Based on this experience, dedicated small muon taggers to be installed in every module have been proposed for the IceCube upgrade.

# EIDESSTATTLICHE ERKLÄRUNG

Ich, Martin Rongen erkläre hiermit, dass diese Dissertation und die darin dargelegten Inhalte die eigenen sind und selbstständig, als Ergebnis der eigenen originären Forschung, generiert wurden.

Hiermit erkläre ich an Eides statt:

1. Diese Arbeit wurde vollständig oder größtenteils in der Phase als Doktorand dieser Fakultät und Universität angefertigt.
2. Sofern irgendein Bestandteil dieser Dissertation zuvor für einen akademischen Abschluss oder eine andere Qualifikation an dieser oder einer anderen Institution verwendet wurde, wurde dies klar angezeigt.
3. Wenn immer andere eigene- oder Veröffentlichungen Dritter herangezogen wurden, wurden diese klar benannt.
4. Wenn aus anderen eigenen- oder Veröffentlichungen Dritter zitiert wurde, wurde stets die Quelle hierfür angegeben. Diese Dissertation ist vollständig meine eigene Arbeit, mit der Ausnahme solcher Zitate.
5. Alle wesentlichen Quellen von Unterstützung wurden benannt.
6. Wenn immer ein Teil dieser Dissertation auf der Zusammenarbeit mit anderen basiert, wurde von mir klar gekennzeichnet, was von anderen und was von mir selbst erarbeitet wurde.
7. Teile dieser Arbeit wurden zuvor veröffentlicht und zwar in:
   - M. Rongen und M. Schaufel (2018). "Design and evaluation of a versatile picosecond light pulser", doi: 10.1088/1748-0221/13/06/P06002
   - Martin Rongen (2016). "Measuring the optical properties of IceCube drill holes", doi: 10.1051/epjconf/201611606011
   - Elisa Resconi, Martin Rongen u. a. (2017). "The Precision Optical CAlibration Module for IceCube-Gen2: First Prototype", doi: 10.22323/1.301.0934
   - Dmitry Chirkin und Martin Rongen (2019b). *Photon propagation through birefringent polycrystals*. IceCube internal report
   - Dmitry Chirkin und Martin Rongen (20. Aug. 2019a). "Light diffusion in birefringent polycrystals and the IceCube ice anisotropy"

*Aachen, October 28, 2019*

Martin Rongen